\documentclass[a4paper]{ThesisStyle}

\usepackage{epic,eepic,amsmath,amsthm,amssymb,epsfig,cite,indentfirst,oldgerm}
\usepackage{multicol,boxedminipage}
\usepackage{epsfig}             % Enable Postscript figure inclusion
\usepackage{graphics}           % Enable Postscript figure inclusion 
\usepackage{subfigure}          % Labels subfigure within a figure environment
\usepackage{wrapfig}            % Wraps text around figure   
\usepackage{lscape}             % The ability to set up landsscape images/pages. 
\usepackage{rotating}

\newcommand{\field}[1]{\mathbb{#1}}

\newcommand{\neweq}{\begin{equation}\small}
\newcommand{\neweqar}{\begin{eqnarray}\small{ }
\newcommand{\neweeqar}{ }\begin{eqnarray}}

\newtheorem{lemma}{Lemma}
\newtheorem{no}{Notation}
\newtheorem{proposition}{Proposition}
\newtheorem{pr}{Proposition}

\newtheorem{theorem}{Theorem}

\def\beqa{\begin{eqnarray}}
\def\eeqa{\end{eqnarray}}
\def\ba{\begin{array}}
\def\ea{\end{array}}

\def\lt({\left(}
\def\rt){\right)}

\def\det{\operatorname{det}}

\begin{document}

%---------------------------------------------------------------------------------------%
%                                      Front Matter                                     %
%---------------------------------------------------------------------------------------%

\title{Studies in Integrable Quantum Lattice Models and Classical Hierarchies}
\author{\textbf{Matthew Luke Zuparic} \\ \\ Under the supervision of Dr. Omar Foda \\ \\ \\ Department of Mathematics and Statistics \\ The University of Melbourne \\ Parkville Victoria 3010 \\ Australia \\ \\ \\ \\ May 2009\\ \\ \\ \\ \\ Submitted in total fulfilment of the requirements of the\\ degree of Doctor of Philosophy}
\date{}

\maketitle

\tableofcontents                 % Contents Page

%---------------------------------------------------------------------------------------%
%                                       Main Body                                       %
%---------------------------------------------------------------------------------------%
\nocite{*}

\newpage
\setcounter{page}{1}\pagenumbering{roman}\pagestyle{plain}\addcontentsline{toc}{chapter}{Abstract}
\noindent\LARGE\textbf{Abstract}\\
\\
\noindent\normalsize The following work is an exploration into certain topics in the broad world of integrable models, both classical and quantum, and consists of two main parts of roughly equal length. The first part, consisting of chapters 1-3, concerns itself \textit{mainly} with correlations between results in classical hierarchies and quantum lattice models. The second part, consisting of chapters 4-6, deals almost entirely with deriving results concerned with quantum lattice models. \\
\\
\noindent\large\textbf{Outline of thesis - including main references used.}\\
\\
\noindent\normalsize
\textbf{Part 1.}\\
\\
\textit{Chapter 1} consists of a detailed, (almost) self contained account into the formulation of the finite 2-Toda classical hierarchy and its corresponding polynomial $\tau$-functions. Readers familiar with this topic can happily skip this section as most of the details can be found in \cite{Toda1,Toda2,Toda3,Toda4,Toda5}, with the slight exception of section 1.5 which details an elementary attempt to extend the polynomial solutions of the $\tau$-function by means of a simple scale transformation. Section 1.6 serves as a reference for fundamental results regarding symmetric polynomials\footnote{All results in this section can be found in chapters I and III of \cite{MacD}.}. The results in this section is assumed intimately throughout the rest of part 1 of the thesis.\\
\\
\textit{Chapter 2} begins with a detailed account of the construction of the quantum phase model using the algebraic Bethe ansatz and the corresponding scalar product. Most of the results in sections 2.1.1-2.1.4 can be found in \cite{phase1,phase2}. Section 2.1.5 details a beautiful and necessary result \cite{Tsilevich} which expands the scalar product as the bilinear sum of Schur polynomials. The first substantial original result in this thesis is found in section 2.1.6 which details the correspondence between the scalar product and the $\tau$-function of the 2-Toda hierarchy. Section 2.2 examines the corresponding Toda wave-vectors for the phase model and shows that they correspond to certain specific classes of correlation functions. We additionally detail how to obtain the single determinant form for these correlation functions. Section 2.3 is a simple observation of the correspondence between the $\tau$-functions of the scale transformed 2-Toda hierarchy detailed in section 1.5 and the scalar product of Hall-Littlewood vertex operators found in \cite{MO2}. The results of these two sections are to appear in \cite{toappear}.\\
\\
\textit{Chapter 3} contains results regarding the correspondence between the six vertex model and the KP hierarchy. Sections 3.1.1-3.1.2 are a general introduction to the model and an overview of the Korepin-Izergin approach \cite{Korepini,23} to deriving the domain wall partition function (DWPF). Section 3.1.3 gives a detailed account, found in \cite{Lascoux1}, of the derivation of a necessary alternative form to the DWPF given by Lascoux. Section 3.1.4 details how an additional form for the DWPF, given by Kirillov and Smirnov \cite{KS}, can be transformed to a form equivalent to that of Lascoux. Section 3.2 serves as a reference for fundamental results regarding free fermions and their corresponding infinite dimensional Fock space. All of these results can be found in \cite{bluebook,6,nicepaper,11}. Section 3.3 details how to express the Lascoux (and Kirillov-Smirnov) DWPF expression in fermionic form. The details of this section form the basis for \cite{P5}. Section 3.4.1 serves as an introduction to the algebraic Bethe ansatz approach to the six vertex model and the definition of the scalar product given this approach. The details of this section can be found in a staggering amount of references, (we give \cite{Lyon5,Lyon6,Lyon2,Lyon1,Lyon3,Lyon4,purplebook} as typical examples), owing to the complexity of the topic and the subsequent interest it has obtained. Section 3.4.2 introduces a necessary form of the scalar product \cite{Slavnov}, given by Slavnov. In this form only one of the sets of quantum variables obeys the necessary Bethe equations. Section 3.4.3 details how Slavnov's result can be re-expressed in a form equivalent to Lascoux's DWPF and section 3.4.4 details the subsequent fermionization of this expression. The details of this section form the basis for \cite{P6}. Section 3.5 serves as a conclusion, detailing how the fermionic forms of the DWPF and scalar product derived in this section automatically satisfy the KP bilinear equation. The details of this section can be found in \cite{bluebook,6}.\\
\\
\textbf{Part 2.}\\
\\
\textit{Chapter 4} marks the beginning of the second part of the thesis, where results concerning classical integrable systems are kept to a minimum, replaced by the consideration of integrable lattice models. The first of these systems is the so called trigonometric coloured Felderhof vertex model \cite{felderhof1,felderhof2,felderhof3}, or alternatively the spin-$\frac{1}{2}$ Deguchi-Akutsu vertex model \cite{DA1,DA2}, which can be thought of as the free fermion six vertex model in the presence of fields. Section 4.1 gives a necessary overview of the model and the corresponding coloured Yang-Baxter equations. In section 4.2.1 we derive the determinant form of the DWPF of the model (with rapidities trivialized) using the corresponding Koerpin-Izergin \cite{Korepini,23} type of proof. In section 4.2.2 we first take the homogeneous limit of the DWPF using a technique detailed in \cite{2}. We then show that the homogeneous DWPF is a $\tau$-function of the 2-Toda molecule equation \cite{11}. A similar observation was also made between homogeneous DWPF of the six vertex model and the 1-Toda molecule equation in \cite{Sog}. The results of section 4.2 were published in \cite{P1}. In section 4.3 we detail how the determinant form of the DWPF can be transformed to a Cauchy determinant, and hence expressed as a product. This simplification is due to the free fermion nature of the model.\\
\\
\textit{Chapter 5} contains results regarding the Baxter solid-on-solid (BSOS) elliptic height model. Sections 5.1.1-5.1.2 detail some of the main differences between height and vertex models, the application of the state variables and the fact the weights are parameterized by elliptic functions rather than trigonometric. We also give a necessary theorem\footnote{In particular, theorem 15.1 of \cite{Baxterbook}.} concerning doubly (anti)periodic functions, of which a majority of elliptic functions are categorized as. Section 5.1.3 introduces the elliptic Boltzmann weights of the model as well as the all important algebraic and pictorial definition(s) of the height Yang-Baxter identities, as detailed in \cite{SOS1,SOS2}. In section 5.1.4 we detail what exactly domain wall boundary conditions (DWBC's) mean for a height model. In section 5.2.1 we express the DWPF recursively using a double summation. The method employed in this section is an elliptic extension of the method first used in \cite{1} on the six vertex model. We drastically simplify the recursive expression for the DWPF in section 5.2.2, again by an elliptic extension of a method first used in \cite{OPCF} on the six vertex model. Finally in section 5.2.3 we express the DWPF as a sum over the permutation group.\\
\\
\textit{Chapter 6} is arguably the most straight-forward chapter of this thesis. In section 6.1.1 we give the definition of the trigonometric Perk-Schultz (PS) vertex model \cite{PSvert2,PSvert3,PSvert4}, the parameterization of the Boltzmann weights and the associated \textit{graded} Yang-Baxter equation. Sections 6.1.2-6.1.3 give the equivalent Korepin-Izergin \cite{Korepini,23} type proof for the DWPF, however, due to an asymmetry in the line permuting vertices, the most natural \textit{guess} for the DWPF turns out to be a product expression, rather than a determinant. The results of these sections were published in \cite{P2}. Section 6.2.1 revisits elliptic functions and offers a necessary result regarding entire doubly \textit{quasi}-periodic functions. The result is a straight-forward generalization of theorems 15.2 and 15.3 in \cite{Baxterbook}. In section 6.2.2 we give the definition of the elliptic PS height model \cite{PSheight1}, the parameterization of the Boltzmann weights and the associated \textit{graded} height Yang-Baxter equation. Sections 6.2.3-6.2.4 give the equivalent elliptic Korepin-Izergin \cite{Korepini,23} type proof for the DWPF, however, again due to an asymmetry in the line permuting faces, the most natural \textit{guess} for the DWPF turns out to be a product expression, rather than a determinant. The results of these sections were published in \cite{P3}. In section 6.2.5 we generate some really nice elliptic identities by performing the same analysis on the PS height model that we performed on the BSOS model in sections 5.2.1-5.2.2, based on the results in \cite{1,OPCF}.

\newpage
\mbox{}
\newpage
\setcounter{page}{5}\addcontentsline{toc}{chapter}{Declaration}\pagestyle{plain}
\noindent\LARGE\textbf{Declaration}\\
\\
\noindent \normalsize
This is to certify that
\begin{itemize}
\item{\textit{the thesis comprises only my original work where indicated in the preface,}}
\item{\textit{due acknowledgement has been made in the text to all other material used,}}
\item{\textit{the thesis is less than 100,000 words in length, exclusive of tables, maps, bibliographies and appendices.\\}}
\end{itemize}
Matthew Zuparic 
\newpage
\mbox{}
\newpage
\setcounter{page}{7}\addcontentsline{toc}{chapter}{Preface}\pagestyle{plain}
\noindent\LARGE\textbf{Preface}\\
\\
\noindent \normalsize This thesis was written under the supervision of
$$
\begin{array}{l}
\textrm{Dr. Omar Foda}\\
\textrm{Department of Mathematics and Statistics}\\
\textrm{The University of Melbourne}\\
\textrm{Victoria 3010, Australia}\\
\textrm{{\tt O.Foda@ms.unimelb.edu.au}}
\end{array}
$$
\noindent \LARGE{\textbf{Chapter 1}}
\normalsize \begin{itemize}
\item{\textbf{1.1-1.4} Review.}
\item{\textbf{1.5} Work done by Matthew Zuparic (MZ).}
\item{\textbf{1.6} Review.}\\
\end{itemize}
\noindent \LARGE{\textbf{Chapter 2}}
\normalsize \begin{itemize}
\item{\textbf{2.1.1-2.1.5} Review.}
\item{\textbf{2.1.6} Work done by Omar Foda (OF) and MZ.}
\item{\textbf{2.2} Work done by MZ.}
\item{\textbf{2.3} Work done by MZ based on previous work by OF and Michael Wheeler (MW).}\\
\end{itemize}
\noindent \LARGE{\textbf{Chapter 3}}
\normalsize \begin{itemize}
\item{\textbf{3.1.1-3.1.3} Review.}
\item{\textbf{3.1.4} Work done by OF and MZ.}
\item{\textbf{3.2} Review.}
\item{\textbf{3.3} Work done by OF, MW and MZ.}
\item{\textbf{3.4.1} Review.}
\item{\textbf{3.4.2} Review.}
\item{\textbf{3.4.3-3.4.4} Work done by OF and MZ.}
\item{\textbf{3.5} Review.}\\
\end{itemize}
\noindent \LARGE{\textbf{Chapter 4}}
\normalsize \begin{itemize}
\item{\textbf{4.1} Review.}
\item{\textbf{4.2.1} Work done by OF, Alex Caradoc (AC) and MZ.}
\item{\textbf{4.2.2} Review and work done by OF and MZ.}
\item{\textbf{4.3} Review.}\\
\end{itemize}
\noindent \LARGE{\textbf{Chapter 5}}
\normalsize \begin{itemize}
\item{\textbf{5.1} Review.}
\item{\textbf{5.2} Work done by OF and MZ.}\\
\end{itemize}
\noindent \LARGE{\textbf{Chapter 6}}
\normalsize \begin{itemize}
\item{\textbf{6.1.1} Review.}
\item{\textbf{6.1.2-6.1.3} Work done by OF and MZ.}
\item{\textbf{6.2.1} Review and work done by OF.}
\item{\textbf{6.2.2} Review.}
\item{\textbf{6.2.3-6.2.4} Work done by OF and MZ.}
\item{\textbf{6.2.5} Work done by MZ.}
\end{itemize}
\newpage  
\setcounter{page}{9}\addcontentsline{toc}{chapter}{Acknowledgements} \pagestyle{plain}
\noindent\LARGE\textbf{Acknowledgements}\\
\\
\noindent \normalsize
First and foremost, to my supervisor Omar Foda. I cannot thank you enough for the countless research hours you have personally invested in this body of work. Needless to say, your enthusiasm for meaningful discovery, integrity, patience, kindness and occasional harsh truths have undoubtedly paved the way for my long journey ahead.\\
\\
To Melli. You have stood by me for the entire time, fixing my (metaphorical) wounds while I constantly banged my forehead against the desk searching for answers. I honestly do not deserve you.\\
\\
To Michael. It is painfully obvious that without your help, especially at the start, this thesis would never have happened. Your professional approach to research is inspirational and I hope that we continue to bounce ideas off each other in the future.\\
\\
To Ellie. You saved me from my shameful computer illiteracy.\\
\\
To the tech support crew, especially Jeff, thanks for your help in maintaining the temperamental dinosaur.\\
\\
To my parents and family, Mira, Branko, Jasna, Stephen, Sarah, Baka, Kerri, Ren\'{e}, Josh, Ana and Adrian. Thankyou for your unconditional support and love through this stressful and trying time.\\
\\
To my friends and colleagues, Daniel, Waven, Kieran, Eliza, Steve, Megan, Gary, Craig, Princess, Catherine, Alex, Fel, Michael, Steve, Heather, Maurice, Paul, Amy, Leigh, Nic, Anita, Anthony, Emily, Alison, Paul, Andy, Diana, Geoffrey, Jan, Richard, Peter, Tony, Chris, Iwan, Ian and Nicholas. Thankyou for your constant encouragement and for instilling a sense of belonging.\\
\\
\textbf{Dishonourable Mention}\\
\\
My computer. Being old is not an excuse.\\
\\
The emergence of random lists on the room 223 whiteboard. Geography is not a team sport.
\newpage 
\indent \textit{Monkey burst into tears. ``Where am I to go?'' he asked.\\
\indent ``Back to where you came from, I should suppose,'' said the Patriarch.\\
\indent ``You don't mean back to the Cave of the Water Curtain in Ao-lai!'' said Monkey.\\
\indent ``Yes,'' said the Patriarch, ``go back as quickly as you can, if you value your life. One thing is certain in any case; you can't stay here.''\\
\indent ``May I point out,'' said Monkey, ``that I have been away from home for twenty years and should be very glad to see my monkey-subjects once more. But I can't consent to go till I have repaid you for all your kindness.''\\
\indent ``I have no desire to be repaid,'' said the Patriarch. ``All that I ask is that if you get into trouble, you should keep my name out of it$\dots$I'm convinced you'll come to no good. So remember, when you get into trouble, I absolutely forbid you say that you are my disciple. If you give a hint of any such thing I shall flay you alive, break all your bones, and banish your soul to the Place of Ninefold Darkness, where it will remain for ten thousand aeons.''\\
\\
Wu Ch'\^{e}ng-\^{e}n, `Journey to the West'}
\newpage
\addcontentsline{toc}{chapter}{List of Figures}
\listoffigures 
\newpage
%%%%%%%%%%%%%%%%%%%%%%%%%%%%%%%%%%%%%%%%%%%%%%%%%%%%%%%%%%%%%%%%%%%%%%%%%%%%%%%%%%%%%%%%%%%%%%%%%%%%%%%%%%%%%%%%%%%%
\chapter{Finite 2-Toda hierarchy}\pagenumbering{arabic}\setcounter{page}{5}\pagestyle{headings}
The discovery and subsequent study of the 2-Toda hierarchy is an obvious landmark in the world of classical integrable hierarchies and the wider study of integrable models. The hierarchy itself is arguably attributed to two separate areas of study, that of the Toda lattice equation and that of the KP hierarchy.\\
\\
\textbf{Integrability of the Toda lattice equation.} The study of the integrable properties of the Toda lattice equation began in 1967 \cite{MToda} when Toda discovered certain \textit{nice} properties of the non linear lattice with an exponential potential. The equation of motion for this system,
\small{\begin{equation}
\partial^2_{x} u_s(x) = e^{u_s(x)-u_{s-1}(x)} - e^{u_{s+1}(x)-u_{s}(x)}  \textrm{  ,  }s \in \field{N}
\label{todarubbish}\end{equation}\normalsize}\normalsize
is obviously referred to as the Toda lattice equation. Inspired by Toda's findings, Ford et. al. \cite{Fordy} deduced that a three particle \textit{spring} system with the Toda potential always has three conserved quantities, thus ensuring integrability of the system. In 1974 both H\'{e}non \cite{Henon} and Flaschka \cite{Flaschka} proved the existence of $N$ conserved quantities for an $N$ particle lattice.\\
\\  
\textbf{The KP hierarchy.} The KP hierarchy arguably began with the landmark paper \cite{GGKM} by Gardner et. al. when they applied the inverse scattering transform method to solve the initial value problem for the celebrated KdV equation. In \cite{Lax} Lax expressed the KdV equation as the compatibility condition of two linear operator equations. Specifically, consider the two differential operators,
\small{\begin{equation*}
L = \partial^2_{x_1} + 2u(x_1,x_2) \textrm{  ,  } B= \partial^3_{x_1} + 3 u(x_1,x_2) \partial_{x_1} +\frac{3}{2} \{\partial_{x_1} u(x_1,x_2) \}
\end{equation*}\normalsize}\normalsize
and the following linear equations,
\small\begin{equation}
L \psi = \lambda \psi \textrm{  ,  } \partial_{x_2} \psi = B \psi
\end{equation}\normalsize
where the eigenvalue $\lambda$ is independent of $\{x_1,x_2\}$. The compatibility condition of these two linear equations,
\small\begin{equation}
\partial_{x_2} L = [B,L]
\end{equation}\normalsize
reduces to the KdV equation. The formal extension to the above compatibility condition was first given in \cite{Zak} by Zakharov and Shabat. Specifically, from the following linear operator equations,
\small\begin{equation}
\partial_{x_m} w(\vec{x}) = B_m w(\vec{x}) \textrm{  ,  } \partial_{x_n} w(\vec{x}) = B_n w(\vec{x}) \textrm{  ,  } \{m,n\} \in \field{N}
\end{equation}\normalsize
(we refer to $w(\vec{x})$ as the \textit{wave-function}), the compatibility conditions yield the Zakharov-Shabat equation,
\small\begin{equation}
\partial_{x_m} B_n - \partial_{x_n} B_m+ [B_n,B_m] =0
\label{ZAK}\end{equation}\normalsize
By choosing suitable differential operators $\{B_m,B_n\}$, one an reduce eq. \ref{ZAK} to any one of an infinite amount of non linear applied partial differential equations, one of which is the KP equation,
\small\begin{equation}
\partial_{x_1} \left( 4 \partial_{x_3}u -12u \partial_{x_1}u  - \partial^3_{x_1}u \right) = 3 \partial^2_{x_2}u
\label{ZAKii}\end{equation}\normalsize
Through defining an appropriate $L$ operator\footnote{$L$ in general is a \textit{pseudo-differential} operator.}, it can also be shown that the Zakharov-Shabat equation is equivalent to the following generalized Lax pair system,
\small\begin{equation}
\partial_{x_m} L = [B_m,L] \textrm{  ,  } m \in \field{N}
\label{Glax}\end{equation}\normalsize
In either form (eq. \ref{ZAK} or \ref{Glax}), the infinite set of differential equations is referred to as the \textit{KP hierarchy}.\\
\\
In this chapter, we shall introduce the 2-Toda hierarchy as it was introduced in the seminal paper \cite{Toda4} by Ueno and Takasaki, as a straight forward generalization of the KP Lax pair system with four sets of operators, $L,M,B_m$ and $C_n$, as opposed to the usual two for KP. From this starting point we shall then feature the results in \cite{Toda1,Toda2,Toda3,Toda4,Toda5}, and demonstrate how the system can be \textit{solved}\footnote{By solved we refer to defining/deriving the quantities $L,M,B_m$ and $C_n$.} by considering a well defined matrix initial value problem. When it comes time to discuss the $\tau$-function of the hierarchy, we shall see the real advantages of this initial value approach.
\section{Definition of the 2-Toda hierarchy}
We begin this section by giving the definition of the 2-Toda hierarchy (with 2 copies of $n-m-1$ time variables) in terms of 4 distinct Lax type systems of first order differential equations. First we give some necessary definitions.\\
\\
Defining the following shift matrices,
\small\begin{equation*}
\begin{split}
& \Lambda^j_{[m,n-1]}= (\delta_{k+j,l})_{k,l \in \{m,\dots,n-1 \}} \\
\left(\Lambda^T_{[m,n-1]} \right)^j =& \Lambda^{-j}_{[m,n-1]} = (\delta_{k-j,l})_{k,l \in \{m,\dots,n-1 \}} 
\end{split}
\end{equation*}\normalsize
we let $E_{kl}$ be the $(k,l)$ unit matrix of size $(n-m)\times (n-m)$,
\small\begin{equation*}
E_{kl} = (\delta_{ik}\delta_{jl})_{i,j,k,l \in \{m,\dots,n-1 \}}
\end{equation*}\normalsize
In this notation the Lie algebra $gl(n-m)$ in naturally generated by the linear combination of all such $(n-m)\times (n-m)$ matrices,
\small\begin{equation*}
gl(n-m) = \left\{ \sum_{i,j \in \{m,\dots,n-1\} }a_{ij}E_{ij} | a_{ij} \in \field{C} \right\}
\end{equation*}\normalsize
The general matrix $A \in gl(n-m)$ is written in the following convenient form,
\small\begin{equation*}
A= \sum^{n-m-1}_{j=-n+m+1}a_j(s)\Lambda^j_{[m,n-1]}
\end{equation*}\normalsize
where $m \le s \le n-1 $ denotes the row of the matrix $A$. The matrix $A\in gl(n-m)$ is said to be a strictly lower triangular matrix if $a_j(s)=0$ for $j \ge 0$ and an upper triangular matrix if $a_j(s)=0$ for $j < 0$. With these distinctions, we label the $(\pm)$ sections of the matrix $A$ as,
\small\begin{equation*}
(A)_+ = \sum^{n-m-1}_{j=0} a_j(s)\Lambda^j_{[m,n-1]} \textrm{  ,  } (A)_- = \sum^{-1}_{j=-n+m+1} a_j(s)\Lambda^j_{[m,n-1]}
\end{equation*}\normalsize
We now define 2 sets of time flows $\vec{x}$ and $\vec{y}$ as,
\small\begin{equation*}
\vec{x} = \{ x_1,x_2,\dots, x_{n-m-1}\} \textrm{  ,  } \vec{y} = \{ y_1,y_2,\dots, y_{n-m-1}\}
\end{equation*}\normalsize
and let $L(\vec{x},\vec{y}), M(\vec{x},\vec{y}), B_n(\vec{x},\vec{y}), C_n(\vec{x},\vec{y}) \in gl(n-m)$ where,
\small\begin{equation*}\begin{array}{lll}
\displaystyle L = \sum^{1}_{-n+m+1} b_j(s,\vec{x},\vec{y}) \Lambda^j_{[m,n-1]} & b_1(s) = 1&  B_n = (L^n)_+\\
\displaystyle M = \sum^{n-m-1}_{j=-1} c_j(s,\vec{x},\vec{y}) \Lambda^j_{[m,n-1]} & c_{-1}(s) \ne 0 & C_n = (M^n)_- 
\end{array}\end{equation*}\normalsize
We now define the 2-Toda hierarchy as the following Lax type system of differential equations,
\small\begin{equation}\begin{array}{ll}
\partial_{x_n} L(\vec{x},\vec{y}) = [B_n(\vec{x},\vec{y}),L(\vec{x},\vec{y})] & \partial_{y_n} L(\vec{x},\vec{y}) = [C_n(\vec{x},\vec{y}),L(\vec{x},\vec{y})] \\
\partial_{x_n} M(\vec{x},\vec{y}) = [B_n(\vec{x},\vec{y}),M(\vec{x},\vec{y})] & \partial_{y_n} M(\vec{x},\vec{y}) = [C_n(\vec{x},\vec{y}),M(\vec{x},\vec{y})]
\end{array}\label{lax}\end{equation}\normalsize
or equivalently (theorem 1.1 of \cite{Toda3}), the Zakharov-Shabat system,
\small\begin{equation}\begin{split}
\partial_{x_{m}} B_{n}-\partial_{x_{n}} B_{m} + \left[ B_{n},B_{m} \right]&=0\\
\partial_{y_{m}} C_{n}-\partial_{y_{n}} C_{m} + \left[ C_{n},C_{m} \right]&=0\\
\partial_{y_{m}} B_{n}-\partial_{x_{n}} C_{m} + \left[ B_{n},C_{m} \right]&=0
\end{split}\label{dog2}\end{equation}\normalsize
\textbf{Compatibility conditions.} It can be shown (theorem 1.2 of \cite{Toda3}) that the above systems are the compatibility conditions of the following linear operator equations,
\small\begin{equation}\begin{array}{ll}
\partial_{x_{m}} W^{(\infty)}(\vec{x},\vec{y}) = B_m W^{(\infty)}(\vec{x},\vec{y})& \partial_{y_{m}} W^{(\infty)}(\vec{x},\vec{y}) = C_m W^{(\infty)}(\vec{x},\vec{y})\\
\partial_{x_{m}} W^{(0)}(\vec{x},\vec{y}) = B_m W^{(0)}(\vec{x},\vec{y})  & \partial_{y_{m}} W^{(0)}(\vec{x},\vec{y}) = C_m W^{(0)}(\vec{x},\vec{y})\\
\end{array}\end{equation}\normalsize
where $W^{(\infty/0)}(\vec{x},\vec{y}) \in GL(n-m)$ are referred to as \textit{wave-matrices}, and the derivation of their exact form forms the basis for the next section of this thesis.\\
\\
\textbf{The Toda lattice equation.} With the following parameterization for $B_1$ and $C_1$,
\small\begin{equation*}
B_1= \delta_{i,j-1} + \delta_{i,j} \partial_{x_1}u(i,\vec{x},\vec{y}) \textrm{  ,  }C_1= \delta_{i,j+1}e^{u(i,\vec{x},\vec{y})-u(i-1,\vec{x},\vec{y})}
\end{equation*}\normalsize
the third Zakharov-Shabat equation for $\{m,n\}=1$ becomes the 2 dimensional Toda lattice equation,
\small\begin{equation}
\partial_{x_1}\partial_{y_1} u(s,\vec{x},\vec{y}) = e^{u(s,\vec{x},\vec{y})-u(s-1,\vec{x},\vec{y})} - e^{u(s+1,\vec{x},\vec{y})-u(s,\vec{x},\vec{y})} 
\label{todarubbish2D}\end{equation}\normalsize
These systems (eq. \ref{lax} and \ref{dog2}) are an obvious parallel to the Lax and Zakharov-Shabat systems that originally defined the KP hierarchy. In what follows, we shall first consider the initial value problem for the 2-Toda hierarchy and hence find explicit values for the entries of the wave-matrices of the hierarchy. If we then define the matrices $L$ and $M$ (and by extension $B_n$ and $C_n$) as specific products of the wave matrices, we shall then show that the wave-matrices allow us to construct 4 distinct sets of first order linear differential equations. Considering the compatibility of these equations then leads us to the definition of the 2-Toda hierarchy found in eq. \ref{lax}. 
\section{The initial value problem}
We begin by defining the constant matrix $A \in GL(n-m) = (a_{i,j})_{i,j = m, \dots,n-1}$, such that $\textrm{det}\left[ a_{ij} \right]_{i,j=m\dots,s-1} \ne 0, m < s \le n$. Our task now is to find wave-matrices $W^{(\infty)}(\vec{x},\vec{y})$ and $W^{(0)}(\vec{x},\vec{y})$ such that,
\small\begin{equation}
W^{(0)}(\vec{x},\vec{y})=W^{(\infty)}(\vec{x},\vec{y}) A
\label{1}\end{equation}\normalsize
where $W^{(\infty)}(\vec{x},\vec{y})$ and $W^{(0)}(\vec{x},\vec{y})$ have the specific form,
\small\begin{equation*} 
\begin{split}
W^{(\infty)}(\vec{x},\vec{y}) = \hat{W}^{(\infty)}(\vec{x},\vec{y}) \exp \left[\sum^{n-m-1}_{k = 1} x_{k} \Lambda^{k}_{[m,n-1]}\right] \\
\hat{W}^{(\infty)}(\vec{x},\vec{y}) = \left(\hat{w}^{(\infty)}_{i-j}(i,\vec{x},\vec{y})\right)^{n-1}_{i,j = m}\\
W^{(0)}(\vec{x},\vec{y}) = \hat{W}^{(0)}(\vec{x},\vec{y}) \exp \left[\sum^{n-m-1}_{k = 1} y_{k} (\Lambda^T_{[m,n-1]})^{k}\right] \\
 \hat{W}^{(0)}(\vec{x},\vec{y}) = \left(\hat{w}^{(0)}_{j-i}(i,\vec{x},\vec{y})\right)^{n-1}_{i,j = m} 
\end{split}
\end{equation*}\normalsize
where the diagonal entries of $\hat{W}^{(\infty)}(\vec{x},\vec{y})$ and $\hat{W}^{(0)}(\vec{x},\vec{y})$ are given by,
\small\begin{equation}
\begin{split}
 \hat{w}^{(\infty)}_j = \left\{ \begin{array}{cc} 
0 & j < 0\\
1 & j =0 \end{array} \right. \\
 \hat{w}^{(0)}_j =  \left\{ \begin{array}{cc}
0 & j < 0\\ 
 \hat{w}^{(0)}_j(\vec{x},\vec{y}) \ne const. & j =0
\end{array}\right.
\end{split}
\label{ogle}\end{equation}\normalsize
\textbf{Origin of the initial value problem.} One immediate consequence of theorem 1.2 of \cite{Toda3} is the following identity\footnote{Eq. 1.2.17 of \cite{Toda3}.},
\small\begin{equation}
\left( \partial^{\{i\}}_{\vec{x}} \partial^{\{j\}}_{\vec{y}} W^{(\infty)}(\vec{x},\vec{y}) \right) \left( W^{(\infty)}(\vec{x},\vec{y}) \right)^{-1} = \left( \partial^{\{i\}}_{\vec{x}} \partial^{\{j\}}_{\vec{y}} W^{(0)}(\vec{x},\vec{y}) \right) \left( W^{(0)}(\vec{x},\vec{y}) \right)^{-1}
\label{poochu}\end{equation}\normalsize
where,
\small\begin{equation*}
\partial^{\{i\}}_{\vec{x}} = \partial^{i_1}_{x_1}\partial^{i_2}_{x_2} \dots \partial^{i_{n-m-1}}_{x_{n-m-1}} \textrm{  ,  } \partial^{\{j\}}_{\vec{y}} = \partial^{j_1}_{y_1}\partial^{j_2}_{y_2} \dots \partial^{j_{n-m-1}}_{y_{n-m-1}}
\end{equation*}\normalsize
and $\{i_k,j_l\} \in \field{N} \bigcup \{0\}$. The generating function of the above equation is given by the following expression\footnote{We shall see in section 1.4.1 that eq. \ref{pilin} is actually the bilinear equation of the 2-Toda hierarchy.},
\small\begin{equation}
W^{(\infty)}(\vec{x}',\vec{y}') \left( W^{(\infty)}(\vec{x},\vec{y}) \right)^{-1} = W^{(0)}(\vec{x}',\vec{y}')  \left( W^{(0)}(\vec{x},\vec{y}) \right)^{-1}
\label{pilin}\end{equation}\normalsize
for general $\{\vec{x},\vec{x}',\vec{y},\vec{y}'  \}$. To see this, consider Taylor expanding the matrices $W^{(\infty/0)}(\vec{x}',\vec{y}')$ at the points $\vec{x}' = \vec{x}$ and $\vec{y}'=\vec{y}$,
\small\begin{equation*}\begin{split}
\prod^{n-m-1}_{k,l=1} \sum^{\infty}_{i_k,j_l=0} \frac{(x'_k-x_k)^{i_k}}{i_k !} \frac{(y'_l-y_l)^{j_l}}{j_l !} \left( \partial^{\{i\}}_{\vec{x}} \partial^{\{j\}}_{\vec{y}} W^{(\infty)}(\vec{x},\vec{y}) \right) \left( W^{(\infty)}(\vec{x},\vec{y}) \right)^{-1}\\
 =\prod^{n-m-1}_{k,l=1} \sum^{\infty}_{i_k,j_l=0} \frac{(x'_k-x_k)^{i_k}}{i_k !} \frac{(y'_l-y_l)^{j_l}}{j_l !} \left( \partial^{\{i\}}_{\vec{x}} \partial^{\{j\}}_{\vec{y}} W^{(0)}(\vec{x},\vec{y}) \right) \left( W^{(0)}(\vec{x},\vec{y}) \right)^{-1}
\end{split}\end{equation*}\normalsize
Collecting the coefficients of the monomials in $(x'_k-x_k)^{i_k}(y'_l-y_l)^{j_l}$, we immediately obtain eq. \ref{poochu} for all values of $\{i_k,j_l\}$.\\
\\
We now express eq. \ref{pilin} in the following equivalent form,
\small\begin{equation*}
 \left( W^{(0)}(\vec{x}',\vec{y}')  \right)^{-1} W^{(\infty)}(\vec{x}',\vec{y}')  =   \left( W^{(0)}(\vec{x},\vec{y}) \right)^{-1} W^{(\infty)}(\vec{x},\vec{y})
\end{equation*}\normalsize
This equation must hold for all $\{\vec{x},\vec{x}',\vec{y},\vec{y}'  \}$, so obviously both sides do not depend on the time variables. Thus the above equation implies eq. \ref{1}.
\subsection{Derivation of the wave-matrices}
\noindent We now derive the remaining non zero entries of the hatted wave-matrices.
\begin{proposition}
\small\begin{equation}\begin{split}
\hat{w}^{(\infty)}_{k}(s,\vec{x},\vec{y}) =& (-1)^k\frac{\textrm{det}\left[a_{ij}(\vec{x},\vec{y}) \right]_{i=m,\dots,\hat{s-k},\dots ,s \atop{ j=m,\dots,s-1}}}{\textrm{det}\left[a_{ij}(\vec{x},\vec{y}) \right]_{i,j=m,\dots,s-1}}, \textrm{  where  } \left\{ \begin{array}{c} 0 \le k \le s-m\\ m < s \le n-1\end{array}\right. \\
\hat{w}^{(0)}_{k}(s,\vec{x},\vec{y}) =& \frac{\textrm{det}\left[a_{ij}(\vec{x},\vec{y}) \right]_{i=m,\dots,s \atop{j=m,\dots,s-1,s+k}}}{\textrm{det}\left[a_{ij}(\vec{x},\vec{y}) \right]_{i,j=m,\dots,s-1}}, \textrm{  where  } \left\{ \begin{array}{c}  0 \le k \le n-s-1\\ m < s \le n-1\end{array}\right.
\end{split}
\label{wavemat1}\end{equation}\normalsize
where,
\small\begin{equation*}\begin{split}
\left( a_{ij}(\vec{x},\vec{y}) \right)^{n-1}_{i,j=m} &= \exp \left[\sum^{\infty}_{k = 1} x_{k} \Lambda^{k}_{[m,n-1]}\right] A \exp \left[-\sum^{\infty}_{k = 1} y_{k} (\Lambda^T_{[m,n-1]})^{k}\right]\\
&= \exp \left[\sum^{n-m-1}_{k = 1} x_{k} \Lambda^{k}_{[m,n-1]}\right] A \exp \left[-\sum^{n-m-1}_{k = 1} y_{k} (\Lambda^T_{[m,n-1]})^{k}\right] 
\end{split}
\end{equation*}\normalsize
\end{proposition}
\textbf{Proof.}
To prove this we begin with eq. \ref{1} and rewrite it as,
\small\begin{equation}
\hat{W}^{(0)}(\vec{x},\vec{y})=\hat{W}^{(\infty)}(\vec{x},\vec{y}) A(\vec{x},\vec{y})
\label{2}\end{equation}\normalsize
where we remember that $\hat{W}^{(0)}(\vec{x},\vec{y})$ is upper triangular and $\hat{W}^{(\infty)}(\vec{x},\vec{y})$ is lower triangular. Therefore, if we consider the zero entries of each individual row on the left hand side of eq. \ref{2}, they can be expressed as,
\small\begin{equation}
\left( \hat{w}^{(\infty)}_{s-m}(s), \dots, \hat{w}^{(\infty)}_{0}(s) \right) \left( a_{ij}(\vec{x},\vec{y}) \right)_{i=m,\dots,s \atop{j=m,\dots,s-1}}  = \underbrace{( 0, \dots, 0)}_{s-m}
\label{3}\end{equation}\normalsize
where $m< s \le n-1$ and $\hat{w}^{(\infty)}_{0}(s) = 1$.\\
\\
At the moment the matrix is of size $(s-m+1)\times(s-m)$. It is possible to add 1 more column appropriately to make the matrix square,
\small\begin{equation*}
\left( \hat{w}^{(\infty)}_{s-m}(s), \dots, \hat{w}^{(\infty)}_{1}(s),1 \right) \left(\begin{array}{ccccc}
a_{m,m} & \dots & \dots & a_{m,s-1} & 0 \\
\vdots & \ddots &  & \vdots& \vdots\\
\vdots &  & \ddots & \vdots & 0\\
a_{s,m} & \dots& \dots & a_{s,s-1} &1 \end{array}\right)
  = \underbrace{( 0, \dots, 0,1)}_{s-m+1}
\end{equation*}\normalsize
In this form we can use Cramer's rule to find the explicit values of $\hat{w}^{(\infty)}_{k}(s)$. For instance, $\hat{w}^{(\infty)}_{s-m}(s)$ is equal to,
\small\begin{equation*}
\frac{\textrm{det}\left(\begin{array}{ccccc}
0 & \dots & \dots & 0 & 1 \\
\vdots & \ddots &  & \vdots& \vdots\\
\vdots &  & \ddots & \vdots & 0\\
a_{s,m} & \dots& \dots & a_{s,s-1} &1 \end{array}\right)}{\textrm{det}\left(\begin{array}{ccccc}
a_{m,m} & \dots & \dots & a_{m,s-1} & 0 \\
\vdots & \ddots &  & \vdots& \vdots\\
\vdots &  & \ddots & \vdots & 0\\
a_{s,m} & \dots& \dots & a_{s,s-1} &1 \end{array}\right)} = (-1)^{s-m}\frac{\textrm{det}\left[a_{ij}(\vec{x},\vec{y}) \right]_{i=m+1,\dots ,s \atop{j=m,\dots,s-1}}}{\textrm{det}\left[a_{ij}(\vec{x},\vec{y}) \right]^{s-1}_{i,j=m}}
\end{equation*}\normalsize
where we have expanded the top determinant along the top row, and the bottom determinant along the rightmost column. Performing the procedure for general $\hat{w}^{(\infty)}_{k}(s)$ immediately leads us to the desired result. For $\hat{w}^{(0)}_{k}(s)$, we consider the non zero entries of each individual row on the left hand side of eq. \ref{2}. These may be expressed as,
\small\begin{equation*}
\left(  \hat{w}^{(0)}_{0}(s), \dots, \hat{w}^{(0)}_{n-s-1}(s) \right) =\left( \hat{w}^{(\infty)}_{s-m}(s), \dots, \hat{w}^{(\infty)}_{0}(s) \right) \left( a_{ij}(\vec{x},\vec{y}) \right)_{i=m,\dots,s \atop{j=s,\dots,n-1}} 
\end{equation*}\normalsize
for $m \le s \le n-1$.\\
\\
What remains is to simply consider each case separately to obtain,
\small\begin{equation*}\begin{split}
 \hat{w}^{(0)}_{p}(s) &= \sum^{s-m}_{k=0}a_{s-k,s+p} \hat{w}^{(\infty)}_{k}(s)\\
 & =  \sum^{s-m}_{k=0}(-1)^k a_{s-k,s+p} \frac{\textrm{det}\left[a_{ij}(\vec{x},\vec{y}) \right]_{i=m,\dots,\hat{s-k},\dots ,s \atop{j=m,\dots,s-1}}}{\textrm{det}\left[a_{ij}(\vec{x},\vec{y}) \right]_{i,j=m,\dots,s-1}}\\
  &= \frac{\textrm{det}\left[a_{ij}(\vec{x},\vec{y}) \right]_{i=m,\dots ,s \atop{j=m,\dots,s-1,s+p}}}{\textrm{det}\left[a_{ij}(\vec{x},\vec{y}) \right]_{i,j=m,\dots,s-1}} \textrm{    $\square$}
\end{split} 
\end{equation*}\normalsize 
\begin{proposition}The entries of the inverse of the 2 hatted wave-matrices, $\left(\hat{W}^{(0)}(\vec{x},\vec{y})\right)^{-1}$ and $\left(\hat{W}^{(\infty)}(\vec{x},\vec{y})\right)^{-1}$, are given as the following,
\small\begin{equation}\begin{split}
\hat{w}^{*(0)}_{k}(s,\vec{x},\vec{y})& = (-1)^k\frac{\textrm{det}\left[a_{ij}(\vec{x},\vec{y}) \right]_{i=m,\dots,s-1 \atop{j=m,\dots,\hat{s-k},\dots,s}}}{\textrm{det}\left[a_{ij}(\vec{x},\vec{y}) \right]_{i,j=m,\dots,s}},\textrm{  where  }\left\{ \begin{array}{c} 0 \le k \le s-m \\ m < s \le n-1 \end{array}\right. \\
\hat{w}^{*(\infty)}_{k}(s,\vec{x},\vec{y}) &= \frac{\textrm{det}\left[a_{ij}(\vec{x},\vec{y}) \right]_{i=m,\dots,s-1,s+k\atop{j=m,\dots,s}}}{\textrm{det}\left[a_{ij}(\vec{x},\vec{y}) \right]_{i,j=m,\dots,s}},\textrm{  where  } \left\{ \begin{array}{c}  0 \le k \le n-s-1\\ m < s \le n-1\end{array}\right.
\end{split}
\label{wavemat2}\end{equation}\normalsize
where $s$ now refers to the column of the entry, as opposed to the row.
\end{proposition}
\textbf{Proof.}
Taking the inverse of eq. \ref{2}, and rearranging accordingly we obtain,
\small\begin{equation}
A(\vec{x},\vec{y}) \left(\hat{W}^{(0)}(\vec{x},\vec{y})\right)^{-1}=\left(\hat{W}^{(\infty)}(\vec{x},\vec{y})\right)^{-1}
\label{4}\end{equation}\normalsize
Where $\left(\hat{W}^{(0)}(\vec{x},\vec{y})\right)^{-1}$ is upper triangular and $\left(\hat{W}^{(\infty)}(\vec{x},\vec{y})\right)^{-1}$ is lower triangular. Therefore, in an exactly analogous method to proposition 1, if we consider the zero entries of each individual column on the right hand side of eq. \ref{4}, they may be expressed as,
\small\begin{equation*}
\left( a_{ij}(\vec{x},\vec{y}) \right)_{i=m, \dots, s-1 \atop{j=m,\dots,s}} \left(\hat{w}^{*(0)}_{s-i}(s)  \right)_{i=m,\dots,s} = (0)_{i=m,\dots,s-1\atop{j=s}}
\end{equation*}\normalsize
where $m < s \le n-1$ and $\hat{w}^{*(0)}_{0}(s) = \frac{1}{\hat{w}^{(0)}_{0}(s)}$.\\
\\
Similarly to proposition 1, we add a row appropriately to the matrix to make it square,
\small\begin{equation*}
\left(\begin{array}{ccccc} a_{m,m} & \dots & \dots & a_{m,s}\\
\vdots & \ddots & & \vdots\\
\vdots & & \ddots &\vdots\\
a_{s-1,m} & \dots & \dots & a_{s-1,s}\\
0 & \dots & 0 & 1 \end{array} \right)
\left(\begin{array}{c} \hat{w}^{*(0)}_{s-m}(s)\\ \vdots \\ \hat{w}^{*(0)}_{1}(s) \\ \frac{1}{\hat{w}^{(0)}_{0}(s)} \end{array}\right) = \left(\begin{array}{c} 0\\ \vdots \\ 0 \\ \frac{1}{\hat{w}^{(0)}_{0}(s)} \end{array}\right) 
\end{equation*}\normalsize
which is the correct form to apply Cramer's rule. Thus we obtain,
\small\begin{equation*}\begin{split}
 \hat{w}^{*(0)}_{k}(s) &= (-1)^k \frac{1}{\hat{w}^{(0)}_{0}(s)} \frac{\textrm{det}\left[a_{ij}(\vec{x},\vec{y}) \right]_{i=m,\dots,s-1 \atop{j=m,\dots,\hat{s-k},\dots,s}}}{\textrm{det}\left[a_{ij}(\vec{x},\vec{y}) \right]_{i,j=m,\dots,s-1}}\\
&=(-1)^k\frac{\textrm{det}\left[a_{ij}(\vec{x},\vec{y}) \right]_{i=m,\dots,s-1 \atop{j=m,\dots,\hat{s-k},\dots,s}}}{\textrm{det}\left[a_{ij}(\vec{x},\vec{y}) \right]_{i,j=m,\dots,s}}
\end{split}
\end{equation*}\normalsize
For $ \hat{w}^{*(\infty)}_{k}(s) $, we consider the non zero entries of each individual column on the right hand side of eq. \ref{4}. These may be conveniently expressed as,
\small\begin{equation*}
\left( a_{ij}(\vec{x},\vec{y}) \right)_{i=s, \dots, n-1 \atop{j=m,\dots,s}} \left(\hat{w}^{*(0)}_{s-i}(s)  \right)_{i=m,\dots,s} = \left(\hat{w}^{*(\infty)}_{i-s}(s)\right)_{i=s,\dots,n-1}
\end{equation*}\normalsize
for $m \le s \le n-1$.\\
\\
Considering each individual entry on the right hand side obtains,
\small\begin{equation*}\begin{split}
\hat{w}^{*(\infty)}_{p}(s) &= \sum^{s-m}_{k=0} a_{s+p,s-k}\hat{w}^{*(0)}_{k}(s) \\
& =  \sum^{s-m}_{k=0} (-1)^k a_{s+p,s-k}\frac{\textrm{det}\left[a_{ij}(\vec{x},\vec{y}) \right]_{i=m,\dots,s-1 \atop{j=m,\dots,\hat{s-k},\dots,s}}}{\textrm{det}\left[a_{ij}(\vec{x},\vec{y}) \right]_{i,j=m,\dots,s}} \\
& = \frac{\textrm{det}\left[a_{ij}(\vec{x},\vec{y}) \right]_{i=m,\dots,s-1,s+p\atop{j=m,\dots,s}}}{\textrm{det}\left[a_{ij}(\vec{x},\vec{y}) \right]_{i,j=m,\dots,s}} \textrm{  $\square$}
\end{split}\end{equation*}\normalsize 
\section{The generalized Lax and Zakharov-Shabat systems}
In direct analogue of the construction of the KP hierarchy in terms of Zakharov-Shabat and generalized Lax systems introduced in the beginning of this chapter, we now detail how the wave-matrices derived in the previous section can be used to construct the necessary operators given in eq. \ref{lax}, which were first used by Ueno and Takasaki to define the 2-Toda hierarchy. 
\begin{lemma}
Consider the matrices,
\small\begin{equation*}\begin{array}{ll}
L = W^{(\infty)}(\vec{x},\vec{y}) \Lambda_{[m,n-1]} \left(  W^{(\infty)}(\vec{x},\vec{y}) \right)^{-1} &B_{k} = \left\{ L^{k} \right\}_+ \\
M = W^{(0)}(\vec{x},\vec{y})\Lambda^T_{[m,n-1]} \left(  W^{(0)}(\vec{x},\vec{y}) \right)^{-1}& C_{k} = \left\{ M^{k} \right\}_-
\end{array}\end{equation*}\normalsize
then we have the following linearization,
\small\begin{equation*}\begin{split}
\partial_{x_{k}} \hat{W}^{(\infty)}(\vec{x},\vec{y}) =& B_{k} \hat{W}^{(\infty)}(\vec{x},\vec{y}) -  \hat{W}^{(\infty)}(\vec{x},\vec{y}) \Lambda^{k}_{[m,n-1]}\\
\partial_{y_{k}} \hat{W}^{(0)}(\vec{x},\vec{y}) =& C_{k} \hat{W}^{(0)}(\vec{x},\vec{y}) -  \hat{W}^{(0)}(\vec{x},\vec{y}) \left(\Lambda^{T}_{[m,n-1]}\right)^{k}\\
\partial_{x_{k}} \hat{W}^{(0)}(\vec{x},\vec{y}) =& B_{k} \hat{W}^{(0)}(\vec{x},\vec{y}) \\
\partial_{y_{k}} \hat{W}^{(\infty)}(\vec{x},\vec{y}) =& C_{k} \hat{W}^{(\infty)}(\vec{x},\vec{y})
\end{split}\end{equation*}\normalsize
Lax type system,
\small\begin{equation*}\begin{array}{ll}
\partial_{x_{k}} L = \left[ B_{k},L \right] &\partial_{x_{k}} M = \left[ B_{k},M \right]\\
\partial_{y_{k}} L = \left[ C_{k},L \right] &\partial_{y_{k}} M = \left[ C_{k},M \right]
\end{array}\end{equation*}\normalsize
and Zakharov-Shabat type system,
\small\begin{equation*}\begin{split}
\partial_{x_{j}} B_{k}-\partial_{x_{k}} B_{j} + \left[ B_{k},B_{j} \right]&=0\\
\partial_{y_{j}} C_{k}-\partial_{y_{k}} C_{j} + \left[ C_{k},C_{j} \right]&=0\\
\partial_{y_{j}} B_{k}-\partial_{x_{k}} C_{j} + \left[ B_{k},C_{j} \right]&=0
\end{split}\end{equation*}\normalsize
\end{lemma}
\textbf{Proof.}
First we note that $L$ and $M$ can be equivalently expressed in terms of the hatted wave-matrices,
\small\begin{equation*}\begin{split}
L =& \hat{W}^{(\infty)}(\vec{x},\vec{y}) \Lambda_{[m,n-1]} \left(  \hat{W}^{(\infty)}(\vec{x},\vec{y}) \right)^{-1} \\
M =& \hat{W}^{(0)}(\vec{x},\vec{y})\Lambda^T_{[m,n-1]} \left(  \hat{W}^{(0)}(\vec{x},\vec{y}) \right)^{-1}
\end{split}\end{equation*}\normalsize
since the following commutators,
\small\begin{equation*}
\left[\exp\left\{ \sum^{n-m-1}_{l=1}x_{l}\Lambda^{l}_{[m,n)} \right\}, \Lambda^{k}_{[m,n)} \right] = \left[\exp\left\{ \sum^{n-m-1}_{l=1}y_{l}\left(\Lambda^{T}_{[m,n)}\right)^{l} \right\}, \left(\Lambda^{T}_{[m,n)}\right)^{k}  \right] 
\end{equation*}\normalsize
give zero. Also, powers of $L$ and $M$ can be conveniently expressed as,
\small\begin{equation*}\begin{split}
L^{j} =& \hat{W}^{(\infty)}(\vec{x},\vec{y}) \Lambda^{j}_{[m,n-1]} \left(  \hat{W}^{(\infty)}(\vec{x},\vec{y}) \right)^{-1} \\
M^{j} =& \hat{W}^{(0)}(\vec{x},\vec{y})\left(\Lambda^T_{[m,n-1]}\right)^{j} \left(  \hat{W}^{(0)}(\vec{x},\vec{y}) \right)^{-1}
\end{split}\end{equation*}\normalsize
For the remainder of this section we shall simply label $W^{(\infty)}(\vec{x},\vec{y})$ as $W^{(\infty)}$ and $W^{(0)}(\vec{x},\vec{y})$ as $W^{(0)}$, and likewise with their inverses.\\
\\
We begin by differentiating eq. \ref{1} with respect to $x_{j}$, and then multiplying $\left(W^{(\infty)} \right)^{-1}$ on the right,
\small\begin{equation}
\left(\partial_{x_{j}} W^{(\infty)}  \right) \left(  W^{(\infty)} \right)^{-1} = \left(\partial_{x_{j}} W^{(0)}  \right) AA^{-1} \left(  W^{(0)} \right)^{-1} = \left(\partial_{x_{j}} W^{(0)}  \right)  \left( W^{(0)} \right)^{-1} 
\label{1.61}\end{equation}\normalsize
and similarly, differentiating with respect to $y_{j}$,
\small\begin{equation}
\left(\partial_{y_{j}}W^{(\infty)}  \right) \left( W^{(\infty)} \right)^{-1}  = \left(\partial_{y_{j}} W^{(0)}  \right)  \left( W^{(0)} \right)^{-1} 
\label{1.62}\end{equation}\normalsize
Decomposing $W^{(\infty)}$ and $W^{(0)}$, and computing the product rule, eq. \ref{1.61} becomes,
\small\begin{equation*}\begin{split}
\left(\partial_{x_{j}} \hat{W}^{(\infty)}  \right) \left(  \hat{W}^{(\infty)} \right)^{-1} + \hat{W}^{(\infty)}  \left( \partial_{x_{j}}\exp\left\{ \sum^{n-m-1}_{k=1}x_{k}\Lambda^{k}_{[m,n-1]} \right\} \right) \left(  W^{(\infty)} \right)^{-1}\\
= \left(\partial_{x_{j}} \hat{W}^{(0)}  \right)  \left( \hat{W}^{(0)} \right)^{-1}
\end{split}\end{equation*}\normalsize
\small\begin{equation}
\Rightarrow \left(\partial_{x_{j}} \hat{W}^{(\infty)}  \right) \left(  \hat{W}^{(\infty)} \right)^{-1} + \hat{W}^{(\infty)} \Lambda^{j}_{[m,n-1]}  \left(  \hat{W}^{(\infty)} \right)^{-1} = \left(\partial_{x_{j}} \hat{W}^{(0)}  \right)  \left( \hat{W}^{(0)} \right)^{-1} 
\label{1.63}\end{equation}\normalsize
and similarly, eq. \ref{1.62} becomes,
\small\begin{equation}
\left(\partial_{y_{j}} \hat{W}^{(\infty)}  \right) \left(  \hat{W}^{(\infty)} \right)^{-1} = \left(\partial_{y_{j}} \hat{W}^{(0)}  \right)  \left( \hat{W}^{(0)} \right)^{-1}  + \hat{W}^{(0)} \left(\Lambda^{T}_{[m,n-1]}  \right)^{j}\left(  \hat{W}^{(0)} \right)^{-1} 
\label{1.64}\end{equation}\normalsize
Since $\hat{W}^{(\infty)}$ and $\left(\hat{W}^{(\infty)}\right)^{-1}$ are lower diagonal matrices, and $\hat{w}^{(\infty)}_0(s,\vec{x},\vec{y})=\hat{w}^{*(\infty)}_0(s,\vec{x},\vec{y})=1$, this means that $\partial_{x_{j}} \hat{W}^{(\infty)} = \left\{ \partial_{x_{j}} \hat{W}^{(\infty)} \right\}_-$. Hence we have that,
\small\begin{equation*}
\left\{\left(\partial_{x_{j}} \hat{W}^{(\infty)}  \right) \left(  \hat{W}^{(\infty)} \right)^{-1} \right\}_+=0
\end{equation*}\normalsize
i.e. the diagonal terms have been eliminated by the differential operator. Similarly, since $\hat{W}^{(0)}$ and $\left(\hat{W}^{(0)}\right)^{-1}$ are upper diagonal matrices, and since $\hat{w}^{(0)}_0(s,\vec{x},\vec{y})=\frac{1}{\hat{w}^{*(0)}_0(s,\vec{x},\vec{y})} \ne \textrm{  constant}$, we have that,
\small\begin{equation*}
\left\{\left(\partial_{y_{j}} \hat{W}^{(0)}  \right) \left(  \hat{W}^{(0)} \right)^{-1} \right\}_-=0
\end{equation*}\normalsize
i.e. the diagonal terms have \emph{not} been eliminated by the differential operator.\\
\\
This implies the following two equalities,
\small\begin{equation*}\begin{split}
\left\{\left(\partial_{x_{j}} \hat{W}^{(\infty)}  \right) \left(  \hat{W}^{(\infty)} \right)^{-1} \right\}_- =& \left(\partial_{x_{j}} \hat{W}^{(\infty)}  \right) \left(  \hat{W}^{(\infty)} \right)^{-1} \\
\left\{\left(\partial_{y_{j}} \hat{W}^{(0)}  \right) \left(  \hat{W}^{(0)} \right)^{-1} \right\}_+ =& \left(\partial_{y_{j}} \hat{W}^{(0)}  \right) \left(  \hat{W}^{(0)} \right)^{-1} 
\end{split}\end{equation*}\normalsize
Considering the $\{\dots \}_+$ part of eq. \ref{1.63} allows us to obtain an alternative definition of $B_{j}$,
\small\begin{equation*}\begin{split}
\left\{ \left(\partial_{x_{j}} \hat{W}^{(\infty)}  \right) \left(  \hat{W}^{(\infty)} \right)^{-1} + \hat{W}^{(\infty)} \Lambda^{j}_{[m,n-1]}  \left(  \hat{W}^{(\infty)} \right)^{-1} = \left(\partial_{x_{j}} \hat{W}^{(0)}  \right)  \left( \hat{W}^{(0)} \right)^{-1} \right\}_+
\end{split}\end{equation*}\normalsize
\small\begin{equation}\begin{split}
\Rightarrow \left\{L^{j} \right\}_+ = B_{j} = \left(\partial_{x_{j}} \hat{W}^{(0)}  \right)  \left( \hat{W}^{(0)} \right)^{-1} 
\label{1.65}\end{split}\end{equation}\normalsize
Similarly, considering the $\{\dots \}_-$ part of eq. \ref{1.64} allows us to obtain an alternative definition of $C_{j}$,
\small\begin{equation*}\begin{split}
\left\{ \left(\partial_{y_{j}} \hat{W}^{(\infty)}  \right) \left(  \hat{W}^{(\infty)} \right)^{-1} = \left(\partial_{y_{j}} \hat{W}^{(0)}  \right)  \left( \hat{W}^{(0)} \right)^{-1}  + \hat{W}^{(0)} \left(\Lambda^{T}_{[m,n-1]}  \right)^{j}\left(  \hat{W}^{(0)} \right)^{-1} \right\}_-
\end{split}\end{equation*}\normalsize
\small\begin{equation}\begin{split}
\Rightarrow \left\{M^{j} \right\}_- = C_{j} = \left(\partial_{y_{j}} \hat{W}^{(\infty)}  \right) \left(  \hat{W}^{(\infty)} \right)^{-1}
\label{1.66}\end{split}\end{equation}\normalsize
\textbf{Linearization equations.} We are now in a position to obtain the 4 linearization equations. Beginning with eq. \ref{1.63} and using eq. \ref{1.65} we obtain,
\small\begin{equation*}\begin{split}
B_{j} & =  \left(\partial_{x_{j}} \hat{W}^{(\infty)}  \right) \left(  \hat{W}^{(\infty)} \right)^{-1} + \hat{W}^{(\infty)} \Lambda^{j}_{[m,n-1]}  \left(  \hat{W}^{(\infty)} \right)^{-1} \\
\Rightarrow B_{j}  \hat{W}^{(\infty)} &=\partial_{x_{j}} \hat{W}^{(\infty)}  + \hat{W}^{(\infty)} \Lambda^{j}_{[m,n-1]}  
\end{split}\end{equation*}\normalsize
Considering the $\{\dots \}_+$ part of eq. \ref{1.63} we obtain,
\small\begin{equation*}\begin{split}
B_{j}& = \left(\partial_{x_{j}} \hat{W}^{(0)}  \right)  \left( \hat{W}^{(0)} \right)^{-1}\\ 
\Rightarrow B_{j} \hat{W}^{(0)}& = \partial_{x_{j}} \hat{W}^{(0)}  
\end{split}\end{equation*}\normalsize
Focusing on eq. \ref{1.64} and in view of eq. \ref{1.66} we obtain,
\small\begin{equation*}\begin{split}
C_{j} &= \left(\partial_{y_{j}} \hat{W}^{(0)}  \right)  \left( \hat{W}^{(0)} \right)^{-1}  + \hat{W}^{(0)} \left(\Lambda^{T}_{[m,n-1]}  \right)^{j}\left(  \hat{W}^{(0)} \right)^{-1} \\
\Rightarrow C_{j} \hat{W}^{(0)}& = \partial_{y_{j}} \hat{W}^{(0)}   + \hat{W}^{(0)} \left(\Lambda^{T}_{[m,n-1]}  \right)^{j}
\end{split}\end{equation*}\normalsize
and finally considering the $\{\dots \}_-$ part of eq. \ref{1.64} we obtain,
\small\begin{equation*}\begin{split}
C_{j} & = \left(\partial_{y_{j}} \hat{W}^{(\infty)}  \right) \left(  \hat{W}^{(\infty)} \right)^{-1}\\
\Rightarrow C_{j}\hat{W}^{(\infty)}&= \partial_{y_{j}} \hat{W}^{(\infty)} 
\end{split}\end{equation*}\normalsize
\textbf{Lax type equations.}
We now obtain the 4 Lax type equations. For the first equation, consider differentiating the initial definition of $L$ with respect to $x_{j}$,
\small\begin{equation*}\begin{split}
\partial_{x_{j}}L = \left(\partial_{x_{j}} \hat{W}^{(\infty)}  \right) \Lambda_{[m,n-1]} \left(  \hat{W}^{(\infty)} \right)^{-1} + \hat{W}^{(\infty)} \Lambda_{[m,n-1]} \left[ \partial_{x_{j}}\left(  \hat{W}^{(\infty)} \right)^{-1}\right]
\end{split}\end{equation*}\normalsize
where,
\small\begin{equation}\begin{split}
 \partial_{x_{j}} \left[ \hat{W}^{(\infty)}\left(  \hat{W}^{(\infty)} \right)^{-1}\right] &= 0  \\
 &=\left( \partial_{x_{j}}\hat{W}^{(\infty)} \right)\left(  \hat{W}^{(\infty)} \right)^{-1} + \hat{W}^{(\infty)}\left[\partial_{x_{j}} \left(  \hat{W}^{(\infty)} \right)^{-1}\right] \\
\Rightarrow \partial_{x_{j}} \left(  \hat{W}^{(\infty)} \right)^{-1} &= - \left(  \hat{W}^{(\infty)} \right)^{-1} \left( \partial_{x_{j}}\hat{W}^{(\infty)} \right)\left(  \hat{W}^{(\infty)} \right)^{-1}\label{h.1}
\end{split}\end{equation}\normalsize
Using the above expression for $\partial_{x_{j}} \left(  \hat{W}^{(\infty)} \right)^{-1}$ on $\partial_{x_{j}}L$ we obtain,
\small\begin{equation*}\begin{split}
\partial_{x_{j}}L &= \underbrace{\left(\partial_{x_{j}} \hat{W}^{(\infty)}  \right)}_{\textrm{use eq. \ref{1.63}}} \Lambda_{[m,n-1]} \left(  \hat{W}^{(\infty)} \right)^{-1}\\
& - \hat{W}^{(\infty)} \Lambda_{[m,n-1]}\left(  \hat{W}^{(\infty)} \right)^{-1} \underbrace{\left( \partial_{x_{j}}\hat{W}^{(\infty)} \right)\left(  \hat{W}^{(\infty)} \right)^{-1}}_{\textrm{use eq. \ref{1.63}}}\\
&= B_{j}  \hat{W}^{(\infty)}  \Lambda_{[m,n-1]} \left(  \hat{W}^{(\infty)} \right)^{-1} -  \hat{W}^{(\infty)}  \Lambda_{[m,n-1]} \left(  \hat{W}^{(\infty)} \right)^{-1}B_{j}\\
\Rightarrow \partial_{x_{j}}L &= \left[ B_{j}, L \right]
\end{split}\end{equation*}\normalsize
The second Lax equation is derived analogously to the first. Consider differentiating $L$ with respect to $y_{j}$,
\small\begin{equation*}\begin{split}
\partial_{y_{j}}L = \underbrace{\left(\partial_{y_{j}} \hat{W}^{(\infty)}  \right)}_{= C_{j}\hat{W}^{(\infty)} } \Lambda_{[m,n-1]} \left(  \hat{W}^{(\infty)} \right)^{-1} + \hat{W}^{(\infty)} \Lambda_{[m,n-1]} \left[ \partial_{y_{j}}\left(  \hat{W}^{(\infty)} \right)^{-1}\right]
\end{split}\end{equation*}\normalsize
where,
\small\begin{equation}\begin{split}
\partial_{y_{j}} \left[ \hat{W}^{(\infty)}\left(  \hat{W}^{(\infty)} \right)^{-1}\right]&= 0\\
&= \underbrace{\left( \partial_{y_{j}}\hat{W}^{(\infty)} \right)\left(  \hat{W}^{(\infty)} \right)^{-1}}_{C_{j}} + \hat{W}^{(\infty)}\left[\partial_{y_{j}} \left(  \hat{W}^{(\infty)} \right)^{-1}\right]\\
\Rightarrow \partial_{y_{j}} \left(  \hat{W}^{(\infty)} \right)^{-1}&= - \left(  \hat{W}^{(\infty)} \right)^{-1} C_{j}\label{h.2}
\end{split}\end{equation}\normalsize
Using the above expression in $\partial_{y_{j}}L$ we obtain,
\small\begin{equation*}\begin{split}
\partial_{y_{j}}L &= C_{j}\hat{W}^{(\infty)}  \Lambda_{[m,n-1]} \left(  \hat{W}^{(\infty)} \right)^{-1} - \hat{W}^{(\infty)} \Lambda_{[m,n-1]}\left(  \hat{W}^{(\infty)} \right)^{-1} C_{j}\\
\Rightarrow \partial_{y_{j}}L &= \left[ C_{j}, L \right]
\end{split}\end{equation*}\normalsize
For the third equation, consider differentiating $M$ with respect to $x_{j}$,
\small\begin{equation*}\begin{split}
\partial_{x_{j}}M = \underbrace{\left(\partial_{x_{j}} \hat{W}^{(0)}  \right)}_{= B_{j}\hat{W}^{(0)} } \Lambda^{T}_{[m,n-1]}  \left(  \hat{W}^{(0)} \right)^{-1} + \hat{W}^{(0)}  \Lambda^{T}_{[m,n-1]}  \left[ \partial_{x_{j}}\left(  \hat{W}^{(0)} \right)^{-1}\right]
\end{split}\end{equation*}\normalsize
where,
\small\begin{equation}\begin{split}
\partial_{x_{j}} \left[ \hat{W}^{(0)}\left(  \hat{W}^{(0)} \right)^{-1}\right] &=0\\
&= \underbrace{\left( \partial_{x_{j}}\hat{W}^{(0)} \right)\left(  \hat{W}^{(0)} \right)^{-1}}_{B_{j}} + \hat{W}^{(0)}\left[\partial_{x_{j}} \left(  \hat{W}^{(0)} \right)^{-1}\right] \\
\Rightarrow \partial_{x_{j}} \left(  \hat{W}^{(0)} \right)^{-1} &= - \left(  \hat{W}^{(0)} \right)^{-1} B_{j}\label{h.3}
\end{split}\end{equation}\normalsize
Using the above expression in $\partial_{x_{j}}M$ we obtain,
\small\begin{equation*}\begin{split}
\partial_{x_{j}}M &=B_{j} \hat{W}^{(0)}  \Lambda^{T}_{[m,n-1]}  \left(  \hat{W}^{(0)} \right)^{-1} - \hat{W}^{(0)}  \Lambda^{T}_{[m,n-1]} \left(  \hat{W}^{(0)} \right)^{-1} B_{j}\\
\Rightarrow \partial_{x_{j}}M &= \left[ B_{j},M \right] 
\end{split}\end{equation*}\normalsize
For the fourth equation, consider differentiating $M$ with respect to $y_{j}$,
\small\begin{equation*}\begin{split}
\partial_{y_{j}}M = \underbrace{\left(\partial_{y_{j}} \hat{W}^{(0)}  \right)}_{\textrm{use eq. \ref{1.64}}}  \Lambda^{T}_{[m,n-1]} \left(  \hat{W}^{(0)} \right)^{-1} + \hat{W}^{(0)}  \Lambda^{T}_{[m,n-1]}  \left[ \partial_{y_{j}}\left(  \hat{W}^{(0)} \right)^{-1}\right]
\end{split}\end{equation*}\normalsize
where,
\small\begin{equation}\begin{split}
\partial_{y_{j}} \left[ \hat{W}^{(0)}\left(  \hat{W}^{(0)} \right)^{-1}\right] &=0\\
&=  \underbrace{\left( \partial_{y_{j}} \hat{W}^{(0)}\right)}_{\textrm{use eq. \ref{1.64}}}\left(  \hat{W}^{(0)} \right)^{-1}+\hat{W}^{(0)} \left[ \partial_{y_{j}}\left(  \hat{W}^{(0)} \right)^{-1}\right]\\
\Rightarrow \partial_{y_{j}}\left(  \hat{W}^{(0)} \right)^{-1}& = - \left(  \hat{W}^{(0)} \right)^{-1} C_{j}+ \left( \Lambda^{T}_{[m,n-1]} \right)^{j}\left(  \hat{W}^{(0)} \right)^{-1}\label{h.4}
\end{split}\end{equation}\normalsize
Using the above expression in $\partial_{y_{j}}M$ we obtain,
\small\begin{equation*}\begin{split}
\partial_{y_{j}}M &= C_{j} \hat{W}^{(0)}  \Lambda^{T}_{[m,n-1]}  \left(  \hat{W}^{(0)} \right)^{-1} - \hat{W}^{(0)}   \Lambda^{T}_{[m,n-1]}  \left(  \hat{W}^{(0)} \right)^{-1} C_{j}\\
\Rightarrow \partial_{y_{j}}M &= \left[ C_{j},M \right] 
\end{split}\end{equation*}\normalsize
\textbf{Zakharov-Shabat type equations.}
For the first Z-S equation, we consider the first linearization equation and compare cross derivatives with respect to $x$, i.e. $\left(\partial_{x_{j}}\partial_{x_{k}} - \partial_{x_{k}}\partial_{x_{j}}\right) \hat{W}^{(\infty)}=0$,
\small\begin{equation*}\begin{split}
\partial_{x_{j}}\left(\partial_{x_{k}} \hat{W}^{(\infty)} \right) &= \partial_{x_{j}}\left(B_{k} \hat{W}^{(\infty)} - \hat{W}^{(\infty)}\Lambda^{k}_{[m,n-1]}  \right)\\
&= \left( \partial_{x_{j}} B_{k} \right)\hat{W}^{(\infty)} + B_{k} B_{j}\hat{W}^{(\infty)}-B_{k} \hat{W}^{(\infty)} \Lambda^{j}_{[m,n-1]}\\
&-B_{j} \hat{W}^{(\infty)} \Lambda^{k}_{[m,n-1]}+\hat{W}^{(\infty)} \Lambda^{j+k}_{[m,n-1]}\\
\Rightarrow \left(\partial_{x_{j}}\partial_{x_{k}} - \partial_{x_{k}}\partial_{x_{j}}\right) \hat{W}^{(\infty)}
 &= \left( \partial_{x_{j}}B_{k} -  \partial_{x_{k}}B_{j}+\left[ B_{k},B_{j} \right] \right) \hat{W}^{(\infty)}
\end{split}\end{equation*}\normalsize
For the second Z-S equation, we consider the second linearization equation and compare cross derivatives with respect to $y$, i.e. $\left(\partial_{y_{j}}\partial_{y_{k}}- \partial_{y_{k}}\partial_{y_{j}}\right) \hat{W}^{(0)}=0$,
\small\begin{equation*}\begin{split}
\partial_{y_{j}}\left(\partial_{y_{k}} \hat{W}^{(0)} \right) &= \partial_{y_{j}}\left(C_{k} \hat{W}^{(0)} - \hat{W}^{(0)}\left(\Lambda^{T}_{[m,n-1]}\right)^{k}  \right)\\
&= \left( \partial_{y_{j}} C_{k} \right)\hat{W}^{(0)} + C_{k} C_{j}\hat{W}^{(0)}-C_{k} \hat{W}^{(0)} \Lambda^{j}_{[m,n-1]}\\
&-C_{j} \hat{W}^{(0)} \left(\Lambda^{T}_{[m,n-1]}\right)^{k}+\hat{W}^{(0)} \left(\Lambda^T_{[m,n-1]}\right)^{j+k}\\
\Rightarrow \left(\partial_{y_{j}}\partial_{y_{k}} - \partial_{y_{k}}\partial_{y_{j}}\right) \hat{W}^{(0)}
 &= \left( \partial_{y_{j}}C_{k} -  \partial_{y_{k}}C_{j}+\left[ C_{k},C_{j} \right] \right) \hat{W}^{(0)}
\end{split}\end{equation*}\normalsize
For the third Z-S equation, we consider cross derivatives with respect to $x$ and $y$ acting on $\hat{W}^{(\infty)}$, $\left(\partial_{x_{j}} \partial_{y_{k}}- \partial_{y_{k}}\partial_{x_{j}}\right) \hat{W}^{(\infty)}=0$,
\small\begin{equation*}\begin{split}
\partial_{x_{j}} \partial_{y_{k}}\hat{W}^{(\infty)} &= \left( \partial_{x_{j}} C_{k} \right) \hat{W}^{(\infty)}+ C_{k}B_{j}\hat{W}^{(\infty)} - C_{k}\hat{W}^{(\infty)}  \Lambda^{j}_{[m,n-1]} \\
\partial_{y_{k}} \partial_{x_{j}} \hat{W}^{(\infty)}&= \left( \partial_{y_{k}} B_{j} \right) \hat{W}^{(\infty)}+ B_{j}C_{k}\hat{W}^{(\infty)} - C_{k}\hat{W}^{(\infty)}  \Lambda^{j}_{[m,n-1]}\\
\Rightarrow \left(\partial_{x_{j}} \partial_{y_{k}}- \partial_{y_{k}}\partial_{x_{j}}\right) \hat{W}^{(\infty)} & =  \left( \partial_{x_{j}}C_{k} -\partial_{y_{k}}B_{j} + \left[ C_{k},B_{j} \right] \right) \hat{W}^{(\infty)}
\end{split}\end{equation*}\normalsize
Thus concluding this section. $\square$
\section{Tau-function of the 2-Toda hierarchy}
\noindent \textbf{Further words on the KP hierarchy.} Continuing with the running example of the KP hierarchy, Hirota \cite{Hirotima} found a systematic method for deriving the $N$ soliton solutions (alternative to the IST method) to the integrable equations produced by the KP hierarchy, known as Hirota's direct method. The heart of the method involves expressing the non linear PDE's in \textit{bilinear form} and employing an \textit{exact} perturbation\footnote{Exact in the sense that even though perturbation is being used, one still obtains an exact analytical result.} argument. For example, the KP equation in bilinear form looks like,
\small\begin{equation}
\left( 4 D_{x_1} D_{x_3}-D^4_{x_1} -3D^2_{x_2} \right) \tau(\vec{x}) . \tau(\vec{x}) = 0
\end{equation}\normalsize
where $u(\vec{x}) =2 \partial^2_{x_1} \log \tau(\vec{x})$, and the Hirota bilinear differential operators, $D_x$, are defined in section 1.4.2. Here the $\tau$-function appears as a simple transformation of the usual function $u$, with no indication of the special properties is possesses. The first time the notion of the $\tau$-function was introduced as an object worthy of study was in the work \cite{tau11}.  Here it was expressed a series of expectation values in the context of holonomic quantum fields.\\
\\
Studying the KP hierarchy, M. and Y. Sato \cite{Sato1,Sato2} discovered some beautiful algebro-geometric properties of the $\tau$-function. Namely, through considering the infinite dimensional Lie algebra $gl(\infty)$, the group orbit of the highest weight vector (the $\tau$-function) is an infinite dimensional Grassmannian manifold. Furthermore, the equations which describe this manifold in function space are the soliton equations. Even more importantly, (as far as this thesis is concerned), was the work \cite{KP1} where the results Sato were expressed in the form of fermionic operators (acting on the Fock space). Additionally, they discovered all of the (infinitely many) non linear PDE's in the KP hierarchy can miraculously be expressed as coefficients of the following \textit{bilinear equation},
\small\begin{equation}
\oint \frac{d \lambda}{2 \pi i} \exp \left\{ \sum^{\infty}_{l=1}(x_{l}-x'_{l}) \lambda^{l} \right\} \frac{\tau\left(\vec{x}-\vec{\epsilon}\left(\frac{1}{\lambda}\right)\right)}{ \tau\left(\vec{x}\right)}\frac{\tau\left(\vec{x}'+\vec{\epsilon}\left(\frac{1}{\lambda}\right)\right)}{\tau\left(\vec{x}'\right)} = 0
\end{equation}\normalsize
For general $\{\vec{x}, \vec{x}' \}$, and where $\vec{\epsilon}\left(\lambda \right) = \left( \lambda, \frac{\lambda^2}{2}, \frac{\lambda^3}{3},\dots \right)$. Thus we see that a $\tau$-function which satisfies one non linear PDE in the hierarchy obviously satisfies \textit{all} equations in the hierarchy. We shall use the results of this line of work extensively in chapter 3. \\
\\
\textbf{Continuing with the 2-Toda hierarchy.} As shown in eq. \ref{lax}, the 2-Toda hierarchy is defined as a series of 4 distinct lax type first order differential equations of the matrices $L,M,B_n$ and $C_n$, which themselves are composed of the wave-matrices $W^{(\infty)},\left( W^{(\infty)} \right)^{-1},W^{(0)}$ and $\left( W^{(0)} \right)^{-1}$. The $\tau$-function is a single function, $\tau(s,\vec{x},\vec{y})$, of the time parameters $\vec{x}$ and $\vec{y}$ and a single parameter $s$, which corresponds to the row number of $W^{(\infty)},W^{(0)}$ or the column number of $\left( W^{(\infty)} \right)^{-1},\left( W^{(0)} \right)^{-1}$. The derivatives of $\tau(s,\vec{x},\vec{y})$ correspond to the entries of the wave-matrices and using this fact, we can express the 2-Toda hierarchy in a single integral bilinear form. \\
\\
\noindent We begin with the following fundamental result.
\begin{lemma} For the function,
\small\begin{equation}
\tau(s,\vec{x},\vec{y}) = \textrm{det}\left[a_{ij}(\vec{x},\vec{y}) \right]_{i,j=m,\dots,s-1}
\label{5}\end{equation}\normalsize
the following four relations hold,
\small\begin{equation}
\hat{w}^{(\infty)}_{k}(s,\vec{x},\vec{y}) = \frac{\zeta_k(-\tilde{\partial}_{\vec{x}})\tau(s,\vec{x},\vec{y})}{\tau(s,\vec{x},\vec{y})},  \textrm{ non zero for } k \in \{0, \dots, s-m\} \label{s1}
\end{equation}\normalsize
\small\begin{equation}
\hat{w}^{(0)}_{k}(s,\vec{x},\vec{y}) = \frac{\zeta_k(-\tilde{\partial}_{\vec{y}})\tau(s+1,\vec{x},\vec{y})}{\tau(s,\vec{x},\vec{y})}, \textrm{ non zero for } k \in \{0, \dots, n-s-1\}\label{s2}
\end{equation}\normalsize
\small\begin{equation}
\hat{w}^{*(\infty)}_{k}(s,\vec{x},\vec{y}) = \frac{\zeta_k(\tilde{\partial}_{\vec{x}})\tau(s+1,\vec{x},\vec{y})}{\tau(s+1,\vec{x},\vec{y})}, \textrm{ non zero for } k \in \{0, \dots, n-s-1\}\label{s3}
\end{equation}\normalsize
\small\begin{equation}
\hat{w}^{*(0)}_{k}(s,\vec{x},\vec{y}) = \frac{\zeta_k(\tilde{\partial}_{\vec{y}})\tau(s,\vec{x},\vec{y})}{\tau(s+1,\vec{x},\vec{y})}, \textrm{ non zero for } k \in \{0, \dots, s-m\}\label{s4}
\end{equation}\normalsize
where,
\small\begin{equation*}
\tilde{\partial}_{\vec{x}} = (\partial_{x_1}, \frac{1}{2} \partial_{x_2}, \frac{1}{3}\partial_{x_3}, \dots ) \textrm{  ,  } \tilde{\partial}_{\vec{y}} = (\partial_{y_1}, \frac{1}{2} \partial_{y_2}, \frac{1}{3}\partial_{y_3}, \dots )
\end{equation*}\normalsize
and the generating function for the $\zeta_k(\vec{x})$'s (referred to as the \textbf{one row character polynomial of order $\mathbf{k}$}) is given by,
\small\begin{equation}\begin{split}
\sum^{\infty}_{k=0} z^k \zeta_k(\vec{x}) &= \exp \left\{ \sum^N_{j=1}z^j x_j \right\} \\
\Rightarrow \zeta_k(\vec{x}) &= \sum_{j_1 + 2j_2 + \dots + k j_k = k} \frac{x^{j_1}_1 \dots x^{j_k}_k}{j_1 ! \dots j_k !} \label{zeta}
\end{split}\end{equation}\normalsize
\end{lemma}
\textbf{Proof.} Assuming eq. \ref{s1} is true, then we have the following,
\small\begin{equation*}\begin{split}
\sum^{s-m}_{k=0} \lambda^k \hat{w}^{(\infty)}_{k}(s,\vec{x},\vec{y})& =\sum^{\infty}_{k=0} \lambda^k \hat{w}^{(\infty)}_{k}(s,\vec{x},\vec{y}) \\
& = \frac{1}{\tau(s,\vec{x},\vec{y})}\sum^{\infty}_{k=0}\lambda^k \zeta_k(-\tilde{\partial}_{\vec{x}}) \tau(s,\vec{x},\vec{y})\\
& = \frac{1}{\tau(s,\vec{x},\vec{y})} \exp\left[-\sum^{\infty}_{j=1}\frac{\lambda^j}{j}\partial_{x_{j}} \right]\tau(s,\vec{x},\vec{y})
\end{split}\end{equation*}\normalsize
\small\begin{equation}
\Rightarrow \sum^{s-m}_{k=0} \lambda^k \hat{w}^{(\infty)}_{k}(s,\vec{x},\vec{y}) = \frac{\tau(s,\vec{x}-\vec{\epsilon}(\lambda),\vec{y})}{\tau(s,\vec{x},\vec{y})}
\label{10}\end{equation}\normalsize
Similarly, equations \ref{s2}-\ref{s4} become,
\begin{eqnarray}
\sum^{n-s-1}_{k=0} \lambda^k \hat{w}^{(0)}_{k}(s,\vec{x},\vec{y}) &= \frac{\tau(s+1,\vec{x},\vec{y}-\vec{\epsilon}(\lambda))}{\tau(s,\vec{x},\vec{y})}\label{s5}\\
\sum^{n-s-1}_{k=0} \lambda^k \hat{w}^{*(\infty)}_{k}(s,\vec{x},\vec{y}) &= \frac{\tau(s+1,\vec{x}+\vec{\epsilon}(\lambda),\vec{y})}{\tau(s+1,\vec{x},\vec{y})}\label{s6}\\
\sum^{s-m}_{k=0} \lambda^k \hat{w}^{*(0)}_{k}(s,\vec{x},\vec{y}) &= \frac{\tau(s,\vec{x},\vec{y}+\vec{\epsilon}(\lambda))}{\tau(s+1,\vec{x},\vec{y})}\label{s7}
\end{eqnarray}
Hence we shall prove lemma 2 by showing that equations \ref{10} - \ref{s7} hold. \\
\\
In the work below we shall require the following relations,
\small\begin{equation}
\sum^{\infty}_{j=1} \frac{\left( \lambda \Lambda_{[m,n-1]} \right)^{j}}{j} = -\log\left( 1 - \lambda \Lambda_{[m,n-1]}\right)\label{s8}
\end{equation}\normalsize
\small\begin{equation}
\exp\left[ -\sum^{\infty}_{j=1} \frac{\left( \lambda \Lambda_{[m,n-1]} \right)^{j}}{j}  \right]= 1- \lambda  \Lambda_{[m,n-1]}\label{s9}
\end{equation}\normalsize
\small\begin{equation}
\exp\left[ \sum^{\infty}_{j=1} \frac{\left( \lambda \Lambda_{[m,n-1]} \right)^{j}}{j}  \right] = \sum^{n-m-1}_{k=0}\lambda^k  \Lambda^k_{[m,n-1]}\label{s10}
\end{equation}\normalsize
where the summation in eq. \ref{s8} obviously truncates at $j=n-m-1$.\\
\\
In addition, we require the \textbf{Cauchy-Binet} identity for expanding the determinant of the product non square matrices,
\small\begin{equation}
\textrm{det} \left( \sum^{p+q}_{l=m} g_{il}h_{lj} \right)^p_{i,j=m} = \sum_{m\le l_m < \dots < l_p \le p+q} \det \left( g_{i l_j} \right)^p_{i,j = m}\det \left( h_{l_i,j} \right)^p_{i,j = m}
\label{17}\end{equation}\normalsize
Let us now consider expanding $\tau(s,\vec{x}-\vec{\epsilon}(\lambda),\vec{y})$
\small\begin{equation*}\begin{split}
=& \textrm{det}\left[a_{ij}(\vec{x}-\vec{\epsilon}(\lambda),\vec{y}) \right]^{s-1}_{i,j=m}\\
=& \textrm{det} \left[ \exp \left\{ \sum^{n-m-1}_{l=1} \left( x_{l} - \frac{\lambda^{l}}{l} \right) \Lambda^{l}_{[m,n-1]} \right\} A \exp \left\{ -\sum^{n-m-1}_{l=1} y_{l} \left( \Lambda^{T}_{[m,n-1]}\right)^{l} \right\}  \right]^{s-1}_{i,j=m}\\
=& \textrm{det} \left[\exp \left\{- \sum^{n-m-1}_{l=1} \frac{\lambda^{l}}{l} \Lambda^{l}_{[m,n-1]} \right\}A(\vec{x},\vec{y}) \right]^{s-1}_{i,j=m}\\
=&\textrm{det} \left[\left( 1- \lambda  \Lambda_{[m,n-1]} \right)A(\vec{x},\vec{y}) \right]^{s-1}_{i,j=m}\\
=& \textrm{det}\left[ \left(\begin{array}{ccccc}
1 & -\lambda & & & \\
 & \ddots & \ddots & & \\
& & \ddots & \ddots &\\
& & & 1 & -\lambda 
\end{array}\right) \left(\begin{array}{cccc}
a_{m,m} &  \dots & a_{m,s-1} \\
\vdots & & \vdots\\
\vdots &  & \vdots\\
\vdots &  & \vdots\\
a_{s,m} & \dots  & a_{s,s-1}\end{array}\right)
\right]\\
=& \textrm{det} \left[\sum^s_{j=m}\left( 1- \lambda  \Lambda_{[m,n-1]} \right)_{ij} \left( a_{jk} (\vec{x},\vec{y}) \right)  \right]^{s-1}_{i,k=m}\\
=& \sum_{m\le j_m < \dots < j_{s-1} \le s} \det \left[ \left( 1- \lambda  \Lambda_{[m,n-1]} \right)_{ij_k} \right]^{s-1}_{i,k = m}\det \left[ a_{j_p l} (\vec{x},\vec{y})\right]^{s-1}_{p,l = m} \\
=& \sum^{s}_{k=m} \det \left[ \left( 1- \lambda  \Lambda_{[m,n-1]} \right)_{ij} \right]_{i = m,\dots,s-1\atop{j=m, \dots, \hat{k}, \dots, s}}\det \left[ a_{ij} (\vec{x},\vec{y})\right]_{i= m,\dots,\hat{k}, \dots,s\atop{j=m, \dots,s-1}}
\end{split}\end{equation*}\normalsize
It is elementary to show that,
\small\begin{equation*}
\det \left[ \left( 1- \lambda  \Lambda_{[m,n-1]} \right)_{ij} \right]_{i = m,\dots,s-1\atop{j=m, \dots, \hat{k}, \dots, s}} = (-1)^{s-k}\lambda^{s-k}
\end{equation*}\normalsize
hence, making the following change of indices, $k \rightarrow s-p$, we obtain,
\small\begin{equation*}
\tau(s,\vec{x}-\vec{\epsilon}(\lambda),\vec{y}) = \sum^{s-m}_{p=0} (-1)^p \lambda^p \det \left[ a_{ij} (\vec{x},\vec{y})\right]_{i= m,\dots,\hat{s-p}, \dots,s\atop{j=m, \dots,s-1}}
\end{equation*}\normalsize
which proves eq. \ref{10}. For eq. \ref{s5} let us consider expanding $\tau(s+1,\vec{x},\vec{y}-\vec{\epsilon}(\lambda))$,
\small\begin{equation*}\begin{split}
=& \textrm{det}\left[a_{ij}(\vec{x},\vec{y}-\vec{\epsilon}(\lambda)) \right]^s_{i,j=m}\\
=& \textrm{det} \left[ \exp \left\{ \sum^{n-m-1}_{l=1}  x_{l}  \Lambda^{l}_{[m,n-1]} \right\} A \exp \left\{ -\sum^{n-m-1}_{l=1}\left( y_{l} - \frac{\lambda^{l}}{l} \right)\left( \Lambda^{T}_{[m,n-1]}\right)^{l} \right\}  \right]^s_{i,j=m}\\
=& \textrm{det} \left[A(\vec{x},\vec{y}) \exp \left\{ \sum^{n-m-1}_{l=1} \frac{\lambda^{l}}{l} \left(\Lambda^{T}_{[m,n-1]} \right)^{l}\right\}\right]^s_{i,j=m}\\
=&\textrm{det} \left[A(\vec{x},\vec{y})  \sum^{n-m-1}_{k=0}\lambda^k  \left(\Lambda^T_{[m,n-1]} \right)^{k} \right]^s_{i,j=m}
\end{split}\end{equation*}\normalsize
\small\begin{equation*}\begin{split}
=& \textrm{det}\left[ \left( \begin{array}{cccc}
a_{m,m} & \dots & \dots & a_{m,n-1} \\
\vdots & & & \vdots\\
a_{s,m} & \dots & \dots & a_{s,n-1} \end{array}\right)\left( \begin{array}{ccccc}
1 & & & & \\
\lambda & 1 & & & \\
\lambda^2 & \lambda & 1 & & \\
\vdots & & & \ddots & \\
\vdots & & & & 1\\
\vdots & & & & \vdots\\
\lambda^{n-m-1} & \dots & \dots & \dots & \lambda^{n-s-2}\end{array}\right)
\right]
\end{split}\end{equation*}\normalsize
\small\begin{equation*}\begin{split}
=&\textrm{det}\left[\sum^{n-m-1}_{j=m} \left( a_{ij}(\vec{x},\vec{y}) \right) \left( \sum^{n-m-1}_{l=0}\lambda^{l}  \left(\Lambda^T_{[m,n-1]} \right)^{l}  \right)_{jk} \right]^s_{i,k=m}\\
=& \sum_{m \le j_m < \dots < j_s \le n-1}\det \left[ a_{ij_l} (\vec{x},\vec{y})\right]_{i,l = m,\dots,s} \textrm{det} \left[  \left(\sum^{n-m-1}_{l=0}\lambda^{l}  \left(\Lambda^T_{[m,n-1]} \right)^{l}  \right)_{j_p,k} \right]^s_{p,k = m}
\end{split}\end{equation*}\normalsize
The term, $\textrm{det} \left[  \left(\sum^{n-m-1}_{l=0}\lambda^{l}  \left(\Lambda^T_{[m,n-1]} \right)^{l} \right)_{j_p,k} \right]^s_{p,k = m}$, simply corresponds to taking out $n-s-1$ rows, leaving $s-m+1$ rows for a proper square matrix. However, if we take out \emph{any} of the rows which contain zeros (of which there are $s-m$ of), the whole determinant is zero. To see this, notice that all the rows without zeros are proportional to each other by varying factors of $\lambda$, i.e. $R_{i+j} = \lambda^{j} R_i$. This means that we have fixed $j_m = m, j_{m+1} = m+1, \dots, j_{s-1} = s-1$, which leads to,
\small\begin{equation*}
= \sum^{n-1}_{k=s}\det \left[ a_{ij} (\vec{x},\vec{y})\right]_{i = m,\dots,s\atop{m,\dots, s-1, k}} \textrm{det} \left[  \left(\sum^{n-m-1}_{l=0}\lambda^{l}  \left(\Lambda^T_{[m,n-1]} \right)^{l}  \right)_{ij} \right]_{i = m,\dots,s-1,k\atop{j=m,\dots,s}}
\end{equation*}\normalsize
It is elementary to show that,
\small\begin{equation*}
\textrm{det} \left[  \left(\sum^{n-m-1}_{l=0}\lambda^{l}  \left(\Lambda^T_{[m,n-1]} \right)^{l}  \right)_{ij} \right]_{i = m,\dots,s-1,k\atop{j=m,\dots,s}} = \lambda^{k-s}
\end{equation*}\normalsize
hence with a change of indices, $k \rightarrow s+p$, we obtain,
\small\begin{equation*}
\tau(s+1,\vec{x},\vec{y}-\vec{\epsilon}(\lambda)) = \sum^{n-s-1}_{p=0} \lambda^p\det \left[ a_{ij} (\vec{x},\vec{y})\right]_{i = m,\dots,s\atop{m,\dots, s-1, s+p}} 
\end{equation*}\normalsize
which proves eq. \ref{s5}.\\
\\
Proving eq. \ref{s6} and \ref{s7} is very similar to proving eq. \ref{10} and \ref{s5}, so we shall not show full details. Expanding $\tau(s+1,\vec{x}+\vec{\epsilon}(\lambda),\vec{y})$
\small\begin{equation*}\begin{split}
=&  \textrm{det}\left[a_{ij}(\vec{x}+\vec{\epsilon}(\lambda),\vec{y}) \right]^s_{i,j=m}\\
=& \textrm{det}\left[ \sum^{n-m-1}_{l=0} \lambda^{l} \Lambda^{l}_{[m,n-1]} A(\vec{x},\vec{y}) \right]^s_{i,j=m}\\
=& \textrm{det}\left[ \sum^{n-1}_{j=m} \left(\sum^{n-m-1}_{l=0} \lambda^{l} \Lambda^{l}_{[m,n-1]} \right)_{ij} \left( a_{jk}(\vec{x},\vec{y})\right) \right]^s_{i,k=m}\\
=& \sum_{m \le j_m < \dots < j_s \le n-1} \textrm{det}\left[ \left( \sum^{n-m-1}_{l=0} \lambda^{l} \Lambda^{l}_{[m,n-1]}\right)_{ij_k} \right]^s_{i,k=m}\det \left[ a_{j_p q} (\vec{x},\vec{y})\right]^s_{p,q = m}
\end{split}\end{equation*}\normalsize
Based on the proof for eq. \ref{s5}, we know that $j_m = m, j_{m+1} = m+1, \dots, j_{s-1}=s-1$, hence,
\small\begin{equation*}
= \sum^{n-1}_{p=s}\textrm{det}\left[ \left( \sum^{n-m-1}_{l=0} \lambda^{l} \Lambda^{l}_{[m,n-1]}\right)_{ij} \right]_{i=m,\dots,s\atop{j=m,\dots, s-1,p}}\det \left[ a_{ij} (\vec{x},\vec{y})\right]_{i = m, \dots, s-1,p\atop{j=m,\dots,s}}
\end{equation*}\normalsize
We also know that $\textrm{det}\left[ \left( \sum^{n-m-1}_{l=0} \lambda^{l} \Lambda^{l}_{[m,n-1]}\right)_{ij} \right]_{i=m,\dots,s\atop{j=m,\dots, s-1,p}} = \lambda^{p-s}$, thus making the following change in indices, $p \rightarrow k+s$, we obtain,
\small\begin{equation*}
\tau(s+1,\vec{x}+\vec{\epsilon}(\lambda),\vec{y}) =  \sum^{n-s-1}_{k=0} \lambda^{k} \det \left[ a_{ij} (\vec{x},\vec{y})\right]_{i = m, \dots, s-1,s+k\atop{j=m,\dots,s}}
\end{equation*}\normalsize
which proves eq. \ref{s6}.\\
\\
For eq. \ref{s7}, we expand $\tau(s,\vec{x},\vec{y}+\vec{\epsilon}(\lambda))$
\small\begin{equation*}\begin{split}
=&  \textrm{det}\left[a_{ij}(\vec{x},\vec{y}+\vec{\epsilon}(\lambda)) \right]^{s-1}_{i,j=m}\\
=&  \textrm{det}\left[A(\vec{x},\vec{y}) \left( 1-\lambda \Lambda^T_{[m,n-1]} \right) \right]^{s-1}_{i,j=m}\\
=& \textrm{det}\left[\sum^{s}_{j=m}\left( a_{ij}(\vec{x},\vec{y})\right) \left( 1-\lambda \Lambda^T_{[m,n-1]} \right)_{jk} \right]^{s-1}_{i,k=m}\\
=& \sum_{m \le j_m < \dots< j_{s-1}\le s} \textrm{det} \left[ a_{ij_k}(\vec{x},\vec{y}) \right]^{s-1}_{i,k = m}  \textrm{det} \left[ \left( 1-\lambda \Lambda^T_{[m,n-1]} \right)_{j_k l}\right]^{s-1}_{k,l = m} \\
=& \sum^{s}_{p=m} \textrm{det} \left[ a_{ij}(\vec{x},\vec{y}) \right]_{i = m,\dots,s-1\atop{j=m,\dots,\hat{p},\dots,s}}  \textrm{det} \left[ \left( 1-\lambda \Lambda^T_{[m,n-1]} \right)_{ij}\right]_{i = m,\dots,\hat{p},\dots,s\atop{j=m, \dots, s-1}}
\end{split}\end{equation*}\normalsize
Since $\textrm{det} \left[ \left( 1-\lambda \Lambda^T_{[m,n-1]} \right)_{ij}\right]_{i = m,\dots,\hat{p},\dots,s\atop{j=m, \dots, s-1}} = (-1)^{s-p}\lambda^{s-p}$, if we make the change of indices, $p \rightarrow s-k$, we obtain,
\small\begin{equation*}
\tau(s,\vec{x},\vec{y}+\vec{\epsilon}(\lambda)) = \sum^{s-m}_{k=0} (-1)^k \lambda^k \textrm{det} \left[ a_{ij}(\vec{x},\vec{y}) \right]_{i = m,\dots,s-1\atop{j=m,\dots,\hat{s-k},\dots,s}} 
\end{equation*}\normalsize
which proves eq. \ref{s7}. $\square$
\subsection{Bilinear relation of the 2-Toda hierarchy}
\begin{lemma}
The function $\tau(s,\vec{x},\vec{y})$ defined in eq. \ref{5} satisfies the following bilinear relationship,
\small\begin{equation}\begin{split}
 \oint \frac{d \lambda}{2 \pi i} \lambda^{s'-s-2} \exp \left\{ \sum^{n-m-1}_{l=1}(y_{l}-y'_{l}) \lambda^{l} \right\} \frac{\tau\left(s+1,\vec{x},\vec{y}-\vec{\epsilon}\left(\frac{1}{\lambda}\right) \right)}{\tau(s,\vec{x},\vec{y})}\frac{\tau \left(s'-1,\vec{x}',\vec{y}'+\vec{\epsilon}\left(\frac{1}{\lambda} \right)\right)}{\tau\left(s',\vec{x}',\vec{y}'\right)}\\
= \oint \frac{d \lambda}{2 \pi i} \lambda^{s-s'} \exp \left\{ \sum^{n-m-1}_{l=1}(x_{l}-x'_{l}) \lambda^{l} \right\} \frac{\tau\left(s,\vec{x}-\vec{\epsilon}\left(\frac{1}{\lambda}\right),\vec{y}\right)}{ \tau\left(s,\vec{x},\vec{y}\right)}\frac{\tau\left(s',\vec{x}'+\vec{\epsilon}\left(\frac{1}{\lambda}\right),\vec{y}'\right)}{\tau\left(s',\vec{x}',\vec{y}'\right)}
\label{biline}\end{split}\end{equation}\normalsize
for general $s,s',\vec{x},\vec{x}',\vec{y},\vec{y}'$. The integration $\oint \frac{d \lambda}{2 \pi i}$ simply refers to the algebraic operation of obtaining the coefficient of $\frac{1}{\lambda}$.
\end{lemma}
\textbf{Proof.} We shall proceed by showing that both sides of this relationship are equivalent to the two sides of another relationship which we know to be true. Let us begin by multiplying eq. \ref{1}, with one set of $\vec{x}$ and $\vec{y}$, on the right by $\left(W^{(0)}(\vec{x}',\vec{y}') \right)^{-1}$, which has a different set of variables $\vec{x}'$ and $\vec{y}'$. Doing so, we obtain the equation,
\small\begin{equation}
 W^{(0)}(\vec{x},\vec{y}) \left(W^{(0)}(\vec{x}',\vec{y}') \right)^{-1} = W^{(\infty)}(\vec{x},\vec{y}) \left(W^{(\infty)}(\vec{x}',\vec{y}') \right)^{-1}
\label{18}\end{equation}\normalsize
Therefore, in order to prove lemma 3, we shall prove two smaller results, showing that the $(s,s')$ entry of the right hand side of eq. \ref{18} is equal to the right hand side of the bilinear relationship for the same choice of $(s,s'+1)$, and similarly for the left hand sides of both equations.\\
\\
\noindent \textbf{Right hand side of bilinear relation.}
\begin{proposition}
\small\begin{equation*}\begin{split}
\oint \frac{d \lambda}{2 \pi i} \lambda^{s-s'-1} \exp \left\{ \sum^{n-m-1}_{l=1}(x_{l}-x'_{l}) \lambda^{l} \right\} \frac{\tau\left(s,\vec{x}-\vec{\epsilon}\left(\frac{1}{\lambda}\right),\vec{y}\right)}{ \tau\left(s,\vec{x},\vec{y}\right)}\frac{\tau\left(s'+1,\vec{x}'+\vec{\epsilon}\left(\frac{1}{\lambda}\right),\vec{y}'\right)}{\tau\left(s'+1,\vec{x}',\vec{y}'\right)} \\
=\left(W^{(\infty)}(\vec{x},\vec{y}) \left(W^{(\infty)}(\vec{x}',\vec{y}') \right)^{-1}\right)_{s,s'}
\end{split}\end{equation*}\normalsize
\end{proposition}
\textbf{Proof.} We begin by expanding the right hand side of proposition 3,
\small\begin{equation*}\begin{split}
\left(W^{(\infty)}(\vec{x},\vec{y}) \left(W^{(\infty)}(\vec{x}',\vec{y}') \right)^{-1}\right)_{s,s'} \\
= \left(\hat{W}^{(\infty)}(\vec{x},\vec{y})\exp\left\{ \sum^{n-m-1}_{l=1}(x_{l}-x'_{l})\Lambda^{l}_{[m,n-1]} \right\} \left(\hat{W}^{(\infty)}(\vec{x}',\vec{y}') \right)^{-1}\right)_{s,s'} 
\end{split}\end{equation*}\normalsize
where,
\small\begin{equation*}
\exp\left\{ \sum^{n-m-1}_{l=1}(x_{l}-x'_{l})\Lambda^{l}_{[m,n-1]} \right\} = \sum^{n-m-1}_{j=0}\zeta_j(\vec{x}-\vec{x}')\Lambda^{j}_{[m,n-1]}
\end{equation*}\normalsize
Recalling that $W^{(\infty)}(\vec{x},\vec{y})$ and $ \left(W^{(\infty)}(\vec{x'},\vec{y'}) \right)^{-1}$ are lower triangular, and\\ $\sum^{n-m-1}_{j=0}\zeta_j(\vec{x}-\vec{x'})\Lambda^{j}_{[m,n-1]}$ is upper triangular, we obtain,
\small\begin{equation*}\begin{split}
=& \left( \hat{w}^{(\infty)}_{i-j}(i,\vec{x},\vec{y}) \right)_{i=s\atop{j \le i}} \left( \zeta_{k-j}(\vec{x}-\vec{x}') \right)_{j \le i \atop{k \ge l}}  \left( \hat{w}^{*(\infty)}_{k-l}(l,\vec{x}',\vec{y}') \right)_{k \ge l \atop{l = s'}}\\
=& \sum^s_{j=m}\sum^{n-1}_{k=s'} \zeta_{k-j}(\vec{x}-\vec{x}') \hat{w}^{(\infty)}_{s-j}(s,\vec{x},\vec{y})\hat{w}^{*(\infty)}_{k-s'}(s',\vec{x}',\vec{y}')
\end{split}\end{equation*}\normalsize
Considering the left hand side of the proposition 3, we recall the following definitions,
\small\begin{equation*}\begin{split}
\exp \left\{ \sum^{n-m-1}_{l=1}(x_{l}-x'_{l}) \lambda^{l} \right\} &= \sum^{\infty}_{j=0} \zeta_{j}(\vec{x}-\vec{x}')\lambda^{j}\\
\frac{\tau\left(s,\vec{x}-\vec{\epsilon}\left(\frac{1}{\lambda}\right),\vec{y}\right)}{ \tau\left(s,\vec{x},\vec{y}\right)} &= \sum^{s-m}_{p=0}\frac{1}{\lambda^p}\hat{w}^{(\infty)}_{p}(s,\vec{x},\vec{y})\\
\frac{\tau\left(s'+1,\vec{x}'+\vec{\epsilon}\left(\frac{1}{\lambda}\right),\vec{y}'\right)}{\tau\left(s'+1,\vec{x}',\vec{y}'\right)}& = \sum^{n-s'-1}_{p=0}\frac{1}{\lambda^p}\hat{w}^{*(\infty)}_{p}(s',\vec{x}',\vec{y}')
\end{split}\end{equation*}\normalsize
Using these definitions the integral becomes,
\small\begin{equation*}\begin{split}
\sum^{\infty}_{\alpha=0} \sum^{s-m}_{\beta=0} \sum^{n-s'-1}_{\gamma=0} \oint \frac{d \lambda}{2 \pi i} \lambda^{\alpha-\beta-\gamma-s'+s-1}\zeta_{\alpha}(\vec{x}-\vec{x}') \hat{w}^{(\infty)}_{\beta}(s,\vec{x},\vec{y})\hat{w}^{*(\infty)}_{\gamma}(s',\vec{x}',\vec{y}')\\
= \sum^{s-m}_{\beta=0} \sum^{n-s'-1}_{\gamma=0} \zeta_{\beta+\gamma+s'-s}(\vec{x}-\vec{x}') \hat{w}^{(\infty)}_{\beta}(s,\vec{x},\vec{y})\hat{w}^{*(\infty)}_{\gamma}(s',\vec{x}',\vec{y}')
\end{split}\end{equation*}\normalsize
making the change of indices $\beta \rightarrow s-j$ and $\gamma \rightarrow k - s'$ we obtain exactly the right hand side of proposition 3. $\square$\\
\\
\noindent \textbf{Left hand side of bilinear relation.}
\begin{proposition}
\small\begin{equation*}\begin{split}
 \oint \frac{d \lambda}{2 \pi i} \lambda^{s'-s-1} \exp \left\{ \sum^{n-m-1}_{l=1}(y_{l}-y'_{l}) \lambda^{l} \right\} \frac{\tau\left(s+1,\vec{x},\vec{y}-\vec{\epsilon}\left(\frac{1}{\lambda}\right) \right)}{\tau(s,\vec{x},\vec{y})}\frac{\tau \left(s',\vec{x}',\vec{y}'+\vec{\epsilon}\left(\frac{1}{\lambda} \right)\right)}{\tau\left(s'+1,\vec{x}',\vec{y}'\right)}  \\
= \left( W^{(0)}(\vec{x},\vec{y}) \left(W^{(0)}(\vec{x'},\vec{y'}) \right)^{-1}\right)_{s,s'}
\end{split}\end{equation*}\normalsize
\end{proposition}
\textbf{Proof.} Expanding the right hand side of proposition 4,
\small\begin{equation*}\begin{split}
\left( W^{(0)}(\vec{x},\vec{y}) \left(W^{(0)}(\vec{x'},\vec{y'}) \right)^{-1}\right)_{s,s'}\\
= \left(\hat{W}^{(0)}(\vec{x},\vec{y})\exp\left\{ \sum^{n-m-1}_{l=1}(y_{l}-y'_{l})\left(\Lambda^{T}_{[m,n-1]}\right)^{l} \right\} \left(\hat{W}^{(0)}(\vec{x}',\vec{y}') \right)^{-1}\right)_{s,s'} 
\end{split}\end{equation*}\normalsize
where,
\small\begin{equation*}
\exp\left\{ \sum^{n-m-1}_{l=1}(y_{l}-y'_{l})\left(\Lambda^{T}_{[m,n-1]}\right)^{l} \right\} = \sum^{n-m-1}_{j=0}\zeta_j(\vec{y}-\vec{y}')\left(\Lambda^{T}_{[m,n-1]}\right)^{j}
\end{equation*}\normalsize
Recalling that $W^{(0)}(\vec{x},\vec{y})$ and $ \left(W^{(0)}(\vec{x'},\vec{y'}) \right)^{-1}$ are upper triangular, and\\ $\sum^{n-m-1}_{j=0}\zeta_j(\vec{y}-\vec{y}')\left(\Lambda^{T}_{[m,n-1]}\right)^{j}$ is lower triangular, we obtain,
\small\begin{equation*}\begin{split}
=& \left( \hat{w}^{(0)}_{j-i}(i,\vec{x},\vec{y}) \right)_{i=s\atop{j \ge i}} \left( \zeta_{j-k}(\vec{y}-\vec{y}') \right)_{j \ge i \atop{k \le l}}  \left( \hat{w}^{*(0)}_{l-k}(l,\vec{x}',\vec{y}') \right)_{k \le l \atop{l = s'}}\\
=& \sum^{n-1}_{j=s}\sum^{s'}_{k=m} \zeta_{j-k}(\vec{y}-\vec{y}') \hat{w}^{(0)}_{j-s}(s,\vec{x},\vec{y})\hat{w}^{*(0)}_{s'-k}(s',\vec{x}',\vec{y}')
\end{split}\end{equation*}\normalsize
Moving on to the left hand side of proposition 4, we recall the following definitions,
\small\begin{equation*}\begin{split}
\exp \left\{ \sum^{n-m-1}_{l=1}(y_{l}-y'_{l}) \lambda^{l} \right\} &= \sum^{\infty}_{j=0} \zeta_{j}(\vec{y}-\vec{y}')\lambda^{j}\\
\frac{\tau\left(s+1,\vec{x},\vec{y}-\vec{\epsilon}\left(\frac{1}{\lambda}\right)\right)}{\tau(s,\vec{x},\vec{y})} &= \sum^{n-s-1}_{k=0} \frac{1}{\lambda^k} \hat{w}^{(0)}_{k}(s,\vec{x},\vec{y})\\
\frac{\tau\left(s',\vec{x}',\vec{y}'+\vec{\epsilon}\left(\frac{1}{\lambda}\right)\right)}{\tau(s'+1,\vec{x}',\vec{y}')} &= \sum^{s'-m}_{k=0} \frac{1}{\lambda^k} \hat{w}^{*(0)}_{k}(s',\vec{x}',\vec{y}') 
\end{split}\end{equation*}\normalsize
Using these definitions the integral becomes,
\small\begin{equation*}\begin{split}
\sum^{\infty}_{\nu=0} \sum^{n-s-1}_{\alpha=0} \sum^{s'-m}_{\beta=0} \oint \frac{d \lambda}{2 \pi i} \lambda^{s'-s-1+\nu-\alpha-\beta}\zeta_{\nu}(\vec{y}-\vec{y}') \hat{w}^{(0)}_{\alpha}(s,\vec{x},\vec{y}) \hat{w}^{*(0)}_{\beta}(s',\vec{x}',\vec{y}')\\
= \sum^{n-s-1}_{\alpha=0} \sum^{s'-m}_{\beta=0}\zeta_{\alpha+\beta-s'+s}(\vec{y}-\vec{y}') \hat{w}^{(0)}_{\alpha}(s,\vec{x},\vec{y}) \hat{w}^{*(0)}_{\beta}(s',\vec{x}',\vec{y}')
\end{split}\end{equation*}\normalsize
Making the change of indices, $\alpha \rightarrow j-s$ and $\beta \rightarrow s'-k$, we obtain exactly the right hand side of proposition 4. $\square$.\\
\\
\textbf{Proof of bilinear identity.} Now we know that the right hand sides of both propositions 3 and 4 are equal. Simply let $s' \rightarrow s'-1$ and we obtain exactly the bilinear relation. $\square$\\
\\
\textbf{Specializing to the mKP and KP hierarchies.}
Specializing to the case $\vec{y} = \vec{y}'$ and $s \ge s'$, we notice that the right hand side of the bilinear relation contains no poles, and hence it reduces to the $(s-s')$th-modified KP (mKP) hierarchy,
\small\begin{equation*}
\oint \frac{d \lambda}{2 \pi i} \lambda^{s-s'} \exp \left\{ \sum^{n-m-1}_{l=1}(x_{l}-x'_{l}) \lambda^{l} \right\} \frac{\tau\left(s,\vec{x}-\vec{\epsilon}\left(\frac{1}{\lambda}\right),\vec{y}\right)}{ \tau\left(s,\vec{x},\vec{y}\right)}\frac{\tau\left(s',\vec{x}'+\vec{\epsilon}\left(\frac{1}{\lambda}\right),\vec{y}\right)}{\tau\left(s',\vec{x}',\vec{y}\right)} = 0
\end{equation*}\normalsize
Specializing again to let $s=s'$, we obtain the KP hierarchy,
\small\begin{equation*}
\oint \frac{d \lambda}{2 \pi i} \exp \left\{ \sum^{n-m-1}_{l=1}(x_{l}-x'_{l}) \lambda^{l} \right\} \frac{\tau\left(s,\vec{x}-\vec{\epsilon}\left(\frac{1}{\lambda}\right),\vec{y}\right)}{ \tau\left(s,\vec{x},\vec{y}\right)}\frac{\tau\left(s,\vec{x}'+\vec{\epsilon}\left(\frac{1}{\lambda}\right),\vec{y}\right)}{\tau\left(s,\vec{x}',\vec{y}\right)} = 0
\end{equation*}\normalsize
\subsection{Extracting non linear partial differential equations from the bilinear relation}
\noindent \textbf{Hirota's bilinear operator.} A necessary definition to proceed in this section is Hirota's bilinear differential operator, $D$, whose operation is defined on the product of two functions, $f(x)$ and $g(x)$. The generating function for $D$ is defined as,
\small\begin{equation}\begin{split}
f(x+y)g(x-y) &=\left( \exp\left\{ y \partial_x \right\} f(x) \right) \exp\left\{ - y \partial_x \right\} g(x)\\
&= \sum^{\infty}_{j=0} \frac{y^j}{j!} D^j_x \left( f(x) . g(x) \right)\\
&= \exp \left\{y D_x \right\} \left( f(x) . g(x) \right),
\label{Hirotaop}\end{split}\end{equation}\normalsize
where the first few explicit examples are, 
\small\begin{equation*}\begin{split}
D_x \left\{ f(x) . g(x) \right\} &=\left\{ \partial_x f(x) \right\} g(x) - f(x)\left\{ \partial_x g(x) \right\}  \\
D^2_x \left\{ f(x) . g(x) \right\} &=\left\{ \partial^2_x f(x) \right\} g(x) -2 \left\{ \partial_x f(x) \right\}\left\{ \partial_x g(x) \right\} + f(x)\left\{ \partial^2_x g(x) \right\}
\end{split}\end{equation*}\normalsize\\
\textbf{Obtaining the non linear PDE's.} The bilinear relation (eq. \ref{biline}) is a deceptively elegant expression which contains an infinite amount of non linear partial differential equations, all of which the $\tau$-function is a solution. To obtain these equations, we make the following change of variables,
\small\begin{equation}\begin{array}{lclcl}
x_i \rightarrow x_i - a_i &,&y_i \rightarrow y_i - b_i &,& i \in \{1,\dots,n-m-1\} \\
x'_i \rightarrow x_i + a_i &,&y'_i \rightarrow y_i + b_i
\end{array}\label{biline2}\end{equation}\normalsize
where the variables $\{a_1,\dots, a_{n-m-1} \}$ and $\{b_1, \dots,b_{n-m-1} \}$ are indeterminants which serve as expansion parameters. Hence eq. \ref{biline} becomes,
\small\begin{equation}\begin{split}
 \oint \frac{d \lambda}{2 \pi i} \lambda^{s'-s-2} e^{\left\{-2 \sum^{n-m-1}_{l=1}b_l \lambda^{l} \right\}} \tau\left(s+1,\vec{x}-\vec{a},\vec{y}-\vec{b}-\vec{\epsilon}\left(\frac{1}{\lambda}\right) \right)\\
\times \tau \left(s'-1,\vec{x}+\vec{a},\vec{y}+\vec{b}+\vec{\epsilon}\left(\frac{1}{\lambda} \right)\right)\\
= \oint \frac{d \lambda}{2 \pi i} \lambda^{s-s'} e^{\left\{-2 \sum^{n-m-1}_{l=1}a_l \lambda^{l} \right\}} \tau\left(s,\vec{x}-\vec{a}-\vec{\epsilon}\left(\frac{1}{\lambda}\right),\vec{y}-\vec{b}\right) \\
\times \tau\left(s',\vec{x}+\vec{a}+\vec{\epsilon}\left(\frac{1}{\lambda}\right),\vec{y} + \vec{b}\right)
\label{rarah}\end{split}\end{equation}\normalsize
Using the definitions given in eq. \ref{Hirotaop}, the bilinears in the $\tau$-functions of the above expression can naturally be re-expressed in terms of Hirota operators. Beginning with the left hand side of eq. \ref{rarah},
\small\begin{equation*}\begin{split}
& \tau\left(s+1,\vec{x}-\vec{a},\vec{y}-\vec{b}-\vec{\epsilon}\left(\frac{1}{\lambda}\right) \right) \tau \left(s'-1,\vec{x}+\vec{a},\vec{y}+\vec{b}+\vec{\epsilon}\left(\frac{1}{\lambda} \right)\right)\\
=& \exp\left\{ \sum^{n-m-1}_{j=1}a_j D_{x_j} \right\}  \tau \left(s'-1,\vec{x},\vec{y}+\vec{b}+\vec{\epsilon}\left(\frac{1}{\lambda} \right)\right) \tau\left(s+1,\vec{x},\vec{y}-\vec{b}-\vec{\epsilon}\left(\frac{1}{\lambda}\right) \right)\\
= &\exp\left\{ \sum^{n-m-1}_{j=1}a_j D_{x_j} \right\}\exp\left\{ \sum^{n-m-1}_{j=1}\left(b_j+\frac{1}{j \lambda^j} \right) D_{y_j} \right\}   \tau \left(s'-1,\vec{x},\vec{y}\right) \tau\left(s+1,\vec{x},\vec{y} \right)\\
=&\sum^{\infty}_{k=0} \frac{1}{\lambda^k}\zeta_k \left( \tilde{D}_{\vec{y}} \right) \exp\left\{ \sum^{n-m-1}_{j=1}\left( a_j D_{x_j} +b_j D_{y_j} \right) \right\}  \tau \left(s'-1,\vec{x},\vec{y}\right) \tau\left(s+1,\vec{x},\vec{y} \right)
\end{split}\end{equation*}\normalsize
where
\small\begin{equation*}
\tilde{D}_{\vec{y}} = \left( D_{y_1}, \frac{1}{2}D_{y_2}, \dots, \frac{1}{n-m-1} D_{y_{n-m-1}} \right)
\end{equation*}\normalsize
Similarly with the right hand side of eq. \ref{rarah},
\small\begin{equation*}\begin{split}
& \tau\left(s,\vec{x}-\vec{a}-\vec{\epsilon}\left(\frac{1}{\lambda}\right),\vec{y}-\vec{b}\right)  \tau\left(s',\vec{x}+\vec{a}+\vec{\epsilon}\left(\frac{1}{\lambda}\right),\vec{y} + \vec{b}\right)\\
=&\sum^{\infty}_{k=0} \frac{1}{\lambda^k}\zeta_k \left( \tilde{D}_{\vec{x}} \right) \exp\left\{ \sum^{n-m-1}_{j=1}\left( a_j D_{x_j} +b_j D_{y_j} \right) \right\}  \tau \left(s',\vec{x},\vec{y}\right) \tau\left(s,\vec{x},\vec{y} \right)
\end{split}\end{equation*}\normalsize
where
\small\begin{equation*}
\tilde{D}_{\vec{x}} = \left( D_{x_1}, \frac{1}{2}D_{x_2}, \dots, \frac{1}{n-m-1} D_{x_{n-m-1}} \right)
\end{equation*}\normalsize
We additionally apply the following exponential expansions,
\small\begin{equation*}
\exp\left\{-2 \sum^{n-m-1}_{l=1}c_l \lambda^{l} \right\} = \sum^{\infty}_{j=0} \lambda^j \zeta_j\left( \left\{ -2c \right\} \right)
\end{equation*}\normalsize
for $c_j = \{ a_j,b_j\}$. In performing the above expansions, obtaining the coefficient of the first order pole in eq. \ref{rarah} is elementary, and thus the bilinear relation becomes,
\small\begin{equation}\begin{split}
\sum^{\infty}_{k=0}  \zeta_{k-t} \left( \left\{ -2b \right\} \right)\zeta_k \left( \tilde{D}_{\vec{y}} \right) \exp\left\{ \sum^{n-m-1}_{j=1}\left( a_j D_{x_j} +b_j D_{y_j} \right) \right\}  \tau \left(t+s,\vec{x},\vec{y}\right) \tau\left(s+1,\vec{x},\vec{y} \right)  \\
= \sum^{\infty}_{k=0} \zeta_{k+t} \left( \left\{ -2a \right\} \right) \zeta_k \left( \tilde{D}_{\vec{x}} \right) \exp\left\{ \sum^{n-m-1}_{j=1}\left( a_j D_{x_j} +b_j D_{y_j} \right) \right\}  \tau \left(t+s+1,\vec{x},\vec{y}\right) \tau\left(s,\vec{x},\vec{y} \right)
\label{rarah2}\end{split}\end{equation}\normalsize
where we have assigned $t = s'-s-1$ for convenience.\\
\\
Expanding the above expression as a polynomial in the variables $\{a\}$ and $\{b\}$, the coefficients of the monomials are the desired non linear PDE's. As an example, consider expanding eq. \ref{rarah2} with the specification $t=-1$. From the coefficient of $b_1$ we obtain the following differential equation,
\small\begin{equation}
D_{x_1}D_{y_1} \tau \left(s,\vec{x},\vec{y}\right) \tau\left(s,\vec{x},\vec{y} \right) + 2 \tau \left(s+1,\vec{x},\vec{y}\right) \tau\left(s-1,\vec{x},\vec{y} \right)=0
\label{todaeq}\end{equation}\normalsize
which is the 2-Toda molecule equation. This equation is used extensively in chapter 4. The 2-Toda molecule equation is related to the lattice equation (eq. \ref{todarubbish}) by the following scale transformation, $\tau(s,\vec{x},\vec{y}) \rightarrow e^{x_1 y_1}\tau(s,\vec{x},\vec{y})$. In bilinear form we obtain,
\small\begin{equation}
D_{x_1}D_{y_1} \tau \left(s,\vec{x},\vec{y}\right) \tau\left(s,\vec{x},\vec{y} \right) + 2 \tau \left(s+1,\vec{x},\vec{y}\right) \tau\left(s-1,\vec{x},\vec{y} \right)= 2 \tau^2\left(s,\vec{x},\vec{y} \right) 
\label{todaeq1}\end{equation}\normalsize
\subsection{Polynomial expressions of the tau-function}
We shall now use eq. \ref{17} to express the $\tau$-function in a more palatable form. Rewriting the following exponentials as,
\small\begin{equation*}\begin{split}
\exp\left\{ \sum^{n-m-1}_{l=1}x_{l}\Lambda^{l}_{[m,n-1]} \right\} &= \sum^{n-m-1}_{j=0}\zeta_j (\vec{x}) \Lambda^{j}_{[m,n-1]}= \left( \zeta_{j-i}(\vec{x}) \right)^{n-1}_{i,j=m}\\
\exp\left\{- \sum^{n-m-1}_{l=1}y_{l}\left(\Lambda^{T}_{[m,n-1]}\right)^{l} \right\} &= \sum^{n-m-1}_{j=0}\zeta_j (-\vec{y}) \left(\Lambda^{T}_{[m,n-1]}\right)^j= \left( \zeta_{i-j}(-\vec{y}) \right)^{n-1}_{i,j=m}
\end{split}\end{equation*}\normalsize
The $\tau$-function now becomes,
\small\begin{equation*}\begin{split}
&\textrm{det}\left[ \sum^{n-1}_{j=m} \zeta_{j-i}(\vec{x}) \left(\sum^{n-1}_{k=m} a_{jk} \zeta_{k-l}(-\vec{y}) \right)\right]^{s-1}_{i,l=m}\\
=& \sum_{m \le j_m < \dots < j_{s-1} \le n-1} \textrm{det} \left[ \zeta_{j_q -i}(\vec{x})\right]^{n-1}_{i,q=m}\textrm{det} \left[  \left(\sum^{n-1}_{k=m} a_{j_i ,k} \zeta_{k -l}(-\vec{y}) \right)\right]^{n-1}_{i,l=m}\\
=& \sum_{m \le j_m < \dots < j_{s-1} \le n-1\atop{m \le k_m < \dots < k_{s-1} \le n-1}} A_{\{j\}\{k\}}  \textrm{det} \left[ \zeta_{j_i -q}(\vec{x})\right]^{s-1}_{i,q=m} \textrm{det} \left[ \zeta_{k_i-q}(-\vec{y}) \right]^{s-1}_{i,q=m}
\end{split}\end{equation*}\normalsize
where we have taken the transpose of the matrix of the second determinant and used the following label,
\small\begin{equation*}
A_{\{j\}\{k\}} = \textrm{det} \left[   a_{j_i ,k_l} \right]^{s-1}_{i,l=m}
\end{equation*}\normalsize
\textbf{Further massaging of the $\tau$-function, character polynomials.} Letting the double summation run from $1, \dots, n-m$, rather than from $m, \dots, n-1$, an immediate simplification of the $\tau$-function is given by,
\small\begin{equation*}\begin{split}
\tau\left(s,\vec{x},\vec{y}\right) &= \sum_{1 \le j_1 < \dots < j_{s-m} \le n-m\atop{1 \le k_1 < \dots < k_{s-m} \le n-m}} A_{\{j'\}\{k'\}}  \textrm{det} \left[ \zeta_{j_i -q}(\vec{x})\right]^{s-m}_{i,q=1} \textrm{det} \left[ \zeta_{k_i-q}(-\vec{y}) \right]^{s-m}_{i,q=1}
\end{split}\end{equation*}\normalsize
where,
\small\begin{equation*}
A_{\{j'\}\{k'\}} = \textrm{det} \left[   a_{j_i+m-1 ,k_l+m-1} \right]^{s-m}_{i,l=1}
\end{equation*}\normalsize
Massaging this expression further, we begin by making the following change in indices,
\small\begin{equation*}
j_{i_1} \rightarrow \lambda_{i_1} + i_1, k_{i_2} \rightarrow \mu_{i_2} + i_2 \textrm{  ,  } \{ i_1,\ i_2\} \in \{1,\dots,  s-m\}
\end{equation*}\normalsize
which transforms the $\tau$-function to the form,
\small\begin{equation*}
\sum_{0 \le \lambda_1 \le \dots \le \lambda_{s-m} \le n-s\atop{0 \le  \mu_1 \le \dots \le  \mu_{s-m} \le n-s}}A_{\{\lambda\}\{\mu\}} \textrm{det} \left[ \zeta_{ \lambda_i+i -q}(\vec{x})\right]^{s-m}_{i,q=1} \textrm{det} \left[ \zeta_{ \mu_i+i -q}(-\vec{y}) \right]^{s-m}_{i,q=1}
\end{equation*}\normalsize
where,
\small\begin{equation*}
 A_{\{ \lambda\}\{ \mu\}} = \textrm{det} \left[   a_{ \lambda_i+i +m-1 , \mu_l+l+ m-1} \right]^{s-m}_{i,l=1} 
\end{equation*}\normalsize
The following operations only apply to the two $\zeta$ determinants. Reversing the ordering of the rows for both matrices,
\small\begin{equation*}\begin{split}
\sum_{0 \le  \lambda_1 \le \dots \le  \lambda_{s-m} \le n-s\atop{0 \le  \mu_1 \le \dots \le  \mu_{s-m} \le n-s}}A_{\{ \lambda\}\{ \mu\}} \textrm{det} \left[ \zeta_{ \lambda_{i}+(s-m+1)-j-i}(\vec{x})\right]^{1}_{i,j=s-m}\\
\times \textrm{det} \left[ \zeta_{ \mu_{i}+(s-m+1)-j-i}(-\vec{y}) \right]^{1}_{i,j=s-m}        
\end{split}\end{equation*}\normalsize
and reversing the ordering of the columns,
\small\begin{equation*}\begin{split}
 \sum_{0 \le  \lambda_1 \le \dots \le  \lambda_{s-m} \le n-s\atop{0 \le  \mu_1 \le \dots \le  \mu_{s-m} \le n-s}}A_{\{ \lambda\}\{ \mu\}} \textrm{det} \left[ \zeta_{ \lambda_{s-m+1-i}+j-i}(\vec{x})\right]_{i,j=1}^{s-m}\textrm{det} \left[ \zeta_{ \mu_{s-m+1-i}+j-i}(-\vec{y})\right]_{i,j=1}^{s-m}
\end{split}\end{equation*}\normalsize
finally, reversing the order of labelling for $ \lambda_i/\mu_j$,
\small\begin{equation*}
 \lambda_i \rightarrow  \lambda_{s-m+1-i} \textrm{  ,  } \mu_j \rightarrow  \mu_{s-m+1-j}
\end{equation*}\normalsize
 we obtain,
\small\begin{equation}\begin{split}
\tau\left(s,\vec{x},\vec{y}\right) &=   \sum_{0 \le  \lambda_{s-m} \le \dots \le  \lambda_{1} \le n-s\atop{0 \le  \mu_{s-m} \le \dots \le  \mu_{1} \le n-s}}A_{\{ \lambda\}\{ \mu\}} \textrm{det} \left[ \zeta_{ \lambda_{i}+j-i}(\vec{x})\right]_{i,j=1}^{s-m}\textrm{det} \left[\zeta_{ \mu_{i}+j-i}(-\vec{y})\right]_{i,j=1}^{s-m} \\
&= \sum_{\{ \lambda\}\{ \mu\} \subseteq (n-s)^{(s-m)}}A_{\{ \lambda\}\{ \mu\}} \chi_{\{ \lambda\}}(\vec{x}) \chi_{\{ \mu\}}(-\vec{y}) \label{H.99}
\end{split}\end{equation}\normalsize
where $\{ \lambda\}$ and $\{ \mu\}$ are partitions contained within the box of dimensions $(n-s)^{(s-m)}$, and $\chi_{\{ \lambda\}}(\vec{x})$ is the character polynomial given by,
\small\begin{equation}
\chi_{\{ \lambda\}}(\vec{x}) = \textrm{det} \left[ \zeta_{ \lambda_{i}+j-i}(\vec{x})\right]_{i,j=1}^{s-m}
\end{equation}\normalsize
\subsection{The restricted tau-function}
\textbf{Restricting the time variables.} In the details above we have assumed that all the times variables are algebraically independent of each other. It is possible to make a restriction on the times variables so that they lose their independence, but the $\tau$-function becomes an element of the symmetric polynomial ring, $\field{C}\{[u_1,\dots,u_{s-m}]^{S_{s-m}},[v_1,\dots,v_{s-m}]^{S_{s-m}}\}$. It is through this process that we are able to match the $\tau$-function of the hierarchy to expressions of extreme interest in statistical mechanics, which is the main topic of the next two chapters.\\
\\
Thus throughout the remainder of this thesis we shall use the following convention,
\begin{itemize}
\item{$\tau(\vec{x},\vec{y})$ denotes that the time variables are algebraically independent, and $\tau(\vec{x},\vec{y})$ is an element of the (non symmetric) polynomial ring\\ $\field{C}[x_1,\dots,x_{n-m-1}, y_1,\dots,y_{n-m-1}]$. We shall refer to the $\tau$-function in this form as unrestricted.}
\item{$\tau(\vec{u},\vec{v})$ denotes that the time variables are not algebraically independent, and $\tau(\vec{u},\vec{v})$ is an element of the (symmetric) polynomial ring\\ $\field{C}\{[u_1,\dots,u_{s-m}]^{S_{s-m}},[v_1,\dots,v_{s-m}]^{S_{s-m}}\}$. We shall refer to the $\tau$-function in this form as restricted.}
\end{itemize}
\textbf{Miwa transformations and Schur polynomials, creating the restricted $\tau$-function.} Performing the following Miwa change of variables from Toda time parameters to symmetric power sums\footnote{We define the symmetric power sums, $p_i(\vec{u})$, and the complete homogeneous symmetric polynomials, $h_i(\vec{u})$, in section 1.6.},
\small\begin{equation*}
x_{k} = \frac{1}{k}p_k(u_1,\dots,u_{s-m}) \textrm{  ,  } - y_{k} = \frac{1}{k} p_k(v_1,\dots,v_{s-m}) \textrm{  ,  }  k \in \{1,\dots, n-m-1\}
\end{equation*}\normalsize
the one row character polynomials become complete homogeneous symmetric polynomials\footnote{Arguably the best method to see this equivalence is through the generating functions of both polynomials. For the complete story see eq. \ref{Mi} and \ref{charac}.},
\begin{eqnarray*}
\zeta_i(\vec{x}) \rightarrow h_i (u_1,\dots,u_{s-m}) &,&\zeta_i(-\vec{y}) \rightarrow h_i (v_1,\dots,v_{s-m}) 
\end{eqnarray*}
Hence the character polynomials in the $\tau$-function expression become Schur polynomials,
\small\begin{equation*}\begin{split}
\tau\left(s,\vec{u},\vec{v}\right) &=   \sum_{0 \le {\lambda}_{s-m} \le \dots \le \lambda_{1} \le n-s\atop{0 \le \mu_{s-m} \le \dots \le \mu_{1} \le n-s}}A_{\{\lambda\}\{\mu\}} \textrm{det} \left[ h_{\lambda_{i}+j-i}(\vec{u})\right]_{i,j=1}^{s-m}\textrm{det} \left[h_{\mu_{i}+j-i}(\vec{v})\right]_{i,j=1}^{s-m}\\
&= \sum_{\{\lambda\}\{\mu\} \subseteq (n-s)^{(s-m)}}A_{\{\lambda\}\{\mu\}} S_{\{\lambda\}}(\vec{u}) S_{\{\mu\}}(\vec{v})
\end{split}\end{equation*}\normalsize
\textbf{A further simplification.} Making the constant matrix $A$ equal to the $(n -m)\times (n-m)$ identity, we immediately obtain the simplified $\tau$-function,
\small\begin{equation*}\begin{split}
\tau\left(s,\vec{x},\vec{y}\right) &=   \sum_{\{ \lambda\}\{ \mu\} \subseteq (n-s)^{(s-m)}}\delta_{\{ \lambda\}\{ \mu\}} \chi_{\{ \lambda\}}(\vec{x}) \chi_{\{ \mu\}}(-\vec{y})\\ 
&= \sum_{\{ \lambda\} \subseteq (n-s)^{(s-m)}} \chi_{\{ \lambda\}}(\vec{x}) \chi_{\{ \lambda\}}(-\vec{y})\\
\tau\left(s,\vec{u},\vec{v}\right) &= \sum_{\{ \lambda\} \subseteq (n-s)^{(s-m)}} S_{\{ \lambda\}}(\vec{u}) S_{\{ \lambda\}}(\vec{v})
\end{split}\end{equation*}\normalsize
The above expression shall be used extensively in the next chapter.
\section{Generating additional symmetric polynomials}
As a small extension of the above results, we present a method of generating additional $\tau$-function expressions by introducing a simple scale transformation of the time variables\footnote{This section is similar to the work \cite{Taka2}, except we consider a scale transformation of the time variables as opposed to a translation.}.
\subsection{A scale transformation of the time variables}
The scale transformations are introduced into the 2-Toda hierarchy by simply multiplying each of the $n-m-1$ time variables $x_{k}/y_{k}$, $k \in \{1,\dots, n-m-1\}$, by a general function $f^{(x/y)}_{k}(t)$,
\small\begin{equation*}
x_{k} \rightarrow f^{(x)}_{k}(t) x_{k} \textrm{  ,  } y_{k} \rightarrow f^{(y)}_{k}(t) y_{k}
\end{equation*}\normalsize
We shall now explicitly show that under these transformations the hierarchy is still well defined.\\
\\
\textbf{The wave-matrices.} Given the constant matrix $A  = (a_{i,j})_{i,j = m, \dots,n-1}\in GL(n-m)$, where $\textrm{det}\left[ a_{ij} \right]_{i,j=m\dots,s-1} \ne 0, m < s \le n$, the wave-matrices $W^{(\infty)}(\vec{x},\vec{y};t)$ and $W^{(0)}(\vec{x},\vec{y};t)$ are defined by the equation,
\small\begin{equation}
W^{(0)}(\vec{x},\vec{y};t)=W^{(\infty)}(\vec{x},\vec{y};t) A
\label{H.1}\end{equation}\normalsize
where $W^{(\infty)}(\vec{x},\vec{y};t)$ and $W^{(0)}(\vec{x},\vec{y};t)$ have the specific form,
\small\begin{equation*} 
\begin{split}
W^{(\infty)}(\vec{x},\vec{y};t) = \hat{W}^{(\infty)}(\vec{x},\vec{y};t) \exp \left[\sum^{n-m-1}_{k = 1}f^{(x)}_{k}(t) x_{k} \Lambda^{k}_{[m,n)}\right] \\
W^{(0)}(\vec{x},\vec{y};t) = \hat{W}^{(0)}(\vec{x},\vec{y};t) \exp \left[\sum^{n-m-1}_{k = 1} f^{(y)}_{k}(t) y_{k} (\Lambda^T_{[m,n)})^{k}\right] 
\end{split}
\end{equation*}\normalsize
where $\hat{W}^{(\infty/0)}(\vec{x},\vec{y};t)$ are lower/upper diagonal respectively, and the diagonal entries of hatted wave-matrices are given by eq. \ref{ogle}.\\
\\
Using the results from propositions 1 and 2, the remaining entries of the hatted wave-matrices, and their inverses, are given by eq. \ref{wavemat1} and \ref{wavemat2}, with the only difference being $a_{ij}(\vec{x},\vec{y}) \rightarrow a_{ij}(\vec{x},\vec{y};t)$ where,
\small\begin{equation*}\begin{split}
\left( a_{ij}(\vec{x},\vec{y};t) \right)^{n-1}_{i,j=m} =&  \exp \left[\sum^{n-m-1}_{k = 1}f^{(x)}_{k}(t) x_{k} \Lambda^{k}_{[m,n-1]}\right] A\\
&\times  \exp \left[-\sum^{n-m-1}_{k = 1}f^{(y)}_{k}(t) y_{k} (\Lambda^T_{[m,n-1]})^{k}\right]
\end{split}\end{equation*}\normalsize
\textbf{The corresponding linear problem.} The various matrix equations (linear, Lax, Zakharov-Shabat) that define this hierarchy are mostly the same except for the inclusion of various factors of $f^{(x)}_{k}(t)$ and $f^{(y)}_{k}(t)$.\\
\\
If we consider the matrices,
\small\begin{equation*}\begin{split}
L = W^{(\infty)}(\vec{x},\vec{y};t) \Lambda_{[m,n-1]} \left(  W^{(\infty)}(\vec{x},\vec{y};t) \right)^{-1} \textrm{  ,  }B_{j} = \left\{ L^{j} \right\}_+ \\
M = W^{(0)}(\vec{x},\vec{y};t)\Lambda^T_{[m,n-1]} \left(  W^{(0)}(\vec{x},\vec{y};t) \right)^{-1} \textrm{  ,  } C_{j} = \left\{ M^{j} \right\}_-
\end{split}\end{equation*}\normalsize
then using the workings/results from lemma 1 we have the following linearization,
\small\begin{equation*}\begin{split}
\partial_{x_{j}} \hat{W}^{(\infty)}(\vec{x},\vec{y};t) &=f^{(x)}_{j}(t) \left\{ B_{j} \hat{W}^{(\infty)}(\vec{x},\vec{y};t) -  \hat{W}^{(\infty)}(\vec{x},\vec{y};t) \Lambda^{j}_{[m,n-1]}\right\}\\
\partial_{y_{j}} \hat{W}^{(0)}(\vec{x},\vec{y};t) &= f^{(y)}_{j}(t) \left\{ C_{j} \hat{W}^{(0)}(\vec{x},\vec{y};t) -   \hat{W}^{(0)}(\vec{x},\vec{y};t) \left(\Lambda^{T}_{[m,n-1]}\right)^{j}\right\}\\
\partial_{x_{j}} \hat{W}^{(0)}(\vec{x},\vec{y};t) &=f^{(x)}_{j}(t) B_{j} \hat{W}^{(0)}(\vec{x},\vec{y};t) \\
\partial_{y_{j}} \hat{W}^{(\infty)}(\vec{x},\vec{y};t)& =f^{(y)}_{j}(t) C_{j} \hat{W}^{(\infty)}(\vec{x},\vec{y};t)
\end{split}\end{equation*}\normalsize
Lax type system,
\small\begin{equation*}\begin{split}
\partial_{x_{j}} L =f^{(x)}_{j}(t) \left[ B_{j},L \right] \textrm{  ,  } \partial_{x_{j}} M = f^{(x)}_{j}(t) \left[ B_{j},M \right]\\
\partial_{y_{j}} L = f^{(y)}_{j}(t) \left[ C_{j},L \right] \textrm{  ,  } \partial_{y_{j}} M = f^{(y)}_{j}(t) \left[ C_{j},M \right]
\end{split}\end{equation*}\normalsize
and Zakharov-Shabat type system,
\small\begin{equation*}\begin{split}
f^{(x)}_{j}(t) \partial_{x_{k}} B_{j}-f^{(x)}_{k}(t)\partial_{x_{j}} B_{k} +f^{(x)}_{j}(t)f^{(x)}_{k}(t) \left[ B_{j},B_{k} \right]=0\\
f^{(y)}_{j}(t)\partial_{y_{k}} C_{j}-f^{(y)}_{k}(t) \partial_{y_{j}} C_{k} +f^{(y)}_{j}(t)f^{(y)}_{k}(t) \left[ C_{j},C_{k} \right]=0\\
f^{(x)}_{j}(t) \partial_{y_{k}} B_{j}-f^{(y)}_{k}(t)\partial_{x_{j}} C_{k} + f^{(x)}_{j}(t)f^{(y)}_{k}(t)\left[ B_{j},C_{k} \right]=0
\end{split}\end{equation*}\normalsize \\
\noindent \textbf{$\tau$-function of the hierarchy.}
\begin{proposition} For the function,
\small\begin{equation}
\tau(s,\vec{x},\vec{y};t) = \textrm{det}\left[a_{ij}(\vec{x},\vec{y};t) \right]^{s-1}_{i,j=m}
\label{H.12}\end{equation}\normalsize
the following 4 relations hold,
\small\begin{equation}\begin{split}
\hat{w}^{(\infty)}_{k}(s,\vec{x},\vec{y};t) = \frac{\zeta_k \left(-\tilde{\partial}^{f^{(x)}}_{\vec{x}}\right)\tau(s,\vec{x},\vec{y};t)}{\tau(s,\vec{x},\vec{y};t)}, & \textrm{ non zero for } k \in \{0, \dots, s-m\} \\
\hat{w}^{(0)}_{k}(s,\vec{x},\vec{y}) = \frac{\zeta_k \left(-\tilde{\partial}^{f^{(y)}}_{\vec{y}}\right)\tau(s+1,\vec{x},\vec{y};t)}{\tau(s,\vec{x},\vec{y};t)},& \textrm{ non zero for } k \in \{0, \dots, n-s-1\}\\
\hat{w}^{*(\infty)}_{k}(s,\vec{x},\vec{y}) = \frac{\zeta_k \left(\tilde{\partial}^{f^{(x)}}_{\vec{x}} \right)\tau(s+1,\vec{x},\vec{y};t)}{\tau(s+1,\vec{x},\vec{y};t)},& \textrm{ non zero for } k \in \{0, \dots, n-s-1\}\\
\hat{w}^{*(0)}_{k}(s,\vec{x},\vec{y}) = \frac{\zeta_k \left(\tilde{\partial}^{f^{(y)}}_{\vec{y}} \right)\tau(s,\vec{x},\vec{y};t)}{\tau(s+1,\vec{x},\vec{y};t)},& \textrm{ non zero for } k \in \{0, \dots, s-m\}
\end{split}\end{equation}\normalsize
where,
\small\begin{equation*}\begin{split}
\tilde{\partial}^{f^{(x)}}_{\vec{x}} &= \left(\frac{1}{f^{(x)}_{1}(t)}\partial_{x_1},  \frac{1}{2 f^{(x)}_{2}(t)}\partial_{x_2}, \frac{1}{3 f^{(x)}_{3}(t)}\partial_{x_3}, \dots \right)\\
\tilde{\partial}^{f^{(y)}}_{\vec{y}} &=  \left(\frac{1}{f^{(y)}_{1}(t)}\partial_{y_1},  \frac{1}{2 f^{(y)}_{2}(t)}\partial_{y_2}, \frac{1}{3 f^{(y)}_{3}(t)}\partial_{y_3}, \dots \right)
\end{split}\end{equation*}\normalsize
\end{proposition}
\textbf{Proof.} As in lemma 1, if the above four equations are true then we have the following, 
\small\begin{equation}\begin{split}
 \sum^{s-m}_{k=0} \lambda^k \hat{w}^{(\infty)}_{k}(s,\vec{x},\vec{y};t) = \frac{\tau(s,\vec{x}-\vec{\epsilon}^{f^{(x)}}(\lambda),\vec{y};t)}{\tau(s,\vec{x},\vec{y};t)}\\
\sum^{n-s-1}_{k=0} \lambda^k \hat{w}^{(0)}_{k}(s,\vec{x},\vec{y};t) = \frac{\tau(s+1,\vec{x},\vec{y}-\vec{\epsilon}^{f^{(y)}}(\lambda);t)}{\tau(s,\vec{x},\vec{y};t)}\\
\sum^{n-s-1}_{k=0} \lambda^k \hat{w}^{*(\infty)}_{k}(s,\vec{x},\vec{y};t) = \frac{\tau(s+1,\vec{x}+\vec{\epsilon}^{f^{(x)}}(\lambda),\vec{y};t)}{\tau(s+1,\vec{x},\vec{y};t)}\\
\sum^{s-m}_{k=0} \lambda^k \hat{w}^{*(0)}_{k}(s,\vec{x},\vec{y};t) = \frac{\tau(s,\vec{x},\vec{y}+\vec{\epsilon}^{f^{(y)}}(\lambda);t)}{\tau(s+1,\vec{x},\vec{y};t)}\label{H.20}
\end{split}\end{equation}\normalsize
where,
\small\begin{equation*}
\vec{\epsilon}^{f^{(x/y)}}(\lambda) = \left(\frac{\lambda}{f^{(x/y)}_{1}(t)}, \frac{\lambda^2 }{2 f^{(x/y)}_{2}(t)},\frac{\lambda^3 }{3 f^{(x/y)}_{3}(t)}, \dots \right)
\end{equation*}\normalsize
By using the methods in lemma 1 we obtain,
\small\begin{equation*}\begin{split}
\tau(s,\vec{x} \mp \vec{\epsilon}^{f^{(x)}}(\lambda),\vec{y};t) = \textrm{det}\left[ (1-\lambda \Lambda_{[m,n)})^{\pm 1} A(\vec{x},\vec{y};t) \right]^{s-1}_{i,j=m}\\
\tau(s,\vec{x} ,\vec{y}\mp \vec{\epsilon}^{f^{(y)}}(\lambda);t) = \textrm{det}\left[  A(\vec{x},\vec{y};t) (1-\lambda \Lambda^T_{[m,n)})^{\mp 1} \right]^{s-1}_{i,j=m}
\end{split}\end{equation*}\normalsize
which upon expanding in terms of $\lambda$ we obtain the required result. $\square$\\
\\
\textbf{Bilinear relation of the hierarchy.} Using the the results of lemma 3, the function $\tau(s,\vec{x},\vec{y};t)$ defined in eq. \ref{H.12} satisfies the following bilinear relationship,
\small\begin{equation}\begin{split}
\oint \frac{d \lambda}{2 \pi i} \lambda^{s-s'} \exp \left\{ \sum^{n-m-1}_{l=1} f^{(x)}_{l}(t) (x_{l}-x'_{l}) \lambda^{l} \right\} \frac{\tau\left(s,\vec{x}-\vec{\epsilon}^{f^{(x)}}\left( \frac{1}{\lambda}\right) ,\vec{y}\right)}{ \tau\left(s,\vec{x},\vec{y};t\right)}\\
\times \frac{\tau\left(s',\vec{x}'+\vec{\epsilon}^{f^{(x)}}\left( \frac{1}{\lambda}\right),\vec{y}';t\right)}{\tau\left(s',\vec{x}',\vec{y}';t\right)} \\
= \oint \frac{d \lambda}{2 \pi i} \lambda^{s'-s-2} \exp \left\{ \sum^{n-m-1}_{l=1} f^{(y)}_{l}(t)(y_{l}-y'_{l}) \lambda^{l} \right\} \frac{\tau\left(s+1,\vec{x},\vec{y}-\vec{\epsilon}^{f^{(y)}}\left( \frac{1}{\lambda} \right);t \right)}{\tau(s,\vec{x},\vec{y};t)} \\
\times \frac{\tau \left(s'-1,\vec{x}',\vec{y}'+\vec{\epsilon}^{f^{(y)}}\left( \frac{1}{\lambda} \right);t\right)}{\tau\left(s',\vec{x}',\vec{y}';t\right)}
\label{bigbilin}\end{split}\end{equation}\normalsize
for general $s,s',\vec{x},\vec{x}',\vec{y},\vec{y}',t$.\\
\\
\textbf{Polynomial form of the $\tau$-function.} Expanding the exponentials,
\small\begin{equation*}\begin{split}
\exp\left\{ \sum^{n-m-1}_{l=1} f^{(x)}_{l}(t) x_{l}\Lambda^{l}_{[m,n)} \right\} &= \sum^{n-m-1}_{j=0}\zeta_j \left(f^{(x)}_{1}(t) x_{1},  f^{(x)}_{2}(t) x_{2}, \dots \right) \Lambda^{j}_{[m,n)}\\
&= \sum^{n-m-1}_{j=0}\zeta_j \left( \left\{ f^{(x)}(t) x \right\}\right) \Lambda^{j}_{[m,n)}\\
&= \left[ \zeta_{j-i} \left( \left\{ f^{(x)}(t) x \right\} \right) \right]^{n-1}_{i,j=m}\\
\exp\left\{- \sum^{n-m-1}_{l=1}f^{(y)}_{l}(t) y_{l}\left(\Lambda^{T}_{[m,n)}\right)^{l} \right\} &= \sum^{n-m-1}_{j=0}\zeta_j \left(\left\{- f^{(y)}(t) y \right\}  \right) \left(\Lambda^{T}_{[m,n)}\right)^j\\
&= \left[ \zeta_{i-j}\left( \left\{- f^{(y)}(t) y \right\}  \right) \right]^{n-1}_{i,j=m}
\end{split}\end{equation*}\normalsize
we obtain the following expression for the $\tau$-function,
\small\begin{equation}\begin{split}
\tau\left(s,\vec{x},\vec{y};t\right) &=  \sum_{\{\lambda\}\{\mu\} \subseteq (n-s)^{(s-m)}}A_{\{\lambda\}\{\mu\}} \chi_{\{\lambda\}}\left(\left\{ f^{(x)}(t) x \right\}  \right) \chi_{\{\mu\}}\left( \left\{- f^{(y)}(t) y \right\}  \right)
\label{msog4}\end{split}\end{equation}\normalsize
where,
\small\begin{equation*}\begin{split}
A_{\{\lambda\}\{\mu\}} &= \textrm{det} \left[   a_{\lambda_{s-m+1-i}+i +m-1 ,\mu_{s-m+1-l}+l+ m-1} \right]^{s-m}_{i,l=1} \\
\chi_{\{\lambda\}}\left(\left\{ f^{(x)}(t) x \right\}  \right)  &= \textrm{det} \left[ \zeta_{\lambda_{i}+j-i}\left( f^{(x)}_1(t) x_1,  f^{(x)}_2(t) x_2, \dots   \right)\right]_{i,j=1}^{s-m}
\end{split}\end{equation*}\normalsize
Additionally, setting the constant matrix $A$ equal to the $(n -m)\times (n-m)$ identity, we immediately obtain the simplified $\tau$-function,
\small\begin{equation}
\tau\left(s,\vec{x},\vec{y};t\right) = \sum_{\{\lambda\} \subseteq (n-s)^{(s-m)}} \chi_{\{\lambda\}}\left( \left\{ f^{(x)}(t) x \right\} \right) \chi_{\{\lambda\}}\left( \left\{ -f^{(y)}(t) y \right\}  \right)
\label{msog3}\end{equation}\normalsize
\subsection{Symmetric polynomials with an additional parameter as (restricted) tau-functions of the hierarchy}
We now present some specific examples of the functions $f^{(x)}_{j}(t)$ and $f^{(y)}_{j}(t)$ which generate some interesting forms for the restricted $\tau$-function. For the examples below, we shall perform the usual Miwa transformations,
$$
x_{k} \rightarrow \frac{1}{k}p_k(u_1,\dots,u_{s-m}) \textrm{  ,  } - y_{k} \rightarrow \frac{1}{k} p_k(v_1,\dots,v_{s-m})
$$
and work in the ring of symmetric functions\\ $\field{C}\{[u_1,\dots,u_{s-m};t]^{S_{s-m}},[v_1,\dots,v_{s-m};t]^{S_{s-m}}\}$.\\
\\
\textbf{Specifying the scaling factors.} For this section we consider the specific values of $f^{(x/y)}_{j}(t)$,
\begin{eqnarray*}
f^{(x)}_{j}(t) = 1 - t^{j}&, & f^{(y)}_{j}(t) = 1
\end{eqnarray*}
and obtain the following form for the restricted $\tau$-function, 
\small\begin{equation}
\tau_s =  \sum_{\{\lambda\} \subseteq (n-s)^{ (s-m)}} \chi_{\{\lambda\}}\left((1-t)p_{1}(\vec{u}),\frac{1-t^2}{2}p_{2}(\vec{u}),\dots  \right) S_{\{\lambda\}}\left(\vec{v} \right)
\end{equation}\normalsize
The function, $\chi_{\{\lambda\}}\left(\left\{ \frac{1-t^{k}}{k}p_{k}(\vec{u}) \right\} \right)$, has a nice form given in terms of the $t$-deformed complete symmetric functions\footnote{These symmetric functions officially have no name, at least none offered in \cite{MacD}. For additional details on symmetric polynomials refer to section 1.6.}, which are labeled $q_{j}(\vec{u};t)$. Hence the restricted $\tau$-function obtains the form, 
\small\begin{equation*}
\tau_s = \sum_{\{\lambda\} \subseteq (n-s)^{(s-m)}} S_{\{\lambda\}}\left( u_1,\dots,u_{s-m};t \right) S_{\{\lambda\}}\left(v_1,\dots,v_{s-m} \right)
\end{equation*}\normalsize
where the function $S_{\{\lambda\}}\left( \vec{u};t \right)$ is given as the $t$-deformed equivalent of the Schur polynomial,
\small\begin{equation*}
S_{\{\lambda\}}\left(u_1, \dots,u_N;t \right) = \textrm{det} \left[ q_{\lambda_i +j-i}(\vec{u};t) \right]^{N}_{i,j=1}
\end{equation*}\normalsize
\textbf{Hall-Littlewood polynomials in the $n \rightarrow \infty$ limit.} We now consider the limit $n \rightarrow \infty$, (the infinite lattice with a free end), to obtain,
\small\begin{equation}\begin{split}
\tau\left(s,\vec{u},\vec{v};t\right) &= \sum_{\{\lambda\} \subseteq (\infty)^{(s-m)}}S_{\{\lambda\}}\left( u_1,\dots,u_{s-m};t \right) S_{\{\lambda\}}\left(v_1,\dots,v_{s-m} \right)\\
&= \prod^{s-m}_{i,j=1} \frac{1-t u_i v_j}{1 - u_i v_j}\\
&= \sum_{\{\lambda\} \subseteq (\infty)^{(s-m)}}P_{\{\lambda\}}\left( u_1,\dots,u_{s-m};t \right) Q_{\{\lambda\}}\left(v_1,\dots,v_{s-m};t \right)
\label{mcog2}\end{split}\end{equation}\normalsize
where $P_{\{\lambda\}}\left( \vec{u};t \right)$ $(Q_{\{\lambda\}}\left( \vec{u};t \right))$ is the celebrated \textbf{Hall-Littlewood} polynomial and the second line in the above expression can be found in eq. 3.4.7 of \cite{MacD}.
\section{Symmetric polynomials}
This section acts as a reference/appendix for the remainder of the thesis, as many results of the remaining chapters assume intimate knowledge of the definitions/results presented in this section\footnote{The results of this section can be found in chapters I and III of \cite{MacD}.}.\\
\\
We begin by stating some very general definitions about polynomials.\\
\\ 
\textbf{Commutative ring.} A commutative ring $\langle R,+,.\rangle$ is a set $R$ with the 2 binary operations addition $(+)$ and multiplication $(.)$ defined on $R$, such that $\langle R,+\rangle$ forms a commutative group.\\
\\
\textbf{Polynomial ring.} The set of all polynomials in $\{u_1, \dots, u_n\}$ with coefficients in a field $k$ (for the remainder of this thesis all fields are $\mathbb{C}$ to avoid confusion) is denoted by $\field{C}[u_1, \dots, u_n]$. $\field{C}[u_1, \dots, u_n]$ forms a commutative ring, which we call a polynomial ring. A subset of this ring is the symmetric polynomial ring which consists of all the polynomials which stay invariant under the action of the symmetric group $S_n$ permuting the variables $\{u_1,\dots,u_n\}$. We label this ring as $\field{C}[u_1, \dots, u_n]^{S_n}$.\\
\\
\textbf{Algebraic independence.} Given a finite set of polynomials, $F=$\\ $\{f_1(\vec{u}),\dots,f_N(\vec{u})\}$, in a finite set of variables, $\{u_1,\dots,u_M\}$, and every non zero polynomial, $\Omega(f_1(\vec{u}),\dots,f_N(\vec{u}))$, constructed entirely from elements of the finite set of polynomials $F$ with coefficients in $\field{C}$, the set $F$ is called algebraically independent if we have:
$$
\Omega(f_1(\vec{u}),\dots,f_N(\vec{u})) \ne 0
$$
for general polynomial $\Omega$.\\
\\
We now concern ourselves with the specific symmetric polynomials that are used within this thesis.\\
\\
\textbf{Symmetric power sums, $\mathbf{p_r(\vec{u})}$.} Where,
\small\begin{equation}
p_r (u_1,\dots,u_N) = \sum^{N}_{i=1}\alpha^{r}_i \textrm{  ,  } 1 \le r < \infty \label{power}
\end{equation}\normalsize
and whose generating function, $P(z;\vec{u})$, is given by,
\small\begin{equation}
P(z;\vec{u})= \sum^{N}_{k=1}\frac{u_k}{1-u_k z} = \sum^{\infty}_{j=0}p_{j+1}(\vec{u}) z^j \label{powergen}
\end{equation}\normalsize
\textbf{Elementary symmetric polynomials, $\mathbf{e_r(\vec{u})}$.} Where,
\small\begin{equation}
e_r(u_1,\dots,u_N) = \sum_{1\le j_1 < \dots < j_r \le N} u_{j_1} \dots u_{j_r} \textrm{  ,  } 0 \le r \le N \label{elementary}
\end{equation}\normalsize
and whose generating function $E(z;\vec{u})$ is given by,
\small\begin{equation}
E(z;\vec{u})= \prod^{N}_{k=1}(1+ u_k z) = \sum^{N}_{j=0}e_{j}(\vec{u}) z^j= \exp\left\{- \sum^{\infty}_{k=1} \frac{(-z)^k}{k} p_k(\vec{u}) \right\} \label{elementarygen}
\end{equation}\normalsize
\textbf{Complete homogeneous symmetric polynomials, $\mathbf{h_r(\vec{u})}$.} Where,
\small\begin{equation}
h_r(u_1,\dots,u_N) = \sum_{1\le j_1 \le \dots \le j_r \le N} u_{j_1} \dots u_{j_r} \textrm{  ,   } 0 \le r < \infty \label{complete}
\end{equation}\normalsize
and whose generating function $H(z;\vec{u})$ is given by,
\small\begin{equation}
H(z;\vec{u})= \prod^{N}_{k=1}\frac{1}{1- u_k z} = \sum^{\infty}_{j=0}h_{j}(\vec{u}) z^j= \exp\left\{ \sum^{\infty}_{k=1} \frac{z^k}{k} p_k(\vec{u}) \right\} \label{completegen}
\end{equation}\normalsize
\textbf{Miwa's change of variables and one row character polynomials.} Performing the following change of variables on the power sums,
\small\begin{equation*}
x_{k} \rightarrow \frac{1}{k} \sum^{N}_{l=1} u^{k}_{l} \textrm{  ,  } k = \{ 1,2, \dots \}
\end{equation*}\normalsize
the generating function of the complete symmetric polynomials becomes that of the one row character polynomials,
\small\begin{equation*}
 \exp\left\{ \sum^{\infty}_{k=1} \frac{z^k}{k} p_k(\vec{u}) \right\}  \rightarrow  \exp\left\{ \sum^{\infty}_{k=1} z^k x_k \right\}
\end{equation*}\normalsize
Thus,
\small\begin{equation*}
h_r(\vec{u})  \rightarrow  \zeta_r(\vec{x}) \label{Mi}
\end{equation*}\normalsize
\textbf{Newton's identities and implications.} It is well known\footnote{Section I.2 of \cite{MacD}.} that when there are only finitely many variables, $\vec{u}=\{u_1,\dots,u_N\}$, the elementary symmetric polynomials, $\{e_1(\vec{u}),\dots,e_N (\vec{u})\}$, are algebraically independent, and form a complete basis for the ring of symmetric polynomials $\field{C}[u_1,\dots,u_N]^{S_N}$.\\
\\
If we consider again the generating function of the symmetric power sums and express it as a log derivative,
\small\begin{equation*}
P(-z;\vec{u}) = \sum^{N}_{k=1}\frac{d}{dz}\log\left(1+u_k z\right)= \frac{d}{dz}\log\prod^{N}_{k=1} \left(1+u_k z \right)
\end{equation*}\normalsize
we obtain a relationship between the generating functions of the symmetric power sums and the elementary symmetric polynomials (a similar relationship also exists between the symmetric power sums and the completely homogeneous symmetric polynomials),
\small\begin{equation*}\begin{split}
P(-z;\vec{u}) = \frac{d}{dz}\log\left(\prod^{N}_{k=1}(1+u_k z)\right) = \frac{d}{dz}\log\left(E(z;\vec{u})\right)\\
\Rightarrow \frac{d}{dz}E(z;\vec{u}) = P(-z;\vec{u}) E(z;\vec{u})
\end{split}\end{equation*}\normalsize
Collecting powers of $z$, we obtain $N$ relationships between symmetric power sums and the elementary symmetric polynomials called Newton's identities,
\small\begin{equation*}
r e_r (\vec{u}) = \sum^{r-1}_{k=0}(-1)^{r-1-k} p_{r-k}(\vec{u}) e_k (\vec{u}) \textrm{  ,  } 1 \le r \le N
\end{equation*}\normalsize
From the above formula it is clear that $p_r \in \field{C}[e_1,\dots,e_r]$ and $e_r \in \field{C}[p_1,\dots,p_r]$. We give the first few examples of both,
\small\begin{equation*}
\begin{array}{clcl}
p_1 = & e_1 & e_1 = & p_1\\
p_2 = & e^2_1 - 2 e_2 & e_2 = & \frac{1}{2}\left(p^2_1 - p_2\right)\\
p_3 = & e^3_1 - 3 e_1 e_2 + 3 e_3 & e_3 = & \frac{1}{6}\left( p^3_1 -3p_1 p_2 + 2 p_3 \right)
\end{array}
\end{equation*}\normalsize
Thus,
\small\begin{equation*}
\field{C}[e_1,\dots,e_N]=\mathbb{C}[p_1,\dots,p_N]
\end{equation*}\normalsize
meaning that the power sum symmetric polynomials, $\{p_1,\dots,p_N\}$, are algebraically independent and form a complete basis for the ring of symmetric polynomials $\field{C}[u_1,\dots,u_N]^{S_N}$. A similar such argument will also show that $\{h_1,\dots,h_N\}$ are algebraically independent and form a complete basis for the ring of symmetric polynomials $\field{C}[u_1,\dots,u_N]^{S_N}$. This leads us to the fundamental theorem of symmetric functions. 
\begin{theorem}{\textbf{The fundamental theorem of symmetric polynomials.}} Assume there exist two polynomial rings $\field{C}[u_1,\dots,u_N]^{S_N}$ and $\field{C}[x_1,\dots,x_N]$, both rings containing necessarily the same number of finite variables. Then there exists an isomorphism between the two polynomial rings, $\field{C}[u_1,\dots,u_N]^{S_N}\cong$\\$ \field{C}[x_1,\dots,x_N]$, with the isomorphism sending $x_j \rightarrow e_j(u_1,\dots,u_N)$, $j \in$\\$ \{1,\dots,N\}$, and vice versa. By the results of Newton's identities, we can also map $x_j \rightarrow p_j(u_1,\dots,u_N)$ or  $x_j \rightarrow h_j(u_1,\dots,u_N)$, $j \in \{1,\dots,N\}$, and the isomorphism still holds.
\end{theorem}
\noindent\textbf{Schur polynomials, $S_{\{\lambda \}}( \vec{u})$.} For a general partition, $\{\lambda \}$, the Schur polynomial, $S_{\{\lambda \}}(u_1,\dots,u_N)$, is given by the following combinatorial definition,
\small\begin{equation}
S_{\{\lambda \}}(u_1,\dots,u_N) = \sum_{T^{\{\lambda\}}_+}u^{t_1}_1 \dots u^{t_N}_N =  \sum_{T^{\{\lambda\}}_-}u^{t_1}_1 \dots u^{t_N}_N \label{Schur3}
\end{equation}\normalsize
where $T^{\{\lambda \}}_{\pm}$ denotes summation over all ascending or descending semi-standard Young tableaux of shape $\{\lambda\}$ respectively. Alternative determinant definitions (which are far more useful for computations) are given by,
\small\begin{equation}\begin{split}
S_{\{\lambda\}}\left(u_1,\dots,u_N \right)& = \frac{\textrm{det}\left[ u^{N-i+\lambda_i}_j \right]^N_{i,j=1}}{\prod_{1\le i<j\le N}(u_i-u_j)}  \\
& =  \textrm{det}\left[ h_{\lambda_i +j-i} (u_1,\dots,u_N) \right]^N_{i,j=1} \\
& =  \textrm{det}\left[ e_{\lambda^{'}_i +j-i} (u_1,\dots,u_N) \right]^N_{i,j=1} \label{Schur1}
\end{split}\end{equation}\normalsize
where the the partition $\{\lambda' \}$ is the conjugate of the partition $\{ \lambda\}$. It is a well known fact that Schur polynomials of $N$ variables provide a complete basis for $\field{C}[u_1, \dots, u_N]^{S_N}$. \\
\\
\textbf{Character polynomials.} From definition \ref{Schur1}, it is elementary to see that under the Miwa change of variables the Schur polynomial becomes the character polynomial,
\small\begin{equation}\begin{split}
\textrm{det}\left[ h_{\lambda_i +j-i} (u_1,\dots,u_N) \right]^N_{i,j=1} & \rightarrow  \textrm{det}\left[ \zeta_{\lambda_i +j-i} (\vec{x}) \right]^N_{i,j=1} \\
&= \chi_{\{ \lambda\}}(\vec{x}) \label{charac}
\end{split}\end{equation}\normalsize
\textbf{Pieri's formula.} Given the partition $\{\lambda \}$, we have the following formula,
\small\begin{equation}
h_{r} (\vec{u}) S_{\{\lambda\}}(\vec{u}) = \sum_{\{\mu\} \supseteq \{ \lambda \}\atop{\{ \mu-\lambda\} \in \field{H}_r }} S_{\{\mu\}}(\vec{u}) \label{pieri}
\end{equation}\normalsize
where $\{ \mu-\lambda\}$ is a skew diagram and $\field{H}_r $ is the set of all horizontal strips of length $r$.\\
\\
\textbf{$t$-deformed symmetric polynomials.} Let us now add an additional parameter, $t \in \field{C}$, to the usual ring of symmetric polynomials, $\field{C}[u_1, \dots,u_N]^{S_N}$. Notice that $t$ is not on the same footing as $\{u_1, \dots, u_N\}$. We label this ring as,
\small\begin{equation*}
\field{C}[u_1, \dots,u_N;t]^{S_N}
\end{equation*}\normalsize
The necessary definitions for the symmetric polynomials with an extra parameter are given below.\\
\\
\textbf{$t$-deformed complete symmetric polynomials, $\mathbf{q(\vec{u};t)}$.} Where,
\small\begin{equation*}\begin{split}
q_0(u_1,\dots,u_N;t) &= 1\\
q_r(u_1,\dots,u_N;t) &= \sum^{N}_{i=1} u^r_{i} \prod_{1 \le j < k \le N} \frac{u_j-t u_k}{u_j-u_k} \textrm{  ,  }1 \le r < \infty
\end{split}\end{equation*}\normalsize
and whose generating function, $Q(z;\vec{u};t)= \frac{H(z; \vec{u})}{H(t z; \vec{u})}$, is given by,
\small\begin{equation}\begin{split}
Q(z;u_1,\dots,u_N;t) = \sum^{\infty}_{j=0} z^j q_j(\vec{u};t) = \exp \left\{\sum^{\infty}_{k = 1} z^{k} \frac{1-t^{k}}{k} p_k(\vec{u}) \right\} \\
=\sum^{\infty}_{j=0} z^j \zeta_j\left((1-t) p_{1}(\vec{u}), \frac{1-t^{2}}{2}p_{2}(\vec{u}),\dots \right) \label{t-complete}
\end{split}\end{equation}\normalsize
\textbf{$t$-deformed Schur polynomials, $\mathbf{S_{\{\lambda \}}( \vec{u};t)}$.} Given the partition $\{ \lambda \}$, we have the $t$-deformed equivalent of the Schur polynomial,
\small\begin{equation}
S_{\{\lambda\}}\left(u_1, \dots,u_N;t \right) = \textrm{det} \left[ q_{\lambda_i +j-i}(\vec{u};t) \right]^{N}_{i,j=1}
\end{equation}\normalsize
\textbf{Hall-Littlewood polynomials, $\mathbf{P_{\{\lambda\}}\left( \vec{u};t \right)}$.} Given the partition $\{ \lambda \}$, the Hall-Littlewood polynomials are defined as,
\small\begin{equation}\begin{split}
P_{\{\lambda\}}\left(\vec{u} ;t \right) &= \frac{1}{v_{\{\lambda\}}(t)} \sum_{\sigma \in S_N} u^{\lambda_1}_{\sigma_1} \dots u^{\lambda_N}_{\sigma_N}  \prod_{1 \le i < j \le N} \frac{u_{\sigma_i} - t u_{\sigma_j}}{u_{\sigma_i} - u_{\sigma_j}} \\
&= \frac{1}{b_{\{\lambda\}}(t)} Q_{\{\lambda\}}\left( \vec{u};t \right)\label{H-Litt}
\end{split}\end{equation}\normalsize
where,
\small\begin{equation*}\begin{split}
v_{\{\lambda\}}(t)   &= \prod_{j \ge 0} \prod^{m_j}_{k=1}\frac{1-t^k}{1-t} \\
b_{\{\lambda\}}(t)   &= v_{\{\lambda\}}(t) \prod_{j \ge 0} \prod^{m_j}_{k=1} (1-t)^{m_j}
\end{split}\end{equation*}\normalsize
for $m_i$ the number of $\lambda_j$ equal to $i$, $i \ge 0$.
%%%%%%%%%%%%%%%%%%%%%%%%%%%%%%%%%%%%%%%%%%%%%%%%%%%%%%%%%%%%%%%%%%%%%%%%%%%%%%%%%%%%%%%%%%%%%%%%%%%%%%%%%%%%%%%%%%%%
\newpage
%%%%%%%%%%%%%%%%%%%%%%%%%%%%%%%%%%%%%%%%%%%%%%%%%%%%%%%%%%%%%%%%%%%%%%%%%%%%%%%%%%%%%%%%%%%%%%%%%%%%%%%%%%%%%%%%%%%%
\chapter{Applications of the 2-Toda hierarchy}
In this chapter we explore the correspondence between the classical 2-Toda hierarchy and the quantum phase model. Sections 2.1.1-2.1.5 form a detailed introduction to the model, the algebraic methods used to construct the scalar product and some additional necessary combinatorial aspects of the model. In section 2.1.6 we detail the aforementioned classical-quantum correspondence and explore some technical details regarding the implications of considering a family of $\tau$-functions with a family of scalar products. In section 2.2 we consider the physical interpretation of the Toda wave-functions with respect to the phase model and reveal a novel method of analyzing certain classes of correlation functions for the model. Section 2.3 is an observation of the correspondence between the 2-Toda hierarchy and the scalar product of the Hall-Littlewood vertex operators.
\section{The phase model}
\subsection{The q-boson algebra}
\noindent In the following we use the notation/results found in \cite{phase1,phase2,phase3,phase4}.\\
\\
\textbf{Operators, commutation relations and Fock space.} The introduction of the phase model customarily begins with the $q$-boson algebra, which is defined by three independent operators $B$, $B^{\dagger}$ and $N$ that satisfy the following commutation relations,
\small\begin{equation*}
[B,B^{\dagger}]=q^{2N} \textrm{  ,  } [N,B]=-B \textrm{  ,  } [N,B^{\dagger}]=B^{\dagger}
\end{equation*}\normalsize
where $q \in \mathbb{C}$. The one dimensional Fock space, $\mathbb{F}$, of the $q$-boson algebra is formed from the state $|n\rangle$, where the label $n \in \mathbb{Z}_+ \cup \{ 0\}$ is called an occupation number. The action of $B^{\dagger}$ and $B$ on elements of the Fock space are given by,
\small\begin{equation*}
B^{\dagger}|n\rangle = [(n+1)_q]^{\frac{1}{2}}|n+1\rangle \textrm{  ,  }B|n\rangle = [n_q]^{\frac{1}{2}}|n-1\rangle 
\end{equation*}\normalsize
where,
\small\begin{equation*}
[n_q]^{\frac{1}{2}} = \frac{1-q^{2n}}{1-q^2}
\end{equation*}\normalsize
The action of operator, $N$, on the Fock space is,
\small\begin{equation*}
N |n\rangle = n|n\rangle
\end{equation*}\normalsize
\textbf{The $\mathbf{q \rightarrow 0}$ limit.} The phase model is constructed from the $q \rightarrow 0$ limit of the $q$-boson algebra. In this limit, $B$ and $B^{\dagger}$ are labeled $\phi$ and $\phi^{\dagger}$ respectively, with $N$ remaining unchanged. The equivalent commutation relations become,
\small\begin{equation*}
[\phi,\phi^{\dagger}]= \pi \textrm{  ,  } [N,\phi]=-\phi \textrm{  ,  } [N,\phi^{\dagger}]=\phi^{\dagger}
\end{equation*}\normalsize
where $\pi = |0\rangle \langle0|$ is the vacuum projector. The Fock states $|n\rangle$ can be constructed from the vacuum state $|0\rangle$, or any such state $|m\rangle$, with repeated operation by the $\phi^{\dagger}$ operator,
\small\begin{equation*}
\left(\phi^{\dagger} \right)^{n-m} |m\rangle = |n\rangle  \textrm{  for  } m < n
\end{equation*}\normalsize
The action of the $\phi$ operator on the vacuum state annihilates it,
\small\begin{equation*}
\phi  |0\rangle = 0
\end{equation*}\normalsize
Note that there is no highest state vector in the Fock space. It is simple to verify that $\phi$ and $\phi^{\dagger}$ can be constructed entirely in terms of Fock states,
\small\begin{equation*}
\phi = \sum^{\infty}_{n=0}|n\rangle \langle n+1| \textrm{  ,  } \phi^{\dagger} = \sum^{\infty}_{n=0}|n+1\rangle \langle n|
\end{equation*}\normalsize
\textbf{M+1 dimensions.} We extend this bosonic algebra and consider the tensor product,
\small\begin{equation*}
\mathbb{F} = \mathbb{F}_0 \otimes \mathbb{F}_1 \otimes \dots \otimes \mathbb{F}_M
\end{equation*}\normalsize
which consists of $M+1$ copies of the one dimensional Fock space. With this extended Fock space we associate $3(M+1)$ independent operators, $\phi_j$, $\phi^{\dagger}_j$ and $N_j$, $0 \le j \le M$, where each operator of index $j$ acts on its respective space,
\small\begin{equation*}
\phi_j = I_0 \otimes I_1 \otimes \dots I_{j-1} \otimes \phi \otimes I_{j+1} \otimes \dots \otimes I_M 
\end{equation*}\normalsize
and similarly for $\phi^{\dagger}_j$ and $N_j$, where $I_j$ is the identity operator in $\mathbb{F}_j$. The corresponding commutation relations are given by,
\small\begin{equation}
[\phi_j,\phi^{\dagger}_k]= \pi_j \delta_{jk} \textrm{  ,  } [N_j,\phi_k]=-\phi_j \delta_{jk} \textrm{  ,  } [N_j,\phi^{\dagger}_k]=\phi^{\dagger}_j \delta_{jk}
\end{equation}\normalsize
Each operator $\phi_j, \phi^{\dagger}_j$ and $N_j$ of index $j$ acts on the corresponding indexed Fock vectors,
\small\begin{equation}\begin{array}{lcl}
\left(\phi_j \right)^{m_j-n_j} |m_j \rangle_j &=& |n_j \rangle_j \textrm{  for  }0 \le n_j < m_j \\
\left(\phi^{\dagger}_j \right)^{n_j-m_j} |m_j \rangle_j &=& |n_j \rangle_j \textrm{  for  } n_j > m_j \ge 0\\
N_j |m_j \rangle_j &=& m_j |m_j \rangle_j
\end{array}\end{equation}\normalsize
where the operator $\phi_j$ annihilates the vacuum state $|0\rangle_j$. The state vectors, $|n_p \rangle_j$, and the corresponding conjugate vectors, $ \langle n_r|_k$, are orthonormal,
\small\begin{equation}
\langle n_r|n_p\rangle_{k,j}= \delta_{pr}\delta_{jk}
\end{equation}\normalsize
We consider the total state vector $|n\rangle$ being made up of the tensor product of the $M+1$ indexed Fock vectors,
\small\begin{equation}
|n\rangle = \bigotimes^{M}_{j=0} |n_j\rangle_j = \prod^M_{k=0} \left(\phi^{\dagger}_k \right)^{n_k} \bigotimes^{M}_{j=0} |0\rangle_j \textrm{  where  } \sum^M_{j=0}n_j = n
\end{equation}\normalsize
The sum, $\sum^M_{j=0}n_j = n$, can be expressed in $p(n,M+1)$ different ways, where $p(n,M+1)$ is the number of possible partitions of a natural number $n$, with the provision that the length of the partition is never greater than $M+1$. For $n \le M+1$, the $p(n,M+1)$ can simply be expressed as $p(n)$, where the length of the partition is unrestricted. $p(n)$ can be calculated by the MacMahon generating function\cite{GEA},
\small\begin{equation*}
\prod^{\infty}_{j=1} \left( \frac{1}{1-q^j} \right) = \sum^{\infty}_{n=0}p(n) q^n
\end{equation*}\normalsize
\subsection{Algebraic Bethe ansatz}
\noindent We define the phase model through the following $L$-operator matrix,
\small\begin{equation}
L_j (u) \equiv \left( 
\begin{array}{cc}
\hat{a}_j (u) & \hat{b}_j (u)\\
\hat{c}_j (u) & \hat{d}_j (u)
\end{array}
 \right)= \left( 
\begin{array}{cc}
\frac{1}{u} & \phi^{\dagger}_j \\
\phi_j & u 
\end{array}
 \right)
\end{equation}\normalsize
where $u \in \mathbb{C}$. Naturally associated with $L_j(u)$ is the $4 \times 4$ matrix $R(v,w)$, $v,w \in \mathbb{C}$, given by,
\small\begin{equation}
R(u,v) = \left( 
\begin{array}{cccc}
f(u,v) &0 & 0 & 0\\
0& g(u,v) & 1 & 0 \\
0 & 0 & g(u,v) & 0\\
0 & 0 & 0 & f(u,v)
\end{array}
 \right)
\end{equation}\normalsize
where $f(u,v) = \frac{u^2}{u^2-v^2}$ and $g(u,v) = \frac{u v}{u^2-v^2}$.\\
\\
\textbf{Intertwining relation and Yang-Baxter equation.} $L$ and $R$ satisfy the following intertwining relation,
\small\begin{equation}
R(u,v)[L_j(u) \otimes L_j(v)] = [L_j(v) \otimes L_j(u)]R(u,v)
\label{inte}\end{equation}\normalsize
where $\otimes$ is the usual tensor product of matrices. The $R$-matrix satisfies the Yang-Baxter equation given by,
\small\begin{equation}\begin{split}
\{ I \otimes R(u,v) \} \{R(u,w)\otimes I \} \{ I \otimes R(v,w) \}\\
= \{ R(v,w) \otimes I \} \{ I \otimes R(u,w)  \}  \{ R(u,v) \otimes I \}
\end{split}\end{equation}\normalsize
where $I$ is the $2 \times 2$ identity matrix.\\
\\
\textbf{The monodromy matrix.} The monodromy matrix, $T(u)$, for the phase model is introduced as the ordered product of all $(M+1)$ $L$-matrices,
\small\begin{equation}
T(u) = L_M(u) L_{M-1}(u) \dots L_0(u) = \left( 
\begin{array}{cc}
A (u) & B (u)\\
C (u) & D (u)
\end{array}
 \right)
\end{equation}\normalsize
As an example, for $M=2$ we have,
\small\begin{equation}
T(u) = \left( 
\begin{array}{cc}
\frac{1}{u^3} + \frac{1}{u}\phi^{\dagger}_1 \phi_0 + \frac{1}{u}\phi^{\dagger}_2 \phi_1 + u \phi^{\dagger}_2 \phi_0 & \frac{1}{u^2} \phi^{\dagger}_0 + \phi^{\dagger}_1 + \phi^{\dagger}_2 \phi_1 \phi^{\dagger}_0+  u^2 \phi^{\dagger}_2\\
\frac{1}{u^2} \phi_2+ \phi_1  + \phi_2 \phi^{\dagger}_1 \phi_0 + u^2 \phi_0 & \frac{1}{u}\phi_2 \phi^{\dagger}_0  + u\phi_2 \phi^{\dagger}_1 + u \phi_1 \phi^{\dagger}_0 + u^3
\end{array}
 \right)
\label{stuufff}\end{equation}\normalsize
Using induction on the intertwining relation (eq. \ref{inte}) the monodromy matrix and the $R$-matrix satisfy an equivalent intertwining relationship,
\small\begin{equation}
R(u,v)[T(u)\otimes T(v)] = [T(v) \otimes T(u)] R(u,v)
\end{equation}\normalsize
which generate sixteen non trivial algebraic relationships. A selection of these expressions include,
\small\begin{equation*}\begin{array}{lcl}
C(u) B(v) & = & g(u,v) \{ A(u) D(v) - A(v) D(u)\}\\
C(u) A(v) & = & f(v,u) A(v) C(u) + g(u,v) A(u) C(v)\\
D(u) B(v) & = & f(v,u) B(v) D(u) + g(u,v) B(u) D(v)\\
\left[ B(u),B(v) \right] &=& \left[C(u),C(v)\right] = 0
\end{array}\end{equation*}\normalsize
\textbf{Creation and annihilation operators.} In order to view $B(u)$ and $C(u)$ as creation and annihilation operators of the phase model respectively, we apply the appropriate commutation relations to the expression, \\$\exp\{\eta N_j\} L_j(u) \exp\left\{ \frac{1}{2} \eta \sigma_z \right\}$, $\eta \in \mathbb{C}$, to receive the following useful identity,
\small\begin{equation}\begin{split}
\exp\{\eta N_j\} L_j(u) \exp\left\{ \frac{1}{2} \eta \sigma_z \right\} &= \left( 
\begin{array}{cc}
\frac{1}{u} \exp \left\{ \eta \left( N_j + \frac{1}{2} \right) \right\} & \exp\left\{ \eta \left( N_j - \frac{1}{2} \right)  \right\}\phi^{\dagger}_j  \\
\exp \left\{ \eta \left( N_j + \frac{1}{2} \right)\right\} \phi_j  & u \exp \left\{ \eta \left( N_j - \frac{1}{2} \right)\right\}
\end{array}
 \right)\\
&= \left( 
\begin{array}{cc}
\frac{1}{u} \exp \left\{ \eta \left( N_j + \frac{1}{2} \right) \right\} & \phi^{\dagger}_j \exp\left\{ \eta \left( N_j + \frac{1}{2} \right)  \right\}  \\
 \phi_j \exp \left\{ \eta \left( N_j - \frac{1}{2} \right)\right\}  & u \exp \left\{ \eta \left( N_j - \frac{1}{2} \right)\right\}
\end{array}
 \right) \\
 \Rightarrow \exp\{\eta N_j\} L_j(u) \exp\left\{ \frac{1}{2} \eta \sigma_z \right\} &= \exp\left\{ \frac{1}{2} \eta \sigma_z \right\} L_j(u) \exp\{\eta N_j\} 
\label{pong1}\end{split}\end{equation}\normalsize
We can extend the following identity to include all the $(M+1)$ vector states. Considering the expression, $\exp\{\eta \hat{N}\} T(u) \exp\left\{ \frac{1}{2} \eta \sigma_z \right\}$, where $\hat{N} = \sum^M_{j=0}N_j$, which measures the total occupation number of the state. Applying eq. \ref{pong1} for individual vector states, we receive a more general identity for all vector states,
\small\begin{equation*}\begin{split}
 &  \exp\{\eta N_M\} L_M(u) \dots \exp\{\eta N_0\} L_0(u) \exp\left\{ \frac{1}{2} \eta \sigma_z \right\}\\
=& \exp\{\eta N_M\} L_M(u) \dots \exp\{\eta N_1\} L_1(u)\exp\left\{ \frac{1}{2} \eta \sigma_z \right\}L_0(u)  \exp\{\eta N_0\} \\
= &\exp\{\eta N_M\} L_M(u) \dots \exp\{\eta N_2\} L_2(u)\exp\left\{ \frac{1}{2} \eta \sigma_z \right\}L_1(u)  \exp\{\eta N_1\} L_0(u)  \exp\{\eta N_0\} \\
\end{split}\end{equation*}\normalsize
the process shown above continues until the $\exp\left\{ \frac{1}{2} \eta \sigma_z \right\}$ term moves to the far left of the expression, and thus we receive the identity,
\small\begin{equation*}
\exp\{\eta \hat{N}\} T(u) \exp\left\{ \frac{1}{2} \eta \sigma_z \right\} = \exp\left\{ \frac{1}{2} \eta \sigma_z \right\} T(u) \exp\{\eta \hat{N}\} 
\end{equation*}\normalsize
If we concentrate on the $B(u)$ (top right hand corner) entry of this identity,
\small\begin{equation*}
\exp\left\{\eta\left( \hat{N} -\frac{1}{2} \right) \right\} B(u) = B(u)\exp\left\{\eta\left( \hat{N} +\frac{1}{2}\right) \right\}
\end{equation*}\normalsize
taking the limit, $\eta \rightarrow 0$, we obtain,
\small\begin{equation}
\hat{N} B(u) = B(u) \left\{ \hat{N}+1\right\}
\end{equation}\normalsize
Performing a similar operation on the $C(u)$ (bottom left hand corner) entry of the matrix identity we obtain,
\small\begin{equation}
\hat{N} C(u) = C(u) \left\{ \hat{N}-1 \right\}
\end{equation}\normalsize
Thus the operator $B(u)$ is a creation operator of the phase model, where one application on a state vector increases the total occupation number by one, while $C(u)$ is the opposing annihilation operator of the phase model, where one application to a state vector decreases the total occupation number by one. We note that $C(u)$ annihilates the total vacuum operator,
\small\begin{equation*}
C(u) |0\rangle = C(u) \bigotimes^{M}_{j=0} |0\rangle_j = 0
\end{equation*}\normalsize
Equivalently, the roles of $B(u)$ and $C(u)$ are reversed when applied to the conjugated vacuum vectors, where $B(u)$ now acts as the annihilation operator and annihilates the conjugate vacuum,
\small\begin{equation*}
\langle 0| B(u) = \bigotimes^{M}_{j=0} \langle 0|_j B(u) = 0
\end{equation*}\normalsize
We also notice that the operators $A(u)$ and $D(u)$ have the following identities,
\small\begin{equation}
[A(u),\hat{N}] = [D(u),\hat{N}] = 0
\end{equation}\normalsize
thus, the vacuum vector is the eigenvector of operators $A(u)$ and $D(u)$,
\small\begin{equation}
A(u)|0\rangle = a_M(u)|0\rangle \textrm{  ,  } D(u)|0\rangle = d_M(u)|0\rangle
\end{equation}\normalsize
where $a_M(u) = \frac{1}{u^{M+1}}$ and $d_M(u) = u^{M+1}$. \\
\\
\textbf{$N$-particle state vector of the phase model.} We construct the $N$-particle vector of the phase model, $|\Psi_M(u_1,\dots,u_N)\rangle$, from the vacuum vector by the following operation,
\small\begin{equation*}
|\Psi_M (u_1,\dots,u_N)\rangle = B(u_1) \dots B(u_N)|0\rangle
\end{equation*}\normalsize
where the total occupation number of $|\Psi_M (u_1,\dots,u_N)\rangle$ is $N$,
\small\begin{equation*}
\hat{N}|\Psi_M (u_1,\dots,u_N)\rangle = N |\Psi_M (u_1,\dots,u_N)\rangle
\end{equation*}\normalsize
The above expression implies that a suitable alternative form for the $N$-particle vector is given by the following,
\small\begin{equation}
|\Psi_M (u_1,\dots,u_N)\rangle = \sum_{0 \le n_0,n_1,\dots,n_M \le N\atop{n_0 + n_1 +\dots + n_M = N}}f_{\{n_0,\dots,n_M \}}(\vec{u}) \prod^M_{k=0} \left(\phi^{\dagger}_k \right)^{n_k} \bigotimes^{M}_{j=0} |0\rangle_j
\end{equation}\normalsize
Alternatively, a more formal statement is obtained by considering the decomposition of the tensor product $\mathbb{F} = \mathbb{F}_0 \otimes \dots \otimes \mathbb{F}_M$ as,
\small\begin{equation*}
\mathbb{F} = \mathbb{F}^0 \oplus \mathbb{F}^1 \oplus \dots \oplus \mathbb{F}^N \oplus \dots
\end{equation*}\normalsize
where $\mathbb{F}^l$ denotes the total Fock space whose number of particles is $l$. Therefore, for general $l$ we have the following statement,
\small\begin{equation}
 \sum_{0 \le n_0,n_1,\dots,n_M \le l\atop{n_0 + n_1 +\dots + n_M =l}} \prod^M_{k=0} \left(\phi^{\dagger}_k \right)^{n_k} \bigotimes^{M}_{j=0} |0\rangle_j = \mathbb{F}^l
\label{decom}\end{equation}\normalsize
\textbf{$N$-particle conjugate state vector of the phase model.} Similarly, the conjugate $N$-particle vector, $\langle \Psi_M (v_1,\dots,v_N)|$, has total occupation number $N$,
\small\begin{equation*}\begin{split}
\langle \Psi_M (v_1,\dots,v_N)| = \langle 0| C(v_N) \dots C(v_1)\\
\langle \Psi_M (v_1,\dots,v_N)| \hat{N} = N \langle \Psi_M (v_1,\dots,v_N)| 
\end{split}\end{equation*}\normalsize
Again, the above expression concerning occupation numbers implies that a suitable alternative form for the conjugate $N$-particle vector is the following,
\small\begin{equation}
\langle \Psi_M (v_1,\dots,v_N)|= \sum_{0 \le n_0,n_1,\dots,n_M \le N\atop{n_0 + n_1 +\dots + n_M = N}}g_{\{n_0,\dots,n_M \}}(\vec{v})  \bigotimes^{M}_{j=0} \langle 0|_j \prod^M_{k=0} \left(\phi_k \right)^{n_k}
\end{equation}\normalsize
where,
\small\begin{equation*}
g_{\{n_0,\dots,n_M \}}(v_1,\dots,v_N) = f_{\{n_0,\dots,n_M \}}(v^{-1}_1,\dots,v^{-1}_N)
\end{equation*}\normalsize
Analogously there exists a decomposition of the conjugate tensor product $\mathbb{F}^* = \mathbb{F}^*_0 \otimes \dots \otimes \mathbb{F}^*_M$ given by,
\small\begin{equation*}
\mathbb{F}^* =\left( \mathbb{F}^* \right)^0 \oplus \left( \mathbb{F}^* \right)^1 \oplus \dots \oplus \left( \mathbb{F}^* \right)^N \oplus \dots
\end{equation*}\normalsize
where $\left( \mathbb{F}^* \right)^l$ denotes the total conjugate Fock space whose number of particles is $l$. Therefore, for general $l$ we have the corresponding conjugate Fock space statement,
\small\begin{equation}
 \sum_{0 \le n_0,n_1,\dots,n_M \le l\atop{n_0 + n_1 +\dots + n_M =l}} \bigotimes^{M}_{j=0} \langle 0|_j \prod^M_{k=0} \left(\phi_k \right)^{n_k} = \left( \mathbb{F}^* \right)^l
\end{equation}\normalsize
\subsection{The scalar product}
\begin{figure}[h!]
\begin{center}
\includegraphics[angle=0,scale=0.16]{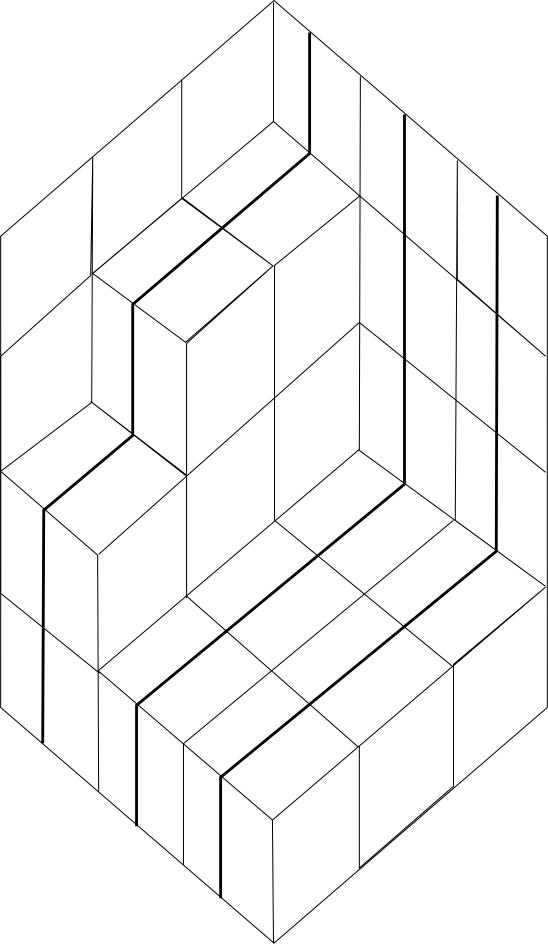}
\caption{\footnotesize{Graphical representation of the plane partition $\pi'$.}}
\label{1.d}
\end{center}
\end{figure}

\begin{figure}[h!]
\begin{center}
\includegraphics[angle=0,scale=0.16]{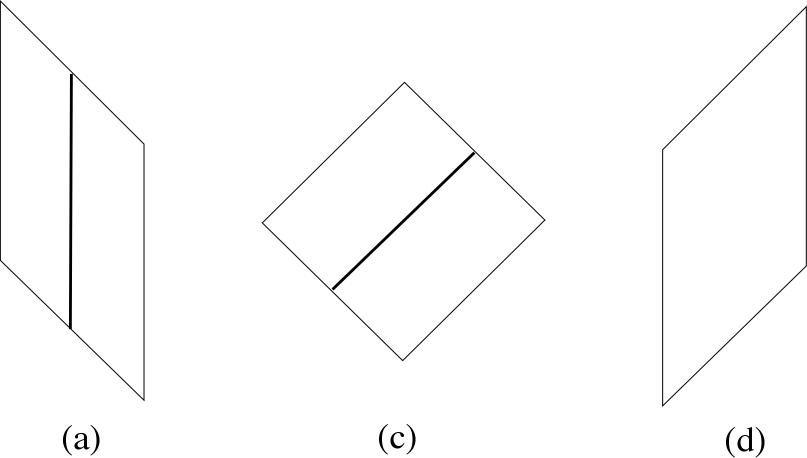}
\caption{\footnotesize{The three types of rhombi used to construct plane partitions.}}
\label{1.e}
\end{center}
\end{figure}

\noindent We now consider the scalar product of the phase model, $\field{S}(N,M| \vec{u},\vec{v})$,
\small\begin{equation*}
\left( \mathbb{F}^* \right)^N \times \mathbb{F}^N \rightarrow \mathbb{C}
\end{equation*}\normalsize
which is defined as the expectation value of the state vectors,
\small\begin{equation}\begin{split}
\field{S}(N,M| \vec{u},\vec{v}) &= \langle \Psi_M (v_1,\dots,v_N)|\Psi_M (u_1,\dots,u_N)\rangle \\
&= \langle 0| C(v_1) \dots C(v_N)B(u_1) \dots B(u_N)|0\rangle \label{scalarfirst}
\end{split}\end{equation}\normalsize
It is possible to obtain a closed form expression for $\field{S}(N,M| \vec{u},\vec{v})$ using the algebraic expressions obtained when we considered the intertwining relation between $R$ and $T$. As an example, consider $\field{S}(1,M| \vec{u},\vec{v})$,
\begin{eqnarray*}
= & \langle0| C(v)B(u) |0\rangle\\
= & g(v,u)\{ \langle0|A(v) D(u)|0\rangle - \langle 0| A(u) D(v)|0\rangle \} \\
= & g(v,u)\{ a_M(v) d_M(u)- a_M(u) d_M(v) \}
\end{eqnarray*}
Building up from the $N=1$ case, it is possible to obtain the expression for general $N$ given as,
\small\begin{equation}
\field{S}(N,M| \vec{u},\vec{v}) = \left\{ \prod_{1 \le j < k \le N} \left( \frac{u_j u_k}{u^2_j-u^2_k}\right) \left( \frac{v_j v_k}{v^2_j-v^2_k}\right)  \right\} \textrm{det} \left[H_{lm}\right]^N_{l,m=1} 
\end{equation}\normalsize
where,
\small\begin{equation}
H_{lm}= \left\{ \left( \frac{u_m}{v_l}\right)^{M+N} -\left( \frac{v_l}{u_m} \right)^{M+N} \right\} \left\{ \frac{1}{\frac{u_m}{v_l} - \frac{v_l}{u_m}} \right\}
\end{equation}\normalsize
Expanding out the geometric series inside the determinant, we obtain the much more useful expression (for the purposes of this work at least),
\small\begin{equation}\begin{split}
H_{lm}& = \left( \frac{1}{u_m v_l}\right)^{M+N-1} \sum_{p_1,p_2\atop{p_1+p_2 = M+N-1}}u^{2 p_1}_m v^{2 p_2}_l\\
& =  \left( \frac{1}{u_m v_l}\right)^{M+N-1} h_{M+N-1}(u^2_m,v^2_l) \label{scalarthird}
\end{split}\end{equation}\normalsize
where $h_{j}(u_1,\dots,u_N)$ are the complete symmetric polynomials of order $j$ and set $\{u_1,\dots,u_N\}$. \\
\\
\textbf{Enumeration of plane partitions.} We now introduce a well known application of the scalar product of the phase model, the enumeration of a certain class of plane partitions \cite{phase2}.\\
\\
A plane partition, $\pi_{j,k}$, is an $r \times s$ array of non negative integers such that,
\small\begin{equation*}
\pi_{j,k} \ge \pi_{j+1,k} \textrm{ and  } \pi_{j,k} \ge \pi_{j,k+1}
\end{equation*}\normalsize
where the integers $\pi_{i,j}$ are referred to as the parts of the plane partition of height $\pi_{i,j}$, and the total sum of the integers, $|\pi|$, is referred to as the volume,
\small\begin{equation*}
|\pi| = \sum^r_{i=1}\sum^s_{j=1}\pi_{i,j}
\end{equation*}\normalsize
If we place a restriction on the maximum height of any integer within the plane partition, $\pi_{i,j} \le t$, the plane partition is said to be contained within a box of side lengths $r \times s \times t$.\\
\\
A typical example of a plane partition within a box of $3 \times 3 \times 4$ is given by the following\footnote{We shall use the following plane partition, $\pi'$, as a running example in this section.}
\small\begin{equation}
\pi'=\left( \begin{array}{ccc}
3 & 1 & 1\\
3 & 1 & 1\\
2 & 1 & 1  
\end{array}\right)
\label{ppexa}\end{equation}\normalsize
\textbf{Graphical representation.} The graphical representation of a plane partition in a $r \times s \times t$ box is given by considering rhombus tilings of a $(r, s, t)$ semiregular hexagon. The plane partition, $\pi'$, is represented by fig. \ref{1.d}, where each representation is constructed entirely from three types of rhombi given in fig. \ref{1.e}.\\
\\
\textbf{q-enumeration.} If we group all plane partitions within a certain $r \times s \times t$ box with the same volume, $\pi$, and give these partitions a weight $q^{|\pi|}$, then the weighted sum of all the plane partitions within an $r \times s \times t$ box is referred to as $q$-enumeration. The integer coefficient of each $q$-weight, $q^{|\pi|}$, tells us how many plane partitions of volume $|\pi|$ exist in that particular $r \times s \times t$ box. The generating function for this enumeration is given by the following,
\small\begin{equation}
\sum_{\textrm{plane}\atop{\textrm{part.}}} q^{|\pi|}_{r \times s \times t}= \prod^r_{i=1}\prod^s_{j=1}\frac{1-q^{t+i+j-1}}{1-q^{i+j-1}}
\end{equation}\normalsize
If we now consider the determinant form of the scalar product of the phase model, and substitute in the following values for $u$ and $v$,
\small\begin{equation*}
u_j = q^{\frac{j-1}{2}} \textrm{  ,  } v_j = q^{-\frac{j}{2}}
\end{equation*}\normalsize
we receive,
\small\begin{equation*}\begin{split}
\field{S}(N,M| \{ q\})&= (-1)^{\frac{N(N-1)}{2}}\left\{ \prod_{1 \le j < k \le N} \left( q^{\frac{j-k}{2}}-q^{-\frac{j-k}{2}} \right)^{-2}  \right\}\\ 
& \times \textrm{det} \left[\frac{\omega^{\frac{l+m-1}{2}}-\omega^{-\frac{l+m-1}{2}}}{q^{\frac{l+m-1}{2}}-q^{-\frac{l+m-1}{2}}}\right]^N_{l,m=1}
\end{split}\end{equation*}\normalsize
where $\omega = q^{M+N}$. \\
\\
Using the following identity \cite{Kuperberg},
\small\begin{equation*}\begin{split}
\textrm{det} \left[\frac{\omega^{\frac{l+m-1}{2}}-\omega^{-\frac{l+m-1}{2}}}{q^{\frac{l+m-1}{2}}-q^{-\frac{l+m-1}{2}}}\right]^N_{l,m=1} &= (-1)^{\frac{N(N-1)}{2}}\left\{ \prod_{1 \le j < k \le N} \left( q^{\frac{j-k}{2}}-q^{-\frac{j-k}{2}} \right)^2  \right\} \\
& \times \prod^N_{l,m=1}\frac{\omega^{\frac{1}{2}} q^{\frac{l-m}{2}} - \omega^{-\frac{1}{2}} q^{-\frac{l-m}{2}}}{q^{\frac{l+m-1}{2}}-q^{-\frac{l+m-1}{2}}}
\end{split}\end{equation*}\normalsize
we obtain the important result,
\small\begin{equation}
\field{S}(N,M| \{ q\}) = q^{-\frac{N^2 M}{2}}\prod^N_{l,m=1} \frac{1-q^{M+l+m-1}}{1-q^{l+m-1}} = q^{-\frac{N^2 M}{2}} \sum_{\textrm{plane}\atop{\textrm{part.}}} q^{|\pi|}_{N \times N \times M}
\label{ppsum}\end{equation}\normalsize
\subsection{Additional combinatorial aspects of the phase model}
\begin{figure}[h!]
\begin{center}
\includegraphics[angle=0,scale=0.30]{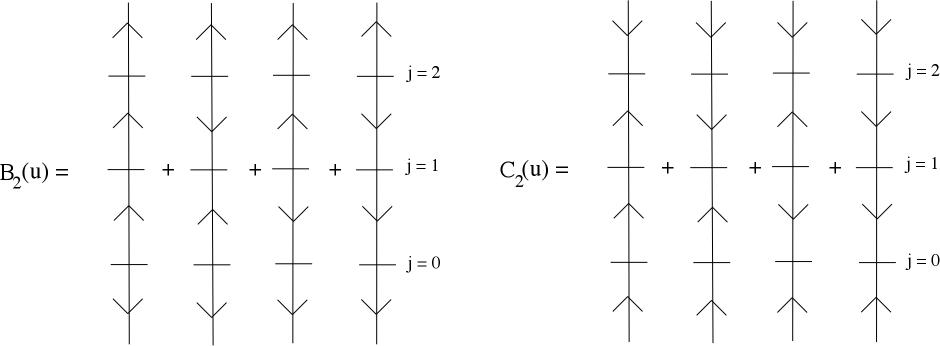}
\caption{\footnotesize{Graphical sum of $B(u)$ and $C(u)$.}}
\label{1.b}
\end{center}
\end{figure}

\begin{figure}[h!]
\begin{center}
\includegraphics[angle=0,scale=0.20]{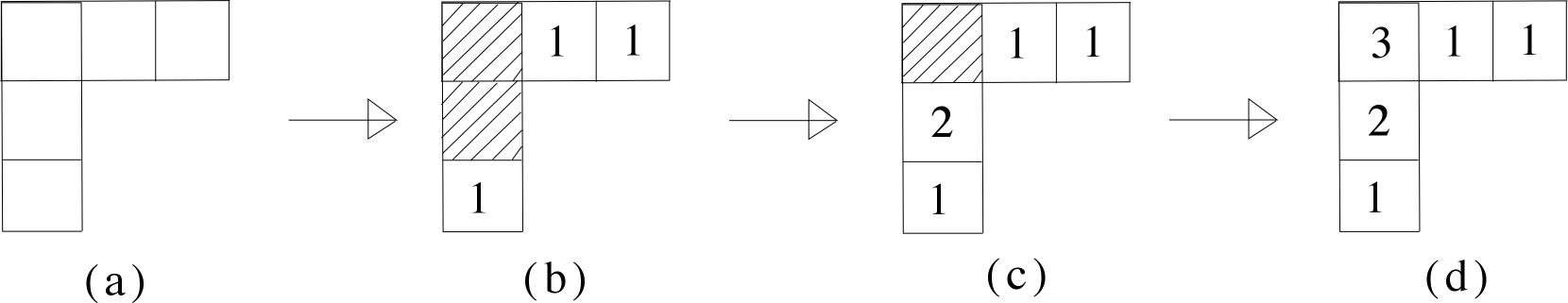}
\caption{\footnotesize{Tableau, $T^{\{\lambda \}}_-$, corresponding to the above upper half array $\pi^{\{\lambda\}}_+$}}
\label{1.f}
\end{center}
\end{figure}

\noindent The next section will enable us to express the state vectors and the scalar product as a weighted sum which counts various objects. This is essential for the analysis presented in the next section. \\
\\
\textbf{Graphical representation of the scalar product and column strict lattice paths.} It is possible to visualize matrix elements of the operator $L_j(u)$ as a vertex with the appropriately attached arrows. 
\begin{figure}[h!]
\begin{center}
\includegraphics[angle=0,scale=0.35]{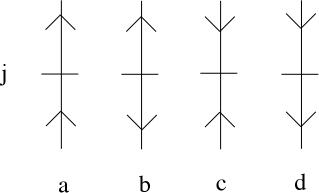}
\caption{\footnotesize{Graphical representation of the 4 matrix operators.}}
\label{1.a}
\end{center}
\end{figure}
\\
To the matrix element, $\hat{a}_j(u) = \frac{1}{u}$, we assign the representation $a$ as above. Similarly the matrix elements $\hat{b}_j(u), \hat{c}_j(u)$ and $\hat{d}_j(u)$ are represented by the vertices corresponding to $b,c$ and $d$ respectively.\\
\\
Elements of the monodromy matrix, $B(u)$ and $C(u)$, are then expressed as the sum over all possible configurations of arrows with corresponding boundary conditions on a one dimensional lattice containing $M+1$ vertical sites. The boundary conditions are as follows, 
\begin{itemize}
\item{For $B(u)$ the top most arrow points north and the bottom most arrow points south.}
\item{For $C(u)$ the top most arrow points south and the bottom most arrow points north.}
\end{itemize}
As a concrete example we give the graphical representation of $B(u)$ and $C(u)$ for $M=2$ (fig. \ref{1.b}) which were calculated earlier in this section (eq. \ref{stuufff}),
\begin{eqnarray*}
B(u) &=&\frac{1}{u} \frac{1}{u} \phi^{\dagger}_0 + \phi^{\dagger}_2 \phi_1 \phi^{\dagger}_0 + \frac{1}{u} \phi^{\dagger}_1 u + \phi^{\dagger}_2 u u\\
C(u) &=&\phi_2 \frac{1}{u} \frac{1}{u}  + u \phi_1 \frac{1}{u} + \phi_2 \phi^{\dagger}_1 \phi_0 + u u \phi_0\\
\end{eqnarray*}
\begin{figure}[h!]
\begin{center}
\includegraphics[angle=0,scale=0.30]{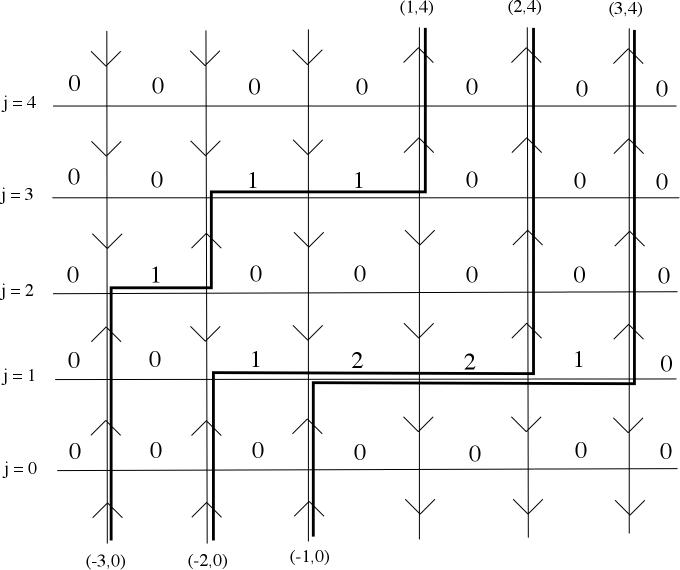}
\caption{\footnotesize{Generic lattice path configuration of $M =4$, $N = 3$, with both occupation numbers and lattice paths given.}}
\label{1.c}
\end{center}
\end{figure}
In order to visualize the scalar product $\field{S}(N,M| \vec{u},\vec{v})$, we now consider a two dimensional lattice of $(M+1)\times 2N$ sites. The first $N$ columns of the lattice are associated with the operators $C(v_j)$ and the remaining $N$ columns with $B(u_j)$.\\
\\
\textbf{Occupation number interpretation.} Each horizontal edge of the lattice is assigned an occupation number indicating the operator ($a,b,c$ or $d$) that is associated with the vertex immediately to the left of the horizontal edge. Explicitly, the $b$ operator increases the occupation number of the horizontal edge by one, the $c$ operator decreases the occupation number by one and the $a$ and $d$ operators leave the corresponding occupation number unchanged. Graphically, the scalar product is then the total sum of allowable configurations of the $(M+1)\times 2N$ lattice, with the appropriate boundary conditions, (the first $N$ top and bottom boundaries point inwards, the remaining $N$ point outwards, and the occupation numbers on the left and right boundaries are zero), taken into consideration.\\
\\
\textbf{Lattice path interpretation.} More interestingly, these configurations can be represented as $N$ non crossing column strict lattice paths on the $(M+1)\times 2N$ lattice. The scalar product is then the allowable configurations of paths beginning from the bottom most first $N$ horizontal edges labelled, $(-N,0),(-N+1,0),\dots,(-1,0)$, and ending at the top most remaining $N$ horizontal edges labelled, $(1,M),(2,M),\dots,(N,M)$. Generally, the $j$th path begins at position $(-N+j-1,0)$ and ends its journey at position $(j,N)$.\\
\\
The direction of the paths is determined by the orientation of the arrows, with paths following the vertical direction of the arrows whilst obeying the rule that paths cannot cross and only one path may exist on each vertical lattice edge. It is also true that the number of paths sharing a particular horizontal edge is equal to the occupation number associated with that horizontal edge. Obviously, the length of each path is $N+M$ units.\\
\\
\textbf{Reading the occupation number of the state vector.} It is important to note that although each column of vertical edges may contain a different set of occupation numbers, the designated occupation number sequence for the particular configuration being described is given by the middle occupation number sequence, between column $-1$ and $1$. Hence, for every set of occupation numbers there is generally more than one lattice path configuration.\\
\\  
As an example, consider a typical lattice path configuration of $M =4$ and $N = 3$ shown in fig. \ref{1.c}. We notice that for the above configuration the occupation number sequence is explicitly,
\small\begin{equation*}
\{n_0,n_1,n_2,n_3,n_4 \} = \{0,2,0,1,0 \}
\end{equation*}\normalsize
\\
\textbf{Correspondence between lattice path configurations of the scalar product and plane partitions.} The $j$th path of the lattice configuration can be thought of as the $j$th column of the array $\pi$. Consider the array, $\pi'$, given as an example earlier in this section. This array, and the lattice path configuration shown in fig. \ref{1.c} are in correspondence with each other.\\
\\
The bottom left entry, $\pi'_{3,1}=2$, corresponds to the first horizontal section of the first path on the second row, and the remaining entries of the column, $(\pi'_{2,1}=3,\pi'_{1,1}=3)$, correspond to the remaining horizontal sections of the first path, on the third row.\\
\\
There obviously exists a similar correspondence between the second and third lattice paths, and the second and third columns of the array $\pi'$ respectively. Therefore, enumerating non crossing column strict lattice paths for a particular occupation number sequence is equivalent to placing that particular occupation number sequence on the diagonal entries of an array and then enumerating all the allowable plane partitions.\\
\\
\textbf{Partition representation of the occupation numbers.} The occupation number sequence, $\{n_0,\dots,n_M \}$, is generally not a partition because it may not, in all generality, contain a sequence of non-negative integers with weakly decreasing order (as the example above shows). Nevertheless, there exists a one to one correspondence between the occupation number sequence $\{n_{0},\dots,n_{M}\}$ of length $M+1$, (provided $\sum^M_{l=0}n_{l} =N$), and some partition $\{\lambda\}=\{\lambda_1,\dots,\lambda_N \}$ of length $N$, as detailed below.\\
\\
Beginning with the occupation number sequence $\{n_{0},\dots,n_{M}\}$, we omit any $n_i$ which are zero, leaving us with a sequence of $k$ non zero numbers. We then order the resulting sequence to give,
\small\begin{equation*}
\{n_{j_1},n_{j_2},\dots,n_{j_k}\} \textrm{  such that  } j_1 > j_2 > \dots > j_k
\end{equation*}\normalsize
and $\sum^k_{l=1}n_{j_k} =N$. Doctoring the original occupation number sequence in this way we then set,
\small\begin{equation*}\begin{array}{lcl}
\lambda_{w_1} = j_1 & \textrm{for} & 1 \le w_1 \le n_{j_1}\\
\lambda_{w_2} = j_2 & \textrm{for} & n_{j_1} + 1 \le w_2 \le \sum^{2}_{l=1}n_{j_l}\\
& \vdots &\\
\lambda_{w_k} = j_k & \textrm{for} & \sum^{k-1}_{l=1}n_{j_l} + 1 \le w_k \le \sum^{k}_{l=1}n_{j_l} = N
\end{array}\end{equation*}\normalsize
From these specifications it is possible to see that the occupation number sequence which corresponds to the partition with the lowest weight, $\{\lambda\}_{L}$, is,
\small\begin{equation*}
\{ n_0,n_1,\dots,n_M\} = \{N, 0,\dots,0\} \Longleftrightarrow \{\lambda\}_{L} = \{0\}
\end{equation*}\normalsize
and, the highest weight partition, $\{\lambda\}_H$, corresponds to the occupation number sequence,
\small\begin{equation*}
\{ n_0,\dots,n_{M-1},n_M\} = \{0,\dots,0,N\} \Longleftrightarrow \{\lambda\}_{H} = \{M^N\} = (M)^{ N}
\end{equation*}\normalsize
Additionally we notice that no parts of any constructed partition are greater than $M$, and no partition is of greater length than $N$. Hence, for all other partitions constructed, $\{\lambda\}$, we have the following statement,
\small\begin{equation*}
\{\lambda\} \subseteq (M)^N
\end{equation*}\normalsize
and hence,
\small\begin{equation}
\sum_{0 \le n_0,n_1,\dots,n_M \le N\atop{n_0 + n_1 +\dots + n_M = N}} \Longleftrightarrow \sum_{\{\lambda\} \subseteq (M)^N}
\end{equation}\normalsize\\
\textbf{Correspondence between upper half plane partitions and semi-standard tableaux of descending order.} It is possible to obtain a one to one correspondence between upper half plane partitions, $\pi^{\{\lambda \}}_{+}$, and semi-standard (column strict) tableaux of descending order\footnote{Semi-standard tableaux are commonly of ascending numerical order, however, descending numerical order is the most convenient convention when we consider skew tableaux.}, $T^{\{\lambda\}}_-$, where the shape of the partition, $\{\lambda \}$, is given by the diagonal entries of the array and the negative subscript denotes descending numerical ordering.\\
\\
We begin by considering a general upper half plane partition, $\pi^{\{\lambda \}}_{+}$, and construct a partition using the diagonal entries,
\small\begin{equation*}
\{ \lambda \} = \{\pi_{1,1},\pi_{2,2},\dots, \pi_{N,N}\}
\end{equation*}\normalsize
Considering the next upper diagonal entries, $\pi_{j,j+1}$, we construct the skew diagram, $\{ \mu_1 \}$,
\small\begin{equation*}
\{ \mu_1 \}= \{\pi_{1,1}-\pi_{1,2},\pi_{2,2}-\pi_{2,3},\dots, \pi_{N-1,N-1}-\pi_{N-1,N},\pi_{N,N}\}
\end{equation*}\normalsize
and place the integer $1$ in the valid regions of the skew diagram\footnote{In ascending tableaux, $N$, would be placed instead of $1$.}. We then consider the upper diagonal entries of the array, $\pi_{j,j+2}$, and construct the skew diagram, $\{ \mu_2 \}$,
\small\begin{equation*}
\{ \mu_2 \} = \{\pi_{1,1}-\pi_{1,3},\pi_{2,2}-\pi_{2,4},\dots, \pi_{N-2,N-2}-\pi_{N,N-2},\pi_{N-1,N-1},\pi_{N,N}\}
\end{equation*}\normalsize
and place the integer $2$ in the valid regions of the skew diagram that have not already been occupied by previous steps in this process. This process continues until the partition contains the numbers $\{1,\dots,N-1 \}$. We then fill the remaining boxes in the partition with the integer $N$, thereby constructing a valid descending semi-standard tableau $T^{\{\lambda \}}_-$ from the upper half plane partition $\pi^{\{\lambda \}}_{+}$.\\
\\
As an example, consider the array given in the past examples,
\small\begin{equation*}
\pi^{\{\lambda\}} = \left(\begin{array}{ccc}
3 & 1&1\\
3&1 & 1\\
2 & 1 &1
\end{array} \right)
\end{equation*}\normalsize
where $\{\lambda \} = \{3,1^2\}$. Based on the construction described above, the corresponding tableau, $T^{\{\lambda\}}_-$, is given by fig. \ref{1.f}. In \textbf{(a)} we construct the partition $\{\lambda\} = (3,1,1)$. In \textbf{(b)} we construct the skew partition $\{ \mu_1\}=(3,1,1)-(1,1,0)$ and place the integer 1 in the valid regions of $\{ \mu_1\}$. The partition $(1,1,0)$ was obtained from the first upper diagonal entries of $\pi^{\lambda}$. In \textbf{(c)} we construct the skew partition $\{ \mu_2\}=(3,1,1)-(1,0,0)$ and place the integer 2 in the valid regions of $\{ \mu_2\}$ that contain no integers. The partition $(1,0,0)$ was obtained from the second upper diagonal entries of $\pi^{\lambda}$. In $\textbf{(d)}$ we place the integer 3 in any remaining entries of $\{\lambda\}$ that don't already contain integers, forming the valid descending semi-standard tableau $T^{\{\lambda \}}_-$ from the upper diagonal plane partition $\pi^{\{\lambda \}}_{+}$.\\
\\
\\
Therefore, the process of enumerating upper half plane partitions for some fixed diagonal, $\{ \lambda \}$, is equivalent to enumerating all the allowable descending tableaux, $T^{\{\lambda \}}_-$, in the shape $\{ \lambda \}$. 
\small\begin{equation*}
\sum_{\textrm{upper plane}\atop{\textrm{part.} \pi^{\{\lambda\}}_+}} \Longleftrightarrow \sum_{T^{\{\lambda \}}_-}
\end{equation*}\normalsize
\subsection{Schur polynomial expansion of the scalar product}
\noindent This section is attributed to Tsilevich \cite{Tsilevich} and details explicitly how the scalar product of the phase model can be expanded as a bilinear sum of Schur functions.\\
\\
\textbf{The $N$ particle state vector as the sum of Schur polynomials.} Let us now return to the $N$-particle state vector of the phase model,
\small\begin{equation}\begin{split}
|\Psi_M (u_1,\dots,u_N)\rangle& = \sum_{0 \le n_{0},n_{1},\dots,n_{M} \le N\atop{n_{0} + n_{1} +\dots + n_{M} = N}}f_{\{n_{0},\dots,n_{M} \}}(\vec{u}) \bigotimes^{M}_{l=0} |n_l\rangle_l\\
&= \sum_{\{\lambda\} \subseteq (M)\times(N)} f_{\{\lambda \}}(\vec{u})  |\lambda \rangle
\end{split}\end{equation}\normalsize
We recall that any individual element of the state vector is an element of the decomposed Fock space (eq, \ref{decom}),
\small\begin{equation*}
\bigotimes^{M}_{l=0} |n_l\rangle_l = | \lambda\rangle  \textrm{    $\in$    } \mathbb{F}^N
\end{equation*}\normalsize
and can also be defined uniquely by the corresponding partition. Using this fact we shall apply a change of basis by realizing the Fock space, $\field{F}^N$, in a linear vector (sub)space of the symmetric polynomials. \\
\\
\textbf{Constructing the symmetric polynomial basis.} Let us first prepare an infinite set of algebraically independent variables $\vec{z} = \{z_1,z_2, \dots \}$. Next, consider the infinite dimensional vector space (ring), $\field{C}[\vec{z}]^S$, of all possible symmetric polynomials consisting of variables $\vec{z}$. As is already known, one linearly independent basis for this vector space is the set of all Schur polynomials consisting of variables $\vec{z}$. Obviously, this vector space is too big for our needs as it consists of Schur polynomials of \textbf{every} partition, whereas the diagrams we are interested in are contained within a finite box. \\
\\
Let us now set the variables $\vec{z}$ such that,
\small\begin{equation*}
h_{M+1}(\vec{z}) = h_{M+2}(\vec{z}) = \dots = 0
\end{equation*}\normalsize
and thus, only $M$ of the $\vec{z}$'s are algebraically independent now. We are assured that,
\small\begin{equation*}
h_1(\vec{z}) \ne 0 \textrm{  ,  } h_2(\vec{z}) \ne 0 \textrm{  , $\dots$ ,  } h_M(\vec{z}) \ne 0
\end{equation*}\normalsize
due to the complete homogeneous symmetric polynomials being algebraically independent. Thus by eq. \ref{Schur1},
\small\begin{equation*}
S_{\{\lambda\}}\left(\vec{z} \right)= \textrm{det}\left[ h_{\lambda_i +j-i} (\vec{z}) \right]_{i,j=1,2, \dots}
\end{equation*}\normalsize
careful inspection shows that whenever $\lambda_1 > M$, we have that $S_{\{\lambda\}}\left(\vec{z} \right) = 0$. Thus no part in the partition $\{ \lambda\}$ will be greater than $M$. We shall call this restricted vector space $\field{C}_M[\vec{z}]^S$ and notice that it still has the properties of a polynomial ring.\\
\\
Nevertheless, a further restriction is necessary, and we now require Schur polynomials whose tableau lengths are no greater than $N$\footnote{Through making this restriction on the vector space we are able to make clear the operation of $\Phi^{-1}$ on the entire vector space. This has the advantage of enabling us to consider the \textit{zero-energy} space in the isomorphism, but has the disadvantage that the actual operation, $\Phi$, is different for each value of $N$.}. Calling this further restricted vector space $\field{C}^N_M[\vec{z}]^S$, we notice that it no longer contains polynomial ring structure. It is this polynomial vector space that we shall use for the change of basis argument. \\
\\
\textbf{Change of basis.} We consider the vector space $\field{F}^N_M$, which is a sub linear vector space of $\field{F}^N$. $\field{F}^N_M$ is spanned by bosonic Fock space elements whose partition representations are of no greater length than $N$ and no single part is greater than $M$. For $c_{\{n_{0},\dots,n_{M} \}} \in \field{C}$, a general element of $\field{F}^N_M$ is given by,
\small\begin{equation*}\begin{array}{lcl}
c_{\{n_{0},\dots,n_{M} \}} \bigotimes^{M}_{l=0} |n_l\rangle_l \in \field{F}^N_M & \textrm{where} & n_{0} + n_{1} +\dots + n_{M} = N\\
& \textrm{and} &\{\lambda\} = \{M^{n_M}, \dots, 2^{n_2}, 1^{n_1},0^{n_0}  \}
\end{array}\end{equation*}\normalsize
It is possible to realize $\field{F}^N_M$ in the algebra of symmetric functions by the following mapping\footnote{See chapter 14 of \cite{Kac} for further details.}, $\Phi = \sum_{\{\mu\} \subseteq \{M^N \}} S_{\{\mu\}}(\vec{z})\langle \mu|$,
\small\begin{equation}
\Phi :  \field{F}^N_M   \rightarrow  \field{C}^N_M[\vec{z}]^S 
\end{equation}\normalsize
where for any $|\alpha \rangle =\sum_{i}  c_{\{\lambda_i \}}| \lambda_i \rangle \in  \field{F}^N_M $ we have,
\small\begin{equation}\begin{split}
\Phi(|\alpha \rangle) &=  \sum_{\{\mu\} \subseteq \{M^N \}} \sum_{i}  c_{\{\lambda_i \}} S_{\{\mu\}}(\vec{z})\langle \mu| \lambda_i \rangle\\
&=\sum_{i}  c_{\{\lambda_i \}} S_{\{\lambda_i \}}(\vec{z})
\end{split}\end{equation}\normalsize
We shall now clear up some ambiguity and show how the linear map $\Phi$ can become an isomorphism. The ambiguity arises due to the Schur functions $S_{\{\mu\}}(\vec{z})$ containing an infinite amount of variables $\vec{z}$. Thus the operation, $\Phi^{-1}$, is made ambiguous when we try to obtain the occupation number $n_0$ of a general vector. However, since we know the number of particles of a state vector of $\field{F}^N_M$, namely $N$, we can always obtain $n_0$ by the simple expression,
\small\begin{equation}
n_0 = N - l(\lambda)
\label{PHI}
\end{equation}\normalsize
where $l(\lambda)$ is given as the length of $\{ \lambda \}$.\\
\\
Thus due to elements in $\field{F}^N_M$ and $\field{C}^N_M[\vec{z}]^S$ being linearly independent basis vectors, an elementary linear space analysis will instantly reveal that $\Phi$ is an isomorphism, hence,
\small\begin{equation*}
\mathbb{F}^N_M  \cong  \field{C}^N_M[\vec{z}]^S
\end{equation*}\normalsize
We now consider the action of $B(u)$ on elements in $\field{C}^N_M[\vec{z}]^S$.
\begin{proposition}
Let $B(u) = u^{-M}\hat{B}(u)$. The operator $\hat{B}(u)$ acts in $\field{C}^N_M[\vec{z}]^S$ as the operator of multiplication by $H_M(u^2;\vec{z})$, where $H_M(t;\vec{z}) = \sum^M_{j=0} t^{j} h_j(\vec{z})$ is the truncated generating function of the complete homogeneous symmetric polynomials.
\end{proposition}
\textbf{Proof.} First we note that the truncated generating function is related to the usual generating function precisely by the condition given for the realization of $\field{C}^N[\vec{z}]^S$,
\small\begin{equation*}
H(t;\vec{z})|_{h_{M+1}(\vec{z})=h_{M+2}(\vec{z}) = \dots = 0} = H_M(t;\vec{z})
\end{equation*}\normalsize
Consider now the operator $\hat{B}(u)$. It is possible to expand the operator as the following polynomial in $u^2$,
\small\begin{equation*}
\hat{B}(u)= \sum^M_{j=0}u^{2j}\hat{B}^{(j)}
\end{equation*}\normalsize
where we now wish to show that the operator $\hat{B}^{(j)}$ acts on $\field{C}^N[\vec{z}]^S$ as the operator of multiplication by $h_j(\vec{z})$. Using the expression,
\small\begin{equation*}
B(u) = \sum^2_{j_1,\dots,j_M =1} \left( L_M (u) \right)_{1,j_M}  \left( L_{M-1} (u) \right)_{j_M,j_{M-1}}  \dots \left( L_1 (u) \right)_{j_2,j_1}  \left( L_{0} (u) \right)_{j_1,2} 
\end{equation*}\normalsize
where we use the following labels,
\small\begin{equation*}
\left( L_{m} (u) \right)_{{1,1}/{2,2}} =u^{-1/+1} \textrm{  ,  } \left( L_{m} (u) \right)_{{1,2}/{2,1}} =\phi^{\dagger /1}_m \textrm{  ,  } 0 \le m \le M
\end{equation*}\normalsize
and defining,
\small\begin{equation*}
\phi_j = \phi^{-1}_j \textrm{  ,  } \phi^{\dagger}_j = \phi^{1}_j \textrm{  ,  } 1= \phi^{0}_j
\end{equation*}\normalsize
we obtain,
\small\begin{equation}
\hat{B}^{(m)} = \sum_{\epsilon_0, \dots, \epsilon_M \in \{ -1,0,1\}^{\dagger} } \phi^{\epsilon_M}_M \dots \phi^{\epsilon_0}_0
\end{equation}\normalsize
So the sum is taken over all $\epsilon_0, \dots, \epsilon_M \in \{ -1,0,1\}$. Associated with this sum are four additional conditions\footnote{It is also possible to generate these conditions quite easily considering the graphical interpretation of the operator $B(u)$. In fact, this is how the author first understood the fourth condition.} on the $\epsilon_i$'s based on the choice of $m$,
\begin{itemize}
\item{ $\boxed{\epsilon_0 \ne -1}$ }
\item{Let  $\epsilon_{j+1}$ and $\epsilon_j $ be two adjacent non zero elements, then $\boxed{\epsilon_{j+1}\epsilon_j \ne 1 \textrm{  $\forall$  } j }$. }
\item{If $\epsilon_M = \dots = \epsilon_{l+1}$=0, and $\epsilon_l$ is non zero, then based on the definition of $B(u)$ given above, $\boxed{\epsilon_l = 1}$.}
\item{For general $m$, assume that the number of $\phi$ operators for a general configuration of $B_M(u)$ is $l$, therefore the number of $\phi^{\dagger}$ operators must be $l+1$. If the $\phi^{\dagger}$ vertices are positioned at $0\le j_1<  j_3 < \dots < j_{2l +1} \le M$, and the $\phi$ vertices are positioned at $j_2 < j_4 < \dots < j_{2l }$, with $j_1 < j_2 < \dots < j_{2l+1}$. The exponent, $2m -M$, is given by,}
\end{itemize}
\small\begin{equation*}\begin{split}
2m -M & = \left(j_1\right) - \sum^{2l+1}_{k=2 }(-1)^{k-1} \left(j_k -j_{k-1} -1 \right) - \left( M - j_{2l+1} \right)\\
& = 2 j_1 - 2j_2 + 2j_3 - \dots - 2j_{2l} + 2j_{2l+1} - M
\end{split}\end{equation*}\normalsize
\small\begin{equation*}
\Rightarrow \boxed{m = \sum^M_{k=0}k \epsilon_k =\sum^M_{k=1}k \epsilon_k}
\end{equation*}\normalsize
We now consider the action of $\phi^{\dagger}_j$ and $\phi_j$ on general $S_{\{ \lambda \}}(\vec{z})$. Based on their effect on the Fock space $\field{F}^N_M$ we have,
\small\begin{equation*}
\begin{array}{cccl}
\phi^{\dagger}_j S_{\{ \lambda \}}(\vec{z}) &=& S_{\{ \mu \}}(\vec{z})  & \textrm{where $\{ \mu \}$ is the partition $\{\lambda\}$ with the}\\
&&&\textrm{row of length $j$ inserted appropriately}
\end{array}
\end{equation*}\normalsize
\small\begin{equation*}
\begin{array}{ccc}
\phi_j S_{\{ \lambda \}}(\vec{z}) &=&
\left\{ \begin{array}{cl}
 S_{\{ \mu \}}(\vec{z})  & \textrm{where $\{ \mu \}$ is the partition $\{\lambda\}$ }\\
&\textrm{with a row of length $j$ deleted}\\
0 & \textrm{if $\{\lambda\}$ does not contain a row of length $j$}
\end{array}\right.
\end{array}
\end{equation*}\normalsize
If we denote $|\nu'_{i}|$ the length of the $i$th column in partition $\{ \nu\}$ and $n_i(\nu)$ as the number of rows of length $i$ in $\{ \nu\}$, then by eq. I.1.4 in \cite{MacD} we have,
\small\begin{equation*}
|\nu'_i| - |\nu'_{i+1}| = n_i(\nu)
\end{equation*}\normalsize
We now consider the action, $\hat{B}^{(m)} S_{\{\lambda\}} (\vec{z}) = S_{\{\mu\}} (\vec{z})$. \\
\\
Denoting $|\theta'_i| = |\mu'_i| - |\lambda'_i|$\footnote{At the moment we are not guaranteed that $\{\mu - \lambda\}$ is a skew diagram as $|\theta'_i|$ could in general be negative for some $i$. Nevertheless, we shall soon show that $|\theta'_i| \ge 0$, meaning that $\{\mu\} \supseteq \{\lambda\} $ and making $\{\mu - \lambda\}$ a skew partition.}, by the action of the $\phi/\phi^{\dagger}$ operators on general $S_{\{\lambda\}} (\vec{z})$ given above, we have,
\small\begin{equation*}
\begin{array}{ccccc}
n_i(\mu) &=& n_i(\lambda) + \epsilon_i\\
\Rightarrow |\theta'_i| &=&|\mu'_i| - |\lambda'_i| & = & |\mu'_{i+1}|+n_i(\mu) +|\lambda'_{i+1}|+n_i(\lambda) \\
&=&|\theta'_{i+1}| + \epsilon_i & \textrm{  for  } &1 \le i \le M
\end{array}
\end{equation*}\normalsize
Beginning from $i=M$ we obtain,
\small\begin{equation*}\begin{array}{clcl}
&|\theta'_M|  &=& \epsilon_M \\
\Rightarrow &|\theta'_{M-1}|  &=& \epsilon_{M-1}+\epsilon_M\\
& & \vdots & \\
\Rightarrow & |\theta'_{j}|  &=& \sum^M_{k=j} \epsilon_k \textrm{    for    } 1 \le j \le M
\end{array}\end{equation*}\normalsize
\textbf{Applying the four conditions on $\mathbf{\epsilon_l}$.} We now consider the four conditions on the $\epsilon_i$'s given above. By condition 1) we have for some $l$, $\epsilon_M = \dots = \epsilon_{l+1} = 0$ and $\epsilon_l = 1.$ Therefore,
\small\begin{equation*}\begin{array}{lcl}
|\theta'_M| = \dots = |\theta'_{l+1}| &=& 0\\
\Rightarrow |\theta'_l| &=& 1
\end{array}\end{equation*}\normalsize
By condition 3) then we have,
\small\begin{equation*}
|\theta'_j| =  \sum^M_{k=j} \epsilon_k  \in  \{ 0,1 \} 
\end{equation*}\normalsize
Since $|\theta'_j| = |\mu'_j| - |\lambda'_j|$, we note that $\{\mu\} \supseteq \{\lambda\}$ and $\{\mu - \lambda\} = \{\mu/\lambda\}$ is indeed a skew partition of no more than one cell in each column. \\
\\
Condition 4) is concealed in the expression sum, $\sum^M_{k=1}|\theta'_k |$,
\small\begin{equation*}
\begin{array}{ccccc}
\sum^M_{k=1}|\theta'_k| &=& \sum^M_{k=1} \sum^M_{j=k} \epsilon_j &= &\sum^M_{k=1} k \epsilon_k\\
\Rightarrow \sum^M_{k=1}|\theta'_k| &=& m
\end{array}
\end{equation*}\normalsize
revealing to us that the skew partition, $\{\mu/\lambda\}$, contains no more than $m$ cells in the form of vertical strips. If we label the set of all vertical strips of no more than length $m$ as $\mathbb{H}_m$, we finally obtain,
\small\begin{equation*}
\hat{B}^{(m)} S_{\{\lambda\}}(\vec{z}) = \sum_{\{\mu\} \supseteq \{ \lambda \}\atop{\{ \mu/\lambda\} \in \mathbb{H}_m }} S_{\{\mu\}}(\vec{z}) \textrm{  ,  } 0 \le m \le M
\end{equation*}\normalsize
The right side of the above equation is nothing more than the right side of Pieri's formula (eq. \ref{pieri}), thus,
\small\begin{equation*}
\hat{B}^{(m)} |\lambda\rangle \cong \hat{B}^{(m)} S_{\{\lambda\}}(\vec{z}) = h_{m}(\vec{z}) S_{\{\lambda\}}(\vec{z}) \textrm{  ,  } 0 \le m \le M \textrm{  $\square$  }
\end{equation*}\normalsize
With the details out of the way we now claim the following prize,
\begin{proposition}
\small\begin{equation*}\begin{split}
|\Psi_M (u_1,\dots,u_N)\rangle &= \left( \frac{1}{u_1\dots u_N}\right)^M\sum_{\{ \lambda\} \subseteq (M)^N} S_{\{\lambda\}} (u^2_1, \dots u^2_N) |\lambda\rangle\\
& \cong \left( \frac{1}{u_1\dots u_N}\right)^M \sum_{\{ \lambda\} \subseteq (M)^N} S_{\{\lambda\}} (u^2_1, \dots u^2_N)S_{\{\lambda\}} (\vec{z})
\end{split}\end{equation*}\normalsize
\end{proposition}
\textbf{Proof.} Using the results from above we have,
\small\begin{equation*}\begin{split}
\left(\prod^N_{j=1}u_j \right)^M B(u_1) \dots B(u_N)|0\rangle &=\hat{B}(u_1) \dots \hat{B}(u_N)|0\rangle\\
&\cong   H_M(u^2_1;\vec{z}) \dots H_M(u^2_N;\vec{z})S_{\{ \phi \}} (\vec{z})\\
&=\lim_{h_l(\vec{z}) \rightarrow 0\atop{l> M}} \prod^{N}_{j=1}\left( \prod^{\infty}_{k=1} \frac{1}{1-u^2_j z_k} \right)\\
&= \sum_{\{ \lambda\} \subseteq (M)^N} S_{\{\lambda\}} (u^2_1, \dots u^2_N)S_{\{\lambda\}} (\vec{z}) \textrm{  $\square$  }
\end{split}\end{equation*}\normalsize
where we recognize the second last line as the limiting case of equation I.4.3 of \cite{MacD}. \\
\\
\textbf{The conjugate $N$-particle state vector as the sum of Schur polynomials.} 
Finding the corresponding form for the conjugate state vector is relatively simple if one is provided with the two results below.
\begin{proposition}
\small\begin{equation*}
B(u) = u A(u)\phi^{\dagger}_0
\end{equation*}\normalsize
\end{proposition}
\textbf{Proof.} The proof of the first result is straightforward,
\small\begin{equation*}\begin{split}
uA(u)\phi^{\dagger}_0 &=  \sum^2_{j_1,\dots,j_M =1} \left( L_M (u) \right)_{1,j_M}  \dots \left( L_1 (u) \right)_{j_2,j_1}\underbrace{\left\{ u \left( L_{0} (u) \right)_{j_1,1}  \phi^{\dagger}_0 \right\}}_{\left( L_{0} (u) \right)_{j_1,2}}\\
&= B(u)  \textrm{    } \square
\end{split}\end{equation*}\normalsize
Which leads us to the second result,
\begin{proposition}
\small\begin{equation*}
C(u) =u^{-1} \phi_0A^{\dagger}_M(u^{-1})
\end{equation*}\normalsize
\end{proposition}
\textbf{Proof.}
\small\begin{equation*}\begin{split}
A^{\dagger}(u^{-1}) &= \sum^2_{j_1,\dots,j_M =1} \left( L^{\dagger}_M (u^{-1}) \right)_{1,j_M}  \dots \left( L^{\dagger}_1 (u^{-1}) \right)_{j_2,j_1} \left( L^{\dagger}_{0} (u^{-1}) \right)_{j_1,1}  \\
&= \sum^2_{j_1,\dots,j_M =1} \left( L_M (u) \right)_{2,j_M}  \dots \left( L_1 (u) \right)_{j_2,j_1} \left( L_{0} (u) \right)_{j_1,2}\\
\Rightarrow u^{-1} \phi_0 A^{\dagger}_M(u^{-1}) &= \sum^2_{j_1,\dots,j_M =1} \left( L_M (u) \right)_{2,j_M}  \dots \left( L_1 (u) \right)_{j_2,j_1} \underbrace{\left\{ u^{-1} \phi_0 \left( L_{0} (u) \right)_{j_1,2} \right\}}_{\left( L_{0} (u) \right)_{j_1,1}} \\
&= C(u) \textrm{    } \square
\end{split}\end{equation*}\normalsize
Therefore, we immediately obtain,
\small\begin{equation*}
\boxed{B^{\dagger}(u^{-1}) = C(u)}
\end{equation*}\normalsize
With these results we are ready to express the conjugate $N$-particle state vector as the sum of Schur polynomials.
\begin{proposition}
\small\begin{equation*}
\langle \Psi_M (v_1,\dots,v_N)| =\left( v_1\dots v_N \right)^M \sum_{\{ \lambda\} \subseteq (M)^N} S_{\{\lambda\}} (v^{-2}_1, \dots v^{-2}_N) \langle \lambda|
\end{equation*}\normalsize
\end{proposition}
\textbf{Proof.}
\small\begin{equation*}\begin{split}
\left(\prod^N_{j=1}\frac{1}{v_j} \right)^M \langle \Psi_M (v_1,\dots,v_N)| &=\langle 0| \hat{C}(v_1) \dots \hat{C}(v_N)\\
&=\left\{ \hat{B}(v^{-1}_1) \dots \hat{B}(v^{-1}_N)|0\rangle \right\}^{\dagger}\\
&= \sum_{\{ \lambda\} \subseteq (M)^N} S_{\{\lambda\}} (v^{-2}_1, \dots v^{-2}_N)\left\{|\lambda\rangle\right\}^{\dagger}\\
&= \sum_{\{ \lambda\} \subseteq (M)^N} S_{\{\lambda\}} (v^{-2}_1, \dots v^{-2}_N)\langle \lambda| \textrm{    } \square
\end{split}\end{equation*}\normalsize
These results allow us to naturally express the scalar product as the bilinear sum of Schur polynomials,
\small\begin{equation}\begin{split}
\field{S}(N,M| \vec{u},\vec{v}) &=\langle \Psi_M (v_1,\dots,v_N)|\Psi_M (u_1,\dots,u_N)\rangle\\
&= \left( \prod^N_{j=1} \frac{v_j}{u_j}  \right)^M  \sum_{\{\lambda\}\{\mu \} \subseteq (M)^N}S_{\{ \lambda \}} \left(\left\{u^{2}_k\right\}\right)S_{\{ \mu \}} \left( \left\{v^{-2}_k\right\} \right) \langle \mu| \lambda\rangle\\
&= \left( \prod^N_{j=1} \frac{v_j}{u_j}  \right)^M \sum_{\{\lambda\} \subseteq (M)^N}S_{\{ \lambda \}} \left(u_1^{2},\dots, u_N^{2}\right)S_{\{ \lambda \}} \left(v_1^{-2},\dots, v_N^{-2}\right)\label{TSI}
\end{split}\end{equation}\normalsize
We now give some additional combinatorial representations for the $N$-particle state vector which we shall use later on.\\
\\
\textbf{Combinatorial definitions of the $N$-particle vector.}
Let us now return to the $N$-particle vector of the phase model,
\small\begin{equation*}
|\Psi_M (u_1,\dots,u_N)\rangle = \sum_{0 \le n_{0},n_{1},\dots,n_{M} \le N\atop{n_{0} + n_{1} +\dots + n_{M} = N}}f_{\{n_{0},\dots,n_{M} \}}(\vec{u}) \bigotimes^{M}_{l=0} |n_l\rangle_l
\end{equation*}\normalsize
Using the \textbf{lattice path} representation of the scalar product, we note that for each occupation number sequence, $\{n_{j_1},\dots,n_{j_k}\}$, we have the following expression,
\small\begin{equation}
f_{\{n_{j_1},\dots,n_{j_k} \}}(\vec{u}) = \sum_{\textrm{allowable paths}\atop{\textrm{in $(M+1)\times N$ lattice}}}u_1^{t^d_1-t^a_1} u_2^{t^d_2-t^a_2} \dots u_N^{t^d_N-t^a_N}
\end{equation}\normalsize
where the sum is taken over all allowable paths in the $(M+1) \times N$ lattice under the conditions,
\begin{itemize}
\item{$n_{j_1}$ paths starting at $(1,j_1)$ and ending at $(1,M),(2,M),\dots,(n_{j_1},M)$}
\item{$n_{j_2}$ paths starting at $(1,j_2)$ and ending at $(n_{j_1} +1,M),\dots,(\sum^2_{l=1}n_{j_l},M)$}
\item{this procedure continues until we have finally $n_{j_k}$ paths starting at $(1,j_k)$ and ending at $(\sum^{k-1}_{l=1}n_{j_l} +1,M),\dots,(N,M)$}
\end{itemize}
The powers $t^d_{l}$ and $t^a_{l}$, $1 \le l \le N$, are equal to the number of $d$ and $a$ vertices respectively in the $l$th column.\\
\\
An alternative form for the above expression when considering upper half \textbf{plane partitions} in an $N\times N \times M$ box is given as,
\small\begin{equation}
f_{\{n_{j_1},\dots,n_{j_k} \}}(\vec{u}) = \sum_{\textrm{upper plane}\atop{\textrm{partitions}}}u_1^{l^d_1-l^a_1} u_2^{l^d_2-l^a_2} \dots u_N^{l^d_N-l^a_N}
\end{equation}\normalsize
where the sum is taken over all allowable plane partitions in the upper half of an $N \times N$ array where the diagonal terms are given by the partition representation of the corresponding occupation number sequence in descending numerical order. Considering the graphical representation of the half plane partition, the powers $l^d_{l}$ and $l^a_{l}$, $1 \le l \le N$, are equal to the number of $d$ and $a$ rhombi respectively in the $l$th column of the half hexagon.\\
\\
Additionally, since there exists a one to one correspondence between the upper plane partition array, $\pi^{\{\lambda \}}_+$, and particular \textbf{semi-standard tableau} of shape $\{ \lambda\}$, another valid combinatorial definition for the function $f_{\{n_{j_1},\dots,n_{j_k} \}}(\vec{u})   = f_{\{\lambda \}}(\vec{u})$ is given by:
\small\begin{equation}
f_{\{\lambda \}}(\vec{u}) = \sum_{T^{\{\lambda \}}_-}u_1^{2t_1 - M} u_2^{2t_2 - M} \dots u_N^{2t_N - M}
\end{equation}\normalsize
where the summation is over all semi-standard Young tableaux of shape $\{\lambda\}$ . The powers, $t_j$, give the weights of $T^{\{\lambda\}}_-$, which count the number of times $j$ appears in the tableau. Note that these powers have been chosen to match the Schur polynomial expression given by eq. \ref{Schur3}.\\
\\
\textbf{Combinatorial definitions of the conjugate $N$-particle vector.} Considering the \textbf{lattice path} representation of the scalar product we obtain,
\small\begin{equation}
g_{\{n_{j_1},\dots,n_{j_k} \}}(\vec{v}) = \sum_{\textrm{allowable paths}\atop{\textrm{in $(M+1)\times N$ lattice}}}v_1^{t^d_1-t^a_1} v_2^{t^d_2-t^a_2} \dots v_N^{t^d_N-t^a_N}
\end{equation}\normalsize
where the sum is taken over all allowable paths in the $(M+1) \times N$ lattice under the conditions,
\begin{itemize}
\item{$n_{j_1}$ paths starting at $(-N,0),(-N-1,0), \dots, (-N+n_{j_1}-1,0)$ and ending at $(-1,n_{j_1})$}
\item{$n_{j_2}$ paths starting at $(-N+n_{j_1} ,0),\dots,(-N+\sum^2_{l=1}n_{j_l}-1,0)$ and ending at $(-1,n_{j_2})$}
\item{this procedure continues until we have finally $n_{j_k}$ paths starting at $(-N+\sum^{k-1}_{l=1}n_{j_l} ,0),\dots,(-1,0)$ and ending at $(-1,n_{j_k})$}
\end{itemize}
Alternatively, when considering lower half \textbf{plane partitions} in an $N\times N \times M$ box we obtain,
\small\begin{equation}
g_{\{n_{j_1},\dots,n_{j_k} \}}(\vec{v}) = \sum_{\textrm{lower plane}\atop{\textrm{partitions}}}v_1^{l^d_1-l^a_1} v_2^{l^d_2-l^a_2} \dots v_N^{l^d_N-l^a_N}
\end{equation}\normalsize
where the sum is taken over all allowable plane partitions in the lower half of an $N \times N$ array where the diagonal terms are given by the partition representation of the corresponding occupation number sequence. \\
\\
When we transform from lower plane partition to \textbf{semi-standard tableau}, we use the usual convention of ascending numerical ordering.\\
\\
As an example, if we consider the array $\pi$ given previously, the corresponding ascending semi-standard tableau is given by fig. \ref{1.d}. In \textbf{(a)} we construct the partition $\{\lambda\} = (3,1,1)$. In \textbf{(b)} we construct the skew partition $\{ \nu_1\}=(3,1,1)-(3,1,0)$ and place the integer 3 in the valid regions of $\{ \nu_1\}$. The partition $(3,1,0)$ was obtained from the first lower diagonal entries of $\pi^{\lambda}$. In \textbf{(c)} we construct the skew partition $\{ \nu_2\}=(3,1,1)-(2,0,0)$ and place the integer 2 in the valid regions of $\{ \nu_2\}$ that contain no integers. The partition $(2,0,0)$ was obtained from the second lower diagonal entries of $\pi^{\lambda}$. In $\textbf{(d)}$ we place the integer 1 in any remaining entries of $\{\lambda\}$ that don't already contain integers, forming the valid ascending semi-standard tableau $T^{\{\lambda \}}_+$ from the lower diagonal plane partition $\pi^{\{\lambda \}}_{-}$.
\begin{figure}[h!]
\begin{center}
\includegraphics[angle=0,scale=0.20]{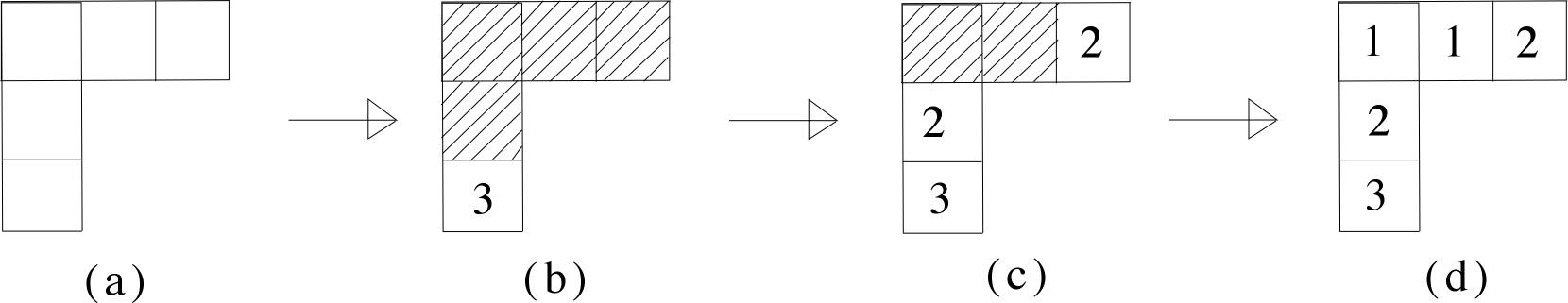}
\caption{\footnotesize{Tableau, $T^{\{\lambda \}}_+$, corresponding to the lower half array $\pi^{\{\lambda\}}_-$}}
\label{1.g}
\end{center}
\end{figure}
Thus, considering the correspondence between the lower plane partition array, $\pi^{\{\lambda \}}_-$, semi-standard tableau of shape $\{ \lambda\}$, another valid combinatorial definition for the function $g_{\{n_{j_1},\dots,n_{j_k} \}}(\vec{v})   = g_{\{\lambda \}}(\vec{v})$ is given by,
\small\begin{equation}
g_{\{\lambda \}}(\vec{v}) = \sum_{T^{\{\lambda \}}_+}v_1^{-2t_1 + M} v_2^{-2t_2 + M} \dots v_N^{-2t_N + M}
\end{equation}\normalsize
where the summation is over all semi-standard Young tableaux of shape $\{\lambda\}$ of ascending numerical order. 
\subsection{Restricting the 2-Toda tau-function to obtain the scalar product}
To begin this section we present the obvious result.
\begin{proposition}
\label{illu}
The scalar product of the phase model for general $N$ and $M$ is, (up to an overall factor of $ \left( \prod^N_{j=1} \frac{v_j}{u_j} \right)^M$), a restricted $\tau$-function of the 2-Toda hierarchy with $A_{\{\lambda\}\{\mu\}} = \delta_{\{\lambda\}\{\mu\}}$, and $s=n-M=m+N$, where $m$ and $n$ are free parameters.
\end{proposition}
\textbf{Proof.} Beginning with the unrestricted $\tau$-function,
\small\begin{equation*}
\tau(s = n-M=m+N, \vec{x},\vec{y}) = \sum_{\{\lambda\} \subseteq (M)^N} \chi_{\{\lambda\} }(\vec{x}) \chi_{\{\lambda\} } (-\vec{y})
\end{equation*}\normalsize
and performing the following change of variables,
\small\begin{equation*}
x_k \rightarrow \frac{1}{k} p_k\left(u^2_1, \dots,u^2_N  \right)\textrm{  ,  } -y_k \rightarrow \frac{1}{k} p_k\left(v^{-2}_1, \dots,v^{-2}_N  \right)\textrm{  ,  } 1 \le k \le N+M-1 
\end{equation*}\normalsize
we obtain,
\small\begin{equation}\begin{split}
\tau\left(s = n-M=m+N, \left\{ u^2_j \right\}, \left\{ v^{-2}_j \right\} \right)& = \sum_{\{\lambda\} \subseteq (M)^N}S_{\{\lambda\} }\left(\left\{ u^2_j \right\} \right) S_{\{\lambda\} }\left(\left\{ v^{-2}_j \right\} \right)\\
&=  \left( \prod^N_{j=1} \frac{u_j}{v_j} \right)^M \field{S}(N,M| \vec{u},\vec{v}) \label{specialtau}
\end{split}\end{equation}\normalsize
which is the required result. $\square$\\
\\
The above result only considers one value of $s$. Let us now consider the family of corresponding restricted $\tau$-functions for other values of $s= m+1, \dots, n$, ($s \ne n-M)$.\\
\\
We begin by clarifying some known facts about the family of unrestricted $\tau$-functions. 
\begin{itemize}
\item{The full family looks like, $\tau_{s=m+1}(\vec{x},\vec{y}), \tau_{s=m+2}(\vec{x},\vec{y}), \dots, \tau_{s=n}(\vec{x},\vec{y})$.}
\item{The valid partitions for each $s$ value are $\{ \lambda\} \subseteq (n-s)^{s-m}$.}
\item{Different values of  $s$ do not change the amount of, (two sets of $n-m-1$), time variables.}
\end{itemize}
We now compare this to the case of the family of restricted $\tau$-functions. 
\begin{itemize}
\item{The initial $\tau$-function, $\tau\left(s = n-M=m+N, \left\{ u^2_j \right\}, \left\{ v^{-2}_j \right\} \right)$, has two sets of $N+M-1$ time variables, but each set is constructed from $N$ symmetric variables.}
\item{The introduction of the  condition $s = n-M=m+N$ means that as $s$ changes, so to do $M$ and $N$.}
\item{By considering the change in the dimensions of the partition, we can obtain how $M$ and $N$ change with $s$.}
\end{itemize}
\small\begin{equation}
s \rightarrow s \pm l \Longleftrightarrow  \left\{ \begin{array}{c} M \rightarrow M \mp l \\ N \rightarrow N \pm l \end{array}\right. 
\label{changeins}\end{equation}\normalsize
\begin{itemize}
\item{Consequently, although the number of time variables does not change with each $s$ value, different values of  $s$ \textbf{do} change the amount of symmetric variables that the time variables are constructed from.}
\end{itemize}
\textbf{An illustrative example.} Consider the complete family of unrestricted $\tau$-functions for $n=5$ and $m=1$. In this case each $\tau$-function contains two sets of $3$ time variables, $\{\vec{x},\vec{y} \}=\{x_1,x_2,x_3,y_1,y_2,y_3\}$,
\small\begin{equation*}\begin{split}
\tau_{s=2} (\vec{x},\vec{y}) = \sum_{\{\lambda\} \subseteq \{3 \}} \chi_{\{\lambda\}}(\vec{x})\chi_{\{\lambda\}}(-\vec{y}), \\
 \tau_{s=3} (\vec{x},\vec{y}) = \sum_{\{\lambda\} \subseteq \{2,2 \}} \chi_{\{\lambda\}}(\vec{x})\chi_{\{\lambda\}}(-\vec{y}),\\
\tau_{s=4} (\vec{x},\vec{y}) = \sum_{\{\lambda\} \subseteq \{1,1,1 \}} \chi_{\{\lambda\}}(\vec{x})\chi_{\{\lambda\}}(-\vec{y}), \\
\tau_{s=5} (\vec{x},\vec{y}) = \sum_{\{\lambda\} = \{\phi \}} \chi_{\{\lambda\}}(\vec{x})\chi_{\{\lambda\}}(-\vec{y})
\end{split}\end{equation*}\normalsize
Consider now restricting $\tau_{s=3}$ of the above family. By proposition \ref{illu} we obtain the $M=N=2$ scalar product. The main question now is, if one $\tau$-function in a family has been restricted to form a scalar product with a certain $M$ and $N$ value, \emph{can the remaining $\tau$-functions of the family also be restricted to form scalar products with valid $M$ and $N$ values?}\\
\\
Naively performing the corresponding restrictions to the family of $\tau$-functions we obtain the following family of scalar products,
\small\begin{equation*}\begin{split}
\tau_{s=2}  = \left( \prod^1_{j=1} \frac{u^i_j}{v^i_j} \right)^3 \field{S}\left(\left.N=1\atop{M=3}\right| \vec{u}^i,\vec{v}^i \right) &\textrm{  ,  } \begin{array}{c}
x_k \rightarrow \frac{1}{k} p_k\left((u^i_1)^2  \right)\\
-y_k \rightarrow \frac{1}{k} p_k\left((v^i_1)^{-2}  \right)
\end{array}   \\
\tau_{s=3}  =  \left( \prod^2_{j=1} \frac{u^{ii}_j}{v^{ii}_j} \right)^2 \field{S}\left(\left.N=2\atop{M=2}\right| \vec{u}^{ii},\vec{v}^{ii}\right)&\textrm{  ,  } \begin{array}{c}
x_k \rightarrow \frac{1}{k} p_k\left((u^{ii}_1)^2, (u^{ii}_2)^2  \right)\\
-y_k \rightarrow \frac{1}{k} p_k\left((v^{ii}_1)^{-2},(v^{ii}_2)^{-2}  \right)
\end{array} \\
\tau_{s=4}  =  \left( \prod^3_{j=1} \frac{u^{iii}_j}{v^{iii}_j} \right)^1 \field{S}\left(\left.N=3 \atop{M=1}\right| \vec{u}^{iii},\vec{v}^{iii} \right) &\textrm{  ,  } \begin{array}{c}
x_k \rightarrow \frac{1}{k} p_k\left((u^{iii}_1)^2, \dots, (u^{iii}_3)^2  \right)\\
-y_k \rightarrow \frac{1}{k} p_k\left((v^{iii}_1)^{-2},\dots ,(v^{iii}_3)^{-2} \right)
\end{array}   \\
\tau_{s=5}  =  \left( \prod^4_{j=1} \frac{u^{iv}_j}{v^{iv}_j} \right)^0 \field{S}\left( \left. N=4\atop{M=0} \right| \vec{u}^{iv},\vec{v}^{iv} \right)&\textrm{  ,  } \begin{array}{c}
x_k \rightarrow \frac{1}{k} p_k\left((u^{iv}_1)^2, \dots,(u^{iv}_4)^2  \right)\\
-y_k \rightarrow \frac{1}{k} p_k\left((v^{iv}_1)^{-2},\dots,(v^{iv}_4)^{-2} \right)
\end{array}   
\end{split}\end{equation*}\normalsize
for $1 \le k \le 3$.\\
\\
This example has illustrated an extremely important issue. We remember that each $\tau$-function contained within a family must contain the same amount of time variables. Furthermore, it is a requirement that these time variables be the same for each value of $s$, if the $\tau$-functions are to obey the bilinear relation, which they obviously must, otherwise we are simply wasting our time.  If this is to be the case for the above example, we have the following set of equations that must be satisfied,
\small\begin{equation*}\begin{split}
 p_k\left((u^i_1)^2  \right)= p_k\left((u^{ii}_1)^2, (u^{ii}_2)^2  \right)=  \dots= p_k\left((u^{iv}_1)^2, \dots,(u^{iv}_4)^2  \right)\\
p_k\left((v^i_1)^{-2}  \right) = p_k\left((v^{ii}_1)^{-2},(v^{ii}_2)^{-2}  \right) =  \dots = p_k\left((v^{iv}_1)^{-2},\dots,(v^{iv}_4)^{-2} \right)
\end{split}\end{equation*}\normalsize
A simple check will reveal that only the trivial solution exists,
\small\begin{equation*}\begin{split}
(u^i_1)^2  = (u^{ii}_{\sigma_{j_1}})^2= (u^{iii}_{\sigma_{j_2}})^2 =(u^{iv}_{\sigma_{j_3}})^2 \\
(v^i_1)^{-2}  = (u^{ii}_{\sigma_{l_1}})^{-2}= (v^{iii}_{\sigma_{l_2}})^{-2} =(v^{iv}_{\sigma_{l_3}})^{-2}
\end{split}\end{equation*}\normalsize
and the remaining variables are set to zero. This obviously trivializes the situation immensely. Thus, at a first glance, the answer to the question is no, due to the fact that \emph{the $\tau$-functions in the family all need to contain the same time variables.}\\
\\
We now generalize the above example.
\begin{proposition}
\label{ref1}
The system of equations, $0 \le l \le M-1$,
\small\begin{equation*}\begin{split}
u^2_1+ \dots +u_{N+l}^{2} & =\mu_1^2+ \dots +\mu_{N}^{2}\\
u_1^4+ \dots +u_{N+l}^{4} & =\mu_1^4+ \dots +\mu_{N}^{4} \\
&\vdots\\
u_1^{2(N+M-1)}+ \dots +u_{N+l}^{2(N+M-1)} & = \mu_1^{2(N+M-1)}+ \dots +\mu_{N}^{2(N+M-1)}
\end{split}\end{equation*}\normalsize
permits only the trivial solution, i.e. $u_{\sigma_j}^2=\mu_j^2$, for $j\in \{1,\dots,N\}$, and the remaining $l$ of the $u_k^2$'s are equal to zero. 
\end{proposition}
\textbf{Proof.} The proof of the above result relies on the fundamental theorem of symmetric functions. We map each symmetric power sum, $p_k\left(\left\{u^2 \right\}\right)$, onto a simpler set of linear polynomials and trivially solve the resulting linear system and show that there are only $(N+l)$ points of intersection in the non symmetric polynomial ring.\\
\\
\textbf{Considering the first $N+l$ equations.} We begin by considering the first $N+l$ equations in the system, the remaining equations will follow easily.\\
\\
\textbf{Applying the fundamental theory of symmetric polynomials.} We note that the left hand side of these polynomial equations exist in the symmetric polynomial ring $\mathbb{C}[u_1^2,\dots,u_{N+l}^2]^{S_{N+l}}$. Consider now the polynomial ring $\mathbb{C}[s_1,\dots,s_{N+l}]$, and recall that the fundamental theorem of symmetric polynomials states that there exists an isomorphism between the two polynomial rings,\\ $\mathbb{C}[u_1^2,\dots,u_{N+l}^2]^{S_{N+l}} \cong \mathbb{C}[s_1,\dots,s_{N+l}]$, with the isomorphism sending\\ $p_j\left(u_1^2,\dots,u_{N+l}^2\right) \rightarrow s_j$, $j=\{1,\dots,N+l\}$. Hence the system of $N+l$ equations in the isomorphic polynomial ring, $\mathbb{C}[s_1,\dots,s_{N+l}]$, has the following form,
\small\begin{equation*}\begin{split}
s_1 & =\mu_1^2+ \dots +\mu_{N}^{2}\\
s_2 & =\mu_1^4+ \dots +\mu_{N}^{4} \\
&\vdots\\
s_{N+l} & = \mu_1^{2(N+l)}+ \dots +\mu_{N}^{2(N+l)}
\end{split}\end{equation*}\normalsize
In this polynomial ring the system is linear and thus trivially only has one solution.\\
\\
Since the two rings are isomorphic, this means that the system in the ring $\mathbb{C}[u_1^2,\dots,u_{N+l}^2]^{S_{N+l}}$ contains one base solution, and every possible permutation of that base solution (since the polynomial ring is symmetric), leading to a total of $(N+l)!$ possible solutions. Since we already trivially know $(N+l)!$ solutions to the system, $u_{\sigma_j}^2= \mu_j^2$ for $j\in \{1,\dots,N\}$, and $u_{\sigma_k}^2= 0$ for $k\in \{N+1,\dots,N+l\}$, this means only the trivial solution exists for the first $N+l$ equations.\\
\\
\textbf{Considering the remaining $M-l-1$ equations.} Since the first $N+l$ equations uniquely solved for the $N+l$ independent variables, any remaining equations of the system are either solved automatically by the solution given by the first $N+l$ equations, or the system has no solution. In this case it is easy to note that the remaining $M-l-1$ equations are solved by the $(N+l)!$ solutions, thus proving the statement.  $\square$\\
\\
We now give the converse result.
\begin{proposition}
\label{ref2}
The system of $N+M-1$ polynomials,
\small\begin{equation*}\begin{split}
u^2_1+ \dots +u_{N}^{2} & =\mu^2_1+ \dots +\mu_{N+l}^{2} \\
u_1^4+ \dots +u_{N}^{4}  & =\mu^4_1+ \dots +\mu_{N+l}^{4} \\
&\vdots\\
u_1^{2(N+M-1)}+ \dots +u_N^{2(N+M-1)}& = \mu^{2(N+M-1)}_1+ \dots +\mu_{N+l}^{2(N+M-1)}
\end{split}\end{equation*}\normalsize
permits no solution unless $l$ of the $\mu_j^2$'s are exactly zero or $l=0$.
\end{proposition}
\textbf{Proof.} This proof of this statement is almost automatic, however, we shall proceed as before and apply the fundamental theorem of symmetric functions. We begin with the first $N$ equations.\\
\\
\textbf{The first $N$ equations.} We notice that the left hand side of the above system exists in the ring $\mathbb{C}[u_1^2,\dots,u_{N}^2]^{S_{N}}$, whereas the right hand side exists in the ring $\mathbb{C}[\mu_1^2,\dots,\mu_{N+l}^2]^{S_{N+l}}$. Using the following isomorphisms,
\small\begin{equation*}\begin{split}
\mathbb{C}[u_1^2,\dots,u_N^2]^{S_N} \cong \mathbb{C}[s_1,\dots,s_N]\\
p_j\left(u_1^2,\dots,u_{N}^2\right) \rightarrow s_j\textrm{  ,  }&j \in \{ 1,\dots, N \}\\
\mathbb{C}[\mu_1^2,\dots,\mu_{N+l}^2]^{S_{N+l}} \cong \mathbb{C}[t_1,\dots,t_{N+l}]\\
p_j\left(\mu_1^2,\dots,\mu_{N+l}^2\right) \rightarrow t_j\textrm{  ,  }& j \in \{ 1,\dots, N+l \}
\end{split}\end{equation*}\normalsize
the first $N$ equations become the simple linear system,
\small\begin{equation*}\begin{split}
s_1 & =t_1\\
s_2 & =t_2 \\
&\vdots\\
s_N & = t_N
\end{split}\end{equation*}\normalsize
which uniquely fixes the $N$ variables, $u_j^2$. To prove the result we need only look at one more equation.\\
\\
\textbf{The $(N+1)$th equation.} To express the left hand side of this equation in the ring $\mathbb{C}[s_1,\dots,s_N]$, we need to construct the $(N+1)$th symmetric power sum of $N$ variables from the previous $N$ symmetric power sums. Some simple examples being,
\small\begin{equation*}\begin{split}
N=1 \textrm{  ,  }& p_2\left( u_1^2 \right) = s^2_1 \\
N=2 \textrm{  ,  }&p_3 \left( u_1^2,u_2^2 \right)  = \frac{1}{2} \left( 3 s_1 s_2 - s^3_1 \right)\\
N=3 \textrm{  ,  }&p_4 \left( u_1^2,u_2^2 ,u_3^2 \right) = \frac{1}{6}\left( s^4_1 - 6 s^2_1 s_2 + 3s^2_2+8 s_1 s_3  \right)
\end{split}\end{equation*}\normalsize
Thus the $(N+1)$th equation looks like,
\small\begin{equation}\begin{split}
f(s_1,\dots, s_N) = t_{N+1} \\
\Rightarrow f(t_1,\dots, t_N) = t_{N+1}
\label{wrong}\end{split}\end{equation}\normalsize
Since $p_{N+1} \left(\mu_1^2,\dots ,\mu_{N+l}^2 \right)$ is algebraically independent of $\{p_1, \dots, p_N\}$, the above expression (eq. \ref{wrong}) is a contradiction, thus proving the proposition. $\square$\\
\\
Using the above results the following lemma comes almost automatically.
\begin{lemma}
Assume we have a particular family of unrestricted $\tau$-functions with particular $m$ and $n$ values,
\small\begin{equation}
\{\tau_{m+1}(\vec{x},\vec{y}), \tau_{m+2}(\vec{x},\vec{y}), \dots, \tau_{n}(\vec{x},\vec{y})\}
\label{taufamily}\end{equation}\normalsize
The process of restricting the entire family so that each $\tau$-function corresponds to a valid scalar product expression,
\small\begin{equation}\begin{split}
\{\gamma^{n-m-1}_1  \field{S}\left(\left.N=1\atop{M=n-m-1}\right| \vec{u}^i,\vec{v}^i \right), \gamma^{n-m-2}_2  \field{S}\left(\left.N=2\atop{M=n-m-2}\right| \vec{u}^{ii},\vec{v}^{ii} \right),\dots \\
 \dots, \gamma^{0}_{n-m}  \field{S}\left(\left.N=n-m\atop{M=0}\right| \vec{\mu},\vec{\nu} \right)\}
\label{scafamily}\end{split}\end{equation}\normalsize
where $\gamma^M_N =\left( \prod^N_{j=1} \frac{u_j}{v_j} \right)^M $, has potentially two (ill) effects.
\begin{itemize}
\item{If each of the above scalar product expressions has two sets of $N$ ($N$ is not constant for each scalar product) symmetric variables, then the 2 sets of $n-m-1=N+M-1$ time variables of the restricted $\tau$-functions are no longer equal, and therefore the bilinear identity is no longer valid.}
\item{If we enforce that the time variables be equal, then we only have two sets of \textbf{one symmetric variable} for each of the scalar product expressions.}
\end{itemize} 
Arguably both scenarios are pointless, so it makes sense to use the results of proposition \ref{illu} and only consider restricting one $\tau$-function in any family.
\end{lemma}
\textbf{Proof.} Using the results of proposition \ref{illu} on all the unrestricted $\tau$-functions in eq. \ref{taufamily} we instantly arrive to the expression in eq. \ref{scafamily}. Additionally, analyzing the scalar product expressions as they are (with two sets of $N$ symmetric variables), the results of propositions \ref{ref1} and \ref{ref2} state that the symmetric power sums, and hence the time variables, cannot be equal. Thus the first point in this lemma becomes obvious. Furthermore, we obtained from propositions \ref{ref1} and \ref{ref2} that the only way for the time variables to be equal is if we trivialize the power sums as indicated in point two of this lemma. $\square$
\section{Analysis of the Toda wave-vectors}
In this section we shall show that the wave-functions associated with the $\tau$-functions that are generated by the scalar product give an alternative method to calculating a certain class of correlation functions, and thus have a natural combinatorial meaning. However, in order to proceed we shall first give the definition of an inner product in the ring of symmetric polynomials which naturally leads to the necessary definition of skew Schur polynomials.\\
\\
\textbf{Orthogonality.} We define the inner product of two symmetric polynomials, $\langle f_1(\vec{u}) , f_2(\vec{u}) \rangle$, in the symmetric polynomial ring, $\mathbb{C}[u_1,\dots,u_N]^{S_N}$, as the following quantity\footnote{See chapter 3 of \cite{Virasoro}.},
\small\begin{equation}\begin{split}
\langle f_1(\vec{u}),f_2(\vec{u}) \rangle & =  \langle f_1(\vec{x}),f_2(\vec{x}) \rangle\\
& \equiv   \lim_{\vec{x} \rightarrow \vec{0}} f_1 (\tilde{\partial}_{\vec{x}}) f_2(\vec{x})
\end{split}\end{equation}\normalsize
where $x_j = \frac{1}{j}p_j(\vec{u})$ and $\tilde{\partial}_{\vec{x}} =\left( \partial_{x_1}, \frac{1}{2}\partial_{x_2}, \frac{1}{3}\partial_{x_3}, \dots\right)$. The Schur polynomials, like the other symmetric polynomials (complete, power sum, elementary) mentioned in this work, form a complete basis for the symmetric polynomial ring $\mathbb{C}[u_1,\dots,u_N]^{S_N}$. The Schur polynomials are special in this regard however as they form an orthonormal basis for the ring,
\small\begin{equation*}
\langle S_{\{\lambda \}}(\vec{u}),S_{\{\mu \}}(\vec{u}) \rangle = \langle \chi_{\{\lambda \}}(\vec{x}),\chi_{\{\mu \}}(\vec{x})\rangle = \delta_{\{\lambda \}\{\mu \}}
\end{equation*}\normalsize
In the following sections we shall call upon an extremely helpful inner product identity given as the following\footnote{See chapter 5 of \cite{Virasoro}.},
\small\begin{equation}
\langle \chi_{\{ \lambda \}}(\vec{x}),k x_k \chi_{\{ \mu \}}(\vec{x}) \rangle = \langle \partial_{x_k} \chi_{\{ \lambda \}}(\vec{x}), \chi_{\{ \mu \}}(\vec{x})\rangle 
\label{helpful}\end{equation}\normalsize\\
\textbf{Skew Schur polynomial.} Given a set of variables $\{u_1,\dots,u_N\}$ and three partitions $\{\lambda\}$, $\{\mu\}$ and $\{\nu \}$ such that $\{\mu\} \subseteq \{\lambda\}$, the skew Schur polynomial $S_{\{\lambda\} / \{ \mu\}} (u_1,\dots,u_N)$ is defined as,
\small\begin{equation}
\langle S_{\{\lambda\} /\{\mu\}}(\vec{u}), S_{\{ \nu \}}(\vec{u}) \rangle = \langle S_{\{\lambda\}}(\vec{u}),S_{\{ \mu \}}(\vec{u}) S_{\{ \nu \}}(\vec{u}) \rangle
\end{equation}\normalsize
It is possible to expand the product of Schur polynomials, $S_{\{ \mu \}}(\vec{u}) S_{\{ \nu \}}(\vec{u})$, as a linear sum of Schur polynomials,
\small\begin{equation}
S_{\{ \mu \}}(\vec{u}) S_{\{ \nu \}}(\vec{u}) = \sum_{\{\gamma \}\atop{|\gamma| = |\mu|+|\nu|}} c^{\{\gamma \}}_{\{ \mu\} \{\nu \}} S_{\{\gamma \}}(\vec{u})
\end{equation}\normalsize
where the positive integers $c^{\{\gamma \}}_{\{ \mu\} \{\nu \}}$ are known as \textit{Littlewood-Richardson coefficients} and they can be derived combinatorially\footnote{For further information see section I.9 of \cite{MacD}.}. The sum then is over all possible partitions which have non zero Littlewood-Richardson coefficients. Putting the above expansion into the expression for skew Schur polynomials, one immediately obtains the expression,
\small\begin{equation}
S_{\{\lambda\} /\{\mu\}}(\vec{u}) = \sum_{\{\gamma \} \subseteq \{ \lambda \} \atop{|\gamma| = |\lambda|-|\mu|}} c^{\{\lambda \}}_{\{ \mu\} \{\gamma \}} S_{\{\gamma \}}(\vec{u})
\end{equation}\normalsize
Again, the sum is over all possible partitions which have non zero Littlewood-Richardson coefficients. The combinatorial definition of the skew Schur polynomial is given by,
\small\begin{equation}
S_{\{\lambda\} /\{\mu\}}(\vec{u})  = \sum_{T^{\{ \lambda - \mu \}}_+} u^{t_1}_1u^{t_2}_2 \dots u^{t_N}_N=\sum_{T^{\{ \lambda - \mu \}}_-} u^{t_1}_1u^{t_2}_2 \dots u^{t_N}_N
\end{equation}\normalsize
where the sum is given over all possible (ascending or descending) semi-standard (column strict) skew tableaux of shape $\{\lambda - \mu \}$, and the $t_j$ give the weights of the tableau (the amount of times $j$ appears in the skew partition).\\
\\
A more convenient expression for the skew Schur polynomials is given by,
\small\begin{equation}
S_{\{\lambda\} /\{\mu\}} (u_1,\dots,u_N) = \textrm{det}[h_{\lambda_i - \mu_j +j-i}(u_1,\dots,u_N)]^N_{i,j=1}
\end{equation}\normalsize
where we notice explicitly that $S_{\{\lambda\} / \{ \mu\}}(\vec{u})=0$ unless $\{\mu \} \subseteq \{\lambda \}$.\\
\\
Performing a Miwa change of variables to the skew Schur polynomial, $\frac{1}{j}p_j(\vec{u}) \rightarrow x_k$, transforms each complete symmetric polynomial to the corresponding one row character polynomial. Hence, we also define the skew character polynomial, $\chi_{\{\lambda\} / \{ \mu\}} (x_1,\dots,x_N)$, as,
\small\begin{equation}
\chi_{\{\lambda\} / \{ \mu\}} (x_1,\dots,x_N)=\textrm{det}[\zeta_{\lambda_i - \mu_j +j-i}(x_1,\dots,x_N)]^N_{i,j=1}
\end{equation}\normalsize
\subsection{Examining the first class of wave-function}
We have two classes of wave-function to consider and we shall begin by considering the $\hat{w}^{(0)}$ class first, as it requires the least amount of work. Using the definitions given in the first section we have,
\small\begin{equation*}\begin{split}
\tau(s) \hat{w}^{(0)}_{k}(s) &= \zeta_k(-\tilde{\partial}_{\vec{y}})\tau(s+1)\\
&= \sum_{\{\lambda \} \subseteq (n-(s+1))^{((s+1)-m)}} \chi_{\{\lambda \}} (\vec{x})\zeta_k(-\tilde{\partial}_{\vec{y}}) \chi_{\{\lambda \}} (-\vec{y}) 
\end{split}\end{equation*}\normalsize
We now require the following result.
\begin{proposition}
\small\begin{equation}
\zeta_j (-\tilde{\partial}_{\vec{y}}) \chi_{\{\lambda \}}(-\vec{y}) = \chi_{\{\lambda \}/\{j\}} (-\vec{y})
\label{nice1}\end{equation}\normalsize
for all partitions $\{\lambda\}$ such that $\{j \} \subseteq \{\lambda\}$.
\end{proposition}
\textbf{Proof.} Consider the inner product of $\zeta_j (-\tilde{\partial}_{\vec{y}}) \chi_{\{\lambda \}}(-\vec{y})$ with a general character polynomial $\chi_{\{\mu \}}(-\vec{y})$. Through the application of eq. \ref{helpful}, the polynomial of differential operators applied to $\chi_{\{\lambda \}}(-\vec{y})$ becomes the equivalent polynomial of simple variables (as opposed to differential operators) multiplied by $\chi_{\{\mu \}}(-\vec{y})$,
\small\begin{equation*}
\langle \zeta_j (-\tilde{\partial}_{\vec{y}}) \chi_{\{\lambda \}}(-\vec{y}) , \chi_{\{\mu \}}(-\vec{y}) \rangle  = \langle \chi_{\{\lambda \}} (-\vec{y}),\zeta_j (-\vec{y}) \chi_{\{\mu \}}(-\vec{y}) \rangle
\end{equation*}\normalsize
The expression $\zeta_j (-\vec{y})$ can be written as a character polynomial with a partition containing a single entry of $j$,
\small\begin{equation*}
\langle \zeta_j (-\tilde{\partial}_{\vec{y}}) \chi_{\{\lambda \}} (-\vec{y}), \chi_{\{\mu \}} (-\vec{y})\rangle  = \langle \chi_{\{\lambda \}}(-\vec{y}) ,\chi_{\{j \}} \chi_{\{\mu \}}(-\vec{y}) \rangle
\end{equation*}\normalsize
The final step simply applies the original definition of a skew Schur polynomial given earlier,
\begin{eqnarray*}
\langle \chi_{\{\lambda \}}(-\vec{y}) ,\chi_{\{j \}}(-\vec{y}) \chi_{\{\mu \}} (-\vec{y})\rangle &=& \langle \chi_{\{\lambda \}/\{j \}} (-\vec{y}), \chi_{\{\mu \}}(-\vec{y}) \rangle\\
\Rightarrow \zeta_j (-\tilde{\partial}_{\vec{y}}) \chi_{\{\lambda \}}(-\vec{y})& =&  \chi_{\{\lambda \}/\{j \}} (-\vec{y}) \textrm{    } \square
\end{eqnarray*}
Therefore, with regards to the first class of wave-functions we have the following result,
\small\begin{equation}
\tau(s) \hat{w}^{(0)}_{k}(s) = \sum_{\{\lambda \} \subseteq (n-(s+1))^{((s+1)-m)}} \chi_{\{\lambda \}} (\vec{x}) \chi_{\{\lambda \}/\{k \}} (-\vec{y}) 
\label{nice2}\end{equation}\normalsize
Thus the upper triangular wave-matrix, $\hat{W}^{(0)}(\vec{x},\vec{y})$, has entries of the form,
\small\begin{equation*}\begin{split}
\hat{W}^{(0)}(\vec{x},\vec{y})&=  \left( \hat{w}^{(0)}_{k-j}(j,\vec{x},\vec{y}) \right)^{n-1}_{j,k= m}\\
&= \left( \frac{1}{\tau(j)}\sum_{\{\lambda \} \subseteq (n-(j+1))^{((j+1)-m)}} \chi_{\{\lambda \}} (\vec{x}) \chi_{\{\lambda \}/\{k-j \}} (-\vec{y}) \right)^{n-1}_{j,k= m}
\end{split}\end{equation*}\normalsize\\
\textbf{The infinite lattice with a free end.} An interesting quirk appears in this result when we let $n \rightarrow \infty$, which is known as dealing with an infinite lattice with a free end\footnote{See section 5.2 of \cite{Toda3}.}. Taking this limit and then expanding the skew polynomial as a linear sum we receive,
\small\begin{equation*}\begin{split}
\tau(s) \hat{w}^{(0)}_{k}(s) & =\sum_{\{\lambda \} \subseteq (\infty)^{((s+1)-m)}} \chi_{\{\lambda \}} (\vec{x}) \chi_{\{\lambda \}/\{k \}} (-\vec{y}) \\
&= \sum_{\{\lambda \} \subseteq (\infty)^{((s+1)-m)}}\chi_{\{\lambda \}} (\vec{x})  \left(\sum_{\{\nu \} \subseteq (\infty)^{((s+1)-m)}} c^{\{\lambda \}}_{\{k \} \{\nu \}}\chi_{\{\nu \}} (-\vec{y}) \right)  \\
&= \sum_{\{\nu \} \subseteq (\infty)^{((s+1)-m)}}  \chi_{\{\nu \}} (-\vec{y}) \left(\sum_{\{\lambda \} \subseteq (\infty)^{((s+1)-m)}} c^{\{\lambda \}}_{\{k \} \{\nu \}}\chi_{\{\lambda \}} (\vec{x}) \right)  \\
& =  \chi_{\{k \}} (\vec{x}) \sum_{\{\nu \} \subseteq (\infty)^{((s+1)-m)}}  \chi_{\{\nu \}} (\vec{x}) \chi_{\{\nu \}} (-\vec{y})\\
& =  \chi_{\{k \}} (\vec{x}) \tau(s+1)
\end{split}\end{equation*}\normalsize
In this case we see that the skew character polynomial decouples and we simply receive $\tau(s+1)$ multiplied by a factor of $\chi_{\{k \}} (\vec{x})$. This case shall be considered as nothing more than an observational quirk, and from now on we shall continue in the finite case where the skew in the partition remains.\\
\\
\textbf{Constructing skew $N$-particle conjugate state vectors.} Consider the following conjugate state vector,
\small\begin{equation*}\begin{split}
\langle 0| \phi_k C(v_2)  \dots C(v_{N}) &=\langle k|  C(v_2)  \dots C(v_{N})\\
&= \langle \Psi^{\{k \}}_M (v_2,\dots,v_{N})|
\end{split}\end{equation*}\normalsize
where the partition $\{k\}$, which consists of one entry, is constructed in the usual manner from the occupation numbers.\\
\\
We now have the following result regarding the allowable partitions of this particular conjugate state vector,
\begin{proposition}
\small\begin{equation*}
\langle \Psi^{\{k \}}_M (v_2,\dots,v_{N})|= \sum_{\{\lambda\} \subseteq \{(M)^{ (N-1)},k\} \atop{\{\lambda\} \supseteq  \{ k \}}} \psi^{(1,k)}_{\{\lambda \}}(v_2,\dots,v_{N}) \langle \lambda|
\end{equation*}\normalsize
\end{proposition}
\textbf{Proof.} Consider the non crossing column strict lattice path interpretation of the state vectors. The operator $\phi_k$ assures us that the first path in the first column makes a directional change from north to east at row $k$. This has the effect that the occupation number sequence will contain at least one entry $n_l$, where $l \ge k$. Transforming the occupation number sequence to a partition $\{ \lambda \}$, we instantly receive the result, $\{ \lambda \} \supseteq \{ k \}$.\\
\\
The fact that the first path in the first column turns east at row $k$ also means that the highest row that the $N$th path can be when it crosses between column $N$ and $N+1$ is $k$. Thus the highest partition obtainable from lattice paths under this restriction are $\{\lambda \} =  \{(M)^{(N-1)},k\}.$   $\square$\\
\\
We now give the following combinatorial definitions of $\psi^{(1,k)}_{\{\lambda \}}(v_2,\dots,v_{N})$. Considering the \textbf{lattice path} interpretation we receive,
\small\begin{equation*}
\psi^{(1,k)}_{\{\lambda \}}(v_2,\dots,v_{N}) = \sum_{\textrm{allowable paths in}\atop{\textrm{$(M+1)\times N$ lattice$^\dagger$}}} v^{t^{d}_2-t^{a}_2}_2 \dots v^{t^{d}_N-t^{a}_N}_N
\end{equation*}\normalsize
where the lattice paths are under the condition that the first path in the first column makes a directional change from north to east at row $k$, and the powers $t^d_j$ and $t^a_j$ give the total amount of $d$ and $a$ vertices in column $j$ respectively.\\
\\
Considering the \textbf{plane partition} interpretation we receive,
\small\begin{equation*}
\psi^{(1,k)}_{\{\lambda \}}(v_2,\dots,v_{N}) = \sum_{\textrm{lower plane part.}\atop{\textrm{in $N \times N\times M$ array$^\dagger$}}}  v^{l^{d}_2-l^{a}_2}_2 \dots v^{l^{d}_N-l^{a}_N}_N
\end{equation*}\normalsize
where the lower plane partitions are under the condition that the entry $\pi_{N,1}$ is equal to $k$, and the powers $l^d_j$ and $l^a_j$ give the total amount of $d$ and $a$ rhombi in column $j$ respectively.\\
\\
Finally, considering the ascending \textbf{Young tableaux} interpretation, we notice that whenever we transform from the lower plane partition to the Young tableau, the fact that $\pi_{N,1}=k$, means that the weight $t_1$ is always equal to $k$. Since the weight $t_1$ does not enter the equation, as $v_1$ is not present, we can simply consider the skew partition $\{ \lambda - k\}$ to generate the tableaux, leading to,
\small\begin{equation*}\begin{split}
\psi^{(1,k)}_{\{\lambda \}}(v_2,\dots,v_{N})  &= \sum_{T^{\{ \lambda -k \}}_+ }  v^{-2 t_2+M}_2 \dots v^{-2 t_N+M}_N\\
&= (v_2 \dots v_N)^M \sum_{T^{\{ \lambda - k \}}_+ }  \left( v^{-2}_2 \right)^{t_2} \dots \left( v^{-2}_N \right)^{t_N}\\
&=  (v_2 \dots v_N)^M S_{\{\lambda \}/\{k \}} (v^{-2}_2, \dots, v^{-2}_N)
\end{split}\end{equation*}\normalsize
Therefore, the skew $N$-particle conjugate state vector is,
\small\begin{equation}
\langle \Psi^{\{k \}}_M (v_2,\dots,v_{N})| =\left( \prod^N_{j=2} v_j \right)^M   \sum_{\{\lambda\} \subseteq \{(M)^{ (N-1)},k\} \atop{\{\lambda\} \supseteq  \{ k \}}}  S_{\{\lambda \}/\{k \}} (v^{-2}_2, \dots, v^{-2}_N)  \langle \lambda|
\end{equation}\normalsize
\textbf{Correlation functions and the wave-vector at $\mathbf{s=n-M-1 = m+N-1}$ as a weighted sum.} Consider then the following correlation function,
\small\begin{equation}\begin{split}
&\langle \Psi^{\{k \}}_M (v_2,\dots,v_{N})|\Psi_M (u_1,\dots,u_{N})\rangle \\
=& \langle 0| \phi_k C(v_2)  \dots C(v_{N}) B(u_1)  \dots B(u_{N})|0\rangle\\
=&\left( \frac{\prod^N_{j=2} v_j}{\prod^N_{j=1} u_j} \right)^M   \sum_{\{\lambda\} \subseteq \{(M)^{ (N-1)},k\} \atop{\{\mu \} \subseteq  (M)^{ (N)}}}  S_{\{\mu \}} (u^{2}_1, \dots, u^{2}_N) S_{\{\lambda \}/\{k \}} (v^{-2}_2, \dots, v^{-2}_N)  \langle \lambda|\mu \rangle\\
=&\left( \frac{\prod^N_{j=2} v_j}{\prod^N_{j=1} u_j} \right)^M   \sum_{\{\lambda\} \subseteq \{(M)^{(N-1)},k\} \atop{\{\lambda\} \supseteq  \{ k \}}}  S_{\{\lambda \}} (u^{2}_1, \dots, u^{2}_N) S_{\{\lambda \}/\{k \}} (v^{-2}_2, \dots, v^{-2}_N)
\end{split}\end{equation}\normalsize
which calculates all the weighted non crossing column strict lattice paths on an $(M+1)\times 2N$ grid with the first path in the first column turning east at row $k$. Compare it now to any of the wave-functions that we calculated earlier,
\small\begin{equation*}
\tau(s) \hat{w}^{(0)}_{k}(s) = \sum_{\{\lambda \} \subseteq (n-(s+1))^{((s+1)-m)}\atop{\{\lambda \} \supseteq \{k \}}} \chi_{\{\lambda \}} (\vec{x}) \chi_{\{\lambda \}/\{k \}} (-\vec{y})
\end{equation*}\normalsize
and concentrate now on the particular row, $s=n-M-1 = m+N-1$, of the wave-matrix. If we restrict the variables as the following,
\small\begin{equation}
x_k \rightarrow \frac{1}{k}p_j\left( \left\{ u^2_j \right\} \right) \textrm{  ,   }-y_k \rightarrow \frac{1}{k}p_j\left( \left\{ v^{-2}_j \right\} \right) \textrm{  ,   } 1 \le k \le N+M-1
\label{restrictq}\end{equation}\normalsize
we immediately obtain,
\small\begin{equation*}\begin{split}
&\tau(n-M-1) \hat{w}^{(0)}_{k}(n-M-1)\\ 
= &\sum_{\{\lambda \} \subseteq (M)^{(N)}\atop{\{\lambda \} \supseteq \{k \}}} S_{\{\lambda \}} (u^2_1,\dots,u^2_N) S_{\{\lambda \}/\{k \}} (v^{-2}_1,\dots,v^{-2}_N).
\end{split}\end{equation*}\normalsize
Now consider the limit $v_1 \rightarrow \infty$. In this limit we obviously obtain,
\small\begin{equation*}
S_{\{\lambda \}/\{k \}} (v^{-2}_1,\dots,v^{-2}_N) \rightarrow S_{\{\lambda \}/\{k \}} (v^{-2}_2,\dots,v^{-2}_N),
\end{equation*}\normalsize
however subtle effects also appear in the summation. Recall the combinatorial definition of the skew Schur polynomial,
\small\begin{equation*}
S_{\{\lambda \}/\{k \}} (v^{-2}_2,\dots,v^{-2}_N) = \sum_{T^{\{\lambda - k \}}_+} \left(v^{-2}_2 \right)^{t_2} \dots \left(v^{-2}_N \right)^{t_N}
\end{equation*}\normalsize
where the sum is given over all possible column strict skew tableaux of shape $\{\lambda - k \}$, and the $t_j$ give the amount of times $j$ appears in the skew partition. In the above case, $j=\{2,\dots,N\}$, thus the total length of any column in the skew partition cannot be greater than $N-1$, otherwise the Young tableau will not be column strict. Therefore, when $\{\lambda\} \subseteq (M)^{ (N)}$, for all the columns in the skew partition $\{\lambda - k\}$, to be no greater than $N-1$ in length we obtain the new restricted condition, $\{\lambda\} \subseteq \{ (M)^{ (N-1)},k\}$.\\
\\
Thus the wave-vector, given by the $s=n-M-1 = m+N-1$ row of the wave-matrix, in the $v_1 \rightarrow \infty$ limit,
\small\begin{equation*}\begin{split}
&\lim_{v_1 \rightarrow \infty} \left( \tau(n-M-1) \hat{w}^{(0)}_{k}(n-M-1) ) \right)^{M}_{k= 0}\\
=& \left(\sum_{\{\lambda \} \subseteq \{(M)^{(N-1)},k\}\atop{\{\lambda\} \supseteq  \{ k\}}} S_{\{\lambda \}} (u^2_1,\dots,u^2_N ) S_{\{\lambda \}/\{k \}} (v^{-2}_2,\dots,v^{-2}_N ) \right)^{M}_{k= 0}\\
=&\left( \frac{\prod^N_{j=1} u_j}{\prod^N_{j=2} v_j} \right)^M  \left( \langle \Psi^{\{k \}}_M (v_2,\dots,v_{N})|\Psi_M (u_1,\dots,u_{N}) \rangle \right)^{M}_{k= 0}
\end{split}\end{equation*}\normalsize
gives exactly (up to a multiplicative factor) all the weighted non crossing column strict lattice paths on an $(M+1)\times 2N$ lattice with the first path in the first column turning right at row $k$, $0 \le k \le M$.\\
\\
\textbf{Single determinant form for the wave-functions.} When initially discussing the scalar product, $\field{S}(N,M| \vec{u},\vec{v})$, it was stated that when using the method of algebraic Bethe ansatz, we could obtain a single determinant form of the scalar product given by,
\small\begin{equation*}\begin{split}
\field{S}(N,M| \vec{u},\vec{v})& = \left\{ \prod_{1 \le j < k \le N} \left( \frac{u_j u_k}{u^2_j-u^2_k}\right) \left( \frac{v_j v_k}{v^2_j-v^2_k}\right)  \right\} \left( \prod^N_{m,l=1} \frac{1}{u_m v_l} \right)^{M+N-1}\\
&  \times \textrm{det} \left[ h_{M+N-1}(u^2_m,v^2_l)\right]^N_{l,m=1}
\end{split}\end{equation*}\normalsize
From this expression, it is possible to obtain a single determinant form for the wave-functions given above\footnote{The details below are given in section VI of \cite{phase2} to obtain single determinant expressions of 1-point correlation functions for the phase model. We expand upon these results in the next subsection to obtain single determinant expressions of $n$-point correlation functions for the model.}.\\
\\
\textbf{Polynomial expansion of the scalar product.} To achieve this, we first examine the operator $C(v)$ briefly. More explicitly, we are interested in the parts of $C(v)$ that contain only $\phi_j$ operators,
\small\begin{equation}
C(v) = \sum^M_{j=0} v^{M-2j} \phi_j + \textrm{terms that contain operators $\phi^{\dagger}_j$}
\end{equation}\normalsize
Thus when $C(v)$ acts on the conjugate vacuum,
\small\begin{equation}
\langle 0|C(v) =v^{M} \sum^M_{j=0} v^{-2j} \langle 0| \phi_j 
\end{equation}\normalsize
we can obtain the scalar product as the following weighted linear sum of correlation functions,
\small\begin{equation}\begin{split}
\field{S}(N,M| \vec{u},\vec{v}) & =  \langle 0|C(v_1)\dots C(v_N)B(u_1) \dots B(u_N)|0 \rangle \\
&=v^{M}_1 \sum^M_{j=0} v^{-2j}_1 \langle 0| \phi_j C(v_2)\dots C(v_N)B(u_1) \dots B(u_N)|0 \rangle \\
& = v^{M}_1 \sum^M_{j=0} v^{-2j}_1  \langle \Psi^{\{j \}}_M (v_2,\dots,v_{N})|\Psi_M (u_1,\dots,u_{N})\rangle 
\end{split}\end{equation}\normalsize
Therefore, if we expand the single matrix form for the scalar product as a polynomial in $v^2_1$, the coefficients will reveal a single matrix form for the correlation functions/wave-functions. The remaining part of this section describes the procedure to do this.\\
\\
\textbf{Polynomial expansion of the determinant.} We begin by relabeling the scalar product as,
\small\begin{equation}
\field{S}(N,M| \vec{u},\vec{v}) = \frac{ \Omega_{\hat{v}_1} }{ v^M_1} \left\{ \prod_{1\le j < k \le N} \frac{1}{v^2_j-v^2_k}\right\}  \textrm{det} \left[ h_{M+N-1}(u^2_k,v^2_j)\right]^N_{j,k=1}
\end{equation}\normalsize
where,
\small\begin{equation}
\Omega_{\hat{v}_1}= \left\{ \prod_{1 \le j < k \le N}  \frac{1}{u^2_j-u^2_k}\right\} \left( \prod^N_{m=1}\prod^N_{l=2} \frac{1}{u_m v_l} \right)^{M}
\end{equation}\normalsize
It is apparent that all the uninteresting multiplicative factors have been bundled into $\Omega_{\hat{v}_1}$. We will now proceed to eliminate the factor $\left\{ \prod_{1\le j < k \le N} \frac{1}{v^2_j-v^2_k}\right\} $.\\
\\
Consider subtracting the $N$th row in the determinant from the $j_1$th row, $1\le j_1 \le N-1$, to obtain,
\small\begin{equation}\begin{split}
&h_{M+N-1}(u^2_k,v^2_{j_1})-h_{M+N-1}(u^2_k,v^2_N)\\
=&  \sum^{M+N-1}_{p=0}\left\{ \left(v^2_{j_1} \right)^{p} - \left(v^2_N \right)^{p} \right\} \left( u^2_k \right)^{M+N-1-p}\\
=&\left(v^2_{j_1} -v^2_N  \right) \sum^{M+N-1}_{p_1=1} \sum^{p_1-1}_{p_2=0}  \left( v^2_{j_1} \right)^{p_2} \left( v^2_N \right)^{p_1-1-p_2} \left( u^2_k \right)^{M+N-1-p_1}\\
=& \left(v^2_{j_1}-v^2_N  \right) \sum_{p_1,p_2,p_3\atop{p_1+p_2+p_3=M+N-2}} \left(v^2_{j_1} \right)^{p_1} \left(v^2_N \right)^{p_2} \left(u^2_m \right)^{p_3}\\
= & \left(v^2_{j_1}-v^2_N  \right) h_{M+N-2}(u^2_k,v^2_{j_1},v^2_N)
\label{idenhhh}\end{split}\end{equation}\normalsize
We then take out the factor of $\prod^{N-1}_{j_1=1}\left(v^2_{j_1}-v^2_N  \right)$ from the determinant and eliminate the corresponding factor on the denominator. Thus the determinant expression becomes,
\small\begin{equation*}\begin{split}
\left\{ \prod_{1\le j < k \le N-1} \frac{1}{v^2_j-v^2_k}\right\} \textrm{det}\left[ \begin{array}{c}
h_{M+N-2}(u^2_k,v^2_{j},v^2_N)\\
h_{M+N-1}(u^2_k,v^2_N)
\end{array} \right]_{j=1,\dots, N-1\atop{k=1,\dots, N}}
\end{split}\end{equation*}\normalsize
\textbf{A necessary identity.} Briefly notice that the identity in eq. \ref{idenhhh} can easily be generalized to the following form,
\small\begin{equation}
h_{p}( \{v^2 \}, v^2_{j})-h_{p}(\{v^2 \},v^2_{k}) =  \left(v^2_{j}-v^2_k  \right) h_{p-1}(\{v^2 \}, v^2_{j},v^2_{k})
\end{equation}\normalsize
where $\{v^2_{j},v^2_{k} \} \notin \{v^2 \}$. We shall use this identity frequently in the work below.\\
\\
We now subtract the $(N-1)$th row in the determinant from the $j_2$th row, $1\le j_2 \le N-2$, to receive,
\small\begin{equation*}\begin{split}
& h_{M+N-2}(u^2_k,v^2_{j_2},v^2_N)-h_{M+N-2}(u^2_k,v^2_{N-1},v^2_N)\\
= & \left(v^2_{j_2}-v^2_{N-1}  \right) h_{M+N-3}(u^2_k,v^2_{j_2},v^2_{N-1},v^2_N)
\end{split}\end{equation*}\normalsize
Eliminating the factor of $\prod^{N-2}_{j_2=1}\frac{1}{\left(v^2_{j_2}-v^2_{N-1}  \right)}$ accordingly we obtain,
\small\begin{equation*}\begin{split}
 \left\{ \prod_{1\le j < k \le N-2} \frac{1}{v^2_j-v^2_k}\right\} \textrm{det}\left[ \begin{array}{c}
h_{M+N-3}(u^2_k,v^2_{j},v^2_{N-1},v^2_N)\\
h_{M+N-2}(u^2_k,v^2_{N-1},v^2_N)\\
h_{M+N-1}(u^2_k,v^2_N)
\end{array} \right]_{j=1,\dots, N-2\atop{k=1,\dots, N}}
\end{split}\end{equation*}\normalsize
By now the general procedure should be crystal clear. Performing this procedure generally an $m$ number of times,  $1 \le m \le N-1$, we eliminate the multiplicative factor, $\left\{\prod^m_{k=1}\prod^{N-m}_{j=1} \frac{1}{(v^2_j-v^2_{N+1-k})}\right\}$, and the determinant reads,
\small\begin{equation*}\begin{split}
 \left\{ \prod_{1\le j < k \le N-m} \frac{1}{v^2_j-v^2_k}\right\} \textrm{det}\left[ \begin{array}{c}
h_{M+N-(m+1)}(u^2_k,v^2_{j},v^2_{N-m},\dots,v^2_N)\\
h_{M+N-m}(u^2_k,v^2_{N-m},\dots,v^2_N)\\
\vdots \\
h_{M+N-2}(u^2_k,v^2_{N-1},v^2_N)\\
h_{M+N-1}(u^2_k,v^2_N)
\end{array} \right]_{j=1,\dots, N-m\atop{k=1,\dots, N}}
\end{split}\end{equation*}\normalsize
Performing the final step, $m = N-1$, we have completely eliminated the multiplicative factor of $\left\{ \prod_{1\le j < k \le N} \frac{1}{v^2_j-v^2_k}\right\} $ and only the top row has terms containing $v_1$. For convenience we relabel this new determinant as,
\small\begin{equation}
\textrm{det}[\Lambda_{j,k}]^N_{j,k=1} =  \textrm{det}\left[ h_{M-1+j}(u^2_k,v^2_{j},v^2_{j+1},\dots,v^2_N) \right]^N_{j,k=1}
\end{equation}\normalsize
It is apparent that only the top row of this determinant contains the variable $v^2_1$. Expressing the complete homogeneous symmetric function(s) in the top row as a polynomial in $v_1$,
\small\begin{equation*}
h_{M}(u^2_k,v^2_{1},\dots,v^2_N) = \sum^M_{q=0} \left(v^2_1 \right)^{q} h_{M-q}(u^2_k,v^2_{2},\dots,v^2_N)
\end{equation*}\normalsize
we expand the determinant along the first row to receive,
\small\begin{equation*}\begin{split}
\textrm{det}[\Lambda_{j,k}]^N_{j,k=1} &=  \textrm{det}\left[ \begin{array}{c}
\sum^M_{q=0} \left(v^2_1 \right)^{q} h_{M-q}(u^2_k,v^2_{2},\dots,v^2_N)\\
h_{M-1+j}(u^2_k,v^2_{j},v^2_{j+1},\dots,v^2_N)
\end{array} \right]_{j=2,\dots,N\atop{k=1,\dots, N}}\\
&= \sum^N_{r=1}(-1)^{r+1} \sum^M_{q=0} \left(v^2_1 \right)^{q} h_{M-q}(u^2_r,v^2_{2},\dots,v^2_N) \textrm{det}[\Lambda_{j,k}]_{j=2 \dots,N \atop{k=1,\dots,\hat{r},\dots,N}}\\
&= \sum^M_{q=0} \left(v^2_1 \right)^{q} \left\{ \sum^N_{r=1}(-1)^{r+1}   h_{M-q}(u^2_r,v^2_{2},\dots,v^2_N) \textrm{det}[\Lambda_{j,k}]_{j=2 \dots,N \atop{k=1,\dots,\hat{r},\dots,N}} \right\}\\
&= \sum^M_{q=0} \left(v^2_1 \right)^{q} \textrm{det}[\Lambda^{(q)}_{j,k}]^N_{j,k=1}
\end{split}\end{equation*}\normalsize
where,
\small\begin{equation}\begin{split}
\textrm{det}[\Lambda^{(q)}_{j,k}]^N_{j,k=1} &= \sum^N_{r=1}(-1)^{r+1}   h_{M-q}(u^2_r,v^2_{2},\dots,v^2_N) \textrm{det}[\Lambda_{j,k}]_{j=2 \dots,N \atop{k=1,\dots,\hat{r},\dots,N}}\\
& =  \textrm{det}\left[ \begin{array}{c}
  h_{M-q}(u^2_k,v^2_{2},\dots,v^2_N)\\
h_{M-1+j}(u^2_k,v^2_{j},v^2_{j+1},\dots,v^2_N)
\end{array} \right]_{j=2,\dots,N\atop{k=1,\dots, N}}
\end{split}\end{equation}\normalsize
Putting everything together, we obtain,
\small\begin{equation}
\field{S}(N,M| \vec{u},\vec{v}) =  v^M_1  \sum^M_{q=0} \left(v^{-2}_1 \right)^{q} \Omega_{\hat{v}_1}  \textrm{det}[\Lambda^{(M-q)}_{j,k}]^N_{j,k=1} 
\label{bigom}\end{equation}\normalsize
which gives us a single determinant form for the (restricted) wave-functions,
\small\begin{equation*}\begin{split}
\Omega_{\hat{v}_1}  \textrm{det}[\Lambda^{(M-q)}_{j,k}]^N_{j,k=1} &= \langle \Psi^{\{q \}}_M (v_2,\dots,v_{N})|\Psi_M (u_1,\dots,u_{N})\rangle\\
&=\left( \frac{\prod^N_{j=2} v_j}{\prod^N_{j=1} u_j} \right)^M  \lim_{v_1 \rightarrow \infty}\tau(n-M-1) \hat{w}^{(0)}_{q}(n-M-1) 
\end{split}\end{equation*}\normalsize\\
\textbf{An alternative form.} It is possible to undo all the operations that have been applied to all the rows of the determinant, save the first row, to obtain the alternative form to eq. \ref{bigom},
\small\begin{equation}
\field{S}(N,M| \vec{u},\vec{v}) =   v^M_1  \sum^M_{q=0} \left(v^{-2}_1 \right)^{q} \tilde{\Omega}_{\hat{v}_1}  \textrm{det}[\tilde{\Lambda}^{(M-q)}_{j,k}]^N_{j,k=1} 
\end{equation}\normalsize
where,
\small\begin{equation}\begin{split}
\textrm{det}[\tilde{\Lambda}^{(M-q)}_{j,k}]^N_{j,k=1} &= \textrm{det}\left[ \begin{array}{c}
h_{q}(u^2_k,v^2_{2},\dots,v^2_N)\\
h_{M-1}(u^2_k,v^2_{j})
\end{array} \right]_{j=2,\dots,N\atop{k=1,\dots, N}}\\
\tilde{\Omega}_{\hat{v}_1} &= \left\{ \prod_{2 \le j < k \le N}  \frac{1}{v^2_j-v^2_k}\right\}  \left\{ \prod_{1 \le j < k \le N}  \frac{1}{u^2_j-u^2_k}\right\} \left( \prod^N_{m=1}\prod^N_{l=2} \frac{1}{u_m v_l} \right)^{M}
\end{split}\end{equation}\normalsize
\subsection{Examining the second class of wave-function} 
We conclude this section by considering the $\hat{w}^{(\infty)}$ wave-functions. Using the definitions given previously we have,
\small\begin{equation*}\begin{split}
\tau(s) \hat{w}^{(\infty)}_k(s) &= \zeta_k (- \tilde{\partial}_{\vec{x}}) \tau(s)\\
& = \sum_{\{ \lambda \} \subseteq (n-s)^{ (s-m)}}\chi_{\{\lambda \}} (-\vec{y}) \zeta_k (- \tilde{\partial}_{\vec{x}})\chi_{\{\lambda \}} (\vec{x})
\end{split}\end{equation*}\normalsize
\begin{proposition}
\small\begin{equation}
\zeta_j (- \tilde{\partial}_{\vec{x}})\chi_{\{\lambda \}} (\vec{x}) = (-1)^j \chi_{\{\lambda \}/ \{1^j \}} (\vec{x})
\end{equation}\normalsize
for all partitions $\{ \lambda \}$ such that $\{ \lambda \} \supseteq \{ 1^j\}$.
\end{proposition}
\textbf{Proof.} 
\small\begin{equation*}\begin{split}
& \langle \zeta_j (-\tilde{\partial}_{\vec{x}}) \chi_{\{\lambda \}}(\vec{x}) , \chi_{\{\mu \}}(\vec{x}) \rangle \\
=&\sum_{\mu_1 +  \dots + j \mu_j = j} (-1)^{\mu_1 + \dots +\mu_j}   \left\langle \frac{\left( \partial_{x_1} \right)^{\mu_1}\left( \frac{1}{2}\partial_{x_2} \right)^{\mu_2} \dots \left( \frac{1}{j}\partial_{x_j} \right)^{\mu_j}}{\mu_1 ! \dots \mu_j !} \chi_{\{\lambda \}}(\vec{x}) , \chi_{\{\mu \}}(\vec{x}) \right\rangle\\
=&\sum_{\mu_1 +  \dots + j \mu_j = j} (-1)^{\mu_1 + \dots +\mu_j}   \left\langle  \chi_{\{\lambda \}}(\vec{x}) ,\frac{\left( x_1 \right)^{\mu_1}\left(x_2 \right)^{\mu_2} \dots \left( x_j \right)^{\mu_j}}{\mu_1 ! \dots \mu_j !} \chi_{\{\mu \}}(\vec{x}) \right\rangle\\
=& \langle \chi_{\{\lambda \}} (\vec{x}),\zeta_j (-\vec{x}) \chi_{\{\mu \}}(\vec{x}) \rangle\\
=& \langle \chi_{\{\lambda \}} (\vec{x}), \chi_{\{j \}}(-\vec{x}) \chi_{\{\mu \}}(\vec{x}) \rangle
\end{split}\end{equation*}\normalsize
where $\{ j \}$ is the partition with the single non zero entry $j$. \\
\\
In order to proceed we now give the following result,
\begin{proposition}
\small\begin{equation}\begin{split}
\chi_{\{\lambda \}} (\vec{x}) &=  \textrm{det} [\zeta_{\lambda_i +j-i} (\vec{x})]^N_{i,j=1}\\
& =  \textrm{det} [(-1)^{\lambda^{'}_i +j-i}\zeta_{\lambda^{'}_i +j-i} (-\vec{x})]^N_{i,j=1}
\end{split}\end{equation}\normalsize
where $\{\lambda^{'}\}$ is the conjugate of the partition $\{ \lambda \}$.
\end{proposition}
\textbf{Proof.}
We recall that under the Miwa transformation the Schur polynomials, $S_{\{\lambda\}} (\vec{u})$, became the character polynomials, $\chi_{\{\lambda \}}(\vec{x})$,
\small\begin{equation*}
\textrm{det}[h_{\lambda_i +j-i} (\vec{u})]^N_{i,j=1} \rightarrow \textrm{det} [\zeta_{\lambda_i +j-i} (\vec{x})]^N_{i,j=1}
\end{equation*}\normalsize
Consider performing the Miwa transformation on the elementary symmetric polynomial definition of the Schur polynomial given by the last line in eq. \ref{Schur1},
\small\begin{equation*}
S_{\{\lambda\}} (\vec{u}) =  \textrm{det} [e_{\lambda^{'}_i +j-i} (\vec{u})]^N_{i,j=1}
\end{equation*}\normalsize
Performing the Miwa transformation we receive,
\small\begin{equation*}\begin{split}
\sum^{N}_{j=0}t^j e_j(\vec{u})& \rightarrow \exp \left\{ \sum^{\infty}_{j=1} (-t)^j (-x_j) \right\} = \sum^{N}_{j=0}t^j (-1)^j \zeta_j(-\vec{x})\\
\Rightarrow e_j(\vec{u}) & \rightarrow  (-1)^j \zeta_j(-\vec{x})
\end{split}\end{equation*}\normalsize
and,
\small\begin{equation*}\begin{split}
\textrm{det} [e_{\lambda^{'}_i +j-i} (\vec{u})]^N_{i,j=1} \rightarrow \textrm{det} [(-1)^{\lambda^{'}_i +j-i}\zeta_{\lambda^{'}_i +j-i} (-\vec{x})]^N_{i,j=1} \textrm{     } \square
\end{split}\end{equation*}\normalsize
\textbf{Continuation of prop. 16.} Thus with this little result we now consider the expression $(-1)^k \chi_{\{k \}}(-\vec{x})$,
\small\begin{equation*}\begin{split}
(-1)^k \chi_{\{k \}}(-\vec{x}) &= \textrm{det} \left[(-1)^{k \delta_{i,1} +j-i}\zeta_{k \delta_{i,1} +j-i} (-\vec{x}) \right]\\
&= \textrm{det} \left[ \zeta_{ \delta_{1,j}+  \delta_{2,j}+ \dots +  \delta_{k,j} +j-i} (\vec{x})\right]\\
&= \chi_{\{1^k \}}(\vec{x})
\end{split}\end{equation*}\normalsize
hence,
\small\begin{equation*}\begin{split}
\langle \chi_{\{\lambda \}} (\vec{x}), \chi_{\{j \}}(-\vec{x}) \chi_{\{\mu \}}(\vec{x}) \rangle &= (-1)^j\langle \chi_{\{\lambda \}} (\vec{x}), \chi_{\{1^j \}}(\vec{x}) \chi_{\{\mu \}}(\vec{x}) \rangle \\
&= (-1)^j \langle \chi_{\{\lambda \}/ \{ 1^j \}} (\vec{x}),  \chi_{\{\mu \}}(\vec{x}) \rangle \\
\Rightarrow \zeta_j (- \tilde{\partial}_{\vec{x}})\chi_{\{\lambda \}} (\vec{x}) &= (-1)^j \chi_{\{\lambda \}/ \{1^j \}} (\vec{x}) \textrm{      } \square
\end{split}\end{equation*}\normalsize
Therefore, with regards to the second class of wave-functions we have the following result,
\small\begin{equation}
\tau(s) \hat{w}^{(\infty)}_{k}(s) =(-1)^k  \sum_{\{\lambda \} \subseteq (n-s)^{(s-m)}} \chi_{\{\lambda \}/\{1^k \}} (\vec{x}) \chi_{\{\lambda \}} (-\vec{y}) ,
\end{equation}\normalsize
and remembering that the wave-matrix $\hat{W}^{(\infty)}(\vec{x},\vec{y})$ is lower triangular, we obtain,
\small\begin{equation*}\begin{split}
\hat{W}^{(\infty)}(\vec{x},\vec{y})&=  \left( \hat{w}^{(\infty)}_{j-k}(j,\vec{x},\vec{y}) \right)^{n-1}_{j,k= m}\\
&= \left( \frac{(-1)^{j-k}}{\tau(j)}\sum_{\{\lambda \} \subseteq (n-j)^{(j-m)}} \chi_{\{\lambda \}/\{1^{j-k} \}} (\vec{x}) \chi_{\{\lambda \}} (-\vec{y}) \right)^{n-1}_{j,k= m}\\
\end{split}\end{equation*}\normalsize\\
\textbf{The infinite lattice with a free end (again).} Analyzing the second class of wave-function in the $n \rightarrow \infty$ limit we receive,
\small\begin{equation*}\begin{split}
\tau(s) \hat{w}^{(\infty)}_{k}(s) & =\sum_{\{\lambda \} \subseteq (\infty)^{(s-m)}} \chi_{\{\lambda \}/\{1^k \}} (\vec{x}) \chi_{\{\lambda \}} (-\vec{y}) \\
&= \sum_{\{\lambda \} \subseteq (\infty)^{(s-m)}}\chi_{\{\lambda \}} (-\vec{y})  \left(\sum_{\{\nu \} \subseteq (\infty)^{(s-m)}} c^{\{\lambda \}}_{\{1^k \} \{\nu \}}\chi_{\{\nu \}} (\vec{x}) \right)  
\end{split}\end{equation*}\normalsize
\small\begin{equation*}\begin{split}
&= \sum_{\{\nu \} \subseteq (\infty)^{(s-m)}}  \chi_{\{\nu \}} (\vec{x}) \left(\sum_{\{\lambda \} \subseteq (\infty)^{(s-m)}} c^{\{\lambda \}}_{\{1^k \} \{\nu \}}\chi_{\{\lambda \}} (-\vec{y}) \right)  \\
& =  \chi_{\{1^k \}} (-\vec{y}) \sum_{\{\nu \} \subseteq (\infty)^{(s-m)}}  \chi_{\{\nu \}} (\vec{x}) \chi_{\{\nu \}} (-\vec{y})\\
& =  \chi_{\{1^k \}} (-\vec{y}) \tau(s)\\
\Rightarrow  \hat{w}^{(\infty)}_{k}(s) &= \chi_{\{1^k \}} (-\vec{y})
\end{split}\end{equation*}\normalsize
Again, we see that the skew character polynomial decouples and we simply receive $\tau(s)$, which is eliminated by the same factor on the denominator, multiplied by a factor given by $\chi_{\{1^k \}} (-\vec{y})$.  As before, this case shall be classed as uninteresting, (maybe even more so than the first case), and from now on we shall continue in the finite case where the skew in the partition remains in the wave-functions, and proceed to uncover their combinatorial meaning.\\
\\
\textbf{Constructing $N$-particle state vectors.} Consider the following state vector,
\small\begin{equation*}\begin{split}
 B(u_1)  \dots B(u_{N-k}) \left( \phi^{\dagger}_1 \right)^k |0\rangle &=B(u_1)  \dots B(u_{N-k})  |1^k\rangle\\
&= |\Psi^{\{1^k\}}_M (u_1,\dots,u_{N-k})\rangle
\end{split}\end{equation*}\normalsize
where the partition $\{1^k\}$ is constructed in the usual manner from the occupation numbers.\\
\\
We now have the following result regarding the allowable partitions of this particular state vector,
\begin{proposition}
\small\begin{equation}
|\Psi^{\{1^k\}}_M (u_1,\dots,u_{N-k})\rangle = \sum_{\{\lambda\} \subseteq \{(M)^{(N-k)},1^k\} \atop{\{\lambda\} \supseteq  \{1^k \}}} \psi^{(2,1^k)}_{\{\lambda \}}(u_1,\dots,u_{N-k}) |\lambda\rangle
\end{equation}\normalsize
\end{proposition}
\textbf{Proof.} Consider again the non crossing column strict lattice path interpretation of the state vectors. The operator(s) $\left( \phi^{\dagger}_1 \right)^k$ assure us that the last $k$ paths, labelled $j_{q}$, $N-k+1 \le q \le N$, make directional changes from east to north at row $1$, column $q$. Thus the largest occupation number sequence can be,
\small\begin{equation*}
\{n_0, n_1, \dots,n_M\} = \{0, k, 0,\dots,0,N-k\} \Rightarrow \{\lambda\} \subseteq \{(M)^{(N-k)},1^k\}
\end{equation*}\normalsize
Also, since columns $\{N-k+1, \dots, N\}$ only contain one $\phi^{\dagger}$ operator each, this means that only columns $\{1, \dots,  N-k\}$ can contain paths in the zeroth row. The fact that the paths are column strict means that the lowest occupation number sequence is,
\small\begin{equation*}
\{n_0, n_1, \dots,n_M\} = \{N-k, k, 0,\dots,0,0\} \Rightarrow \{\lambda\} \supseteq \{1^k\} \textrm{    } \square
\end{equation*}\normalsize
We now give the following combinatorial definitions of $\psi^{(2,1^k)}_{\{\lambda \}}(u_1,\dots,u_{N-k})$. Considering the \textbf{lattice path} interpretation we obtain,
\small\begin{equation*}
\psi^{(2,1^k)}_{\{\lambda \}}(u_1,\dots,u_{N-k})= \sum_{\textrm{allowable paths in}\atop{\textrm{$(M+1)\times N$ lattice$^\dagger$}}} u^{t^{d}_1-t^{a}_1}_1 \dots u^{t^{d}_{N-k}-t^{a}_{N-k}}_{N-k}
\end{equation*}\normalsize
where the lattice paths are under the condition that the last $k$ paths, labelled $j_{q}$, $N-k+1 \le q \le N$, make directional changes from east to north at row $1$, column $q$, and only columns $\{1, \dots,  N-k\}$ can contain paths in the zeroth row. The powers $t^d_j$ and $t^a_j$ give the total amount of $d$ and $a$ vertices in column $j$ respectively.\\
\\
Considering the \textbf{plane partition} interpretation we obtain,
\small\begin{equation*}
\psi^{(2,1^k)}_{\{\lambda \}}(u_1,\dots,u_{N-k})= \sum_{\textrm{upper plane part.}\atop{\textrm{in $N \times N \times M$ array$^\dagger$}}} u^{l^{d}_1-l^{a}_1}_1 \dots u^{l^{d}_{N-k}-l^{a}_{N-k}}_{N-k}
\end{equation*}\normalsize
where the upper plane partitions are under the condition that the top-right most $k \times k$ entries are equal to one. This obviously places restrictions on the remaining entries, as per the conditions of a plane partition. For example, the remaining $(N-k)\times (N-k)$ bottom-right entries can only either be zero or one accordingly,
\small\begin{equation*}
\pi^{\{\lambda \}}_+ = \left( \begin{array}{ccccccccc}
\pi_{1,1} & \dots  & \pi_{1,N-k} & 1 & \dots& \dots&\dots &\dots &  1\\
 	      & \ddots & \vdots         & \vdots  & && & &\vdots \\
& 			&\pi_{N-k,N-k} & 1 &\dots  & \dots&\dots &\dots & 1\\
& 			& & 1 &\dots  &\dots & \dots&\dots & 1\\
& 			& &  &\ddots  && && \vdots\\
& 			& &  &  &1&\dots &\dots& 1\\
& 			& &  &  && \pi_{k+1,k+1}& \dots& \pi_{k+1,N}\\
& 			& &  &  && & \ddots& \vdots\\
& 			& &  &  && && \pi_{N,N}
\end{array}\right)
\end{equation*}\normalsize
The powers $l^d_j$ and $l^a_j$ give the total amount of $d$ and $a$ rhombi in column $j$ respectively.\\
\\
Finally, considering the descending \textbf{Young tableaux} interpretation, we notice that whenever we transform from the upper plane partition to the Young tableau, the weights $t_{N-k+1}= \dots = t_{N} = 1$ and their position in the tableau is exactly $\{t_N = T^{\{\lambda\}}_{1,1},t_{N-1} = T^{\{\lambda\}}_{2,1},\dots,t_{N-k+1} = T^{\{\lambda\}}_{k,1}  \}$. Since these weights do not enter the equation, due to $u_{N-k+1},\dots,u_N$ not being present, we can simply consider the skew partition $\{\lambda - 1^k\}$ to generate the tableaux\footnote{Incidentally, it is at this point the reason we considered the tableaux in descending order becomes apparent. Had we considered ascending order we would need to invert the numbers to obtain the required results.}. Thus we obtain,
\small\begin{equation}\begin{split}
\psi^{(2,1^k)}_{\{\lambda \}}(u_1,\dots,u_{N-k})  &= \sum_{T^{\{ \lambda -1^k \}}_- }  u^{2 t_1-M}_1 \dots u^{2 t_{N-k}-M}_{N-k}\\
&= \left(\frac{1}{u_1 \dots u_{N-k}}\right)^M \sum_{T^{\{ \lambda -1^k \}}_- }  \left(u^{2 }_1\right)^{t_1} \dots \left(u^{2 }_{N-k}\right)^{t_{N-k}}\\
&=   \left(\frac{1}{u_1 \dots u_{N-k}}\right)^M  S_{\{\lambda \}/\{1^k \}} (u^{2}_1, \dots, u^{2}_{N-k})
\end{split}\end{equation}\normalsize
meaning the $N$-particle state vector is given by the following,
\small\begin{equation}\begin{split}
&|\Psi^{\{1^k\}}_M (u_1,\dots,u_{N-k})\rangle\\
=& \left(\prod^{N-k}_{j=1}\frac{1}{u_j}\right)^M   \sum_{\{\lambda\} \subseteq \{(M)^{ (N-k)},1^k\} \atop{\{\lambda\} \supseteq  \{1^k \}}}  S_{\{\lambda \}/\{1^k \}} (u^{2}_1, \dots, u^{2}_{N-k}) |\lambda\rangle
\end{split}\end{equation}\normalsize
\textbf{Correlation functions and the wave-vector at $\mathbf{s=n-M = m+N}$ as a weighted sum.} Consider then the correlation function,
\small\begin{equation}\begin{split}
&\left( \frac{\prod^{N-k}_{j=1} u_j}{\prod^{N}_{j=1} v_j} \right)^M \langle\Psi_M (v_1,\dots,v_{N})|\Psi^{\{1^k\}}_M (u_1,\dots,u_{N-k})\rangle\\
=&\left( \frac{\prod^{N-k}_{j=1} u_j}{\prod^{N}_{j=1} v_j} \right)^M \langle0| C(v_1)  \dots C(v_{N}) B(u_1)  \dots B(u_{N-k})\left(\phi^{\dagger}_1 \right)^{k}|0\rangle\\
=&   \sum_{\{\lambda\} \subseteq \{(M)^{ (N-k)},1^k\} \atop{\{\mu \} \subseteq  (M)^{(N)}}}  S_{\{\lambda \}/\{1^k \}} (u^{2}_1, \dots, u^{2}_{N-k}) S_{\{\mu \}} (v^{-2}_1, \dots, v^{-2}_N)  \langle\mu|\lambda\rangle\\
=&  \sum_{\{\lambda\} \subseteq \{(M)^{(N-k)},1^k\} \atop{\{\lambda \} \supseteq  \{1^k\}}}  S_{\{\lambda \}/\{1^k \}} (u^{2}_1, \dots, u^{2}_{N-k}) S_{\{\lambda \}} (v^{-2}_1, \dots, v^{-2}_N)
\end{split}\end{equation}\normalsize
which calculates all the weighted non crossing column strict lattice paths on an $(M+1)\times 2N$ grid with the final $k$ paths, labelled $j_q$, $N-k+1 \le q \le N$, turning north at row 1, column $N-k+1 \le q \le N$. Additionally, only columns $1 \le q \le N-k$ can contain paths in the zeroth row. Compare the above result now with any of the wave-functions that we calculated earlier,
\small\begin{equation*}
\tau(s) \hat{w}^{(\infty)}_{k}(s) =(-1)^k  \sum_{\{\lambda \} \subseteq (n-s)^{(s-m)}\atop{\{\lambda \} \supseteq \{1^k \}}} \chi_{\{\lambda \}/\{1^k \}} (\vec{x}) \chi_{\{\lambda \}} (-\vec{y}) 
\end{equation*}\normalsize
and concentrate on the particular row, $s=n-M = m+N$, of the wave-matrix, restricting the variables as before (eq. \ref{restrictq}), to obtain,
\small\begin{equation*}\begin{split}
&\tau(n-M) \hat{w}^{(\infty)}_{k}(n-M)\\ 
= &(-1)^k \sum_{\{\lambda \} \subseteq (M)^{(N)}\atop{\{\lambda \} \supseteq \{1^k \}}} S_{\{\lambda \}/\{1^k \}} (u^2_1,\dots,u^2_N) S_{\{\lambda \}} (v^{-2}_1,\dots,v^{-2}_N)
\end{split}\end{equation*}\normalsize
Now consider the limit $u_{N-k+1} = \dots = u_N = 0$,
\small\begin{equation*}
\lim_{ u_{j} \rightarrow 0 \atop{N-k+1 \le j \le N} } S_{\{\lambda \}/\{1^k \}} (u^2_1,\dots,u^2_N) = S_{\{\lambda \}/\{1^k \}} (u^2_1,\dots,u^2_{N-k}).
\end{equation*}\normalsize
Unsurprisingly however we again witness subtle effects in the summation. Recalling the combinatorial definition of the skew Schur polynomial,
\small\begin{equation*}
S_{\{\lambda \}/\{1^k \}} (u^2_1,\dots,u^2_{N-k}) = \sum_{T^{\{\lambda - 1^k \}}_-} \left(u^{2}_1 \right)^{t_1} \dots \left(u^{2}_{N-k} \right)^{t_{N-k}}
\end{equation*}\normalsize
where the sum is given over all possible descending column strict skew tableaux of shape $\{\lambda - 1^k \}$, and the $t_j$ give the amount of times $j$ appears in the skew partition. In the above case, $j=\{1,\dots,N-k\}$, thus the total length of any column in the skew partition cannot be greater than $N-k$, otherwise the Young tableau will not be column strict. Therefore, when $\{\lambda\} \subseteq (M)^{(N)}$, for all the columns in the skew partition $\{\lambda - 1^k\}$, to be no greater than $N-k$ in length we obtain the new restricted condition, $\{\lambda\} \subseteq \{ (M)^{(N-k)},1^k\}$.\\
\\
Thus the wave-vector, given by the $s=n-M = m+N$ row of the wave-matrix, in the $u_{N-k+1} = \dots = u_N = 0$ limit,
\small\begin{equation}\begin{split}
&\lim_{ u_{j} \rightarrow 0 \atop{N-k+1 \le j \le N} }  \left( \tau(n-M) \hat{w}^{(\infty)}_{k}(n-M) ) \right)^{M}_{k= 0}\\
=& \left((-1)^k \sum_{\{\lambda \} \subseteq \{(M)^{(N-k)},1^k\}\atop{\{\lambda \} \supseteq \{1^k \}}} S_{\{\lambda \}/\{1^k \}} (u^2_1,\dots,u^2_{N-k}) S_{\{\lambda \}} (v^{-2}_1,\dots,v^{-2}_N)  \right)^{M}_{k= 0}\\
=&\left( \frac{\prod^{N-k}_{j=1} u_j}{\prod^N_{j=1} v_j} \right)^M  \left( \langle \Psi_M (v_1,\dots,v_{N})|\Psi^{\{1^k\}}_M (u_1,\dots,u_{N-k})\rangle \right)^{M}_{k= 0}
\end{split}\end{equation}\normalsize
gives exactly (up to a multiplicative factor) all the weighted non crossing column strict lattice paths on an $(M+1)\times 2N$ with the final $k$ paths, $1< k \le N$, labelled $j_q$, $N-k+1 \le q \le N$, turning north at row 1, column $N-k+1 \le q \le N$ and only the first $N-k$ columns can contain paths in the zeroth row.\\
\\
\textbf{Single determinant form for the wave-functions.}\\
\\
\textbf{Comment.} For both classes of wave-function all the results up to this point have been a mirror image of each other up to a slight variation. In the following result however, the mirroring ceases. Obtaining the single determinant form for the second class of wave-functions is a long process that displays a surprising asymmetry with the first class of wave-functions. \\
\\
\textbf{Polynomial expansion (in $\mathbf{u_N}$) of the scalar product.} We begin by examining the operator $B(u)$, as we are interested in the parts of $B(u)$ that contain only $\phi^{\dagger}_j$ and $\phi_1$ operators,
\small\begin{equation}\begin{split}
B(u) &=u^{-M} \left\{ \sum^M_{j=0} u^{2j} \phi^{\dagger}_j +\sum^{M-2}_{j=0} u^{2j+2} \phi^{\dagger}_0 \phi_1 \phi^{\dagger}_{j+2} \right\} \\
& \textrm{$+$ terms that contain operators $\phi_j$   ,   $j \in \{2,\dots,N\}$} 
\end{split}\end{equation}\normalsize
Thus when $B(u)$ acts on the vacuum,
\small\begin{equation}
B(u)|0\rangle =u^{-M} \sum^M_{j=0} u^{2j}  \phi^{\dagger}_j |0\rangle
\end{equation}\normalsize
we can obtain the scalar product as the following weighted linear sum of correlation functions,
\small\begin{equation}\begin{split}
\field{S}(N,M| \vec{u},\vec{v}) & =  \langle0|C(v_1)\dots C(v_N)B(u_1) \dots B(u_N)|0\rangle\\
& = u^{-M}_N \sum^M_{j=0} u^{2j}_N  \langle\Psi_M (v_1,\dots,v_{N})|\Psi^{\{j\}}_M (u_1,\dots,u_{N-1})\rangle\\
& = u^{-M}_N \sum^M_{j=0} u^{2j}_N \field{S}^N_M(\{j \})
\end{split}\end{equation}\normalsize
where we have defined,
\small\begin{equation}\begin{split}
\field{S}^N_M(\{j \})=  \langle\Psi_M (v_1,\dots,v_{N})|\Psi^{\{j\}}_M (u_1,\dots,u_{N-1})\rangle
\end{split}\end{equation}\normalsize
for notational convenience.\\
\\
Our ultimate goal is to find the single determinant expression of,
\small\begin{equation}\begin{split}
\field{S}^N_M(\{1^k \}) &= \field{S}^N_M(\{\underbrace{1, \dots, 1}_{k} \})\\
&= \langle\Psi_M (v_1,\dots,v_{N})|\Psi^{\{1^k\}}_M (u_1,\dots,u_{N-k})\rangle
\end{split}\end{equation}\normalsize
which shall be achieved through many steps. Nevertheless, we begin this process by explicitly finding the required expression for small $k$, and then using induction to fill in the gaps.\\
\\
\textbf{Deriving the coefficient, $\mathbf{\field{S}^N_M(\{q \})}$, $\mathbf{0 \le q \le M}$.} Expanding the scalar product as a series in $u^2_N$ involves exactly the same procedure as expanding it as a series in $v^2_1$. We begin by relabeling the scalar product as,
\small\begin{equation*}
\field{S}(N,M| \vec{u},\vec{v}) = \Omega_{\hat{u}_N} \frac{1}{ u^M_N} \left\{ \prod_{1 \le j < k \le N}  \frac{1}{u^2_j-u^2_k}\right\}  \textrm{det} \left[ h_{M+N-1}(u^2_k,v^2_j)\right]^N_{j,k=1}
\end{equation*}\normalsize
where,
\small\begin{equation}
\Omega_{\hat{u}_s}  = \left\{\prod_{1 \le j < k \le N} \frac{1}{v^2_j-v^2_k} \right\} \left( \prod^{s-1}_{m=1}\prod^N_{l=1} \frac{1}{u_m v_l} \right)^{M} \textrm{  ,  } 1 \le s \le N
\end{equation}\normalsize
In the corresponding section for the first wave-functions, we used a series of row operations to eliminate the factor of $\prod_{1 \le j < k \le N}\left(  v^2_j-v^2_k \right)$ on the denominator. Using the corresponding column operations to eliminate the factor of $\prod_{1 \le j < k \le N}\left( u^2_j-u^2_k \right)$ in the denominator of the above expression we obtain,
\small\begin{equation*}
\field{S}(N,M| \vec{u},\vec{v}) =  \Omega_{\hat{u}_N} \frac{1}{ u^M_N}  \textrm{det} \left[ h_{M+N-k}(u^2_1,\dots, u^2_{k},v^2_j) \right]^N_{j,k=1}
\end{equation*}\normalsize
and expressing the entries of the final column as a polynomial in $u^2_N$,
\small\begin{equation*}
h_{M}(u^2_1,\dots, u^2_N,v^2_j) = \sum^M_{q=0} u^{2q}_N h_{M-q}(u^2_1,\dots, u^2_{N-1},v^2_j),
\end{equation*}\normalsize
we receive,
\small\begin{equation}\begin{split}
\field{S}^N_M(\{q \}) = \Omega_{\hat{u}_N} \textrm{det} \left[ h_{M+N-k}(\{u^2\}_{k},v^2_j), h_{M-q}(\{u^2\}_{N-1},v^2_j)\right]_{j=1,\dots,N\atop{k=1,\dots,N-1 }}
\label{nice33}\end{split}\end{equation}\normalsize
where $\{u^2\}_{k} = \{u^2_1, \dots, u^2_k  \}$.\\
\\
\textbf{Polynomial expansion (in $\mathbf{u_{N-1}}$) of $\mathbf{\field{S}^N_M(\{1 \})}$.} We now build upon eq. \ref{nice33} and consider the quantity,
\small\begin{equation}
B(u_{N-1})\phi^{\dagger}_1|0\rangle =u^{-M}_{N-1} \sum^M_{j=0} u^{2j}_{N-1}  \phi^{\dagger}_j \phi^{\dagger}_1 |0\rangle + u^{-(M+2)}_{N-1} \sum^{M-2}_{j=2} u^{2j}_{N-1}  \phi^{\dagger}_0  \phi^{\dagger}_{j} |0\rangle
\end{equation}\normalsize
Hence,
\small\begin{equation}\begin{split}
\field{S}^N_M(\{1 \}) &= u^{-M}_{N-1} \sum^M_{j=0} u^{2j}_{N-1} \field{S}^N_M(\{j,1 \}) + u^{-(M+2)}_{N-1} \sum^{M}_{j=2} u^{2j}_{N-1}   \field{S}^N_M(\{0,j \})\\
&= u^{-M}_{N-1}\left\{ \field{S}^N_M(\{0,1 \}) + u^{2M}_{N-1} \field{S}^N_M(\{M,1 \})\right\}   \\
&+  \textrm{   } u^{-M}_{N-1}\left\{ \sum^{M-1}_{j=1} u^{2j}_{N-1}\left[ \field{S}^N_M(\{j,1 \}) + \field{S}^N_M(\{0,j+1 \}) \right] \right\}
\end{split}\end{equation}\normalsize
where we recognize that,
\small\begin{equation}
\field{S}^N_M(\{m,n \}) = \field{S}^N_M(\{n,m \})
\end{equation}\normalsize
Thus, if we expand $\field{S}^N_M(\{1 \})$ as a series in $u_{N-1}$, the coefficient of $u^{-M+2}_{N-1}$ is,
\small\begin{equation*}
\field{S}^N_M(\{1,1 \}) + \field{S}^N_M(\{0,2 \}) 
\end{equation*}\normalsize
At this point we run into a potential problem. In order to obtain $\field{S}^N_M(\{1,1 \}) =\field{S}^N_M(\{1^2 \}) $, we need to first find $\field{S}^N_M(\{0,2 \})$.\\
\\
\textbf{Deriving $\mathbf{\field{S}^N_M(\{0,2 \})}$ from the polynomial expansion of $\mathbf{\field{S}^N_M(\{0 \})}$.} Luckily, this can be achieved by expanding $\field{S}^N_M(\{0 \})$ as a series in $u_{N-1}$,
\small\begin{equation*}
\field{S}^N_M(\{0 \}) = u^{-M}_{N-1} \sum^M_{j=0}u^{2j}_{N-1} \field{S}^N_M(\{j,0 \})
\end{equation*}\normalsize
Substituting $q=0$ into eq. \ref{nice33} we have,
\small\begin{equation*}
 \field{S}^N_M(\{0 \}) =  \frac{\Omega_{\hat{u}_{N-1}}}{u^M_{N-1}} \textrm{det} \left[ h_{M+N-k}(\{u^2\}_{k},v^2_j), h_{M}(\{u^2\}_{N-1},v^2_j)\right]_{j=1,\dots,N\atop{k=1,\dots,N-1 }}.
\end{equation*}\normalsize
where $\Omega_{\hat{u}_{N}}=\frac{\Omega_{\hat{u}_{N-1}}}{u^M_{N-1}}$. We now rewrite the entries of the $N$th column, multiplied by $u^2_{N-1}$, $u^2_{N-1}h_{M}(\{u^2\}_{N-1},v^2_j)$, as,
\small\begin{equation*}\begin{split}
& \sum^{M+1}_{q=1} u^{2q }_{N-1} h_{M+1-q}(\{u^2\}_{N-2},v^2_j)\\
=& \sum^{M+1}_{q=0} u^{2q }_{N-1} h_{M+1-q}(\{u^2\}_{N-2},v^2_j) - h_{M+1}(\{u^2\}_{N-2},v^2_j)\\
=& h_{M+1}(\{u^2\}_{N-1},v^2_j)- h_{M+1}(\{u^2\}_{N-2},v^2_j),
\end{split}\end{equation*}\normalsize
to obtain,
\small\begin{equation*}\begin{split}
 \field{S}^N_M(\{0 \}) =  \frac{\Omega_{\hat{u}_{N-1}}}{u^{M+2}_{N-1}} \textrm{det} \left[ c_{jk}, h_{M+1}(\{u^2\}_{N-1},v^2_j)- h_{M+1}(\{u^2\}_{N-2},v^2_j)\right]_{j=1,\dots,N\atop{k=1,\dots,N-1 }},
\end{split}\end{equation*}\normalsize
where,
\small\begin{equation*}\begin{split}
c_{jk} = h_{M+N-k}(\{u^2\}_{k},v^2_j).
\end{split}\end{equation*}\normalsize
Subtracting column $N-1$ from column $N$ we receive,
\small\begin{equation*}\begin{split}
\field{S}^N_M(\{0 \}) = -\frac{\Omega_{\hat{u}_{N-1}}}{u^{M+2}_{N-1}} \textrm{det} \left[ c_{jk},  h_{M+1}(\{u^2\}_{N-2},v^2_j)\right]_{j=1,\dots,N\atop{k=1,\dots,N-1 }}.
\end{split}\end{equation*}\normalsize
In the above form of $\field{S}^N_M(\{0 \})$, only column $N-1$ is a function of $u^2_{N-1}$. Expanding the $(N-1)$th column as a polynomial in $u_{N-1}$, we obtain,
\small\begin{equation*}\begin{split}
&\frac{u^{M}_{N-1}}{\Omega_{\hat{u}_{N-1}}} \field{S}^N_M(\{0 \}) \\
=&- \sum^{N}_{r=1}(-1)^{N-1+r} \sum^{M+1}_{q=0}u^{2q-2}_{N-1} h_{M+1-q}(\{u^2\}_{N-2},v^2_j) \\
& \times \textrm{det} \left[ c_{jk},h_{M+1}(\{u^2\}_{N-2},v^2_j) \right]_{j=1, \dots, \hat{r}, \dots, N\atop{k=1,\dots,N-2 }}\\
=& - \sum^{M+1}_{q=0}u^{2q-2}_{N-1} \textrm{det} \left[ c_{jk},h_{M+1-q}(\{u^2\}_{N-2},v^2_j),h_{M+1}(\{u^2\}_{N-2},v^2_j) \right]_{j=1, \dots, N \atop{k=1,\dots,N-2 }}\\
=& \sum^{M}_{q=0}u^{2q}_{N-1} \textrm{det} \left[ c_{jk},h_{M+1}(\{u^2\}_{N-2},v^2_j),h_{M-q}(\{u^2\}_{N-2},v^2_j) \right]_{j=1, \dots, N \atop{k=1,\dots,N-2 }}
\end{split}\end{equation*}\normalsize
where the $q = 0$ case on the second last line is eliminated due to column $N-1$ and $N$ being equal. \\
\\
Thus we obtain the result,
\small\begin{equation}\begin{split}
\field{S}^N_M(\{q,0 \}) &= \Omega_{\hat{u}_{N-1}} \textrm{det} \left[ c_{j,k},h_{M+1}(\{u^2\}_{N-2},v^2_j),h_{M-q}(\{u^2\}_{N-2},v^2_j) \right]_{j=1, \dots, N \atop{k=1,\dots,N-2 }}
\label{already}\end{split}\end{equation}\normalsize
which means we now have the necessary results to obtain $\field{S}^N_M(\{1^2 \})$.\\
\\
\textbf{Deriving $\mathbf{\field{S}^N_M(\{1^2 \})}$ from the polynomial expansion of $\mathbf{\field{S}^N_M(\{1 \})}$.}  We now consider $\field{S}^N_M(\{1 \})$,
\small\begin{equation*}
 \field{S}^N_M(\{1 \}) =  \frac{ \Omega_{\hat{u}_{N-1}}}{u^M_{N-1}} \textrm{det} \left[ c_{jk}, h_{M-1}(\{u^2\}_{N-1},v^2_j)\right]_{j=1,\dots,N\atop{k=1,\dots,N-1 }}
\end{equation*}\normalsize
Labelling the individual columns of the matrix as $C_{j}$, $1\le j \le N$, we consider the quantity, $u^4_{N-1} C_{N} - C_{N-1} $,
\small\begin{equation*}\begin{split}
&u^4_{N-1} h_{M-1}(\{u^2\}_{N-1},v^2_j) - h_{M+1}(\{u^2\}_{N-1},v^2_j)\\
=&\sum^{M-1}_{q=0}u^{2q+4}_{N-1} h_{M-1-q}(\{u^2\}_{N-2},v^2_j) - \sum^{M+1}_{q=0}u^{2q}_{N-1}h_{M+1-q}(\{u^2\}_{N-2},v^2_j)\\
=& -\sum^{1}_{q=0}u^{2q}_{N-1}h_{M+1-q}(\{u^2\}_{N-2},v^2_j) \\
=&  -\sum^{1}_{q=0}u^{2q}_{N-1}\varrho^{N-2,j}_{q-1}
\end{split}\end{equation*}\normalsize
where we label the symmetric polynomials as,
\small\begin{equation}\begin{split}
\varrho^{\alpha,\beta}_{\gamma}=h_{M-\gamma}(\{u^2\}_{\alpha},v^2_{\beta})
\end{split}\end{equation}\normalsize
for notational convenience. Thus continuing with the expansion we obtain,
\small\begin{equation*}\begin{split}
\field{S}^N_M(\{1 \})&=-\frac{\Omega_{\hat{u}_{N-1}}}{u^{M+4}_{N-1}}\textrm{det}\left[ c_{jk}, \sum^{M+1}_{q_1=0}u^{2q_1}_{N-1} \varrho^{N-2,j}_{q_1-1},\sum^{1}_{q_2=0}u^{2q_2}_{N-1} \varrho^{N-2,j}_{q_2-1} \right]_{j=1,\dots,N\atop{k=1,\dots,N-2 }}\\
&=u^{-M}_{N-1} \sum^{M+1}_{q_1=2}\sum^{1}_{q_2=0}u^{2q_1+2 q_2-4}_{N-1}\Omega_{\hat{u}_{N-1}} \textrm{det}\left[ c_{jk},  \varrho^{N-2,j}_{q_2-1},\varrho^{N-2,j}_{q_1-1} \right]_{j=1,\dots,N\atop{k=1,\dots,N-2 }}
\end{split}\end{equation*}\normalsize
where we are interested in the indices $(q_1,q_2) = (r,1)$, and $(r+1,0)$, $2 \le r \le M$,
\small\begin{equation*}\begin{split}
\field{S}^N_M(\{r-1,1 \})+\field{S}^N_M(\{r,0 \}) &=\Omega_{\hat{u}_{N-1}} \left( \textrm{det}\left[ c_{jk},  \varrho^{N-2,j}_{0},\varrho^{N-2,j}_{r-1} \right]_{j=1,\dots,N\atop{k=1,\dots,N-2 }} \right. \\
&\left.  + \textrm{det}\left[ c_{jk},  \varrho^{N-2,j}_{-1},\varrho^{N-2,j}_{r} \right]_{j=1,\dots,N\atop{k=1,\dots,N-2 }}\right)
\end{split}\end{equation*}\normalsize
Since we already have the explicit form of $\field{S}^N_M(\{r,0 \})$, given in eq. \ref{already}, this leaves us with the result,
\small\begin{equation}
\field{S}^N_M(\{r-1,1 \}) = \Omega_{\hat{u}_{N-1}} \textrm{det}\left[ c_{jk}, \varrho^{N-2,j}_{0},\varrho^{N-2,j}_{r-1} \right]_{j=1,\dots,N\atop{k=1,\dots,N-2 }}
\end{equation}\normalsize\\
\textbf{Towards the general result.} We now have enough knowledge to conclude this section with an inductive proof of the following result.
\begin{proposition}
\small\begin{equation}
\field{S}^N_M(\{r_1, \dots, r_p \}) = \Omega_{\hat{u}_{N+1-p}} \textrm{det}\left[ c_{j,k}, \varrho^{N-p,j}_{r_p+1-p},\varrho^{N-p,j}_{r_{p-1}+2-p},\dots,\varrho^{N-p,j}_{r_{1}} \right]_{j=1,\dots,N\atop{k=1,\dots,N-p }}
\label{tochi}\end{equation}\normalsize
where,
\small\begin{equation*}\begin{split}
r_1 \in \{ 0,1, \dots,M \} \textrm{  ,  } r_2 \in \{0,1\} \textrm{  , \dots ,  } r_p \in \{ 0,1\}\\
r_1 \ge  r_2 \ge \dots \ge r_p \textrm{   ,   } 1 \le p \le N
\end{split}\end{equation*}\normalsize
\end{proposition}
\textbf{Proof.} We have shown that the above statement is true for $p = 1,2$. Let us assume that the general case is true up to $p$, and show that the $p+1$ case follows naturally from this assumption.\\
\\
\textbf{Polynomial expansion (in $\mathbf{u^2_{N-p}}$) of $\mathbf{\field{S}^N_M(\{1^{p-r},0^{r}  \})}$.} Using the expansion of $B(u_{N-p})$ we obtain the expression, $\field{S}^N_M(\{1^{p-r},0^{r} \})$, $0 \le  r \le p-1$, as the usual weighted sum,
\small\begin{equation}\begin{split}
\field{S}^N_M(\{1^{p-r},0^{r}  \})   &= u^{-M}_{N-p} \sum^M_{j=0} u^{2j}_{N-p}\field{S}^N_M(\{j,1^{p-r},0^{r}  \}) \\
&+ u^{-(M+2)} _{N-p}\sum^{M}_{j=2} u^{2j}_{N-p}   \field{S}^N_M(\{j,1^{p-r-1},0^{r+1} \})\\
&= u^{-M}_{N-p}\left\{ \field{S}^N_M(\{1^{p-r},0^{r+1} \}) + u^{2M}_{N-p} \field{S}^N_M(\{M,1^{p-r},0^{r} \})\right\} +u^{-M}_{N-p}  \\
&\times \left\{ \sum^{M-1}_{j=1} u^{2j}_{N-p}\left[ \field{S}^N_M(\{j,1^{p-r},0^{r}  \}) + \field{S}^N_M(\{j+1,1^{p-r-1},0^{r+1} \}) \right] \right\}
\label{machma}\end{split}\end{equation}\normalsize
Additionally for $\field{S}^N_M(\{0^{p} \})$, $(r=p)$, we have,
\small\begin{equation}
\field{S}^N_M(\{0^{p} \})=u^{-M}_{N-p}\sum^M_{j=0} u^{2j}_{N-p} \field{S}^N_M(\{j,0^{p} \})
\label{ache2}\end{equation}\normalsize
In order to verify the proposed result we need to derive (using the polynomial expansion method) the explicit forms for the following expressions,
\begin{itemize}
\item{$\field{S}^N_M(\{j,0^{p} \})$, this is the most elementary calculation.}
\item{$\field{S}^N_M(\{1^{p-r},0^{r+1} \})$ and $\field{S}^N_M(\{M,1^{p-r},0^{r} \})$, the coefficients of $u^{-M}_{N-p}$ and $u^{M}_{N-p}$ in eq. \ref{machma}.}
\item{$\field{S}^N_M(\{j,1^{p-r},0^{r}  \})$ and $\field{S}^N_M(\{j+1,1^{p-r-1},0^{r+1} \})$,  the coefficients of $u^{-M+2j}_{N-p}$ in eq. \ref{machma}. This case will obviously involve a seperating argument.}
\end{itemize}
\textbf{Deriving $\mathbf{\field{S}^N_M(\{q, 0^p \})}$, $\mathbf{0 \le q \le M}$, from the polynomial expansion of $\mathbf{\field{S}^N_M(\{ 0^p \})}$.} The assumed form (eq. \ref{tochi}) of $\field{S}^N_M(\{0^{p} \})$ is explicitly given as,
\small\begin{equation*}\begin{split}
\field{S}^N_M(\{0^p \}) = \frac{\Omega_{\hat{u}_{N-p}}}{u^M_{N-p}} \textrm{det}\left[ \varrho^{k,j}_{k-N}, \varrho^{N-p,j}_{-p}, \varrho^{N-p,j}_{1-p},\dots,\varrho^{N-p,j}_{-1},\varrho^{N-p,j}_{0} \right]_{j=1,\dots,N\atop{k=1,\dots,N-p-1 }}
\end{split}\end{equation*}\normalsize
Using the following symmetric polynomial identity, for $\varrho^{N-p,j}_{s}$, $-p \le s \le 0$,
\small\begin{equation}\begin{split}
\varrho^{N-p,j}_{s}& =  \sum^{M+s}_{q=0}u^{2q}_{N-p}h_{M+s-q} (\{u^2 \}_{N-p-1},v^2_j)\\
\Rightarrow \varrho^{N-p,j}_{s} - u^{2}_{N-p} \varrho^{N-p,j}_{s+1} &=  \varrho^{N-p-1,j}_{s} 
\label{niceiden}\end{split}\end{equation}\normalsize
we apply the following column operations (in order),
\small\begin{equation*}\begin{array}{lcl}
C_{N-p} &\rightarrow &C_{N-p} - u^2_{N-p}C_{N-p+1}\\
& \vdots&\\
C_{N-1} &\rightarrow& C_{N-1} - u^2_{N-p}C_{N}
\end{array}\end{equation*}\normalsize
to obtain,
\small\begin{equation*}\begin{split}
\field{S}^N_M(\{0^p \}) &= \frac{\Omega_{\hat{u}_{N-p}}}{u^M_{N-p}} \textrm{det}\left[  \varrho^{k,j}_{k-N},\varrho^{N-p-1,j}_{-p},\dots,\varrho^{N-p-1,j}_{-1},\varrho^{N-p,j}_{0} \right]_{j=1,\dots,N\atop{k=1,\dots,N-p-1 }}
\end{split}\end{equation*}\normalsize
Realizing that the final column solely contains terms of $u^2_{N-p}$, we expand along this column to obtain,
\small\begin{equation}\begin{split}
&\field{S}^N_M(\{0^p \}) \\
=&\frac{\Omega_{\hat{u}_{N-p}}}{u^M_{N-p}} \sum^M_{q=0}u^{2q}_{N-p} \textrm{det}\left[  \varrho^{k,j}_{k-N},\varrho^{N-p-1,j}_{-p},,\dots,\varrho^{N-p-1,j}_{-1},\varrho^{N-p-1,j}_{q} \right]_{j=1,\dots,N\atop{k=1,\dots,N-p-1 }}
\label{ache}\end{split}\end{equation}\normalsize
and comparing eq. \ref{ache} with \ref{ache2} we have,
\small\begin{equation}
\field{S}^N_M(\{q, 0^p \}) = \Omega_{\hat{u}_{N-p}}  \textrm{det}\left[  \varrho^{k,j}_{k-N},\varrho^{N-p-1,j}_{-p},\dots,\varrho^{N-p-1,j}_{-1},\varrho^{N-p-1,j}_{q} \right]_{j=1,\dots,N\atop{k=1,\dots,N-p-1 }}
\label{prepatory}\end{equation}\normalsize
With this prepatory case completed, we now move on to expand the more general expression, $\field{S}^N_M(\{1^r,0^{p-r} \})$.\\
\\
\textbf{Deriving $\mathbf{\field{S}^N_M(\{1^{p-r},0^{r+1}  \}) }$ and $\mathbf{\field{S}^N_M(\{M,1^{p-r},0^{r}  \})}$ from the polynomial expansion of $\mathbf{\field{S}^N_M(\{1^r,0^{p-r} \})}$.} The assumed form (eq. \ref{tochi}) of $\field{S}^N_M(\{1^r,0^{p-r} \})$ is explicitly given as,
\small\begin{equation*}\begin{split}
&\field{S}^N_M(\{1^r,0^{p-r} \}) \\
=&\frac{\Omega_{\hat{u}_{N-p}}}{u^M_{N-p}}  \textrm{det}\left[  \varrho^{k,j}_{k-N},\underbrace{\varrho^{N-p,j}_{-p},\dots, \varrho^{N-p,j}_{r-p}}_{r+1},\underbrace{\varrho^{N-p,j}_{r-p+2} \dots , \varrho^{N-p,j}_{1}}_{p-r} \right]_{j=1,\dots,N\atop{k=1,\dots,N-p-1 }}
\end{split}\end{equation*}\normalsize
Applying the column operations (in order),
\small\begin{equation*}\begin{array}{lcl}
C_{N-p} &\rightarrow & C_{N-p} - u^2_{N-p}C_{N-p+1}   \\
& \vdots &  \\
C_{N-p+r-1} &\rightarrow & C_{N-p+r-1} - u^2_{N-p}C_{N-p+r}\\
C_{N-p+r+1} &\rightarrow & C_{N-p+r+1} - u^2_{N-p}C_{N-p+r+2}\\
& \vdots & \\
C_{N-1} &\rightarrow & C_{N-1} - u^2_{N-p}C_{N}
\end{array}\end{equation*}\normalsize
in conjunction with the symmetric polynomials identities in eq. \ref{niceiden}, we obtain,
\small\begin{equation}\begin{split}
&\field{S}^N_M(\{1^r,0^{p-r} \})\\
=&\frac{\Omega_{\hat{u}_{N-p}}}{u^M_{N-p}}  \textrm{det}\left[  \varrho^{k_1,j}_{k_1-N},\varrho^{N-p-1,j}_{-k_2}, \varrho^{N-p,j}_{r-p},\varrho^{N-p-1,j}_{-k_3},\varrho^{N-p,j}_{1} \right]^{j=1,\dots,N \atop{ k_1=1,\dots,N-p-1} }_{k_2=p,\dots,p-r+1\atop{  k_3=p-r-1,\dots,0}}
\label{nonconv}\end{split}\end{equation}\normalsize
where $k_2$ and $k_3$ are in descending order.\\
\\ 
Realizing that only columns $N-p+r$ and $N$ contain the variable $u^2_{N-p}$, we now proceed to suppress all columns except $N-p+r$ and $N$ from eq. \ref{nonconv} for notational convenience, 
\small\begin{equation}\begin{split}
& \textrm{det}\left[\mathcal{P}, \varrho^{N-p,j}_{r-p},\varrho^{N-p,j}_{1} \right]\\
=&  \textrm{det}\left[  \varrho^{k_1,j}_{k_1-N},\varrho^{N-p-1,j}_{-k_2}, \varrho^{N-p,j}_{r-p},\varrho^{N-p-1,j}_{-k_3},\varrho^{N-p,j}_{1} \right]^{j=1,\dots,N \atop{ k_1=1,\dots,N-p-1} }_{k_2=p,\dots,p-r+1\atop{  k_3=p-r-1,\dots,0}}
\label{conv}\end{split}\end{equation}\normalsize
Expanding the entries of column $N-p+r$,
\small\begin{equation}
\varrho^{N-p,j}_{r-p} = \sum^{M+p-r}_{q=0} u^{2q}_{N-p} \varrho^{N-p-1,j}_{r-p+q}
\label{summy}\end{equation}\normalsize
we notice that at index $q=2, 3, \dots, p-r$, eq. \ref{summy} is proportional to column $N-p+r+1, N-p+r+2, \dots, N-1$ respectively, thus we can delete these indices from the sum.\\
\\
Additionally, for the indices, $q=p-r+1, \dots, M+p-r$, we have,
\small\begin{equation*}
\sum^{M+p-r}_{q=p-r+1} u^{2q}_{N-p} \varrho^{N-p-1,j}_{r-p+q} = u^{2(p-r+1)}_{N-p} \varrho^{N-p,j}_{1}
\end{equation*}\normalsize
which is proportional to column $N$. Taking advantage of the above results, the entries of column $N-p+r$ can be reduced to,
\begin{eqnarray*}
\sum^1_{q=0} u^{2q}_{N-p} \varrho^{N-p-1,j}_{r-p+q}
\end{eqnarray*}
without affecting the value of the determinant.\\
\\
Thus expanding $\field{S}^N_M(\{1^r,0^{p-r} \})$ as a polynomial in $u^2_{N-p}$ we obtain,
\small\begin{equation}\begin{split}
 \frac{1}{\Omega_{\hat{u}_{N-p}}} \field{S}^N_M(\{1^r,0^{p-r} \}) = \sum^{M-1}_{q_1=0}\sum^1_{q_2=0} u^{2(q_1+q_2)-M}_{N-p} \textrm{det}\left[ \mathcal{P}, \varrho^{N-p-1,j}_{r-p+q_2},\varrho^{N-p-1,j}_{1+q_1} \right]\\
= u^{-M}_{N-p}  \textrm{det}\left[ \mathcal{P}, \varrho^{N-p-1,j}_{r-p},\varrho^{N-p-1,j}_{1} \right] + u^{M}_{N-p}\textrm{det}\left[ \mathcal{P}, \varrho^{N-p-1,j}_{r-p+1},\varrho^{N-p-1,j}_{M} \right]\\
+ \sum^{M-1}_{q=1} u^{-M+2q}_{N-p} \left\{  \textrm{det}\left[ \mathcal{P}, \varrho^{N-p-1,j}_{r-p},\varrho^{N-p-1,j}_{1+q} \right]+\textrm{det}\left[ \mathcal{P}, \varrho^{N-p-1,j}_{r-p+1},\varrho^{N-p-1,j}_{q} \right]  \right\}
\label{nicenice}\end{split}\end{equation}\normalsize
Comparing eq. \ref{nicenice} with eq. \ref{machma} we instantly obtain the sought after expressions,
\small\begin{equation}\begin{split}
 \field{S}^N_M(\{1^{p-r},0^{r+1}  \}) &= \Omega_{\hat{u}_{N-p}} \textrm{det}\left[ \mathcal{P}, \varrho^{N-p-1,j}_{r-p},\varrho^{N-p-1,j}_{1} \right] \\
\field{S}^N_M(\{M,1^{p-r},0^{r}  \}) &= \Omega_{\hat{u}_{N-p}} \textrm{det}\left[ \mathcal{P}, \varrho^{N-p-1,j}_{r-p+1},\varrho^{N-p-1,j}_{M} \right] 
\label{biggest}\end{split}\end{equation}\normalsize\\
\textbf{Disentangling the remainder by considering the overlap of terms.} Additionally from comparing eq. \ref{nicenice} with eq. \ref{machma} we obtain the entangled expressions,
\small\begin{equation}\begin{split}
 \field{S}^N_M(\{q,1^{p-r},0^{r}  \}) + \field{S}^N_M(\{q+1,1^{p-r-1},0^{r+1} \})  \\
= \Omega_{\hat{u}_{N-p}} \left\{  \textrm{det}\left[ \mathcal{P}, \varrho^{N-p-1,j}_{r-p},\varrho^{N-p-1,j}_{1+q} \right]+\textrm{det}\left[ \mathcal{P}, \varrho^{N-p-1,j}_{r-p+1},\varrho^{N-p-1,j}_{q} \right]  \right\}
\label{tangle}\end{split}\end{equation}\normalsize
for $1 \le q \le M-1$, $0 \le r \le p-1$.\\
\\
In order to disentangle this expression, we consider strategic $r$ values where one term in eq. \ref{tangle} is already known from a previous result. To begin, consider $\mathbf{r=p-1}$, $1 \le q \le M-1$,
\small\begin{equation}\begin{split}
 \field{S}^N_M(\{q,1,0^{p-1}  \}) + \underbrace{\field{S}^N_M(\{q+1,0^{p} \})}_{\textrm{use eq. \ref{prepatory}}}  \\
= \Omega_{\hat{u}_{N-p}} \left\{  \textrm{det}\left[ \mathcal{P}, \varrho^{N-p-1,j}_{-1},\varrho^{N-p-1,j}_{1+q} \right]+\textrm{det}\left[ \mathcal{P}, \varrho^{N-p-1,j}_{0},\varrho^{N-p-1,j}_{q} \right]  \right\}\\
\Rightarrow \field{S}^N_M(\{q,1,0^{p-1}  \}) =  \Omega_{\hat{u}_{N-p}} \textrm{det}\left[ \mathcal{P}, \varrho^{N-p-1,j}_{-1},\varrho^{N-p-1,j}_{1+q} \right]
\label{tangle2}\end{split}\end{equation}\normalsize
The above result allows us to similarly consider $\textbf{r=p-2}$,
\small\begin{equation}\begin{split}
 \field{S}^N_M(\{q,1^2,0^{p-2}  \}) + \underbrace{\field{S}^N_M(\{q+1,1,0^{p-1} \})}_{\textrm{use eq. \ref{tangle2} for $1\le q \le M-2$}\atop{\textrm{use eq. \ref{biggest} for $q=M-1$}} }  \\
= \Omega_{\hat{u}_{N-p}} \left\{  \textrm{det}\left[ \mathcal{P}, \varrho^{N-p-1,j}_{-2},\varrho^{N-p-1,j}_{1+q} \right]+\textrm{det}\left[ \mathcal{P}, \varrho^{N-p-1,j}_{-1},\varrho^{N-p-1,j}_{q} \right]  \right\}\\
\Rightarrow  \field{S}^N_M(\{q,1^2,0^{p-2}  \})  =  \Omega_{\hat{u}_{N-p}}  \textrm{det}\left[ \mathcal{P}, \varrho^{N-p-1,j}_{-2},\varrho^{N-p-1,j}_{1+q} \right]
\label{tangle3}\end{split}\end{equation}\normalsize
Thus applying the above algorithm a general number of times we are able to fully disentangle eq. \ref{tangle} for general $r$ and $q$,
\small\begin{equation}
 \field{S}^N_M(\{q,1^r,0^{p-r}  \})  =  \Omega_{\hat{u}_{N-p}}  \textrm{det}\left[ \mathcal{P}, \varrho^{N-p-1,j}_{-r},\varrho^{N-p-1,j}_{1+q} \right]
\label{tangle4}\end{equation}\normalsize
Thus putting everything together, we receive,
\small\begin{equation}\begin{split}
\field{S}^N_M(\{r_1, \dots, r_{p+1} \})\\ 
= \Omega_{\hat{u}_{N-p}} \textrm{det}\left[ c_{jk},\varrho^{N-p-1,j}_{r_{p+1}-p}, \varrho^{N-p-1,j}_{r_p+1-p},\dots,\varrho^{N-p-1,j}_{r_{1}} \right]_{j=1,\dots,N\atop{k=1,\dots,N-p-1 }}\\
r_1 \in \{ 0,1, \dots,M \} \textrm{  ,  } r_2 \in \{0,1\} \textrm{  , \dots ,  } r_{p+1} \in \{ 0,1\}\\
r_1 \ge  r_2 \ge \dots \ge r_{p+1} \textrm{   ,   } 1 \le p \le N-1
\end{split}\end{equation}\normalsize
which completes our inductive proof. $\square$\\
\\
\textbf{Final result for the second class of wave-function.} Thus, letting $r_1 = \dots = r_{p}=1$ in eq. \ref{tochi}, we obtain the single determinant form for the second class of wave-function,
\small\begin{equation}\begin{split}
\field{S}^N_M(\{1^p\}) &= \langle \Psi_M (v_1,v_2,\dots,v_{N})|\Psi^{\{1^p\}}_M (u_1,\dots,u_{N-q})\rangle  \\
&=\Omega_{\hat{u}_{N-p+1}} \textrm{det}\left[ \varrho^{k,j}_{k-N},\varrho^{N-p,j}_{2-p}, \varrho^{N-p,j}_{3-p},\dots,\varrho^{N-p,j}_{1} \right]_{j=1,\dots,N\atop{k=1,\dots,N-p }}\\
&=\lim_{u_r \rightarrow 0\atop{N-p+1 \le r \le N}} \left( \frac{\prod^N_{j=1}v_j}{\prod^{N-p}_{j=1}u_j} \right)^M \tau(n-M) \hat{w}^{(\infty)}_{p}(n-M)  
\end{split}\end{equation}\normalsize
for $1 \le p \le N$.
\section{Hall-Littlewod plane partitions}
Unlike the previous 2 sections of this chapter, this section functions more as an observation of the correspondence of the results obtained in \cite{MO2} and the scale transformed 2-Toda hierarchy shown in section 1.5 of this thesis. Due to the limited nature of the results obtained, we leave most definitions of this section to a minimum. 
\subsection{Charged t-fermions}
For a more complete introduction to charged $t$-fermions, see \cite{Jing1,Jing2}. A comprehensive introduction to $t=0$ free fermions and their associated Fock space is given in section 3.2 of this thesis and the references contained therein. \\
\\
\textbf{$t$-anti commutation relations. } The following model is defined by the non commutative operators $\psi_i$ and $\psi^*_j$, $i,j\in \field{Z}$, whose anti commutation relations are given by,
\small\begin{equation}\begin{split}
\{ \psi_m, \psi_n \}_+ &= t \psi_{m+1} \psi_{n-1} + t\psi_{n+1} \psi_{m-1}\\
\{ \psi^*_m, \psi^*_n \}_+ &= t \psi^*_{m-1} \psi^*_{n+1} + t\psi^*_{n-1} \psi^*_{m+1}\\
\{ \psi_m, \psi^*_n \}_+ &= t \psi_{m-1} \psi^*_{n-1} + t\psi^*_{n+1} \psi_{m+1}+(1-t)^2 \delta_{mn}
\end{split}\end{equation}\normalsize
where $t \in \field{C}$.\\
\\
\textbf{$t$-Heisenberg generators.} Additionally, we define the $t$-Heisenberg generators, $H^{(t)}_m$, $m \in \field{Z} \backslash \{0\}$,
\small\begin{equation}
H^{(t)}_m = \left\{ \begin{array}{cc}
\frac{1}{1-t} \sum_{j \in \field{Z}} \psi_j \psi^*_{j+m} & m \ge 1 \\
\frac{1}{(1-t)(1-t^{-m})} \sum_{j \in \field{Z}} \psi_j \psi^*_{j+m} & m \le -1 
\end{array}\right.
\end{equation}\normalsize
whose commutation relation is given by,
\small\begin{equation*}
[ H^{(t)}_m, H^{(t)}_n] = \frac{m}{1-t^{|m|}}\delta_{m,-n}
\end{equation*}\normalsize
\textbf{$t$-vertex operators.}  Lastly we define the $t$-vertex operators, $\Gamma_{\pm}(u,t)$, as exponentials of weighted sums of the $t$-Heisenberg generators,
\small\begin{equation}\begin{split}
\Gamma_+(u,t) &= \exp \left\{ - \sum^{\infty}_{m=1} \frac{1-t^m}{m}\frac{1}{u^m}H^{(t)}_m \right\} \\
\Gamma_-(u,t) &= \exp \left\{ - \sum^{\infty}_{m=1} \frac{1-t^m}{m}u^m H^{(t)}_{-m} \right\}
\end{split}\end{equation}\normalsize
\subsection{Vertex operator expectation value} In \cite{Onky}, Okounkov et. al. observed that ($t=0$) vertex operator expectation values taken at special limits of the $u_i$'s and $v_j$'s generate random plane partitions. This process was generalized in \cite{MO1} for neutral free fermions, and in the corresponding special limits of the $u_i$'s and $v_j$'s diagonally strict plane partitions were generated. For general $u$ and $v$ values, it is known \cite{NimmoBKP} that the expectation value, given as the bilinear sum of $Q$-Schur polynomials, is a (restricted) $\tau$-function of the BKP hierarchy with two sets of time variables. \\
\\
We now give the results of \cite{MO2} and show that the expectation value for general $t$ is a $\tau$-function of the hierarchy detailed in section 1.5.
\small\begin{equation}\begin{split}
\field{S}_{N} (u_1,\dots,u_N,v_1,\dots,v_N;t) = \langle 0| \Gamma_+(u^{-1}_N,t) \dots \Gamma_+(u^{-1}_1,t) \Gamma_-(v_1,t) \dots \Gamma_-(v_N,t) |0 \rangle
\end{split}\end{equation}\normalsize
is given explicitly as,
\small\begin{equation}\begin{split}
\field{S}_{N} (\vec{u},\vec{v};t)&= \prod^N_{j,k =1} \frac{1-t u_j v_k}{1-u_j v_k}\\
&= \sum_{\{\lambda\} \subseteq (\infty)^{(N)}}P_{\{\lambda\}}\left( u_1,\dots,u_{N};t \right) Q_{\{\lambda\}}\left(v_1,\dots,v_{N};t \right)
\end{split}\end{equation}\normalsize
where $P_{\{\lambda\}}\left( \vec{u};t \right) = \frac{1}{b_{\{\lambda \}}}Q_{\{\lambda\}}\left( \vec{u};t \right)$, is the Hall-Littlewood polynomial of partition $\{ \lambda \}$.\\
\\
\textbf{The expectation value as a restricted, scale transformed, 2-Toda $\tau$-function.}
We now consider the unrestricted, scale transformed, $\tau$-function constructed in section 1.5,
\small\begin{equation}\begin{split}
\tau(s=N+m, \vec{u},\vec{v};t) &= \field{S}_{N} (u_1,\dots,u_N,v_1,\dots,v_N;t) \\
&= \sum_{\{\lambda\} \subseteq (\infty)^{(N)}}P_{\{\lambda\}}\left( u_1,\dots,u_{N};t \right) Q_{\{\lambda\}}\left(v_1,\dots,v_{N};t \right)
\end{split}\end{equation}\normalsize
we obtain the observation that the restricted $\tau$-function is equal to the finite scalar product of the $t$-vertex operators.
%%%%%%%%%%%%%%%%%%%%%%%%%%%%%%%%%%%%%%%%%%%%%%%%%%%%%%%%%%%%%%%%%%%%%%%%%%%%%%%%%%%%%%%%%%%%%%%%%%%%%%%%%%%%%%%%%%%%
\newpage
%%%%%%%%%%%%%%%%%%%%%%%%%%%%%%%%%%%%%%%%%%%%%%%%%%%%%%%%%%%%%%%%%%%%%%%%%%%%%%%%%%%%%%%%%%%%%%%%%%%%%%%%%%%%%%%%%%%%
\chapter{The six vertex model and KP}
\section{Domain wall partition function (DWPF)}
In this chapter we introduce the well studied six vertex model. This model is statistical in nature, and as such, most of the interesting quantities  consist of weighted sums of all allowable configurations. Within the framework of this chapter, we are interested in two main quantities, the domain wall partition function (DWPF) and the associated scalar product. The main aim of this chapter is to show that both quantities can be fermionized in a specific form \cite{bluebook, 6} that automatically means that these quantities are KP $\tau$-functions with restricted time variables. \\
\\
We begin this section with a detailed introduction to the statistics of the six vertex model under domain wall boundary conditions (DWBC's). The literature on the following model, for both periodic and DW boundary conditions, is \textit{immense} and we offer chap. 8 of \cite{Baxterbook} and sections VI-VII of \cite{purplebook} as typical examples for the model under respective boundary conditions.
\subsection{Overview of the model}
\begin{figure}[h!]
\begin{center}
\includegraphics[angle=0,scale=0.25]{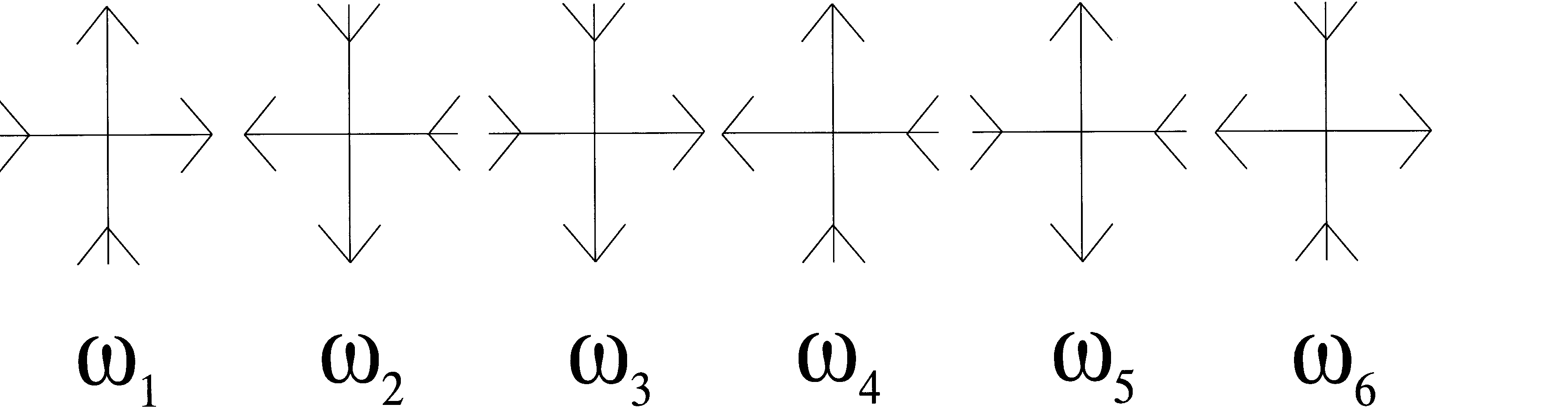}
\caption{\footnotesize{The six allowable vertex configurations.}}
\label{6Ve.b}
\end{center}
\end{figure}

\begin{figure}[h!]
\begin{center}
\includegraphics[angle=0,scale=0.25]{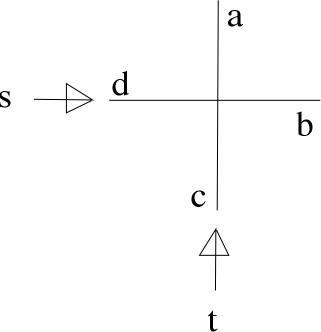}
\caption{\footnotesize{Labeling of the vertex $X(s,t)^{a.b}_{d,c}$.}}
\label{6Ve.c}
\end{center}
\end{figure}

\noindent \textbf{The $N \times N$ lattice and rapidity flows.} Consider a square lattice with $N$ horizontal lines (rows) and $N$ vertical lines (columns) that intersect at $N^2$ points (vertices). To each row we associate a horizontal rapidity flow, $s_i \in \field{C}$, $1\le i \le N$, which is oriented from left to right. Similarly to each column we associate a vertical rapidity, $t_j \in \field{C}$, $1 \le j \le N$, which is oriented from bottom to top.
\begin{figure}[h!]
\begin{center}
\includegraphics[angle=0,scale=0.25]{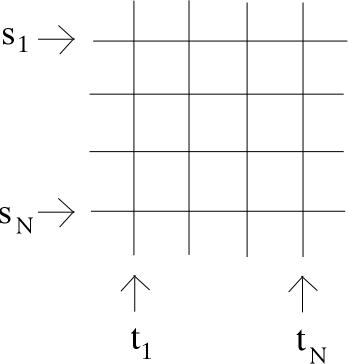}
\caption{\footnotesize{The $N\times N$ lattice with horizontal and vertical rapidity flows.}}
\label{6Ve.a}
\end{center}
\end{figure}\\
\\
\noindent \textbf{State variables and vertex weights.} With each of the $N^2$ vertices are associated four state variables, represented as arrows pointing in or out of the intersection. This obviously leads to $2^4 = 16$ distinct configurations for each vertex. We now impose that only those vertices with two arrows pointing in and two arrows pointing out are allowed, thus restricting the amount of allowable configurations to six. These allowable configurations are shown in fig. \ref{6Ve.b}.
\\
With each allowable vertex is an associated Boltzmann weight, specified by the difference of the the horizontal and vertical rapidities, $s_i-t_j$, $1\le i,j \le N$, and a global crossing parameter, $\lambda \in \field{C}$,
\small\begin{equation}\begin{split}
X(s,t)^{1,1}_{1,1} = \omega_1(s-t) &= \sinh \left( \lambda( -s+t +1) \right) \\
X(s,t)^{2,2}_{2,2} = \omega_2(s-t) &= \sinh \left( \lambda( -s+t +1) \right) \\
X(s,t)^{2,1}_{1,2} = \omega_3(s-t) &= \sinh \left( \lambda( -s+t ) \right) \\
X(s,t)^{1,2}_{2,1} = \omega_4(s-t) &= \sinh \left( \lambda( -s+t ) \right) \\
X(s,t)^{1,2}_{1,2} = \omega_5(s-t) &= \sinh \left( \lambda \right) \\
X(s,t)^{2,1}_{2,1} = \omega_6(s-t) &= \sinh \left( \lambda \right)
\label{Boltz.6V}\end{split}\end{equation}\normalsize
Fig. \ref{6Ve.c} specifies the convention used for the assignment of the state variables.
\\
\textbf{Yang Baxter equations.} Using this specific parameterization of the six allowable weights, we are assured that the Yang Baxter equations, 
\small\begin{equation}\begin{split}
&\sum_{g_1, g_2, g_3 \in \{1,2\}}X(s_1-s_2)^{h_1 h_2}_{g_2 g_1} X(s_1-s_3)^{g_1 h_3}_{g_3 q_1}  X(s_2-s_3)^{g_2 g_3}_{q_3 q_2} \\
=& \sum_{g_1, g_2, g_3 \in \{1,2\}}X(s_2-s_3)^{h_2 h_3}_{g_3 g_2} X(s_1-s_3)^{h_1 g_3}_{q_3 g_1}  X(s_1-s_2)^{g_1 g_2}_{q_2 q_1}  
\label{YangB.6V}\end{split}\end{equation}\normalsize
are valid.\\
\\
\textbf{Domain wall boundary conditions (DWBC's).} For the remainder of this section we specify that the outer-most left and right arrows point outwards, and the outer-most top and bottom arrows point inwards, while the inside bulk remains free.
\begin{figure}[h!]
\begin{center}
\includegraphics[angle=0,scale=0.25]{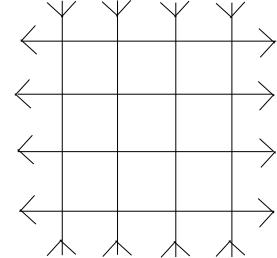}
\caption{\footnotesize{Typical example of DWBC's.}}
\label{6Ve.d}
\end{center}
\end{figure}\\
\\
\textbf{Domain wall partition function (DWPF).} The DWPF, $Z_N$, is defined as the weighted sum over all valid lattice configurations under DWBC's.
\small\begin{equation}
Z_N \left( \vec{s},\vec{t},\lambda \right) =Z_N \left( \vec{s},\vec{t}\right) = \sum_{\textrm{allowable}\atop{\textrm{configurations}}} \left\{ \prod_{\textrm{all}\atop{\textrm{vertices}}} X^{a,b}_{c,d}(s_i-v_j) \right\} 
\label{weightedsum.6V}\end{equation}\normalsize
\textbf{Korepin's conditions for $Z_N$.} In \cite{Korepini} Korepin obtained four conditions that uniquely determine the expression for $Z_N \left( \vec{s},\vec{t}\right)$. These are,
\begin{itemize}
\item{$Z_N \left( \vec{s},\vec{t}\right)$ is an order $N-1$ trigonometric polynomial in any of the rapidities $\{s\}$ or $\{t\}$.}
\item{$Z_N \left( \vec{s},\vec{t}\right)$ is a symmetric polynomial in the set $\{s\}$ and the set $\{t\}$.}
\item{Setting the rapidity variables, $s_1 = t_1+1$, we obtain the recursion relation,}
\end{itemize}
\small\begin{equation*}\begin{split}
Z_N|_{s_1 = t_1+1} = & \left( \prod^N_{i=2} \sinh \left( \lambda (-s_i+t_1) \right) \right) \sinh \left( \lambda \right) \left( \prod^N_{j=2} \sinh \left( \lambda (-s_1+t_j) \right) \right) \\
& \times Z_{N-1}(\hat{s}_1,\hat{t}_1)
\end{split}\end{equation*}\normalsize
\begin{itemize}
\item{The initial condition is given by $Z_1(s_1,t_1) = \sinh  \left( \lambda \right)$.}
\end{itemize}
We shall generate similar conditions for alternative vertex/height models in chapters 4 and 6 of this work.\\
\\
We are now ready to begin considering the determinant expressions of the DWPF that adhere to the above four conditions.
\subsection{Izergin's determinant expression}
Performing the change of variables,
\small\begin{equation}\begin{split}
e^{2 \lambda s_i} &=  u_i \textrm{  ,  } 1 \le i \le N  \\
 e^{2 \lambda t_i} &=  v_i \textrm{  ,  } 1 \le i \le N  \\
e^{-2 \lambda } &=  q 
\label{p.2}\end{split}\end{equation}\normalsize
Izergin's determinant expression the DWPF is given by,
\small\begin{equation}
Z^{I}_{N} (\vec{u},\vec{v}) =\Upsilon_N \frac{\prod^N_{i,j=1}(u_i-v_j)(q u_i - v_j)}{\prod_{1 \le i < j \le N} (u_i - u_j)(v_j-v_i)} \textrm{det}\left[\frac{1}{(u_i-v_j)(q u_i-v_j)}\right]^N_{i,j=1} 
\label{p.1} \end{equation}\normalsize
where $\Upsilon_N = 2^{N(N-1)} q^{\frac{1}{2}N(N-1)} \left( \prod^N_{i=1}u_i v_i \right)^{\frac{N-1}{2}} $.\\
\\
It is a relatively straightforward process \cite{23} to show that Izergin's determinant expression satisfies the four conditions of Korepin.
\subsection{Lascoux's determinant expression} An equivalent expression of the partition function using basis (elementary and homogeneous) symmetric polynomials, due to Lascoux \cite{Lascoux1}, is given by,
\small\begin{equation}
Z^L_N(\vec{u},\vec{v}) =\Upsilon_N \textrm{det} \left[ \left( h_{j-i}(\vec{u}) \right)^{2N-1}_{i,j=1}\left( \kappa_{j,k}(\vec{v})  \right)^{2N-1}_{j,k=1} \right]^N_{i,k=1},
\label{p.3}
\end{equation}\normalsize
where,
\small\begin{equation}
\kappa_{j,k}(\vec{v}) = \frac{q^{j-k+1}-q^{k-1}}{q-1} (-1)^{N-j+k-1}e_{N-j+k-1}(\vec{v}) 
\end{equation}\normalsize\\
\textbf{Expanding $Z^L_N$ as a polynomial in Schur/character polynomials.} Using the Cauchy-Binet formula to expand the determinant of the product of non square matrices, the above expression becomes,
\small\begin{equation*}
Z^{L}_{N  } (\vec{u},\vec{v}) =  \Upsilon_N  \sum_{1 \le j_1 < \dots < j_{N} \le 2N-1} \textrm{det} \left[ h_{j_l -i}(\vec{u})\right]^N_{i,l=1}\textrm{det} \left[ \kappa_{j_l,k}(\vec{v}) \right]^N_{l,k=1}
\end{equation*}\normalsize
Performing the change of variable, $j_{\alpha} \rightarrow \lambda_{\alpha}+ \alpha, 1 \le \alpha \le N$, and applying the workings from section 1.4.3 to express the determinant of the complete homogeneous symmetric polynomials as Schur polynomials, we obtain the following expression,
\small\begin{equation}\begin{split}
Z^{L}_{N  } (\vec{u},\vec{v})& =\Upsilon_N \sum_{0 \le \lambda_1 \le  \dots \le \lambda_{N} \le N-1} \textrm{det} \left[ h_{\lambda_i + i -l}(\vec{u})\right]^N_{i,l=1}\textrm{det} \left[ \kappa_{\lambda_l +l,k}(\vec{v}) \right]^N_{l,k=1}\\
&=\Upsilon_N \sum_{\{\lambda\} \subseteq (N-1)^{N}} c^{(N)}_{\{\lambda\}}(\vec{v})S_{\{\lambda\}}(\vec{u}) 
\label{fermone}\end{split}\end{equation}\normalsize
where we have the usual identities, 
\small\begin{equation*}\begin{split}
\sum_{\{\lambda\} \subseteq (N-1)^{N}} &=  \sum_{0 \le \lambda_N \le  \dots \le \lambda_{1} \le N-1} \\
 S_{\{\lambda\}}(\vec{u}) &= \textrm{det} \left[ h_{\lambda_i + j -i}(\vec{u})\right]^N_{i,j=1}
\end{split}\end{equation*}\normalsize
and the coefficients, $c^{(N)}_{\{\lambda\}} (\vec{v})$, are given by,
\small\begin{equation}
c^{(N)}_{\{\lambda\}} (\vec{v})= \textrm{det} \left[ \kappa_{\lambda_{N+1-i} +i,j}(\vec{v}) \right]^N_{i,j=1}
\label{p.18} \end{equation}\normalsize
In \cite{Strog1,Strog2,Lascoux2} it was shown that when the crossing parameter is equal to a third root of unity, the partition function is symmetric between \textit{both} sets of rapidities, and can be expressed as a \textit{single} Schur polynomial in both sets of rapidities.\\
\\
We shall show that the above expression for the DWPF can be fermionized via the boson-fermion correspondence. By definition, this form is a restricted $\tau$-function of the KP hierarchy. For more details of this statement, see section 3.5 of this work. First however, we shall explicitly derive Lascoux's form  (eq. \ref{p.3}) starting from Izergin's (eq. \ref{p.1}). Additionally, we also consider one more alternative expression for the DWPF which involves basis symmetric polynomials, given by Kirillov and Smirnov \cite{KS}.\\
\\
\textbf{Derivation of Lascoux's expression.} We shall present the derivation of Lascoux's result for two reasons. Firstly Lascoux's result, like Tsilevich's result considered in section 2.1.5, is quite pretty and demands respect. Secondly and more importantly, contained within this derivation is a series of row operations that are \textit{extremely} helpful in section 3.4.2.\\
\\
\textbf{Necessary definitions.} The most important object in this derivation is the so called \textit{divided difference operator}, $\partial_i$, which acts on functions involving pairs of variables $\{u_i,z_i \}$, $i \in \field{N}$,
\small\begin{equation}
\partial_i : f(\{u_i,z_i \}) \rightarrow \frac{f(\{u_i,z_i \}) - f(\{u_i \leftrightarrow z_i \})}{u_i - z_i}
 \end{equation}\normalsize
We also define the symmetric polynomials $w_k(\{u_1, \dots, u_M\}|\{v_1, \dots v_N\})$, which generally consist of two sets of variables which do not necessarily have the same cardinality. The generating function, $W(z;\vec{u}|\vec{v})$, is given by the multiplication of the generating functions of the elementary symmetric polynomials and the complete homogeneous symmetric polynomials,
\small\begin{equation}\begin{split}
W(z;\vec{u}|\vec{v}) =\sum^{\infty}_{j=0} z^j w_j(\vec{u}|\vec{v}) &= \frac{\prod^N_{j_1=1}(1-z v_{j_1})}{\prod^M_{j_2=1}(1-z u_{j_2})}  \\
\Rightarrow w_j(\vec{u}|\vec{v})=&\sum^j_{k=0} (-1)^k e_k(\vec{v}) h_{j-k}(\vec{u})
\label{l.sym}\end{split}\end{equation}\normalsize
where $e(\vec{v})$ and $h(\vec{u})$ are given by the usual elementary symmetric polynomials and the complete homogeneous symmetric polynomials respectively. Obviously, when the first and second sets are empty we obtain,
\small\begin{equation*}\begin{split}
 w_j(\phi | \vec{v})&= (-1)^j e_j(\vec{v}) \\
 w_j(\vec{u} | \phi)&=h_{j}(\vec{u})
\end{split}\end{equation*}\normalsize
Lastly, we define the two row symmetric function, $w_{(j,k)}(\{\vec{u}|\vec{v}\}, \{\vec{\mu}|\vec{\rho}\} )$, (which contains four sets of variables) given by the $2 \times 2$ determinant expression,
\small\begin{equation}
w_{(j,k)}(\{\vec{u}|\vec{v}\},\{\vec{\mu}| \vec{\rho}\}) = \textrm{det} \left[ \begin{array}{cc}
w_j(\vec{u}|\vec{v}) & w_{j+1} (\vec{u}|\vec{v}) \\
w_{k-1} (\vec{\mu}| \vec{\rho}) & w_{k} (\vec{\mu}| \vec{\rho}) 
\end{array}\right]
\label{l.2row}\end{equation}\normalsize\\
\textbf{Comment on necessary results.} The derivation of Lascoux's form relies on the following four (seemingly unmotivated) necessary results regarding the divided difference operator and the symmetric polynomials $w_k(\vec{u}|\vec{v})$. We obtain these four results and explicitly show how they are applied for the derivation of $Z^{L}_{N  } (\vec{u},\vec{v})$. \\
\\
\textbf{Result (1).}
\small\begin{equation}
w_{(N,j)}(\{u_i|\vec{v} \}, \{u_i,z_i|\phi\})= z^j_i \prod^N_{k=1}(u_i-v_k)
\label{l.res1}
\end{equation}\normalsize
To verify eq. \ref{l.res1} we expand $w_{(N,j)}(\{u_i|\vec{v} \}, \{u_i,z_i|\phi\})$ explicitly,
\small\begin{equation*}
w_{(N,j)}(\{u_i|\vec{v} \}, \{u_i,z_i|\phi\})= w_N(u_i|\vec{y}) h_j(u_i,z_i) - w_{N+1}(u_i| \vec{v})h_{j-1}(u_i,z_i)
\end{equation*}\normalsize
where\footnote{We remember that $h_j(u_i) = u^j_i.$},
\small\begin{equation*}\begin{split}
w_N(u_i|\vec{v}) &= \sum^N_{k=0}(-1)^k u^{N-k}_ie_{k}(\vec{v})\\
h_j(u_i,z_i)  &= \sum^j_{k=0}u^k_{i} z^{j-k}_i\\
w_{N+1}(u_i|\vec{v}) &= \sum^N_{k=0}(-1)^k u^{N+1-k}_i e_{k}(\vec{v}) + (-1)^{N+1}  \underbrace{e_{N+1}(\vec{v})}_{=0}
\end{split}\end{equation*}\normalsize
Putting this all together we obtain,
\small\begin{equation*}\begin{split}
w_{(N,j)}(\{u_i|\vec{v} \}, \{u_i,z_i|\phi\})&=u^N_i \underbrace{\sum^{N}_{k=0}\left(-\frac{1}{u_i}\right)^k e_{k}(\vec{v})}_{E\left(-\frac{1}{u_i};\vec{v} \right)} \underbrace{\left\{ \sum^j_{p=0}u^p_{i} z^{j-p}_i- u_i \sum^{j-1}_{p=0}u^p_{i} z^{j-1-p}_i \right\}}_{z^j_i}\\
&=z^j_i u^N_i \prod^N_{k=1} \left( 1-\frac{v_k}{u_i} \right),
\end{split}\end{equation*}\normalsize
which verifies eq. \ref{l.res1}. Taking the determinant of the above result, (for $1 \le i,j \le N$), we have,
\small\begin{equation}\begin{split}
\textrm{det}\left[ w_{(N,j)}(\{u_i|\vec{v} \}, \{u_i,z_i|\phi\}) \right]^N_{i,j=1} &= \textrm{det}\left[ z^{j-1}_i \prod^N_{k=1} \left( u_i-v_k \right)  \right]^N_{i,j=1} \\
&=\prod^N_{i,k=1} \left( u_i-v_k \right) \textrm{det}\left[ z^{j-1}_i  \right]^N_{i,j=1} \\
&=\prod^N_{i,k=1} \left( u_i-v_k \right) \prod_{1\le i < j \le N} (z_j-z_i)
\label{l.res2}\end{split}
\end{equation}\normalsize\\
\textbf{Result (2), the action of $\mathbf{\partial_i}$ on $\mathbf{w_j (u_i|\vec{v})}$.} Considering the action of $\partial_i$ on $w_j (u_i|\vec{v})$, where the first set contains only one element $u_i$, we obtain,
\small\begin{equation}\begin{split}
\partial_i \left(w_j (u_i|\vec{v}) \right) &= \sum^j_{k=0}(-1)^k e_{k}(\vec{v}) \partial_i \left( u^{j-k}_i \right) \\
&= \sum^j_{k=0}(-1)^k e_{k}(\vec{v}) \frac{ u^{j-k}_i - z^{j-k}_i}{u_i - z_i}\\
&= \sum^j_{k=0}(-1)^k e_{k}(\vec{v})\underbrace{ \sum^{j-k-1}_{p=0} u^p_i z^{j-k-1-p}_i}_{h_{j-k-1}(u_i,z_i)}\\
&= w_{j-1}( u_i,z_i|\vec{v}) ,
\end{split}\end{equation}\normalsize
since on the second last line the $k = j$ term, $h_{-1}(u_i,z_i)$, is zero.\\
\\
\textbf{Result (3), the action of $\mathbf{\partial_i}$ on $\mathbf{w_{(N,j)}(\{u_i|\vec{v} \}, \{u_i,z_i|\phi\})}$.} Expanding the term $w_{(N,j)}(\{u_i|\vec{v} \}, \{u_i,z_i|\phi\})$, and applying the operator we have,
\small\begin{equation}
\partial_i \left( w_{(N,j)}(\{u_i|\vec{v} \}, \{u_i,z_i|\phi\}) \right) =\partial_i \left( w_{N}(u_i|\vec{v}) h_j(u_i,z_i) \right) - \partial_i \left( w_{N+1}(u_i|\vec{v}) h_{j-1}(u_i,z_i) \right)
\label{l.imp}
\end{equation}\normalsize
we notice that the homogeneous symmetric polynomials in this expression are invariant under the action of $u_i \leftrightarrow z_i$. Thus eq. \ref{l.imp} can be expressed more conveniently as,
\small\begin{equation}
\partial_i \left( w_{(N,j)}(\{u_i|\vec{v} \}, \{u_i,z_i|\phi\}) \right) =\partial_i \left( w_{N}(u_i|\vec{v})  \right)h_j(u_i,z_i) - \partial_i \left( w_{N+1}(u_i|\vec{v}) \right) h_{j-1}(u_i,z_i)
\label{l.imp2}
\end{equation}\normalsize
Using result (2) from above, eq. \ref{l.imp2} becomes,
\small\begin{equation}\begin{split}
\partial_i \left( w_{(N,j)}(\{u_i|\vec{v} \}, \{u_i,z_i|\phi\}) \right) &= w_{N-1}(u_i,z_i|\vec{v})  h_j(u_i,z_i) - w_{N}(u_i,z_i|\vec{v}) h_{j-1}(u_i,z_i)\\
&= w_{(N-1,j)} (\{u_i,z_i|\vec{v}\},\{ u_i,z_i| \phi \}) 
\label{l.mid}
\end{split}\end{equation}\normalsize
\textbf{Result (4), expanding $\mathbf{w_{(N-1,j)} (\{u_i,z_i|\vec{v}\},\{ u_i,z_i| \phi \})} $.} We shall now provide the derivation of two alternative forms for $w_{(N-1,j)} (\{u_i,z_i|\vec{v}\},\{ u_i,z_i| \phi \}) $, given by,
\small\begin{equation}\begin{split}
&w_{(N-1,j)} (\{u_i,z_i|\vec{v}\},\{ u_i,z_i| \phi \})\\
 =& \sum^{N-1}_{p=-1}(-1)^{N-1-p} w_{(p,j)} (\{u_i,z_i|\phi\} ,\{u_i,z_i|\phi\}) e_{N-1-p}(\vec{v}) 
\label{l.alt1}
\end{split}\end{equation}\normalsize
and,
\small\begin{equation}\begin{split}
&w_{(N-1,j)} (\{u_i,z_i|\vec{v}\},\{ u_i,z_i| \phi \})\\
=& \sum^{N-1}_{p=j} (-1)^{N-1-p}w_{(p,j)}(\{u_i,z_i|\phi\},\{ u_i,z_i| \phi \}) e_{N-1-p}(\vec{v}) \\
&-  \sum^{j-1}_{p=0} (-1)^{N-p}w_{(j-1,p)}(\{u_i,z_i|\phi\},\{ u_i,z_i| \phi \}) e_{N-p}(\vec{v}) 
\label{l.alt2}
\end{split}\end{equation}\normalsize
Using the definition of $w_j(\vec{u}|\vec{v})$ in terms of a summation of $e(\vec{v})$'s and $h(\vec{u})$'s, eq. \ref{l.mid} becomes,
\small\begin{equation*}\begin{split}
w_{(N-1,j)} (\{u_i,z_i|\vec{v}\},\{ u_i,z_i| \phi \}) = w_{N-1}(u_i,z_i|\vec{v})  h_j(u_i,z_i) - w_{N}(u_i,z_i|\vec{v}) h_{j-1}(u_i,z_i)\\
= \sum^{N-1}_{p_1=-1} (-1)^{N-1-p_1}h_{p_1}(u_i,z_i) e_{N-1-p_1}(\vec{v})  h_j(u_i,z_i)\\
 -\sum^{N}_{p_2=0} (-1)^{N-p_2}h_{p_2}(u_i,z_i) e_{N-p_2} (\vec{v}) h_{j-1}(u_i,z_i)
\end{split}\end{equation*}\normalsize
where we have added the $p_1 =-1$ term in the summation since it produces a zero. If we make the change of indices $p_2 \rightarrow p_2+1$, we obtain, 
\small\begin{equation*}\begin{split}
&w_{(N-1,j)} (\{u_i,z_i|\vec{v}\},\{ u_i,z_i| \phi \})\\
=& \sum^{N-1}_{p=-1}(-1)^{N-1-p}\underbrace{\left( h_p(u_i,z_i)  h_j(u_i,z_i) -h_{p+1}(u_i,z_i)h_{j-1}(u_i,z_i)  \right)}_{w_{(p,j)} (\{u_i,z_i|\phi\} ,\{u_i,z_i|\phi\})} e_{N-1-p}(\vec{v})
\end{split}\end{equation*}\normalsize
which verifies eq. \ref{l.alt1}. To verify eq. \ref{l.alt2} we begin by separating eq. \ref{l.alt1} into the two values of $p$, $j \le p \le N-1$ and $-1 \le p \le j-2$, where we note that the $p=j-1$ term is zero,
\small\begin{equation*}\begin{split}
&w_{(N-1,j)} (\{u_i,z_i|\vec{v}\},\{ u_i,z_i| \phi \})\\
=&\sum^{N-1}_{p=j} (-1)^{N-1-p}  w_{(p,j)} (\{u_i,z_i|\phi\} ,\{u_i,z_i|\phi\})  e_{N-1-p}(\vec{v})\\
&+ \sum^{j-2}_{p=-1}\left( h_p(u_i,z_i)  h_j(u_i,z_i) -h_{p+1}(u_i,z_i)h_{j-1}(u_i,z_i)  \right)e_{N-1-p}(\vec{v})
\end{split}\end{equation*}\normalsize
We notice that the first term is in the correct form. For the second term, we make the change of indices $p \rightarrow p-1$, and take out an overall negative factor to obtain,
\small\begin{equation*}
- \sum^{j-1}_{p=0} (-1)^{N-p}\textrm{det} \left[ \begin{array}{cc}
h_{j-1}(u_i,z_i) & h_j(u_i,z_i) \\
h_{p-1}(u_i,z_i) & h_p(u_i,z_i) 
\end{array}\right] e_{N-p}(\vec{v})
\end{equation*}\normalsize
thus verifying eq. \ref{l.alt2}.\\
\\
\textbf{Main section of the derivation.} We now come to the main section of the derivation of Lascoux's result, which relies on applying a series of divided difference operators on a known determinant expression (Cauchy's identity) and obtaining the Izergin-Korepin determinant. We then use the four results results to reconstruct this determinant in terms of symmetric polynomials.\\
\\
\textbf{Obtaining Izergin's determinant from Cauchy's determinant using divided difference operators.} We now consider the product of operators, $\partial_1 \dots \partial_N$, acting on Cauchy's determinant in the form, \small\begin{equation*}
\textrm{det}\left[\frac{1}{z_i-v_j} \right]^N_{i,j=1} = \frac{\prod_{1 \le i < j \le N}(z_i-z_j)(v_j-v_i) }{\prod_{1 \le i < j \le N}(z_i-v_j)}
\end{equation*}\normalsize
To simplify the situation, we notice that $z_i$ only appears in row $i$, thus each operator, $\partial_k$, only acts on a single row. Taking this into account we obtain for a single operator,
\small\begin{equation*}\begin{split}
\partial_k \left(\textrm{det}\left[\frac{1}{z_i-v_j} \right]^N_{i,j=1}\right) &= \sum^{N}_{p=1} (-1)^{p-k}\frac{\frac{1}{z_k-v_p}-\frac{1}{u_k-v_p}}{u_k-z_k} \textrm{det}\left[\frac{1}{z_i-v_j} \right]_{i=1,\dots,\hat{k}, \dots,N \atop{j=1,\dots,\hat{p},\dots,N}}\\
&= \sum^{N}_{p=1} (-1)^{p-k}\frac{1}{(u_k-v_p)(z_k-v_p)} \textrm{det}\left[\frac{1}{z_i-v_j} \right]_{i=1,\dots,\hat{k}, \dots,N \atop{j=1,\dots,\hat{p},\dots,N}}\\
&= \textrm{det}\left[\begin{array}{c}
\frac{1}{z_{i_1}-v_j}\\
\frac{1}{(u_k-v_p)(z_k-v_p)}\\
\frac{1}{z_{i_2}-v_j}
\end{array}\right]^{i_1=1,\dots,k-1 \atop{i_2=k+1,\dots,N}}_{j=1,\dots,N}
\end{split}\end{equation*}\normalsize
Thus applying the product of $N$ divided difference operators we obtain,
\small\begin{equation}
 \partial_1 \dots \partial_N \left(\textrm{det}\left[\frac{1}{z_i-v_j} \right]^N_{i,j=1}\right) = \textrm{det}\left[\frac{1}{(u_i-v_j)(z_i-v_j)} \right]^N_{i,j=1}
\end{equation}\normalsize
which is Izergin's determinant expression in the limit $z_i \rightarrow q u_i$, $1\le i \le N$. We now concentrate on obtaining the whole of Izergin's DWPF, $Z^I_N(\vec{u},\vec{v})$, up to the factor $\Upsilon_N$.\\
\\
\textbf{Obtaining $\mathbf{Z^I_N}$ in terms of basis symmetric polynomials.} Consider now the product form of the Cauchy determinant and massage it to obtain,
\small\begin{equation}\begin{split}
& \frac{\prod_{1\le i < j \le N}(v_j-v_i)(z_i-z_j) }{\prod^N_{i,j=1}(z_i-v_j)}\\
=& \left\{ \frac{\prod_{1\le i < j \le N}(v_j-v_i) }{\prod^N_{i,j=1}(u_i-v_j)(z_i-v_j)} \right\} \underbrace{ \left\{\prod_{1\le i < j \le N}(z_i-z_j)\prod^N_{i,j=1}(u_i-v_j)  \right\} }_{\textrm{use eq. \ref{l.res2}}}\\
=&(-1)^{\frac{N(N-1)}{2}}\left\{ \frac{\prod_{1\le i < j \le N}(v_j-v_i) }{\prod^N_{i,j=1}(u_i-v_j)(z_i-v_j)} \right\} \textrm{det}\left[ w_{(N,j-1)}(\{u_i|\vec{v} \}, \{u_i,z_i |\phi\}) \right]^N_{i,j=1}
\label{before2}\end{split}\end{equation}\normalsize
It is elementary to see that the factor, $\frac{\prod_{1\le i < j \le N}(v_j-v_i) }{\prod^N_{i,j=1}(u_i-v_j)(z_i-v_j)}$, is invariant under the action $u_i \leftrightarrow z_i$. Thus applying the product of divided difference operators  we obtain,
\small\begin{equation*}\begin{split}
&\frac{1}{\Upsilon_N}Z^I_N(\vec{u},\vec{v})\\
=& \left\{ \frac{\prod^N_{i,j=1}(u_i-v_j)(q u_i-v_j)}{\prod_{1\le i < j \le N}(u_i-u_j)(v_j-v_i)}\right\} \lim_{z_i \rightarrow q u_i \atop{i \in \{1,\dots,N \}}} \underbrace{\partial_1 \dots \partial_N \left( \textrm{det}\left[\frac{1}{(z_i-v_j)} \right]^N_{i,j=1} \right)}_{\textrm{use eq. \ref{before2}}} \\
=& \frac{(-1)^{\frac{N(N-1)}{2}}}{\prod_{1\le i < j \le N}(u_i-u_j)} \lim_{z_i \rightarrow q u_i \atop{i \in \{1,\dots,N \}}} \underbrace{\partial_1 \dots \partial_N\left(  \textrm{det}\left[ w_{(N,j-1)}(\{u_i|\vec{v} \}, \{u_i,z_i |\phi\}) \right]^N_{i,j=1}\right)}_{\textrm{use eq. \ref{l.mid}}} \\
=&  \frac{(-1)^{\frac{N(N-1)}{2}}}{\prod_{1\le i < j \le N}(u_i-u_j)}  \textrm{det}\left[ w_{(N-1,j-1)}(\{u_i, qu_i|\vec{v} \}, \{u_i, qu_i |\phi\}) \right]^N_{i,j=1}
\end{split}\end{equation*}\normalsize\\
\textbf{Expanding the matrix entry $\mathbf{w_{(N-1,j-1)}(\{u_i, qu_i|\vec{v} \}, \{u_i, qu_i |\phi\})}$.} We now apply eq. \ref{l.alt2} on the entries, $w_{(N-1,j-1)}(\{u_i, qu_i|\vec{v} \}, \{u_i, qu_i |\phi\})$, of the matrix to obtain,
\small\begin{equation*}\begin{split}
&w_{(N-1,j-1)}(\{u_i, qu_i|\vec{v} \}, \{u_i, qu_i |\phi\})\\
=&\sum^{N-1}_{p=j-1} (-1)^{N-1-p}w_{(p,j-1)}(\{u_i, qu_i|\phi\},\{ u_i, qu_i|\phi \}) e_{N-1-p}(\vec{v})\\
 &-  \sum^{j-2}_{p=0} (-1)^{N-p}w_{(j-2,p)}(\{u_i, qu_i|\phi\},\{ u_i, qu_i|\phi \}) e_{N-p}(\vec{v})
\end{split}\end{equation*}\normalsize
where (for $p_1 \ge p_2$),
\small\begin{equation}\begin{split}
&w_{(p_1,p_2)}(\{u_i, qu_i|\phi\},\{ u_i, qu_i|\phi \})\\
= & h_{p_1}(u_i,qu_i)h_{p_2}(u_i,qu_i)-h_{p_1+1}(u_i,qu_i)h_{p_2-1}(u_i,qu_i)\\
=& u^{p_1+p_2}_i \left\{ \sum^{p_1}_{s_1=0}\sum^{p_2}_{s_2=0}q^{p_1+p_2-s_1-s_2} - \sum^{p_1+1}_{s_1=0}\sum^{p_2-1}_{s_2=0}q^{p_1+p_2-s_1-s_2}\right\}\\
=&  u^{p_1+p_2}_i [q]^{p_1}_{p_2}
\end{split}\end{equation}\normalsize
where we have used the label $\left(\sum^{p_1}_{s=p_2}q^s \right) =  [q]^{p_1}_{p_2}$. Thus using the above result we obtain,
\small\begin{equation}\begin{split}
w_{(p,j-1)}(\{u_i, qu_i|\phi\},\{ u_i, qu_i|\phi \}) &= u^{p+j-1}_i [q]^{p}_{j-1}\\
w_{(j-2,p)}(\{u_i, qu_i|\phi\},\{ u_i, qu_i|\phi \}) &= u^{j-2+p}_i [q]^{j-2}_{p}
\end{split}\end{equation}\normalsize
and the partition function expression, $\frac{1}{\Upsilon_N}Z^I_N(\vec{u},\vec{v})$, becomes,
\small\begin{equation}
\frac{(-1)^{\frac{N(N-1)}{2}}}{\displaystyle\prod_{1\le i < j \le N}(u_i-u_j)}  \textrm{det}\left[ \begin{array}{c}
 \sum^{N-1}_{p=j-1}(-1)^{N-1-p} u^{p+j-1}_i[q]^{p}_{j-1}e_{N-1-p}(\vec{v}) \\
- \sum^{j-3}_{p=-1}(-1)^{N-p-1}u^{p+j-1}_i[q]^{j-2}_{p+1} e_{N-1-p}(\vec{v})
 \end{array} \right]^N_{i,j=1}
\label{temp1}
\end{equation}\normalsize\\
\textbf{Eliminating the removable poles.} It is now necessary to eliminate the removable poles, $\frac{1}{\prod_{1\le i < j \le N}(u_i-u_j)}$. Using the following symmetric function identity,
\small\begin{equation}
h_m(\{u\},u_j ) - h_m(\{u\}, u_k ) = (u_j-u_k)h_{m-1}(\{u\},u_j,u_k )
\label{completeid}
\end{equation}\normalsize 
where $ u_j, u_k \nsubseteq \{u \}$, we perform the $N-1$ row operations (in order), 
\small\begin{equation}\begin{array}{lcl}
R_1& \rightarrow & R_1 - R_{2}\\
R_2 &\rightarrow & R_2-R_3\\
&\vdots&\\
R_{N-1}& \rightarrow & R_{N-1}-R_{N}
\end{array}\label{rowop1}\end{equation}\normalsize
Under these operations eq. \ref{temp1} becomes,
\small\begin{equation*}\begin{split}
&\frac{1}{\Upsilon_N}Z^I_N(\vec{u},\vec{v}) = \frac{(-1)^{\frac{N(N-1)}{2}}}{ \prod_{1\le i < j \le N\atop{i \ne j+1}}(u_i-u_j)}\\
\times& \textrm{det}\left[ \begin{array}{l}
 \sum^{N-1}_{p=j-1}(-1)^{N-1-p} h_{p+j-2}(u_i,u_{i+1})[q]^{p}_{j-1}e_{N-1-p}(\vec{v}) \\
- \sum^{j-3}_{p=-1}(-1)^{N-1-p}h_{p+j-2}(u_i,u_{i+1})[q]^{j-2}_{p+1} e_{N-1-p}(\vec{v})\\
 \sum^{N-1}_{p=j-1}(-1)^{N-1-p} u^{p+j-1}_N [q]^{p}_{j-1}e_{N-1-p}(\vec{v}) \\
-  \sum^{j-3}_{p=-1}(-1)^{N-1-p}u^{p+j-1}_N [q]^{j-2}_{p+1} e_{N-1-p}(\vec{v})
\end{array} \right]_{i=1, \dots, N-1\atop{j=1,\dots,N }}
\end{split}\end{equation*}\normalsize
Continuing with this pattern and applying the $N-2$ operations (in order),
\small\begin{equation}
R_i \rightarrow R_i - R_{i+2} \textrm{  ,  } 1 \le i \le N-2
\label{rowop2}\end{equation}\normalsize
followed by the $N-3$ operations (in order),
\small\begin{equation}
R_i \rightarrow R_i - R_{i+3} \textrm{  ,  } 1 \le i \le N-3
\label{rowop3}\end{equation}\normalsize
until finally we have just the single operation,
\small\begin{equation}
R_1 \rightarrow R_1 - R_{N}
\label{rowopN-1}\end{equation}\normalsize
eq. \ref{temp1} becomes,
\small\begin{equation}\begin{split}
&\frac{1}{\Upsilon_N}Z^I_N(\vec{u},\vec{v}) = (-1)^{\frac{N(N-1)}{2}}\\
\times& \textrm{det}\left[ \begin{array}{l}
\sum^{N-1}_{p=j-1}(-1)^{N-1-p} h_{p+j-1-(N-i)}(\hat{u}_1,\dots,\hat{u}_{i-1})[q]^{p}_{j-1}e_{N-1-p}(\vec{v}) \\
- \sum^{j-3}_{p=-1}(-1)^{N-1-p}h_{p+j-1-(N-i)}(\hat{u}_1,\dots,\hat{u}_{i-1})[q]^{j-2}_{p+1} e_{N-1-p}(\vec{v})
\end{array} \right]^N_{i,j=1}
\label{temp3}\end{split}
\end{equation}\normalsize
which completely eliminates the poles.\\
\\
\textbf{Clearing up the homogeneous symmetric polynomials.} In their current form (eq. \ref{temp3}), most of the polynomials, $h_{p+j-1-(N-i)}(\hat{u}_1,\dots,\hat{u}_{i-1})$, contain incomplete sets of the variables $\{u\}$. We now detail the required algorithm to make all of these polynomials contain the complete set $\{u\}$. For this task we require the following identity\footnote{The identity can be derived easily enough using an inductive argument.},
\small\begin{equation}
h_j(u_1,\dots,u_k) = \sum^l_{p=1}h_1(u_p)h_{j-1}(u_p,\dots,u_k ) + h_j(u_{l+1},\dots,u_k) \textrm{  ,  } 1\le  l \le k-1
\label{completeid2}
\end{equation}\normalsize
Using the above identity, if we apply the following (ordered) row operations,
\small\begin{equation}\begin{array}{lcl}
R_N& \rightarrow& R_N + h_1(u_{N-1}) R_{N-1}\\
R_{N-1} &\rightarrow & R_{N-1} + h_1(u_{N-2}) R_{N-2}\\
& \vdots & \\
R_{2} &\rightarrow & R_{2} + h_1(u_{1}) R_{1}
\end{array}\label{rowopother1}
\end{equation}\normalsize
This increases the number of variables in each homogeneous symmetric polynomial by one (except for those in $R_1$ which already have $N$ variables). Continuing this process (in order),
\small\begin{equation}
\left.\begin{array}{lcl}
R_N & \rightarrow& R_N + h_1(x_{N-j}) R_{N-j}\\
& \vdots &\\
R_{1+j}& \rightarrow& R_{1+j} + h_1(x_{1}) R_{1}
\end{array} \right\} \textrm{  for  } j = 2,3, \dots, N-1
\label{rowopother2}\end{equation}\normalsize
eq. \ref{temp3} becomes,
\small\begin{equation*}
(-1)^{\frac{N(N-1)}{2}} \textrm{det}\left[ \begin{array}{l}
\sum^{N-1}_{p=j-1}(-1)^{N-1-p} h_{p+j-1-(N-i)}(\vec{u})[q]^{p}_{j-1}e_{N-1-p}(\vec{v}) \\
- \sum^{j-3}_{p=-1}(-1)^{N-1-p}h_{p+j-1-(N-i)}(\vec{u})[q]^{j-2}_{p+1} e_{N-1-p}(\vec{v})
\end{array} \right]^N_{i,j=1}
\end{equation*}\normalsize
Finally, reordering the rows as follows,
\small\begin{equation}
R_i \rightarrow R_{N+1-i} \textrm{  ,  } i = 1, \dots, N
\end{equation}\normalsize
we obtain,
\small\begin{equation}
\frac{1}{\Upsilon_N}Z^I_N(\vec{u},\vec{v})  = \textrm{det}\left[ \begin{array}{l}
\sum^{N-1}_{p=j-1}(-1)^{N-1-p} h_{p+j-i}(\vec{u})[q]^{p}_{j-1}e_{N-1-p}(\vec{v}) \\
- \sum^{j-3}_{p=-1}(-1)^{N-1-p}h_{p+j-i}(\vec{u})[q]^{j-2}_{p+1} e_{N-1-p}(\vec{v})
\end{array} \right]^N_{i,j=1}
\label{perm}
\end{equation}\normalsize
which is the expanded form of Lascoux's determinant expression, (eq. \ref{p.3}).
\subsection{Kirillov-Smirnov determinant expression}
For completeness we now analyze the alternative form for the DWPF expressed in terms of basis symmetric polynomials given by Kirillov-Smirnov in \cite{KS},
\small\begin{equation}
Z^{KS}_{N} (\vec{u},\vec{v}) =  \Upsilon_N \frac{\textrm{det} \left[ T^{(0)}_{(j,0)}(\{ k \} | \vec{u})  , T^{(1)}_{(j,0)}(\{ k \} | \vec{v})\right]_{j = 0, \dots, 2N-1 \atop{k = 1, \dots,N}}}{q^{\frac{N}{2}}\left( q^{\frac{1}{2}}-q^{-\frac{1}{2}} \right)^N \left[ \prod_{1 \le i < j \le N} (u_i-u_j)(v_i-v_j) \right] e_N(\vec{v}) } \label{p.ks1}
\end{equation}\normalsize
where,
\small\begin{equation}
T^{(z)}_{(j,p)}(\{ I \} | \vec{u}) = \frac{1}{q^{\frac{p}{2}}} e_j \left(  q^{\frac{1}{2}} u_{n \in \{ 1, \dots, N \} }, \frac{1}{q^{\frac{1}{2}}} u_{n \notin  \{I\}} \right) - z q^{\frac{p}{2}} e_j \left(  q^{\frac{1}{2}} u_{n \notin  \{I\} }, \frac{1}{q^{\frac{1}{2}}} u_{n \in \{ 1, \dots, N \} } \right)
\end{equation}\normalsize
As with Izergin's determinant form, the advantage of the above expression is that it is relatively easy to show that it adheres to Korepin's four conditions. We shall consider the above form and show that hidden inside is an similar form to Lascoux's.\\
\\
To proceed we first eliminate the Vandermonde determinants in the denominator.\\
\\
\textbf{Elimination of the removable poles.}
\begin{proposition}
\small\begin{equation}
Z^{KS}_{N } (\vec{u},\vec{v}) =\Upsilon_N \frac{\textrm{det} \left[ T^{(0)}_{(j-k,k-1)}(\{ 1,\dots,k \} | \vec{u})  , T^{(1)}_{(j-k,k-1)}(\{ 1,\dots,k \} | \vec{v})\right]_{j = 1, \dots, 2N \atop{k = 1, \dots,N}}}{q^{\frac{N}{2}}\left( q^{\frac{1}{2}}-q^{-\frac{1}{2}} \right)^N  e_N(\vec{v}) } \label{p.ks2} 
\end{equation}\normalsize
\end{proposition}
\textbf{Proof.}
We notice that we can refer naturally to the left hand and right hand side of the determinant in eq. \ref{p.ks1} as they have $z$ values 0 and 1 respectively. Now consider the $j$th column, labeled $C_j$, $1 \le j \le N$, in both sides of this determinant, (we shall keep $z$ general in the below calculations).\\
\\
Using the following relation between the elementary symmetric polynomials,
\small\begin{equation}
e_j(w_1, \dots, w_N) = e_j(w_1, \dots,\hat{w}_k,\dots, w_N) + w_k e_{j-1}(w_1, \dots,\hat{w}_k,\dots, w_N)
\label{p.ks5}
\end{equation}\normalsize
we first note the following result,
\small\begin{equation}\begin{split}
T^{(z)}_{(j,p)}(\{ I \} \cup \{k_1\} | \vec{w})  - T^{(z)}_{(j,p)}(\{ I \}  \cup \{k_2\}| \vec{w}) \\
= q^{-\frac{p}{2}} \left[ e_j \left(  q^{\frac{1}{2}} w_{n \in \{ 1, \dots, N \} }, q^{-\frac{1}{2}} w_{n \notin  \{I\} \cup \{k_1 \}} \right) -e_j \left(  q^{\frac{1}{2}} w_{n \in \{ 1, \dots, N \} }, q^{-\frac{1}{2}} w_{n \notin  \{I\} \cup \{k_2 \}} \right) \right ]\\
- z q^{\frac{p}{2}} \left[  e_j \left(  q^{\frac{1}{2}} w_{n \notin  \{I\} \cup \{k_1 \}}, q^{-\frac{1}{2}} w_{n \in \{ 1, \dots, N \} } \right) -e_j \left(  q^{\frac{1}{2}} w_{n \notin  \{I\} \cup \{k_2 \}}, q^{-\frac{1}{2}} w_{n \in \{ 1, \dots, N \} } \right)  \right] \\
= \left(w_{k_2} - w_{k_1}  \right) \left[ q^{-\frac{p+1}{2}} e_{j-1} \left(  q^{\frac{1}{2}} w_{n \in \{ 1, \dots, N \} }, q^{-\frac{1}{2}} w_{n \notin  \{I\} \cup \{k_1,k_2 \}} \right)\right.\\
 \left. - z q^{\frac{p+1}{2}} e_{j-1} \left(  q^{\frac{1}{2}} w_{n \notin  \{I\}\cup \{k_1,k_2\} }, q^{-\frac{1}{2}} w_{n \in \{ 1, \dots, N \} } \right)  \right] \\
=\left(w_{k_2} - w_{k_1}  \right) T^{(z)}_{(j-1,p+1)}(\{ I \} \cup \{k_1,k_2\} | \vec{w})
\label{p.ks3}\end{split}\
\end{equation}\normalsize
Thus performing the column operations,
\small\begin{equation*}
C_j \rightarrow C_j-C_1 \textrm{  ,  } 2 \le j \le N
\end{equation*}\normalsize
in both sides of the determinant we eliminate a factor of $\prod^N_{ j=2} (u_1-u_j)(v_1-v_j)$ from the denominator and eq. \ref{p.ks1} becomes,
\small\begin{equation*}\begin{split}
\Upsilon_N \frac{\textrm{det} \left[ T^{(0)}_{(j,0)}(\{ 1 \} | \vec{u}) ,T^{(0)}_{(j-1,1)}(\{1, k \} | \vec{u})  , T^{(1)}_{(j,0)}(\{ 1 \} | \vec{v}),T^{(1)}_{(j-1,1)}(\{ 1,k \} | \vec{v})\right]_{j = 0, \dots, 2N-1 \atop{k = 2, \dots,N}}}{q^{\frac{N}{2}}\left( q^{\frac{1}{2}}-q^{-\frac{1}{2}} \right)^N \left[ \prod_{2 \le i < j \le N} (u_i-u_j)(v_i-v_j) \right]  e_N(\vec{v})   }
\end{split}\end{equation*}\normalsize
Continuing this process we perform the following column operations (in order),
\small\begin{equation*}\begin{array}{lclc}
C_{j_3} & \rightarrow &C_{j_3}-C_2 & 3 \le j_3 \le N \\
C_{j_4}& \rightarrow &C_{j_4}-C_3 & 4 \le j_4 \le N\\
& \vdots&   \\
C_{j_n}& \rightarrow &C_{j_n}-C_{n-1} & n \le j_n \le N
\end{array}\end{equation*}\normalsize
for $3 \le n \le N$ in both sides of the determinant. Doing so we eliminate a factor of $\prod^{n}_{i=2}\prod^N_{ j=i+1} (u_i-u_j)(v_i-v_j)$ from the denominator and obtain,
\small\begin{equation*}
Z^{KS}_{N  } (\vec{u},\vec{v}) =\Upsilon_N \frac{\textrm{det} \left[\field{T}^{(0)}_{j,n\atop{k_1,k_2}} (\vec{u}) ,\field{T}^{(1)}_{j,n\atop{k_1,k_2}} (\vec{v})\right]^{j = 0, \dots, 2N-1}_{k_1 = 1,\dots,n \atop{k_2 = n+1, \dots,N}}}{q^{\frac{N}{2}}\left( q^{\frac{1}{2}}-q^{-\frac{1}{2}} \right)^N \left[ \prod_{n \le i < j \le N} (u_i-u_j)(v_i-v_j) \right] e_N(\vec{v}) }
\end{equation*}\normalsize
where,
\small\begin{equation*}\begin{split}
\field{T}^{(z)}_{j,n\atop{k_1,k_2}} (\vec{w})= \left\{ T^{(z)}_{(j-k_1+1,k_1-1)}(\{ 1,\dots,k_1 \} | \vec{w}) ,T^{(z)}_{(j-(n-1),n-1)}(\{1, \dots,n-1 ,k_2 \} | \vec{w}) \right\} 
\end{split}\end{equation*}\normalsize
Setting $n = N$, we obtain eq. \ref{p.ks2}. $\square$\\
\\
We now wish to massage eq. \ref{p.ks2} into a form that contains elementary symmetric polynomials in one \emph{complete} set of variables only, as opposed to the mixed state that they presently exist. \\
\\
\textbf{Clearing up the symmetric polynomials.}
\begin{proposition}
\small\begin{equation}
Z^{KS}_{N} (\vec{u},\vec{v}) = \Upsilon_N \frac{\textrm{det} \left[ b^{(0)}_{(j,k)}( \vec{u}) , b^{(1)}_{(j,k)}(\vec{v})\right]_{j = 1, \dots, 2N \atop{k = 1, \dots,N}}}{q^{\frac{N}{2}}\left( q^{\frac{1}{2}}-q^{-\frac{1}{2}} \right)^N e_N(\vec{v})  } 
\end{equation}\normalsize
where,
\small\begin{equation}
b^{(z)}_{(j,k)}( \vec{w}) = q^{-\frac{k-1}{2}} e_{j-k} \left( q^{\frac{1}{2}} w_{n \in \{ 1, \dots, N \} }\right) - z q^{\frac{k-1}{2}} e_{j-k} \left( q^{-\frac{1}{2}} w_{n \in \{ 1, \dots, N \} } \right)
\label{p.ks4}
\end{equation}\normalsize
\end{proposition}
\textbf{Proof.} For the following proof we keep the $z$ value general. Naturally referring to the left hand and right hand side of the determinant in eq. \ref{p.ks2}, we notice that column $N$ of both sides of the determinant is already of the required form. Thus for $z = 0,1$, we call this term $b^{(z)}_{(j,N)}$,
\small\begin{equation*}\begin{split}
T^{(z)}_{(j-N,N-1)}(\{ 1,\dots,N  \} | \vec{w}) = b^{(z)}_{(j,N)}( \vec{w}) =& q^{-\frac{N-1}{2}} e_{j-N} \left(   q^{\frac{1}{2}} w_{n \in \{ 1, \dots, N \} } \right) \\
& - z  q^{\frac{N-1}{2}} e_{j-N} \left(    q^{-\frac{1}{2}} w_{n \in \{ 1, \dots, N \} } \right) 
\end{split}\end{equation*}\normalsize
Using the symmetric polynomial identity in eq. \ref{p.ks5}, $C_{N-1}$ on both sides of eq. \ref{p.ks2} to be massaged the following way,
\small\begin{equation*}\begin{split}
C_{j,N-1} = q^{-\frac{N-2}{2}} e_{j-(N-1)} \left(  q^{\frac{1}{2}} w_{n \in \{ 1, \dots, N \} } ,q^{-\frac{1}{2}}w_N \right) \\
- z q^{\frac{N-2}{2}} e_{j-(N-1)} \left(  q^{-\frac{1}{2}} w_{n \in \{ 1, \dots, N \} },q^{\frac{1}{2}} w_N \right)\\
= q^{-\frac{N-1}{2}} w_N e_{j-N} \left(  q^{\frac{1}{2}} w_{n \in \{ 1, \dots, N \} }  \right) + q^{\frac{N-2}{2}} e_{j-(N-1)} \left(  q^{\frac{1}{2}} w_{n \in \{ 1, \dots, N \} } \right)\\
 -z \left[ q^{\frac{N-1}{2}}w_N e_{j-N} \left(   q^{-\frac{1}{2}} w_{n \in \{ 1, \dots, N \} }\right) +q^{\frac{N-2}{2}} e_{j-(N-1)} \left( q^{-\frac{1}{2}} w_{n \in \{ 1, \dots, N \} } \right) \right]\\
\Rightarrow C_{j,N-1} = w_N  b^{(z)}_{(j,N)}( \vec{w}) +  b^{(z)}_{(j,N-1)}( \vec{w})
\end{split}\end{equation*}\normalsize 
We wish to express the remaining columns in a similar manner. To do this we use the following extended version of eq. \ref{p.ks5},
\small\begin{equation}
e_p \left( w_{n \in \{ I\} \cup \{k_1,\dots,k_m\} } \right) = \sum^m_{l =0} e_l(w_{n \in \{ k_1,\dots,k_m \}}) e_{p-l} \left( w_{n \in \{ I\} } \right)
\label{p.ks6}
\end{equation}\normalsize
The proof of the above formula can be obtained through an elementary induction argument, where we apply eq. \ref{p.ks5} to $e_p \left( w_{n \in \{ I\} \cup \{k_1,\dots,k_m\} } \right)$ a general number of times. The first few cases are given explicitly as, 
\small\begin{equation*}\begin{split}
e_p \left( w_{n \in \{ I\} \cup \{k_1,\dots,k_m\} } \right) \\
= e_1(w_{k_m}) e_{p-1} \left( w_{n \in \{ I\} \cup \{k_1,\dots,k_{m-1}\} } \right) + e_{p} \left( w_{n \in \{ I\} \cup \{k_1,\dots,k_{m-1}\} } \right) \\
=e_2(w_{k_{m-1},k_m}) e_{p-2} \left( w_{n \in \{ I\} \cup \{k_1,\dots,k_{m-2}\} } \right)  \\
+e_1(w_{k_{m-1},k_m}) e_{p-1} \left( w_{n \in \{ I\} \cup \{k_1,\dots,k_{m-2}\} } \right)  + e_{p} \left( w_{n \in \{ I\} \cup \{k_1,\dots,k_{m-2}\} } \right) 
\end{split}\end{equation*}\normalsize 
and so forth. Using the above result we can now express the remaining columns, $C_m$, $1 \le m \le N-2$, as a linear sum of terms involving the remaining $b^{(z)}_{(j,k)}$'s,
\small\begin{equation}\begin{split}
C_{j,m} =  T^{(z)}_{(j-m,m-1)}(\{ 1,\dots,m  \} |\vec{w}) \\
=   q^{-\frac{m-1}{2}} e_{j-m} \left(   q^{\frac{1}{2}} w_{n \in \{ 1, \dots, N \} },  q^{-\frac{1}{2}} w_{n \in  \{m+1, \dots,N \}} \right)\\
 - z  q^{\frac{m-1}{2}} e_{j-m} \left(   q^{\frac{1}{2}} w_{n \in  \{m+1, \dots,N\} },  q^{-\frac{1}{2}} w_{n \in \{ 1, \dots, N \} } \right) \\
= q^{-\frac{m-1}{2}} \sum^{N-m}_{l=0} q^{-\frac{l}{2}}  e_{l} \left( w_{n \in  \{m+1, \dots,N \}} \right) e_{j-m -l} \left(   q^{\frac{1}{2}} w_{n \in \{ 1, \dots, N \} } \right)\\
 -z  q^{\frac{m-1}{2}} \sum^{N-m}_{l=0}  q^{\frac{l}{2}} e_l \left(   w_{n \in  \{m+1, \dots,N\} }\right) e_{j-m-l} \left(    q^{-\frac{1}{2}} w_{n \in \{ 1, \dots, N \} } \right)\\
= \sum^{N-m}_{l=0} e_l \left(   w_{n \in  \{m+1, \dots,N\} }\right) \left[    q^{-\frac{m+l-1}{2}}  e_{j-m -l} \left(  q^{\frac{1}{2}} w_{n \in \{ 1, \dots, N \} } \right)\right.\\
\left.- z q^{\frac{m+l-1}{2}} e_{j-m-l} \left(   q^{-\frac{1}{2}}w_{n \in \{ 1, \dots, N \} } \right) \right]\\\
\Rightarrow C_{j,m} = \sum^{N-m}_{l=0} e_l \left(   w_{n \in  \{m+1, \dots,N\} }\right) b^{(z)}_{(j,m+l)}(\vec{w})
\label{p.ks7}
\end{split}\end{equation}\normalsize
Thus, performing the column operations (in order),
\small\begin{equation*}\begin{array}{lcl}
C_{N-1}& \rightarrow & C_{N-1} - e_1(u_N) C_N\\
C_{N-2}& \rightarrow & C_{N-2} - e_1(u_N,u_{N-1}) C_{N-1} - e_2(u_N,u_{N-1}) C_N\\
&\vdots \\
C_m& \rightarrow& C_m - \sum^{N-m}_{l=1} e_l \left(   u_{n \in  \{m+1, \dots,N\} }\right) C_{l+m}\\
&\vdots & \\
C_1& \rightarrow& C_1 - \sum^{N-1}_{l=1} e_l \left(   u_{n \in  \{2, \dots,N\} }\right) C_{l+1}
\end{array}\end{equation*}\normalsize 
and similarly for the right hand side of the determinant, eq. \ref{p.ks2} becomes eq. \ref{p.ks4}, completing the proposition. $\square$\\
\\
Extracting the $q$'s out of the elementary symmetric polynomials present in the $b^{(z)}_{(j,k)}$'s,
\small\begin{equation*}
b^{(z)}_{(j,k)}(\vec{w}) = \left\{q^{-\frac{2k-j-1}{2}} - z q^{\frac{2k-j-1}{2}}  \right\} e_{j-k}(\vec{w})
\end{equation*}\normalsize
eq.  \ref{p.ks4} becomes,
\small\begin{equation*}\begin{split}
 \frac{\Upsilon_N}{q^{\frac{N}{2}} e_N(\vec{v})  }\textrm{det} \left[  q^{-\frac{2k-j-1}{2}}e_{j-k}(\vec{u})  , \frac{ \left\{ q^{-\frac{2k-j-1}{2}} -  q^{\frac{2k-j-1}{2}}  \right\} }{\left( q^{\frac{1}{2}}-q^{-\frac{1}{2}} \right)}e_{j-k}(\vec{v})\right]_{j = 1, \dots, 2N \atop{k = 1, \dots,N}}\\
= \frac{\Upsilon_N}{ e_N(\vec{v})  }\textrm{det} \left[e_{j-k}(\vec{u})  , \frac{q^{\frac{N}{2}} \left\{ q^{1-k} -  q^{k-j}  \right\} }{ \left( q-1 \right)}e_{j-k}(\vec{v})\right]_{j = 1, \dots, 2N \atop{k = 1, \dots,N}}
\end{split}\end{equation*}\normalsize 
We now rearrange the ordering of the left and right hand side columns respectively as $k \rightarrow N- k+1$, $1 \le k \le N$, and the rows as $j \rightarrow 2N -j +1$, $1 \le j \le 2N$, to obtain,
\small\begin{equation}
Z^{KS}_{N } (\vec{u},\vec{v})=   \frac{\Upsilon_N}{ e_N(\vec{v})  }\textrm{det} \left[e_{N-j+k}(\vec{u})  , \frac{\left\{ q^{ k} -  q^{j-k}  \right\} }{ \left( 1-q \right)}e_{N-j+k}(\vec{v})\right]_{j = 1, \dots, 2N \atop{k = 1, \dots,N}}
\label{p.ks8}
\end{equation}\normalsize
where we notice that the bottom row on the right hand side are all zeros. We consider expanding the above $2N \times 2N$ determinant as a bilinear sum of $N \times N$ determinants using the Laplace expansion.\\
\\
\textbf{Laplace expansion of determinants.} Let $D_{2N} = \textrm{det}[d_{jk}]^{2N}_{j,k=1}$ be a $2N$th order determinant. It can be expressed as a bilinear sum of $N \times N$ order determinants by either of the following expressions,
\small\begin{equation*}\begin{split}
D_{2N} &=\sum_{\sigma \in S^*_{2N}} (-1)^{\sum^N_{l=1}\left( j_{\sigma_l} + k_l \right) } D_N\left( \begin{array}{ccc}
j_{\sigma_1} & \dots & j_{\sigma_N}\\
k_1 & \dots & k_N
\end{array}  \right) D_N\left( \begin{array}{ccc}
j_{\sigma_{N+1}}  & \dots & j_{\sigma_{2N}}\\
k_{N+1}  & \dots & k_{2N}
\end{array}  \right)  \\
&=\sum_{\sigma \in S^*_{2N}} (-1)^{\sum^N_{l=1}\left( j_{l} + k_{\sigma_l} \right) } D_N\left( \begin{array}{ccc}
j_1 & \dots & j_N\\
k_{\sigma_1}  & \dots & k_{\sigma_N}
\end{array}  \right) D_N\left( \begin{array}{ccc}
j_{N+1}  & \dots &j_{2N}\\
k_{\sigma_{N+1}}  & \dots & k_{\sigma_{2N}}
\end{array}  \right)
\end{split}\end{equation*}\normalsize 
where for the first expression,
\small\begin{equation*}\begin{split}
j_{\sigma_1} < j_{\sigma_2} < \dots < j_{\sigma_N} &\textrm{  ,  } j_{\sigma_{N+1}} < j_{\sigma_{N+2}} < \dots < j_{\sigma_{2N}} \\
k_1 < k_2 < \dots < k_N &\textrm{  ,  } k_{N+1} < k_{N+2} < \dots < k_{2N}
\end{split}\end{equation*}\normalsize 
and for the second expression,
\small\begin{equation*}\begin{split}
j_1 < j_2 < \dots < j_N &\textrm{  ,  }j_{N+1} < j_{N+2} < \dots <j_{2N}\\
k_{\sigma_1} < k_{\sigma_2} < \dots < k_{\sigma_N} &\textrm{  ,  } k_{\sigma_{N+1}} < k_{\sigma_{N+2}} < \dots < k_{\sigma_{2N}}
\end{split}\end{equation*}\normalsize 
Additionally, $D_N$ denotes the $N$th order determinant,
\small\begin{equation*}
D_N\left( \begin{array}{ccc}
j_{1} & \dots & j_{N}\\
k_1 & \dots & k_N
\end{array}  \right) = \textrm{det} [d_{j_m,k_n}]^N_{m,n=1}
\end{equation*}\normalsize \\
\textbf{Laplace expanding $\mathbf{Z^{KS}_N}$.} Applying the first Laplace expansion to eq. \ref{p.ks8}, we set columns,
\small\begin{equation*}
k_1 = 1, \dots, k_N = N , k_{N+1} = N+1, \dots , k_{2N} =2N
\end{equation*}\normalsize
to immediately obtain,
\small\begin{equation*}\begin{split}
Z^{KS}_{N } (\vec{u},\vec{v}) =&\frac{(-1)^{\frac{N}{2}(N+1)}\Upsilon_N}{ e_N(\vec{v})  } \sum_{\sigma \in S^*_{2N}\atop{j_{\sigma_N} = 2N}} (-1)^{\sum^N_{l=1} j_{\sigma_l}  } \textrm{det} \left[  e_{N-j_{\sigma_m }+k}(\vec{u}) \right]^N_{m,k=1}\\
&\times \textrm{det} \left[ \frac{\left\{ q^{k} -  q^{j_{\sigma_{N+m}}-k}  \right\} }{ \left( 1-q \right)}e_{N-j_{\sigma_{N+m}}+k}(\vec{v})    \right]^{N}_{m,k=1}
\end{split}\end{equation*}\normalsize
where the condition, $j_{\sigma_N} = 2N$, is due to the entries in the bottom most right hand row in eq. \ref{p.ks8} being zero, i.e. all terms with $j_{\sigma_{2N}} = 2N$ are zero.\\
\\
Using the following relation,
\small\begin{equation*}
 (-1)^{\sum^{N}_{l=1}j_{\sigma_l}} =  (-1)^{\frac{2N(2N+1)}{2}} (-1)^{\sum^{N}_{l=1}j_{\sigma_{N+l}}} =  (-1)^N(-1)^{\sum^{N}_{l=1}j_{\sigma_{N+l}}} 
\end{equation*}\normalsize 
the DWPF becomes,
\small\begin{equation}
Z^{KS}_{N } (\vec{u},\vec{v}) =  \frac{\Upsilon_N}{ e_N(\vec{v})  } \sum_{\sigma \in S^*_{2N}\atop{j_{\sigma_N} = 2N}}  \textrm{det} \left[  e_{N-j_{\sigma_m }+k}(\vec{u}) \right]^N_{m,k=1}\textrm{det} \left[  \varphi_{j_{\sigma_{N+m}},k} \right]^{N}_{m,k=1}
\label{p.ks9}
\end{equation}\normalsize
where,
\small\begin{equation}
\varphi_{j,k} = (-1)^{\frac{(N-1)}{2}+j}\frac{\left\{ q^{k} -  q^{j-k}  \right\} }{ \left( 1-q \right)}e_{N-j+k}(\vec{v}) \label{p.ks9'}
\end{equation}\normalsize
We now turn our attention to the summation, which in the current form is quite unruly.\\
\\
\textbf{Expressing the (restricted) sum over the symmetric group as the sum of partitions.} Expanding it out explicitly, it is possible (and advantageous) to express the (restricted) summation over the symmetric group as the following,
\small\begin{equation}
\sum_{\sigma \in S^*_{2N}\atop{j_{\sigma_N} = 2N}} = \sum_{1 \le j_{\sigma_1} < j_{\sigma_2} < \dots < j_{\sigma_N-1} \le 2N-1 \atop{ j_{\sigma_N} = 2N}}\sum_{1 \le j_{\sigma_{N+1}} < j_{\sigma_{N+2}} < \dots < j_{\sigma_{2N}} \le 2N-1 \atop{\ne  j_{\sigma_1} , j_{\sigma_2} , \dots, j_{\sigma_{N-1}} }}
\label{p.ks10}
\end{equation}\normalsize
where we view the sum involving $j_{\sigma_{N+l}}$, $1\le l \le N$, as uniquely fixed depending on the value of $j_{\sigma_{l}}$.\\
\\
Making the following convenient change of variables,
\small\begin{equation*}
\begin{array}{lcllcl}
j_{\sigma_1} &= & N - \lambda_{1} +1&j_{\sigma_{N+1}} & = & \mu_N +1 \\
j_{\sigma_2} &= & N - \lambda_{2} +2 &j_{\sigma_{N+2}} & = & \mu_{N-1} +2\\
&\vdots & & & \vdots  &\\
j_{\sigma_{N-1}} &=&  N - \lambda_{N-1} +N-1 & j_{\sigma_{2N-1}} & = & \mu_{2} +N-1\\
\lambda_N &=&0 &  j_{\sigma_{2N}} & = & \mu_{1} +N
\end{array}
\end{equation*}\normalsize
the summation becomes,
\small\begin{equation*}
\sum_{\sigma \in S^*_{2N}\atop{j_{\sigma_N} = 2N}}= \sum_{0 \le \lambda_{N-1} \le  \dots \le \lambda_1 \le N \atop{ \lambda_N  =0}}\sum_{0 \le \mu_N \le \dots \le \mu_1 \le  N-1 \atop{\mu_{N+1-l} + l \ne N+1 - \lambda_{1}  , N+2 - \lambda_{2}, \dots, 2N-1 - \lambda_{N-1} }}
\end{equation*}\normalsize
where we appreciate that,
\small\begin{equation*}
\sum_{0 \le \lambda_{N-1} \le  \dots \le \lambda_1 \le N \atop{ \lambda_N  =0}} = \sum_{\{\lambda \} \subseteq (N)^{N-1}} \textrm{  ,  } \sum_{0 \le \mu_N \le \dots \le \mu_1 \le  N-1 } = \sum_{\{ \mu \} \subseteq (N-1)^N}
\end{equation*}\normalsize
We now focus on the summation of the $\mu$'s. Based on its initial form in eq. \ref{p.ks10} we know that it only has one unique configuration for every partition $\{ \lambda_1, \dots, \lambda_{N-1}\} = \{\lambda\}$. We shall proceed to show that the specific configuration in question is the conjugate of $\{ \lambda\}$.
\begin{proposition}
\small\begin{equation*}
\sum_{\{\mu \} \subseteq (N-1)^N  \atop{\mu_{N+1-l} + l \ne N+1 - \lambda_{1}  , N+2 - \lambda_{2}, \dots, 2N-1 - \lambda_{N-1} }} = \sum_{\{\mu \}= \{ \lambda' \}} 
\end{equation*}\normalsize
for all partitions $ \{ \lambda \} \subseteq (N)^{N-1} $.
\end{proposition}
\textbf{Proof.}
Expanding the conditions on the parts of $\{\mu\}$, we obtain a comprehensive list of their forbidden values,
\small\begin{equation}
\begin{array}{lcllcl}
\mu_N & \ne & N - \lambda_1 & N+1 - \lambda_2 & \dots & 2N-2 - \lambda_{N-1} \\
\mu_{N-1} & \ne & N-1 - \lambda_1 & N - \lambda_2 & \dots & 2N-3 - \lambda_{N-1} \\
& \vdots &\\
\mu_1 & \ne & 1 - \lambda_1 & 2 - \lambda_2 & \dots & N-1 - \lambda_{N-1} \\
\end{array}
\label{p.ks11}
\end{equation}\normalsize
As stated earlier, we know as a fact that there is only one valid partition that satisfies each of these conditions. Thus to complete the proposition we shall proceed to verify that the conjugate of $\{ \lambda\}$ indeed adheres to all of these conditions.\\
\\
We recall that the partition $\{ \lambda \} \subseteq (N)^{N-1} $ can be expressed in the following convenient form,
\small\begin{equation*}
(\lambda_1, \dots, \lambda_{N-1}, \lambda_N =0) = (N^{m_N}, (N-1)^{m_{N-1}}, \dots, 1^{m_1} )
\end{equation*}\normalsize
for $ m_1 + \dots + m_N \le N-1$. Using the above expression we obtain,
\small\begin{equation}\begin{split}
m_j = \lambda'_j - \lambda'_{j+1}  \textrm{  for  }  1 \le j \le N-1  \\
m_N = \lambda'_N \label{p.ks12}
\end{split}\end{equation}\normalsize
Labeling, $l(\lambda) = \lambda'_1$, as the length of $\{ \lambda \}$, (the sum of the parts), we now let $\mu_j = \lambda'_j$, $1 \le j \le N$. Using eq. \ref{p.ks12}, the conditions in eq. \ref{p.ks11} can be expressed conveniently as,
\small\begin{equation}
\begin{array}{lcl}
l (\lambda ) + \lambda_j & \ne& j  \\
l (\lambda ) + \lambda_j & \ne& j +1 +m_1 \\
l(\lambda) + \lambda_j & \ne & j + 2 + m_1 + m_2  \\
& \vdots &\\
l(\lambda) + \lambda_j & \ne & j + N-1 + \sum^{N-1}_{i=1} m_i 
\end{array}\label{p.ks13}
\end{equation}\normalsize
for $1 \le j \le N-1$. Let us verify the first few of these conditions, and in doing so, the method to verify these conditions generally is made apparent. \\
\\
\textbf{Verifying $\mathbf{l (\lambda ) + \lambda_j  \ne j}$, $\mathbf{1 \le j \le N-1}$.} This is the simplest of the conditions. There are two values of $\lambda_j$ that we must examine to verify the condition.
\small\begin{equation*}
\begin{array}{lcl}
( \lambda_j =  0) && l(\lambda) < j\\
 & \Rightarrow & l (\lambda ) + \lambda_j  < j  \\
(\lambda_j> 0)&&  l(\lambda) \ge j\\
&\Rightarrow & l (\lambda ) + \lambda_j  > j
\end{array}
\end{equation*}\normalsize
Hence the condition is verified.\\
\\
\textbf{Verifying $\mathbf{l (\lambda ) + \lambda_j  \ne j+1 + m_1}$, $\mathbf{1 \le j \le N-1}$.} Three values of $\lambda_j$ must be examined for this condition to be verified.
\small\begin{equation*}
\begin{array}{lcll}
(\lambda_j =  0) && l(\lambda) < j  \\
 & \Rightarrow & m_1 \le  l(\lambda)   \\
 & \Rightarrow & l (\lambda ) + \lambda_j  <m_1+ j+1  \\
(\lambda_j =  1) && l(\lambda) = j + s & 0\le s \le N-j-1\\
 & \Rightarrow & m_1 \ge s+1  \\
 & \Rightarrow & l (\lambda ) + \lambda_j  <m_1+ j+1  \\
(\lambda_j > 1) && m_1 \le  l(\lambda) - j\\
&\Rightarrow & m_1+j \le  l(\lambda)\\
&\Rightarrow & l (\lambda ) + \lambda_j  > m_1 + j +1
\end{array}
\end{equation*}\normalsize
Explicitly verifying one more condition is enough to make the general case transparent.\\
\\
\textbf{Verifying $\mathbf{ l(\lambda ) + \lambda_j  \ne j+2 + m_1+m_2}$, $\mathbf{1 \le j \le N-1}$.} Unsurprisingly, four values of $\lambda_j$ must be examined for this condition to be verified.
\small\begin{equation*}
\begin{array}{lcll}
(\lambda_j =  0)& & l(\lambda) < j &\\
 & \Rightarrow & m_1+m_2 \le  l(\lambda) & \\
 & \Rightarrow & l (\lambda ) + \lambda_j  <m_1+m_2+ j+2 &\\
(\lambda_j =  1 )& & l(\lambda) = j + s & 0\le s \le N-j-1\\
 & \Rightarrow & m_1  \ge s+1  \\
 & \Rightarrow & l (\lambda ) + \lambda_j  <m_1+m_2+ j+2  &\\
(\lambda_j =  2) & & l(\lambda) = j + s_1 + s_2 & 0\le s_1+s_2 \le N-j-1\\
 & \Rightarrow & m_1+m_2 \ge s_1+s_2+1  \\
 & \Rightarrow & l (\lambda ) + \lambda_j  <m_1+m_2 +j+2  \\
( \lambda_j > 2)& & m_1+m_2 \le  l(\lambda) - j\\
&\Rightarrow & m_1+m_2+j \le  l(\lambda) \\
&\Rightarrow & l (\lambda ) + \lambda_j  > m_1+m_2 + j +2
\end{array}
\end{equation*}\normalsize
Which leaves us to verify the general case in an obvious fashion.\\
\\
\textbf{Verifying $\mathbf{l (\lambda ) + \lambda_j  \ne j+n + m_1+\dots+m_n }$, $\mathbf{1 \le j \le N-1}$, $\mathbf{3 \le n \le N-1}$.} For this case we systematically show that each value of $\lambda_j$ adheres to the condition.
\small\begin{equation*}
\begin{array}{lcll}
(\lambda_j =  0) & & l(\lambda) < j &\\
 & \Rightarrow & m_1+\dots + m_n \le  l(\lambda) & \\
 & \Rightarrow & l (\lambda ) + \lambda_j  <m_1+\dots + m_n+ j+n &\\
(\lambda_j =  r) & & l(\lambda) = j + s_1+ \dots + s_r & 0\le s_1+ \dots +s_r \le N-j-1\\
1\le r \le n & \Rightarrow &  m_1+\dots+ m_r \ge s_1+ \dots +s_r+1  \\
 & \Rightarrow & l (\lambda ) + \lambda_j  <m_1+\dots + m_n+ j+n  &\\
( \lambda_j > n)&  & m_1+\dots + m_n \le  l(\lambda) - j\\
&\Rightarrow & m_1+\dots + m_n+j \le  l(\lambda) \\
&\Rightarrow & l (\lambda ) + \lambda_j  > m_1+\dots + m_n + j +n
\end{array}
\end{equation*}\normalsize
Thus the proposition is verified. $\square$\\
\\
Applying the above result to eq. \ref{p.ks9} we immediately obtain,
\small\begin{equation*}\begin{split}
Z^{KS}_{N } (\vec{u},\vec{v}) &=\frac{\Upsilon_N}{ e_N(\vec{v})  } \sum_{0 \le \lambda_N \le \dots \le \lambda_1 \le N\atop{\lambda_N = 0}} \textrm{det} \left[ e_{\lambda_j +k-j} (\vec{u}) \right]^N_{j,k=1} \textrm{det} \left[ \varphi_{ \lambda'_{N-j+1}+ j,k} (\vec{v}) \right]^N_{j,k=1} \\
&= \frac{\Upsilon_N}{ e_N(\vec{v})  } \sum_{\{\lambda \} \subseteq (N)^{N-1}} g^{(N)}_{\{ \lambda' \}} (\vec{v}) S_{\{ \lambda' \}} (\vec{u})\\
& = \frac{\Upsilon_N}{ e_N(\vec{v})  } \sum_{\{\lambda \} \subseteq (N-1)^{N}} g^{(N)}_{\{ \lambda \}} (\vec{v}) S_{\{ \lambda \}} (\vec{u}) \\
&=\frac{\Upsilon_N}{ e_N(\vec{v})  } \sum_{0 \le \lambda_N \le \dots \le \lambda_1 \le N-1} \textrm{det} \left[ h_{\lambda_j +k-j} (\vec{u}) \right]^N_{j,k=1} \textrm{det} \left[ \varphi_{ \lambda_{N-j+1}+ j,k} (\vec{v}) \right]^N_{j,k=1}
\end{split}\end{equation*}\normalsize
where,
\small\begin{equation*}
g^{(N)}_{\{ \lambda \}} (\vec{v}) = \textrm{det} \left[ \varphi_{ \lambda_{N-j+1}+ j,k} (\vec{v}) \right]^N_{j,k=1}
\end{equation*}\normalsize
Using the Cauchy-Binet formula, or simply comparing this expression to \ref{p.18} we see that,
\small\begin{equation}
Z^{KS}_{N } (\vec{u},\vec{v}) =\frac{\Upsilon_N}{ e_N(\vec{v})  } \textrm{det} \left[ \left( h_{j-i}(\vec{u}) \right)^{2N-1}_{i,j=1}\left( \varphi_{j,k}(\vec{v}) \right)^{2N-1}_{j,k=1} \right]^N_{i,k=1}
\label{p.ks10'}
\end{equation}\normalsize
\section{Charged free fermions}
\begin{figure}[h!]
\begin{center}
\includegraphics[angle=0,scale=0.30]{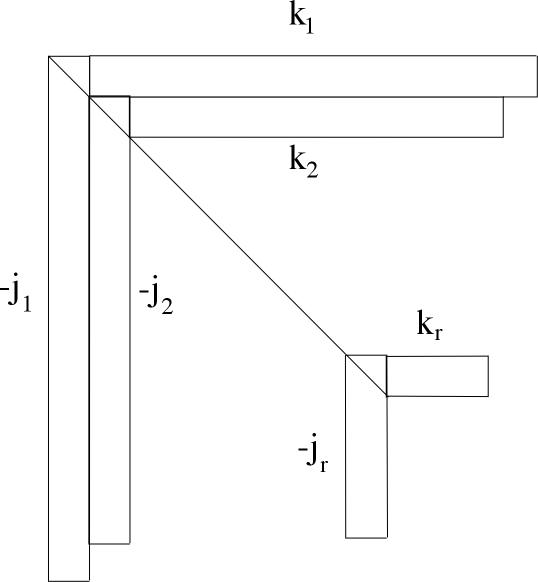}
\caption{\footnotesize{Partition $\{ \lambda\}$, consisting of $r$ hooks, which corresponds to the fermionic expression in eq. \ref{p.8}.}}
\label{p.a}
\end{center}
\end{figure}

In order to give the main results of this chapter we recall some necessary definitions/results regarding Clifford algebras, charged free fermions and their corresponding Fock space. This section, much like section 1.6 serves as an appendix of necessary definitions. In the following we use the integer labeling conventions for fermions found in \cite{6} as opposed to the $\frac{1}{2}$-integer labeling found in \cite{bluebook}. \\
\\
\textbf{The Clifford algebra and free fermion operators.} We define two infinite sets of generators, $\psi^*_i, \psi_i$, $i \in \field{Z}$, over $\field{C}$, which form a Clifford algebra, $\field{A}$, and satisfy the following anti-commutation relations,
\small\begin{equation}
\left\{ \psi_i, \psi_j \right\}_+ = 0 \textrm{  ,  } \left\{ \psi^*_i, \psi^*_j \right\}_+ = 0 \textrm{  ,  }  \left\{ \psi_i, \psi^*_j \right\}_+ = \delta_{ij}
\label{p.p}
\end{equation}\normalsize
We refer to a free fermion as an element of the (infinite) set of all linear combinations of the Clifford algebra, $\field{W}$,
\small\begin{equation*}
\field{W} = \left( \bigoplus_{m \in \field{Z}} \field{C} \psi_m \right) \oplus  \left( \bigoplus_{n \in \field{Z}} \field{C} \psi^*_n \right)
\end{equation*}\normalsize
Within $\field{W}$ there exist two subsets which form the creation and annihilation operators,
\small\begin{equation*}
\field{W}_{cr} = \left( \bigoplus_{m \ge 0} \field{C} \psi_m \right) \oplus  \left( \bigoplus_{n < 0} \field{C} \psi^*_n \right) \textrm{  ,  } \field{W}_{ann} = \left( \bigoplus_{m < 0} \field{C} \psi_m \right) \oplus  \left( \bigoplus_{n \ge 0} \field{C} \psi^*_n \right)
\end{equation*}\normalsize
We refer to $\field{A}_{cr}$ and $\field{A}_{ann}$ as the set of all possible ordered strings of creation and annihilation operators. A typical element of $\field{A}_{cr}$ and $\field{A}_{ann}$ are given below,
\small\begin{equation*}
\begin{array}{lcl}
a_{cr } \in \field{A}_{cr} = \psi^*_{j_1} \dots  \psi^*_{j_r} \psi_{k_s} \dots  \psi_{k_1} &\textrm{where}& j_1 < \dots < j_r < 0 \le k_s < \dots < k_1 \\
a_{ann } \in \field{A}_{ann} = \psi^*_{k_1} \dots  \psi^*_{k_s} \psi_{j_r} \dots  \psi_{j_1}  &\textrm{where}& j_1 < \dots < j_r < 0 \le k_s < \dots < k_1
\end{array}
\end{equation*}\normalsize
\textbf{The Fock space $\mathbf{\field{F}}$.} The (infinite dimensional) Fock space associated with the Clifford algebra $\field{A}$, referred to as $\field{F}$\footnote{A beautiful interpretation of the Fock space is given as an infinite one dimensional Maya diagram of black and white stones where the action of the operators move the stones around in a very specific way. For further details see chap. 4 of \cite{bluebook}.}, is characterized by the two properties,
\begin{itemize}
\item{All elements of $\field{W}_{ann}$ annihilate the vacuum,}
\end{itemize}
\small\begin{equation*}
 w_{ann}|0 \rangle = 0 \textrm{  $\forall$  }   w_{ann} \in \field{W}_{ann} 
\end{equation*}\normalsize
\begin{itemize}
\item{The entire Fock space can be generated by applying elements of $\field{A}_{cr}$ to the vacuum,}
\end{itemize}
\small\begin{equation*}
 \field{F} \equiv \field{A}_{cr}|0 \rangle
\end{equation*}\normalsize
Additionally, if we view a typical element of $\field{F}$,
\small\begin{equation}\begin{split}
 \psi^*_{j_1} \dots  \psi^*_{j_r} \psi_{k_s} \dots  \psi_{k_1} |0\rangle \textrm{  where  }& j_1 < \dots < j_r < 0 \le k_s < \dots < k_1\\
&s-r = l
\label{p.9}
\end{split}\end{equation}\normalsize
we refer to $l \in \field{Z}$ as the \emph{charge} of the element. Elements of $\field{F}$ with the same charge form a subspace, $\field{F}_l$, and hence $\field{F}$ decomposes into the following direct sum of vector spaces,
\small\begin{equation*}
\field{F} = \dots \oplus \field{F}_{-1} \oplus \field{F}_0 \oplus \field{F}_{1} \oplus \dots
\end{equation*}\normalsize
\textbf{The conjugate Fock space $\mathbf{\field{F}^*}$.} The conjugate Fock space, $\field{F}^*$, is defined similarly.
\begin{itemize}
\item{All elements of $\field{W}_{cr}$ annihilate the conjugate vacuum,}
\end{itemize}
\small\begin{equation*}
 \langle 0| w_{cr} = 0 \textrm{  $\forall$  }  w_{cr} \in \field{W}_{cr} 
\end{equation*}\normalsize
\begin{itemize}
\item{$\field{F}^*$ can be generated by applying elements of $\field{A}_{ann}$ to the conjugate vacuum,}
\end{itemize}
\small\begin{equation*}
 \field{F}^* \equiv \langle 0 | \field{A}_{ann}
\end{equation*}\normalsize
Additionally, viewing a typical element of $\field{F}^*$,
\small\begin{equation}\begin{split}
\langle 0| \psi^*_{k_1} \dots  \psi^*_{k_s} \psi_{j_r} \dots  \psi_{j_1} \textrm{  where  }& j_1 < \dots < j_r < 0 \le k_s < \dots < k_1 \\
&r-s = l
\label{p.10}
\end{split}\end{equation}\normalsize
elements of $\field{F}^*$ with the same charge, $l \in \field{Z}$, form a conjugate subspace, $\field{F}^*_l$, and hence $\field{F}^*$ decomposes into the following direct sum of conjugate vector spaces,
\small\begin{equation*}
\field{F}^* = \dots \oplus \field{F}^*_{-1} \oplus \field{F}^*_0 \oplus \field{F}^*_{1} \oplus \dots
\end{equation*}\normalsize
\textbf{The inner product.} We consider the inner product of dual vector spaces,
\small\begin{equation*}
\field{F}^* \times \field{F} = \langle 0|\left( \prod^{n_1}_{i=1} w^i_{ann}\right)  \left( \prod^{n_2}_{j=1}  w^j_{cr} \right) |0 \rangle \rightarrow \field{C}
\end{equation*}\normalsize
for all $w^j_{cr}/ w^i_{ann} \in  \field{W}_{cr} / \field{W}_{ann}$. The quantity $\langle 0| \dots|0 \rangle$ is referred to as the vacuum expectation value, defined by,
\begin{equation}\begin{split}
\langle 0|0 \rangle = 1 \textrm{  ,  } \langle 0| \psi_i \psi_j |0 \rangle = \langle 0| \psi^*_i \psi^*_j |0 \rangle = 0 \\
\langle 0| \psi_i \psi^*_j |0 \rangle = \left\{ \begin{array}{cc}
\delta_{ij} & i = j < 0 \\
0 & \textrm{otherwise}
\end{array}\right.
  \textrm{  ,  } \langle 0| \psi^*_i \psi_j |0 \rangle =\left\{ \begin{array}{cc}
\delta_{ij} & i = j \ge 0 \\
0 & \textrm{otherwise}
\end{array}\right.
\label{p.11}
\end{split}\end{equation}\normalsize
Using the above definitions and the anti-commutation relations, the expectation value of a general string of free fermions, $w_1 \dots w_r $, can be calculated, known as \textit{Wick's theorem},
\small\begin{equation}\begin{split}
\langle 0|w_1 \dots w_r  |0 \rangle =& \left\{ \begin{array}{cc} 
0 & r \textrm{  odd}\\
\sum_{\sigma \in S^*_r} sgn( \sigma) \langle 0|w_{\sigma_1}  w_{\sigma_2}  |0 \rangle \dots  \langle 0|w_{\sigma_{r-1}}  w_{\sigma_r}  |0 \rangle &  r \textrm{  even}
\end{array}\right.  \\
&  \sigma_1 < \sigma_2,  \dots,  \sigma_{r-1} < \sigma_r  \\
& \sigma_1 < \sigma_3 <  \dots < \sigma_{r-1}
\label{p.12}
\end{split}\end{equation}\normalsize
\textbf{Fermionic representation of the Lie algebra $\mathbf{gl(\infty)}$.} The fermionic representation of the algebra $gl(\infty)$ is given by the following bilinear sum,
\small\begin{equation}
gl(\infty) = \left\{ \sum_{i,j \in \field{Z}} a_{ij} :\psi_i \psi^*_j : \right\} \oplus \field{C} \label{p.15} 
\end{equation}\normalsize
where,
\small\begin{equation}
: \psi_i \psi^*_j : \equiv \psi_i \psi^*_j - \langle 0| \psi_i \psi^*_j  |0 \rangle 
\label{p.13}
\end{equation}\normalsize
and the coefficients, $a_{ij} \in \field{C}$, satisfy the following finiteness condition,
\small\begin{equation}
\exists \textrm{  } N \in \field{N}, \textrm{  such that  } a_{ij} =0 \textrm{  ,  } \forall \textrm{  } |i-j| > N
\label{p.14}
\end{equation}\normalsize
Additionally, for some $X_A \in gl(\infty)$, we have following helpful commutation relations,
\small\begin{equation}
[X_A, \psi_j] = \sum_{i \in \field{Z}}a_{ij} \psi_i \textrm{  ,  } [X_A, \psi^*_j] = \sum_{i \in \field{Z}}(-a_{ji}) \psi^*_i \label{p.46}
\end{equation}\normalsize\\
\textbf{Heisenberg subalgebras of $\mathbf{gl(\infty)}$.} We now consider important subalgebras of this Lie algebra, labeled $H_m \in gl(\infty)$, by setting $a_{ij} = \delta_{j,i+m}$, $m \in \field{Z}$,
\small\begin{equation}
H_m = \sum_{i \in \field{Z}} : \psi_i \psi^*_{i+m}:
\label{p.16}
\end{equation}\normalsize
which satisfy the following commutation relations,
$$
[H_m, H_n] = m \delta_{mn}
$$
Thus $H_n$, $n \ne 0$, along with central element 1 span a Heisenberg subalgebra $\field{H}$ in $gl(\infty)$. We additionally define the generating function for this subalgebra as follows,
\small\begin{equation}
H_{\pm}(\vec{x}) \equiv \sum^{\infty}_{m = 1} x_{\pm m} H_{\pm m}
\label{p.17}
\end{equation}\normalsize
Notice that the generating function only contains fermionic terms of charge zero.\\
\\
\textbf{Boson-fermion correspondence.} It is possible to realize expressions in fermionic Fock space as elements in the polynomial ring $\field{C} [x_1,x_2, \dots]$ (bosons) by applying the following theorem.
\begin{theorem}
The following map,
\small\begin{equation*}
\Phi: \field{F} = \oplus_l \field{F}_l  \longrightarrow \field{C} [z,z^{-1}, x_1,x_2, \dots] = \oplus_l z^{l} \field{C} [x_1,x_2, \dots] 
\end{equation*}\normalsize
where,
\small\begin{equation*}
\Phi \left[ \left( \prod^n_{j=1} w^j_{cr} \right) |0 \rangle \right] = \oplus_l z^{l} \langle 0| \psi^*_0 \dots  \psi^*_{l-1} \exp \{H_+(\vec{x}) \} \left( \prod^n_{j=1} w^j_{cr} \right) |0 \rangle
\end{equation*}\normalsize
$\forall$ $w^j_{cr} \in \field{W}_{cr}$, is an isomorphism of vector spaces.
\end{theorem}
In what follows, we shall only have to consider fermionic expressions of zero charge.\\
\\
\textbf{Character polynomials.} The following formula uses the vector space isomorphism to generate character polynomials (which serve as a basis for $\field{C}[\vec{x}]$) from zero charge fermionic expressions. 
\begin{equation}
\langle 0| \exp\{H_+(\vec{x})\} \psi^*_{j_1} \dots  \psi^*_{j_r} \psi_{k_r} \dots  \psi_{k_1} |0\rangle = (-1)^{j_1 + \dots + j_r}\chi_{\{\lambda\}}(\vec{x}) 
\label{p.8}
\end{equation}\normalsize
where $j_1 < \dots < j_r < 0 \le k_r < \dots < k_1$, and $r$ corresponds to the amount of hooks in the corresponding partition. The correspondence between the partition $\{ \lambda\}$ and the integers $\{j_1,\dots,j_r,k_1,\dots, k_r\}$ is explained in diagram \ref{p.a}.
\section{Fermionic expression of the DWPF}
\label{ferm6V}
We now come to the first of the two main results in this chapter, given by the following lemma.
\begin{lemma}
Eq. \ref{fermone} can be expressed as the bosonization of the following fermionic Fock space expression,
\small\begin{equation}
\exp\left\{ X^{(N)}_{0} \right\}\exp\left\{ X^{(N)}_{1} \right\} \dots \exp\left\{ X^{(N)}_{N-2} \right\} |0\rangle \Upsilon_N c^{(N)}_{\{\phi\}}
\label{p.6}
\end{equation}\normalsize
where,
\small\begin{equation}
\begin{array}{lcl}
X^{(N)}_{0} &=& -\tilde{c}^{(N)}_{\{1 \}} \psi^*_{-1}\psi_0 + \tilde{c}^{(N)}_{\{1^2 \}} \psi^*_{-2}\psi_0 + \dots + (-1)^N\tilde{c}^{(N)}_{\{1^N \}} \psi^*_{-N}\psi_0  \\
X^{(N)}_{1} &=& -\tilde{c}^{(N)}_{\{2 \}} \psi^*_{-1}\psi_1 + \tilde{c}^{(N)}_{\{2,1 \}} \psi^*_{-2}\psi_1 + \dots + (-1)^N\tilde{c}^{(N)}_{\{2,1^{N-1} \}} \psi^*_{-N}\psi_1 \\
& \vdots&  \\
X^{(N)}_{N-2} &=& \sum^{N}_{j=1}(-1)^{j} \tilde{c}^{(N)}_{\{N-1,1^{j-1} \}} \psi^*_{-j}\psi_{N-2} \end{array}
\label{p.7}
\end{equation}\normalsize
and the coefficients, $\tilde{c}^{(N)}_{\{\lambda \}} = \frac{c^{(N)}_{\{\lambda \}}}{c^{(N)}_{\{\phi \}}}$, are given by eq. \ref{p.18}. We shall refer to the group elements, $\exp\left\{ X^{(N)}_{j} \right\} \in GL(\infty)$, as generators.
\end{lemma}
\textbf{Example, $\mathbf{N=3}$.} Before we give a proof of the above lemma, we first give the simplest non trivial example.\\
\\
Using the following definitions,
\small\begin{equation*}
c^{(3)}_{\{\lambda\}} = \textrm{det}  \left[ \frac{q^{\lambda_{4-i} +i-j+1}-q^{j-1}}{q-1} (-1)^{2-(\lambda_{4-i} +i-j)}e_{2-(\lambda_{4-i} +i-j)}(\vec{v}) \right]^3_{i,j=1}
\end{equation*}\normalsize
and,
\small\begin{equation*}\begin{split}
X^{(3)}_{0} = &-\tilde{c}^{(3)}_{\{1 \}} \psi^*_{-1}\psi_0 + \tilde{c}^{(3)}_{\{1^2 \}} \psi^*_{-2}\psi_0 -\tilde{c}^{(3)}_{\{1^3 \}} \psi^*_{-3}\psi_0  \\
X^{(3)}_{1}=& -\tilde{c}^{(3)}_{\{2 \}} \psi^*_{-1}\psi_1 + \tilde{c}^{(3)}_{\{2,1 \}} \psi^*_{-2}\psi_1 -\tilde{c}^{(3)}_{\{2,1^2 \}} \psi^*_{-3}\psi_1
\end{split}\end{equation*}\normalsize
we obtain the expansion,
\small\begin{equation*}\begin{split}
\exp\left\{ X^{(3)}_{0} \right\}\exp\left\{ X^{(3)}_{1} \right\}\\
 = 1- \tilde{c}^{(3)}_{\{1 \}} \psi^*_{-1}\psi_0 + \tilde{c}^{(3)}_{\{1^2 \}} \psi^*_{-2}\psi_0  +\left(\tilde{c}^{(3)}_{\{1 \}} \tilde{c}^{(3)}_{\{2,1^2 \}} -\tilde{c}^{(3)}_{\{2 \}} \tilde{c}^{(3)}_{\{1^3 \}} \right) \psi^*_{-3} \psi^*_{-1}\psi_0 \psi_1 \\
+ \tilde{c}^{(3)}_{\{2,1 \}} \psi^*_{-2}\psi_1 -\tilde{c}^{(3)}_{\{2,1^2 \}} \psi^*_{-3}\psi_1 + \left(\tilde{c}^{(3)}_{\{2,1 \}} \tilde{c}^{(3)}_{\{1^3 \}} -\tilde{c}^{(3)}_{\{1^2 \}} \tilde{c}^{(3)}_{\{2,1^2 \}} \right) \psi^*_{-3} \psi^*_{-2}\psi_0 \psi_1 \\
-\tilde{c}^{(3)}_{\{1^3 \}} \psi^*_{-3}\psi_0 - \tilde{c}^{(3)}_{\{2 \}} \psi^*_{-1}\psi_1+ \left(\tilde{c}^{(3)}_{\{2 \}} \tilde{c}^{(3)}_{\{1^2 \}} -\tilde{c}^{(3)}_{\{1 \}} \tilde{c}^{(3)}_{\{2,1 \}} \right) \psi^*_{-2} \psi^*_{-1}\psi_0 \psi_1
\end{split}\end{equation*}\normalsize
We now detail the method required to simplify the three bilinear expressions in the above coefficients. For further details on Pl\"{u}cker relations see \cite{nicepaper}.\\
\\
\textbf{Generation of Pl\"{u}cker relations.} To simplify the three bilinear expressions of the coefficients contained above we label $\gamma_{\mu}$ as the following column vector,
\small\begin{equation*}
\gamma_{\mu} = \left(\frac{q^{\mu-j+2}-q^{j-1}}{q-1} (-1)^{1-\mu +j}e_{1-\mu +j}(\vec{v}) \right)^T_{j=1,2,3} = \left( \begin{array}{c}
\kappa_{\mu+1,1}\\
\kappa_{\mu+1,2}\\
\kappa_{\mu+1,3}
\end{array} \right)
\end{equation*}\normalsize
Using this notation the coefficients $c^{(3)}_{\{\lambda\}}$ can be constructed in the following convenient way,
\small\begin{equation*}
c^{(3)}_{\{\lambda\}}  = \textrm{det} \left[ \left( \gamma_{\lambda_3},\gamma_{\lambda_2 +1},\gamma_{\lambda_1+2} \right)^T  \right] =\left| \gamma_{\lambda_3},\gamma_{\lambda_2 +1},\gamma_{\lambda_1+2}  \right|
\end{equation*}\normalsize
We now consider the following $6 \times 6$ determinant expression,
$$
 \left| \begin{array}{cccccc}
\gamma_{\mu_1} & \gamma_{\mu_2} & \gamma_{\nu_1} & \gamma_{\nu_2}& \gamma_{\nu_3}& \gamma_{\nu_4}\\
0 & 0& \gamma_{\nu_1} & \gamma_{\nu_2}& \gamma_{\nu_3}& \gamma_{\nu_4}
\end{array} \right| = 0
$$
and use Laplace expansion to obtain a bilinear sum of $3 \times 3$ determinants,
\begin{eqnarray*}
\left| \gamma_{\mu_1} , \gamma_{\mu_2} , \gamma_{\nu_1} \right| \left| \gamma_{\nu_2} , \gamma_{\nu_3} , \gamma_{\nu_4} \right| + \left| \gamma_{\mu_1} , \gamma_{\mu_2} , \gamma_{\nu_3} \right| \left| \gamma_{\nu_1} , \gamma_{\nu_2} , \gamma_{\nu_4} \right| &&\\
-\left| \gamma_{\mu_1} , \gamma_{\mu_2} , \gamma_{\nu_2} \right| \left| \gamma_{\nu_1} , \gamma_{\nu_3} , \gamma_{\nu_4} \right| - \left| \gamma_{\mu_1} , \gamma_{\mu_2} , \gamma_{\nu_4} \right| \left| \gamma_{\nu_1} , \gamma_{\nu_2} , \gamma_{\nu_3} \right| &=&0
\end{eqnarray*}
where we have a total of six arbitrary indices, $(\mu_1,\mu_2,\nu_1,\nu_2,\nu_3,\nu_4)$. In order to derive the three necessary Pl\"{u}cker relations we input the following three sets of values for the indices, $(\mu_1,\mu_2,\nu_1,\nu_2,\nu_3,\nu_4)$, to obtain,
\small\begin{equation*}
\begin{array}{lcl}
(2,3,4,0,1,2) &:& c^{(3)}_{\{2,1 \}}c^{(3)}_{\{1^3 \}} - c^{(3)}_{\{1^2 \}}c^{(3)}_{\{2,1^2 \}} = - c^{(3)}_{\{\phi \}}c^{(3)}_{\{2^3 \}}\\
(1,3,4,0,1,2) &:& c^{(3)}_{\{1 \}}c^{(3)}_{\{2,1^2 \}} - c^{(3)}_{\{2 \}}c^{(3)}_{\{1^3 \}} = c^{(3)}_{\{\phi \}}c^{(3)}_{\{2^2,1 \}}\\
(0,3,4,0,1,2) &:& c^{(3)}_{\{2 \}}c^{(3)}_{\{1^2 \}} - c^{(3)}_{\{1 \}}c^{(3)}_{\{2,1 \}} =  -c^{(3)}_{\{\phi \}}c^{(3)}_{\{2^2 \}}
\end{array}
\end{equation*}\normalsize
Thus the polynomial expansion of the product of the two generators becomes the following,
\small\begin{equation*}\begin{split}
\exp\left\{ X^{(3)}_{0} \right\}\exp\left\{ X^{(3)}_{1} \right\}\\
 = 1- \tilde{c}^{(3)}_{\{1 \}} \psi^*_{-1}\psi_0 + \tilde{c}^{(3)}_{\{1^2 \}} \psi^*_{-2}\psi_0 + \tilde{c}^{(3)}_{\{2,1 \}} \psi^*_{-2}\psi_1 -\tilde{c}^{(3)}_{\{2,1^2 \}} \psi^*_{-3}\psi_1-\tilde{c}^{(3)}_{\{1^3 \}} \psi^*_{-3}\psi_0   \\
- \tilde{c}^{(3)}_{\{2 \}} \psi^*_{-1}\psi_1 +\tilde{c}^{(3)}_{\{2^2,1 \}} \psi^*_{-3} \psi^*_{-1}\psi_0 \psi_1 - \tilde{c}^{(3)}_{\{2^3 \}} \psi^*_{-3} \psi^*_{-2}\psi_0 \psi_1 - \tilde{c}^{(3)}_{\{2^2 \}} \psi^*_{-2} \psi^*_{-1}\psi_0 \psi_1
\end{split}\end{equation*}\normalsize
where every bilinear term in the fermions contains a coefficient with a partition containing a single hook, and every term containing the product of four fermions contains a coefficient with a partition containing two hooks. \\
\\
\textbf{Bosonization.} Applying eq. \ref{p.8} on the inner product expression we obtain,
\small\begin{equation*}
\langle 0| \exp\{H_+(\vec{x})\}e^{ X^{(3)}_{0}}e^{ X^{(3)}_{1}} |0\rangle \Upsilon_3 c^{(3)}_{\{ \phi \}}(\vec{v}) = \Upsilon_3 \sum_{\{ \lambda \} \subseteq (2)^3} c^{(3)}_{\{ \lambda \}} (\vec{v}) \chi_{\{ \lambda \}}(\vec{x})
\end{equation*}\normalsize
and restricting the time variables in the usual manner,
\small\begin{equation*}
x_j \rightarrow \frac{1}{j}p_j(u_1,u_2,u_3) \textrm{  ,  } j \in \{ 1, 2, \dots \}
\end{equation*}\normalsize
we have,
\small\begin{equation*}\begin{split}
\langle 0| \exp\left\{H_+\left( \left\{ \frac{1}{j}p_j(\vec{u}) \right\} \right)\right\}e^{ X^{(3)}_{0}}e^{ X^{(3)}_{1}} |0\rangle \Upsilon_3 c^{(3)}_{\{ \phi \}}(\vec{v}) &= \Upsilon_3 \sum_{\{ \lambda \} \subseteq (2)^3} c^{(3)}_{\{ \lambda \}} (\vec{v}) S_{\{ \lambda \}}(\vec{u})\\
&= Z^L_3(\vec{u},\vec{v})
\end{split}\end{equation*}\normalsize
which completes the $N=3$ example.\\
\\
It should be apparent by now that proving the lemma of this section requires us to verify that the necessary Pl\"{u}cker relations are generated appropriately. We shall proceed slowly and show that this is the case for the multiplication of two general generators. This shall serves as the base case for the inductive proof of the lemma that shall follow.
\subsection{Multiplication of two generators}
\label{2oper}
\noindent For $0 \le l_1 < l_2 \le N-2$, consider the following elements of $gl( \infty )$,
\small\begin{equation*}\begin{split}
X^{(N)}_{l_1} = \sum^{N}_{j=1}(-1)^{j} \tilde{c}^{(N)}_{\{l_1+1,1^{j-1} \}} \psi^*_{-j}\psi_{l_1} \\
 X^{(N)}_{l_2} = \sum^{N}_{j=1}(-1)^{j} \tilde{c}^{(N)}_{\{l_2+1,1^{j-1} \}} \psi^*_{-j}\psi_{l_2} 
\end{split}\end{equation*}\normalsize
These two sums of fermionic bilinears contain coefficients that are labeled by partitions containing a single hook of varying dimensions. When we multiply these two sums, we will obtain a sum of a product of four fermions, with bilinear terms in the coefficients. The main crux of this section is to detail, through the application of Pl\"{u}cker relations, the method of simplifying these bilinear sums of coefficients (whose partitions are labeled by single hooks) into single coefficients (whose partition is labeled by two hooks).\\
\\
We begin by multiplying the exponentiation of the above bilinear sums of fermions,
\small\begin{equation*}
\exp \left\{ X^{(N)}_{l_1}  \right\}\exp \left\{ X^{(N)}_{l_2}  \right\} = 1 + X^{(N)}_{l_1} + X^{(N)}_{l_2} + X^{(N)}_{l_1} X^{(N)}_{l_2}
\end{equation*}\normalsize
where we notice immediately that non linear terms in either $X^{(N)}_{l_1}$ or $X^{(N)}_{l_2}$ do not survive due to the anti-commutation relations.\\
\\
\textbf{Obtaining the bilinear terms in the coefficients.} It is obvious that the linear terms, $X^{(N)}_{l_1}$ and $X^{(N)}_{l_2}$, in the above expression do not require any work as they do not contain any bilinear terms in the coefficients. Concentrating then on the cross term, $X^{(N)}_{l_1} X^{(N)}_{l_2}$, we have,
\small\begin{equation}
X^{(N)}_{l_1} X^{(N)}_{l_2} = \sum^N_{j_1=1}\sum^N_{j_2=1\atop{j_2 \ne j_1}} (-1)^{j_1+j_2} \tilde{c}^{(N)}_{\{l_1+1,1^{j_1-1} \}}\tilde{c}^{(N)}_{\{l_2+1,1^{j_2-1} \}}\psi^*_{-j_1}\psi_{l_1}\psi^*_{-j_2}\psi_{l_2}
\label{objert1}\end{equation}\normalsize
Commuting the fermions to the following desired form,
\small\begin{equation*}
\psi^*_{-j_2}\psi^*_{-j_1}\psi_{l_1}\psi_{l_2} \textrm{  ,  } 1 \le j_1 < j_2 \le N
\end{equation*}\normalsize
we obtain,
\small\begin{equation*}\begin{split}
& \sum^N_{j_1=1}\sum^N_{j_2=1\atop{j_2 \ne j_1}} (-1)^{j_1+j_2+1} \tilde{c}^{(N)}_{\{l_1+1,1^{j_1-1} \}}\tilde{c}^{(N)}_{\{l_2+1,1^{j_2-1} \}}\psi^*_{-j_1}\psi^*_{-j_2}\psi_{l_1}\psi_{l_2} \\
=&\left(\sum_{1 \le j_1 < j_2 \le N} + \sum_{1 \le j_2 < j_1 \le N}   \right)(-1)^{j_1+j_2+1} \tilde{c}^{(N)}_{\{l_1+1,1^{j_1-1} \}}\tilde{c}^{(N)}_{\{l_2+1,1^{j_2-1} \}}\psi^*_{-j_1}\psi^*_{-j_2}\psi_{l_1}\psi_{l_2}
\end{split}\end{equation*}\normalsize
\small\begin{equation*}\begin{split}
=&\sum_{1 \le j_1 < j_2 \le N}(-1)^{j_1+j_2} \left(\tilde{c}^{(N)}_{\{l_1+1,1^{j_1-1} \}}\tilde{c}^{(N)}_{\{l_2+1,1^{j_2-1} \}} - \tilde{c}^{(N)}_{\{l_1+1,1^{j_2-1} \}}\tilde{c}^{(N)}_{\{l_2+1,1^{j_1-1} \}} \right)\\
& \times \psi^*_{-j_2}\psi^*_{-j_1}\psi_{l_1}\psi_{l_2}
\end{split}\end{equation*}\normalsize
In order to simplify the above bilinear relationship with the coefficients we proceed much the same as we did for the example with $N=3$, but on a much larger scale.\\
\\
\textbf{Generation of the necessary Pl\"{u}cker relations.} To begin we label $\gamma_{\mu}$ as the following column vector,
\small\begin{equation}
\gamma_{\mu} = \left(\frac{q^{\mu-j+2}-q^{j-1}}{q-1} (-1)^{N-2-\mu +j}e_{N-2-\mu +j}(\vec{v}) \right)^T_{j=1,\dots,N } = \left( \begin{array}{c}
\kappa_{\mu+1,1}\\
\kappa_{\mu+1,2}\\
\vdots\\
\kappa_{\mu+1,N}
\end{array} \right)
\label{5.19}\end{equation}\normalsize
and hence the coefficient, $c^{(N)}_{\{\lambda\}}$, can be expressed as the determinant of the following length $N$ vector of column vectors $\gamma_{\mu}$,
\begin{equation}
c^{(N)}_{\{\lambda\}}  =  \left| \gamma_{\lambda_N},\gamma_{\lambda_{N-1} +1},\dots, \gamma_{\lambda_2+N-2}, \gamma_{\lambda_1+N-1}  \right|
\label{5.20}\end{equation}\normalsize
Now we consider the following $2N \times 2N$ determinant expression,
$$
 \left| \begin{array}{ccccccc}
\gamma_{\mu_1} & \dots & \gamma_{\mu_{N-1}} & \gamma_{\nu_1}& \gamma_{\nu_2}&\dots & \gamma_{\nu_{N+1}}\\
0 & \dots & 0 & \gamma_{\nu_1}& \gamma_{\nu_2}&\dots & \gamma_{\nu_{N+1}}
\end{array} \right| = 0
$$
and Laplace expand it to obtain a bilinear sum of $N \times N$ determinants,
\small\begin{equation}
\sum^{N+1}_{p=1} (-1)^{p+1}\left| \begin{array}{ccccc}
\gamma_{\nu_1} & \dots & \hat{\gamma}_{\nu_{p}} & \dots & \gamma_{\nu_{N+1}}
\end{array} \right| \left| \begin{array}{cccc}
\gamma_{\mu_1} & \dots & \gamma_{\mu_{N-1}} & \gamma_{\nu_p}
\end{array} \right|   =0
\label{p.21}
\end{equation}\normalsize
where we have $2N$ arbitrary indices. We now consider inputting the following specific values for $(\mu_1, \dots, \mu_{N-1}, \nu_1, \dots \nu_{N+1})$,
\small\begin{equation*}
\begin{array}{lcl}
(\mu_1, \dots, \mu_{N-j_2}) &=& (0, \dots, N-j_2-1)\\
(\mu_{N-j_2+1},\dots, \mu_{N-j_1-1}) &=& (N-j_2 +1 , \dots , N-j_1 -1)\\
(\mu_{N-j_1},\dots, \mu_{N-2}) &=& (N-j_1 +1 , \dots , N-1)\\
(\mu_{N-1}, \nu_{1}) &=& (N+l_1,N+l_2)\\
(\nu_2, \dots, \nu_{N+1}) &=& (0, \dots, N-1)
\end{array}
\end{equation*}\normalsize
to obtain the following bilinear sum of determinants,
\small\begin{equation*}\begin{split}
\left|\gamma_{0} , \dots , \gamma_{N-1} \right| 
\left| \gamma_{0} ,\dots , \hat{\gamma}_{N-j_2} , \dots , \hat{\gamma}_{N-j_1} , \dots , \gamma_{N-1} , \gamma_{N+l_1} , \gamma_{N+l_2} \right|&\\
+(-1)^{N-j_2+1} \left| \gamma_{N+l_2} , \gamma_{0} , \dots , \hat{\gamma}_{N-j_2} , \dots , \gamma_{N-1}  \right|&\\
 \times\left| \gamma_{0} , \dots , \hat{\gamma}_{N-j_2} , \dots , \hat{\gamma}_{N-j_1} , \dots , \gamma_{N-1} , \gamma_{N+l_1}  ,\gamma_{N-j_2} \right| &\\
+(-1)^{N-j_1+1}\left| \gamma_{N+l_2} , \gamma_{0} , \dots , \hat{\gamma}_{N-j_1} , \dots , \gamma_{N-1}  \right| &\\
  \times \left| \gamma_{0} , \dots , \hat{\gamma}_{N-j_2} , \dots , \hat{\gamma}_{N-j_1} , \dots , \gamma_{N-1} , \gamma_{N+l_1}  ,\gamma_{N-j_1} \right|&=0
\end{split}\end{equation*}\normalsize
Ordering the columns of the above determinant expressions so that indices of a higher integer are placed to the right, we obtain the required Pl\"{u}cker relations,
\small\begin{equation*}\begin{split}
c^{(N)}_{\{l_1+1,1^{j_1-1} \}}c^{(N)}_{\{l_2+1,1^{j_2-1} \}} - c^{(N)}_{\{l_1+1,1^{j_2-1} \}}c^{(N)}_{\{l_2+1,1^{j_1-1} \}}\\
=  c^{(N)}_{\{ \phi \}}c^{(N)}_{\{l_2+1,l_1+2,2^{j_1-1},1^{j_2-j_1-1} \}}
\end{split}\end{equation*}\normalsize
Thus eq. \ref{objert1} reduces to the form,
\small\begin{equation}
X^{(N)}_{l_1}  X^{(N)}_{l_2} =  \sum_{1 \le j_1 < j_2 \le N}(-1)^{j_1+j_2} \tilde{c}^{(N)}_{\{l_2+1,l_1+2,2^{j_1-1},1^{j_2-j_1-1} \}}  \psi^*_{-j_2}\psi^*_{-j_1}\psi_{l_1} \psi_{l_2}
\label{p.22}
\end{equation}\normalsize
Referring back to fig. \ref{p.a}, we see that the partition of each coefficient, expressible as a double hook, correctly corresponds to the partition generated by the product of four fermions.\\
\\
With the base case now complete, we shall now use induction to prove that multiplying a general number of orbit operators produces the required coefficients.
\subsection{Multiplication of an arbitrary number of generators}
\noindent We shall now generalize the above result, that the multiplication of $k$ sums of bilinear fermions with coefficients labeled by single hook partitions simplifies, through the use of Pl\"{u}cker relations, into the sum of a product of $2k$ fermions, where each fermionic expression is accompanied by the required coefficient labeled by the partition consisting of necessarily $k$ hooks. This result is proven using induction by the following proposition.
\label{koper}
\begin{proposition}
For $0 \le l_1 < \dots < l_{k} \le N-2$,
\small\begin{equation}
X^{(N)}_{l_1} \dots X^{(N)}_{l_{k}} =  \sum_{1 \le j_1 < \dots < j_{k} \le N}(-1)^{j_1+\dots + j_{k}} \tilde{c}^{(N)}_{\{ \lambda\}_k}   \psi^*_{-j_k} \dots \psi^*_{-j_1}\psi_{l_1}\dots  \psi_{l_k} \label{p.23}
\end{equation}\normalsize
where $\{ \lambda\}_k$ is the partition consisting of $k$ hooks given explicitly as,
\small\begin{equation}
\{ \lambda\}_k =\{l_{k}+1,l_{k-1}+2,\dots, l_{1}+k,k^{j_1-1},(k-1)^{j_2-j_1-1},\dots, 1^{j_{k}-j_{k-1}-1} \}
\label{particl}\end{equation}\normalsize
\label{korbit}
\end{proposition}
\textbf{Proof.} We begin by noting that we have proven the above formula for $k=2$. Let us now assume that eq. \ref{p.23} holds for some value of $k$, we shall now show explicitly that it also holds for $k+1$. Hence we naturally consider the multiplication of $k+1$ bilinear sums of fermions and generate the bilinear terms in the coefficients.\\
\\
\textbf{Generating the bilinear terms in the coefficients.} For $0 \le l_1 < \dots < l_{k+1} \le N-2$,
\small\begin{equation*}\begin{split}
X^{(N)}_{l_1} \dots X^{(N)}_{l_{k+1}} =  \sum_{1 \le j_1 < \dots < j_{k} \le N} \sum^N_{j_{k+1}=1\atop{j_{k+1} \ne j_1,\dots, j_k}}(-1)^{j_1+\dots + j_{k+1}} \tilde{c}^{(N)}_{\{ \lambda\}_k} \hat{c}^{(N)}_{\{l_{k+1}+1,1^{j_{k+1}-1}\}}\\ 
 \times \psi^*_{-j_k} \dots \psi^*_{-j_1}\psi_{l_1}\dots  \psi_{l_k}\psi^*_{-j_{k+1}}\psi_{l_{k+1}}
\end{split}\end{equation*}\normalsize
where we break up the summation in the convenient form,
\small\begin{equation*}\begin{split}
 \sum_{1 \le j_1 < \dots < j_{k} \le N} \sum^N_{j_{k+1}=1\atop{j_{k+1} \ne j_1,\dots, j_k}} =  \sum_{1 \le j_1 < \dots < j_{k+1} \le N} +  \sum_{1 \le j_1 < \dots <j_{k-1} < j_{k+1} < j_{k} \le N} + \dots\\
\dots +   \sum_{1 \le j_1 < j_{k+1}< j_2 < \dots < j_{k} \le N} +  \sum_{1 \le j_{k+1}< j_1 < \dots  < j_{k} \le N} 
\end{split}\end{equation*}\normalsize
Ordering the fermions appropriately in each summation and reassigning indices so that only one summation is necessary, we obtain the following expression,
\small\begin{equation}\begin{split}
X^{(N)}_{l_1} \dots X^{(N)}_{l_{k+1}} = \sum_{1 \le j_1 < \dots < j_{k+1} \le N} (-1)^{j_1+\dots + j_{k+1}}\\
\times \left[ \sum^k_{p=0}(-1)^{p} \tilde{c}^{(N)}_{\sigma_p \left( \{\lambda\}_k \right)}  \tilde{c}^{(N)}_{\{l_{k+1}+1, 1^{j_{k+1-p}-1} \}} \right] \psi^*_{-j_{k+1}} \dots \psi^*_{-j_1}\psi_{l_1}\dots\psi_{l_{k+1}}
\end{split}\end{equation}\normalsize
The partition $\sigma_p \left( \{\lambda\}_k\right)$, $1 \le p \le k$, is obtained by rearranging the indices of partition $\{\lambda\}_k$ appropriately, where $\sigma_0 \left( \{\lambda\}_k\right)$ denotes that there is no change to the indices. As a concrete example, consider $k=4$. Labeling $\field{L}_4= \{l_{4}+1,l_{3}+2,l_{2}+3, l_{1}+4 \}$ we have,
\small\begin{equation*}
\sigma_0 \left( \{\lambda\}_4\right)=  \{\lambda\}_4 =  \{\field{L}_4,4^{j_1-1},3^{j_2-j_1-1},2^{j_3-j_2-1}, 1^{j_{4}-j_{3}-1} \} 
\end{equation*}\normalsize
$\sigma_1 \left( \{\lambda\}_4\right)$ is obtained from the summation $ \sum_{1 \le j_1 < j_2 < j_3 < j_5 < j_{4} \le N}$. Performing the relabeling $j_5 \leftrightarrow j_4$ to the indices of this summation expresses it in the required form. Thus the partition $\sigma_1 \left( \{\lambda\}_4\right)$ is simply $\{\lambda\}_4$ with the aforementioned index relabeling,
\small\begin{equation*}
\sigma_1 \left( \{\lambda\}_4\right)=   \{\field{L}_4,4^{j_1-1},3^{j_2-j_1-1},2^{j_3-j_2-1}, 1^{j_{5}-j_{3}-1} \} 
\end{equation*}\normalsize
Similarly, $\sigma_2 \left( \{\lambda\}_4\right)$ is obtained from the summation $ \sum_{1 \le j_1 < j_2 < j_5 < j_3 < j_{4} \le N}$. Performing the relabelings (in order) $j_5 \leftrightarrow j_3$, $j_5 \leftrightarrow j_4$, to the indices of this summation expresses it in the required form. Thus the partition $\sigma_2 \left( \{\lambda\}_4\right)$ is explicitly,
\small\begin{equation*}
\sigma_2 \left( \{\lambda\}_4\right)=    \{\field{L}_4,4^{j_1-1},3^{j_2-j_1-1},2^{j_4-j_2-1}, 1^{j_{5}-j_{4}-1} \}
\end{equation*}\normalsize
$\sigma_3 \left( \{\lambda\}_4\right)$ and $\sigma_4 \left( \{\lambda\}_4\right)$ are obviously obtained in an equivalent manner.\\
\\
For the case with general $k$, labelling $\field{L}_k = \{l_{k}+1,l_{k-1}+2,\dots, l_{1}+k \}$ we have explicitly,
\small\begin{equation}\begin{array}{lcl}
\sigma_0 \left( \{\lambda\}_k\right)& =& \{\field{L}_k,k^{j_1-1},\dots, 1^{j_{k}-j_{k-1}-1} \} \\
\sigma_1 \left( \{\lambda\}_k\right) &=&  \{\field{L}_k,k^{j_1-1},\dots, 1^{j_{k+1}-j_{k-1}-1} \}\\
\sigma_2 \left( \{\lambda\}_k\right) &=&  \{\field{L}_k,k^{j_1-1},\dots, 2^{j_{k}-j_{k-2}-1},1^{j_{k+1}-j_{k}-1} \}\\
&\vdots&\\
\sigma_p \left(\{\lambda\}_k \right) &=&\{\field{L}_k,k^{j_1-1},\dots,p^{j_{k-(p-2)}-j_{k-p}-1},\dots ,1^{j_{k+1}-j_{k}-1} \}\\
&\vdots&\\
\sigma_k \left(\{\lambda\}_k \right) &=&\{\field{L}_k,k^{j_2-1},(k-1)^{j_3-j_2-1},\dots,1^{j_{k+1}-j_{k}-1} \}
\end{array}\label{lotsofcrap}\end{equation}\normalsize
Our next step is to simplify the following bilinear sum of coefficients,
\small\begin{equation}
\sum^k_{p=0}(-1)^{p} \tilde{c}^{(N)}_{\sigma_p \left( \{\lambda\}_k \right)}  \tilde{c}^{(N)}_{\{l_{k+1}+1, 1^{j_{k+1-p}-1} \}}
\end{equation}\normalsize
using appropriate Pl\"{u}cker relations.\\
\\
\textbf{Generation of the necessary Pl\"{u}cker relations.} We again consider the bilinear sum of $N \times N$ determinants given in eq. \ref{p.21}. This time however we input the following (more general) values for the $2N$ indices $(\mu_1, \dots, \mu_{N-1}, \nu_1, \dots \nu_{N+1})$,
\small\begin{equation*}
\begin{array}{lcl}
(\mu_1, \dots, \mu_{N-j_{k+1}}) &=& (0, \dots, N-j_{k+1}-1)\\
(\mu_{N-j_{k+1}+1},\dots, \mu_{N-j_{k}-1}) &=& (N-j_{k+1} +1 , \dots , N-j_{k} -1)\\
(\mu_{N-j_{k}},\dots, \mu_{N-j_{k-1}-2}) &=& (N-j_{k} +1 , \dots , N-j_{k-1} -1)\\
(\mu_{N-j_{k-1}-1},\dots, \mu_{N-j_{k-2}-3}) &=& (N-j_{k-1} +1 , \dots , N-j_{k-2} -1)\\
& \vdots& \\
(\mu_{N-j_{1}-(k-1)},\dots, \mu_{N-(k+1)}) &=& (N-j_{1} +1 , \dots , N-1)\\
(\mu_{N-k}, \dots,  \mu_{N-1}, \nu_1) &=& (N+l_1,\dots, N+l_{k},N+l_{k+1})\\
(\nu_2, \dots, \nu_{N+1}) &=& (0, \dots, N-1)
\end{array}\end{equation*}\normalsize
Doing so, eq. \ref{p.21} becomes,
\small\begin{equation}\begin{split}
\left|\gamma_{0} , \dots , \gamma_{N-1} \right|  \left| \Gamma^{(-)}_{j}  ,\Gamma^{(+)}_l, \gamma_{N+l_{k+1}} \right|  \\
+(-1)^{N-j_{k+1}+1}  \left| \gamma_{N+l_{k+1}} , \gamma_{0} , \dots , \hat{\gamma}_{N-j_{k+1}} , \dots , \gamma_{N-1}  \right| \left|\Gamma^{(-)}_{j},  \Gamma^{(+)}_l  ,\gamma_{N-j_{k+1}} \right| \\
+(-1)^{N-j_k+1} \left| \gamma_{N+l_{k+1}} , \gamma_{0} , \dots , \hat{\gamma}_{N-j_k} , \dots , \gamma_{N-1}  \right| \left|\Gamma^{(-)}_{j}, \Gamma^{(+)}_l ,\gamma_{N-j_k} \right| \\
+ \dots + (-1)^{N-j_1+1} \left| \gamma_{N+l_{k+1}} , \gamma_{0} , \dots , \hat{\gamma}_{N-j_1} , \dots , \gamma_{N-1}  \right| \left|\Gamma^{(-)}_{j}, \Gamma^{(+)}_l  ,\gamma_{N-j_1} \right| &=0  \label{6VA}
\end{split}\end{equation}\normalsize
where we have used the following labels,
\small\begin{equation}\begin{split}
\Gamma^{(-)}_{j} =& \{ \gamma_{0} ,\dots ,\gamma_{N-1-j_{k+1}}, \hat{\gamma}_{N-j_{k+1}},\gamma_{N+1-j_{k+1}} , \dots \\
& \dots,\gamma_{N-1-j_{k}}, \hat{\gamma}_{N-j_{k}},\gamma_{N+1-j_{k}} , \dots ,\gamma_{N-1-j_1}, \hat{\gamma}_{N-j_1} ,\gamma_{N+1-j_1}, \dots , \gamma_{N-1} \}\\
\Gamma^{(+)}_l  =& \{ \gamma_{N+l_1} , \gamma_{N+l_2}, \dots, \gamma_{N+l_{k}} \}
\end{split}\end{equation}\normalsize
As with the $k=2$ case, ordering the columns in eq. \ref{6VA} so that the indices of a higher integer are placed to the right, we obtain the required Pl\"{u}cker relations,
\small\begin{equation*}
 \sum^k_{p=0}(-1)^{p} c^{(N)}_{\sigma_p \left( \{\lambda\}_k \right)}  c^{(N)}_{\{l_{k+1}+1, 1^{j_{k+1-p}-1} \}} =  c^{(N)}_{ \{\phi\}} c^{(N)}_{ \{\lambda\}_{k+1} }
\end{equation*}\normalsize
where $ \{\lambda\}_{k+1}$ is the partition consisting of $k+1$ hooks given explicitly as,
\small\begin{equation*}
 \{ \lambda\}_{k+1} =\{l_{k+1}+1,l_{k}+2,\dots, l_{1}+(k+1),(k+1)^{j_1-1},k^{j_2-j_1-1},\dots, 1^{j_{k+1}-j_{k}-1} \}
\end{equation*}\normalsize
Thus we obtain,
\small\begin{equation*}\begin{split}
X^{(N)}_{l_1} \dots X^{(N)}_{l_{k+1}} =&  \sum_{1 \le j_1 < \dots < j_{k+1} \le N}(-1)^{j_1+\dots + j_{k+1}} \\
& \times \tilde{c}^{(N)}_{\{ \lambda\}_{k+1}}   \psi^*_{-j_{k+1}} \dots \psi^*_{-j_1}\psi_{l_1}\dots  \psi_{l_{k+1}} 
\end{split}\end{equation*}\normalsize
which completes the proof of the proposition. $\square$\\
\\
\textbf{Proving the lemma.} Now consider the multiplication of all the $N-1$ generators,
\small\begin{equation}\begin{split}
e^{ X^{(N)}_{0} } \dots e^{ X^{(N)}_{N-2}} &= \sum^{N-1}_{k=0} e_k \left(  X^{(N)}_{0}, \dots,  X^{(N)}_{N-2} \right)\\
&= 1 + \sum^{N-1}_{k=1} \sum_{0 \le l_1 < \dots < l_k \le N-2} X^{(N)}_{l_1} \dots  X^{(N)}_{l_k}
\label{p.25}
\end{split}\end{equation}\normalsize 
Applying eq. \ref{p.23} the above expression becomes,
\small\begin{equation}\begin{split}
e^{ X^{(N)}_{0} } \dots e^{ X^{(N)}_{N-2}} =& 1 + \sum^{N-1}_{k=1} \sum_{0 \le l_1 < \dots < l_k \le N-2}\sum_{1 \le j_1 < \dots < j_{k} \le N}(-1)^{j_1+\dots + j_{k}}\\ 
& \times \tilde{c}^{(N)}_{\{ \lambda\}_{k}}   \psi^*_{-j_{k}} \dots \psi^*_{-j_1}\psi_{l_1}\dots  \psi_{l_{k}} 
\label{p.26}
\end{split}\end{equation}\normalsize\\
\textbf{Bosonization.} Referring to fig. \ref{p.a}, we can see that every value of $k$ in eq. \ref{p.26} generates every possible fermionic expression that corresponds to a partition consisting of $k$ hooks, contained within the partition $\{ (N-1)^{N} \}$. Additionally, eq. \ref{p.23} shows that the fermionic expressions are accompanied by the required coefficient (and sign). Thus, performing the inner product, 
\small\begin{equation*}\begin{split}
& \langle 0|  \exp\{H_+(\vec{x})\} e^{ X^{(N)}_0 } \dots e^{ X^{(N)}_{N-2}}  | 0 \rangle c^{(N)}_{\{ \phi \}}(\vec{v}) \Upsilon_N \\
=& \Upsilon_N\left(  c^{(N)}_{\{ \phi \}}(\vec{v}) \chi_{\{ \phi \}} (\vec{x}) +  \sum^{N-1}_{k=1} \sum_{0 \le l_1 < \dots < l_k \le N-2}\sum_{1 \le j_1 < \dots < j_k \le N} c^{(N)}_{\{ \lambda\}_k}(\vec{v}) \chi_{\{ \lambda\}_k} (\vec{x}) \right)\\
=& \Upsilon_N \sum_{\{\lambda\} \subseteq (N-1)^N} c^{(N)}_{\{\lambda\}}(\vec{v})\chi_{\{\lambda\}}(\vec{x})
\end{split}\end{equation*}\normalsize
and restricting the time variables in the usual way,
\small\begin{equation*}
x_j \rightarrow \frac{1}{j}p_j(u_1,\dots,u_N) \textrm{  ,  } j \in \{ 1, 2, \dots \}
\end{equation*}\normalsize
we obtain,
\small\begin{equation*}\begin{split}
\langle 0| \exp\left\{H_+\left( \left\{ \frac{1}{j}p_j(\vec{u}) \right\} \right)\right\}e^{ X^{(N)}_0 } \dots e^{ X^{(N)}_{N-2}}  |0\rangle c^{(N)}_{\{ \phi \}}(\vec{v})  \Upsilon_N\\
= \Upsilon_N \sum_{\{ \lambda \} \subseteq (N-1)^N} c^{(N)}_{\{ \lambda \}} (\vec{v}) S_{\{ \lambda \}}(\vec{u})\\
= Z^L_N(\vec{u},\vec{v})
\end{split}\end{equation*}\normalsize
which proves the lemma of this section. $\square$
\section{Scalar product of the six vertex model}
Having finished fermionizing the DWPF, we now consider the next fundamental quantity of the six vertex model, the scalar product. In order to proceed however we need to introduce the algebraic Bethe ansatz (ABA). The ABA admits a more formal construction\footnote{As opposed to Korepin's four properties for the DWPF.} of fundamental quantities of the six vertex model than has previously been considered. For more details regarding the methods and results of the ABA, refer to \cite{Lyon5,Lyon6,Lyon2,Lyon1,Lyon3,Lyon4,purplebook} and the further references contained therein.
\subsection{Algebraic Bethe ansatz}
\noindent \textbf{XXZ Hamiltonian.} To begin, we consider the $M$ identical vector spaces, $\field{V}_i \cong \field{C}^2$, $i \in \{1,\dots,M\}$, and their tensor product, $V_1 \otimes V_2 \otimes \dots \otimes V_M$. We define the Hamiltonian of the zero field $XXZ$ spin-$\frac{1}{2}$ chain with $M$ sites as the following,
\small\begin{equation}
H = \sum^M_{j=1} \left\{ \sigma^x_j \sigma^x_{j+1} + \sigma^y_j \sigma^y_{j+1} + \Delta \left( \sigma^z_j \sigma^z_{j+1} -1 \right)\right\}
\label{Hammm}\end{equation}\normalsize
where $\sigma^{x,y,z}_j \in \textrm{End}(V_j)$ are the usual spin-$\frac{1}{2}$ Pauli matrices, $\sigma^{x,y,z}_1 = \sigma^{x,y,z}_{M+1}$ and $-1 < \Delta \le 1 $. The (systematic) process of finding the eigenvalues and eigenvectors of the above Hamiltonian is achieved through the algebraic Bethe ansatz which we now introduce.\\
\\
\textbf{Algebraic Bethe ansatz.} The most fundamental object in the algebraic Bethe ansatz construction of the six vertex model, the $R$-matrix, is given as,
\small\begin{equation*}
 R_{ab}(s,t) = \left( \begin{array}{cccc}
[s-t+1] & 0 & 0 & 0\\
0 & [s-t] & [1] & 0\\
0 & [1] & [s-t] & 0\\
0 & 0 & 0 & [s-t+1]
\end{array}\right)_{ab}
\end{equation*}\normalsize
where $[s] = \sinh (\lambda s)$ and $s,t,\lambda \in \field{C}$. The subscripts, $a,b \in \{1,\dots,M\}$, referred to as \textit{quantum indices}, denote that the corresponding $R$-matrix acts in the tensor product $V_a \otimes V _b$, that is, $R_{ab}(s,t) \in \textrm{End}(V_a \otimes V _b)$. \\
\\
\textbf{A note on constructing the $R$-matrices.} For $b = a+1$ we have explicitly,
\small\begin{equation*}
R_{a,a+1}(s,t) = \underbrace{\field{I}_2 \otimes \dots \otimes  \field{I}_2}_{a-1} \otimes R(s,t) \otimes \underbrace{\field{I}_2 \otimes \dots \otimes  \field{I}_2}_{M-a-1}
\end{equation*}\normalsize
where $\field{I}_2$ is the $2 \times 2$ identity matrix. For $b > a+1$, we require the use of the permutation matrix, $\Pi_{ab}$,
\small\begin{equation*}
 \Pi_{ab} = \left( \begin{array}{cccc}
1 & 0 & 0 & 0\\
0 & 0 & 1 & 0\\
0 & 1& 0 & 0\\
0 & 0 & 0 & 1
\end{array}\right)_{ab}
\end{equation*}\normalsize
where the operation of $\Pi_{ab}$ on the tensor product $V_a \otimes V_b$ permutes the ordering of the quantum spaces,
\small\begin{equation*}
\Pi_{ab}\left\{ V_a \otimes V_b \right\} = V_b \otimes V_a
\end{equation*}\normalsize
Taking $M=4$ in the following example, we can construct $R_{13}(s,t)$ explicitly as $\Pi_{23}R_{12}(s,t) \Pi_{23}$. To see this, consider its action on $V_1 \otimes V_2 \otimes V_3 \otimes V_4$,
\small\begin{equation*}\begin{split}
R_{13}(s,t) \{ V_1 \otimes V_2 \otimes V_3 \otimes V_4\} &= \Pi_{23}R_{12}(s,t) \Pi_{23} \{  V_1 \otimes V_2 \otimes V_3 \otimes V_4\}\\
&=  \Pi_{23}R_{12}(s,t)  \{  V_1 \otimes V_3 \otimes V_2 \otimes V_4\}\\
&=  \Pi_{23}  \{  R(s,t) (V_1 \otimes V_3 ) \otimes V_2 \otimes V_4\}
\end{split}\end{equation*}\normalsize
which is the required expression. Obviously we can construct $R_{24}(s,t)$ similarly, i.e. $R_{24}(s,t) = \Pi_{34}R_{23}(s,t) \Pi_{34}$, and $R_{14}(s,t)$ can be constructed recursively from $R_{13}(s,t) $ or $R_{24}(s,t) $,
\small\begin{equation*}\begin{split}
R_{14}(s,t) &= \Pi_{34}R_{13}(s,t) \Pi_{34}\\
&=  \Pi_{12}R_{24}(s,t)  \Pi_{12}
\end{split}\end{equation*}\normalsize
Extending this procedure to general $M$ should be clear.\\
\\
\textbf{Intertwining relations.} The $R$-matrix satisfies the Yang-Baxter equation in the product of vector spaces $V_a \otimes V_b \otimes V_c$, $a,b,c \in \{1,\dots,M \}$,
\small\begin{equation}
R_{ab}(s_1, s_2) R_{a c}(s_1, s_3)R_{b c}(s_2 , s_3) = R_{b c}(s_2, s_3) R_{a c}(s_1, s_3)R_{a b}(s_1,  s_2)
\label{ybslav}\end{equation}\normalsize 
Defining the separate \textit{auxiliary vector spaces}, $V_{\alpha_i} \cong \field{C}^2$, $i \in \{1,2\}$, we now define the $L$-operator,
\small\begin{equation}
L_{\alpha a}(s,t) = R_{\alpha a}(s,t) \textrm{  ,  } a \in \{ 1, \dots, M \}
\label{lopera}\end{equation}\normalsize
where $L_{\alpha a} (s,t) \in \textrm{End}(V_{\alpha} \otimes V_a)$, is referred to as a \textit{local} operator.\\
\\
\textbf{Remark.} In the following we define vector spaces with greek indices as auxiliary, and those with latin as quantum.\\
\\
By virtue of the Yang-Baxter equation we have the following intertwining relation in $V_{\alpha_1}\otimes V_{\alpha_2} \otimes V_a$,
 \small\begin{equation}
R_{\alpha_1 \alpha_2}(s_1, s_2) L_{\alpha_1 a}(s_1, s_3)L_{\alpha_2 a}(s_2 , s_3) = L_{\alpha_2 a}(s_2 , s_3)  L_{\alpha_1 a}(s_1, s_3)R_{\alpha_1 \alpha_2}(s_1, s_2)
\label{intert}\end{equation}\normalsize 
Using the $L$-matrices we now define the \textit{global} monodromy matrix, $T_{\alpha}(s,\vec{t}) \in \textrm{End}(V_{\alpha} \otimes V_1 \otimes \dots \otimes V_M)$, as,
 \small\begin{equation}
T_{\alpha}(s,\vec{t}) =  \left(  \begin{array}{cc}
A(s,\vec{t}) & B(s,\vec{t})\\
C(s,\vec{t}) & D(s,\vec{t})
\end{array}\right)_{\alpha} = L_{\alpha 1}(s,t_1) \dots L_{\alpha M}(s,t_M)
\label{monodrom}\end{equation}\normalsize 
It is customary to suppress the quantum rapidities, $\{t_1, \dots, t_M\}$, in the expression of $T_{\alpha}(s)$, and in the expression of the operators $A(s),B(s),C(s),D(s) \in \textrm{End}(V_1 \otimes \dots \otimes V_M)$. Using the local intertwining relation, eq. \ref{intert}, it is possible to apply a simple inductive argument to obtain the following global intertwining relation,
 \small\begin{equation}
R_{\alpha_1 \alpha_2}(s_1, s_2) T_{\alpha_1 }(s_1)T_{\alpha_2 }(s_2 ) = T_{\alpha_2 }(s_2 )  T_{\alpha_1}(s_1)R_{\alpha_1 \alpha_2}(s_1, s_2)
\label{globintert}\end{equation}\normalsize 
As a simple but illustrative example, consider the $M=2$ case of the left hand side of the above equation (with suppressed rapidities). Realizing that $L$-operators with different indices commute, the proof is almost automatic,
 \small\begin{equation*}\begin{split}
R_{\alpha_1 \alpha_2}  T_{\alpha_1 }T_{\alpha_2 } &= R_{\alpha_1 \alpha_2}  L_{\alpha_1 1 } \underbrace{ L_{\alpha_1 2 }L_{\alpha_2 1 }}_{= L_{\alpha_2 1 }L_{\alpha_1 2 }} L_{\alpha_2 2 }\\
&=\underbrace{ R_{\alpha_1 \alpha_2}  L_{\alpha_1 1 }L_{\alpha_2 1 }}_{= L_{\alpha_2 1 }L_{\alpha_1 1 }R_{\alpha_1 \alpha_2} } L_{\alpha_1 2 } L_{\alpha_2 2 }\\
&= L_{\alpha_2 1 }L_{\alpha_1 1 }\underbrace{R_{\alpha_1 \alpha_2}  L_{\alpha_1 2 } L_{\alpha_2 2 }}_{L_{\alpha_2 2 }L_{\alpha_1 2 }R_{\alpha_1 \alpha_2}}\\
&=L_{\alpha_2 1 }\underbrace{L_{\alpha_1 1 }L_{\alpha_2 2 }}_{=L_{\alpha_2 2 }L_{\alpha_1 1 }}L_{\alpha_1 2 }R_{\alpha_1 \alpha_2}\\
&= T_{\alpha_2 } T_{\alpha_1 }R_{\alpha_1 \alpha_2} 
\end{split}\end{equation*}\normalsize 
The proof for general $M$ involves almost no more work.\\
\\
\textbf{Algebraic relations.} Expanding eq. \ref{globintert} in matrix form in the (auxiliary) space $V_{\alpha_1} \otimes V_{\alpha_2}$,
 \small\begin{equation}\begin{split}
&R(s_1,s_2)  \left(  \begin{array}{cccc}
A_1A_2 &A_1 B_2 & B_1A_2 & B_1B_2 \\
A_1C_2 & A_1D_2 & B_1C_2&B_1D_2\\
C_1A_2 &C_1 B_2 & D_1A_2 & D_1B_2 \\
C_1C_2 & C_1D_2 & D_1C_2&D_1D_2
\end{array}\right)\\
= &\left(  \begin{array}{cccc}
A_1A_2 &B_1 A_2 & A_1B_2 & B_1B_2 \\
C_1A_2 & D_1A_2 & C_1B_2&D_1B_2\\
A_1C_2 &B_1 C_2 & A_1D_2 & B_1D_2 \\
C_1C_2 & D_1C_2 & C_1D_2&D_1D_2
\end{array}\right)R(s_1,s_2)
\end{split}\label{algeq}\end{equation}\normalsize 
we obtain no less than sixteen algebraic relations between the operators $A,B,C$ and $D$.\\
\\
\textbf{Simultaneous eigenvectors.} We now consider the eigenvectors of the $XXZ$ Hamiltonian, labeled $|\Psi^M \rangle$, which are simultaneous eigenvectors of the trace of the monodromy matrix, $\textrm{tr}_{\alpha}\left\{ T_{\alpha}(s)\right\}  = A(s)+ D(s)$, due to the following commutation relation,
 \small\begin{equation}
[H, \textrm{tr}_{\alpha}\left\{ T_{\alpha}(s)\right\} ] = 0
\label{commmmmu}\end{equation}\normalsize 
Thus finding the sought after eigenvector, $|\Psi^M \rangle$, obviously hinges on our ability to solve the following eigenvalue equation,
 \small\begin{equation}
(A(s)+ D(s))|\Psi^M \rangle = \beta(s) |\Psi^M \rangle
\end{equation}\normalsize 
where the eigenvalue, $ \beta(s)$, is a general function involving $s$.\\
\\
\textbf{The ansatz.} The ansatz for the above eigenvalue equation is to set the eigenvector as,
 \small\begin{equation}
|\Psi^M \rangle =|\Psi^M_N(\vec{s}) \rangle = B(s_1) \dots B(s_N)|0 \rangle \textrm{  ,  } N \le M 
\label{ansatz}\end{equation}\normalsize
where,
 \small\begin{equation*}
|0 \rangle = \underbrace{\left( 1 \atop{0} \right) \otimes \dots \otimes \left( 1 \atop{0} \right) }_{M}
\end{equation*}\normalsize\\
\textbf{The Bethe equations.} Using the following formulas,
 \small\begin{equation*}\begin{split}
A(s)|0 \rangle = a(s)| 0 \rangle = \prod^M_{j=1}[s- t_j +1] | 0 \rangle\\
 D(s)|0 \rangle = d(s)| 0 \rangle = \prod^M_{j=1}[s- t_j ] | 0 \rangle
\end{split}\end{equation*}\normalsize
and the algebraic relations obtained from eq. \ref{algeq}, it is possible to commute $A(s)$ and $D(s)$ through the product of $B$ operators to obtain that $|\Psi^M_N(\vec{s}) \rangle$ is only an eigenvector if the rapidities, $\{s_1, \dots,s_N\}$, satisfy the system of transcendental \textit{Bethe equations},
 \small\begin{equation}
(-1)^{N-1} \frac{a(s_i)}{d(s_i)} \prod^N_{j=1\atop{\ne i}} \frac{[s_j-s_i+1]}{[s_i-s_j+1]} = 1 \textrm{  ,  } 1 \le i \le N
\label{BETHEEQ}\end{equation}\normalsize
Specifying $|\Psi^M_N(\vec{s})_{\beta} \rangle$ as the eigenvector whose rapidities satisfy the Bethe equations, we have the following eigenvalue equation,
\small\begin{equation*}
(A(s)+ D(s))|\Psi^M_N(\vec{s})_{\beta} \rangle = \left( a(s) \prod^N_{j=1}[s-t_j+1] + d(s) \prod^N_{j=1}[t_j-s+1]   \right) |\Psi^M_N(\vec{s})_{\beta} \rangle
\end{equation*}\normalsize
We are now ready to construct various fundamental objects of the six-vertex model using the algebraic Bethe ansatz notation.\\
\\
\textbf{A familiar example, the DWPF.} Defining the conjugate vector, $\langle 1|$, as,
 \small\begin{equation*}
\langle 1| = \underbrace{(0,1) \otimes \dots \otimes (0,1)}_{M} 
\end{equation*}\normalsize
and fixing $M=N$, the DWPF of the six vertex model, $Z_N(\vec{s},\vec{t})$, as defined in eq. \ref{p.1} is also given by the following expectation value expression,
\small\begin{equation*}
Z_N(\vec{s},\vec{t}) = \langle 1| \Psi^M_N(\vec{s}) \rangle =\langle 1|  B(s_1) \dots B(s_N)|0 \rangle
\end{equation*}\normalsize
where the rapidities $\{s_1,\dots, s_N\}$ are not required to satisfy the Bethe equations. Note that this expression does not give a systematic way of deriving the determinant solution for the partition function.\\
\\
\textbf{The scalar product.} We now define the conjugate eigenvector, $\langle \Psi^M_N(\vec{r})|$ as the following product of $C$ operators,
\small\begin{equation}
\langle \Psi^M_N(\vec{r})| = \langle 0|   C(r_1) \dots C(r_N)
\label{conjvec}\end{equation}\normalsize
where,
\small\begin{equation*}
\langle 0| = \underbrace{(1,0) \otimes \dots \otimes (1,0)}_{M}
\end{equation*}\normalsize
The scalar product, $\field{S}^M_N(\vec{r},\vec{s},\vec{t})$, is given as the expectation value of the general eigenvector and its conjugate,
 \small\begin{equation}\begin{split}
\field{S}^M_N(\vec{r},\vec{s},\vec{t}) &= \langle \Psi^M_N(\vec{r})| \Psi^M_N(\vec{s}) \rangle \\
 &=  \langle 0|   C(r_1) \dots C(r_N) B(s_1) \dots B(s_N)|0 \rangle
\label{nonslav}\end{split}\end{equation}\normalsize
Generally, such expressions are quite hard to calculate exactly as they involve sums of $\left( 2 N \atop{N} \right)$ terms\footnote{For exact details of this summation expression see eq. (IX.1.3) in \cite{purplebook}.}. In the following, due to Slavnov \cite{Slavnov} we give a determinant form for the scalar product when one set of rapidities satisfies the Bethe equations.
\subsection{Slavnov's determinant expression}
If the rapidities in the set, $\{s_1, \dots, s_N\}$, satisfy the system of Bethe equations given in eq. \ref{BETHEEQ}, then the expression for the scalar product (eq. \ref{nonslav}) simplifies to a manageable determinant form, given by,
 \small\begin{equation}\begin{split}
\field{S}^M_N(\vec{r},\vec{s}_{\beta},\vec{t}) &= \langle \Psi^M_N(\vec{r})| \Psi^M_N(\vec{s})_{\beta} \rangle \\
&= \frac{[\lambda]^N \prod^N_{i,j=1}[r_i-s_j +1] }{\prod_{1 \le i < j \le N} [r_i-r_j][s_j-s_i]}\left\{\prod^N_{k=1} \prod^M_{l=1} [r_k-t_l][s_k-t_l]  \right\} \textrm{det}\left( M_{ij}\right)^N_{ij=1}
\label{slavio}\end{split}\end{equation}\normalsize
where the entries of the determinant are given by,
 \small\begin{equation}\begin{split}
M_{ij} =& \frac{1}{[r_i-s_j][r_i-s_j+1]} \\
&- \frac{(-1)^N}{[s_j-r_i][s_j-r_i+1]}\left\{ \prod^M_{k=1}\frac{[r_i-t_k+1]}{[r_i-t_k]} \prod^N_{l=1}\frac{[s_l-t_i+1]}{[r_i-s_l+1]} \right\}
\label{slaviodet}\end{split}\end{equation}\normalsize
Setting the variables as follows,
 \small\begin{equation*}
u_i = e^{2 \lambda r_i} \textrm{  ,  } v_i = e^{2 \lambda s_i} \textrm{  ,  } w_i = e^{2 \lambda t_i} \textrm{  ,  } q = e^{2 \lambda} 
\end{equation*}\normalsize
and absorbing the numerator of eq. \ref{slavio} into the determinant, we obtain the following, more useful form for the Slavnov scalar product,
\small\begin{equation}
\field{S}^M_N(\vec{u},\vec{v}_{\beta},\vec{w})=\frac{\Upsilon^M_N }{\prod_{1\le i < j \le N}(u_i-u_j)}\textrm{det}\left[ \field{M}_{ij}(u_i,\vec{v},\vec{w}) \right]^N_{ij=1}
\label{slav.1}
\end{equation}\normalsize
where the multiplicative factor is given by,
\small\begin{equation}
\Upsilon^M_N =\frac{(-1)^{N^2} q^{-N\left(\frac{M}{2}+N-1\right)}}{ \left\{ \prod^N_{i=1}u_i v_i \right\}^{\frac{M-1}{2}}\left\{ \prod^M_{i=1}w_i \right\}^{N}} \frac{(q^{\frac{1}{2}} - q^{-\frac{1}{2}})^N}{\prod^N_{i,j=1\atop{i \ne j}} (q^{\frac{1}{2}} v_i - q^{-\frac{1}{2}}v_j)}\frac{1}{\prod_{1\le i< j \le N}(v_j-v_i)}
\label{multfac}\end{equation}\normalsize
and the entries of the determinant are,
\small\begin{equation}\begin{split}
\field{M}_{ij} = \frac{1}{-u_i+v_j} \left\{ \prod^M_{k=1}(u_i-w_k)\prod^N_{k=1\atop{\ne j}} (q u_i -v_k)\prod^M_{k=1}(q v_j-w_k)\prod^N_{k=1\atop{\ne j}} (q v_k -v_j)\right.  \\
- \left. \prod^M_{k=1}(q u_i-w_k)\prod^N_{k=1\atop{\ne j}} (q v_k -u_i)\prod^M_{k=1}( v_j-w_k)\prod^N_{k=1\atop{\ne j}} (q v_j -v_k) \right\} \label{slav.2}
\end{split}\end{equation}\normalsize
We now proceed to show that the scalar product, normalized appropriately, is a $\tau$-function of the KP hierarchy with restricted time variables as power sums in the rapidities $\{u_1,\dots,u_N\}$.
\subsection{Schur polynomial expansion of the scalar product}
\begin{lemma}
Using a method detailed explicitly below, an equivalent form to Slavnov's expression for the scalar product is given by,
\small\begin{equation}
\field{S}^M_N(\vec{u},\vec{v}_{\beta},\vec{w})=\Upsilon^{'M}_N  \textrm{det}\left[ \left(h_{k-i} (\vec{u}) \right)^{N+M-1}_{i,k=1} \left( \rho_{k,j}(\vec{v},\vec{w}) \right)^{N+M-1}_{k,j=1} \right]^N_{i,j=1}
\label{slav.3}
\end{equation}\normalsize
where $\Upsilon^{'M}_N = (-1)^{\frac{N(N-1)}{2}} \Upsilon^{M}_N$, and, 
\small\begin{equation}\begin{split}
\rho_{k,j} = v^{-k}_{j} \sum^{N+M-1}_{\xi=k} \sum^{k-1}_{\eta=0} \sum^{\textrm{min}\{M,\xi \} }_{\alpha= \textrm{max}\{0,\xi-N+1 \}} \sum^{\textrm{min}\{M,\eta \}}_{\zeta=0} (-1)^{N+\xi+\eta} q^{N-1}  \\
\times\left\{ q^{\xi-\eta+2 \zeta-\alpha} - q^{\eta-\xi+2 \alpha- \zeta} \right\} e_{M-\alpha}(\vec{w})e_{M-\zeta}(\vec{w}) e_{N-1-\eta+\zeta}(\vec{v},\hat{v}_j)e_{N-1-\xi+\alpha}(\vec{v},\hat{v}_j) v^{\xi+\eta}_j \label{slav.4}
\end{split}\end{equation}\normalsize
\end{lemma}
\textbf{Proof.} We begin by expanding the entries of the determinant, $\field{M}_{ij}(u_i,\vec{v},\vec{w})$, as (symmetric) polynomials in $\vec{v}$ and $\vec{w}$. In the workings below we label $e_n(\hat{v}_j)=e_n(\vec{v},\hat{v}_j)$ for notational convenience.
\scriptsize\begin{equation*}\begin{array}{ll}
\displaystyle \prod^M_{k=1}(u_i-w_k) = \sum^M_{n=0}(-1)^n u^{M-n}_i e_n(\vec{w}) &\displaystyle \prod^N_{k=1\atop{\ne j}} (q u_i -v_k) = \sum^{N-1}_{n=0}(-1)^n (qu_i)^{N-1-n} e_n(\hat{v}_j)\\
\displaystyle\prod^M_{k=1}(q v_j-w_k) = \sum^M_{n=0}(-1)^n (qv_j)^{M-n} e_n(\vec{w}) & \displaystyle\prod^N_{k=1\atop{\ne j}} (q v_k -v_j) = \sum^{N-1}_{n=0} q^n(-v_j)^{N-1-n} e_n(\hat{v}_j)\\
\displaystyle \prod^M_{k=1}(q u_i-w_k) = \sum^M_{n=0}(-1)^n (qu_i)^{M-n}e_n(\vec{w})&\displaystyle  \prod^N_{k=1\atop{\ne j}} (q v_k -u_i) = \sum^{N-1}_{n=0} q^n(-u_i)^{N-1-n} e_n(\hat{v}_j)\\
\displaystyle\prod^M_{k=1}( v_j-w_k) = \sum^M_{n=0}(-1)^n (v_j)^{M-n} e_n(\vec{w})&\displaystyle \prod^N_{k=1\atop{\ne j}} (q v_j -v_k) =\sum^{N-1}_{n=0} (-1)^n(q v_j)^{N-1-n} e_n(\hat{v}_j)
\end{array}
\end{equation*}\normalsize
Using the above polynomial expansions, $(-u_i+v_j)\field{M}_{ij}(u_i,\vec{v},\vec{w})$ becomes,
\small\begin{equation*}\begin{split}
\sum^M_{m_1,m_2=0} \sum^{N-1}_{n_1,n_2=0} (-1)^{N-1+m_1+m_2+n_1+n_2} q^{M+N-1-m_2+n_2-n_1}e_{m_1}(\vec{w})e_{m_2}(\vec{w}) e_{n_1}(\hat{v}_j)\\
\times e_{n_2}(\hat{v}_j) \left( u^{M+N-1-m_1-n_1}_i v^{M+N-1-m_2-n_2}_j-u^{M+N-1-m_2-n_2}_i v^{M+N-1-m_1-n_1}_j \right)
\end{split}\end{equation*}\normalsize
Performing the following change of indices,
\small\begin{equation*}\begin{array}{ll}
m_1 \rightarrow M - \alpha & m_2 \rightarrow M - \zeta\\
n_1 \rightarrow N-1 - \beta & n_2 \rightarrow N-1 - \delta
\end{array}
\end{equation*}\normalsize
we obtain,
\small\begin{equation*}\begin{split}
\field{M}_{ij}=& \frac{1}{-u_i+v_j}\sum^M_{\alpha,\zeta=0} \sum^{N-1}_{\beta,\delta=0} (-1)^{N-1+\alpha+\beta+\zeta+\delta} q^{N-1+\zeta-\delta+\beta} \\
&\times e_{M-\alpha}(\vec{w}) e_{M-\zeta}(\vec{w}) e_{N-1-\beta}(\hat{v}_j) e_{N-1-\delta}(\hat{v}_j) \left( u^{\alpha+\beta}_i v^{\zeta+\delta}_j-u^{\zeta+\delta}_i v^{\alpha+\beta}_j \right)
\end{split}\end{equation*}\normalsize
Making the additional change in indices, $\alpha+\beta=\xi$ and $\zeta+\delta=\eta$ for obvious convenience we obtain,
\small\begin{equation*}\begin{split}
\field{M}_{ij}=&\sum^M_{\alpha,\zeta=0} \sum^{N-1+\alpha}_{\xi= \alpha} \sum^{N-1+\zeta}_{\eta= \zeta}  q^{N-1-\eta+\xi-\alpha+2\zeta} \\
&\times \underbrace{(-1)^{N+\xi+\eta} e_{M-\alpha}(\vec{w}) e_{M-\zeta}(\vec{w}) e_{N-1-\xi+\alpha}(\hat{v}_j) e_{N-1-\eta+\zeta}(\hat{v}_j)}_{E_{\alpha,\zeta,\xi,\eta}(\vec{v},\hat{v}_j,\vec{w})} \frac{ \left( u^{\xi}_i v^{\eta}_j-u^{\eta}_i v^{\xi}_j \right)}{u_i-v_j}
\end{split}\end{equation*}\normalsize
To deal with the denominator we consider the 2 cases, $\xi > \eta$ and $\xi < \eta$, the case $\xi=\eta$ is trivially zero.\\
\\
Hence for $\xi > \eta$,
\small\begin{equation}\begin{split}
&\sum^M_{\alpha,\zeta=0} \sum^{N-1+\alpha}_{\xi= \alpha \atop{\xi > \eta}} \sum^{N-1+\zeta}_{\eta= \zeta} q^{N-1-\eta+\xi-\alpha+2\zeta} E_{\alpha,\zeta,\xi,\eta}(\vec{v},\hat{v}_j,\vec{w}) u^{\eta}_i v^{\eta}_j \frac{ \left( u^{\xi-\eta}_i - v^{\xi-\eta}_j \right)}{u_i-v_j}\\
=& \sum^M_{\alpha,\zeta=0} \sum^{N-1+\alpha}_{\xi= \alpha } \sum^{\xi-1}_{\eta= \zeta} q^{N-1-\eta+\xi-\alpha+2\zeta} E_{\alpha,\zeta,\xi,\eta}(\vec{v},\hat{v}_j,\vec{w}) \left( \sum^{\xi-\eta-1}_{\nu=0 } u^{\eta+\nu}_i v^{\xi-1-\nu}_j \right)\label{slav.5}
\end{split}\end{equation}\normalsize
and similarly for $\xi < \eta$,
\small\begin{equation}\begin{split}
&-\sum^M_{\alpha,\zeta=0} \sum^{N-1+\alpha}_{\xi= \alpha } \sum^{N-1+\zeta}_{\eta= \zeta \atop{\eta > \xi}} q^{N-1-\eta+\xi-\alpha+2\zeta} E_{\alpha,\zeta,\xi,\eta}(\vec{v},\hat{v}_j,\vec{w}) u^{\xi}_i v^{\xi}_j \frac{ \left( u^{\eta-\xi}_i - v^{\eta-\xi}_j \right)}{u_i-v_j}\\
=&- \sum^M_{\alpha,\zeta=0} \sum^{\eta-1}_{\xi= \alpha } \sum^{N-1+\zeta}_{\eta= \zeta} q^{N-1-\eta+\xi-\alpha+2\zeta} E_{\alpha,\zeta,\xi,\eta}(\vec{v},\hat{v}_j,\vec{w}) \left( \sum^{\eta-\xi-1}_{\nu =0 } u^{\xi+\nu}_i v^{\eta-1-\nu}_j \right) \label{slav.6}
\end{split}\end{equation}\normalsize
Exchanging the index labelling $\alpha \leftrightarrow \zeta$ and $\xi \leftrightarrow \eta$ in eq. \ref{slav.6} and adding this with eq. \ref{slav.5}, the matrix entry $\field{M}_{ij}(u_i,\vec{v},\vec{w})$ becomes the following,
\small\begin{equation}
 \underbrace{\sum^M_{\alpha,\zeta=0} \sum^{N-1+\alpha}_{\xi= \alpha } \sum^{\xi-1}_{\eta= \zeta} \sum^{\xi-\eta}_{\nu=1 }}_{\sum^{(1)}_{\alpha,\zeta,\xi,\eta,\nu}} \underbrace{q^{N-1}\left\{ q^{\xi-\eta+2\zeta-\alpha} -q^{\eta-\xi+2\alpha-\zeta} \right\} E_{\alpha,\zeta,\xi,\eta}(\vec{v},\hat{v}_j,\vec{w}) v^{\xi-\nu}_j}_{\field{E}^{(j)}_{\alpha,\zeta,\xi,\eta,\nu}(\vec{v},\vec{w})}  u^{\eta+\nu-1}_i  
\label{slav.7}
\end{equation}\normalsize\\
\textbf{Eliminating the Vandermonde in $\{ u\}$.} We are now in a position to eliminate the removable poles (Vandermonde) in the $u$'s. To complete this task we employ eq. \ref{completeid} and perform the same row operations that eliminated the equivalent poles in the derivation of Lascoux's result, i.e.
\small\begin{equation*}\begin{array}{lcl}
R_i \rightarrow R_i - R_{i+1} &,& i=1,2, \dots, N-1\\
R_i \rightarrow R_i - R_{i+2} &,& i=1,2, \dots, N-2\\
 &\vdots & \\
R_i \rightarrow R_i - R_{i+N-2} &,& i=1,2\\
R_1 \rightarrow R_1 - R_{N}
\end{array}\end{equation*}\normalsize
Hence the scalar product (eq. \ref{slav.1}) becomes,
\small\begin{equation}
\Upsilon^M_N\textrm{det} \left[ \begin{array}{c}
\sum^{(1)}_{\alpha,\zeta,\xi,\eta,\nu} \field{E}^{(j)}_{\alpha,\zeta,\xi,\eta,\nu}(\vec{v},\vec{w})  h_{\eta+\nu -N}(u_1,\dots,u_N) \\
\sum^{(1)}_{\alpha,\zeta,\xi,\eta,\nu} \field{E}^{(j)}_{\alpha,\zeta,\xi,\eta,\nu}(\vec{v},\vec{w})  h_{\eta+\nu -(N-1)} (u_2,\dots,u_N)\\
\vdots\\
\sum^{(1)}_{\alpha,\zeta,\xi,\eta,\nu} \field{E}^{(j)}_{\alpha,\zeta,\xi,\eta,\nu}(\vec{v},\vec{w})  h_{\eta+\nu -1} (u_N) \end{array}
  \right]_{j=1,\dots,N} \label{slav.8}
\end{equation}\normalsize\\
\textbf{Clearing up the homogenous symmetric polynomials in $\{u\}$.} Additionally, we wish to make all of the homogeneous symmetric polynomials functions of all the $u$ variables. To achieve this we employ eq. \ref{completeid2} and again perform the same row operations that cleared up the homogeneous symmetric polynomials in the derivation of Lascoux's result,
\small\begin{equation*}\begin{array}{lcl}
R_i \rightarrow R_i +h_1(u_{i-1}) R_{i-1} &,& i=N,N-1,\dots, 2\\
R_i \rightarrow R_i +h_1(u_{i-2}) R_{i-1} &,& i=N,N-1,\dots, 3\\
 &\vdots & \\
R_i \rightarrow R_i +h_1(u_{i-(N-2)}) R_{i-1} &,& i=N,N-1\\
R_1 \rightarrow R_N +h_1(u_1) R_{N-1}
\end{array}\end{equation*}\normalsize
Performing these row operations eq. \ref{slav.8} becomes,
\small\begin{equation}\begin{split}
\field{S}^M_N(\vec{u},\vec{v},\vec{w}) &= \Upsilon^M_N \textrm{det} \left[ \sum^{(1)}_{\alpha,\zeta,\xi,\eta,\nu} \field{E}^{(j)}_{\alpha,\zeta,\xi,\eta,\nu}(\vec{v},\vec{w})  h_{\eta+\nu -N -1+i}(\vec{u})   \right]^N_{i,j=1}\\
&= \underbrace{(-1)^{\frac{1}{2}N(N-1)} \Upsilon^M_N}_{\Upsilon^{'M}_N} \textrm{det} \left[ \sum^{(1)}_{\alpha,\zeta,\xi,\eta,\nu} \field{E}^{(j)}_{\alpha,\zeta,\xi,\eta,\nu}(\vec{v},\vec{w})  h_{\eta+\nu -i}(\vec{u})   \right]^N_{i,j=1} \label{slav.9}
\end{split}\end{equation}\normalsize
where we have exchanged rows $i $ and $N -i +1$, $1 \le i \le N$, to obtain the second line from the first. \\
\\
Performing the convenient change of index, $k = \eta+\nu$, we obtain the following expression for the scalar product, $\field{S}^M_N(\vec{u},\vec{v}_{\beta},\vec{w})$,
\small\begin{equation}
 \Upsilon^{'M}_N \textrm{det} \left[ \begin{array}{c}
\sum^M_{\alpha,\zeta=0} \sum^{N-1+\alpha}_{\xi= \alpha } \sum^{\xi-1}_{\eta= \zeta} \sum^{\xi}_{k=1+\eta } q^{N-1}\left\{ q^{\xi-\eta+2\zeta-\alpha} -q^{\eta-\xi+2\alpha-\zeta} \right\} \\
\times E_{\alpha,\zeta,\xi,\eta}(\vec{v},\hat{v}_j,\vec{w}) v^{\xi+\eta-k}_j h_{k-i}(\vec{u}) \end{array}
  \right]^N_{i,j=1} \label{slav.10}
\end{equation}\normalsize
In order to complete the lemma we need one last result given by the following proposition.
\begin{proposition}
\small\begin{equation*}
\sum^M_{\alpha,\zeta=0} \sum^{N-1+\alpha}_{\xi= \alpha } \sum^{\xi-1}_{\eta= \zeta} \sum^{\xi}_{k=1+\eta } = \sum^{N+M-1}_{k=1} \sum^{N+M-1}_{\xi=k} \sum^{k-1}_{\eta=0} \sum^{\textrm{min}\{M,\xi \} }_{\alpha= \textrm{max}\{0,\xi-N+1 \}} \sum^{\textrm{min}\{M,\eta \}}_{\zeta=0}
\end{equation*}\normalsize
\end{proposition}
\textbf{Proof.} 
We begin verifying this result by making $k$ (instead of $\alpha$ and $\zeta$) an independent variable. Since the largest value of $\xi$ is $N+M-1$ and the lowest value of $\eta$ is $0$, we immediately see that the allowable values of $k$ as an independent variable are $1 \le k \le N+M-1$. Additionally, analyzing the final summation on the left hand side, $\sum^{\xi}_{k=1+\eta}$, we can obtain the allowable values of $\xi$ and $\eta$ for each value of $k$,
\small\begin{equation*}\begin{array}{lll}
&&k = \xi, k = \xi-1, k = \xi -2, \dots \\
&\Rightarrow & \xi = k, \xi = k+1, \dots, \xi = N+M-1\\
&& k = \eta+1, k = \eta+2, k = \eta +3, \dots \\
&\Rightarrow & \eta = k-1, \eta = k-2, \dots, \eta = 0\\
&\Rightarrow &\sum^M_{\alpha,\zeta=0} \sum^{N-1+\alpha}_{\xi= \alpha } \sum^{\xi-1}_{\eta= \zeta} \sum^{\xi}_{k=1+\eta } = \sum^{N+M-1}_{k=1}\sum^{N+M-1}_{\xi=k} \sum^{k-1}_{\eta=0} \sum^{(*)}_{\alpha,\zeta} 
\end{array}\end{equation*}\normalsize
To discern the allowable values of $\alpha$ and $\zeta$ we proceed in the same manner. Analyzing $\sum^{N-1+\alpha}_{\xi= \alpha }$ and $ \sum^{\xi-1}_{\eta= \zeta}$ respectively,
\small\begin{equation*}\begin{array}{lll}
&&\xi = \alpha, \xi = \alpha+1, \dots, \xi = N-1+\alpha\\
&\Rightarrow& \alpha= \xi, \alpha = \xi -1, \dots, \alpha = \xi -N+1\\
&&\eta = \zeta, \eta = \zeta+1, \dots \\
&\Rightarrow& \zeta= \eta, \zeta = \eta -1, \dots, \zeta = 0
\end{array}\end{equation*}\normalsize
Taking into account that $0 \le \alpha,\zeta \le M$, we obtain the forms,
\small\begin{equation*}
\textrm{max}\{0,\xi-N+1 \}\le \alpha \le \textrm{min}\{M,\xi \} \textrm{  ,  } 0 \le \zeta \le \textrm{min}\{M,\eta \}
\end{equation*}\normalsize
which completes the proposition. $\square$\\
\\
Thus eq. \ref{slav.10} becomes,
\small\begin{equation}\begin{split}
& \Upsilon^{'M}_N\textrm{det} \left[ \begin{array}{c}
 \sum^{N+M-1}_{k=1}h_{k-i}(\vec{u}) \left( v^{-k}_j  \sum^{N+M-1}_{\xi=k} \sum^{k-1}_{\eta=0} \sum^{\textrm{min}\{M,\xi \} }_{\alpha= \textrm{max}\{0,\xi-N+1 \}}  \right. \\
\left.  \sum^{\textrm{min}\{M,\eta \}}_{\zeta=0} q^{N-1}\left\{ q^{\xi-\eta+2\zeta-\alpha} -q^{\eta-\xi+2\alpha-\zeta} \right\} E_{\alpha,\zeta,\xi,\eta}(\vec{v},\hat{v}_j,\vec{w}) v^{\xi+\eta}_j \right) \end{array}
  \right]^N_{i,j=1} \\
= & \Upsilon^{'M}_N\textrm{det} \left[  \sum^{N+M-1}_{k=1}h_{k-i}(\vec{u}) \rho_{k,j}(\vec{v},\vec{w})  \right]^N_{i,j=1} \\
=& \Upsilon^{'M}_N \textrm{det}\left[ \left(h_{k-i} (\vec{u}) \right)^{N+M-1}_{i,k=1} \left( \rho_{k,j}(\vec{v},\vec{w}) \right)^{N+M-1}_{k,j=1} \right]^N_{i,j=1}
\label{slav.11}
\end{split}\end{equation}\normalsize
where $\rho_{k,j}(\vec{v},\vec{w})$ is given in eq. \ref{slav.4}. $\square$\\
\\
Applying the Cauchy-Binet formula to expand the above expression in terms of Schur polynomials in $\vec{u}$ we obtain,
\small\begin{equation}\begin{split}
\field{S}^M_N(\vec{u},\vec{v}_{\beta},\vec{w}) &=\Upsilon^{'M}_N  \sum_{0 \le \lambda_1 \le  \dots \le \lambda_{N} \le M-1}\textrm{det} \left[ h_{\lambda_i +k -i}(\vec{u})\right]^N_{i,l=1} \\
& \times \textrm{det} \left[\rho_{\lambda_{N+1-k} +k,j}(\vec{v},\vec{w}) \right]^N_{k,j=1}\\
&=\Upsilon^{'M}_N  \sum_{\{\lambda\} \subseteq (M-1)^{N}} g^{(M,N)}_{\{\lambda\}}(\vec{v},\vec{w})S_{\{\lambda\}}(\vec{u}) \label{slav.12}
\end{split}\end{equation}\normalsize
where,
\small\begin{equation}
g^{(M,N)}_{\{\lambda\}} (\vec{v},\vec{w})= \textrm{det} \left[ \rho_{\lambda_{N+1-i} +i,j}(\vec{v},\vec{w}) \right]^N_{i,j=1} \label{slav.13}
\end{equation}\normalsize
It is this form of the scalar product that we shall fermionize.
\subsection{Fermionic form of the scalar product}
As the above expression for the scalar product is an equivalent expression to Lascoux's form for the DWPF, we have the following result.
\begin{lemma}
Eq. \ref{slav.12} is the bosonization of the following fermionic expression,
\small\begin{equation}
\exp\left\{ Y^{(M,N)}_{0} \right\}\exp\left\{ Y^{(M,N)}_{1} \right\} \dots \exp\left\{ Y^{(M,N)}_{M-2} \right\} |0\rangle g^{(M,N)}_{\{\phi\}} \Upsilon^{'M}_N
\label{slav.15}
\end{equation}\normalsize
where the $Y^{(M,N)}$'s are given as,
\small\begin{equation}\begin{array}{lcl}
Y^{(M,N)}_{0} &=& -\tilde{g}^{(M,N)}_{\{1 \}} \psi^*_{-1}\psi_0 + \tilde{g}^{(M,N)}_{\{1^2 \}} \psi^*_{-2}\psi_0 + \dots + (-1)^N\tilde{g}^{(M,N)}_{\{1^N \}} \psi^*_{-N}\psi_0  \\
Y^{(M,N)}_{1} &=& -\tilde{g}^{(M,N)}_{\{2 \}} \psi^*_{-1}\psi_1 + \tilde{g}^{(M,N)}_{\{2,1 \}} \psi^*_{-2}\psi_1 + \dots + (-1)^N\tilde{g}^{(M,N)}_{\{2,1^{N-1} \}} \psi^*_{-N}\psi_1 \\
& \vdots& \\
Y^{(M,N)}_{M-2} &=& \sum^{N}_{j=1}(-1)^{j} \tilde{g}^{(M,N)}_{\{M-1,1^{j-1} \}} \psi^*_{-j}\psi_{M-2} 
\label{slav.16}
\end{array}\end{equation}\normalsize
and the coefficients, $\tilde{g}^{(M,N)}_{\{\lambda \}} = \frac{g^{(M,N)}_{\{\lambda \}}}{g^{(M,N)}_{\{\phi \}}}$, are given by eq. \ref{slav.13}. 
\end{lemma}
\noindent \textbf{Proof.} It is obvious that this result is analogous, but more general, to the equivalent result for the partition function. It is not surprising then that the method of proof will also be analogous. Before we start with the details however, it would be wise to address the issues that make this result slightly different from the result regarding the partition function.
\begin{itemize}
\item{The coefficients, $g^{(M,N)}_{\{\lambda \}}(\vec{v},\vec{w})$, are more complicated.}
\item{The allowable dimensions of the partition, $\{\lambda \} \subseteq (M-1)^N$, are more general.}
\end{itemize}
In order to prove eq. \ref{slav.15} we shall explicitly address these two issues, and show how they can be overcome. \\
\\
\textbf{The necessary Pl\"{u}cker relations do not change.} The main result from section \ref{ferm6V} was arguably that bilinear sums of the coefficients, $c^{(N)}_{\{ \lambda\}}(\vec{v})$, simplified into the required coefficient term using the appropriate Pl\"{u}cker identities. The Pl\"{u}cker identity results were possible due to the coefficients, $c^{N}_{\{\lambda\}}(\vec{v})$, being determinants of $N\times N$ submatrices of a larger $(2N-1) \times N$ \textit{master} matrix, given by $\left(\kappa_{j,k}(\vec{v})  \right)_{j=1,\dots,2N-1\atop{k=1,\dots,N}}$, 
\small\begin{equation*}
\kappa_{j,k} = \frac{q^{j-k+1}-q^{k-1}}{q-1} (-1)^{N-j+k-1}e_{N-j+k-1}(\vec{v}) 
\end{equation*}\normalsize
Our current situation with the new coefficients is obviously not terribly different from section \ref{ferm6V}, as the coefficients, $g^{(M,N)}_{\{\lambda \}}(\vec{v},\vec{w})$, are also determinants of $N\times N$ submatrices constructed from the larger $(N+M-1) \times N$ \textit{master} matrix, given by $\left( \rho_{j,k}(\vec{v},\vec{w}) \right)_{j=1,\dots,N+M-1\atop{k=1,\dots,N}}$, 
\small\begin{equation*}\begin{split}
\rho_{j,k}= v^{-j}_{k} \sum^{N+M-1}_{\xi=j} \sum^{j-1}_{\eta=0} \sum^{\textrm{min}\{M,\xi \} }_{\alpha= \textrm{max}\{0,\xi-N+1 \}} \sum^{\textrm{min}\{M,\eta \}}_{\zeta=0} (-1)^{N+\xi+\eta} q^{N-1}  \\
 \times \left\{ q^{\xi-\eta+2 \zeta-\alpha} - q^{\eta-\xi+2 \alpha- \zeta} \right\} e_{M-\alpha}(\vec{w})e_{M-\zeta}(\vec{w}) e_{N-1-\eta+\zeta}(\vec{v},\hat{v}_k)e_{N-1-\xi+\alpha}(\vec{v},\hat{v}_k) v^{\xi+\eta}_k
\end{split}\end{equation*}\normalsize
Thus although the individual entries are obviously more complex, and the dimensions of the master matrix are more general, we expect to be able to generate all the necessary Pl\"{u}cker relations. This can be seen explicitly by labeling $\gamma_{\mu}$ as the $N\times 1$ column vector,
\small\begin{equation}
\gamma_{\mu} = \left( \begin{array}{c}
\rho_{\mu+1,1}(\vec{v},\vec{w}) \\
\rho_{\mu+1,2}(\vec{v},\vec{w})\\
\vdots\\
\rho_{\mu+1,N}(\vec{v},\vec{w}) \end{array}\right)
\label{slav.1.1}\end{equation}\normalsize
which allows us to generate the coefficients through the determinant expression,
\small\begin{equation}
g^{(M,N)}_{\{\lambda \}}(\vec{v},\vec{w}) = \left| \gamma_{\lambda_N}, \gamma_{\lambda_{N-1}+1}, \dots, \gamma_{\lambda_2 +N-2}, \gamma_{\lambda_1 + N-1}  \right|
\label{slav.1.2}
\end{equation}\normalsize
Using this notation allows us to instantly generate the required Pl\"{u}cker identities. Simply replacing the coefficient $c^{N}_{\{\lambda\}}(\vec{v})$ by the corresponding coefficient $g^{(M,N)}_{\{\lambda \}}(\vec{v},\vec{w})$, the argument shown in section \ref{2oper} is essentially exactly the same, leading to the base case result for the simplification of the sum of two bilinears in the coefficients,
\small\begin{equation}\begin{split}
g^{(M,N)}_{\{l_1+1,1^{j_1-1} \}}g^{(M,N)}_{\{l_2+1,1^{j_2-1} \}} - g^{(M,N)}_{\{l_1+1,1^{j_2-1} \}}g^{(M,N)}_{\{l_2+1,1^{j_1-1} \}}\\
=  g^{(M,N)}_{\{ \phi \}}g^{(M,N)}_{\{l_2+1,l_1+2,2^{j_1-1},1^{j_2-j_1-1} \}}
\label{slav.1.3}
\end{split}\end{equation}\normalsize
for $0 \le l_1 < l_2 \le M-2$ and $1 \le j_1  < j_{2} \le N$. Thus the multiplication of two general $Y^{(M,N)}$'s gives the following required form,
\small\begin{equation}
Y^{(M,N)}_{l_1}  Y^{(M,N)}_{l_2} =  \sum_{1 \le j_1 < j_2 \le N}(-1)^{j_1+j_2} \tilde{g}^{(M,N)}_{\{l_2+1,l_1+2,2^{j_1-1},1^{j_2-j_1-1} \}}  \psi^*_{-j_2}\psi^*_{-j_1}\psi_{l_1} \psi_{l_2}
\label{slav.1.4}
\end{equation}\normalsize
Using these results, we can again replace the coefficient $c^{N}_{\{\lambda\}}(\vec{v})$ by the corresponding coefficient $g^{(M,N)}_{\{\lambda \}}(\vec{v},\vec{w})$ to obtain the equivalent general result given in section \ref{koper} concerning the simplification of a sum of a general number of bilinears in the coefficients,
\small\begin{equation}
 \sum^{k-1}_{p=0}(-1)^{p} g^{(M,N)}_{\sigma_p \left( \{\lambda\}_{k-1} \right)}  g^{(M,N)}_{\{l_{k}+1, 1^{j_{k-p}-1} \}} = g^{(M,N)}_{ \{\phi\}} g^{(M,N)}_{ \{\lambda\}_{k} }
\label{slav.1.5}
\end{equation}\normalsize
for $0 \le l_1 < \dots < l_k \le M-2$ and $1 \le j_1  <\dots <  j_{k} \le N$, and the partition labels, $\sigma_p \left( \{\lambda\}_{k-1} \right)$, are given by eq. \ref{lotsofcrap}.\\
\\
An immediate consequence of eq. \ref{slav.1.5} is the following result concerning the multiplication of a general numbers of $Y^{(M,N)}$'s,
\small\begin{equation}
Y^{(M,N)}_{l_1} \dots Y^{(M,N)}_{l_{k}} =  \sum_{1 \le j_1 < \dots < j_{k} \le N}(-1)^{j_1+\dots + j_{k}} \tilde{g}^{(M,N)}_{\{ \lambda\}_k}   \psi^*_{-j_k} \dots \psi^*_{-j_1}\psi_{l_1}\dots  \psi_{l_k} \label{slav.17}
\end{equation}\normalsize
which shows explicitly that despite the more general nature of the coefficients, the required Pl\"{u}cker relations are still generated. This allows us to consider the next issue.\\
\\
\textbf{Generalizing the dimensions of the partition.} Realizing that all the required Pl\"{u}cker relations are still intact, we are now in a condition to consider the multiplication of all the generators,
\small\begin{equation*}\begin{split}
e^{Y^{(M,N)}_{0}} \dots e^{ Y^{(M,N)}_{M-2}} &= \sum^{M-1}_{k=0}e_k \left(  Y^{(M,N)}_{0} ,\dots, Y^{(M,N)}_{M-2}  \right)\\
&= 1 + \sum^{M-1}_{k=1}\sum_{0 \le l_1 < \dots < l_k \le M-2} Y^{(M,N)}_{l_1} \dots Y^{(M,N)}_{l_{k}}
\end{split}\end{equation*}\normalsize
Applying eq. \ref{slav.17}, the above expression becomes,
\small\begin{equation}
1 + \sum^{M-1}_{k=1}\sum_{0 \le l_1 < \dots < l_k \le M-2} \sum_{1 \le j_1 < \dots < j_{k} \le N} (-1)^{j_1+\dots + j_{k}} \tilde{g}^{(M,N)}_{\{ \lambda\}_k}   \psi^*_{-j_k} \dots \psi^*_{-j_1}\psi_{l_1}\dots  \psi_{l_k} \label{MAS}
\end{equation}\normalsize
Examining eq. \ref{MAS} we notice that considering $M-1$ generators (as opposed to $N-1$) has accomplished two things.
\begin{itemize}
\item{The summation term $\sum_{0 \le l_1 < \dots < l_k \le M-2} \sum_{1 \le j_1 < \dots < j_{k} \le N}$ generates every possible fermionic expression that corresponds to a partition with $k$ hooks, contained within the partition $\{(M-1)^N \}$.}
\item{Since $1 \le k \le M-1$, this immediately means that all the partitions within the rectangle $\{(M-1)^N \}$ are generated.}
\end{itemize}
Thus we have successfully generalized the dimensions of the partition.\\
\\
\textbf{Bosonization.} Having explicitly addressed the issues of generalization of the coefficients and the dimensions of the allowable partitions, we now apply the boson-fermion correspondence to eq. \ref{slav.15}. We recall that eq. \ref{slav.17} assures us that each fermionic expression is accompanied by the required coefficient and sign, thus we obtain the result,
\small\begin{equation*}\begin{split}
&\Upsilon^{'M}_N g^{(M,N)}_{\{\phi\}}  \langle 0| \exp\left\{H_+(\vec{x})\right\}e^{Y^{(M,N)}_{0}} \dots e^{ Y^{(M,N)}_{M-2}} |0\rangle \\
=&\displaystyle \Upsilon^{'M}_N \left( g^{(M,N)}_{\{\phi\}} \chi_{\{\phi\}}(\vec{x}) + \sum^{M-1}_{k=1}\sum_{0 \le l_1 < \dots < l_k \le M-2} \sum_{1 \le j_1 < \dots < j_{k} \le N} (-1)^{j_1+\dots + j_{k}} \right.\\
& \times \left. g^{(M,N)}_{\{ \lambda\}_k} \chi_{\{\lambda \}_k}( \vec{x})  \right)\\
=&  \displaystyle\Upsilon^{'M}_N  \sum_{\{\lambda\} \subseteq (M-1)^{N}} g^{(M,N)}_{\{\lambda\}}(\vec{v},\vec{w})\chi_{\{\lambda\}}(\vec{x}) 
\end{split}\end{equation*}\normalsize
Restricting the time variables appropriately,
\small\begin{equation*}
x_j \rightarrow \frac{1}{j}p_j(u_1,\dots,u_N) \textrm{  ,  } j \in \{1,2,\dots\}
\end{equation*}\normalsize
we obtain,
\small\begin{equation*}\begin{split}
&\Upsilon^{'M}_N g^{(M,N)}_{\{\phi\}}  \langle 0| \exp\left\{H_+\left( \frac{1}{j}p_j(\vec{u}) \right)\right\}e^{Y^{(M,N)}_{0}} \dots e^{ Y^{(M,N)}_{M-2}} |0\rangle \\
=&  \displaystyle\Upsilon^{'M}_N  \sum_{\{\lambda\} \subseteq (M-1)^{N}} g^{(M,N)}_{\{\lambda\}}(\vec{v},\vec{w})S_{\{\lambda\}}(\vec{u}) \\
=&  \field{S}^M_N(\vec{u},\vec{v}_{\beta},\vec{w})
\end{split}\end{equation*}\normalsize
which completes the lemma. $\square$
\section{KP tau-functions and fermions}
This section contains classical results (found in \cite{bluebook,6}) that algebraically show that the above form of the DWPF and scalar product, as a fermionic inner product, is by construction, a $\tau$-function of the KP hierarchy.\\
\\
\textbf{Further fermionic definitions.} We first define the following generating sums of the free fermions,
\small\begin{equation*}
\psi(k) = \sum_{j \in \mathbb{Z}} \psi_j k^j \textrm{  ,  } \psi^*(k) = \sum_{j \in \mathbb{Z}} \psi^*_j k^{-j} 
\end{equation*}\normalsize
Applying the anti-commutation relations, and the Baker-Campbell-Hausdorff formula\footnote{$e^{H(\vec{x})} X e^{-H(\vec{x})} = X + [H(\vec{x}), X]+ \frac{1}{2!}[H(\vec{x}),[H(\vec{x}), X]] + \frac{1}{3!}[H(\vec{x}),[H(\vec{x}),[H(\vec{x}),X]]] + \dots $} we obtain the following,
\small\begin{equation}\begin{split}
e^{H_+(\vec{x})}\psi(k)e^{-H_+(\vec{x})} &= \exp\left\{\sum^{\infty}_{n=1}k^n x_n \right\} \psi(k) \\
 e^{H_+(\vec{x})}\psi^*(k)e^{-H_+(\vec{x})} &= \exp\left\{-\sum^{\infty}_{n=1}k^x t_n \right\} \psi^*(k)
\label{p.42}
\end{split}\end{equation}\normalsize
Considering the inner product expression, $\langle 0| \psi(k_1) \psi^*(k_2) |0 \rangle$, we obtain the following geometric simplification, 
\small\begin{equation*}\begin{split}
\langle 0| \psi(k_1) \psi^*(k_2) |0 \rangle &= \sum^{-1}_{n_1,n_2=-\infty} k^{n_1}_1 k^{-n_2}_2 \langle 0| \psi_{n_1} \psi^*_{n_2} |0 \rangle\\
&= \sum^{-1}_{n_1=-\infty} \left( \frac{k_1}{ k_2} \right)^{n_1} = \sum^{\infty}_{n_1=0} \left( \frac{k_2}{ k_1} \right)^{n_1}-1\\
&= \frac{k_2}{k_1-k_2}
\end{split}\end{equation*}\normalsize
Generalizing the above expression using Wick's theorem we obtain the following determinant expression,
\small\begin{equation*}\begin{split}
 &\langle 0| \psi(k_1)\dots \psi(k_p)  \psi^*(l_1)\dots \psi^*(l_p) |0 \rangle \\
=& \sum_{\sigma \in S_{p}} sgn( \sigma) \langle 0|\psi(k_{1})  \psi^*(l_{\sigma_{1}})  |0 \rangle \dots  \langle 0|\psi(k_{p})  \psi^*(l_{\sigma_{p}})  |0 \rangle \\
=& \textrm{det}\left(  \langle 0|\psi(k_{i})  \psi^*(l_{\sigma_{j}})  |0 \rangle \right)^p_{i,j=1}=\textrm{det}\left(  \frac{l_j}{k_i-l_j}   \right)^p_{i,j=1} \\
=& \left( \prod^p_{i=1} l_i \right) \left( \frac{\prod_{1 \le i < j \le p} (k_i-k_j)(l_j-l_i)}{\prod_{1 \le i < j \le p} (k_i-l_j)} \right)
\end{split}\end{equation*}\normalsize
where the last line is due to \textit{Cauchy's identity}. By an analogous argument we have,
\small\begin{equation*}
\langle 0| \psi^*(k_1) \psi(k_2) |0 \rangle = \frac{k_1}{k_1-k_2}
\end{equation*}\normalsize
which has the following generalization,
\small\begin{equation*}\begin{split}
& \langle 0| \psi^*(k_1)\dots \psi^*(k_p)  \psi(l_1)\dots \psi(l_p) |0 \rangle \\
=& \sum_{\sigma \in S_{p}} sgn( \sigma) \langle 0|\psi^*(k_{1})  \psi(l_{\sigma_{1}})  |0 \rangle \dots  \langle 0|\psi^*(k_{p})  \psi(l_{\sigma_{p}})  |0 \rangle \\
=& \left( \prod^p_{i=1} k_i \right) \left( \frac{\prod_{1 \le i < j \le p} (k_i-k_j)(l_j-l_i)}{\prod_{1 \le i < j \le p} (k_i-l_j)} \right)
\end{split}\end{equation*}\normalsize
This leads us to the first of two necessary results.
\begin{proposition}
\begin{eqnarray}
\langle 0| \psi^*_0 \psi(\lambda) =& \langle 0| \exp \left\{- \sum^{\infty}_{n=1}\frac{1}{n \lambda^n} H_n \right\} =&\langle 0| e^{-h(\lambda)} \label{p.43}\\
\langle 0| \psi_{-1} \psi^*(\lambda) =&\lambda \langle 0| \exp \left\{ \sum^{\infty}_{n=1}\frac{1}{n \lambda^n} H_n \right\}=&\lambda \langle 0|e^{h(\lambda)} \label{p.44}
\end{eqnarray}
\end{proposition}
\textbf{Proof.} In order to verify eq. \ref{p.43}, it is enough to show that,
\small\begin{equation*}\begin{split}
&\langle 0| \psi^*_0 \psi(\lambda) \psi(k_1)\dots \psi(k_p)  \psi^*(l_1)\dots \psi^*(l_p) |0 \rangle\\
 =&\langle 0| e^{-h(\lambda)}\psi(k_1)\dots \psi(k_p)  \psi^*(l_1)\dots \psi^*(l_p) |0 \rangle 
\end{split}\end{equation*}\normalsize
for general $p \in \field{N}$.\\
\\
Focusing on the right hand side of the above expression and inserting $e^{h(\lambda)}e^{-h(\lambda)}$ in between the generating sums we obtain the following rational expression,
\small\begin{equation*}\begin{split}
& \langle 0| \underbrace{e^{-h(\lambda)}\psi(k_1)e^{h(\lambda)}}_{\exp{\left\{ -\sum^{\infty}_{n=1}\frac{1}{n}\left( \frac{k_1}{\lambda} \right)^n \right\}}\psi(k_1)} \underbrace{e^{-h(\lambda)}\psi(k_2)e^{h(\lambda)}}_{\exp\left\{ -\sum^{\infty}_{n=1}\frac{1}{n}\left( \frac{k_2}{\lambda} \right)^n \right\}\psi(k_2)}\dots   \underbrace{e^{-h(\lambda)} \psi^*(l_p)e^{h(\lambda)}}_{\exp\left\{ \sum^{\infty}_{n=1}\frac{1}{n}\left( \frac{l_p}{\lambda} \right)^n \right\}\psi^*(l_p)}\\
& \times \underbrace{e^{-h(\lambda)} |0 \rangle}_{|0 \rangle}\\
=&\left( \prod^p_{j=1} \frac{\lambda-k_j}{\lambda-l_j } \right) \langle 0| \psi(k_1)\dots \psi(k_p)  \psi^*(l_1)\dots \psi^*(l_p) |0 \rangle\\
=& \left( \prod^p_{j=1} \frac{\lambda-k_j}{\lambda-l_j } \right) \left( \prod^p_{i=1} l_i \right) \left( \frac{\prod_{1 \le i < j \le p} (k_i-k_j)(l_j-l_i)}{\prod^p_{i , j = 1} (k_i-l_j)} \right)
\end{split}\end{equation*}\normalsize
Focusing on the left hand side, we use the fact that $\psi^*_0 \psi(k) = 1 -\psi(k) \psi^*_0$ to commute the $\psi^*_0$ operator to the right hand side of the inner product expression. We label $\lambda = k_0$ in the workings below for notational convenience,
\small\begin{equation*}\begin{split}
& \langle 0| \psi^*_0 \psi( k_0) \psi(k_1)\dots \psi(k_p)  \psi^*(l_1)\dots \psi^*(l_p) |0 \rangle \\
=& \langle 0| \psi(k_1)\dots \psi(k_p)  \psi^*(l_1)\dots \psi^*(l_p) |0 \rangle - \langle 0| \ \psi(k_0) \psi^*_0 \psi(k_1)\dots \psi(k_p)  \psi^*(l_1)\dots \psi^*(l_p) |0 \rangle\\
 \vdots\\
=&  \sum^{p}_{j=0} (-1)^j \langle 0| \psi(k_0)\dots \psi(k_{j-1})\psi(k_{j+1}) \dots  \psi(k_p)  \psi^*(l_1)\dots \psi^*(l_p) |0 \rangle \\
=&  \left( \prod^p_{i=1} l_i \right)  \prod_{1 \le i < j \le p}(l_j-l_i) \left\{  \left( \frac{\prod_{1 \le i < j \le p} (k_i-k_j)}{\prod^p_{i , j = 1} (k_i-l_j)} \right) \right.\\
& \left.+ \sum^{p}_{r=1} (-1)^r  \left( \frac{\prod^p_{j=1\ne{r}}(k_0 -k_j)\prod_{1 \le i < j \le p \atop{\ne r}}(k_i-k_j) }{\prod^p_{j=1}(k_0-l_j)\prod^p_{i=1\atop{\ne r}}\prod^p_{j=1} (k_i-l_j)} \right) \right\}\\
=& \frac{ \left( \prod^p_{i=1} l_i \right)  \prod_{1 \le i < j \le p}(l_j-l_i)}{ \prod^p_{j=1}(\lambda-l_j)\prod^p_{i , j = 1} (k_i-l_j)} \left\{  \sum^{p}_{r=0} (-1)^r \left[ \prod_{0 \le i < j \le p \atop{\ne r}}(k_i-k_j) \right] \left[ \prod^p_{j=1} (k_r-l_j) \right]  \right\}
\end{split}\end{equation*}\normalsize
We now concentrate on the term contained within the curly brackets. Expanding the $\{l_1,\dots,l_p\}$ variables in terms of elementary symmetric polynomials we obtain,
\small\begin{equation*}\begin{split}
&\sum^{p}_{r=0} (-1)^r \left[ \prod_{0 \le i < j \le p \atop{\ne r}}(k_i-k_j) \right] \left[ \prod^p_{j=1} (k_r-l_j) \right]\\
=& \sum^p_{s=0}e_s(\vec{l}) \sum^{p}_{r=0} (-1)^r (-k_r)^{p-s} \left[ \prod_{0 \le i < j \le p \atop{\ne r}}(k_i-k_j) \right] 
\end{split}\end{equation*}\normalsize
We now claim that all terms in the above sum for $s \ne 0$ are equal to zero. To show this we express the product as a Vandermonde determinant,\\
$\left[ \prod_{0 \le i < j \le p \atop{\ne r}}(k_i-k_j) \right] = \textrm{det} \left( k^{i}_j \right)_{i=0, \dots, p-1 \atop{j=0,\dots,\hat{r}, \dots,p }}$, and the sum becomes the following determinant expansion expression,
\small\begin{equation*}
 \sum^{p}_{r=0} (-1)^r (k_r)^{p-s} \textrm{det} \left( k^{i}_j \right)_{i=0, \dots, p-1 \atop{j=0,\dots,\hat{r}, \dots,p }} =  \left\{ \begin{array}{cc}
0 & 1 \le s \le p\\
 \textrm{det} \left( k^{i}_j \right)_{i,j=0,\dots,p} & s=0 
\end{array} \right. 
\end{equation*}\normalsize
Hence we obtain,
\small\begin{equation}\begin{split}
 \sum^{p}_{r=0} (-1)^r \left[ \prod_{0 \le i < j \le p \atop{\ne r}}(k_i-k_j) \right] \left[ \prod^p_{j=1} (k_r-l_j) \right]  &=  \textrm{det} \left( k^{i}_j \right)_{i,j=0,\dots,p} \\
 &=  \left[ \prod_{1 \le i < j \le p} (k_i-k_j) \right] \left[ \prod^p_{j=1} (\lambda-k_j) \right] \label{p.45}
\end{split}\end{equation}\normalsize
which completes the verification of eq. \ref{p.43}.\\
\\
To verify eq. \ref{p.44} we wish to verify the following expression,
\small\begin{equation*}\begin{split}
&\langle 0| \psi_{-1} \psi^*(\lambda) \psi^*(k_1)\dots \psi^*(k_p)  \psi(l_1)\dots \psi(l_p) |0 \rangle\\
 =& \lambda \langle 0| e^{h(k)}\psi^*(k_1)\dots \psi^*(k_p)  \psi(l_1)\dots \psi(l_p) |0 \rangle
\end{split}\end{equation*}\normalsize
Focusing on the right hand side as before we obtain,
\small\begin{equation*}\begin{split}
&\ \lambda \langle 0| \underbrace{e^{h(\lambda)}\psi^*(k_1)e^{-h(\lambda)}}_{\exp\left\{ -\sum^{\infty}_{n=1}\frac{1}{n}\left( \frac{k_1}{\lambda} \right)^n \right\}\psi^*(k_1)} \underbrace{e^{h(\lambda)}\psi^*(k_2)e^{-h(\lambda)}}_{\exp\left\{ - \sum^{\infty}_{n=1}\frac{1}{n}\left( \frac{k_2}{\lambda} \right)^n \right\}\psi^*(k_2)}\dots   \underbrace{e^{h(\lambda)} \psi(l_p)e^{-h(\lambda)}}_{\exp\left\{ \sum^{\infty}_{n=1}\frac{1}{n}\left( \frac{l_p}{\lambda} \right)^n \right\}\psi(l_p)}\\
& \times \underbrace{e^{h(\lambda)} |0 \rangle}_{|0 \rangle}\\
=&  \lambda \left( \prod^p_{j=1} \frac{\lambda-k_j}{\lambda-l_j } \right) \left( \prod^p_{i=1} k_i \right) \left( \frac{\prod_{1 \le i < j \le p} (k_i-k_j)(l_j-l_i)}{\prod^p_{i , j = 1} (k_i-l_j)} \right)
\end{split}\end{equation*}\normalsize
Focusing on the left hand side, we use the fact that $\psi_{-1} \psi^*(k) = k -\psi^*(k) \psi_{-1}$ to commute the $\psi_{-1}$ operator to the right hand side of the inner product expression. We label $\lambda = k_0$ in the workings below for notational convenience,
\small\begin{equation*}\begin{split}
& \langle 0| \psi_{-1} \psi^*( k_0) \psi^*(k_1)\dots \psi^*(k_p)  \psi(l_1)\dots \psi(l_p) |0 \rangle \\
=& k_0 \langle 0| \psi^*(k_1)\dots \psi^*(k_p)  \psi(l_1)\dots \psi(l_p) |0 \rangle\\
& - \langle 0| \ \psi(k_0) \psi_{-1} \psi^*(k_1)\dots \psi^*(k_p)  \psi(l_1)\dots \psi(l_p) |0 \rangle\\
\vdots& \\
=&  \sum^{p}_{j=0} (-1)^j k_j \langle 0| \psi^*(k_0)\dots \psi^*(k_{j-1})\psi^*(k_{j+1}) \dots  \psi^*(k_p)  \psi(l_1)\dots \psi(l_p) |0 \rangle \\
=&  \left( \prod^p_{i=0} k_i \right)  \prod_{1 \le i < j \le p}(l_j-l_i) \left\{  \left( \frac{\prod_{1 \le i < j \le p} (k_i-k_j)}{\prod^p_{i , j = 1} (k_i-l_j)} \right)\right.\\
&\left. + \sum^{p}_{r=1} (-1)^r  \left( \frac{\prod^p_{j=1\ne{r}}(k_0 -k_j)\prod_{1 \le i < j \le p \atop{\ne r}}(k_i-k_j) }{\prod^p_{j=1}(k_0-l_j)\prod^p_{i=1\atop{\ne r}}\prod^p_{j=1} (k_i-l_j)} \right) \right\}\\
=& \lambda \frac{ \left( \prod^p_{i=1} k_i \right)  \prod_{1 \le i < j \le p}(l_j-l_i)}{ \prod^p_{j=1}(\lambda-l_j)\prod^p_{i , j = 1} (k_i-l_j)} \left\{  \sum^{p}_{r=0} (-1)^r \left[ \prod_{0 \le i < j \le p \atop{\ne r}}(k_i-k_j) \right] \left[ \prod^p_{j=1} (k_r-l_j) \right]  \right\}\\
=& \lambda \left( \prod^p_{j=1} \frac{\lambda-k_j}{\lambda-l_j } \right) \left( \prod^p_{i=1} k_i \right) \left( \frac{\prod_{1 \le i < j \le p} (k_i-k_j)(l_j-l_i)}{\prod^p_{i , j = 1} (k_i-l_j)} \right)
\end{split}\end{equation*}\normalsize
where we have applied eq. \ref{p.45} to proceed from the second last line to the last line. This completes the proof of the proposition. $\square$\\
\\
We now focus on the second of the two necessary results.
\begin{proposition}
For any $X \in gl(\infty)$ and $|\alpha \rangle$, $|\beta \rangle$ $\in \field{F}$, we have the following relation,
\small\begin{equation}
\sum_{n \in \field{Z}}e^{X} \psi_n  |\alpha \rangle \otimes e^{X} \psi^*_n  |\beta \rangle = \sum_{n \in \field{Z}} \psi_n e^{X} |\alpha \rangle \otimes \psi^*_n e^{X} |\beta \rangle    \label{p.47}
\end{equation}\normalsize
\end{proposition}
\textbf{Proof.}
Let us focus on the left hand expression and commute $\psi_n / \psi^*_n$ to the left using eq. \ref{p.46} and the Baker-Campbell-Hausdorff formula,
\small\begin{equation*}\begin{split}
e^{X} \psi_n &= \left\{  \psi_n + \sum_{m_1 \in \field{Z}} a_{m_1 n} \psi_{m_1}+ \frac{1}{2!} \sum_{m_1 m_2 \in \field{Z}} a_{m_2 m_1}a_{m_1 n} \psi_{m_2}  + \dots \right\}e^{X}\\
e^{X} \psi^*_n &= \left\{  \psi^*_n - \sum_{m_1 \in \field{Z}} a_{ n m_1} \psi^*_{m_1}+\frac{1}{2!}  \sum_{m_1 m_2 \in \field{Z}} a_{m_1 m_2}a_{n m_1} \psi^*_{m_2}  - \dots \right\}e^{X}
\end{split}\end{equation*}\normalsize
We now group the powers of the coefficient, $a$, to obtain,
\small\begin{equation*}\begin{split}
\sum_{n \in \field{Z}}e^{X} \psi_n  \otimes e^{X} \psi^*_n = \sum_{n \in \field{Z}} \psi_n e^{X}  \otimes \psi^*_n e^{X}\\
  +  \sum_{n m_1 \in \field{Z}} \left\{ a_{m_1 n} \psi_{m_1}  \otimes \psi^*_{n} -a_{ n m_1} \psi_{n}  \otimes \psi^*_{m_1} \right\} e^{X}  \otimes e^{X} \\
+ \sum_{n m_1 m_2 \in \field{Z}} \left\{ \frac{a_{m_2 m_1}a_{m_1 n}}{2!}\psi_{m_2}  \otimes \psi^*_{n}-a_{m_1 n}a_{n m_2}\psi_{m_1}  \otimes \psi^*_{m_2}  \right.\\
\left. + \frac{a_{m_1 m_2}a_{n m_1}}{2!}\psi_{n}  \otimes \psi^*_{m_2} \right\} e^{X}  \otimes e^{X}  + \dots 
\end{split}\end{equation*}\normalsize
\small\begin{equation}\begin{split}
\Rightarrow \sum_{n \in \field{Z}}e^{X} \psi_n  \otimes e^{X} \psi^*_n=\sum_{n \in \field{Z}} \psi_n e^{X}  \otimes \psi^*_n e^{X}\\
+ \sum^{\infty}_{j=1} \sum^{j}_{k=0} \frac{1}{k!} \frac{(-1)^{j-k}}{(j-k)!}  \sum_{m_1 \dots m_k l_1 \dots l_{j-k} \in \field{Z}} \left( \prod^k_{r=1}a_{m_r m_{r-1}} \right)\\
\times \left( \prod^{j-k}_{r=1}a_{l_{r-1} l_{r}} \right) \psi_{m_{k}} e^{X}  \otimes \psi^*_{l_{j-k}} e^{X}  \label{p.48}
\end{split}\end{equation}\normalsize
where we have labeled $m_0 = l_0 = n$ in the above equation. It is obvious that we need to verify that the summation over the dummy index, $1 \le j < \infty$, is zero.\\
\\
To accomplish this we perform the following change of indices for each individual value of $j$ and $k$,
\small\begin{equation*}
\begin{array}{lclllcl}
n & \leftrightarrow & l_{j-k}  &&m_1 & \leftrightarrow & l_{j-k+1} \\
l_1 & \leftrightarrow & l_{j-k-1} && m_2 & \leftrightarrow & l_{j-k+2} \\
l_2 & \leftrightarrow & l_{j-k-2}  && & \vdots & \\ 
& \vdots & && m_k & \leftrightarrow & l_{j}\\
l_{\frac{j-k}{2} -1} & \leftrightarrow & l_{\frac{j-k}{2} +1} & j-k \textrm{  even  }\\
l_{\frac{j-k-1}{2} } & \leftrightarrow & l_{\frac{j-k+1}{2} } &  j-k \textrm{  odd  }
\end{array}
\end{equation*}\normalsize
Doing this, the aforementioned summation over $j$ in eq. \ref{p.48} becomes,
\small\begin{equation*}
\sum^{\infty}_{j=1} \underbrace{\left(  \sum^{j}_{k=0} \frac{1}{k!} \frac{(-1)^{j-k}}{(j-k)!}  \right)}_{=0 \textrm{  for all  } j \ne 0} \sum_{l_1 \dots l_j  \in \field{Z}}\left( \prod^{j}_{r=1}a_{l_{r} l_{r-1}} \right) \psi_{l_j} e^{X}  \otimes \psi^*_{n} e^{X} 
\end{equation*}\normalsize
We can see immediately that the above summation equals zero for all $j \ne 0$ by expanding the series, $e^s . e^{-s} = 1$,
\small\begin{equation*}
e^s . e^{-s} = 1 + \sum^{\infty}_{j=1} s^j \sum^{j}_{k=0} \frac{1}{k!} \frac{(-1)^{j-k}}{(j-k)!} 
\end{equation*}\normalsize
knowing that all terms $s^j$, $j \ge 1$, are equal to zero, which proves the proposition. $\square$\\
\\
The above proposition also implies that for $X_{1}, X_{2}, \dots, X_{N} \in gl(\infty)$, and $g = e^{X_{1}}e^{X_{2}} \dots e^{X_{N}}$, then we have,
\small\begin{equation}
\sum_{n \in \field{Z}}g \psi_n  |\alpha \rangle \otimes g\psi^*_n  |\beta \rangle = \sum_{n \in \field{Z}} \psi_n g |\alpha \rangle \otimes \psi^*_n g |\beta \rangle  \label{p.49}
\end{equation}\normalsize
Additionally, with the choice $|\alpha \rangle = |\beta \rangle = |0 \rangle$, then the above expression becomes zero,
\small\begin{equation}
\sum_{n \in \field{Z}}g \psi_n  |0 \rangle \otimes g\psi^*_n |0 \rangle =  \sum_{n \in \field{Z}} \psi_n g |0 \rangle \otimes \psi^*_n g |0 \rangle = 0  \label{p.50}
\end{equation}\normalsize
due to either $\psi_n  |0 \rangle = 0$, or $\psi^*_n  |0 \rangle = 0$, for all values of $n \in \field{Z}$. \\
\\
We now put all the results together to show that a $\tau$-function of the form,
\small\begin{equation}
\tau (\vec{x}) = \langle 0| \exp\{ H_+ (\vec{x}) \} g |0 \rangle \label{p.51} 
\end{equation}\normalsize
satisfies the KP bilinear hierarchy.
\begin{proposition}
Any inner product expression of the form \ref{p.51} obeys the following bilinear relation,
\small\begin{equation*}
\oint \frac{d \lambda}{2 \pi i} \exp \left\{ \sum^{\infty}_{n=1}(x_{n}-x'_{n}) \lambda^{n} \right\} \tau\left(\vec{x}-\vec{\epsilon}\left(\frac{1}{\lambda}\right)\right)  \tau\left(\vec{x}'+\vec{\epsilon}\left(\frac{1}{\lambda}\right)\right) = 0
\end{equation*}\normalsize
where $\vec{\epsilon}\left(\frac{1}{\lambda}\right) = \left( \frac{1}{\lambda}, \frac{1}{2 \lambda^2}, \dots \right)$.
\end{proposition}
\textbf{Proof.} Beginning with the right hand side of eq. \ref{p.50}, and applying $\langle 0| \psi^*_0 e^{H_+(\vec{x})} \otimes \langle 0| \psi_{-1} e^{H_+(\vec{x}')}$ we obtain the bilinear inner product expression,
\small\begin{equation*}\begin{split}
0 =& \sum_{n \in \field{Z}}\langle 0| \psi^*_0 e^{H_+(\vec{x})}  \psi_n g |0 \rangle \langle 0| \psi_{-1} e^{H_+(\vec{x}')}\psi^*_n g |0 \rangle \\
=& \oint \frac{d \lambda}{2 \pi i \lambda} \langle 0| \psi^*_0 e^{H_+(\vec{x})}  \psi (\lambda) g |0 \rangle \langle 0| \psi_{-1} e^{H_+(\vec{x}')}\psi^*(\lambda) g |0 \rangle
\end{split}\end{equation*}\normalsize
Commuting the generating sums, $ \psi (\lambda)/ \psi^* (\lambda)$, with the operators $e^{H_+(\vec{x})}/e^{H_+(\vec{x}')}$, using eq. \ref{p.42}, we obtain,
\small\begin{equation*}\begin{split}
0=& \oint \frac{d \lambda}{2 \pi i \lambda} \exp \left\{ \sum^{\infty}_{n=1}(x_{n}-x'_{n})\lambda^{n} \right\}  \underbrace{\langle 0| \psi^*_0   \psi (\lambda) }_{\textrm{use eq. \ref{p.43}} }e^{H_+(\vec{x})} g |0 \rangle \underbrace{\langle 0| \psi_{-1}\psi^*(\lambda)}_{\textrm{use eq. \ref{p.44}}} e^{H_+(\vec{x}')} g |0 \rangle\\
=& \oint \frac{d \lambda}{2 \pi i}\exp \left\{ \sum^{\infty}_{n=1}(x_{n}-x'_{n})\lambda^{n} \right\}  \langle 0|  \exp \left\{ H_+ \left( \vec{x} - \vec{\epsilon}\left(\frac{1}{\lambda} \right)  \right) \right\}   g |0 \rangle \\
& \times  \langle 0| \exp \left\{ H_+ \left( \vec{x}' + \vec{\epsilon}\left(\frac{1}{\lambda}\right)  \right) \right\} g |0 \rangle \textrm{   $\square$}
\end{split}\end{equation*}\normalsize
%%%%%%%%%%%%%%%%%%%%%%%%%%%%%%%%%%%%%%%%%%%%%%%%%%%%%%%%%%%%%%%%%%%%%%%%%%%%%%%%%%%%%%%%%%%%%%%%%%%%%%%%%%%%%%%%%%%%
\newpage
%%%%%%%%%%%%%%%%%%%%%%%%%%%%%%%%%%%%%%%%%%%%%%%%%%%%%%%%%%%%%%%%%%%%%%%%%%%%%%%%%%%%%%%%%%%%%%%%%%%%%%%%%%%%%%%%%%%%
\chapter{The trigonometric Felderhof model}
In \cite{felderhof1} Felderhof diagonalized the transfer matrix of Baxter's \cite{Bax1,Bax2} free-fermion elliptic eight vertex model\footnote{The eight-vertex model is a generalization of the six-vertex model, where, in order for the transfer matrices of the model to commute (to ensure integrability), one must parameterize the weights by elliptic functions rather than the usual trigonometric, due to the additional two allowable vertices. For a detailed introduction to this model see Chap. 10 of \cite{Baxterbook}}. The aforementioned transfer matrix is expressed in terms of fermionic operators, which leads to an easy survey of eigenvalues and eigenvectors. The model in question was studied earlier by Fan and Wu \cite{FanWu1,FanWu2} in the context of deriving exact and approxiamte solutions for the free energy of the model under periodic boundary conditions. The analysis by Fan and Wu relied on the earlier work of Kastelyn \cite{Kast} where the periodic partition function can be expressed as the summation of dimers on a lattice.\\
\\
Further studies were conducted by Felderhof in \cite{felderhof2,felderhof3} which generalized the model and placed it in the presence of fields (where the fields were effectively parameterized by colours, in the same sense that temperature is effectively parameterized by the rapidities), whilst remaining free-fermion, hence the fermionic operator methods found in \cite{felderhof1} still applied. In \cite{DA1} the trigonometric limit of the Felderhof model was found to be one of a hierarchy of coloured vertex models. In \cite{DA2}, the hierarchy was extended to coloured elliptic height models.\\
\\
In this section we focus solely on the trigonometric limit of the Felderhof coloured vertex model found in \cite{felderhof2,felderhof3}. This corresponds to the spin-$\frac{1}{2}$ vertex model of the hierarchy found in \cite{DA1}, and as such, we shall use the convenient parameterization found in this paper. Being a spin-$\frac{1}{2}$ model this section shares much similarity with the six-vertex model considered in the last section. This should come as no surprise as taking a specific limit of the colour variables recovers the free-fermion six-vertex model.\\
\\
The analysis conducted by Felderhof was under periodic boundary conditions (PBC's). We now concern ourselves with DWBC's and perform a similar Korepin-Izergin analysis found in \cite{Korepini,23} for the derivation of the DWPF.
\section{Defining the model}
\begin{figure}[h!]
\begin{center}
\includegraphics[angle=0,scale=0.25]{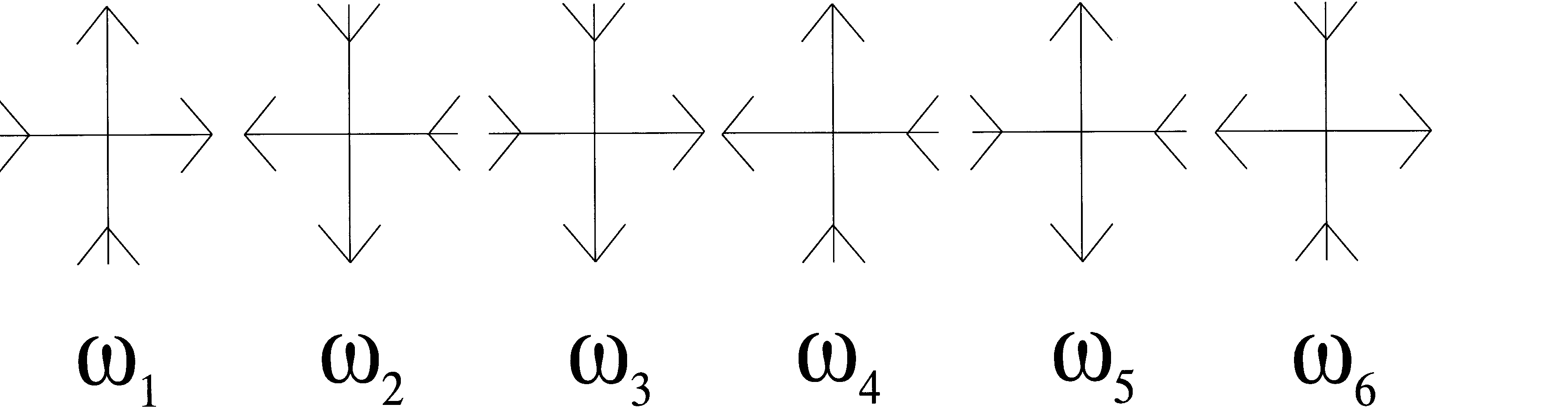}
\caption{\footnotesize{Labelling of the 6 vertices}}
\label{nonfin}
\end{center}
\end{figure}

\noindent \textbf{Lattice lines - rapidities and colours.} Consider an $N \times N$ lattice of vertices, with horizontal rapidity flows $u_i \in \field{C}$, $1 \le i \le N$, which flow from left to right, and vertical rapidity flows $v_j\in \field{C}$, $1 \le j \le N$, which flow from bottom to top. Additionally, associated with each horizontal lattice line is the colour variable $\alpha_i\in \field{C}$, $1 \le i \le N$, and associated with each vertical lattice line is the colour variable $\beta_j\in \field{C}$, $1 \le j \le N$.
\begin{figure}[h!]
\begin{center}
\includegraphics[angle=0,scale=0.25]{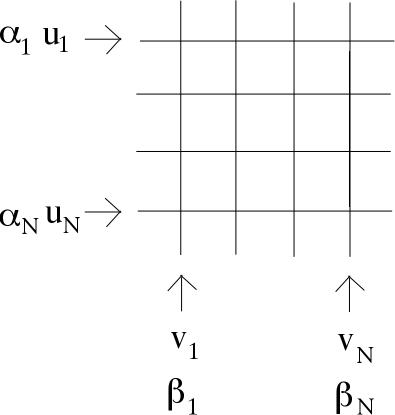}
\caption{\footnotesize{The $N \times N$ lattice with rapidity and colour flows.}}
\label{colour.1}
\end{center}
\end{figure}\\
\\
\noindent \textbf{Allowable vertices.} Each of the $N^2$ vertices contains 4 arrows (state variables) either pointing up or down (left or right). We define the allowable vertices as those shown in fig. \ref{nonfin}, which are of the same configuration as the six vertex model.\\
\\
\\
\\
\noindent \textbf{Boltzmann weights.} As usual we assign a specific algebraic Boltzmann weight to each vertex, labelled as $\omega_i$, $i = 1,\dots,6$, which for an inhomogeneous lattice, the algebraic weights are dependent on the horizontal and vertical rapidities, $\{u_i,v_j\}$, in addition to the horizontal and vertical colours, $\{\alpha_i,\beta_j \}$.\\
\begin{figure}[h!]
\begin{center}
\includegraphics[scale=0.26]{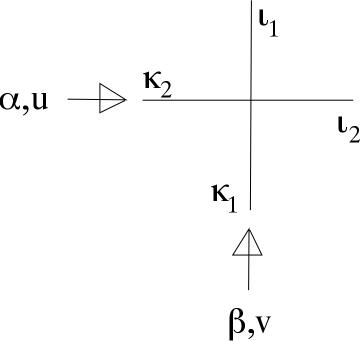}
\caption{\footnotesize{Parameter labelling for vertex $X_{\alpha,\beta} (u-v)^{\iota_1,\iota_2}_{\kappa_2,\kappa_1}$}. }
\label{nonfin2}
\end{center}
\end{figure}\\
Generally the colour variables appear in a non trivial manner in the weights, unlike the rapidities which always appear in the form $u_i - v_j$. \\
\\
The specific parameterizations of the six weights are given as the following,
\begin{equation}\begin{array}{lll}
X_{\alpha,\beta}(u-v)^{11}_{11}=&\omega_1(\alpha,\beta;u-v) =& 1 - \alpha \beta e^{u-v}\\
X_{\alpha,\beta}(u-v)^{22}_{22}=&\omega_2(\alpha,\beta; u-v) =& e^{u-v} - \alpha \beta \\
X_{\alpha,\beta}(u-v)^{21}_{12}=&\omega_3(\alpha,\beta; u-v) =&   \beta - \alpha e^{u-v}\\
X_{\alpha,\beta}(u-v)^{12}_{21}=&\omega_4(\alpha,\beta; u-v) =& \alpha - \beta e^{u-v}\\
X_{\alpha,\beta}(u-v)^{12}_{12}=&\omega_5(\alpha,\beta; u-v) =& e^{u-v} \sqrt{1-\alpha^2} \sqrt{1-\beta^2}\\
X_{\alpha,\beta}(u-v)^{21}_{21}=&\omega_6(\alpha,\beta; u-v) =& \sqrt{1-\alpha^2} \sqrt{1-\beta^2}
\end{array}\end{equation}\normalsize\\
\textbf{Free fermion model.} By definition, the model that we are dealing with is considered a free fermion model as the (homogeneous) weights satisfy the following algebraic expression,
\small\begin{equation*}
\omega_1 \omega_2 +\omega_3 \omega_4 = \omega_5 \omega_6 
\end{equation*}\normalsize
This has long standing implications for the inherent complexity of the model \cite{Bax3}\footnote{In the aforementioned work, Baxter showed that the free-fermion six-vertex model is equivalent to variations of the well studied Ising model and as such fundamental quantities (partition functions, etc.) of the six-vertex model can be expressed in terms of those of the regular square lattice Ising model.} as we shall see shortly. Additionally, when $\alpha = \beta = i$ we obtain the usual free fermion six vertex model in the absence of external fields. \\
\\
\textbf{Coloured Yang-Baxter equation.} These weights satisfy the following coloured Yang-Baxter equation,
\small\begin{equation*}\begin{split}
&\sum_{g_1, g_2, g_3 \in \{1,2\}}X_{\alpha_1,\alpha_2}(u_1-u_2)^{h_1 h_2}_{g_2 g_1} X_{\alpha_1,\alpha_3}(u_1-u_3)^{g_1 h_3}_{g_3 q_1}  X_{\alpha_2,\alpha_3}(u_2-u_3)^{g_2 g_3}_{q_3 q_2} \\
=& \sum_{g_1, g_2, g_3 \in \{1,2\}}X_{\alpha_2,\alpha_3}(u_2-u_3)^{h_2 h_3}_{g_3 g_2} X_{\alpha_1,\alpha_3}(u_1-u_3)^{h_1 g_3}_{q_3 g_1}  X_{\alpha_1,\alpha_2}(u_1-u_2)^{g_1 g_2}_{q_2 q_1}  ,
\end{split}\end{equation*}\normalsize
which we shall employ in the following sections. \\
\\
\textbf{Domain wall boundary conditions.} DWBC's, as in the last section, correspond to the top and bottom-most arrows pointing inward, and the left and right-most arrows pointing outward.\\
\begin{figure}[h!]
\begin{center}
\includegraphics[angle=0,scale=0.25]{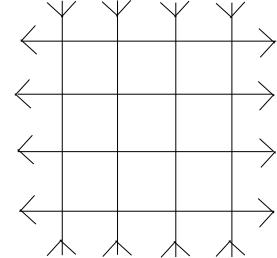}
\caption{\footnotesize{Typical example of DWBC's}}
\label{colourDWBC}
\end{center}
\end{figure}\\
\textbf{Domain wall partition function.} The DWPF $Z_{N}$, as always, is defined as the sum over all allowable weighted configurations of the $N \times N$ lattice that satisfy the required DWBC's,
\small\begin{equation*}
Z_{N}(\vec{u},\vec{v},\vec{\alpha},\vec{\beta})  = \sum_{\textrm{allowable}\atop{\textrm{configurations}}}\left( \prod_{\textrm{vertices}} X_{\alpha_{i},\beta_j} (u_{i}-v_j)\right)
\end{equation*}\normalsize\\
\textbf{Condition on rapidities for remainder of the chapter.} For the remainder of this chapter we require that the difference of rapidities, $u_{i}-v_{j}$, is equal to an integer multiple of $2 \pi i$,
\small\begin{equation*}
u_{i}-v_{j} = 2 n \pi i \textrm{  ,  } i, j \in \{ 1, \dots, N\} \textrm{  ,  } n \in \mathbb{Z}
\end{equation*}\normalsize
Given this condition, the dependence on the rapidities for the weights drops away and we are left with weights dependent solely on colour variables,
\begin{equation*}\begin{array}{lll}
\omega_1(\alpha,\beta;2 n \pi i ) =&\omega_2(\alpha,\beta;2 n \pi i ) =& 1 - \alpha \beta \\
\omega_3(\alpha,\beta; 2 n \pi i ) =&-\omega_4(\alpha,\beta; 2 n \pi i ) =&   \beta - \alpha \\
\omega_5(\alpha,\beta; 2 n \pi i) =&\omega_6(\alpha,\beta; 2 n \pi i) =& \sqrt{1-\alpha^2} \sqrt{1-\beta^2}
\end{array}\end{equation*}\normalsize
Therefore, for notational convenience we make the allocations,
\begin{equation}\begin{array}{lll}
\omega_1(\alpha,\beta;2 n \pi i ) =&\omega_2(\alpha,\beta;2 n \pi i ) =& a(\alpha,\beta )\\
\omega_3(\alpha,\beta; 2 n \pi i ) =&-\omega_4(\alpha,\beta; 2 n \pi i ) =& b(\alpha,\beta) \\
\omega_5(\alpha,\beta; 2 n \pi i) =&\omega_6(\alpha,\beta; 2 n \pi i) =& c(\alpha,\beta)
\end{array}\end{equation}\normalsize
We now concern ourselves with the partition function. Firstly we derive the determinant form of the DWPF for the model using the method given in \cite{Korepini,23}, and then use a standard technique devised in \cite{2} to find the homogeneous limit of the DWPF. We then give some interesting properties of the homogeneous DWPF involving the 2-Toda molecule equation. Lastly however, we show that the determinant form ultimately exists as a Cauchy determinant, and hence we obtain a product form for the DWPF.
\section{Determinant form of the DWPF}
We now follow the work of \cite{Korepini} by presenting the corresponding four properties which uniquely determines the closed form determinant expression for the DWPF.
\subsection{Korepin-like properties and derivation.}
\begin{figure}[h!]
\begin{center}
\includegraphics[angle=0,scale=0.19]{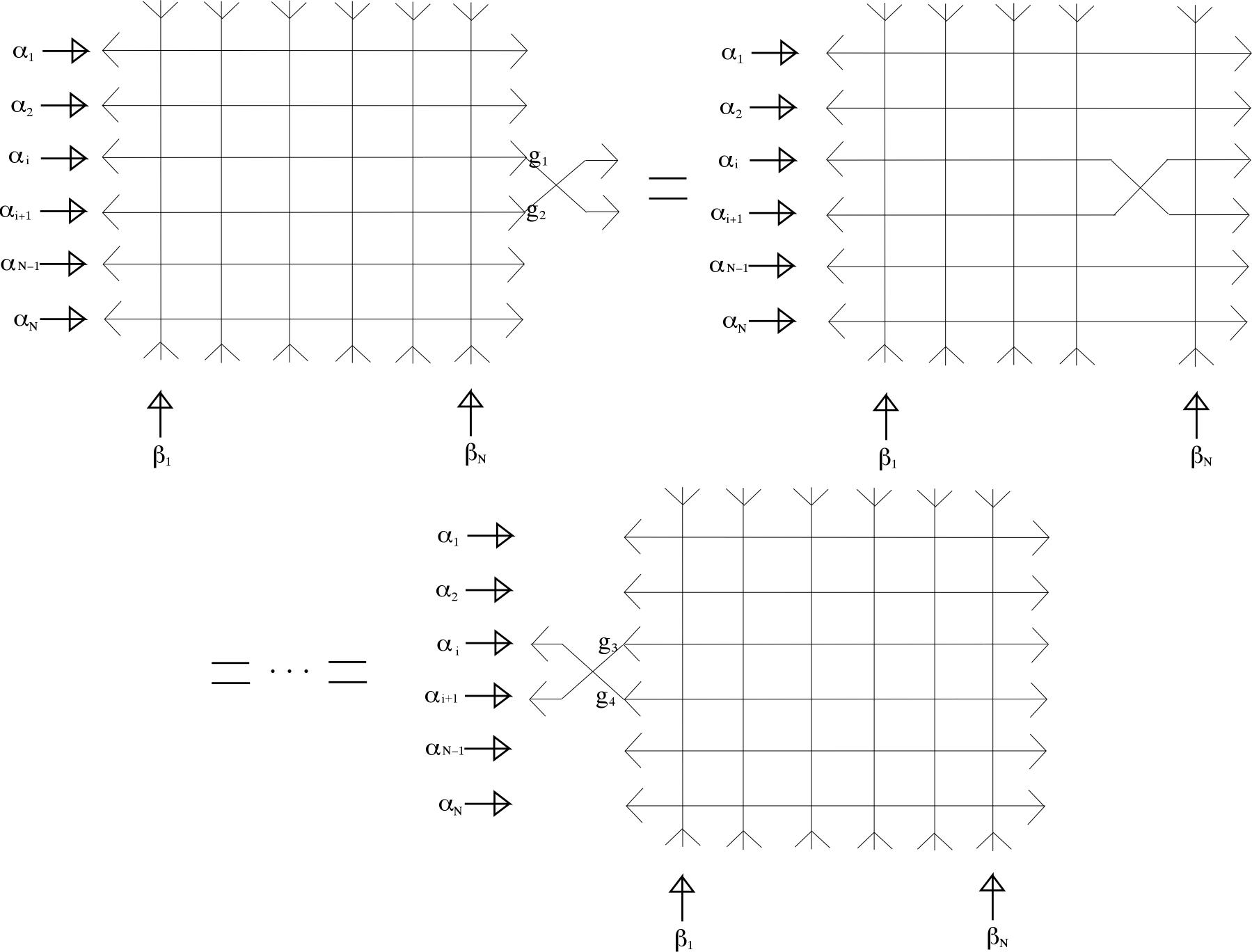}
\caption{\footnotesize{Graphical representation of $ Z_N(\vec{\alpha},\vec{\beta})  \omega_1(\alpha_{i}, \alpha_{i+1}) $ and the subsequent Yang-Baxter procedure}}
\label{sym}
\end{center}
\end{figure}

\noindent \textbf{Property 1.} The initial condition is given as,
\small\begin{equation*}
Z_1(\alpha_1,\beta_1) = c(\alpha_1,\beta_1) = \sqrt{1-\alpha^2_1}\sqrt{1-\beta^2_1}
\end{equation*}\normalsize
\textbf{Proof.} Simply let $N=1$ and we see that the DWBC's demand that the arrangement of the entire lattice is a single $\omega_6$ vertex. $\square$\\
\\
\textbf{Property 2.} $Z_N(\vec{\alpha},\vec{\beta})$ is a polynomial of order $N-1$ in $\alpha_{i}$ and $\beta_{j}$, $i, j \in \{1, \dots, N \}$, up to a factor of $\sqrt{1-\alpha^2_{i}}$ and $\sqrt{1-\beta^2_{j}}$ respectively. \\
\\
\textbf{Proof.}  It is elementary to notice that the DWBC's force each row (column) of an allowable configuration to contain an odd number of $c$ vertices. $\square$ \\
\\
\textbf{Property 3.} $Z_N(\vec{\alpha},\vec{\beta})$ is a symmetric function in each set of colours, $\{\alpha\}$ and $\{\beta\}$.\\
\\
\textbf{Proof.} Consider the graphical representation of $Z_N(\vec{\alpha},\vec{\beta})  \omega_1(\alpha_{i}, \alpha_{i+1}) $, $1 \le i \le N-1$, as shown in fig. \ref{sym}. We notice that since state variables $g_1$ and $g_2$ are fixed (all other configurations produce non allowable vertices), we can use the Yang-Baxter equation to shift the intersection of $\alpha_{i}$ and $\alpha_{i+1}$ through to the left side of the lattice as shown in the diagram. \\
\\
When this process is complete we notice that what remains is the partition function with colours $\alpha_{i}$ and $\alpha_{i+1}$ exchanged, (since state variables $g_3$ and $g_4$ are fixed), multiplied by $\omega_2(\alpha_{i },\alpha_{i+1})$. We can now achieve this result for any permutation of the $\{\alpha\}$ colours by performing this process the required number of times. The method of the proof for the $\{\beta\}$ colours is equivalent and involves applying the Yang-Baxter equation along the columns instead of the rows. $\square$
\\
\\
\textbf{Property 4.} Fixing the colours such that $\alpha_1 \beta_1 = 1$, we obtain the following recursion relation,
\small\begin{equation}\begin{split}
Z_N(\vec{\alpha},\vec{\beta})|_{\alpha_1 \beta_1 = 1} =& (-1)^{N-1}c(\alpha_1,\beta_1)|_{\alpha_1 \beta_1 = 1} \left\{ \prod^N_{j=2}b(\alpha_1,\beta_j) \right\} \left\{\prod^N_{i =2}b(\alpha_{i},\beta_1)\right\} \\
&\times Z_{N-1}(\{\alpha\}_{i \in \{2,\dots,N\}} , \{\beta\}_{j \in \{2,\dots,N\}})
\label{c.other6}
\end{split}\end{equation}\normalsize
\textbf{Proof.} First we notice that the vertex at position $(1,1)$ is forced to either be an $\omega_2$ or an $\omega_6$ due to DWBC. The condition, $\alpha_1 \beta_1 = 1$, further specializes this vertex to an $\omega_6$, as an $\omega_2$ vertex under this condition is zero. With the vertex at $(1,1)$ forced to be $\omega_6$, we notice that due to the DWBC's, the entire first row is fixed into one string of $\omega_3$ vertices and the entire first column is fixed into one string of $\omega_4$ vertices. The remaining $(N-1)^2$ vertices are arranged (almost miraculously) into exactly the $Z_{N-1}$ configuration with colours $\alpha_1$ and $\beta_1$ missing. $\square$ 
\begin{figure}[h!]
\begin{center}
\includegraphics[angle=0,scale=0.21]{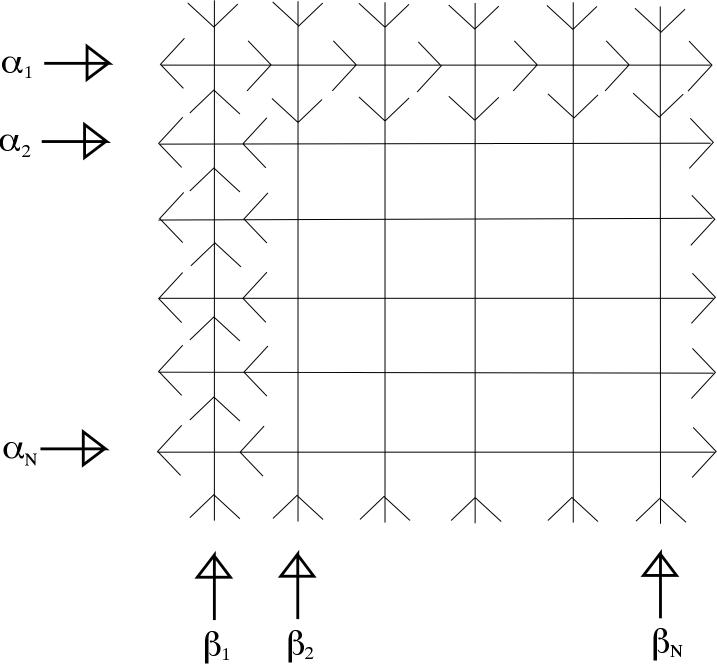}
\caption{\footnotesize{$Z_N$ under the condition $\alpha_1 \beta_1 = 1$}}
\label{prop4}
\end{center}
\end{figure}\\
\\
We now show that the above properties uniquely determine the DWPF.\\
\\
\textbf{A result regarding the Korepin-like properties of $Z_N$.}
\begin{proposition}
The above four properties uniquely determine the DWPF of the trigonometric coloured Felderhof vertex model.
\end{proposition}
\textbf{Proof.} We proceed by assuming that there exist two expressions which satisfy the above four properties, the partition function $Z_N(\vec{\alpha},\vec{\beta})$ and an entirely different function $\field{Z}^F_N(\vec{\alpha},\vec{\beta})$. By property 1 we obtain the base case,
\small\begin{equation*}
Z_1(\alpha_1,\beta_1) = \field{Z}^F_1(\alpha_1,\beta_1) =  c(\alpha_1,\beta_1)
\end{equation*}\normalsize
Let us now assume that the two expressions are equal up to some integer $N$, and prove the $N+1$ case.\\
\\
From property 3 both expressions are symmetric in the $\{ \alpha\} = \{\alpha_1, \dots, \alpha_{N+1}\}$ variables and the $\{ \beta\}=\{\beta_1, \dots, \beta_{N+1}\}$ variables. From property 2 both expressions are order $N$ polynomials in $\alpha_1$ and $\beta_1$, up to a factor of $\sqrt{1-\alpha^2_{1}} \sqrt{1-\beta^2_{1}}$. From property 4, and the fact that $Z_N = \field{Z}^F_N$, we can obtain the value of this polynomial at the $N+1$ points,
\small\begin{equation*}
b(\alpha_1, \beta_j) = 0 \textrm{  ,  }i,j \in \{ 1, \dots,N+1\}
\end{equation*}\normalsize
and similarly with $\beta_1$, which provide the necessary equations to obtain the coefficients of the determinant. $\square$\\
\\
\textbf{Determinant form for $\mathbf{Z_N}$.} Following the work of \cite{23} we now present the following determinant solution for the  $N \times N$ DWPF,
\small\begin{equation}\begin{split}
Z_N(\vec{\alpha},\vec{\beta}) = &\left\{\frac{\prod^N_{i,j =1}(\alpha_{i}-\beta_j) (1-\alpha_{i}\beta_j)}{\prod_{1 \le i < j \le N} (\alpha_{i}-\alpha_{j}) (\beta_j - \beta_i)}\right\}   \\
&\times \left\{ \prod^N_{i=1}\sqrt{1-\alpha^2_i}\sqrt{1-\beta^2_i} \right\} \textrm{det}\left[ \frac{1}{(\alpha_{i}-\beta_j)(1-\alpha_{i}\beta_j)} \right]^N_{i,j=1}
\label{c.18}\end{split}\end{equation}\normalsize
and show that it satisfies all four of the required properties.\\
\\
\textbf{Verification that $\mathbf{Z_N}$ satisfies the Korepin-like properties.}\\
\\
\textbf{Verification of property 1.} This is the most obvious case, simply taking $N=1$ in eq. \ref{c.18} is sufficient.\\
\\
\textbf{Verification of property 2.} We wish to show that $Z_N$ is a polynomial of degree $N-1$ in $\alpha_1$ with a factor of $\sqrt{1-\alpha^2_1}$, (we only have to verify for $\alpha_1$ due to condition 3).\\
\\
First we note the explicit factor of $\sqrt{1-\alpha^2_1}$ in eq. \ref{c.18}. In order to show that the remaining part of eq. \ref{c.18} is a polynomial of $\alpha_1$ and not a rational function, it suffices to show that the poles have zero residues. The first poles are located at the denominator of the determinant,
\small\begin{equation*}
\lim_{\alpha_{1}\rightarrow \beta_j \atop{\textrm{  or  } \alpha_{1}\beta_j \rightarrow 1}}(\alpha_{1}-\beta_j)(1-\alpha_{1}\beta_j)
\end{equation*}\normalsize
However, it is quite obvious that these poles are always cancelled by the numerator of the partition function in the limit. \\
\\
The second pole comes from the denominator of the partition function,
\small\begin{equation*}
\lim_{\alpha_{1} \rightarrow \alpha_{i}\atop{\textrm{$2 \le i \le N$}}}\prod^{N}_{i = 2}  (\alpha_{1}-\alpha_{i})
\end{equation*}\normalsize
However, a close examination of the determinant reveals that this pole would indeed be cancelled by the zero that would occur from rows $1$ and $i$ being exactly the same in the determinant. Thus, $Z_N(\vec{\alpha},\vec{\beta})$ is indeed a polynomial in $\alpha_1$ (with a factor of $\sqrt{1-\alpha^2_1}$) as opposed to a rational function.\\
\\
To find the degree of this polynomial we note that the numerator is of order $2N$ in $\alpha_1$, while the denominator is of order $(N-1)$. In the determinant, we note that the only place that $\alpha_1$ appears is in the first row, hence the determinant is a polynomial in the denominator. Thus eq. \ref{c.18} is a polynomial in $\alpha_1$ of order $2N - (N-1) - 2 = N-1$. A similar analysis can be done for $\beta_1$.\\
\\
\textbf{Verification of property 3.}  To see that $Z_N$ is symmetric in $\{\alpha\}$, we simply exchange $\alpha_{i}$ with $\alpha_{j}$, $i \neq j$, in eq. \ref{c.18}. The numerator is invariant under this process, but the denominator obtains up a minus sign. To the determinant however, this process is equivalent to exchanging two rows. When we interchange these two rows back in their original order we obtain an additional minus sign, thus leaving eq. \ref{c.18} invariant. It is an equivalent process to show that $Z_N$ is symmetric in $\{\beta\}$, but this time we obviously switch the columns of the determinant.\\
\\
\textbf{Verification of property 4.}   Finally, we wish to show that eq. \ref{c.18} obeys the recursion relation. In order to do this, we shall split the multiplicative factor of eq. \ref{c.18} into those parts that contain $\alpha_1$ and $\beta_1$ and those that do not,
\small\begin{equation*}\begin{split}
Z_N(\vec{\alpha},\vec{\beta}) =& (\alpha_1 -\beta_1 )(1 - \alpha_1 \beta_1) \sqrt{1-\alpha^2_1}\sqrt{1-\beta^2_1} \\
 & \times \left\{ \frac{\prod^N_{i =2}(\alpha_{i} - \beta_1) (1-\alpha_{i}\beta_1)(\alpha_{1}-\beta_i) (1-\alpha_{1}\beta_i)}{\prod^N_{j = 2} (\alpha_{1}-\alpha_{j})(\beta_j - \beta_1)} \right\} 
\end{split}\end{equation*}\normalsize
\small\begin{equation*}\begin{split}
&\times \left\{ \prod^N_{i=2}\sqrt{1-\alpha^2_i}\sqrt{1-\beta^2_i} \right\} \left\{ \frac{\prod^N_{i,j  =2} (\alpha_{i}-\beta_j)(1-\alpha_{i}\beta_j)}{\prod_{2 \le i < j \le N} (\alpha_{i}-\alpha_{j}) (\beta_j - \beta_i)} \right\} \\
&\times \textrm{det}\left[ \frac{1}{(\alpha_{i}-\beta_j)(1-\alpha_{i} \beta_j)} \right]^N_{i,j =1}
\end{split}\end{equation*}\normalsize
Absorbing $(1 - \alpha_1 \beta_1)$ into the first column (equivalently row) of the determinant and taking the limit $\alpha_1 \beta_1 \rightarrow 1$, the entries of the first column of the determinant evaluate to zero except for the first entry,
\small\begin{equation*}\begin{split}
\lim_{\alpha_1 \beta_1 \rightarrow 1} (1 - \alpha_1 \beta_1)  \textrm{det}\left[ \phi(\alpha_{\gamma},\beta_k) \right]^N_{\gamma,k =1} =&\left| \begin{array}{cccc}
\frac{1}{(\alpha_1-\beta_1)} &\phi(\alpha_{1},\beta_2) & \dots & \phi(\alpha_{1},\beta_N)\\
0 &\vdots& & \vdots\\
\vdots & \vdots & & \vdots \\
0 &\phi(\alpha_{N},\beta_2) & \dots & \phi(\alpha_{N},\beta_N)
\end{array} \right| \\
 =&\frac{1}{(\alpha_1-\beta_1)}\textrm{det}\left[\phi(\alpha_{i},\beta_j) \right]^N_{i,j=2 } 
\end{split}\end{equation*}\normalsize
where,
\small\begin{equation*}
\phi(\alpha_{i},\beta_j) = \frac{1}{(\alpha_{i}-\beta_j)(1 - \alpha_{i} \beta_j)}
\end{equation*}\normalsize
Additionally, taking the limit in the multiplicative factor we obtain,
\small\begin{equation*}\begin{split}
&\lim_{\alpha_1 \beta_1 \rightarrow 1} \left\{ \frac{\prod^N_{i =2}(\alpha_{i}-\beta_1)(1-\alpha_{i}\beta_1)}{\prod^N_{i = 2} (\alpha_{1}-\alpha_{i})} \right\} \left\{ \frac{\prod^N_{j =2} (\alpha_{1}-\beta_j)(1-\alpha_{1}\beta_j)}{\prod^N_{j=2} (\beta_j - \beta_1)}\right\} \\
=& (-1)^{N-1} \left\{ \frac{1}{\alpha^{N-1}_1}\prod^N_{i =2}(\alpha_{i}-\beta_1)\right\} \left\{ \frac{1}{\beta^{N-1}_1} \prod^N_{j =2} (\alpha_{1}-\beta_j)\right\}\\
=& (-1)^{N-1} \left\{\prod^N_{i=2}b(\alpha_{i},\beta_1) \right\} \left\{\prod^N_{j=2}b(\alpha_1,\beta_j) \right\}
\end{split}\end{equation*}\normalsize
Therefore putting everything together we obtain exactly eq. \ref{c.other6}, which verifies that eq. \ref{c.18} satisfies the four properties.
\subsection{Homogeneous lattice and the 2-Toda molecule equation}
In order to find the homogeneous partition function, we let the vertical and horizontal colours be parameterized by the the same variable respectively,
\small\begin{equation}
\alpha_{i} \rightarrow \alpha \textrm{  ,  } \beta_j \rightarrow \beta \textrm{  ,  } 1 \le i, j \le N
\label{c.21}\end{equation}\normalsize
The method in which we do this requires a little explaining, because simply taking the limit leads to some obvious unresolved singularities. We will first deal with the $\alpha$'s in each row.\\
\\
Setting $\alpha_1 \rightarrow \alpha$ we obtain for the partition function,
\small\begin{equation*}\begin{split}
Z_N =& \frac{\prod^N_{i = 2}\prod^N_{j=1}(\alpha_{i}-\beta_{j})(1-\alpha_{i}\beta_{j})}{\prod_{2 \le i < j \le N} (\alpha_{j} - \alpha_{i})\prod_{1 \le i < j \le N} (\beta_{i} - \beta_{j})} \left\{ \prod^N_{i=2}\sqrt{1-\alpha^2_{i}} \right\} \left\{ \prod^N_{j=1}\sqrt{1-\beta^2_j} \right\}  \\
& \times \sqrt{1-\alpha^2} \left\{ \frac{\prod^N_{j=1} (\alpha - \beta_j)(1 - \alpha \beta_j )}{\prod^N_{i=2 }(\alpha_{i}-\alpha)} \right\} \textrm{det}\left[\begin{array}{c}
\phi(\alpha,\beta_j) \\
\phi(\alpha_{i},\beta_j) 
\end{array} \right]_{i=2,\dots,N\atop{j=1,\dots,N}}
\end{split}\end{equation*}\normalsize\\
\textbf{Eliminating poles by expanding entries of the determinant.} Using $\alpha_2 = \alpha + (\alpha_2 - \alpha)$, entries in the second row of the determinant can be expanded using the translation operator,
\small\begin{equation}\begin{split}
\phi(\alpha + (\alpha_2 - \alpha),\beta_j) &= \exp \left\{ (\alpha_2 - \alpha)\partial_{\alpha} \right\}\phi(\alpha ,\beta_j)  \\
&= \sum^{\infty}_{n=0} \frac{1}{n!}(\alpha_2 - \alpha)^n \partial^n_{\alpha} \phi(\alpha,\beta_j)
\label{c.24} \end{split}
\end{equation}\normalsize
hence the first two rows of the determinant are now,
\small\begin{equation*}
\left[ \begin{array}{c}
\phi(\alpha,\beta_j) \\
\phi(\alpha,\beta_j) + (\alpha_2-\alpha)\partial_{\alpha}\phi(\alpha,\beta_j) + \dots 
\end{array}\right]_{j=1,\dots,N}
\end{equation*}\normalsize
Subtracting the first row from the second row and taking out a common factor of $(\alpha_2-\alpha)$ from the second row, the entries of the second row become,
\small\begin{equation*}
\sum^{\infty}_{n=1} \frac{1}{n!}(\alpha_2 - \alpha)^{n-1} \partial^n_{\alpha}\phi(\alpha,\beta_j)
\end{equation*}\normalsize
Noticing that the denominator of the partition function contains one factor of $(\alpha_2-\alpha)$, we can now eliminate this potential pole with the factor that has been extracted from the determinant, thus allowing us to take the limit $\alpha_2 \rightarrow \alpha$,
\small\begin{equation*}\begin{split}
Z_N =& \frac{\prod^N_{i = 3}\prod^N_{j=1}(\alpha_{i}-\beta_{j})(1-\alpha_{i}\beta_{j})}{\prod_{3 \le i < j \le N} (\alpha_{j} - \alpha_{k})\prod_{1 \le i <j \le N} (\beta_{i} - \beta_{j})} \left\{ \prod^N_{i=3}\sqrt{1-\alpha^2_{i}} \right\} \left\{ \prod^N_{j=1}\sqrt{1-\beta^2_j} \right\}  \\
& \times \left\{ \frac{\sqrt{1-\alpha^2} \prod^N_{j=1} (\alpha - \beta_j)(1 - \alpha \beta_j )}{\prod^N_{i=3 }(\alpha_{i}-\alpha)} \right\}^2 \textrm{det}\left[\begin{array}{c}
\partial^{i_1-1}_{\alpha}\phi(\alpha,\beta_k) \\
\phi(\alpha_{i_2},\beta_k) 
\end{array} \right]^{i_1=1,2\atop{i_2=3,\dots,N} }_{j=1,\dots,N}
\end{split}\end{equation*}\normalsize
Following the same procedure for the third row of the determinant, the first 3 rows of the determinant have the form,
\small\begin{equation*}
\left[\begin{array}{c}
\phi(\alpha,\beta_j) \\
\partial_{\alpha}\phi(\alpha,\beta_j)\\
\phi(\alpha,\beta_j) + {(\alpha_3-\alpha)}\partial_{\alpha}\phi(\alpha,\beta_j)+ \frac{(\alpha_3-\alpha)^2}{2!}\partial^2_{\alpha}\phi(\alpha,\beta_j) + \dots 
\end{array} \right]_{j=1,\dots,N}
\end{equation*}\normalsize
Subtracting the first and second rows (with appropriate factors) from the third row, and then taking out a common factor of $\frac{(\alpha_3-\alpha)^2}{2!}$ and eliminating it with the same factor on the denominator, the partition function in the limit $\alpha_3 \rightarrow \alpha$ becomes,
\small\begin{equation*}\begin{split}
Z_N =&\frac{1}{2!} \frac{\prod^N_{i = 4}\prod^N_{j=1}(\alpha_{i}-\beta_{j})(1-\alpha_{i}\beta_{j})}{\prod_{4 \le i < j \le N} (\alpha_{j} - \alpha_{k})\prod_{1 \le i <j \le N} (\beta_{i} - \beta_{j})} \left\{ \prod^N_{i=4}\sqrt{1-\alpha^2_{i}} \right\} \left\{ \prod^N_{j=1}\sqrt{1-\beta^2_j} \right\}  \\
& \times \left\{ \frac{\sqrt{1-\alpha^2} \prod^N_{j=1} (\alpha - \beta_j)(1 - \alpha \beta_j )}{\prod^N_{i=4 }(\alpha_{i}-\alpha)} \right\}^3 \textrm{det}\left[\begin{array}{c}
\partial^{i_1-1}_{\alpha}\phi(\alpha,\beta_k) \\
\phi(\alpha_{i_2},\beta_k) 
\end{array} \right]^{i_1=1,2,3\atop{i_2=4,\dots,N} }_{j=1,\dots,N}
\end{split}\end{equation*}\normalsize
Continuing this procedure now to row $N$ and eliminating all of the poles in $\{ \alpha\}$, we obtain,
\small\begin{equation*}\begin{split}
Z_N(\alpha,\vec{\beta}) =& \frac{1}{\left[\prod^{N-1}_{n=1}n!\right]} \left\{\sqrt{1-\alpha^2} \right\}^N  \left\{ \prod^N_{j=1}\sqrt{1-\beta^2_j} \right\} \\
& \times \frac{\prod^N_{j=1}(\alpha-\beta_{j})^N(1-\alpha \beta_{j})^N}{\prod_{1 \le i < j \le N} (\beta_{i} - \beta_{j})} \textrm{det}\left[ \partial^{i-1}_{\alpha}\phi(\alpha,\beta_j) \right]_{i,j=1,\dots,N}
\end{split}\end{equation*}\normalsize
which takes care of the limit $\alpha_{i} \rightarrow \alpha \textrm{  ,  } 1 \le i \le N$. The limit $\beta_{j} \rightarrow \beta \textrm{  ,  }  1 \le j \le N$, can obviously be performed in exactly the same manner, but this time instead of dealing with rows, we deal with columns. First however, we need to take out the negatives in the Vandermonde expression,
\small\begin{equation*}
\frac{1}{\prod_{1 \le i < j \le N} (\beta_{i} - \beta_{j})} = \frac{(-1)^{\frac{N(N-1)}{2}}}{\prod_{1 \le i < j \le N} (\beta_{j} - \beta_{i})}
\end{equation*}\normalsize
Therefore, going through the exact same procedure in order to take the limit, but this time with columns instead of rows, we obtain the DWPF with homogeneous weights,
\small\begin{equation}\begin{split}
Z_N(\alpha,\beta) = &\frac{(-1)^{\frac{N(N-1)}{2}}}{\left[\prod^{N-1}_{n=1} n! \right]^2}\left\{\sqrt{1-\alpha^2}\sqrt{1-\beta^2} \right\}^N  \\ 
&\times \{(\alpha-\beta)(1-\alpha \beta) \}^{N^2} \textrm{det}\left[ \partial^{i-1}_{\alpha} \partial^{j-1}_{\beta} \phi(\alpha,\beta) \right]_{i,j=1,\dots,N}
\label{goatrage}\end{split}\end{equation}\normalsize\\
\textbf{Properties of the homogeneous lattice.} It was shown in \cite{Sog} that the determinant solution of the homogenous six vertex model, which contains one parameter, is a $\tau$-function of the 1-Toda molecule equation. We now proceed to show that our current homogeneous determinant solution is a $\tau$-function of the 2-Toda molecule equation.\\
\\
In order to make the connection however, we need to introduce some additional definitions which ultimately lead to the \textit{bilinear Jacobi determinant identity} \cite{11,TOM}.\\
\\
\textbf{Cofactors.} Consider a matrix $A = (a_{ij})_{1 \le i,j \le n}$ whose determinant is $D$. The cofactor $\Delta_{ij}$ with respect to $a_{ij}$ is the determinant of the matrix obtained by eliminating the $i$th row and the $j$th column from $A$, multiplied by $(-1)^{i+j}$. Given this definition, single row and single column Laplace expansion of $D$ can be respectively expressed as,
\small\begin{equation}\begin{split}
D = &\sum^n_{i=1} a_{ij} \Delta_{ij} \textrm{  ,  } j \in \{ 1, \dots, n\}\\
 =& \sum^n_{j=1} a_{ij} \Delta_{ij} \textrm{  ,  } j \in \{ 1, \dots, n\}
\label{1.23}
\end{split}\end{equation}\normalsize
These are special cases of the orthogonality relations,
\small\begin{equation}\begin{split}
\sum^n_{i=1} a_{ij} \Delta_{ik} = \delta_{jk} D\\
\sum^n_{j=1} a_{ij} \Delta_{kj} = \delta_{ik} D
\label{1.24}\end{split}\end{equation}\normalsize
In order to prove the orthogonality conditions, note that if $j \ne k$ or $i \ne k$ respectively, then the corresponding determinant has repeated rows or columns.\\
\\
\textbf{Additional notation.} The $(n-1)$th-order determinant obtained by by eliminating the $j$th row and the $k$th column from an $n$th-order determinant $D=$ det$(a_{ij})_{ i,j = 1,\dots, n}$ is called the $(j,k)$th minor of $D$, which we shall denote as $D\left[ j \atop{k} \right]$. As defined above, the cofactor $\Delta_{jk}$ equals $D\left[ j \atop{k} \right]$ multiplied by the signature $(-1)^{j+k}$. That is,
\small\begin{equation}
\Delta_{jk} = (-1)^{j+k} D\left[ j \atop{k} \right] 
\label{1.19}\end{equation}\normalsize
where,
\small\begin{equation}
D\left[ j \atop{k} \right] = \textrm{det}\left(a_{ij} \right)_{ i= 1 ,\dots, \hat{j},\dots,n \atop{j=1,\dots,\hat{k},\dots,n}}
\label{1.20}\end{equation}\normalsize
In the same way, we denote the $(N-2)$nd-order determinant obtained by eliminating the $j$th and $k$th rows and the $l$th and $m$th columns from the determinant $D$ as \scriptsize$D\left[ \begin{array}{cc} j & k \\ l & m \end{array}  \right]$\normalsize. This notation naturally leads to the bilinear Jacobi determinant identity given below.\\
\\
\textbf{Bilinear Jacobi determinant identity.}
\begin{lemma}
\small\begin{equation}
D\left[ \begin{array}{c} n-1\\ n-1 \end{array} \right] D\left[ \begin{array}{c} n\\ n \end{array} \right] - D\left[ \begin{array}{c} n-1\\ n \end{array} \right]D\left[ \begin{array}{c} n\\ n-1 \end{array} \right] = D\left[ \begin{array}{cc} n-1 & n \\ n-1 & n \end{array} \right] D
\label{1.21}\end{equation}\normalsize
\end{lemma}
\textbf{Proof.} We begin by considering the product of the general $n$th order determinant $D$, and a peculiar $n$th order determinant of cofactors, which we denote by $\left|\begin{array}{cc}
\field{I}_r & \Delta^{(12)}\\
0_{(n-r)\times r} & \Delta^{(22)}
\end{array}\right|$,
\small\begin{equation}
D \left|\begin{array}{cc}
\field{I}_r & \Delta^{(12)}\\
0_{(n-r)\times r} & \Delta^{(22)}
\end{array}\right| = \left|\begin{array}{cc}
A_{11} & A_{12}\\
A_{21} & A_{22}
\end{array}\right|\left|\begin{array}{cc}
\field{I}_r & \Delta^{(12)}\\
0_{(n-r)\times r} & \Delta^{(22)}
\end{array}\right|
\label{1.22}\end{equation}\normalsize
where $\field{I}_r$ is the $r \times r$ identity matrix, $0_{(n-r)\times r}$ is the $(n-r) \times r$ zero matrix and,
\small\begin{equation*}\begin{array}{ll}
A_{11} =   \left( a_{ij} \right)_{i,j=1\dots,r}
& A_{12} =  \left( a_{ij} \right)_{i=1\dots,r\atop{j=r+1,\dots,n}}\\
 A_{21} =   \left( a_{ij} \right)_{i=r+1\dots,n\atop{j=1,\dots,r}}
&A_{22} =  \left( a_{ij} \right)_{i,j=r+1\dots,n} \\
\Delta^{(12)} = \left( \begin{array}{ccc}
\Delta_{r+1,1} & \dots & \Delta_{n,1}\\
\vdots & & \vdots\\
\Delta_{r+1,r} & \dots & \Delta_{n,r}
\end{array}\right) 
& \Delta^{(22)} = \left( \begin{array}{ccc}
\Delta_{r+1,r+1} & \dots & \Delta_{n,r+1}\\
\vdots & & \vdots\\
\Delta_{r+1,n} & \dots & \Delta_{n,n}
\end{array}\right)
\end{array}\end{equation*}\normalsize
Therefore, expanding eq. \ref{1.22} we obtain,
\small\begin{equation}
\left|\begin{array}{cc}
A_{11} & A_{12}\\
A_{21} & A_{22}
\end{array}\right|\left|\begin{array}{cc}
\field{I}_r & \Delta^{(12)}\\
0_{(n-r)\times r} & \Delta^{(22)}
\end{array}\right| = \left|\begin{array}{cc}
A_{11} & A_{11} \Delta^{(12)} + A_{12} \Delta^{(22)}\\
A_{21} & A_{21} \Delta^{(12)} + A_{22} \Delta^{(22)}
\end{array}\right| 
\label{1.25}\end{equation}\normalsize
Focusing on the entries in the top right hand corner we have the following,
\small\begin{equation*}\begin{split}
A_{11} \Delta^{(12)} = &\left( \begin{array}{ccc}
\sum^r_{j=1} a_{1,j} \Delta_{r+1,j}  & \dots & \sum^r_{j=1} a_{1,j} \Delta_{n,j} \\
\vdots  & & \vdots\\
\sum^r_{j=1} a_{r,j} \Delta_{r+1,j}  & \dots & \sum^r_{j=1} a_{r,j} \Delta_{n,j}
\end{array}\right)\\
A_{12} \Delta^{(22)} = &\left( \begin{array}{ccc}
\sum^n_{j=r+1} a_{1,j} \Delta_{r+1,j}  & \dots & \sum^n_{j=r+1} a_{1,j} \Delta_{n,j} \\
\vdots & & \vdots\\
\sum^n_{j=r+1} a_{r,j} \Delta_{r+1,j}  & \dots & \sum^n_{j=r+1} a_{r,j} \Delta_{n,j}
\end{array}\right)
\end{split}\end{equation*}\normalsize
Thus considering their sum we obtain, 
\small\begin{equation*}\begin{split}
A^{(11)} \Delta_{12} + A^{(22)} \Delta_{22} &= \left( \begin{array}{ccc}
\sum^n_{j=1} a_{1,j} \Delta_{r+1,j} & \dots & \sum^n_{j=1} a_{1,j} \Delta_{n,j} \\
\vdots  & & \vdots\\
\sum^n_{j=1} a_{r,j} \Delta_{r+1,j}  & \dots & \sum^n_{j=1} a_{r,j} \Delta_{n,j}
\end{array}\right)\\
&= \left( \begin{array}{cccc}
\delta_{1,r+1} & \delta_{1,r+2} & \dots & \delta_{1,n} \\
\vdots & \vdots & & \vdots\\
\delta_{r,r+1} &\delta_{r,r+2} & \dots & \delta_{r,n}
\end{array}\right) D \\
&= 0_{r\times (n-r)}
\end{split}\end{equation*}\normalsize
Similarly considering the entries in the bottom right hand corner we have,
\small\begin{equation*}\begin{split}
A_{21} \Delta^{(12)} =& \left( \begin{array}{cccc}
\sum^r_{j=1} a_{r+1,j} \Delta_{r+1,j}  & \dots & \sum^r_{j=1} a_{r+1,j} \Delta_{n,j} \\
\vdots  & & \vdots\\
\sum^r_{j=1} a_{n,j} \Delta_{r+1,j}  & \dots & \sum^r_{j=1} a_{n,j} \Delta_{n,j}
\end{array}\right)\\
A_{22} \Delta^{(22)} = & \left( \begin{array}{ccc}
\sum^n_{j=r+1} a_{r+1,j} \Delta_{r+1,j} &  \dots & \sum^n_{j=r+1} a_{r+1,j} \Delta_{n,j} \\
\vdots & & \vdots\\
\sum^n_{j=r+1} a_{n,j} \Delta_{r+1,j} & \dots & \sum^n_{j=r+1} a_{n,j} \Delta_{n,j}
\end{array}\right)
\end{split}\end{equation*}\normalsize
whose sum is,
\small\begin{equation*}\begin{split}
A_{21} \Delta^{(12)} + A^{(22)} \Delta_{22}& = \left( \begin{array}{ccc}
\sum^n_{j=1} a_{r+1,j} \Delta_{r+1,j}  & \dots & \sum^n_{j=1} a_{r+1,j} \Delta_{n,j} \\
\vdots & & \vdots\\
\sum^n_{j=1} a_{n,j} \Delta_{r+1,j}  & \dots & \sum^n_{j=1} a_{n,j} \Delta_{n,j}
\end{array}\right)\\
&= \left( \begin{array}{ccc}
\delta_{r+1,r+1}  & \dots & \delta_{r+1,n} \\
\vdots  & & \vdots\\
\delta_{n,r+1}  & \dots & \delta_{n,n}
\end{array}\right) D \\
& = \field{I}_{n-r} D
\end{split}\end{equation*}\normalsize
Thus, eq. \ref{1.25} becomes,
\small\begin{equation}
D \left|\begin{array}{cc}
\field{I}_r & \Delta^{(12)}\\
0_{(n-r)\times r} & \Delta^{(22)}
\end{array}\right|=  \left|\begin{array}{cc}
A_{11} & A_{11} \Delta^{(12)} + A_{12} \Delta^{(22)}\\
A_{21} & A_{21} \Delta^{(12)} + A_{22} \Delta^{(22)}
\end{array}\right| = \left|\begin{array}{cc}
A_{11} &0_{r\times (n-r)}\\
A_{21} & \field{I}_{n-r} D
\end{array}\right| 
\label{1.26}\end{equation}\normalsize
where $0_{r\times (n-r)}$ is the $r \times (n-r)$ zero matrix and $\field{I}_{n-r}$ is the $(n-r) \times (n-r)$ identity matrix. \\
\\
Using the following simplifications,
\small\begin{equation*}\begin{array}{lll}
 \left|\begin{array}{cc}
\field{I}_r & \Delta^{(12)}\\
0_{(n-r)\times r} & \Delta^{(22)}
\end{array}\right| &= \textrm{det}\left(\field{I}_r \right)\textrm{det}\left( \Delta^{(22)} \right)& = \textrm{det}\left( \Delta^{(22)} \right)\\
\left|\begin{array}{cc}
A_{11} &0_{r\times (n-r)}\\
A_{21} & \field{I}_{n-r} D
\end{array}\right|  &= \textrm{det}\left( A_{11} \right)\textrm{det}\left( \field{I}_{n-r} D \right)& = \textrm{det}\left(  A_{11} \right) D^{n-r}
\end{array}\end{equation*}\normalsize
eq. \ref{1.26} becomes simply, $ \textrm{det}( \Delta^{(22)})D =  \textrm{det}(A_{11}) D^{n-r}$. Writing the above out explicitly,
\small\begin{equation}
\left| \begin{array}{ccc}
\Delta_{r+1,r+1} & \dots & \Delta_{n,r+1}\\
\vdots & & \vdots\\
\Delta_{r+1,n} & \dots & \Delta_{n,n}
\end{array}\right| = \left| \begin{array}{ccc}
a_{1,1} & \dots & a_{1,r}\\
\vdots & & \vdots\\
a_{r,1} & \dots & a_{r,r}
\end{array}\right| D^{n-r-1}
\label{1.28}\end{equation}\normalsize
and fixing $r = n-2$ we receive,
\small\begin{equation*}
\underbrace{\left| \begin{array}{cc}
\Delta_{n-1,n-1} & \Delta_{n,n-1} \\
\Delta_{n-1,n} &  \Delta_{n,n}
\end{array}\right|}_{=\Delta_{n,n}\Delta_{n-1,n-1} - \Delta_{n-1,n}\Delta_{n,n-1}} = \underbrace{\left| \begin{array}{ccc}
a_{1,1} & \dots & a_{1,n-2}\\
\vdots & & \vdots\\
a_{n-2,1} & \dots & a_{n-2,n-2}
\end{array}\right|}_{=  D\left[ n-1 \atop{n-1} \right. \left. n \atop{n} \right]} D
\end{equation*}\normalsize
Recognizing that, 
\small\begin{equation*}\begin{array}{lcl}
\Delta_{n,n} =D\left[ \begin{array}{c} n\\ n \end{array} \right]&,& \Delta_{n-1,n-1} = D\left[ \begin{array}{c} n-1\\ n-1 \end{array} \right]\\ \Delta_{n-1,n}= D\left[ \begin{array}{c} n-1\\ n \end{array} \right]&, & \Delta_{n,n-1}= D\left[ \begin{array}{c} n\\ n-1 \end{array} \right] 
\end{array}\end{equation*}\normalsize
the above expression becomes,
\small\begin{equation}
D\left[ \begin{array}{c} n-1\\ n-1 \end{array} \right] D\left[ \begin{array}{c} n\\ n \end{array} \right] - D\left[ \begin{array}{c} n-1\\ n \end{array} \right]D\left[ \begin{array}{c} n\\ n-1 \end{array} \right] = D\left[ \begin{array}{cc} n-1 & n \\ n-1 & n \end{array} \right] D
\label{1.29}\end{equation}\normalsize
which is the required result. $\square$\\
\\
The Jacobi bilinear identity itself is only one half of the process of showing that the determinant expression obtained in eq. \ref{goatrage} is a $\tau$-function of the 2-Toda molecule equation. In what is follow, we introduce the molecule equation, and show that any determinant in \textit{bi-directional Wronskian}\footnote{A general bi-directional Wronskian determinant is of the form $ \textrm{det}\left[ \partial^{i-1}_{x} \partial^{j-1}_{y} \Psi(x,y) \right]^N_{i,j=1}$, where $\Psi(x,y)$ is a general function.} form is a $\tau$-function to such an equation.\\
\\
\textbf{2-dimensional Toda molecule equation.} The 2-dimensional Toda equation is defined as,
\small\begin{equation}
\partial_x \partial_y Q_s (x,y) = V_{s+1} (x,y) - 2V_s (x,y) + V_{s-1} (x,y)
\label{1.30}\end{equation}\normalsize
where,
\small\begin{equation}
Q_s (x,y) = \left\{ \begin{array}{cc}
 \log\{V_s (x,y)\} & \textrm{  for Toda molecule}\\
 \log\{1 + V_s (x,y)\} & \textrm{  for Toda lattice}
\end{array} \right. \label{garbage}
\end{equation}\normalsize
where $s \in \{ 0,1,\dots \}$.\\
\\
Through the convenient transformation,
\small\begin{equation*}
V_s (x,y)= \partial_x \partial_y \log \{\tau_s(x,y) \}
\end{equation*}\normalsize
the 2-dimensional Toda molecule equation becomes,
\small\begin{displaymath}
\partial_x \partial_y \log \left[\partial_x \partial_y  \log\{\tau_s(x,y)\}  \right] =\partial_x \partial_y \log \left\{ \frac{\tau_{s+1}(x,y) \tau_{s-1}(x,y)}{\tau^2_s(x,y)} \right\}
\end{displaymath}\normalsize
where if we complete the integrals with respect to $x$ and $y$ and take the integration constants to be zero, we receive the following bilinear differential equation,
\small\begin{equation}
 \left\{\partial_x \partial_y \tau_s(x,y) \right\} \tau_s(x,y)  - \left\{ \partial_x  \tau_s(x,y) \right\} \left\{  \partial_y \tau_s(x,y) \right\}  = \tau_{s+1}(x,y) \tau_{s-1}(x,y)
\label{1.34}\end{equation}\normalsize
Or equivalently, using Hirota's bilinear operators we obtain the compact form,
\small\begin{equation}
D_x D_y \tau_s (x,y). \tau_s(x,y) = 2 \tau_{s+1} (x,y)\tau_{s-1}(x,y)
\label{1.35}\end{equation}\normalsize
In section 1.4.2 it was shown that this is one of the non linear PDE's that can be obtained from the 2-Toda hierarchy bilinear relation.\\
\\
\textbf{Bi-directional wronskian solutions to the 2-Toda molecule equation.} The solution, $\tau_s$, of the above bilinear equation can be expressed by means of an $s \times s$ bi-wronskian,
\small\begin{equation}
\tau_s =\left\{ \begin{array}{cc}
1 & s=0\\
 \textrm{det} \left[\partial^{i-1}_x  \partial^{j-1}_y  \Psi (x,y) \right]_{ i,j =1,\dots, s} & s \ne 0
\end{array}\right.
\label{1.36}\end{equation}\normalsize
where $\Psi(x,y)$ is, for now, an arbitrary function of $\{x,y\}$ and the natural number, $s$, is not only the position of the Toda molecule, but also the degree of the wronskian determinant. \\
\\
In order to prove that $\tau_s$ given by eq. \ref{1.36} solves the bilinear 2-Toda molecule equation we introduce the $(s+1) \times (s+1)$, $s \times s$ and $(s-1) \times (s-1)$ determinants, $D$, $D\left[ a_1 \atop{b_1} \right]$ and \scriptsize$D \left[ \begin{array}{cc} a_1 & a_2 \\ b_1 & b_2 \end{array}  \right]$\normalsize respectively, 
\small\begin{equation}
\begin{array}{lcl}
D &=&  \textrm{det} \left[\partial^{i-1}_x  \partial^{j-1}_y  \Psi (x,y) \right]_{ i,j =1,\dots, s+1}  \\
D\left[ \begin{array}{c} a_1  \\ b_1  \end{array} \right]&= & \textrm{det} \left[\partial^{i-1}_x  \partial^{j-1}_y  \Psi (x,y) \right]_{ i=1,\dots,\hat{a_1},\dots, s+1\atop{j=1,\dots,\hat{b_1},\dots, s+1}}\\
D\left[ \begin{array}{cc}  a_1 & a_2 \\ b_1 & b_2 \end{array} \right] &= & \textrm{det} \left[\partial^{i-1}_x  \partial^{j-1}_y  \Psi (x,y) \right]_{ i=1,\dots,\hat{a_1},\dots,\hat{a_2},\dots, s+1\atop{j=1,\dots,\hat{b_1},\dots,\hat{b_2},\dots, s+1}}
\end{array}
\label{1.37}\end{equation}\normalsize
If we use the label,
\small\begin{equation}
D = \tau_{s+1}
\label{tatau}\end{equation}\normalsize
then we have the following convenient expressions,
\begin{equation}
\tau_s = D\left[ s+1 \atop{s+1} \right] \textrm{  ,  } \tau_{s-1} = D\left[ \begin{array}{cc}  s & s+1 \\ s & s+1 \end{array} \right] 
\label{1.38}\end{equation}\normalsize\\
\textbf{Maya diagrams.} We now consider how to express $\partial_x  \tau_s$, $\partial_y  \tau_s$ and $\partial_x \partial_y \tau_s$ in a form similar to eqs. \ref{tatau} and \ref{1.38}. To do so it is advantageous to view $\tau_s$ as the following Maya diagram,
\small\begin{displaymath}
\tau_s = \overbrace{\dots \underbrace{\bullet}_{i=s-2,} \underbrace{\bullet}_{i=s-1,} \underbrace{\bullet}_{i=s,} \underbrace{\circ}_{i=s+1,} \dots}^{x} \textrm{  ,  } \overbrace{\dots \underbrace{\bullet}_{j=s-2,} \underbrace{\bullet}_{j=s-1,} \underbrace{\bullet}_{j=s,} \underbrace{\circ}_{j=s+1,} \dots}^{y}
\end{displaymath}\normalsize
In the above notation, a black dot in position $i$ in the $x$ section represents the row,
\small\begin{equation}
\left( \partial^{i-1}_x  \Psi(x,y), \partial^{i-1}_x \partial_y \Psi(x,y),  \dots, \partial^{i-1}_x \partial^{s-1}_y \Psi(x,y) \right)
\label{1.39}\end{equation}\normalsize
and a black dot in position $j$ in the $y$ section represents the column,
\small\begin{equation}
\left( \begin{array}{c}
\partial^{j-1}_y  \Psi(x,y)\\
\partial^{j-1}_y \partial_x \Psi(x,y)\\
\vdots\\
\partial^{i-1}_y \partial^{s-1}_x \Psi(x,y)
\end{array} \right)
\label{1.40}\end{equation}\normalsize
When considering $\partial_x  \tau_s$ it is best to differentiate row by row with respect to $x$ rather than column by column. Using elementary multilinear differentiation, we see that we have a sum of $s$ terms. Thinking of $\tau_s$ as a Maya diagram, it is elementary to see that differentiating one specific row simply moves its corresponding black dot up one position. Thus, all but one of the $s$ Maya diagrams will have two black dots in the same position. Having two black dots in the same position corresponds to having repeated rows, hence the only Maya diagram that survives is the one that doesn't have two stones in the same position,
\small\begin{equation}
\partial_x  \tau_s = \overbrace{\dots \underbrace{\bullet}_{i=s-2,} \underbrace{\bullet}_{i=s-1,} \underbrace{\circ}_{i=s,} \underbrace{\bullet}_{i=s+1,} \dots}^{x}\textrm{  ,  }  \overbrace{\dots \underbrace{\bullet}_{j=s-2,} \underbrace{\bullet}_{j=s-1,} \underbrace{\bullet}_{j=s}, \underbrace{\circ}_{j=s+1,} \dots}^{y} = D\left[ s \atop{s+1} \right]
\label{1.41}\end{equation}\normalsize
Applying the same procedure to $\partial_y \tau_s$, except differentiating each column separately with respect to $y$, we obtain,
\small\begin{equation}
\partial_y  \tau_s = \overbrace{\dots \underbrace{\bullet}_{i=s-2,} \underbrace{\bullet}_{i=s-1,} \underbrace{\bullet}_{i=s,} \underbrace{\circ}_{i=s+1,} \dots}^{x} \textrm{  ,  } \overbrace{\dots \underbrace{\bullet}_{i=s-2,} \underbrace{\bullet}_{i=s-1,} \underbrace{\circ}_{i=s,} \underbrace{\bullet}_{i=s+1,} \dots}^{y} = D\left[ s+1 \atop{s} \right]
\label{1.42}\end{equation}\normalsize
Applying both procedures for $\partial_x \partial_y  \tau_s$, we have
\small\begin{equation}
\partial_x \partial_y  \tau_s = \overbrace{\dots \underbrace{\bullet}_{i=s-2,} \underbrace{\bullet}_{i=s-1,} \underbrace{\circ}_{i=s,} \underbrace{\bullet}_{i=s+1,} \dots}^{x} \textrm{  ,  } \overbrace{\dots \underbrace{\bullet}_{i=s-2,} \underbrace{\bullet}_{i=s-1,} \underbrace{\circ}_{i=s,} \underbrace{\bullet}_{i=s+1,} \dots}^{y} = D\left[ \begin{array}{c}  s \\ s  \end{array} \right] 
\label{1.43}\end{equation}\normalsize\\
Putting everything from this section together now, if we re-express the 2-Toda molecule equation (eq. \ref{1.34}), 
\small\begin{displaymath}
\left\{\partial_x \partial_y \tau_s(x,y) \right\} \tau_s(x,y)  - \left\{ \partial_x  \tau_s(x,y) \right\} \left\{  \partial_y \tau_s(x,y) \right\}  =\tau_{s-1}(x,y) \tau_{s+1}(x,y) 
\end{displaymath}\normalsize
using the Maya diagram notation we obtain,
\small\begin{displaymath}
 D\left[ \begin{array}{c} s\\ s \end{array} \right] D\left[ \begin{array}{c} s+1\\ s+1 \end{array} \right] - D\left[ \begin{array}{c} s\\ s+1 \end{array} \right]D\left[ \begin{array}{c} s+1\\ s \end{array} \right] = D\left[ \begin{array}{cc} s & s+1 \\ s & s+1 \end{array} \right] D
\end{displaymath}\normalsize
which is the Jacobi bilinear identity for determinants, hence verifying that the determinant expression for $\tau_s$ (eq. \ref{1.36}) is a solution of the 2-Toda molecule equation. \\
\\
\textbf{The homogeneous DWPF is a $\mathbf{\tau}$-function.} Hence, comparing eq. \ref{goatrage} with eq. \ref{1.36}, we see immediately that the determinant expression of the homogeneous partition function is a $\tau$-function that satisfies the 2-Toda molecule equation.\\
\\
\textbf{A note on the free energy.} In a similar process as shown above, it was shown in \cite{Sog} that the homogeneous six-vertex DWPF is a $\tau$-function of the 1-Toda molecule equation. In \cite{8,3} this property was used to extract information about the free energy of the model. This same method was unsurprisingly applied to the homogeneous Felderhof DWPF presented here. However, the results that were obtained were very murky (and thus shall not be presented here). The reasoning behind this murkiness can be attributed to the model being free-fermion. Specifically, considering different values of the variable, $\Delta = \frac{1}{2}(\omega_5 \omega_6 - \omega_3 \omega_4 - \omega_1 \omega_2)$, for the different phases (ferro-electric, disordered, etc.), as is usually done in this kind of analysis, obviously will not work. In the next section we shall see that the DWPF trivializes and the free energy can be taken directly. 
\section{Product form of the DWPF}
Due to the model being a free fermion model, it is expected that the inherent complexities that exist with the general non free fermion model are somehow dwindled down. This is true with the free fermion six vertex model under both domain wall and periodic boundary conditions. In the case of the 6V-DWPF\footnote{This can easily be verified.}, the determinant exists in Cauchy form and hence can be expressed as a product, and in the case of the 6V-PPF, the horribly complex Bethe equations are trivialized \cite{Baxterbook}.\\
\\
Unsurprisingly, given the statement in the last paragraph, the determinant form for the DWPF of the current free fermion model also exists in Cauchy form. We give the details below, limited as they are.\\
\\
\textbf{Manipulating the determinant entries to Cauchy form.} We begin with the entries of the determinant, $\phi(\alpha_i,\beta_j)$, of the inhomogeneous DWPF and rearrange them as the following,
\small\begin{equation*}\begin{split}
\phi(\alpha_i,\beta_j) &= \frac{1}{(\alpha_i - \beta_j)(1-\alpha_i \beta_j)}\\
&=  \frac{1}{\alpha_i(1-\beta^2_j) - \beta_j(1-\alpha^2_i)}\\
&=  \frac{(1-\alpha^2_i)(1-\beta^2_j) }{\frac{\alpha_i}{1-\alpha^2_i} - \frac{\beta_j}{1-\beta^2_j}}
\end{split}\end{equation*}\normalsize
Thus considering the determinant we obtain,
\small\begin{equation*}\begin{split}
\textrm{det} \left[\phi(\alpha_i,\beta_j) \right]^N_{i,j=1} &= \textrm{det} \left[ \frac{(1-\alpha^2_i)(1-\beta^2_j) }{\frac{\alpha_i}{1-\alpha^2_i} - \frac{\beta_j}{1-\beta^2_j}} \right]^N_{i,j=1}\\
&=\left( \prod^N_{i=1} (1-\alpha^2_i)(1-\beta^2_i) \right)\textrm{det} \left[ \frac{1 }{\frac{\alpha_i}{1-\alpha^2_i} - \frac{\beta_j}{1-\beta^2_j}} \right]^N_{i,j=1}
\end{split}\end{equation*}\normalsize
which is obviously of Cauchy type. Expanding the determinant as a product we see immediately that,
\small\begin{equation*}\begin{split}
\textrm{det} \left[\phi(\alpha_i,\beta_j) \right]^N_{i,j=1} =& \left( \prod_{1\le i<j \le N} (1-\alpha_i\alpha_j )(1-\beta_i \beta_j) \right)\\
& \times  \frac{\prod_{1\le i<j \le N} (\alpha_i-\alpha_j )(\beta_j-\beta_i )}{\prod^N_{i,j=1}(\alpha_i - \beta_j)(1-\alpha_i \beta_j)}
\end{split}\end{equation*}\normalsize
and hence, the inhomogeneous DWPF simplifies quite dramatically,
\small\begin{equation}
Z_N\left(\vec{\alpha},\vec{\beta} \right) = \prod^N_{i,j=1} \sqrt{1-\alpha_i \alpha_j}\sqrt{1-\beta_i \beta_j}
\label{trivialsh}\end{equation}\normalsize
Taking the homogeneous limit we obtain the even more simplified expression,
\small\begin{equation}\begin{split}
Z_N\left(\alpha,\beta \right)& =\left( \sqrt{1-\alpha^2}\sqrt{1-\beta^2} \right)^{N^2}\\
&= \left( c\left(\alpha,\beta \right)  \right)^{N^2}
\label{trivialsh2}\end{split}\end{equation}\normalsize
%%%%%%%%%%%%%%%%%%%%%%%%%%%%%%%%%%%%%%%%%%%%%%%%%%%%%%%%%%%%%%%%%%%%%%%%%%%%%%%%%%%%%%%%%%%%%%%%%%%%%%%%%%%%%%%%%%%%
\newpage
%%%%%%%%%%%%%%%%%%%%%%%%%%%%%%%%%%%%%%%%%%%%%%%%%%%%%%%%%%%%%%%%%%%%%%%%%%%%%%%%%%%%%%%%%%%%%%%%%%%%%%%%%%%%%%%%%%%%
\chapter{Baxter's solid on solid (BSOS) model}
In \cite{Bax4} Baxter introduced the BSOS model, which originated through work on the eight-vertex model. In the aforementioned work, the BSOS model was introduced through the vertex-SOS correspondence, where weights of the eight-vertex model are linked to the weights of the BSOS model through intertwining vectors. A peculiarity with the BSOS model is that it actually bears closer resemblance to the six-vertex model, as we shall see shortly. \\
\\
In the following introduction to the model we shall use the notation presented in \cite{SOS1,SOS2}.
\section{Definition of the BSOS model}
\subsection{State variables - vertex and height models}
In the previous chapter we dealt with a vertex model whose configuration was given solely by state variables on each of the four sides of the vertex, designated by the variables $1,2$, or graphically as arrows pointing in or out, with the addition of rapidity and colour flows.  An alternative method of describing such configurations exists by replacing the vertex with a square face, where the state variables (heights) are now placed on the corners of each face. The rapidity and/or colour flows are left unchanged. This new model is called an interaction-round-a-face (IRF) or a solid-on-solid (SOS) model, or simply a height model. Thus the equivalent of the $N \times N$ vertex lattice with horizontal rapidities $\{u\}$ and vertical rapidities $\{v\}$ is the $N \times N$ face lattice with the same horizontal and vertical rapidities.
\begin{figure}[h!]
\begin{center}
\includegraphics[angle=0,scale=0.30]{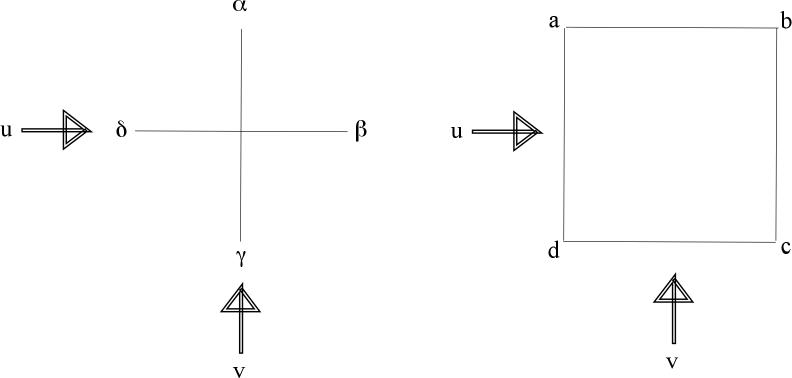}
\caption{\footnotesize{To the left a vertex configuration defined by state variables (heights) $\alpha,\beta,\gamma,\delta$ and rapidities $u,v$, and to the right a height configuration defined by state variables $a,b,c,d$ and same rapidities.}}
\label{5.a}
\end{center}
\end{figure}\\
\\
As an introduction to the particular model we are about to use, the best course of action would be to first introduce the definitions of various elliptic functions and some necessary properties, followed then by the definition of the weights the BSOS model and the Yang-Baxter equation(s). Following this we then define what is meant by DWBC's in the sense of a height model.
\subsection{Elliptic functions}
We define the half period magnitudes, $K_1, K_2$ as the quantities,
\small\begin{equation*}\begin{split}
K_1 &= \frac{1}{2}\pi \prod^{\infty}_{n=1} \left( \left\{ \frac{1 + q^{2n-1}}{1-q^{2n-1}} \right\}\left\{  \frac{1 - q^{2n}}{1+q^{2n}}\right\} \right)^2\\
K_2 &= -\frac{1}{\pi} K_1 \log (q) 
\end{split}\end{equation*}\normalsize
where $q$ is known as the elliptic nome and typically has a value between zero and one. Given $K_1$ and $K_2$, the elliptic theta functions are then defined as,
\small\begin{equation*}\begin{split}
H(u) &= 2 q^{\frac{1}{4}} \sin \left( \frac{\pi u}{2 K_1} \right) \prod^{\infty}_{n=1}\left\{ 1 - 2 q^{2n} \cos \left( \frac{\pi u}{K_1} \right)+ q^{4n}  \right\} \left\{ 1-q^{2n} \right\}\\
H_1(u) &= 2 q^{\frac{1}{4}} \cos \left( \frac{\pi u}{2 K_1} \right) \prod^{\infty}_{n=1}\left\{ 1 + 2 q^{2n} \cos \left( \frac{\pi u}{K_1}\right) + q^{4n}\right\} \left\{ 1-q^{2n} \right\}\\
&= H(u+K_1)\\
\Theta(u) &=  \prod^{\infty}_{n=1}\left\{ 1 - 2 q^{2n-1} \cos \left( \frac{\pi u}{K_1} \right)+ q^{4n-2} \right\} \left\{ 1-q^{2n} \right\}\\
\Theta_1(u) &=  \prod^{\infty}_{n=1}\left\{ 1 +2 q^{2n-1} \cos \left( \frac{\pi u}{K_1} \right)+ q^{4n-2}  \right\} \left\{ 1-q^{2n} \right\}\\
&= \Theta(u+K_1)\\
\end{split}\end{equation*}\normalsize
where $u \in \mathbb{C}$. The (simple) zeroes of theta functions are given by,
\small\begin{equation*}\begin{array}{lll}
H(u_{mn})=0 & \textrm{for  } &u_{mn} = 2m K_1 + 2 i n K_2\\
\Theta(u_{mn})=0 & \textrm{for  } &u_{mn} = 2m K_1 + 2i \left( n+\frac{1}{2} \right) K_2
\end{array}\end{equation*}\normalsize
where $m,n \in \mathbb{Z}$.  We also have the important quasi-periodic relations,
\small\begin{equation*}\begin{split}
H(u + 2 m K_1) =& (-1)^m H(u)\\
H(u + 2 i n K_2) =& (-1)^n q^{-n^2} \exp \left( -\frac{i n \pi u}{K_1} \right) H(u)
\end{split}\end{equation*}\normalsize
where $m,n \in \mathbb{Z}$. A function which satisfies both of these conditions (up to some constant) are referred to as doubly quasi-periodic. We now present an elementary (but nonetheless necessary) result (theorem 15.1 of \cite{Baxterbook}) regarding doubly (anti) periodic functions.
\begin{theorem}
\label{Louv}
If a function is doubly (anti) periodic and is analytic inside and on a period rectangle, then it is a constant.
\end{theorem}
\textbf{Proof.} The proof is elementary. Since the function is analytic in and on the period rectangle, it is bounded in and on the rectangle. The double (anti) periodicity assures us that the function is analytic and bounded everywhere. Hence by Liouville's theorem in complex variable theory, the function is a constant. $\square$\\
\\
\textbf{Comment.} Using the above result it is possible to verify various elliptic identities, (the simplest non trivial example being),
\small\begin{equation}
\begin{split}
H(x-y)H(x+y)H(u+v)H(u-v) =& H(u+x)H(u-x)H(v+y)H(v-y) \\
&-H(u+y)H(u-y)H(v+x)H(v-x)
\end{split}
\label{simple}
\end{equation}\normalsize
without using the explicit definition of $H(u)$\footnote{It is necessary however to use $H(-u) = -H(u)$.}. The usual method one would use to prove the above identity is to consider the right hand side divided by the left hand side, which we shall call $P(u)$. We then show that the zeros of the denominator are at the same positions of those of the numerator and that $P(u)$ satisfies necessary doubly (anti) periodic conditions. Thus by the above theorem, $P(u)$ is a constant. All that remains is to show that the constant is equal to one, by evaluating $P(u)$ at some obvious value of $u$. The reason we give the theorem here is because it is the only result necessary to verify the height Yang-Baxter identities which shall be given shortly.\\
\\
In the following chapter we shall rely heavily on other results regarding genuinely quasi-periodic functions (as opposed to simply doubly (anti) periodic). The results of this chapter rely entirely on the fact that the weights of the model obey the height Yang-Baxter equation. Thus we shall leave any further results regarding quasi-periodic functions for the relevant section of the next chapter.
\subsection{Weights of the model and the Yang-Baxter equation}
We begin by labelling the face configuration in figure \ref{5.a} with state variables $a,b,c,d\in \field{Z}$ and rapidities $u,v \in \field{C}$ by,
\small\begin{equation*}
W \left( \left. \begin{array}{cc} a & b \\ d & c \end{array} \right|  u -v  \right)
\end{equation*}\normalsize
and the only restriction on the state variables being,
\small\begin{equation*}
|a-b| = |b-d| = |d-c| = |c-a| = 1
\end{equation*}\normalsize
This leaves six classes of non zero weights. Labelling,
\small\begin{equation*}
H(\lambda u) \Theta (\lambda u) = [u]
\end{equation*}\normalsize
where $\lambda \in \mathbb{C}$, the six classes of non zero weights are parameterized by,
\begin{equation}\begin{array}{lclcl}
W \left( \left. \begin{array}{cc} l & l \pm 1 \\ l \pm 1 & l \pm 2 \end{array} \right|  u-v  \right) &= & W_A (u-v)& = & \frac{[u-v+1]}{[1] } \\
W \left( \left. \begin{array}{cc} l & l \pm 1 \\ l \mp 1 & l \end{array} \right|  u-v  \right) &= & W^l_{B,\pm} (u-v) &= & \frac{[u-v]}{[1] }\frac{[\zeta + l \pm 1]}{[\zeta+l ] } \\
W \left( \left. \begin{array}{cc} l & l \pm 1 \\ l \pm 1 & l \end{array} \right|  u-v  \right) &= & W^l_{C,\pm} (u-v) &= & \frac{[\zeta + l \mp (u-v)]}{[\zeta+l ] }
\end{array}
\label{11.2}\end{equation}\normalsize
where $\zeta \in \mathbb{C}$. With this parameterization the Yang-Baxter equation looks like,
\small\begin{equation*}\begin{split}
&\sum_{g \in \mathbb{Z}} W \left( \left. \begin{array}{cc} f & g \\ a & b \end{array} \right|  u_1 -u_3  \right) W \left( \left. \begin{array}{cc} e & d \\ f & g \end{array} \right|  u_2 - u_3 \right) W \left( \left. \begin{array}{cc} d & c \\ g & b \end{array} \right|  u_2 -u_1  \right)\\
=&\sum_{g \in \mathbb{Z}} W \left( \left. \begin{array}{cc} e & d \\ g & c \end{array} \right|  u_1 -u_3  \right) W \left( \left. \begin{array}{cc} g & c \\ a & b \end{array} \right|  u_2 - u_3 \right) W \left( \left. \begin{array}{cc} e & g \\ f & a \end{array} \right|  u_2 -u_1  \right)
\end{split}\end{equation*}\normalsize 
for $a,b,c,d,e,f \in \mathbb{Z}$ and $u_1,u_2,u_3 \in \mathbb{C}$.
\begin{figure}[h!]
\begin{center}
\includegraphics[angle=0,scale=0.20]{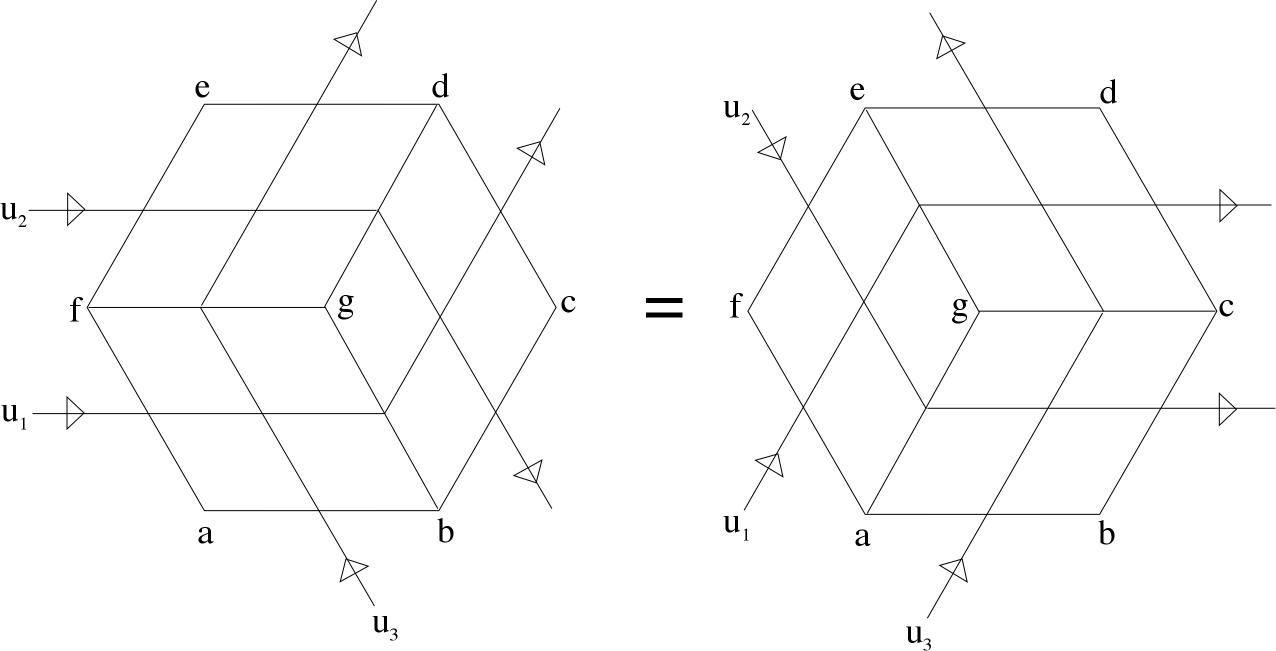}
\caption{\footnotesize{Graphical description of the height Yang-Baxter equation}}
\label{5.d}
\end{center}
\end{figure}
\subsection{DWBC's}
\begin{wrapfigure}[8]{r}{44mm}
\begin{center}
\includegraphics[scale=0.20, angle=0]{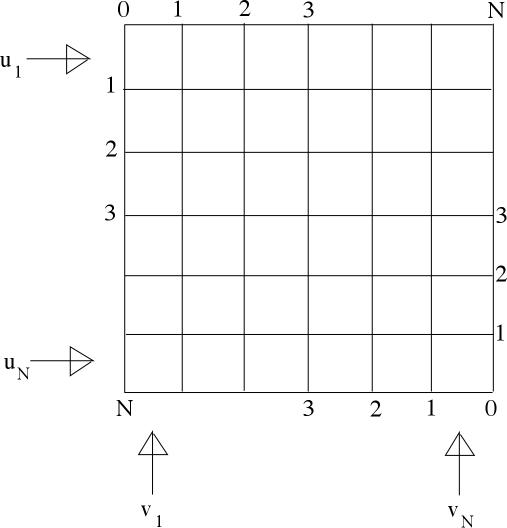}
\caption{\footnotesize{DWBC for the $N \times N$ BSOS model}}
\label{5.e}
\end{center}
\end{wrapfigure}
We define the $N \times N$ DWBC's for the BSOS model as the $N \times N$ face lattice with the top left height equal to zero, with subsequent heights to the right and south increasing by increments of one until they equal $N$. The remaining heights then decrease by increments of one until they meet at the bottom right corner, which is equal to zero. \\
\\
\\
\\
\\
\\
\section{Properties of the DWPF}
In this section we examine properties of the DWPF for this particular model using techniques applied to the six-vertex model to derive one-point correlation functions \cite{1, OPCF}. The overall goal of this section was obviously to derive an Izergin-like expression for the DWPF, but this was not to be the case as Rosengren \cite{rosen} was to publish his admirable result while this work was being conducted.\\
\\  
As usual, the $N \times N$ DWPF is defined as the weighted sum of all allowable $N \times N$ face configurations with rapidities $\{u\}$ and $\{ v\}$ given DWBC,
\small\begin{equation*}
Z_{N}\left(\vec{u},\vec{v}\right) = \sum_{\textrm{allowable}\atop{\textrm{configurations}}} \left\{ \prod_{\textrm{faces}} W \left( \left. \begin{array}{cc} a  & b  \\ d  &c \end{array} \right|  u_{i}-v_j  \right) \right\}
\end{equation*}\normalsize
\subsection{Deriving the recurrence relation for the partition function}
\begin{figure}[h!]
\begin{center}
\includegraphics[angle=0,scale=0.30]{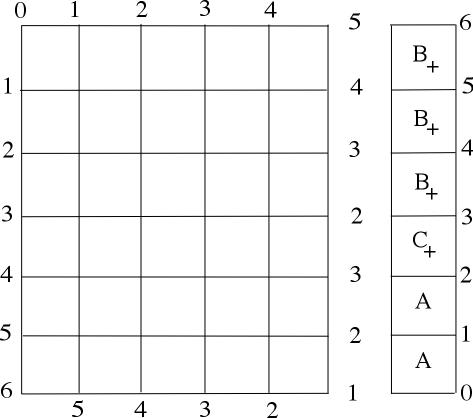}
\caption{\footnotesize{Example involving a $6 \times 6$ lattice with a $W_{C,+}$ face at row 4 of the right most column}}
\label{5.h}
\end{center}
\end{figure}

\begin{figure}[h!]
\begin{center}
\includegraphics[angle=0,scale=0.20]{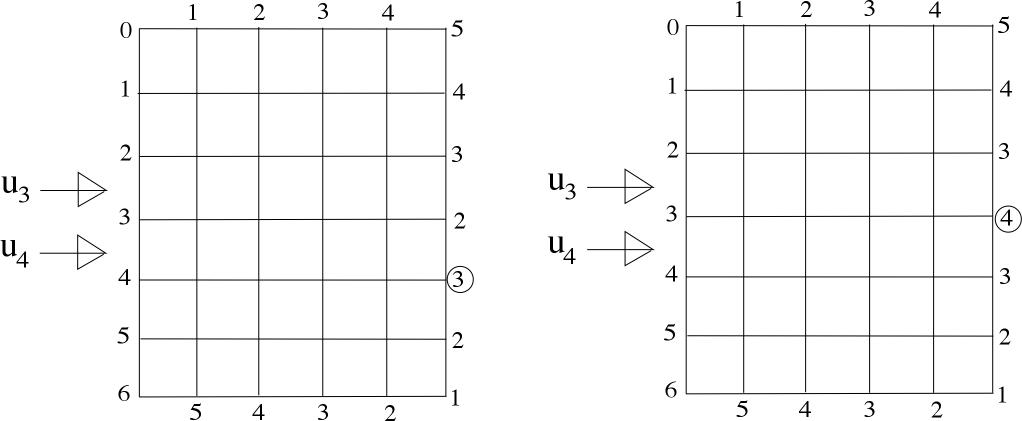}
\caption{\footnotesize{A typical example portraying the difference between $L[r_n,u_n]$ (on the left) and $L[r_{n-1},u_{n-1}]$ (on the right) for $N=6$, $n=4$.}}
\label{5.j}
\end{center}
\end{figure}

\begin{figure}[h!]
\begin{center}
\includegraphics[angle=0,scale=0.25]{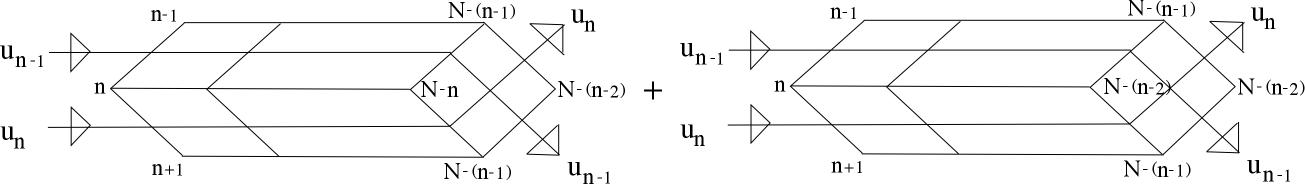}
\caption{\footnotesize{Multiplying $L[r_n,u_n]$ by $W^{N-(n-1)}_{B,+}(u_{n-1}-u_{n})$ (on the left) and $L[r_{n-1},u_{n-1}]$ by $W^{N-(n-1)}_{C,+}(u_{n-1}-u_{n})$ (on the right).}}
\label{5.k}
\end{center}
\end{figure}

\begin{figure}[h!]
\begin{center}
\includegraphics[angle=0,scale=0.25]{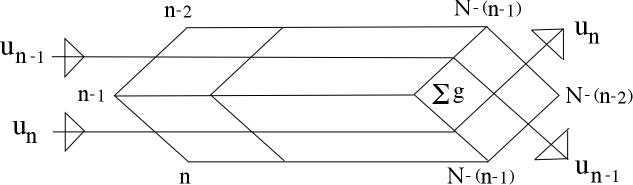}
\caption{\footnotesize{The internal height being summed over all allowable values}}
\label{5.l}
\end{center}
\end{figure}

\begin{figure}[h!]
\begin{center}
\includegraphics[angle=0,scale=0.25]{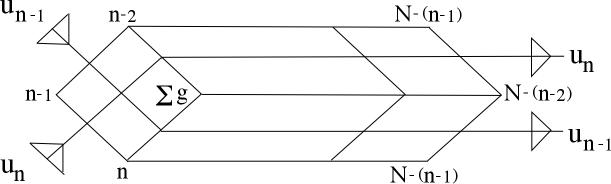}
\caption{\footnotesize{Applying the Yang-Baxter equation to shift the intertwining of horizontal rapidities to the left hand side of the lattice}}
\label{5.m}
\end{center}
\end{figure}

\begin{figure}[h!]
\begin{center}
\includegraphics[angle=0,scale=0.16]{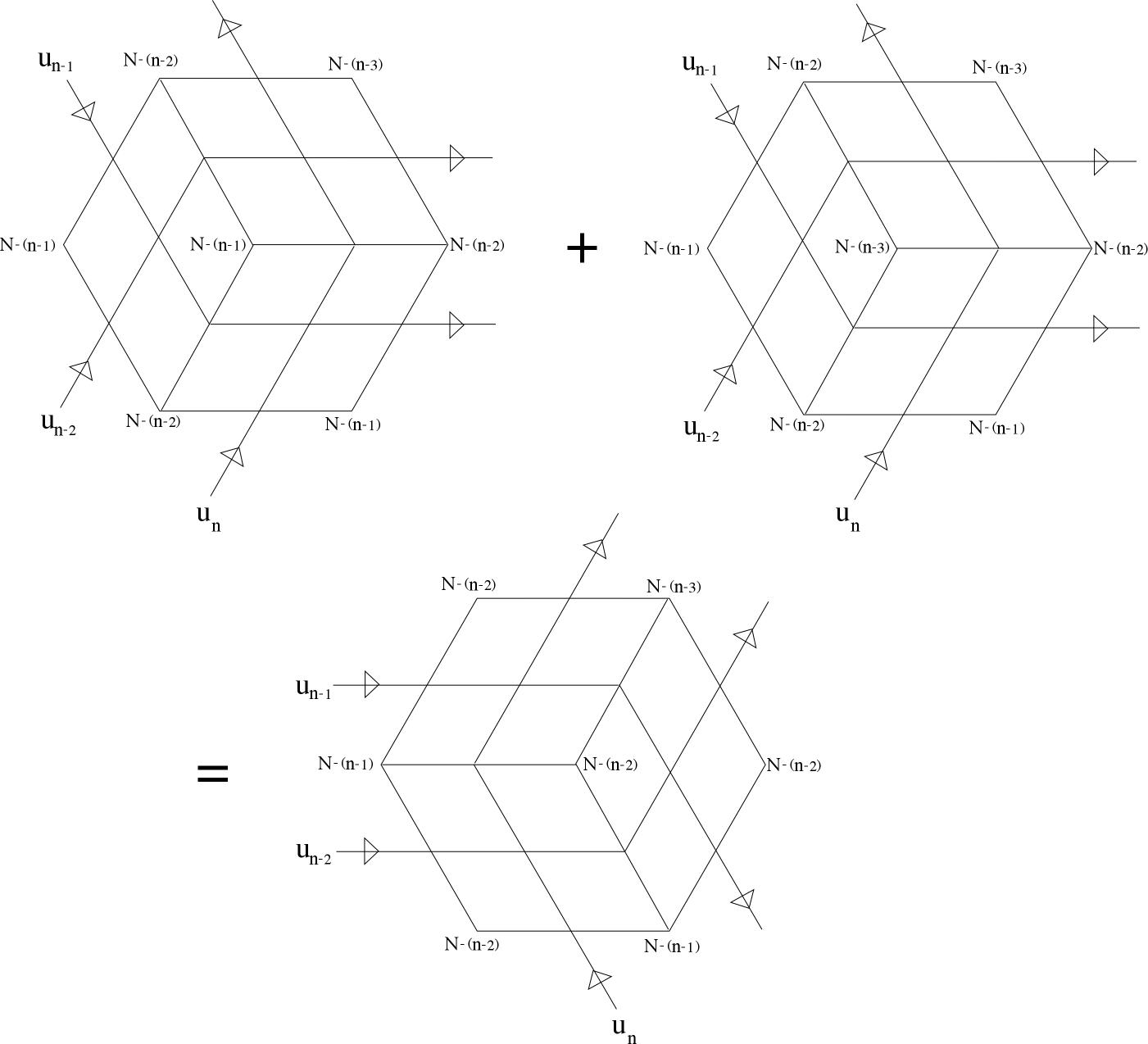}
\caption{\footnotesize{Graphical representation of the Y-B identity in consideration}}
\label{5.n}
\end{center}
\end{figure}

\begin{figure}[h!]
\begin{center}
\includegraphics[angle=0,scale=0.25]{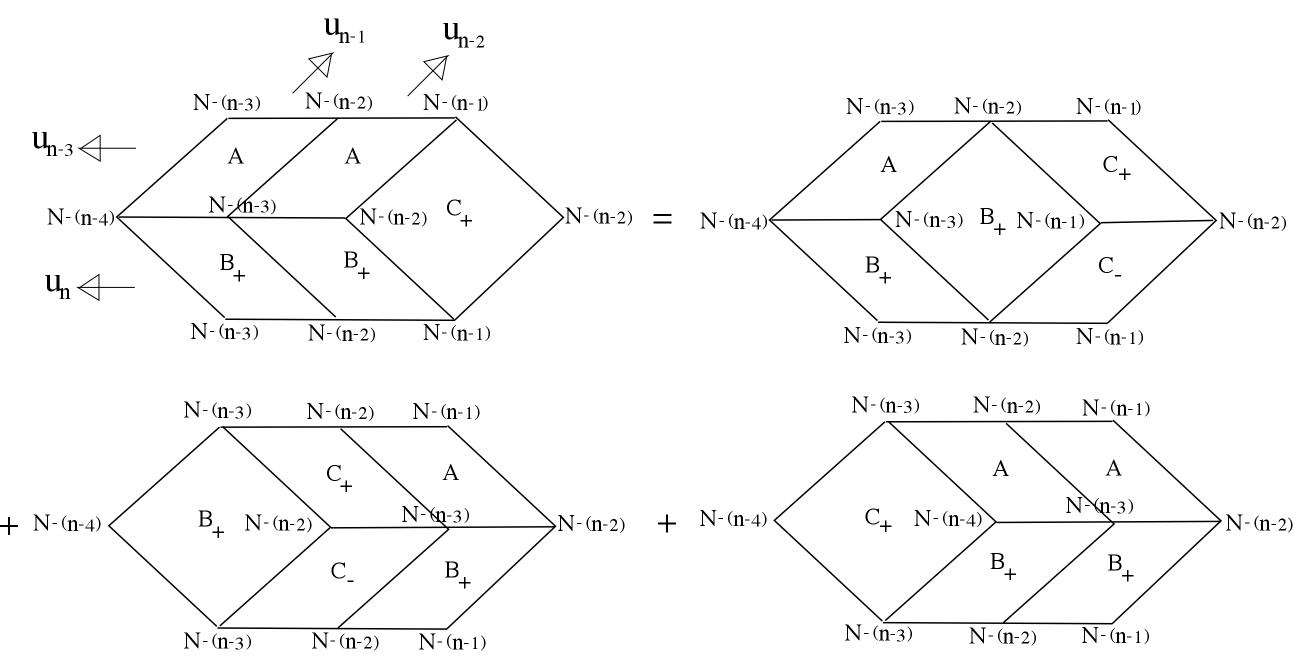}
\caption{\footnotesize{Yang-Baxter equation that simplifies the coefficient of $L[r_{n-3},u_{n-3}]$.}}
\label{5.o}
\end{center}
\end{figure}

\begin{figure}[h!]
\begin{center}
\includegraphics[angle=0,scale=0.25]{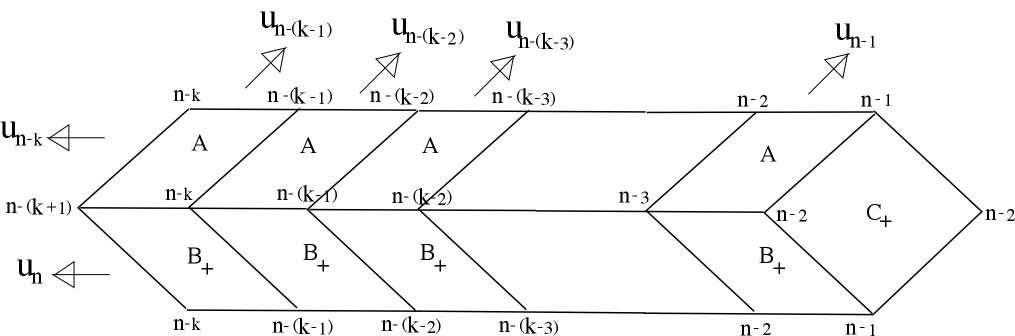}
\caption{\footnotesize{Graphical representation of the right hand side of proposition \ref{bigprop}. Notice that state variables $n-j$, $1 \le j \le k+1$, actually correspond to height $N-(n-j)$.}}
\label{5.p}
\end{center}
\end{figure}

\begin{figure}[h!]
\begin{center}
\includegraphics[angle=0,scale=0.25]{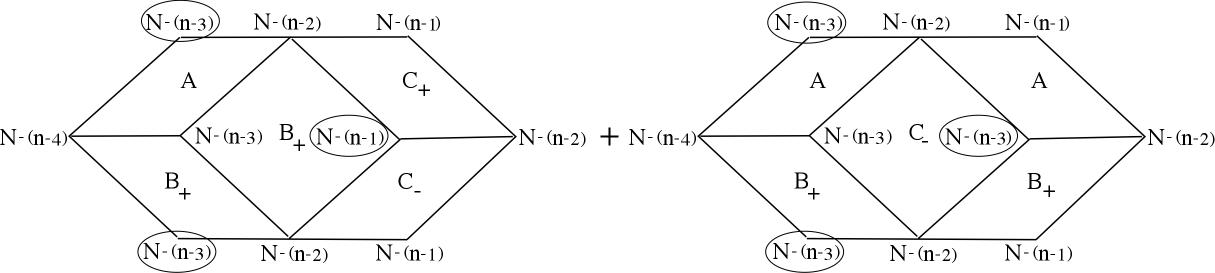}
\caption{\footnotesize{Applying the Yang-Baxter operation once to the configuration in fig. \ref{5.p} to move the diamond face one unit to the left. Notice that the far left of both configurations has been omitted.}}
\label{5.q}
\end{center}
\end{figure}

\begin{figure}[h!]
\begin{center}
\includegraphics[angle=0,scale=0.25]{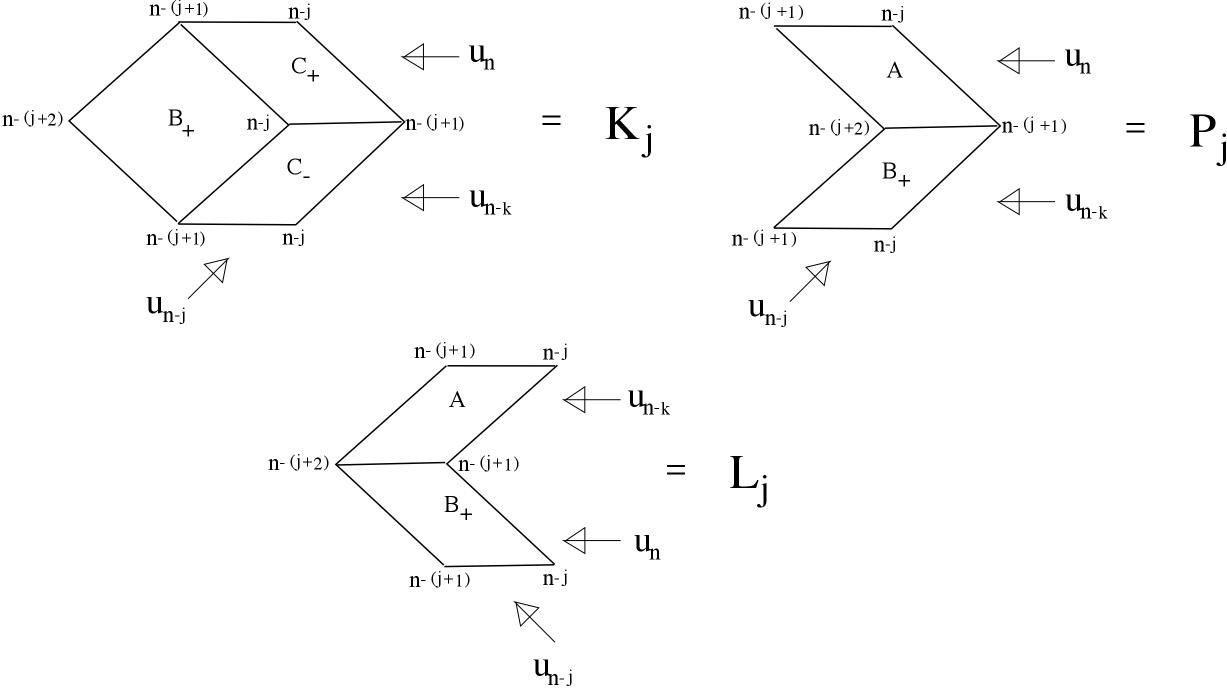}
\caption{\footnotesize{Graphical notation used to verify proposition 22. Notice that state variables $n-j$, $1 \le j \le k+1$, actually correspond to height $N-(n-j)$.}}
\label{5.r}
\end{center}
\end{figure}

In this section we shall use the results in \cite{1}, which were used as an alternative method to \cite{OPCF} for calculating one-point correlation functions for the six-vertex model. We shall show how this method can be used on the BSOS model to derive a complete recursive form for the DWPF, and from this point, derive the closed form expression for the DWPF involving sums over the symmetric group.\\
\\ 
\textbf{The right most column.} We begin by considering the right most column of the $N \times N$ lattice. For any allowable configuration of the model, the presence of DWBC's means that no $W_{C,-}$ faces are allowable and only one $W_{C,+}$ face is (necessarily) present in the right most column. It is then easy to see (fig. \ref{5.h}) that all the faces above the $W_{C,+}$ face are of type $W_{B,+}$, and all faces below the $W_{C,+}$ face are of type $W_{A}$. Hence, if the $W_{C,+}$ weight occurs at row $n$, $1 \le n \le N$, then the right most column has weight,
\small\begin{equation*}
\left( \prod^{n-1}_{j=1}W^{N-j}_{B,+}(u_j-v_N) \right) W^{N-n}_{C,+}(u_n-v_N) \left( \prod^{N}_{j=n+1} W_A(u_j-v_N) \right)
\end{equation*}\normalsize\\
\\
\\
\textbf{The remaining $N \times (N-1)$ lattice.} We label the remaining $N \times (N-1)$ lattice as $L[r_n,u_n]$. Performing a summation for $n = 1, \dots, N$, we obtain the entire DWPF in the form,
\small\begin{equation}\begin{split}
Z^{(0)}_{N }\left(\vec{u},\vec{v}\right) =& \sum^N_{n=1} L[r_n,u_n] \left( \prod^{n-1}_{j=1}W^{N-j}_{B,+}(u_j-v_N) \right) W^{N-n}_{C,+}(u_n-v_N) \\
&\times \left( \prod^{N}_{j=n+1} W_A(u_j-v_N) \right)
\label{N.3}\end{split}\end{equation}\normalsize
where the superscript $(0)$ in the expression $Z^{(0)}_{N }\left(\vec{u},\vec{v}\right)$ denotes the value of the top left height.\\
\\
\textbf{Freezing the top row.} 
\begin{figure}[h!]
\begin{center}
\includegraphics[angle=0,scale=0.30]{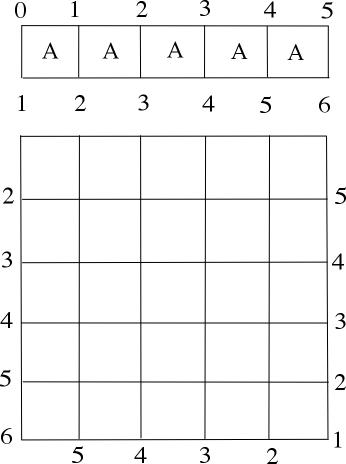}
\caption{\footnotesize{An example of $L[r_1,u_1]$. Notice the entire top row is frozen into $W_A$ faces leaving behind an $(N-1)\times(N-1)$ lattice with DWBC}}
\label{5.i}
\end{center}
\end{figure}
Consider the $N \times (N-1)$ configuration $L[r_1,u_1]$. It is elementary to recognize that the entire top row of this particular configuration is frozen into $W_A$ faces. If we extract these faces, what remains is an $(N-1)\times(N-1)$ lattice with DWBC's. However, it must be noted that the lowest height on the boundary is no longer zero but one. The highest height is still $N$.
\small\begin{equation*}
\Rightarrow L[r_1,u_1] =\left( \prod^{N-1}_{j=1}W_A(u_1 - v_j)\right) Z^{(1)}_{N-1} \left(\vec{u},\vec{v},\hat{u}_1,\hat{v}_N \right)
\end{equation*}\normalsize
The crux of the work that is to follow consists of using the Yang-Baxter equation(s) to express a general $N \times (N-1)$ configuration $L[r_n,u_n]$ as a sum of configurations whose top rows are frozen into the aforementioned position, which in the end shall give a recursive relation for the partition function of the BSOS model.\\
\\
The main tool which we have at our disposal (which boils down to applying the Yang-Baxter equation strategically) shall be referred to as rolling. \\
\\
\textbf{Rolling once.} We begin by considering the general $N \times (N-1)$ configuration, $L[r_n,u_n]$, whose progression of right most heights is interrupted at the $n$th row. We additionally consider the general $N \times (N-1)$ configuration, $L[r_{n-1},u_{n-1}]$, which is the same configuration as $L[r_n,u_n]$, except that the progression of the right most heights is interrupted at row $n-1$. We give an example of the difference of these two configurations in fig. \ref{5.j}.
\newpage
\noindent We proceed the rolling procedure by multiplying the configuration $L[r_n,u_n]$ by the face $W^{N-(n-1)}_{B,+}(u_{n-1}-u_{n})$ and the configuration $L[r_{n-1},u_{n-1}]$ by the face $W^{N-(n-1)}_{C,+}(u_{n-1}-u_{n})$, as shown in fig. \ref{5.k}.
\\
\\
Considering the sum of these configurations, we notice that the internal height is conveniently being summed over all allowable values, thus the sum can be explicitly written as in fig. \ref{5.l}.
\\
\\
We are now in a position to apply the Yang-Baxter equation to the above configuration and shift the intertwining of the $u_{n-1}$ and $u_n$ rapidities to the left hand side of the $N \times (N-1)$ lattice, as shown in fig. \ref{5.m}.
\\
\\
Notice however that sum of state variables, $\sum g$, is actually fixed at $n-1$. Thus we obtain the following algebraic expression,
\small\begin{equation}\begin{split}
 L[r_n,u_n] =& \frac{W_A(u_{n-1}-u_{n})}{W^{N-(n-1)}_{B,+}(u_{n-1}-u_{n})}  L[r_{n-1},u_{n}]\\
&- \frac{W^{N-{n-1)}}_{C,+}(u_{n-1}-u_{n})}{W^{N-(n-1)}_{B,+}(u_{n-1}-u_{n})}  L[r_{n-1},u_{n-1}]\label{N.4}
\end{split}\end{equation}\normalsize
Using the explicit form for the weights we know the following identities hold,
\small\begin{equation*}\begin{split}
W^{k}_{C,+}(u_{i}-u_{j}) &= W^{k}_{C,-}(u_j-u_{i})\\
W^{k}_{B,+}(u_{i}-u_{j}) &= - W^{k}_{B,+}(u_j-u_i)
\end{split}\end{equation*}\normalsize
Hence eq. \ref{N.4} can be expressed in the more palatable form,
\small\begin{equation}
 L[r_n,u_n] = f^N_{n-1} \left( u_{n-1}\atop{u_n} \right)  L[r_{n-1},u_{n}] +g^N_{n-1} \left( u_{n}\atop{u_{n-1}} \right)  L[r_{n-1},u_{n-1}]\label{N.5}
\end{equation}\normalsize
where,
\small\begin{equation*}
 f^l_{k} \left( u_i \atop{u_j} \right)  =  \frac{W_A(u_{i}-u_{j})}{W^{l-k}_{B,+}(u_{i}-u_{j})} \textrm{  ,  } g^l_{k} \left( u_{i}\atop{u_{j}} \right)  =  \frac{W^{l-k}_{C,-}(u_{i}-u_{j})}{W^{l-k}_{B,+}(u_{i}-u_{j})}  
\end{equation*}\normalsize
Equation \ref{N.5} is the conclusion of the first rolling procedure on a general $N \times (N-1)$ lattice configuration whose progression of right most heights is interrupted at row $n-1$. We have succeeded in shifting the interruption up by one row, but at the cost of producing two configurations instead of one. It is at this point that one should obtain a slight feeling of dread, as we can now see that for every rolling procedure we double the amount of configurations. This doubling shall be taken care of however by applying the Yang-Baxter equation at strategic times in the below algorithm to make the number of configurations manageable. We shall demonstrate this process by rolling an additional time.\\
\\
\textbf{Rolling twice.} We now consider what happens when we apply the rolling procedure to the configurations $L[r_{n-1},u_{n}]$ and $L[r_{n-1},u_{n-1}]$. Using the above procedure it is immediate that we obtain the following results,
\small\begin{equation}\begin{split}
L[r_{n-1},u_{n}] &= f^N_{n-2} \left( u_{n-2}\atop{u_n} \right)  L[r_{n-2},u_{n}] +g^N_{n-2} \left( u_{n}\atop{u_{n-2}} \right)  L[r_{n-2},u_{n-2}]\\
L[r_{n-1},u_{n-1}] &= f^N_{n-2} \left( u_{n-2}\atop{u_{n-1}} \right)  L[r_{n-2},u_{n-1}] +g^N_{n-2} \left( u_{n-1}\atop{u_{n-2}} \right)  L[r_{n-2},u_{n-2}]\\
\Rightarrow L[r_{n},u_{n}] &= \left\{  f^N_{n-1} \left( u_{n-1}\atop{u_{n}} \right)g^N_{n-2} \left( u_{n}\atop{u_{n-2}} \right)+ g^N_{n-1} \left( u_{n}\atop{u_{n-1}} \right)g^N_{n-2} \left( u_{n-1}\atop{u_{n-2}} \right)  \right\} \\
&\times L[r_{n-2},u_{n-2}]  +  g^N_{n-1} \left( u_{n}\atop{u_{n-1}} \right)f^N_{n-2} \left( u_{n-2}\atop{u_{n-1}} \right) L[r_{n-2},u_{n-1}] \\
&+f^N_{n-1} \left( u_{n-1}\atop{u_n} \right)f^N_{n-2} \left( u_{n-2}\atop{u_n} \right) L[r_{n-2},u_{n}]\label{N.6}
\end{split}\end{equation}\normalsize
Let us now analyze the coefficient of $L[r_{n-2},u_{n-2}]$ carefully with the intention of reducing it by applying some Yang-Baxter identity. \\
\\
We begin by multiplying the coefficient by the following factor,
\small\begin{equation*}
W^{N-(n-1)}_{B,+}(u_{n-1}-u_n) W^{N-(n-2)}_{B,+}(u_{n}-u_{n-2}) W^{N-(n-2)}_{B,+}(u_{n-1}-u_{n-2})
\end{equation*}\normalsize
to obtain,
\small\begin{equation}\begin{split}
&W_A(u_{n-1}-u_n)W^{N-(n-2)}_{C,+}(u_{n-2}-u_{n}) W^{N-(n-2)}_{B,+}(u_{n-1}-u_{n-2}) \\
+& W^{N-(n-1)}_{C,+}(u_{n-1}-u_n) W^{N-(n-2)}_{B,+}(u_{n-2}-u_{n}) W^{N-(n-2)}_{C,-}(u_{n-1}-u_{n-2})\label{N.7}
\end{split}\end{equation}\normalsize
The coefficient, in the form of eq. \ref{N.7} can indeed be recognized as the left hand side of a Yang-Baxter identity, whose diagram is given in fig. \ref{5.n}.
\\
\\
Hence the desired reduced form for the coefficient of $L[r_{n-2},u_{n-2}]$ is,
\small\begin{equation}\begin{split}
&\frac{W^{N-(n-2)}_{B,+}(u_{n-1}-u_n)}{W^{N-(n-1)}_{B,+}(u_{n-1}-u_n)} \frac{W_{A}(u_{n-1}-u_{n-2})}{W^{N-(n-2)}_{B,+}(u_{n-1}-u_{n-2})} \frac{ W^{N-(n-1)}_{C+}(u_{n-2}-u_{n})}{ W^{N-(n-2)}_{B,+}(u_{n}-u_{n-2})}\\
 =&f^N_{n-2} \left( u_{n-1}\atop{u_{n-2}} \right)g^N_{n-1} \left( u_{n}\atop{u_{n-2}} \right) \label{N.8}
\end{split}\end{equation}\normalsize
where we have used the following identity,
\small\begin{equation*}
W^{k}_{B,+}(u_{i}-u_{j})W^{l}_{B,+}(u_{m}-u_{n}) = W^{l}_{B,+}(u_{i}-u_{j})W^{k}_{B,+}(u_{m}-u_{n})
\end{equation*}\normalsize
Substituting eq. \ref{N.8} in eq. \ref{N.6} we obtain the twice rolled, Yang-Baxter reduced form of $L[r_{n},u_{n}]$,
\small\begin{equation}\begin{split}
L[r_{n},u_{n}]=&f^N_{n-1} \left( u_{n-1}\atop{u_n} \right)f^N_{n-2} \left( u_{n-2}\atop{u_n} \right) L[r_{n-2},u_{n}] \\
&+  g^N_{n-1} \left( u_{n}\atop{u_{n-1}} \right)f^N_{n-2} \left( u_{n-2}\atop{u_{n-1}} \right) L[r_{n-2},u_{n-1}]  \\
&+  g^N_{n-1} \left( u_{n}\atop{u_{n-2}} \right)f^N_{n-2} \left( u_{n-1}\atop{u_{n-2}} \right) L[r_{n-2},u_{n-2}] \label{N.9}
\end{split}\end{equation}\normalsize\\
\textbf{Remark.} Guessing the result for rolling a general number of times should now be quite obvious. Nevertheless, in order to provide a proper proof of the result the Yang-Baxter process involved for rolling three times is \textit{highly illuminating} and instantly shows the method required for the general proof.\\
\\
\textbf{Rolling three times.} The results for rolling three times are as follows,
\small\begin{equation*}\begin{split}
L[r_{n},u_{n}] =& f^N_{n-1} \left( u_{n-1}\atop{u_n} \right)f^N_{n-2} \left( u_{n-2}\atop{u_n} \right) f^N_{n-3} \left( u_{n-3}\atop{u_n} \right)L[r_{n-3},u_{n}] \\
&+ g^N_{n-1} \left( u_{n}\atop{u_{n-1}} \right)f^N_{n-2} \left( u_{n-2}\atop{u_{n-1}} \right) f^N_{n-3} \left( u_{n-3}\atop{u_{n-1}} \right)L[r_{n-3},u_{n-1}] \\
&+ g^N_{n-1} \left( u_{n}\atop{u_{n-2}} \right)f^N_{n-2} \left( u_{n-1}\atop{u_{n-2}} \right) f^N_{n-3} \left( u_{n-3}\atop{u_{n-2}} \right)L[r_{n-3},u_{n-2}] 
\end{split}\end{equation*}\normalsize
\small\begin{equation*}\begin{split}
&+ \left\{ f^N_{n-1} \left( u_{n-1}\atop{u_{n}} \right)f^N_{n-2} \left( u_{n-2}\atop{u_{n}} \right) g^N_{n-3} \left( u_{n}\atop{u_{n-3}} \right) \right.\\
& + g^N_{n-1} \left( u_{n}\atop{u_{n-1}} \right)f^N_{n-2} \left( u_{n-2}\atop{u_{n-1}} \right) g^N_{n-3} \left( u_{n-1}\atop{u_{n-3}} \right) \\
& \left. + g^N_{n-1} \left( u_{n}\atop{u_{n-2}} \right)f^N_{n-2} \left( u_{n-1}\atop{u_{n-2}} \right) g^N_{n-3} \left( u_{n-2}\atop{u_{n-3}} \right) \right\} L[r_{n-3},u_{n-3}]
\end{split}\end{equation*}\normalsize
The coefficient of $L[r_{n-3},u_{n-3}]$ can be seen as one half of a Yang-Baxter identity, where each term consists of five faces, as given in fig. \ref{5.o}. 
\\
Using the above identity the coefficient of $L[r_{n-3},u_{n-3}]$ immediately simplifies to,
\small\begin{equation*}
g^N_{n-1} \left( u_{n}\atop{u_{n-3}} \right)f^N_{n-2} \left( u_{n-1}\atop{u_{n-3}} \right) f^N_{n-3} \left( u_{n-2}\atop{u_{n-3}} \right)
\end{equation*}\normalsize
In order to provide a framework for the case of rolling a general number of times, we need to break up $f^N_{k} \left( u_{i}\atop{u_{j}} \right)$ as the product of two functions, one involving rapidities exclusively and one involving heights exclusively,
\small\begin{equation*}\begin{split}
f^N_{k} \left( u_{i}\atop{u_{j}} \right) &= \frac{[u_i-u_j+1]}{[u_i-u_j]} \frac{[\zeta + N-k]}{[\zeta + N-k+1]}\\
&= f_{(1)} \left( u_{i}\atop{u_{j}} \right) f_{(2)}\left( N\atop{k} \right) 
\end{split}\end{equation*}\normalsize
Using this separation of $f^N_{k} \left( u_{i}\atop{u_{j}} \right)$, we can express $L[r_{n},u_{n}]$ in the following highly suggestive form,
\small\begin{equation}\begin{split}
L[r_{n},u_{n}] =&  \left\{   \prod^3_{l=2} f_{(2)}\left( N\atop{n-l} \right) \right\}  \sum^3_{k=1}g^N_{n-1} \left( u_{n}\atop{u_{n-k}} \right)\left\{\prod^3_{j=1\atop{\ne k}}   f_{(1)} \left( u_{n-j}\atop{u_{n-k}} \right) L[r_{n-3},u_{n-k}]\right\} \\
&+\left\{ \prod^3_{j=1}f_{(1)} \left( u_{n-j}\atop{u_{n}} \right) f_{(2)}\left( N\atop{n-j} \right)  \right\} L[r_{n-3},u_{n}] \label{N.10}
\end{split}\end{equation}\normalsize\\
\textbf{Rolling many times.} In order to give the general formula for rolling $L[r_{n},u_{n}]$ a general number of times, we first give the little result,
\begin{proposition}
\label{bigprop}
\small\begin{equation}\begin{split}
&\left\{ \prod^{k-1}_{j=1}f_{(1)}\left( u_{n-j} \atop{u_n} \right)f_{(2)}\left( N \atop{n-j} \right)  \right\} g^N_{n-k}\left( u_{n} \atop{u_{n-k}} \right) \\
&+ \left\{ \prod^{k-1}_{j=2} f_{(2)}\left( N \atop{n-j} \right)  \right\} \sum^{k-1}_{p=1}g^N_{n-1}\left( u_{n} \atop{u_{n-p}} \right)  \left\{ \prod^{k-1}_{j=1\atop{\ne p}}f_{(1)}\left( u_{n-j} \atop{u_{n-p}} \right)  \right\} g^N_{n-k}\left( u_{n-p} \atop{u_{n-k}} \right)\\
=& g^N_{n-1}\left( u_{n} \atop{u_{n-k}} \right) \left\{ \prod^{k-1}_{j=1} f_{(1)}\left( u_{n-j} \atop{u_{n-k}} \right)  f_{(2)}\left( N \atop{n-(j+1)} \right)  \right\} 
\label{notu}\end{split}\end{equation}\normalsize
for $2 \le k \le n-1$.
\end{proposition}
\textbf{Comment.} The left hand side of eq. \ref{notu} is obviously the coefficient of the configuration $L[r_{n-k},u_{n-k}]$, after rolling $k$ times. We now verify the above expression using the obvious generalization of the Yang-Baxter diagram shown when we considered rolling three times\\
\\
\textbf{Proof of proposition \ref{bigprop}.} We first consider the single term on the right hand side of proposition \ref{bigprop}, shown pictorially as fig. \ref{5.p}.
\\
\\
It is our goal now to use the Yang-Baxter relations to shift the \textbf{diamond} face from the far right to the far left. However, we use the rule that when any additional use of the Yang-Baxter operation to a particular configuration yields only one configuration as opposed to two, we leave that particular configuration and move on. This way we generate the desired number of configurations. Consider the example of applying the Y-B operation once to the initial configuration (fig. \ref{5.p}) shown in fig. \ref{5.q}.
For the configurations shown in fig. \ref{5.q} notice that one more operation of Yang-Baxter on the left configuration will only yield one configuration, whereas one more operation of Yang-Baxter on the right configuration will yield two configurations. The underlying mechanism of which configuration to choose relies on the three circled heights being equal. \\
\\
We now expand the configuration shown in fig. \ref{5.p} totally, using the aforementioned rule by introducing the graphical notation shown in fig. \ref{5.r}.
 \\
\\
\noindent Using these graphical assignments, the configuration in fig. \ref{5.p} can be expanded immediately to give,
\small\begin{equation*}\begin{split}
&\left\{ \prod^{k-1}_{q=1} L_{q} \right\} W^{N-(n-1)}_{C,-}(u_{n}-u_{n-k})= \sum^{k-1}_{l=1} \left\{ \prod^{k-1}_{q_1 = l+1} L_{q_1} \right\} K_l  \left\{ \prod^{l-1}_{q_2 = 1} P_{q_2} \right\}\\
 +& W^{N-(n-k)}_{C,+}(u_{n-k}-u_n) \left\{ \prod^{k-1}_{q = 1} P_{q} \right\}
\end{split}\end{equation*}\normalsize
Dividing both sides of this expression by the following multiplicative product,
\small\begin{equation*}\begin{split}
W^{N-(n-2)}_{C,-}(u_{n-k}-u_n) \left\{ \prod^{k-1}_{q_1=2} W^{N-(n-q_1)}_{B,+}(u_{n-(k-q_1)}-u_{n-k}) \right\}\\
 \times \left\{ \prod^{k-1}_{q_2=2} W^{N-(n-1-q_2)}_{B,+}(u_{n-(k-q_2)}-u_{n}) \right\}
\end{split}\end{equation*}\normalsize
instantly verifies proposition \ref{bigprop}. $\square$\\
\\
Using proposition \ref{bigprop}, the (reduced) result of rolling a general number of times immediately becomes,
\small\begin{equation}\begin{split}
L[r_{n},u_{n}] =& \left\{   \prod^{n-q}_{l=2} f_{(2)}\left( N\atop{n-l} \right) \right\}  \sum^{n-q}_{k=1}g^N_{n-1} \left( u_{n}\atop{u_{n-k}} \right)\left\{\prod^{n-q}_{j=1\atop{\ne k}}   f_{(1)} \left( u_{n-j}\atop{u_{n-k}} \right) L[r_{q},u_{n-k}]\right\}  \\
&+\left\{ \prod^{n-q}_{j=1}f_{(1)} \left( u_{n-j}\atop{u_{n}} \right) f_{(2)}\left( N\atop{n-j} \right)  \right\} L[r_{q},u_{n}] \label{N.11}
\end{split}\end{equation}\normalsize\\
\textbf{Putting everything together.} Thus letting $q = 1$ in the above expression we obtain,
\small\begin{equation}\begin{split}
L[r_{n},u_{n}] &= \left\{   \prod^{n-1}_{l=2} f_{(2)}\left( N\atop{n-l} \right) \right\}  \sum^{n-1}_{k=1}g^N_{n-1} \left( u_{n}\atop{u_{n-k}} \right)\left\{\prod^{n-1}_{j=1\atop{\ne k}}   f_{(1)} \left( u_{n-j}\atop{u_{n-k}} \right) L[r_{1},u_{n-k}]\right\}  \\
&+\left\{ \prod^{n-1}_{j=1}f_{(1)} \left( u_{n-j}\atop{u_{n}} \right) f_{(2)}\left( N\atop{n-j} \right)  \right\} L[r_{1},u_{n}]\\
&= \sum^n_{k=1}\left\{ \frac{g^{N}_{n-1} \left( u_n \atop{u_k} \right)}{f^{N}_{n-1} \left( u_n \atop{u_k} \right)}  \frac{f_{(2)} \left( N \atop{k} \right)}{f_{(2)} \left( N \atop{n} \right)} \prod^n_{j=1\atop{\ne k}}f^{N}_{j} \left( u_j \atop{u_k} \right)  \right\} L[r_{1},u_{k}]\label{N.12}
\end{split}\end{equation}\normalsize
As stated earlier, the configuration $L[r_{1},u_{k}]$ consists of a top row completely frozen into $W_A$ faces, leading to the expression,
\small\begin{equation}
 L[r_1,u_k] =\left\{ \prod^{N-1}_{j=1}W_A(u_k - v_j)\right\} Z^{(1)}_{N-1} \left( \vec{u},\vec{v},\hat{u}_k,\hat{v}_N \right) \label{N.13}
\end{equation}\normalsize
Substituting eq. \ref{N.13} into eq. \ref{N.12} we obtain,
\small\begin{equation}\begin{split}
L[r_{n},u_{n}] =&  \sum^n_{k=1}\left\{ \frac{g^{N}_{n-1} \left( u_n \atop{u_k} \right)}{f^{N}_{n-1} \left( u_n \atop{u_k} \right)}  \frac{f_{(2)} \left( N \atop{k} \right)}{f_{(2)} \left( N \atop{n} \right)} \prod^n_{j=1\atop{\ne k}}f^{N}_{j} \left( u_j \atop{u_k} \right)  \right\} \left\{ \prod^{N-1}_{j=1}W_A(u_k - v_j)\right\} \\
& \times Z^{(1)}_{N-1} \left( \vec{u},\vec{v},\hat{u}_k,\hat{v}_N \right) \label{N.14}
\end{split}\end{equation}\normalsize
Finally, substituting eq. \ref{N.14} into eq. \ref{N.3} we receive the complete recurrence relation for the DWPF of the BSOS model,
\small\begin{equation}\begin{split}
Z^{(h)}_{N }(\{u\},\{v\}) =& \sum^N_{n=1}\sum^n_{k=1}  \left( \prod^{n-1}_{j=1}W^{N-j}_{B,+}(u_j-v_N) \right) W^{N-n}_{C,+}(u_n-v_N) \\
&\times \left( \prod^{N}_{j=n+1} W_A(u_j-v_N) \right)\left\{ \frac{g^{N}_{n-1} \left( u_n \atop{u_k} \right)}{f^{N}_{n-1} \left( u_n \atop{u_k} \right)}  \frac{f_{(2)} \left( N \atop{k} \right)}{f_{(2)} \left( N \atop{n} \right)} \prod^n_{j=1\atop{\ne k}}f^{N}_{j} \left( u_j \atop{u_k} \right)  \right\}\\
&\times \left\{ \prod^{N-1}_{j=1}W_A(u_k - v_j)\right\}Z^{(1)}_{N-1} \left( \vec{u},\vec{v},\hat{u}_k,\hat{v}_N \right) \label{N.15}
\end{split}\end{equation}\normalsize
Note that the above expression involves a double summation. In what follows we shall apply a method originally devised in \cite{OPCF} for the six-vertex model, to simplify the above expression into a summation over a single index.
\subsection{Simplifying the recurrence relation}
\label{STRL}
\begin{figure}[h!]
\begin{center}
\includegraphics[angle=0,scale=0.15]{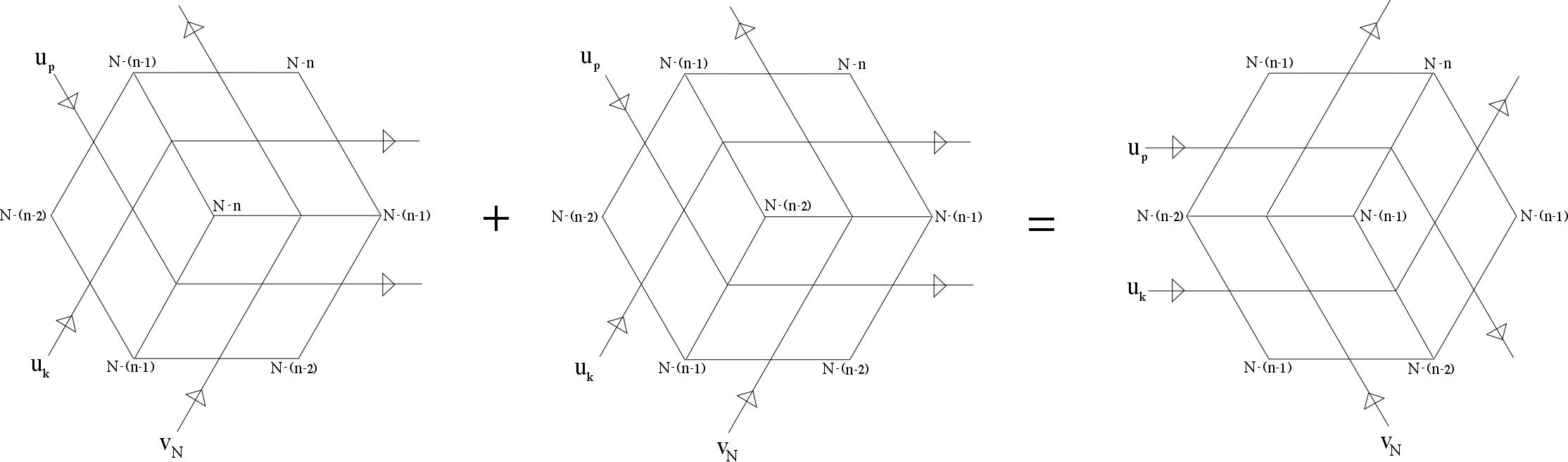}
\caption{\footnotesize{Yang-Baxter identity necessary for this section.}}
\label{5.s}
\end{center}
\end{figure}

We now carefully consider each specific case of the value of $n$, $1 \le n \le N$, for all valid values of $k$, $1 \le k \le n$, and show how we can combine both summation variables to transform eq. \ref{N.15} into a single summation. The goal of this process is to carefully combine all the coefficients of each $Z^{(1)}_{N-1} \left( \vec{u},\vec{v},\hat{u}_k,\hat{v}_N \right)$ term, $1 \le k \le N$, using various Yang-Baxter identities.\\
\\
\textbf{Comment on Y-B identities.} It is fortunate that in the details below, only one Y-B identity is necessary, given in fig. \ref{5.s}.
\\
\\
For the above diagram, the necessary values of $n,p$ and $k$ can easily be inferred from the workings below.\\
\\
\textbf{Step (1i): $\mathbf{(n=1, k=1)+(n=2, k=1)}$.} Combining these two expressions we obtain,
\small\begin{equation}\begin{split}
&\underbrace{\left\{ W^{N-1}_{C,+}(u_1-v_N) W_A(u_2-v_N) + W^{N-1}_{B,+}(u_1-v_N) W^{N-2}_{C,+}(u_2-v_N) g^N_1 \left( u_2\atop{u_1} \right) \right\}}_{\textrm{use Y-B identity}} \\
&\times \left( \prod^{N}_{j=3} W_A(u_j-v_N) \right)  \left\{ \prod^{N-1}_{j=1}W_A(u_1 - v_j)\right\}Z^{(1)}_{N-1} \left(\hat{u}_1,\hat{v}_N \right) \\
=& W^{N-2}_{C,+}(u_1-v_N) W^{N-1}_{B,+}(u_2-v_N) f^N_1 \left( u_2\atop{u_1} \right) \left( \prod^{N}_{j=3} W_A(u_j-v_N) \right)\\
&\times  \left\{ \prod^{N-1}_{j=1}W_A(u_1 - v_j)\right\} Z^{( 1)}_{N-1}  \left(\hat{u}_1,\hat{v}_N \right) 
\label{N.16}
\end{split}\end{equation}\normalsize
Eq. \ref{N.16} simplifies the first two expressions containing $k=1$. Our next step is to simplify this expression with the next $k=1$ term, and also simplify the first two $k=2$ terms. \\
\\
\textbf{Step (2i): $\mathbf{(n=3,k=1)}$ + eq. \ref{N.16}}
\small\begin{equation}\begin{split}
&\underbrace{\left\{ W^{N-2}_{C,+}(u_1-v_N) W_A(u_3-v_N) + W^{N-2}_{B,+}(u_1-v_N) W^{N-3}_{C,+}(u_3-v_N) g^N_2 \left( u_3\atop{u_1} \right) \right\}}_{\textrm{use Y-B identity}} \\
&\times W^{N-1}_{B,+}(u_2-v_N) f^N_1 \left( u_2\atop{u_1} \right)  \left( \prod^{N}_{j=4} W_A(u_j-v_N) \right)  \left\{ \prod^{N-1}_{j=1}W_A(u_1 - v_j)\right\}\\
& \times Z^{(1)}_{N-1} (\hat{u}_1,\hat{v}_N ) \\
=& W^{N-3}_{C,+}(u_1-v_N) \left( \prod^3_{j=2} W^{N-(j-1)}_{B,+}(u_j-v_N)  f^N_{j-1} \left( u_j\atop{u_1} \right)\right) \left( \prod^{N}_{j=4} W_A(u_j-v_N) \right) \\
 & \times \left\{ \prod^{N-1}_{j=1}W_A(u_1 - v_j)\right\}Z^{(1)}_{N-1} (\hat{u}_1,\hat{v}_N )\\
=& W^{N-3}_{C,+}(u_1-v_N) \left( \prod^3_{j=2} W_{B}(u_j-v_N)  f_{(1)} \left( u_j\atop{u_1} \right)\right) \left( \prod^{N}_{j=4} W_A(u_j-v_N) \right) \\
 & \times \left\{ \prod^{N-1}_{j=1}W_A(u_1 - v_j)\right\}Z^{(1)}_{N-1} (\hat{u}_1,\hat{v}_N )\label{N.17}
\end{split}\end{equation}\normalsize
where we have noted that the following expression,
\small\begin{equation*}\begin{split}
W^{N-(j-1)}_{B,+}(u_j-v_N)  f^N_{j-1} \left( u_j\atop{u_1} \right) &= \frac{[u_j-v_N]}{[1]}\frac{[u_j - u_1+1]}{[u_j - u_1]}\\
&= W_{B}(u_j-v_N)  f_{(1)} \left( u_j\atop{u_1}\right)
\end{split}\end{equation*}\normalsize
has no height dependence.\\
\\
\textbf{Step (2ii): $\mathbf{(n=2,k=2)+(n=3,k=2)}$.} 
\small\begin{equation}\begin{split}
&\underbrace{\left\{ W^{N-2}_{C,+}(u_2-v_N) W_A(u_3-v_N) + W^{N-2}_{B,+}(u_2-v_N) W^{N-3}_{C,+}(u_3-v_N) g^N_2 \left( u_3\atop{u_2} \right) \right\}}_{\textrm{use Y-B identity}} \\
&\times W_{B}(u_1-v_N) f_{(1)} \left( u_1\atop{u_2} \right)  \left( \prod^{N}_{j=4} W_A(u_j-v_N) \right)  \left\{ \prod^{N-1}_{j=1}W_A(u_2 - v_j)\right\}\\
& \times Z^{(1)}_{N-1} (\hat{u}_2,\hat{v}_N ) \\
=& W^{N-3}_{C,+}(u_2-v_N) \left( \prod^3_{j=1\atop{\ne 2}} W_{B}(u_j-v_N)  f_{(1)} \left( u_j\atop{u_2} \right)\right) \left( \prod^{N}_{j=4} W_A(u_j-v_N) \right) \\
 & \times \left\{ \prod^{N-1}_{j=1}W_A(u_2 - v_j)\right\}Z^{(1)}_{N-1} (\hat{u}_2,\hat{v}_N ) \label{N.18}
\end{split}\end{equation}\normalsize
Thus using the logic of steps one and two, step three consists of three obvious stages.\\
\\
\textbf{Step (3i): $\mathbf{(n=4,k=1)}$ + eq. \ref{N.17}}
\small\begin{equation}\begin{split}
&\underbrace{\left\{ W^{N-3}_{C,+}(u_1-v_N) W_A(u_4-v_N) + W^{N-3}_{B,+}(u_1-v_N) W^{N-4}_{C,+}(u_4-v_N) g^N_3 \left( u_4\atop{u_1} \right) \right\}}_{\textrm{use Y-B identity}}\\
&\times \left( \prod^3_{j=2} W_{B}(u_j-v_N) f_{(1)} \left( u_j\atop{u_1} \right) \right)  \left( \prod^{N}_{j=5} W_A(u_j-v_N) \right)  \left\{ \prod^{N-1}_{j=1}W_A(u_1 - v_j)\right\}\\
& \times Z^{(1)}_{N-1} (\hat{u}_1,\hat{v}_N ) \\
=& W^{N-4}_{C,+}(u_1-v_N) \left( \prod^4_{j=2} W_{B}(u_j-v_N)  f_{(1)} \left( u_j\atop{u_1} \right)\right) \left( \prod^{N}_{j=5} W_A(u_j-v_N) \right)\\
&  \times \left\{ \prod^{N-1}_{j=1}W_A(u_1 - v_j)\right\}Z^{(1)}_{N-1} (\hat{u}_1,\hat{v}_N )\label{N.19}
\end{split}\end{equation}\normalsize
We proceed similarly for steps $(3ii)$ and $(3iii)$.\\
\\
\textbf{Step (3ii): $\mathbf{(n=4,k=2)}$ + eq. \ref{N.18}}
\small\begin{equation}\begin{split}
W^{N-4}_{C,+}(u_2-v_N) \left( \prod^4_{j=1\atop{\ne 2}} W_{B}(u_j-v_N)  f_{(1)} \left( u_j\atop{u_2} \right)\right) \left( \prod^{N}_{j=5} W_A(u_j-v_N) \right) \\
 \times \left\{ \prod^{N-1}_{j=1}W_A(u_2 - v_j)\right\}Z^{(1)}_{N-1} (\hat{u}_2,\hat{v}_N ) \label{N.20}
\end{split}\end{equation}\normalsize
\textbf{Step (3iii): $\mathbf{(n=3,k=3) + (n=4,k=3)}$ }
\small\begin{equation}\begin{split}
W^{N-4}_{C,+}(u_3-v_N) \left( \prod^4_{j=1\atop{\ne 3}} W_{B}(u_j-v_N)  f_{(1)} \left( u_j\atop{u_3} \right)\right) \left( \prod^{N}_{j=5} W_A(u_j-v_N) \right) \\
  \times \left\{ \prod^{N-1}_{j=1}W_A(u_3 - v_j)\right\}Z^{(h+1)}_{N-1} (\hat{u}_3,\hat{v}_N ) \label{N.21}
\end{split}\end{equation}\normalsize
Therefore given these intermediate steps, we can express the partition function in the following suggestive form,
\small\begin{equation}\begin{split}
Z^{(0)}_N (\vec{u},\vec{v}) =& \sum^3_{r=1}W^{N-4}_{C,+}(u_r-v_N) \left( \prod^4_{j=1\atop{\ne r}} W_{B}(u_j-v_N)  f_{(1)} \left( u_j\atop{u_r} \right)\right)  \\
&  \times\left( \prod^{N}_{j=5} W_A(u_j-v_N) \right)   \left\{ \prod^{N-1}_{j=1}W_A(u_r - v_j)\right\}Z^{(1)}_{N-1} (\hat{u}_r,\hat{v}_N )\\
& + (n=4,k=4) + (n=\{5,\dots,N\}, k=\{ 1,\dots,n\}) 
\end{split}\end{equation}\normalsize
Inspired from the above expression we now propose a general simplified\footnote{In the sense that there is only one summation.} form for the partition function and use inductive techniques for a proof.\\
\\
\textbf{Step $\mathbf{(m)}$, $\mathbf{1 \le m \le N-1}$.} 
\begin{proposition}
\small\begin{equation}\begin{split}
Z^{(0)}_N (\vec{u},\vec{v}) =& \sum^m_{r=1}W^{N-(m+1)}_{C,+}(u_r-v_N) \left( \prod^{m+1}_{j=1\atop{\ne r}} W_{B}(u_j-v_N)  f_{(1)} \left( u_j\atop{u_r} \right)\right)  \\
&  \times   \left( \prod^{N}_{j=m+2} W_A(u_j-v_N) \right)  \left\{ \prod^{N-1}_{j=1}W_A(u_r - v_j)\right\}Z^{(1)}_{N-1} (\hat{u}_r,\hat{v}_N )  \\
&+ (n=m+1,k=m+1) + (n=\{m+2,\dots,N\}, k=\{ 1,\dots,n\}) \label{N.24}
\end{split}\end{equation}\normalsize
\end{proposition}
\textbf{Proof.} We notice that the above expression is valid for $m = 1,2,3$. Let us now assume that it is valid for general $m$, and analyze the situation for $m+1$. Note that this proof requires that $m$ in the above expression is not equal to $N-1$.\\
\\
\textbf{Step $\mathbf{((m +1) \underbrace{\mathbf{i \dots i}}_{r})}$, $\mathbf{1 \le r \le m}$.}\\
\\
\textbf{Comment.} Note that the first $m$ steps of this proof can be accomplished in the following \textit{one} procedure by keeping the $r$ variable general.\\
\\
Beginning with the $r$th component of the above summation, we add this to the $(n=m+2, k = r)$ component of eq. \ref{N.15}. 
\small\begin{equation}\begin{split}
&\underbrace{\left\{\begin{array}{c} W^{N-(m+1)}_{C,+}(u_r-v_N) W_A(u_{m+2}-v_N) \\+ W^{N-(m+1)}_{B,+}(u_r-v_N) W^{N-(m+2)}_{C,+}(u_{m+2}-v_N) g^N_{m+1} \left( u_{m+2}\atop{u_r} \right) \end{array} \right\}}_{\textrm{use Y-B identity}} \\
&\times \left( \prod^{m+1}_{j=1\atop{\ne r}} W_{B}(u_j-v_N) f_{(1)} \left( u_j\atop{u_r} \right) \right)  \left( \prod^{N}_{j=m+3} W_A(u_j-v_N) \right) \\
& \times  \left\{ \prod^{N-1}_{j=1}W_A(u_r - v_j)\right\} Z^{(1)}_{N-1} (\hat{u}_r,\hat{v}_N ) \\
=&W^{N-(m+2)}_{C,+}(u_r-v_N) \left( \prod^{m+2}_{j=1\atop{\ne r}} W_{B}(u_j-v_N)  f_{(1)} \left( u_j\atop{u_r} \right)\right) \\
&  \times   \left( \prod^{N}_{j=m+3} W_A(u_j-v_N) \right)  \left\{ \prod^{N-1}_{j=1}W_A(u_r - v_j)\right\}Z^{(1)}_{N-1} (\hat{u}_r,\hat{v}_N ) \label{N.22}
\end{split}\end{equation}\normalsize
This leads us to the final stage of step $m+1$.\\
\\
\textbf{Step $\mathbf{((m+1)\underbrace{\mathbf{i \dots i}}_{m+1}): (n=m+1,k=m+1)+(n=m+2,k=m+1)}$.}
\small\begin{equation}\begin{split}
&\underbrace{\left\{\begin{array}{c} W^{N-(m+1)}_{C,+}(u_{m+1}-v_N) W_A(u_{m+2}-v_N)\\ + W^{N-(m+1)}_{B,+}(u_{m+1}-v_N) W^{N-(m+2)}_{C,+}(u_{m+2}-v_N) g^N_{m+1} \left( u_{m+2}\atop{u_{m+1}} \right) \end{array} \right\}}_{\textrm{use Y-B identity}} \\
&\times \left( \prod^{m}_{j=1} W_{B}(u_j-v_N) f_{(1)} \left( u_j\atop{u_{m+1}} \right) \right)  \left( \prod^{N}_{j=m+3} W_A(u_j-v_N) \right)\\
& \times   \left\{ \prod^{N-1}_{j=1}W_A(u_{m+1} - v_j)\right\}  Z^{(1)}_{N-1} (\hat{u}_{m+1},\hat{v}_N ) \\
=&W^{N-(m+2)}_{C,+}(u_{m+1}-v_N) \left( \prod^{m+2}_{j=1\atop{\ne m+1}} W_{B}(u_j-v_N)  f_{(1)} \left( u_j\atop{u_{m+1}} \right)\right)  \\
&  \times   \left( \prod^{N}_{j=m+3} W_A(u_j-v_N) \right)  \left\{ \prod^{N-1}_{j=1}W_A(u_{m+1} - v_j)\right\}Z^{(1)}_{N-1} (\hat{u}_{m+1},\hat{v}_N ) \label{N.23}
\end{split}\end{equation}\normalsize
Thus adding eqs. \ref{N.22} and \ref{N.23} we obtain the required expression for $m+1$,
\small\begin{equation*}\begin{split}
Z^{(0)}_N (\vec{u},\vec{v}) =& \sum^{m+1}_{r=1}W^{N-(m+2)}_{C,+}(u_r-v_N) \left( \prod^{m+2}_{j=1\atop{\ne r}} W_{B}(u_j-v_N)  f_{(1)} \left( u_j\atop{u_r} \right)\right)   \\
&  \times   \left( \prod^{N}_{j=m+3} W_A(u_j-v_N) \right)  \left\{ \prod^{N-1}_{j=1}W_A(u_r - v_j)\right\}Z^{(1)}_{N-1} (\hat{u}_r,\hat{v}_N )\\
&+ (n=m+2,k=m+2) + (n=\{m+3,\dots,N\}, k=\{ 1,\dots,n\}) \textrm{    } \square
\end{split}\end{equation*}\normalsize
Therefore, substituting $m = N-1$ in eq. \ref{N.24} and evaluating the $(n=N, k=N)$ term we receive the following simplified recursion relation expression for the BSOS DWPF,
\small\begin{equation}\begin{split}
Z^{(0)}_N (\vec{u},\vec{v}) =& \sum^N_{r=1}W^{0}_{C,+}(u_r-v_N) \left( \prod^{N}_{j=1\atop{\ne r}} W_{B}(u_j-v_N)  f_{(1)} \left( u_j\atop{u_r} \right)\right)   \\
&  \times    \left\{ \prod^{N-1}_{j=1}W_A(u_r - v_j)\right\}Z^{(1)}_{N-1} (\hat{u}_r,\hat{v}_N )  \label{N.25}
\end{split}\end{equation}\normalsize
\subsection{Sum over the symmetric group}
Using eq. \ref{N.25} we can express the $N \times N$ DWPF, not as a recursion relation, but as a sum over all possible permutations of the string $[1,2,\dots,N]$. The derivation of this result begins with considering the first few terms, $Z_1$ and $Z_2$, which allows us to guess an appropriate form for $Z_N$. This form is then proven by using eq. \ref{N.25}.\\
\\
\textbf{Case 1, $\mathbf{N=1}$.} We have immediately that,
\small\begin{equation*}
Z^{(0)}_1(u_1,v_1) = W^{0}_{C,+}(u_1-v_1) 
\end{equation*}\normalsize
\textbf{Case 2, $\mathbf{N=2}$.} Using the above result we obtain,
\small\begin{equation*}\begin{split}
Z^{(0)}_2(\vec{u},\vec{v}) &=   \sum_{\sigma \in S_2} \left\{ \prod_{1 \le i<j\le 2} W_B(u_{\sigma_i}-v_j )   f_{(1)} \left( u_{\sigma_i}\atop{u_{\sigma_j}} \right) W_A(u_{\sigma_j}-v_i )  \right\} \\
& \times \left( \prod^2_{k=1} W^{2-k}_{C,+}(u_{\sigma_k}-v_k) \right)
\end{split}\end{equation*}\normalsize
This suggestive form of $N=2$ brings us to the obvious guess for the general $N$ result.
\begin{proposition}
\small\begin{equation}\begin{split}
Z^{(0)}_N(\vec{u},\vec{v}) =&\sum_{\sigma \in S_N} \left\{ \prod_{1 \le i<j\le N} W_B(u_{\sigma_i}-v_j )   f_{(1)} \left( u_{\sigma_i}\atop{u_{\sigma_j}} \right) W_A(u_{\sigma_j}-v_i )  \right\}\\
& \times  \left( \prod^N_{k=1} W^{N-k}_{C,+}(u_{\sigma_k}-v_k) \right)
\label{N.26}\end{split}
\end{equation}\normalsize
\end{proposition}
\textbf{Proof.} We notice that the above expression holds for $N=1,2$. Assuming the above form holds for some $N$, we now consider the $N+1$ form of eq. \ref{N.26},
\small\begin{equation*}\begin{split}
Z^{(0)}_{N+1} (\vec{u},\vec{v}) =& \sum^{N+1}_{r=1}W^{0}_{C,+}(u_r-v_{N+1}) \left( \prod^{N+1}_{j=1\atop{\ne r}} W_{B}(u_j-v_{N+1})  f_{(1)} \left( u_j\atop{u_r} \right)\right)  \\
& \times \left\{ \prod^{N}_{j=1}W_A(u_r - v_j)\right\}Z^{(1)}_{N} (\hat{u}_r,\hat{v}_{N+1} ) 
\end{split}
\end{equation*}\normalsize
and substitute into this recursive expression the assumed form for $Z_N$ in eq. \ref{N.25} to obtain,
\small\begin{equation}\begin{split}
Z^{(0)}_{N+1} (\vec{u},\vec{v}) =& \sum^{N+1}_{r=1}  W^{0}_{C,+}(u_r-v_{N+1}) \left( \prod^{N+1}_{j=1\atop{\ne r}} W_{B}(u_j-v_{N+1})  f_{(1)} \left( u_j\atop{u_r} \right)\right)\\
& \times \left\{ \prod^{N}_{j=1}W_A(u_r - v_j)\right\}     \sum_{\sigma \in S^{(r)}_{N+1}}  \left( \prod^{N+1}_{k=1\atop{\ne r}} W^{N+1-\beta^{(r)}_k}_{C,+}\left(u_{\sigma_k}-v_{\beta^{(r)}_k} \right) \right)  \\
& \times\left\{ \prod_{1 \le i<j\le N+1\atop{\ne r}} W_B \left(u_{\sigma_i}-v_{\beta^{(r)}_j} \right)   f_{(1)} \left( u_{\sigma_i}\atop{u_{\sigma_j}} \right) W_A \left(u_{\sigma_j}-v_{\beta^{(r)}_i} \right)  \right\} \label{N.27}
\end{split}
\end{equation}\normalsize
where,
\small\begin{equation*}
\beta^{(r)}_{k} = \left\{ \begin{array}{ccc}
k &,& 1 \le k \le r-1\\
k-1 &,& r+1 \le k \le N+1
\end{array}\right.
\end{equation*}\normalsize
and the sum, $\sum_{\sigma \in S^{(r)}_{N+1} }$, is the sum over all possible permutations of the length $N$ string $[1,\dots,\hat{r},\dots,N+1]$. Hence,
\small\begin{equation*}
\sum_{\sigma \in S^{(r)}_{N+1} } = \sum_{\sigma_1,\dots,\hat{\sigma}_r,\dots,\sigma_{N+1}=\{1,\dots, \hat{r},\dots,N+1\} \atop{\sigma_1 \ne \dots \ne \sigma_{N+1}}}
\end{equation*}\normalsize
In order to complete the proof we use the following labels,
\small\begin{equation*}\begin{split}
\sigma_{j} \rightarrow \sigma_{j-1} & \textrm{  ,  }  j \in \{ r+1, \dots N+1 \} \\
r \rightarrow \sigma_{N+1}
\end{split}
\end{equation*}\normalsize
Under these convenient change of labels eq. \ref{N.27} becomes,
\small\begin{equation*}\begin{split}
Z^{(0)}_{N+1} (\vec{u},\vec{v}) = \sum^{N+1}_{\sigma_{N+1}=1}  W^{0}_{C,+}(u_{\sigma_{N+1}}-v_{N+1}) \left( \prod^{N}_{j=1} W_{B}(u_{j}-v_{N+1}) W_A(u_{\sigma_{N+1}} - v_j) \right.  \\
\times \left. f_{(1)} \left( u_j\atop{u_{\sigma_{N+1}}} \right)  \right) \sum_{\sigma \in S^{(\sigma_{N+1})}_{N+1}} \left\{ \prod_{1 \le i<j\le N} W_B(u_{\sigma_i}-v_{j} )   f_{(1)} \left( u_{\sigma_i}\atop{u_{\sigma_j}} \right)W_A(u_{\sigma_j}-v_{j} )  \right\}  \\
\times \left( \prod^{N}_{k=1} W^{N+1-k}_{C,+} (u_{\sigma_k}-v_{k}) \right)  \\
=\sum_{\sigma \in S_{N+1}} W^{0}_{C,+}(u_{\sigma_{N+1}}-v_{N+1}) \left( \prod^{N}_{j=1} W_{B}(u_{\sigma_j}-v_{N+1}) W_A(u_{\sigma_{N+1}} - v_j) \right. \\
\times \left. f_{(1)} \left( u_{\sigma_j}\atop{u_{\sigma_{N+1}}} \right) \right)  \left\{ \prod_{1 \le i<j\le N} W_B(u_{\sigma_i}-v_{j} )   f_{(1)} \left( u_{\sigma_i}\atop{u_{\sigma_j}} \right) W_A(u_{\sigma_j}-v_{j} )  \right\}\\
\times \left( \prod^{N}_{k=1} W^{N+1-k}_{C,+} (u_{\sigma_k}-v_{k}) \right)
\end{split}
\end{equation*}\normalsize
\small\begin{equation*}\begin{split}
\Rightarrow Z^{(0)}_{N+1} (\vec{u},\vec{v}) = \sum_{\sigma \in S_{N+1}} \left\{ \prod_{1 \le i<j\le N+1} W_B(u_{\sigma_i}-v_j )   f_{(1)} \left( u_{\sigma_i}\atop{u_{\sigma_j}} \right) W_A(u_{\sigma_j}-v_i )  \right\}\\
\times \left( \prod^{N+1}_{k=1} W^{N+1-k}_{C,+}(u_{\sigma_k}-v_k) \right) \textrm{   $\square$}
\end{split}
\end{equation*}\normalsize
%%%%%%%%%%%%%%%%%%%%%%%%%%%%%%%%%%%%%%%%%%%%%%%%%%%%%%%%%%%%%%%%%%%%%%%%%%%%%%%%%%%%%%%%%%%%%%%%%%%%%%%%%%%%%%%%%%%%
\newpage
%%%%%%%%%%%%%%%%%%%%%%%%%%%%%%%%%%%%%%%%%%%%%%%%%%%%%%%%%%%%%%%%%%%%%%%%%%%%%%%%%%%%%%%%%%%%%%%%%%%%%%%%%%%%%%%%%%%%
\chapter{The Perk-Schultz (PS) models}
\section{The trigonometric PS vertex model}
In \cite{PSvert1}, J. Perk and C. Schultz discovered a new family of vertex models with commuting transfer matrices, making the models integrable. In more recent years \cite{slrs}, these vertex models have been associated with the Lie \textit{superalgebras} $sl(r+1|s+1)$, due to the obvious asymmetry in the Boltzmann weights when the state variables are conjugated. In \cite{PSvert2} Zhao et. al. derived the determinant representation of the DWPF for the $sl(1|1)$ and $sl(2|1)$ PS vertex models. The $sl(1|1)$ results were then applied in \cite{PSvert3} to derive analytic expressions of the one and two-point correlation functions. \\
\\
In this section of the thesis we use a modified Korepin-Izergin argument and the fact that the weights are asymmetrical under state variable conjugation to derive the \textbf{factorised} form for the DWPF of the model under general $r$ and $s$.
\subsection{Definition  of the model}
\noindent In the remainder of this section we use the notation found in \cite{PSvert4}.\\
\\
\noindent \textbf{Two distinct sets of state variables.} We define the two sets $\field{B}_{-} = \{1,   \ldots, s+1  \}$, $\field{B}_{+} = \{s+2, \ldots, r+s+2 = L\}$, and their union $\field{B}$,
\small\begin{equation*}
\field{B}=\{ \underbrace{1,   \dots,   s+1}_{\field{B}_-}, \underbrace{s+2, \dots, r+s+2 = L}_{\field{B}_+} \}
\end{equation*}\normalsize
It is from these two sets that the allowable weights of the model obtain their state variables. \\
\\
\textbf{The $N \times N$ lattice playground.} Being a 2-dimensional vertex model, our main arena is an $N \times N$ lattice, where each horizontal line in the lattice is labeled by a horizontal rapidity $u_i \in \field{C}$, and each vertical line in the lattice is labeled by a vertical rapidity $v_j\in \field{C}$.
\begin{figure}[h!]
\begin{center}
\includegraphics[angle=0,scale=0.25]{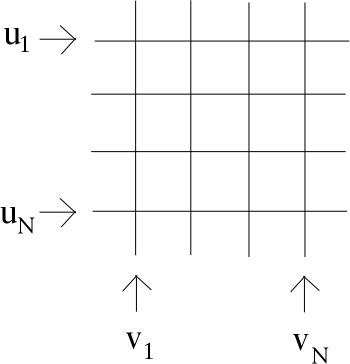}
\caption{\footnotesize{The $N\times N$ lattice with horizontal and vertical rapidity flows.}}
\label{PS.a}
\end{center}
\end{figure}\\
\\
\textbf{Parameterization of the vertices.} Each of the $N^2$ vertices are labeled by the difference of the corresponding horizontal and vertical rapidities, $u_i-v_j$, and additionally by a set of four state variables, $\{a,b,c,d\} \in \field{B}$.
\begin{figure}[h!]
\begin{center}
\includegraphics[angle=0,scale=0.25]{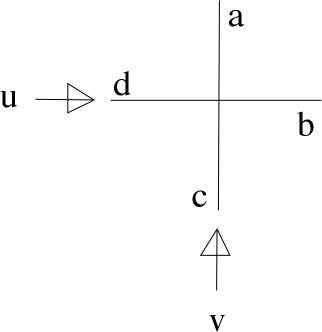}
\caption{\footnotesize{Labeling of the vertex $X(u-v)^{a.b}_{d,c}$.}}
\label{PS.b}
\end{center}
\end{figure}
\newpage
\noindent \textbf{Allowable weights of the model.} Let $a,b \in \field{B}$. The non-vanishing weights of the $sl(r+1|s+1)$ Perk-Schultz (PS) models are given the following parameterizations,
\small\begin{equation*}\begin{array}{lll}
X^{a,a}_{a,a}(u,v) &=&
\left\{
\begin{array}{cc}
\frac{\sinh \left( \eta(1-u+v)\right)}{\sinh\left(\eta\right)}, & a \in \field{B}_- \\
\frac{\sinh \left(\eta(1+u-v)\right)}{\sinh\left(\eta\right)},  & a \in \field{B}_+
\end{array}\right. \\
X^{a,b}_{b,a}(u,v) &=&
\left\{
\begin{array}{cc}
-\frac{\sinh \left(\eta (u-v) \right)}{\sinh\left(\eta\right)},  &
a,b \in{\field{B}_-}\   \textrm{  or  }   a,b \in{\field{B}_+}\ \\
\frac{\sinh \left(\eta (u-v)\right)}{\sinh\left(\eta\right)},  &
{\rm otherwise}
\end{array}\right.\\
X^{a,b}_{a,b}(u,v)& =& \left\{ \begin{array}{cc} e^{\eta (u-v)},  &
a <  b   \\
e^{-\eta (u-v)}, &
a >  b
\end{array}\right.
\end{array}\end{equation*}\normalsize
where $\eta \in \field{C}$ plays the role of the global crossing parameter.\\
\\
\noindent \textbf{Yang-Baxter equation.} Given state variables $\{h_1,h_2,h_3,q_1,q_2,q_3\} \in \field{B}$ and rapidities $\{u_1,u_2,u_3\} \in \field{C}$, the above weights of the PS model obey the following Yang-Baxter identities,
\small\begin{equation*}\begin{split}
&\sum_{g_1, g_2, g_3 \in \field{B}}X^{h_1 h_2}_{g_2 g_1} (u_1-u_2)X^{g_1 h_3}_{g_3 q_1}(u_1-u_3)  X^{g_2 g_3}_{q_3 q_2}(u_2-u_3) \\
=& \sum_{g_1, g_2, g_3 \in \field{B}}X^{h_2 h_3}_{g_3 g_2}(u_2-u_3) X^{h_1 g_3}_{q_3 g_1}(u_1-u_3)  X^{g_1 g_2}_{q_2 q_1}(u_1-u_2)  ,
\end{split}\end{equation*}\normalsize
\noindent \textbf{Non-invariance of the weights under state conjugation.} Unlike the weights of the six vertex model in chapter 3 of this work, it is clear by inspection that the weights of the PS vertex models are not invariant under conjugation of state variables. This is an important running theme in this section as we use this property extensively to obtain the product form of the domain wall partition function.\\
\\
\textbf{Weight symmetry and equivalence of the PS models.} From the definition of the weights it is clear that they are symmetric in the state variables in set $\field{B}_{-}$ and set $\field{B}_{+}$ respectively. This fact makes every choice for DWBC's equivalent, effectively making all DWPF's for general $r$ and $s$ variables equivalent to the $sl(1|1)$ partition function. \\
\\
\noindent \textbf{DWBC's.} We define the $sl(r+1|s+1) $ PS DWBC's as follows:
\begin{itemize}
\item{The state variables on all bonds on the right most and bottom most boundaries are equal to $L=r+s+2 \in \field{B}_{+}$.}
\item{The state variables on all bonds on the left most and top most boundaries are equal to $1 \in \field{B}_{-}$.}
\end{itemize}
\begin{figure}[h!]
\begin{center}
\includegraphics[angle=0,scale=0.25]{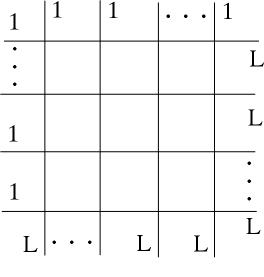}
\caption{\footnotesize{The $N\times N$ lattice with DWBC's.}}
\label{PS.c}
\end{center}
\end{figure}
\noindent Due to the symmetry that exists in the weights we could have taken any single representative in $\field{B}_{+}$ and $\field{B}_{-}$ for DWBC's. Ultimately, we expect no difference between the results we obtain from the $sl(r+1|s+1)$ model, and what has already been obtained from the $sl(1|1)$ model.\\
\\
\textbf{Non-invariance of the line permuting vertices.} Arguably, one of the most useful properties of the $sl(r+1|s+1)$ PS vertex models with DWBC's is the non-invariance of the $X^{a,a}_{a,a}(u,v)$ vertex when $a$ changes from $L$ (an element of $\field{B}_+$) to $1$ (an element of $\field{B}_-$), and vice versa,
\small\begin{equation*}\begin{split}
X^{L,L}_{L,L}(u,v) &= \frac{\sinh \left(\eta(1+u-v)\right)}{\sinh\left(\eta\right)} \\
X^{1,1}_{1,1}(u,v) &= \frac{\sinh \left(\eta(1-u+v)\right)}{\sinh\left(\eta\right)} 
\end{split}\end{equation*}\normalsize
So useful is this property that through using it alone one can uncover the form of the DWPF up to a multiplicative constant. We give the details of this property below.
\subsection{Equivalent trigonometric Korepin properties of the DWPF.} 
\begin{figure}[h!]
\begin{center}
\includegraphics[angle=0,scale=0.25]{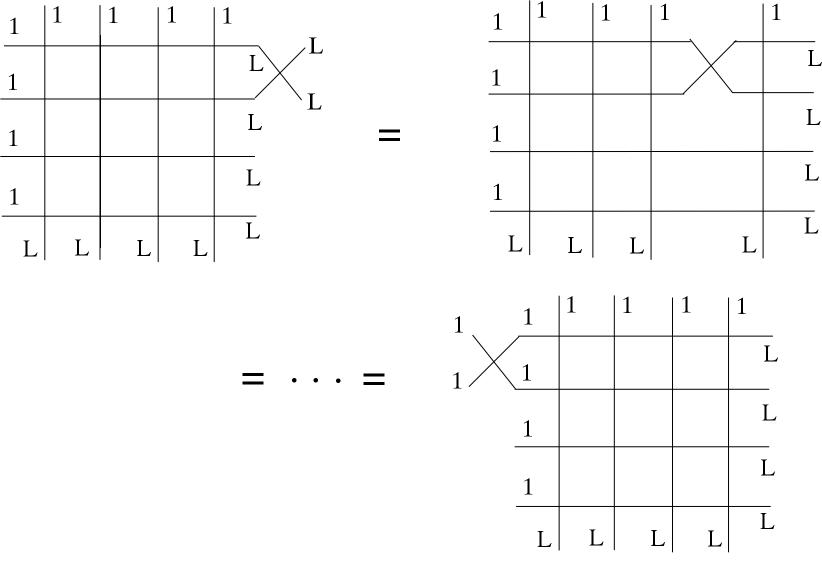}
\caption{\footnotesize{Graphical derivation of eq. \ref{PS.1}.}}
\label{PS.rAR}
\end{center}
\end{figure}

As usual, the DWPF of the $sl(r+1|s+1)$ PS vertex model, $Z^{(r,s)}_N(\vec{u},\vec{v})$, is given as the sum over all allowable weighted configurations of the $N \times N$ lattice under DWBC's,
\small\begin{equation*}
Z^{(r,s)}_N(\vec{u},\vec{v}) = \sum_{\textrm{allowable}\atop{\textrm{configurations}}} \left( \prod_{\textrm{vertices}} X^{a,b}_{d,c} (u_i-v_j) \right)
\end{equation*}\normalsize
In what is to follow we shall suppress the $(r,s)$ superscripts.\\
\\
\textbf{Korepin-like properties of the DWPF.} In direct analogy with \cite{Korepini}, we shall propose four properties that the $sl(r+1|s+1)$ PS DWPF should satisfy. We shall then show that these four properties uniquely determine the partition function, which allows us to simply postulate a valid expression and show that the four properties are satisfied.\\
\\ 
{\bf Property 1.}  The partition function is a polynomial of order $N$ in $e^{\eta u_1}$ and $e^{-\eta v_1}$. Additionally one of the zeroes of the order $N$ polynomial(s) exists in the trivial form $e^{\eta u_1}= 0$ and $e^{-\eta v_1}=0$ respectively.\\
\\
\textbf{Proof.} An elementary analysis of the top row of vertices reveals that all valid configurations are of the form,
\small\begin{equation*}\begin{split}
\left( \prod^{l-1}_{j=1}X^{1,1}_{1,1}(u_1,v_j)  \right) X^{1,L}_{1,L}(u_1,v_l)\left( \prod^{N}_{j=l+1} X^{1,L}_{L,1}(u_1,v_j)  \right) &\textrm{  ,  } 1 \le l \le N\\
= \left( \prod^{l-1}_{j=1} \frac{\sinh \left(\eta(1-u_1+v_j)\right)}{\sinh\left(\eta\right)} \right) e^{\eta (u_1-v_l)} \left( \prod^{N}_{j=l+1}\frac{\sinh \left(\eta (u_1-v_j)\right)}{\sinh\left(\eta\right)}  \right) 
\end{split}\end{equation*}\normalsize
A similar analysis of the left most column reveals that all valid configurations are of the form:
\small\begin{equation*}\begin{split}
\left( \prod^{l-1}_{j=1}X^{1,1}_{1,1}(u_j,v_1)  \right) X^{1,L}_{1,L}(u_l,v_1)\left( \prod^{N}_{j=l+1} X^{L,1}_{1,L}(u_j,v_1)  \right) \textrm{  ,  }  1 \le l \le N\\
=(-1)^{N-l} \left( \prod^{l-1}_{j=1} \frac{\sinh \left(\eta(1-u_j+v_1)\right)}{\sinh\left(\eta\right)} \right) e^{\eta (u_l-v_1)} \left( \prod^{N}_{j=l+1}\frac{\sinh \left(\eta (u_j-v_1)\right)}{\sinh\left(\eta\right)}  \right) 
\end{split}\end{equation*}\normalsize
{\bf Property 2.} From the Yang-Baxter equation it is possible to show that,
\small\begin{equation}
Z_{N} \left( \vec{u}, \vec{v} \right) = \prod^N_{j=2} \frac{X^{1,1}_{1,1}(u_1-u_j)}{X^{L,L}_{L,L}(u_1-u_j)} Z_{N} \left( u_2, \dots, u_N, u_1, \vec{v}\right)
\label{PS.1}
\end{equation}\normalsize
\noindent which gives exactly $N-1$ zeroes of the polynomial in the form $X^{1,1}_{1,1}(u_1 - u_j) = 0$, $j=2, \dots, N$. Performing the equivalent technique on the $\vec{v}$'s we obtain,
\small\begin{equation}
Z_{N} \left( \vec{u},\vec{v}\right) = \prod^{N}_{j=2} \frac{X^{L,L}_{L,L}(v_1 - v_j)}{X^{1,1}_{1,1}(v_1 - v_j)} Z_{N} \left(\vec{u}, v_2, \dots, v_N, v_1\right)
\label{PS.2}
\end{equation}\normalsize
which gives exactly $N-1$ zeroes of the polynomial in the form $X^{L,L}_{L,L}(v_1 - v_j) = 0$, $j=2, \dots, N$.\\
\\
\textbf{Proof.} We consider placing an $X^{L,L}_{L,L}(u_1 - u_2)$ vertex\footnote{Remember that in earlier discussions we label this vertex, and its conjugate, $X^{1,1}_{1,1}(u_1 - u_2)$, as line permuting vertices.} on the right hand side of $Z_{N} \left(\vec{u},\vec{v}\right)$. This process is displayed in the first diagram of fig. \ref{PS.rAR}. We notice that there are no other valid internal state variables for this vertex, thus we can apply the Yang-Baxter equation and shift it through to the left hand side of the lattice. Once at the left side of the lattice, the vertex is fixed to a $X^{L,L}_{L,L}(u_1 - u_2)$ configuration, as these are the only valid internal state variables, and the rapidities $u_1$ and $u_2$ are switched. Thus we obtain,
\small\begin{equation*}
Z_{N} \left( \vec{u}, \vec{v} \right) =  \frac{X^{1,1}_{1,1}(u_1-u_2)}{X^{L,L}_{L,L}(u_1-u_2)} Z_{N} \left( u_2,u_1,u_3, \dots, u_N, \vec{v}\right)
\end{equation*}\normalsize
Applying this process an additional $N-2$ times so that $u_1$ is the bottom-most rapidity, we obtain eq. \ref{PS.1}.\\
\\
The method for the verification of eq. \ref{PS.2} is obviously entirely analogous. We begin by applying the line permuting vertex, $X^{1,1}_{1,1}(v_1 - v_2)$, to the top of $Z_{N} \left(\vec{u},\vec{v}\right)$, and apply the Yang-Baxter equation repeatedly until we obtain,
\small\begin{equation*}
Z_{N} \left( \vec{u},\vec{v}\right) =  \frac{X^{L,L}_{L,L}(v_1 - v_2)}{X^{1,1}_{1,1}(v_1 - v_2)} Z_{N} \left(\vec{u}, v_2,v_1,v_3, \dots, v_N\right)
\end{equation*}\normalsize
Repeating this procedure an additional $N-2$ times so that $v_1$ is the right-most rapidity, we obtain eq. \ref{PS.2}.
\\
\\
{\bf Property 3.} $Z_{N}$ satisfies two recursion relations. The first equation can be derived by freezing rapidities $u_1$ and $v_1$ so that the top left hand corner vertex is always $X^{1,L}_{1,L}$,
\small\begin{equation}\begin{split}
Z_{N}|_{u_1=v_1+1} =& X^{1,L}_{1,L}(1) \left( \prod^N_{k=2} X^{1,L}_{L,1}(u_1 - v_k)X^{L,1}_{1,L}(u_k - v_1) \right) \\
& \times Z_{N-1}\left(\vec{u},\vec{v},\hat{u}_1,\hat{v}_1\right)
\label{PS.3}
\end{split}\end{equation}\normalsize
The second equation can be derived by freezing rapidities $u_N$ and $v_1$ so that the bottom left hand corner vertex is always $X^{1,L}_{1,L}$,
\small\begin{equation}\begin{split}
Z_{N}|_{u_N = v_1} =& X^{1,L}_{1,L}(0) \left( \prod^{N - 1}_{k = 1} X^{1,1}_{1,1}(u_k - v_1) \right) \left( \prod^{N}_{k = 2} X^{L,L}_{L,L}(u_N - v_j) \right)  \\
& \times  Z_{N-1} \left(\vec{u},\vec{v},\hat{u}_N,\hat{v}_1\right)  \label{PS.tt}
\end{split}\end{equation}\normalsize
\textbf{Proof.} Simply freezing the top left vertex to $X^{1,L}_{1,L}$ by fixing the the rapidities, $u_1=v_1+1$, we see immediately that the first row and first column of the $N \times N$ lattice are frozen in the configuration given by eq. \ref{PS.3}. We additionally notice that the remaining vertices in the $(N-1)\times(N-1)$ bulk are (miraculously) under DWBC's, hence verifying eq. \ref{PS.3}. The verification of \ref{PS.tt} is an exactly equivalent process.
\begin{figure}[h!]
\begin{center}
\includegraphics[angle=0,scale=0.20]{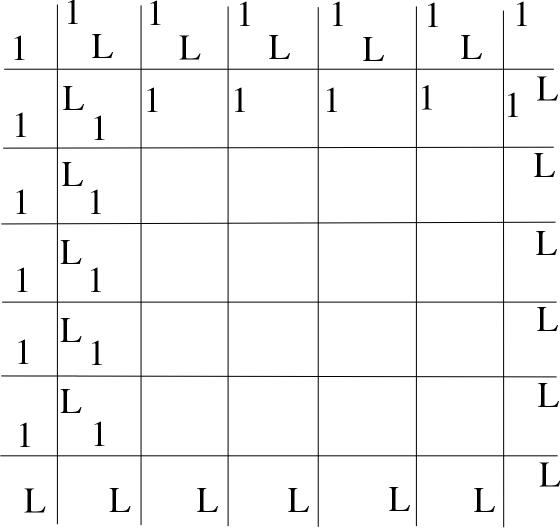}
\caption{\footnotesize{Graphical derivation of eq. \ref{PS.3}.}}
\label{PS.d}
\end{center}
\end{figure}\\
\\
{\bf Property 4.} The initial condition is given by:
\small\begin{equation*}
Z_{1}(u_1,v_1)=  X^{1,N}_{1,N}(u_1-v_1)
\end{equation*}\normalsize
\textbf{Proof.} Simply setting the DWBC's for a single vertex produces the desired result.\\
\\
\noindent \textbf{An inductive result regarding the four properties of the DWPF.}
\begin{lemma}
The above four properties uniquely determine the DWPF of the $sl(r+1|s+1)$ PS vertex models.
\end{lemma}
\noindent \textbf{Proof.}  We begin by assuming that there exist two expressions which satisfy the above four properties. We refer to $Z_{N} \left(\vec{u},\vec{v}\right)$ as the actual partition function, and $\field{Z}^{PS}_{N} \left(\vec{u},\vec{v}\right)$, which is an altogether different expression. By condition $4$ we obtain our base case,
\small\begin{equation*}
Z_{1} \left(u_1,v_1\right) = \field{Z}^{PS}_{1} \left(u_1,v_1\right) = X^{1,N}_{1,N}(u_1-v_1)
\end{equation*}\normalsize
Let us now assume that the two expressions are equal up to some integer $N$ and use induction for the $N+1$ case. From condition $1$ both $Z_{N+1}$ and $\field{Z}^{PS}_{N+1}$ are order $N+1$ polynomials in $e^{\eta u_1}$ and $e^{-\eta v_1}$, and both have zeros of the form $e^{\eta u_1}= 0$ and $e^{-\eta v_1}=0$. From condition $2$ we know both expressions additionally share the $N$ zeroes of $u_1$,
\small\begin{equation*}
X^{1,1}_{1,1}(u_1 - u_j) = 0 \textrm{  ,  } j=2, \dots, N+1
\end{equation*}\normalsize
and the $N$ zeroes of $v_1$,
\small\begin{equation*}
X^{L,L}_{L,L}(v_1 -v_j) = 0 \textrm{  ,  } j=2, \dots, N+1
\end{equation*}\normalsize
making a grand total of $N+1$ shared zeroes. Thus by conditions $1$ and $2$ we know that,
\small\begin{equation}
Z_{N+1}= \mathcal{C} \field{Z}^{PS}_{N+1}
\label{PS.6}
\end{equation}\normalsize
where $\mathcal{C}$ is a multiplicative constant independent of rapidities $u_1$ and $v_1$. To find $\mathcal{C}$ we apply eq. \ref{PS.3} (or \ref{PS.tt}) of condition $3$ to eq. \ref{PS.6} to obtain,
\small\begin{equation*}\begin{split}
&X^{1,L}_{1,L}(1) \left( \prod^{N+1}_{k=2} X^{1,L}_{L,1}(u_1 - v_k)X^{L,1}_{1,L}(u_k - v_1) \right) Z_{N}\left(\vec{u},\vec{v},\hat{u}_1,\hat{v}_1\right)\\
=& \mathcal{C} X^{1,L}_{1,L}(1) \left( \prod^{N+1}_{k=2} X^{1,L}_{L,1}(u_1 - v_k)X^{L,1}_{1,L}(u_k - v_1) \right) \field{Z}^{PS}_{N}\left(\vec{u},\vec{v},\hat{u}_1,\hat{v}_1\right)
\end{split}\end{equation*}\normalsize 
and using the inductive assumption that $Z_{N}= \field{Z}^{PS}_{N}$, we obtain $\mathcal{C} = 1$. $\square$ 
\subsection{Product form for the DWPF}
\begin{proposition}
\small\begin{equation}\begin{split}
Z_{N} \left( \vec{u},\vec{v}\right) = \left( \prod^{N}_{k=1} X^{1,L}_{1,L} (u_k - v_k) \right) \left( \prod_{1 \le i < j \le N} X^{1,1}_{1,1}(u_i - u_j) X^{L,L}_{L,L}(v_i - v_j) \right)  \\
= \left( \prod^{N}_{k=1} e^{ \eta (u_k - v_k)} \right) \left( \prod_{1 \le i < j \le N} \frac{\sinh  \eta(1- u_i + u_j)  }{\sinh \eta } \frac{\sinh \eta(1+ v_i - v_j)  }{\sinh \eta}\right)
\label{PS.5}
\end{split}\end{equation}\normalsize
\end{proposition}
\noindent \textbf{Proof.} Verifying the above proposition obviously relies on showing that eq. \ref{PS.5} obeys the four conditions.\\
\\
\textbf{Verification of property 1.} All dependence on $u_1$ and $v_1$ in eq. \ref{PS.5} can be written immediately as,
\small\begin{equation}
 e^{ \eta u_1} \prod^N_{ j= 2} \sinh  \eta(1- u_1 + u_j) \textrm{  ,  } e^{ -\eta v_1} \prod^N_{ j= 2} \sinh  \eta(1+ v_1 - v_j) \label{PS.7}
\end{equation}\normalsize
which are degree $N$ polynomials in $ e^{ \eta u_1}$ and $ e^{ -\eta v_1}$ and contain zeros of the form $e^{ \eta u_1} = e^{ -\eta v_1}=0$ respectively.\\
\\
\textbf{Verification of property 2.} From eq. \ref{PS.7} it is immediate that the remaining $N-1$ zeros are of the required form.\\
\\
\textbf{Verification of property 3.}  Substituting the values $u_1 = v_1 +1$, eq. \ref{PS.5} becomes,
\small\begin{equation*}\begin{split}
Z_{N} \left( \vec{u},\vec{v}\right)|_{u_1 = v_1 +1} = e^{\eta} (-1)^{N-1}\left( \prod^N_{ j= 2} \frac{ \sinh  \eta(u_j-v_1)}{ \sinh  \eta} \frac{ \sinh  \eta(u_1-v_j)}{ \sinh  \eta} \right) \\
\times  \left( \prod^{N}_{k=2} X^{1,L}_{1,L} (u_k - v_k) \right) \left( \prod_{2 \le i < j \le N} X^{1,1}_{1,1}(u_i - u_j) X^{L,L}_{L,L}(v_i - v_j) \right)
\end{split}\end{equation*}\normalsize
which is exactly eq. \ref{PS.3}. Verifying eq. \ref{PS.tt} is an equivalent process.\\
\\
\textbf{Verification of property 4.} Simply substituting $N=1$ in eq. \ref{PS.5} obtains the desired result.\\
\\
Thus we have verified the proposition for the product form of the partition function. $\square$
\section{The elliptic PS height model}
\noindent In \cite{PSheight1}, Deguchi and Martin introduced the elliptic height equivalent of the trigonometric vertex model considered in the previous section. In this section of the thesis we introduce the height model and derive the product form of the DWPF in an equivalent process considered in the previous section. Additionally, due to the DWPF being sufficiently simple, we can use the methods of the previous chapter to derive non trivial elliptic identities of a general number of terms.
\subsection{Elliptic functions revisited}
\noindent As promised in the previous chapter, we shall now present some additional necessary results regarding quasi-periodic functions. \\
\\
\textbf{Useful theorem regarding quasi-periodic functions (continued).} 
The following result is similar to theorems 15(b) and 15(c) in section 15.3 of \cite{Baxterbook}. The aforementioned theorems consider results regarding meromorphic (anti)periodic functions, whereas the result below concerns entire quasi-periodic functions.
\begin{theorem}
\label{quas}
Consider $f(u)$ which is an entire quasi-periodic function satisfying the relations,
\small\begin{equation}
f\left(u+\frac{2K_1}{\lambda}\right) = (-1)^N f(u)\label{PS.b.1}
\end{equation}\normalsize 
\small\begin{equation}
f\left(u+\frac{2K_2}{\lambda}\right) = (-1)^{N}\left( \frac{1}{q} \right)^N \exp \left\{- \frac{i \pi \lambda \left(N u - \sum^N_{j=1}\kappa_j \right) }{K_1} \right\} f(u) \label{PS.b.2}
\end{equation}\normalsize
Given the above relations the following forms for $f(u)$ apply,
\small\begin{equation}\begin{split}
f(u)& = \mathcal{C} \prod^N_{j=1} H\left( \lambda(u- \kappa_j) \right) \\
&= \mathcal{C} \left( \prod^{N-1}_{j=1} H\left( \lambda(u- \kappa_j )\right)  \right) H\left( \lambda(u- \eta + \sum^{N-1}_{j=1}\kappa_j )\right)
\label{PS.b.3}
\end{split}\end{equation}\normalsize
where $\{ \kappa_1,\dots,\kappa_N,\mathcal{C}, \lambda\} \in \field{C}$ and $\eta = \sum^{N}_{j=1} \kappa_j$.
\end{theorem}
\textbf{Proof.} 
\begin{figure}[h!]
\begin{center}
\includegraphics[angle=0,scale=0.21]{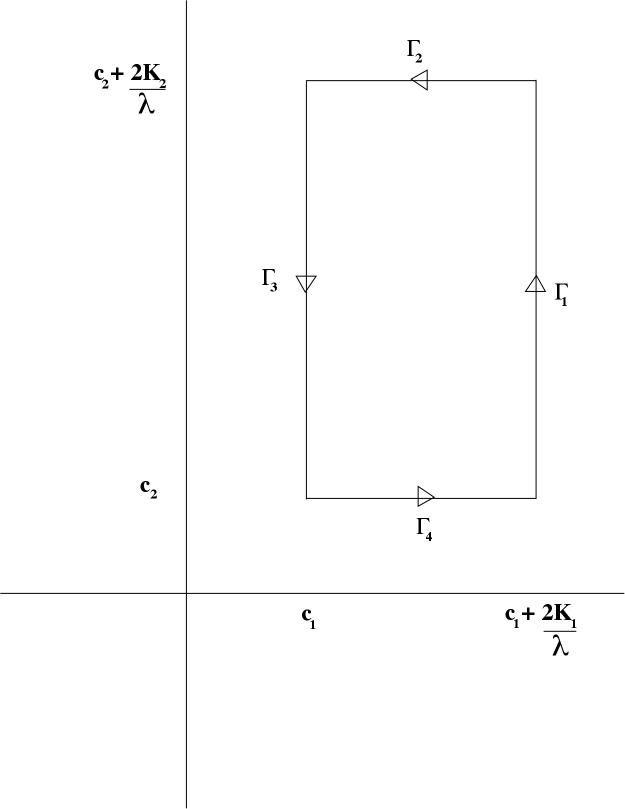}
\caption{\footnotesize{A typical period rectangle in the complex plane of width $\frac{2K_1}{\lambda}$ and height $\frac{2K_2}{\lambda}$}}
\label{5.b}
\end{center}
\end{figure}
We can choose a period rectangle in the $\field{C}$ plane such that the isolated zeroes of $f(u)$ are not on the boundary. We then consider the following integral,
\small\begin{equation*}
\int_{\Gamma_1+\Gamma_2+\Gamma_3+\Gamma_4} \frac{ f'(u)}{f(u)} du = \int_{\Gamma_1+\Gamma_3} \frac{ f'(u)}{f(u)} du +\int_{\Gamma_2+\Gamma_4} \frac{ f'(u)}{f(u)} du 
\end{equation*}\normalsize
Performing the integral of $\Gamma_3$ and $\Gamma_4$ we obtain,
\small\begin{equation*}\begin{split}
 \int_{\Gamma_3} \frac{ f'(u)}{f(u)} du &= \int_{u= c_1+i \{c_2 + 2K_2 / \lambda \}}^{u=c_1 +ic_2 } \frac{d}{du} \log f(u) du \\
&= \log \left( \frac{f(c_1 +ic_2)}{f(c_1+i \{c_2 + 2K_2 /  \lambda \})} \right) \\
 \int_{\Gamma_4} \frac{ f'(u)}{f(u)} du &= \int^{u= c_1+2K_1 /  \lambda+i c_2}_{u=c_1+ic_2} \frac{d}{du}\log f(u) du \\
&=\log \left( \frac{f( c_1+2K_1 / \lambda +i c_2)}{f(c_1+ic_2)} \right)
\end{split}\end{equation*}\normalsize
For the remaining two integrals we simply apply the quasi-periodic conditions of $f(u)$. For the integral along $\Gamma_1$ we receive,
\small\begin{equation*}\begin{split}
 \int_{\Gamma_1} \frac{ f'(u)}{f(u)} du &= - \int_{\Gamma_3} \frac{ f'(u+2K_1/ \lambda)}{f(u+2K_1 /\lambda)} du \\
&=- \log \left( \frac{f(c_1+2K_1 / \lambda+ic_2)}{f(c_1+2K_1/ \lambda+i \{c_2 + 2K_2/  \lambda\})} \right)\\
&= - \log \left( \frac{(-1)^N f(c_1 +ic_2)}{(-1)^N f(c_1+i \{c_2 + 2K_2/ \lambda \})} \right)\\
&=  - \int_{\Gamma_3} \frac{ f'(u)}{f(u)} du
\end{split}\end{equation*}\normalsize
Thus we obtain the result $\int_{\Gamma_1+\Gamma_3} \frac{ f'(u)}{f(u)} du =0$. The integral along $\Gamma_2$ is slightly different however,
\small\begin{equation*}\begin{split}
 \int_{\Gamma_2} \frac{ f'(u)}{f(u)} du &=- \int_{\Gamma_4} \frac{ f'(u+i2K_2/ \lambda )}{f(u+i 2K_2/ \lambda)} du  \\
&=-\log \left( \frac{f( c_1+2K_1 /\lambda +i \{c_2+2K_2/ \lambda\})}{f(c_1+i\{c_2+2K_2/ \lambda\})} \right)
\end{split}\end{equation*}\normalsize
\small\begin{equation*}\begin{split}
&=- \log \left( \frac{(-1)^N q^{-N}\exp\{\alpha\} \exp\left\{ - i2\pi N \right\} f( c_1+2K_1/ \lambda +ic_2)}{(-1)^N q^{-N}\exp\{\alpha\} f(c_1+ic_2)} \right)\\
&= i2\pi N -  \int_{\Gamma_4} \frac{ f'(u )}{f(u)} du
\end{split}\end{equation*}\normalsize
where we have labeled $\alpha = -\frac{i\pi \lambda}{K_1} \left(N c_1 + i N c_2 -\eta \right)$. Thus we obtain\\$\int_{\Gamma_1+\Gamma_2+\Gamma_3+\Gamma_4} \frac{ f'(u)}{f(u)} du =i2\pi N$, and since $f(u)$ is an entire function, this tells us that $f(u)$ contains exactly $N$ zeros in the period rectangle, which we shall label as $\{\kappa_1,\kappa_2, \dots, \kappa_N\}$.\\
\\
Let us now consider the function,
\small\begin{equation*}
\phi(u) = \mathcal{C} \prod^N_{j=1} H\left( \lambda( u- \kappa_j )\right) 
\end{equation*}\normalsize
where $\phi(u)$ obeys the same quasi-periodic conditions as $f(u)$. Additionally, we consider the expression,
\small\begin{equation}
 \frac{d}{du} \log\left( \frac{f(u)}{\phi(u)}\right) = \frac{f'(u)}{f(u)} - \frac{\phi'(u)}{\phi(u)} 
\label{PS.b.4}
\end{equation}\normalsize
By construction eq. \ref{PS.b.4} is doubly (anti)-periodic and analytic inside the period rectangle, hence by Liouville's theorem it is a constant, which we shall label as $\kappa$.\\
\\
Integrating eq. \ref{PS.b.4} with respect to $u$ we obtain,
\small\begin{equation}
f(u) = \mathcal{C} e^{\kappa u} \prod^N_{j=1} H\left( \lambda( u- \kappa_j) \right) 
\label{PS.b.5}
\end{equation}\normalsize 
Finally, by considering the quasi-periodic conditions we fix the constant $\kappa$ to equal zero. To obtain the final expression in eq. \ref{PS.b.3} we simply let $\eta = \sum^N_{j=1} \kappa_j$ without loss of generality. $\square$\\
\\
The above result shall be used to obtain the DWPF of the elliptic $gl(P|M)$ PS height model. 
\subsection{Definition of the model} 
\begin{figure}[h!]
\begin{center}
\includegraphics[angle=0,scale=0.15]{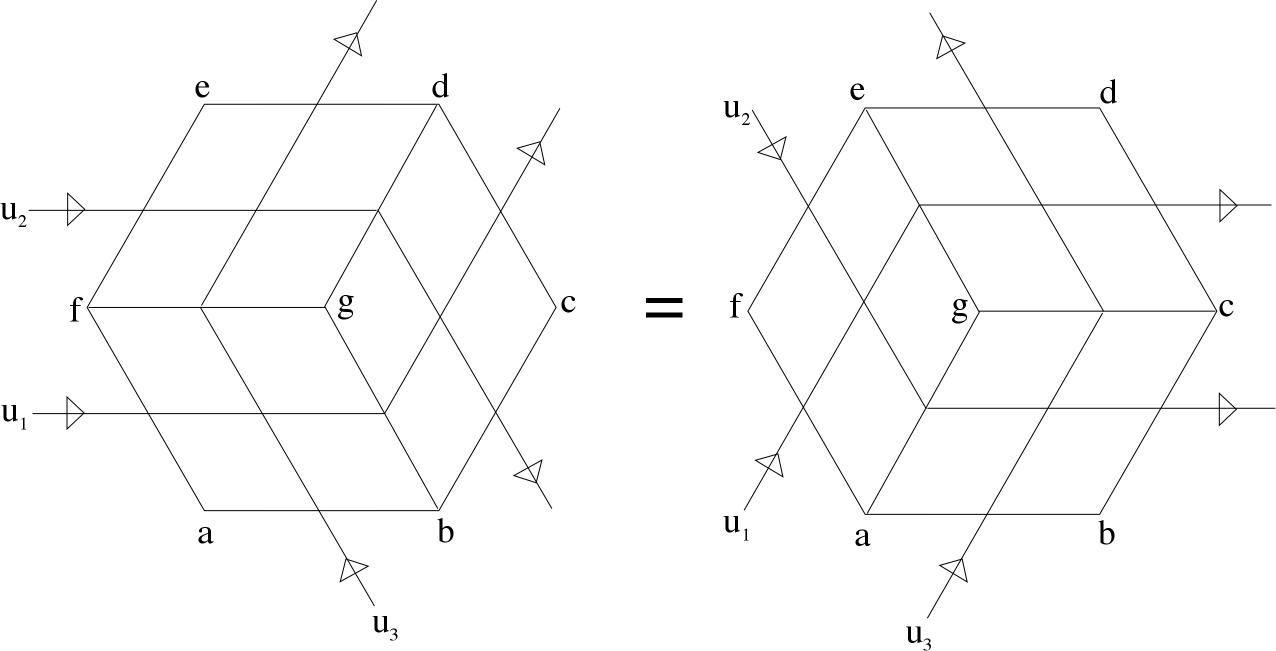}
\caption{\footnotesize{Graphical representation of the Yang-Baxter identities.}}
\label{5.fff}
\end{center}
\end{figure}

\noindent The following definitions are given in section 2.6 of \cite{PSheight1}.\\
\\
\noindent \textbf{Comment.} Due to the $gl(P|M)$ PS IRF model being the solid on solid equivalent to the $sl(r+1|s+1)$ PS vertex model, this section shall read very similarly to the first section of this chapter. To begin the similarities we now label $P+M = L$ in the remainder of this section.\\
\\
\textbf{State vectors from $\field{Z}^{L}$ and additional definitions.}
We introduce the notation,
\small\begin{equation}
 \hat{e}_k = \{ \underbrace{0, \dots,0}_{k-1}, 1,\underbrace{0 ,\dots,0}_{L-k}\} \textrm{  ,  } k \in \{ 1, \dots, L\}
\label{PS.b.5'}
\end{equation}\normalsize
as a unit vector in the field $\field{Z}^L = \underbrace{\field{Z} \times \dots \times \field{Z}}_{L}$. Additionally we introduce the, (as yet unmotivated), $L \times L$ matrix $\omega$ and length $L$ vector $\vec{\epsilon}$, where $\omega$ is an arbitrary constant antisymmetric matrix,
\small\begin{equation}\begin{array}{lcl}
(\omega_{ij})^L_{i,j=1} & =& - (\omega_{ji})^L_{i,j=1}\\
\omega_{ij} \in \field{C}& \textrm{  ,  } & i,j \in \{ 1,\dots, L\} 
\end{array}\label{PS.b.6}\end{equation}\normalsize
and $\epsilon_k$ is given simply as,
\small\begin{equation}
\epsilon_{k} = \left\{ \begin{array}{ccc}
+1 & \textrm{for} & 1 \le k \le P \\
-1 & \textrm{for} & P+1 \le k \le L \end{array}  \right.
\label{PS.b.7}
\end{equation}\normalsize
With these definitions we are now properly equipped to define our solid on solid (SOS) playing field and the parameterization of the face weights.\\
\\
\textbf{The $N \times N$ SOS playground.} As with the BSOS model, the $N \times N$ faces for the $gl(P|M)$ PS IRF model have both horizontal rapidity flows, $u_i \in \field{C}$, $1 \le i \le N$ and vertical rapidity flows, $v_j \in \field{C}$, $1 \le j\le N$.\\
\\
\begin{figure}[h!]
\begin{center}
\includegraphics[angle=0,scale=0.21]{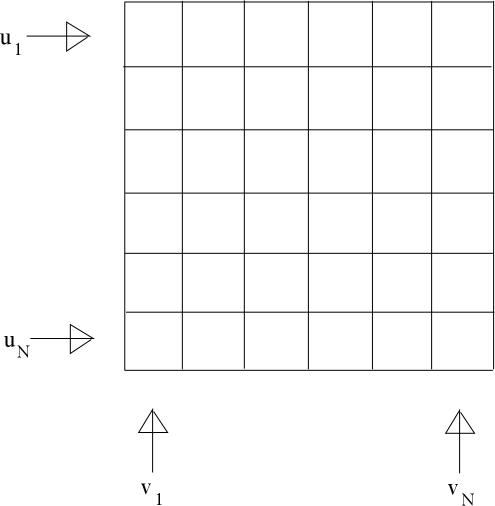}
\caption{\footnotesize{The $N \times N$ SOS lattice with vertical and horizontal rapidity flows.}}
\label{PS.b.a}
\end{center}
\end{figure}\\
\\
However, unlike the BSOS model, whose state variables were elements of $\field{Z}$, the state variables of $gl(P|M)$ PS IRF model include additional generality in that they are vectors which are strictly elements of $\field{Z}^L$. 
\begin{figure}[h!]
\begin{center}
\includegraphics[angle=0,scale=0.21]{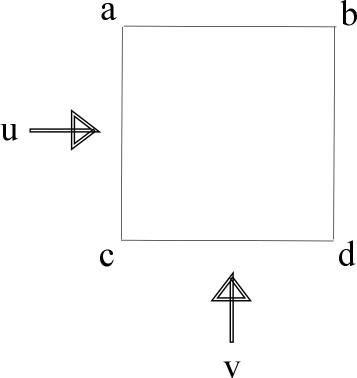}
\caption{\footnotesize{A typical face of the model with state variable vectors $\{\vec{a},\vec{b},\vec{c},\vec{d}\}\in \field{Z}^L$.}}
\label{PS.b.b}
\end{center}
\end{figure}\\
\\
\textbf{Labeling of the faces.} Each of the $N^2$ faces are labeled by the difference of the corresponding horizontal and vertical rapidities, $u_i-v_j$, and obviously by the set of four state vectors, $\{\vec{a},\vec{b},\vec{c},\vec{d}\}\in \field{Z}^L$. Hence we label the weight in fig. \ref{PS.b.b} as $W \left( \left. \begin{array}{cc} \vec{a} & \vec{b} \\ \vec{c} & \vec{d} \end{array} \right|  u-v  \right)$.\\
\\
\textbf{Allowable face configurations and weights.} We remark that the Boltzmann weights of the model are set to zero unless the differences $\vec{b}-\vec{a},\vec{d}-\vec{a}, \vec{c}-\vec{d}$ and $\vec{c}-\vec{b}$ are, up to a sign, equal to some unit vector $\hat{e}_k$. In the following we use the notation,
\small\begin{equation*}
[u] = H(\lambda u)
\end{equation*}\normalsize
where $\lambda \in \field{C}$ plays the role of the global crossing parameter. Thus, given an initial state vector $\vec{a} \in \field{Z}^L$, the non zero weights are parameterized by,
\small\begin{equation*}\begin{array}{lcll}
W \left( \left. \begin{array}{cc} \vec{a} & {\vec{a}+ \hat{e}_{j}} \\ {\vec{a}+ \hat{e}_{j}} & {{\vec{a}+ 2 \hat{e}_{j}}} \end{array} \right|  u-v  \right) &=&\displaystyle \frac{[ 1+ \epsilon_{j} (u-v) ]}{[1]} \\
W \left( \left. \begin{array}{cc} \vec{a}& {\vec{a}+ \hat{e}_{k}} \\ {\vec{a}+ \hat{e}_{j}} & \vec{a}+ \hat{e}_{j}+ \hat{e}_{k} \end{array} \right|  u -v \right)& =&\displaystyle \frac{[u-v][a_{jk}-1]}{[1][a_{jk}]}, & j \ne k\\
W \left( \left. \begin{array}{cc} {\vec{a}} & {\vec{a}+ \hat{e}_{j}} \\ {\vec{a}+ \hat{e}_{j}} & {{\vec{a}+ \hat{e}_{j}+ \hat{e}_{k}}} \end{array} \right|  u-v  \right) &=&\displaystyle \frac{[a_{jk}- (u-v) ]}{[a_{jk}]},& j \ne k
\end{array}\end{equation*}\normalsize
where $a_{jk}= \epsilon_{j} a_{j}-\epsilon_{k} a_{k}+\omega_{jk}$, ($a_j$ being the $j$th component of the state vector $\vec{a}$).  \\
\\
\textbf{Yang-Baxter equation.} Given state vectors $\{\vec{a},\vec{b},\vec{c},\vec{d},\vec{e},\vec{f}\} \in \field{Z}^L$ and rapidities $\{u_1,u_2,u_3\} \in \field{C}$, the above weights of the $gl(P|M)$ PS IRF model obey the following Yang-Baxter identities \cite{SOS1,SOS2},
\small\begin{equation*}\begin{split}
&\displaystyle \sum_{\vec{g} \in \field{Z}^L} W \left( \left. \begin{array}{cc} \vec{f} & \vec{g} \\ \vec{a} & \vec{b} \end{array} \right|  u_1-u_3  \right) W \left( \left. \begin{array}{cc} \vec{e} & \vec{d} \\ \vec{f} & \vec{g} \end{array} \right|  u_2-u_3 \right) W \left( \left. \begin{array}{cc} \vec{d} & \vec{c} \\ \vec{g} & \vec{b} \end{array} \right|  u_2-u_1  \right)\\
=&\displaystyle \sum_{\vec{g} \in \field{Z}^L} W \left( \left. \begin{array}{cc} \vec{e} & \vec{d} \\ \vec{g} & \vec{c} \end{array} \right|  u_1-u_3  \right) W \left( \left. \begin{array}{cc} \vec{g} & \vec{c} \\ \vec{a} & \vec{b} \end{array} \right| u_2-u_3  \right) W \left( \left. \begin{array}{cc} \vec{e} & \vec{g} \\ \vec{f} & \vec{a} \end{array} \right|  u_2-u_1  \right)
\end{split}\end{equation*}\normalsize
whose graphical representation is given in fig. \ref{5.fff}.
\\
\\
\textbf{DWBC's.} The DWBC's for the $gl(P|M)$ IRF model follow naturally from the DWBC's of the previous $sl(r+1|s+1)$ vertex model. Firstly, we place an arbitrary element of $\field{Z}^L$, labeled $\vec{a}_{(0)}$, as the top left height of the $N \times N$ faces. Each subsequent outermost height to the right and south of the top left corner increase by one unit of $\hat{e}_1$. Then south of the top right corner, and right of the bottom left corner, each subsequent outermost height increase by one unit of $\hat{e}_{L}$ until the bottom right height is $\vec{a}_{(0)} +N \hat{e}_1 + N \hat{e}_{L}$.  
\begin{figure}[h!]
\begin{center}
\includegraphics[angle=0,scale=0.24]{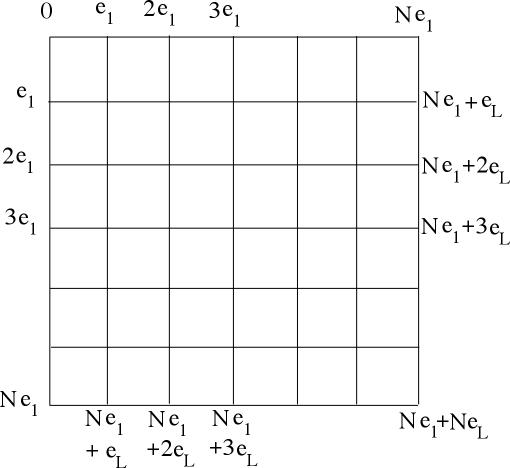}
\caption{\footnotesize{DWBC's for the model, notice that the presence of $\vec{a}_{(0)}$ from all state vectors has been suppressed.}}
\label{5.ffg}
\end{center}
\end{figure}\\
\\
Due to the DWBC's, there are only six types of faces that we need to consider. If we label $\hat{e}_1 = \hat{e}_+$ and $\hat{e}_{L} = \hat{e}_-$, the necessary weights are given by,
\small\begin{equation*}\begin{array}{lclcl}
W \left( \left. \begin{array}{cc} \vec{a} & {\vec{a}+ \hat{e}_{\pm}} \\ {\vec{a}+ \hat{e}_{\pm}} & {{\vec{a}+ 2 \hat{e}_{\pm}}} \end{array} \right|  u-v  \right) &=& W_{A,\pm}(u-v) &=& \frac{[ 1\pm (u-v) ]}{[1]}\\
W \left( \left. \begin{array}{cc} {\vec{a}} & {\vec{a}+ \hat{e}_{\pm}} \\ {\vec{a}+ \hat{e}_{\mp}} & \vec{a}+ \hat{e}_{+}+ \hat{e}_{-} \end{array} \right|  u-v  \right) &=&W^{(\vec{a})}_{B,\pm}(u-v)&=& \frac{[u-v][ a_{1,L} \pm 1]}{[1][ a_{1,L}]} \\
W \left( \left. \begin{array}{cc} {\vec{a}} & {\vec{a}+ \hat{e}_{\pm}} \\ {\vec{a}+ \hat{e}_{\pm}} & {{\vec{a}+ \hat{e}_{+}+ \hat{e}_{-}}} \end{array} \right|  u -v \right)&=&W^{(\vec{a})}_{C,\pm}(u-v) &=& \frac{[ a_{1,L} \mp (u-v) ]}{[ a_{1,L}]}
\end{array}\end{equation*}\normalsize
It is possible to consider the rapidity and height section of the $B$ weights separately. Labelling $W^{(\vec{a})}_{B,\pm}(u-v) = W^{(\vec{a})}_{B,\pm}W_{B}(u-v)$ We shall use the convention,
\small\begin{equation}\begin{split}
W^{(\vec{a})}_{B,\pm} &= \frac{[ a_{1,L} \pm 1]}{[ a_{1,L}]}\\
W_{B}(u-v)&= \frac{[u-v]}{[1]} \label{PSdecouple}
\end{split}\end{equation}\normalsize
\textbf{Non-invariance of the line permuting vertices.} As with the PS vertex models, the $gl(P|M)$ height models with DWBC's display a non-invariance of the $W_{A,\pm}(u-v)$ vertex when $\hat{e}_i$ changes from $1$ to $L$, and vice versa,
\small\begin{equation*}\begin{split}
W \left( \left. \begin{array}{cc} \vec{a} & {\vec{a}+ \hat{e}_{1}} \\ {\vec{a}+ \hat{e}_{1}} & {{\vec{a}+ 2 \hat{e}_{1}}} \end{array} \right|  u-v  \right) =&\frac{[ 1+ (u-v) ]}{[1]} \\
W \left( \left. \begin{array}{cc} \vec{a} & {\vec{a}+ \hat{e}_{L}} \\ {\vec{a}+ \hat{e}_{L}} & {{\vec{a}+ 2 \hat{e}_{L}}} \end{array} \right|  u-v  \right) =& \frac{[ 1- (u-v) ]}{[1]}
\end{split}\end{equation*}\normalsize
We shall exploit this property in much the same way that we did for the vertex model.
\subsection{Equivalent elliptic Korepin properties and the DWPF} 
\begin{figure}[h!]
\begin{center}
\includegraphics[angle=0,scale=0.25]{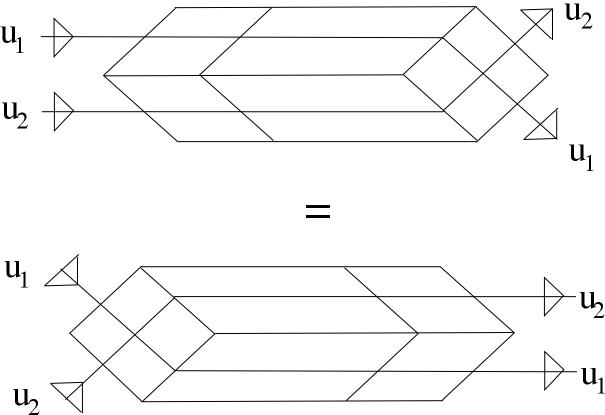}
\caption{\footnotesize{Graphical derivation of eq. \ref{PS.b.11}, notice that state vectors have been suppressed.}}
\label{PSelip2}
\end{center}
\end{figure}

\begin{figure}[h!]
\begin{center}
\includegraphics[angle=0,scale=0.24]{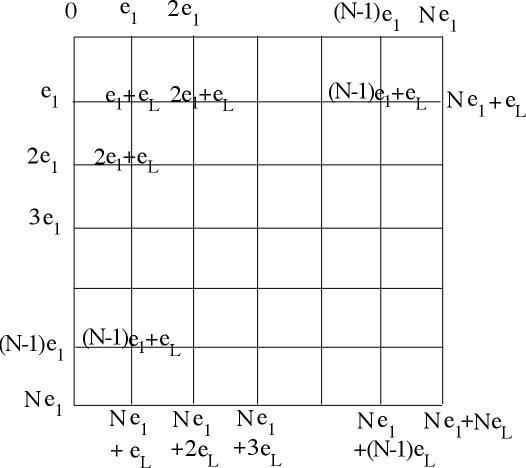}
\caption{\footnotesize{Graphical derivation of eq. \ref{PS.b.13}, notice $\vec{a}_{(0)}$ has been suppressed from all stater vectors.}}
\label{PS.9999}
\end{center}
\end{figure}

The DWPF of the $gl(P|M)$ PS height model, $Z_N\left(\vec{u},\vec{v}\right)$, is given as the sum over all allowable weighted configurations of the $N \times N$ faces under DWBC's,
\small\begin{equation*}
Z_N\left(\vec{u},\vec{v}\right) = \sum_{\textrm{allowable}\atop{\textrm{configurations}}} \left( \prod_{\textrm{faces}} W \left( \left. \begin{array}{cc} \vec{a} & \vec{b} \\ \vec{c} & \vec{d} \end{array} \right|  u_i-v_j  \right) \right)
\end{equation*}\normalsize
\textbf{Korepin-like properties of the DWPF.} We now list four properties which the DWPF of the model satisfies. We shall then show that the DWPF is necessarily determined uniquely by satisfying these properties.\\
\\ 
{\bf Property 1.}  The partition function satisfies the following quasi-periodic conditions in $u_1$,
\small\begin{equation}
Z_N\left(\vec{u} ,\vec{v}\right)|_{u_1 \rightarrow u_1+2\frac{K_1}{\lambda}} =(-1)^N Z_N\left(\vec{u},\vec{v}\right) \label{PS.b.8}
\end{equation}\normalsize
\small\begin{equation}
Z_N\left(\vec{u} ,\vec{v}\right)|_{u_1 \rightarrow u_1+2\frac{K_2}{\lambda}}  = \left(- \frac{1}{q} \right)^N \exp\left\{ - \frac{i \pi \lambda }{K_1}\left(Nu_1 - \sum^N_{j=1}v_j - \left( \vec{a}_{(0)} \right)_{1,L} \right)  \right\}
Z_N\left(\vec{u},\vec{v}\right) \label{PS.b.9}
\end{equation}\normalsize
and in $v_1$,
\small\begin{equation}
Z_N\left(\vec{u} ,\vec{v}\right)|_{v_1 \rightarrow v_1+2\frac{K_1}{\lambda}} =(-1)^N Z_N\left(\vec{u},\vec{v}\right) \label{PS.bq}
\end{equation}\normalsize
\small\begin{equation}
Z_N\left(\vec{u} ,\vec{v}\right)|_{v_1 \rightarrow v_1+2\frac{K_2}{\lambda}}  = \left(- \frac{1}{q} \right)^N \exp\left\{ - \frac{i \pi \lambda }{K_1}\left(Nv_1 - \sum^N_{j=1}u_j + \left( \vec{a}_{(0)} \right)_{1,L} \right)  \right\} Z_N\left(\vec{u},\vec{v}\right) \label{PS.bs}
\end{equation}\normalsize
\textbf{Proof.} An elementary analysis of the top row reveals that, for $1 \le l \le N$, all valid configurations are of the form,
\small\begin{equation}\begin{split}
& \left\{ \displaystyle \prod^{l-1}_{j=1} W \left( \left. \begin{array}{cc} \vec{a}_{(0)}+(j-1) \hat{e}_{+} & {\vec{a}_{(0)}+ j \hat{e}_{+}} \\ {\vec{a}_{(0)}+ j \hat{e}_{+}} & {{\vec{a}_{(0)}+ (j+1) \hat{e}_{+}}} \end{array} \right|  u_1-v_j  \right)  \right\} \\
\times & W \left( \left. \begin{array}{cc} \vec{a}_{(0)}+(l-1) \hat{e}_{+} & {\vec{a}_{(0)}+ l \hat{e}_{+}} \\ {\vec{a}_{(0)}+ l \hat{e}_{+}} & {{\vec{a}_{(0)}+ l \hat{e}_{+}+\hat{e}_{-}}} \end{array} \right|  u_1-v_l  \right) \\
\times & \left\{ \displaystyle \prod^{N}_{j=l+1}  W \left( \left. \begin{array}{cc} \vec{a}_{(0)}+(j-1) \hat{e}_{+} & {\vec{a}_{(0)}+ j \hat{e}_{+}} \\ {\vec{a}_{(0)}+ (j-1) \hat{e}_{+}+\hat{e}_{-}} & {{\vec{a}_{(0)}+ j \hat{e}_{+}+\hat{e}_{-}}} \end{array} \right|  u_1-v_j  \right)  \right\}  \\
=& \displaystyle \left( \prod^{l-1}_{j=1} \frac{[1+u_1-v_j]}{[1]} \right) \frac{[\left(\vec{a}_{(0)}+(l-1) \hat{e}_{+}\right)_{1,L}-u_1+ v_l]}{[\left(\vec{a}_{(0)}+(l-1) \hat{e}_{+}\right)_{1,L}]} \\
\times & \displaystyle \left( \prod^{N}_{j=l+1}\frac{[u_1-v_j][\left(\vec{a}_{(0)}+(j-1) \hat{e}_{+}\right)_{1,L}+1]}{[1][\left(\vec{a}_{(0)}+(j-1) \hat{e}_{+}\right)_{1,L}]} \right)  \label{PS.b.10}
\end{split}\end{equation}\normalsize
Verifying the first quasi-periodic condition is elementary. To verify the second condition we consider the coefficients that appear due to $u_1 \rightarrow u_1 + \frac{2 K_2}{\lambda}$ in the $j \in \{1,\dots,l-1\}$, $j=l$ and $j \in \{l+1,\dots,N\}$ expressions in eq. \ref{PS.b.10},
\small\begin{equation*}\begin{array}{lcl}
1 \le j \le l-1 &,& \displaystyle \left(- \frac{1}{q} \right)^{l-1} \exp\left\{ - \frac{i \pi \lambda}{K_1} \left( (l-1)(1+u_1) - \sum^{l-1}_{j=1}v_j \right) \right\}\\
j=l &,& \displaystyle \left(- \frac{1}{q} \right) \exp\left\{ - \frac{i \pi \lambda}{K_1} \left( u_1- \left(\vec{a}_{(0)}+(l-1) \hat{e}_{+}\right)_{1,L} - v_l \right) \right\}\\
l+1 \le j \le N&,& \displaystyle \left(- \frac{1}{q} \right)^{N-l} \exp\left\{ - \frac{i \pi \lambda}{K_1} \left( (N-l)u_1 - \sum^{N}_{j=l+1}v_j \right) \right\}
\end{array}\end{equation*}\normalsize
where $ \left(\vec{a}_{(0)}+(l-1) \hat{e}_{+}\right)_{1,L}  =  \left(\vec{a}_{(0)}\right)_{1,L} +l-1$. Multiplying these three terms together we obtain eq. \ref{PS.b.9}.\\
\\
Verifying eqs. \ref{PS.bq} and \ref{PS.bs} consists of much of the same process.\\
\\
{\bf Property 2.} From the Yang-Baxter equation it is possible to show that,
\small\begin{equation}
Z_{N} \left( \vec{u}, \vec{v} \right) = \prod^N_{j=2} \frac{[1+u_1-u_j]}{[1-(u_1-u_j)]} Z_{N} \left( u_2, \dots, u_N, u_1, \vec{v}\right)
\label{PS.b.11}
\end{equation}\normalsize
\noindent which gives exactly $N-1$ zeroes per period rectangle in the form $u_1 = u_j -1 + \frac{2m K_1}{\lambda}+ \frac{2 i n K_2}{\lambda} $, $j=2, \dots, N$, $\{m,n\} \in \field{Z}$. Performing the equivalent technique on the $\vec{v}$'s we obtain,
\begin{equation}
Z_{N} \left( \vec{u},\vec{v}\right) = \prod^{N}_{j=2} \frac{[1-(v_1-v_j)]}{[1+v_1-v_j]} Z_{N} \left(\vec{u}, v_2, \dots, v_N, v_1\right)
\label{PS.b.12}
\end{equation}\normalsize
which gives exactly $N-1$ zeroes of the form $v_1 = v_j+1+\frac{2m K_1}{\lambda}+ \frac{2 i n K_2}{\lambda} $, $j=2, \dots, N$, $\{m,n\} \in \field{Z}$.\\
\\
\textbf{Proof.} We consider placing a \scriptsize$W \left( \left. \begin{array}{cc} \vec{a}_{(0)}+N \hat{e}_{+} & \vec{a}_{(0)}+ N \hat{e}_{+}+  \hat{e}_{-} \\ \vec{a}_{(0)}+ N \hat{e}_{+}+\hat{e}_{-} & \vec{a}_{(0)}+ N \hat{e}_{+}+2\hat{e}_{-} \end{array} \right|  u_1-u_2  \right)$\normalsize face\footnote{Remember that in earlier discussions we label this face, and its conjugate as line permuting face.} on the right hand side of $Z_{N} \left(\vec{u},\vec{v}\right)$. This process is displayed in the first diagram of fig. \ref{PSelip2}. We notice that there are no other valid internal state vectors for this face, thus we can apply the Yang-Baxter equation and shift it through to the left hand side of the bulk. Once at the left side of the lattice, the face is fixed to a \scriptsize$W \left( \left. \begin{array}{cc} \vec{a}_{(0)} & \vec{a}_{(0)}+ \hat{e}_{+} \\ \vec{a}_{(0)}+ \hat{e}_{+}& \vec{a}_{(0)}+ 2 \hat{e}_{+} \end{array} \right|  u_1-u_2  \right)$\normalsize configuration, as these are the only valid internal state vectors, and the rapidities $u_1$ and $u_2$ are switched. Thus we obtain,
\small\begin{equation*}
Z_{N} \left( \vec{u}, \vec{v} \right) = \frac{[1+u_1-u_2]}{[1-(u_1-u_2)]} Z_{N} \left( u_2,u_1,u_3, \dots, u_N, \vec{v}\right)
\end{equation*}\normalsize
Applying this process an additional $N-2$ times so that $u_1$ is the bottom-most rapidity, we obtain eq. \ref{PS.b.11}.\\
\\
The method for the verification of eq. \ref{PS.b.12} is obviously entirely analogous. We begin by applying the line permuting face, \scriptsize$W \left( \left. \begin{array}{cc} \vec{a}_{(0)} & \vec{a}_{(0)}+ \hat{e}_{+} \\ \vec{a}_{(0)}+ \hat{e}_{+}& \vec{a}_{(0)}+ 2 \hat{e}_{+} \end{array} \right|  v_1-v_2  \right)$\normalsize, to the top of $Z_{N} \left(\vec{u},\vec{v}\right)$, and apply the Yang-Baxter equation repeatedly until we obtain,
\small\begin{equation*}
Z_{N} \left( \vec{u},\vec{v}\right) =  \frac{[1-(v_1-v_2)]}{[1+v_1-v_2]} Z_{N} \left(\vec{u}, v_2,v_1,v_3, \dots, v_N\right)
\end{equation*}\normalsize
Repeating this procedure an additional $N-2$ times so that $v_1$ is the right-most rapidity, we obtain eq. \ref{PS.b.12}.
\\
\\
{\bf Property 3.} $Z_{N}$ satisfies two recursion relations per period rectangle. The first equation can be derived by freezing rapidities $u_1= v_1-1$, so that the top left hand corner face is always of a $W_{C,+}$ configuration,
\small\begin{equation}\begin{split}
Z^{\vec{a}_{(0)}}_{N}|_{u_1=v_1-1}  = & W^{\vec{a}_{(0)}}_{C,+}\left(-1  \right) \left( \prod^N_{k=2} W^{\vec{a}_{(0)}+(j-1)\hat{e}_+}_{B,-}(u_j - v_1) \right)  \\
& \times \left( \prod^N_{k=2} W^{\vec{a}_{(0)}+(j-1)\hat{e}_+}_{B,+}(u_1 - v_j) \right) Z^{\vec{a}_{(0)}+\hat{e}_+ + \hat{e}_- }_{N-1}\left(\vec{u},\vec{v},\hat{u}_1,\hat{v}_1\right)
\label{PS.b.13}
\end{split}\end{equation}\normalsize
The second equation can be derived by freezing rapidities $u_N$ and $v_1$ so that the bottom left hand corner vertex is always a $W_{C,+}$ configuration,
\small\begin{equation}\begin{split}
Z^{\vec{a}_{(0)}}_{N}|_{u_N = v_1} =& W^{\vec{a}_{(0)} +(N-1)\hat{e}_+ }_{C,+}(0) \left( \prod^{N - 1}_{k = 1} W_{A,+}(u_k - v_1) \right)  \\
& \times   \left( \prod^{N}_{k = 2} W_{A,-}(u_N - v_j) \right) Z^{\vec{a}_{(0)} +\hat{e}_+ }_{N-1} \left(\vec{u},\vec{v},\hat{u}_N,\hat{v}_1\right)  \label{PS.b.14}
\end{split}\end{equation}\normalsize
Equivalent results exist for other period rectangles.\\
\\
\textbf{Proof.} Simply freezing the top left face to a $W_{C,+}$ configuration by fixing the the rapidities, $u_1=v_1-1$, we see immediately (fig. \ref{PS.9999}) that the faces of the first row and first column are frozen in the configuration given by eq. \ref{PS.b.13}. We additionally notice that the remaining faces in the $(N-1)\times(N-1)$ bulk are under DWBC's, with the top left hand state vector equal to $\vec{a}_{(0)}+\hat{e}_+ + \hat{e}_- $, hence verifying eq. \ref{PS.b.13}. The verification of \ref{PS.b.14} is an exactly equivalent process.
\\
\\
{\bf Property 4.} The initial condition is given by,
\small\begin{equation*}
Z^{\vec{a}_{(0)}}_{1}(u_1,v_1)=  W^{\vec{a}_{(0)}}_{C,+} (u_1-v_1)
\end{equation*}\normalsize
\textbf{Proof.} Simply setting the DWBC's for a single face produces the desired result.\\
\\
\noindent \textbf{An inductive result regarding the four properties of the DWPF.}
\begin{lemma}
The above four properties uniquely determine the DWPF of the $gl(P|M)$ PS height models.
\end{lemma}
\noindent \textbf{Proof.} We begin by assuming that there exist two expressions which satisfy the above four properties. We refer to $Z^{\vec{a}_{(0)}}_{N} \left(\vec{u},\vec{v}\right)$ as the actual partition function, and $\field{Z}_{N} \left(\vec{u},\vec{v}\right)$, which is an altogether different expression. By condition $4$ we obtain our base case,
\small\begin{equation*}
Z^{\vec{a}_{(0)}}_{1} \left(u_1,v_1\right) = \field{Z}_{1} \left(u_1,v_1\right) = W^{\vec{a}_{(0)}}_{C,+} (u_1-v_1).
\end{equation*}\normalsize
Let us now assume that the two expressions are equal up to some integer $N$ and use induction for the $N+1$ case. From condition $1$ and theorem \ref{quas}, both $Z^{\vec{a}_{(0)}}_{N+1}$ and $\field{Z}_{N+1}$ are order $N+1$ polynomials per period rectangle in rapidity $ u_1$. Additionally for both expressions, by condition 2, $N$ of those zeros are of the form,
\small\begin{equation*}
u_1 = u_j -1 + \frac{2m K_1}{\lambda}+ \frac{2 i n K_2}{\lambda} \textrm{  ,  } j=2, \dots, N+1\textrm{  ,  } \{m,n\} \in \field{Z}
\end{equation*}\normalsize
From condition $1$ we also know that the remaining zero in $u_1$ must be of the form,
\small\begin{equation*}
u_1 = \sum^{N+1}_{j=2} \left(v_j-u_j \right)  +\left( \vec{a}_{(0)} \right)_{1,L} +N-1+ \frac{2m K_1}{\lambda}+ \frac{2 i n K_2}{\lambda}  \textrm{  ,  } \{m,n\} \in \field{Z}
\end{equation*}\normalsize
making a grand total of $N+1$ shared zeroes per period rectangle. Thus by conditions $1$ and $2$ we know that:
\small\begin{equation}
Z^{\vec{a}_{(0)}}_{N+1}= \mathcal{C}(\hat{u}_1) \field{Z}_{N+1},
\label{PS.777}
\end{equation}\normalsize
where $ \mathcal{C}(\hat{u}_1)$ is a multiplicative constant independent of rapidity $u_1$. To find $ \mathcal{C}(\hat{u}_1)$ we apply the first (or equivalently second) recursion relation of condition $3$ (and the inductive assumption that $Z^{\vec{a}}_{N}= \field{Z}_{N}$), to eq. \ref{PS.777} to immediately obtain $\mathcal{C}(\hat{u}_1)  = 1$. $\square$ \\
\\
Note that an alternative form of the above proof can be equivalently carried out using the zeros of $v_1$.
\subsection{Product form for the DWPF.}
\begin{proposition}
The following product form for the DWPF satisfies the four Izergin-like properties,
\small\begin{equation}\begin{split}
Z^{\vec{a}_{(0)}}_{N}(\vec{u},\vec{v}) =& \frac{[\left( \vec{a}_{(0)} \right)_{1,L} +(N-1) - \sum^N_{k=1}(u_k-v_k)]}{[\left( \vec{a}_{(0)} \right)_{1,L} +(N-1)]}\\
&\times  \prod_{1\le i < j \le N} \frac{[1+u_i-u_j]}{[1]} \frac{[1-(v_i-v_j)]}{[1]}
\label{mas}\end{split}\end{equation}\normalsize
\end{proposition}
\noindent \textbf{Proof.} As is required, we explicitly verify each of the four properties.\\
\\
\textbf{Verification of property 1.} Verifying eq. \ref{PS.b.8} is elementary. To verify eq. \ref{PS.b.9} we consider first the coefficient that appears due to applying $u_1 \rightarrow u_1 + \frac{2K_2}{\lambda}$ to the term $[\left( \vec{a}_{(0)} \right)_{1,L} +(N-1) - \sum^N_{k=1}(u_k-v_k)]$,
\small\begin{equation*}
- \frac{1}{q} \exp \left\{ - \frac{i \pi \lambda}{K_1} \left(  \sum^N_{k=1}(u_k-v_k) -\left( \vec{a}_{(0)} \right)_{1,L} -(N-1) \right) \right\}
\end{equation*}\normalsize
and similarly to  $\prod^N_{j=2} [1+u_1-u_j]$,
\small\begin{equation*}
\left(- \frac{1}{q} \right)^{N-1} \exp \left\{ - \frac{i \pi \lambda}{K_1} \left( (N-1)u_1 - \sum^N_{k=2}u_k  +(N-1) \right) \right\}
\end{equation*}\normalsize
Multiplying these terms together we obtain eq. \ref{PS.b.9}. Verifying eqs. \ref{PS.bq} and \ref{PS.bs} is an equivalent process.\\
\\
\textbf{Verification of property 2.} From eqn. \ref{mas} it is immediate that $N-1$ of the zeros, per period rectangle, exist in the form,
\small\begin{equation*}
u_1 = u_j -1 + \frac{2m K_1}{\lambda}+ \frac{2 i n K_2}{\lambda} \textrm{  ,  } j=\{2, \dots, N\} \textrm{  ,  } \{m,n\} \in \field{Z},
\end{equation*}\normalsize
and similarly with $v_1$.\\
\\
\textbf{Verification of property 3.} Freezing the rapidities $u_1$ and $v_1$ such that $u_1 = v_1 -1$, eq. \ref{mas} becomes,
\small\begin{equation}\begin{split} 
& \frac{[\left( \vec{a}_{(0)} \right)_{1,L} +N]}{[\left( \vec{a}_{(0)} \right)_{1,L} +(N-1)]} \prod^N_{j=2} \frac{[u_j-v_1]}{[1]} \frac{[u_1-v_j]}{[1]} \\ 
\times & \frac{[\left( \vec{a}_{(0)} \right)_{1,L} +N - \sum^N_{k=2}(u_k-v_k)]}{[\left( \vec{a}_{(0)} \right)_{1,L} +N]} \prod_{2\le i < j \le N} \frac{[1+u_i-u_j]}{[1]} \frac{[1-(v_i-v_j)]}{[1]}\label{PS.www}
\end{split}\end{equation}\normalsize
where we recognize that,
\small\begin{equation*}\begin{split} 
& \frac{[\left( \vec{a}_{(0)} \right)_{1,L} +N]}{[\left( \vec{a}_{(0)} \right)_{1,L} +(N-1)]} \prod^N_{j=2} \frac{[u_j-v_1]}{[1]} \frac{[u_1-v_j]}{[1]}\\
=&W^{\vec{a}_{(0)}}_{C,+}\left(-1  \right) \left( \prod^N_{k=2} W^{\vec{a}_{(0)}+(j-1)\hat{e}_+}_{B,-}(u_j - v_1)  W^{\vec{a}_{(0)}+(j-1)\hat{e}_+}_{B,+}(u_1 - v_j) \right)
\end{split}\end{equation*}\normalsize
and,
\small\begin{equation*}\begin{split} 
 \frac{[\left( \vec{a}_{(0)}+\hat{e}_+ + \hat{e}_-  \right)_{1,L} +(N-2) - \sum^N_{k=2}(u_k-v_k)]}{[\left( \vec{a}_{(0)}+\hat{e}_+ + \hat{e}_-  \right)_{1,L} +(N-2)]} \prod^N_{i,j=2\atop{i<j}} \frac{[1+u_i-u_j]}{[1]} \frac{[1-(v_i-v_j)]}{[1]}\\
=Z^{\vec{a}_{(0)}+\hat{e}_+ + \hat{e}_- }_{N-1}\left(\vec{u},\vec{v},\hat{u}_1,\hat{v}_1\right)
\end{split}\end{equation*}\normalsize
which verifies the first recursion relation. Verifying the second is an equivalent process.\\
\\
\textbf{Verification of property 4.} Simply let $N =1$ in eq. \ref{mas} to obtain the required result.\\
\\
Since eq. \ref{mas} obeys the four Izergin-type properties, it is uniquely the DWPF of the $gl(P|M)$ PS IRF model. $\square$\\
\subsection{Generation of elliptic identities using the product form of the DWPF}
\begin{figure}[h!]
\begin{center}
\includegraphics[angle=0,scale=0.14]{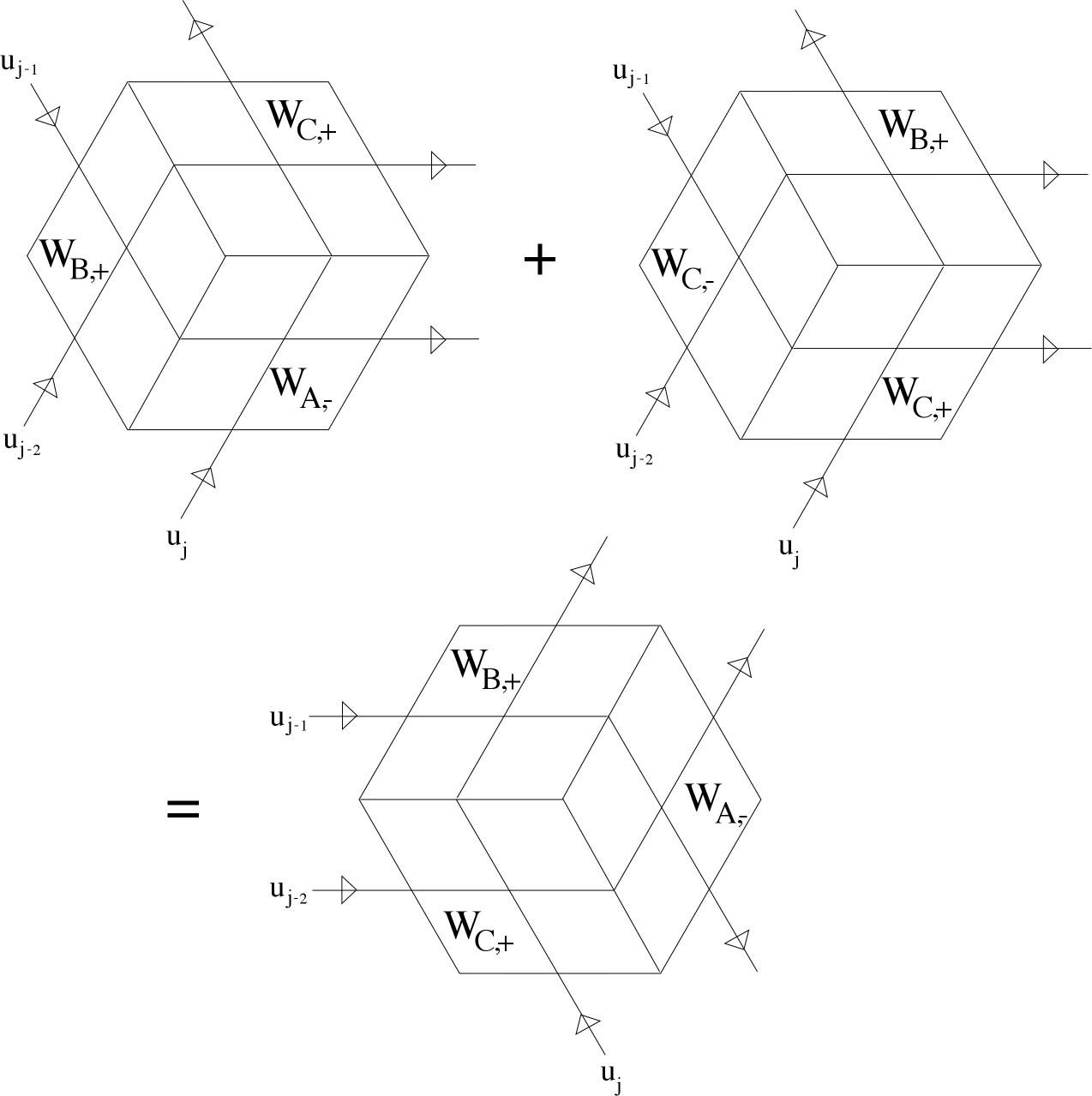}
\caption{\footnotesize{Graphical representation of the Y-B identity under consideration. Notice that the state vectors have been omitted and can easily be inferred from eq. \ref{PS.L.3}.}}
\label{PS.YBident}
\end{center}
\end{figure}

Given that we have obtained a closed form product expression for the DWPF of this elliptic model, we are now in a position to combine these results with the results of the previous chapter, (that is, obtaining a recursive formula for the partition function), to generate some non trivial elliptic identities. Since the identities are valid by construction, we are not required to verify them \textit{traditionally} using theorem 3. Obviously this is an advantage as applying such a theorem to verify large identities can be an extremely involved and painful process.\\
\\
\textbf{Remark.} The process of applying the rolling technique to this model initially holds several differences and complications, and as such we shall begin slowly. Nevertheless, all the proofs of the main results necessary are exactly the same, hence they shall not be verified explicitly.\\
\\
\textbf{Rolling revisited.} In order to generate the recursive formula for the partition function, we need to play the same game that we did in the previous chapter, that is, applying the rolling technique \cite{1} multiple times so that the top row is frozen into $(N-1)$ $W_{A,+}$ faces.\\
\\
We begin by considering the right most column of the $N \times N$ bulk with DWBC's. We recognise that the boundary conditions force the partition function to contain only one single $W_{C,+}$ configuration in the right most column.
\begin{figure}[h!]
\begin{center}
\includegraphics[angle=0,scale=0.25]{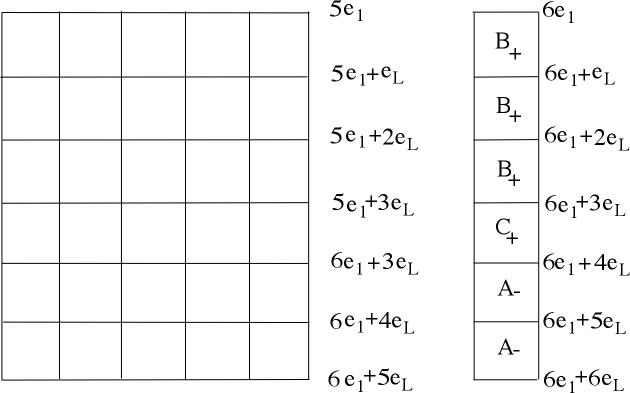}
\caption{\footnotesize{Example involving $6 \times 6$ bulk SOS. $6 \times 5$ bulk on the left hand side labeled as $L[r_4,u_4]$.}}
\label{PSrightcoli}
\end{center}
\end{figure}\\
\\
As for the remaining weights, those above the $W_{C,+}$ are of a $W_{B,+}$ configuration, and those below are of a $W_{A,-}$ configuration. Thus we obtain,
\small\begin{equation}\begin{split}
Z^{\vec{a}_0}_{N } = \sum^N_{j=1}\left(\prod^{j-1}_{k=1} W^{\vec{a}_0+(N-1)\hat{e}_1\atop{+(k-1)\hat{e}_L}}_{B,+}(u_k-v_N) \right) W^{\vec{a}_0+(N-1)\hat{e}_1\atop{+(j-1)\hat{e}_L}}_{C,+}(u_j-v_N)\\
\times \left(\prod^{N}_{k=j+1}W_{A,-}(u_k-v_N) \right) L[r_j,u_j] 
\label{L.1}\end{split}\end{equation}\normalsize
where as before, $L[r_j,u_j]$ is defined as the remaining $N \times (N-1)$ lattice where the lone outermost  $W_{C,+}$ face \textit{was} present on the $j$th row. The main goal of this section is to use the Yang-Baxter equation to force the remaining $N \times (N-1)$ bulk into a configuration similar to that of $L[r_1,u_1]$, thus forcing the remaining $(N-1)$ faces of the top row into $W_{A,+}$'s. Thus we obtain,
\small\begin{equation}
L[r_1,u_1] = \left( \prod^{N-1}_{j=1} W_{A,+}(u_1-v_j) \right)Z^{\vec{a}_{(0)}+\hat{e}_1 }_{N-1}(\hat{v}_N) \label{L!!!}
\end{equation}\normalsize
\begin{figure}[h!]
\begin{center}
\includegraphics[angle=0,scale=0.25]{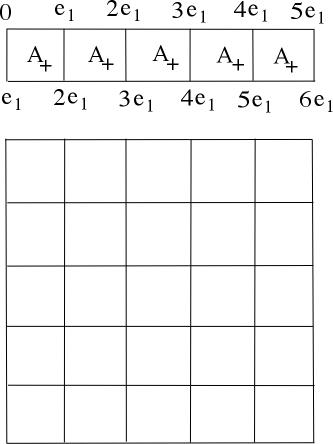}
\caption{\footnotesize{An example of $L[r_1,u_1]$. Notice the entire top row is frozen into $W_{A,+}$ faces leaving behind an $(N-1)\times (N-1)$ bulk with DWBC's, with $\vec{a}_{(0)}+\hat{e}_1$ as the top left state vector.}}
\label{PSfrozeni}
\end{center}
\end{figure}\\
\\
\textbf{Rolling Once.} We begin this procedure by considering a general $N \times (N-1)$ configuration, $L[r_j,u_j]$, whose right-most state vector jumps from $\vec{a}_{(0)} + (N-1) \hat{e}_1 + (j-1)\hat{e}_L$ to $\vec{a}_{(0)} + N \hat{e}_1 + (j-1)\hat{e}_L$ whilst crossing the horizontal rapidity flow, $u_j$, in a downward direction. In addition we also consider $L[r_{j-1},u_{j-1}]$, whose right-most state vector jumps from $\vec{a}_{(0)} + (N-1) \hat{e}_1 + (j-2)\hat{e}_L$ to $\vec{a}_{(0)} + N \hat{e}_1 + (j-2)\hat{e}_L$ whilst crossing $u_{j-1}$ in a downward direction.
\begin{figure}[h!]
\begin{center}
\includegraphics[angle=0,scale=0.25]{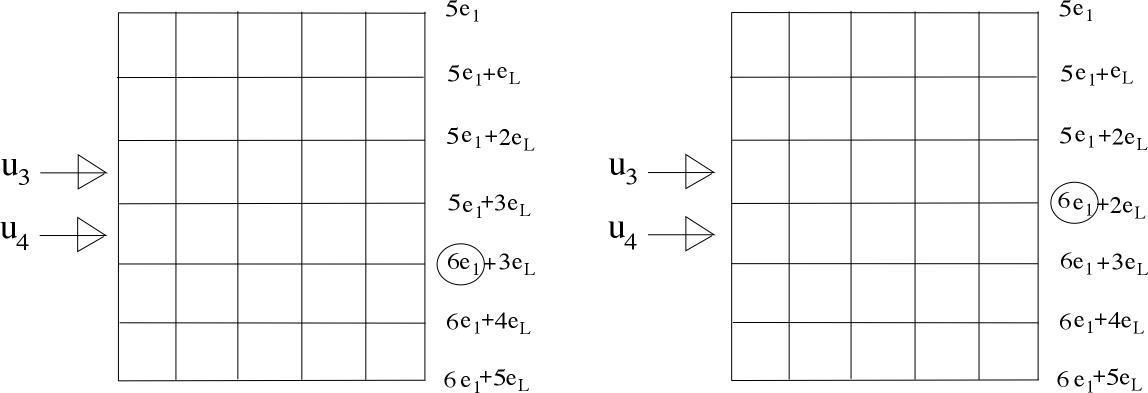}
\caption{\footnotesize{Typical example portraying the difference between $L[r_j,u_j]$ (left) and $L[r_{j-1},u_{j-1}]$ (right) for $j=4, N=6$.}}
\label{PSLdiffi}
\end{center}
\end{figure}\\
We now multiply the configuration $L[r_j,u_j]$ by the face $W^{\vec{a}_{(0)} + (N-1)\hat{e}_1\atop{+ (j-2)\hat{e}_L}}_{B,+}(u_{j-1}-u_{j})$ and the configuration $L[r_{j-1},u_{j-1}]$ by the face $W^{\vec{a}_{(0)} + (N-1)\hat{e}_1\atop{+ (j-2)\hat{e}_L}}_{C,+}(u_{j-1}-u_{j})$. As with the BSOS model, we notice that this sum of the two configurations means that the internal height is being summed over all allowable values.
\newpage
\begin{figure}[h!]
\begin{center}
\includegraphics[angle=0,scale=0.23]{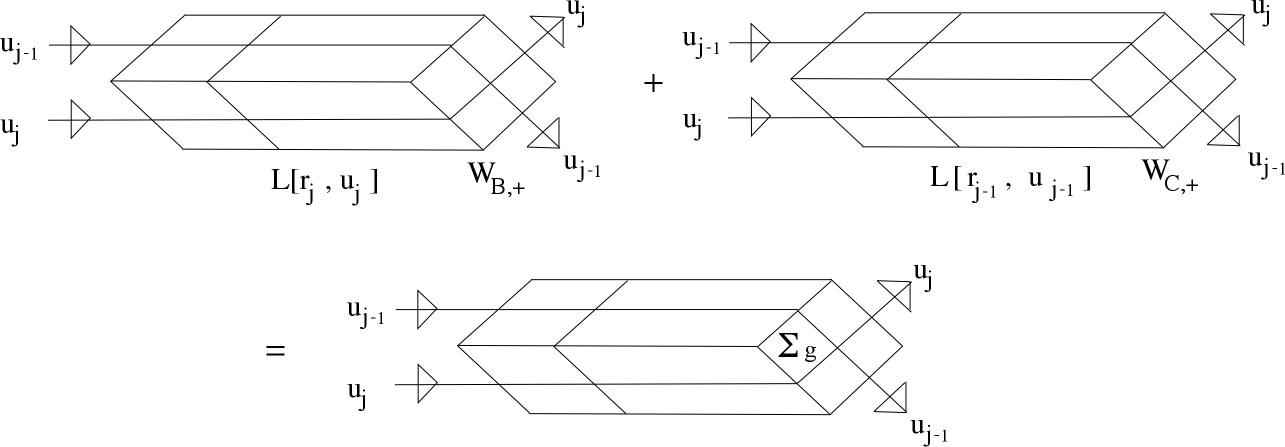}
\caption{\footnotesize{Multiplying $L[r_j,u_j]$ by $W^{\vec{a}_{(0)} + (N-1)\hat{e}_1\atop{+ (j-2)\hat{e}_L}}_{B,+}(u_{j-1}-u_{j})$ (on the left) and $L[r_{j-1},u_{j-1}]$ by $W^{\vec{a}_{(0)} + (N-1)\hat{e}_1\atop{+ (j-2)\hat{e}_L}}_{C,+}(u_{j-1}-u_{j})$ (on the right).}}
\label{PS.angryfish}
\end{center}
\end{figure}
\noindent We are now in a position to apply the Yang-Baxter equation to the above configuration and shift the intertwining of the $u_{j-1}$ and $u_j$ rapidities to the left hand side of the $N \times (N-1)$ bulk.
\begin{figure}[h!]
\begin{center}
\includegraphics[angle=0,scale=0.25]{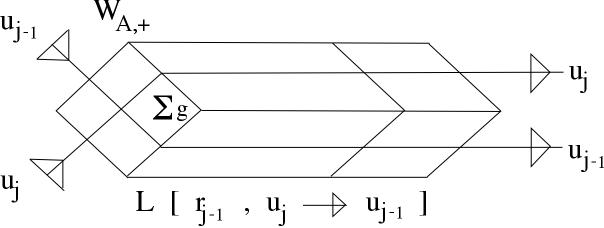}
\caption{\footnotesize{Applying the Yang-Baxter equation to shift the intertwining of horizontal rapidities to the left hand side of the bulk.}}
\label{PS.internalsum}
\end{center}
\end{figure}\\
\\
Where we notice that the sum of state vectors, $\sum g$, is actually fixed at $\vec{a}_{(0)}+(j-2)\hat{e}_1$. Thus we obtain the algebraic expression,
\small\begin{equation}
 L[r_j,u_j] = f^{(j-2)}_{+} \left( u_{j-1} \atop{u_j} \right)  L[r_{j-1},u_{j} \rightarrow u_{j-1}] - g^{(j-2)} \left( u_{j} \atop{u_{j-1}} \right)  L[r_{j-1},u_{j-1}]
\label{PS.L.1}
\end{equation}\normalsize
where we have used the following identities,
\small\begin{equation*}\begin{array}{ll}
W^{\vec{a}}_{C,+}(u_i-u_j) = W^{\vec{a}}_{C,-}(u_j-u_i)& W^{\vec{a}}_{B,+}(u_i-u_j) = - W^{\vec{a}}_{B,+}(u_j-u_i)\\
 f^{(n)}_{\pm} \left( u_{j-1} \atop{u_j} \right) = \frac{W_{A,\pm}(u_{j-1}-u_j)}{W^{\vec{a}_{(0)} + (N-1)\hat{e}_1\atop{ +n \hat{e}_L}  }_{B,+}(u_{j-1}-u_{j})} &  g^{(n)} \left( u_{j} \atop{u_{j-1}} \right) =\frac{W^{\vec{a}_{(0)} + (N-1)\hat{e}_1\atop{ + n \vec{e}_L} }_{C,-}(u_{j}-u_{j-1})}{W^{\vec{a}_{(0)} + (N-1)\hat{e}_1\atop{ +n\hat{e}_L} }_{B,+}(u_{j}-u_{j-1})}
\end{array}\end{equation*}\normalsize
\textbf{A note on exchanging rows in the bulk.} In the previous chapter it was not necessary to keep track of any exchanges in the horizontal rapidities due to the invariance in the line permuting faces, i.e.
\small\begin{equation*}
W_{A,+}(u_i-u_j) = W_{A,-}(u_i-u_j)= W_{A}(u_i-u_j).
\end{equation*}\normalsize
We obviously do not have such invariance in this chapter. Thus it shall become necessary very shortly to keep track of such exchanges, such as the term $L[r_{j-1},u_{j} \rightarrow u_{j-1}]$ in eq. \ref{PS.L.1}.\\
\\
\textbf{Rolling twice.} We now apply the rolling procedure to the configurations $L[r_{j-1},u_{j} \rightarrow u_{j-1}]$ and $L[r_{j-1},u_{j-1}]$. Using the above procedure we obtain the results,
\small\begin{equation*}\begin{split}
L[r_{j-1},u_{j}\rightarrow u_{j-1}] =& f^{ (j-3)}_+ \left( u_{j-2} \atop{u_j} \right) L[r_{j-2},u_{j}\rightarrow u_{j-1} \rightarrow u_{j-2}] \\
&+ g^{ (j-3)} \left( u_{j} \atop{u_{j-2}} \right)   L[r_{j-2},u_{j-2}, u_{j}\rightarrow u_{j-1}] \\
L[r_{j-1},u_{j-1}] =& f^{ (j-3)}_+ \left( u_{j-2} \atop{u_{j-1}} \right)  L[r_{j-2},u_{j-1}\rightarrow u_{j-2}] \\
&+ g^{ (j-3)} \left( u_{j-1} \atop{u_{j-2}} \right)   L[r_{j-2},u_{j-2}]
\end{split}\end{equation*}\normalsize
\textbf{An additional note on exchanging rows in the bulk.} In the previous chapter, due to the invariance of the rapidity permuting faces we had the following result,
\small\begin{equation*}
L[r_{j-2},u_{j-2}, u_{j}\rightarrow u_{j-1}] = L[r_{j-2},u_{j-2}].
\end{equation*}\normalsize
With the current definition of the non invariant rapidity permuting faces, the above result now looks comparable to property 2 (eq. \ref{PS.b.11}) of the DWPF:
\small\begin{equation}
L[r_{j-2},u_{j-2}, u_{j}\rightarrow u_{j-1}] =\frac{W_{A,-}(u_{j-1}-u_j)}{W_{A,+}(u_{j-1}-u_j)} L[r_{j-2},u_{j-2}].
\label{PS.L.4}
\end{equation}\normalsize
Using the above result, and collecting the coefficients of the $N \times (N-1)$ bulk terms we obtain,
\small\begin{equation}\begin{split}
L[r_{j},u_{j}] =f^{(j-2)}_+ \left( u_{j-1}\atop{u_j} \right)f^{(j-3)}_+ \left( u_{j-2}\atop{u_j} \right) L[r_{j-2},u_{j}\rightarrow u_{j-1} \rightarrow u_{j-2}]  \\
+  g^{(j-2)} \left( u_{j}\atop{u_{j-1}} \right)f^{(j-3)}_+ \left( u_{j-2}\atop{u_{j-1}} \right) L[r_{j-2},u_{j-1} \rightarrow u_{j-2}] \\
+\left\{  f^{ (j-2)}_- \left( u_{j-1}\atop{u_{j}} \right)g^{ (j-3)} \left( u_{j}\atop{u_{j-2}} \right)+ g^{ (j-2)} \left( u_{j}\atop{u_{j-1}} \right)g^{ (j-3)} \left( u_{j-1}\atop{u_{j-2}} \right)  \right\} L[r_{j-2},u_{j-2}] \label{PS.L.2}
\end{split}\end{equation}\normalsize
As before, we now analyze the coefficient of $L[r_{j-2},u_{j-2}]$ carefully with the intention of reducing it using some Yang-Baxter identity. \\
\\
We begin by multiplying the coefficient by a factor of,
\small\begin{equation*}
W^{(j-2)}_{B,+}(u_{j-1}-u_j) W^{(j-3)}_{B,+}(u_{j}-u_{j-2}) W^{(j-3)}_{B,+}(u_{j-1}-u_{j-2})
\end{equation*}\normalsize
to obtain,
\small\begin{equation}\begin{split}
W_{A,-}(u_{j-1}-u_j)W^{(j-3)}_{C,+}(u_{j-2}-u_{j}) W^{(j-3)}_{B,+}(u_{j-1}-u_{j-2})\\
+ W^{(j-2)}_{C,+}(u_{j-1}-u_j) W^{(j-3)}_{B,+}(u_{j-2}-u_{j}) W^{(j-3)}_{C,-}(u_{j-1}-u_{j-2})\label{PS.L.3}
\end{split}\end{equation}\normalsize
In the form of eq. \ref{PS.L.3}, the coefficient can be recognized as the left hand side of a Yang-Baxter identity, whose diagram is given in fig. \ref{PS.YBident}.
\\
\\
Hence the reduced form for the coefficient of $L[r_{j-2},u_{j-2}]$ is given by,
\small\begin{equation}\begin{split}
&\frac{W^{\vec{a}_{(0)}+(N-1) \hat{e}_1\atop{+(j-3)\hat{e}_L} }_{B,+}(u_{j-1}-u_j)}{W^{\vec{a}_{(0)}+(N-1) \hat{e}_1\atop{+(j-2)\hat{e}_L}}_{B,+}(u_{j-1}-u_j)} \frac{W_{A,-}(u_{j-1}-u_{j-2})}{W^{\vec{a}_{(0)}+(N-1) \hat{e}_1\atop{+(j-3)\hat{e}_L}}_{B,+}(u_{j-1}-u_{j-2})} \frac{ W^{\vec{a}_{(0)}+N \hat{e}_1\atop{+(j-3)\hat{e}_L}}_{C,+}(u_{j-2}-u_{j})}{ W^{\vec{a}_{(0)}+(N-1) \hat{e}_1\atop{+(j-3)\hat{e}_L}}_{B,+}(u_{j}-u_{j-2})}\\
 =&f^{(j-3)}_- \left( u_{j-1}\atop{u_{j-2}} \right)g^{(j-2)} \left( u_{j}\atop{u_{j-2}} \right) \label{PS.L.666}
\end{split}\end{equation}\normalsize
where we have used the results that the height and rapidity parts of the $W_{B,\pm}$ weights decouple (eq. \ref{PSdecouple}), and,
\small\begin{equation}\begin{split}
\left( \vec{a}_{(0)} + j_1 \hat{e}_1+ j_2 \hat{e}_L \right)_{1,L} = \left(\vec{a}_{(0)} + k_1 \hat{e}_1+ k_2 \hat{e}_L \right)_{I,L}  \textrm{  for  } j_1+j_2=k_1+k_2 \\
\Rightarrow W^{\vec{a}_{(0)}+N \hat{e}_1\atop{+(j-3)\hat{e}_L}}_{C,+}(u_{j-2}-u_{j}) = W^{\vec{a}_{(0)}+(N-1) \hat{e}_1\atop{+(j-2)\hat{e}_L}}_{C,+}(u_{j-2}-u_{j})  \label{PSpower}
\end{split}\end{equation}\normalsize
Thus after applying the rolling procedure twice we obtain the following for $L[r_{j},u_j]$,
\small\begin{equation*}\begin{split}
L[r_{j},u_j] = g^{(j-2)} \left( u_{j}\atop{u_{j-2}} \right) f^{(j-3)}_- \left( u_{j-1}\atop{u_{j-2}} \right) L[r_{j-2},\field{U}_{j-2}] +  g^{(j-2)} \left( u_{j}\atop{u_{j-1}} \right)\\
 \times f^{(j-3)}_+ \left( u_{j-2}\atop{u_{j-1}} \right) L[r_{j-2},\field{U}_{j-1}] +f^{(j-2)}_+ \left( u_{j-1}\atop{u_j} \right)f^{(j-3)}_+ \left( u_{j-2}\atop{u_j} \right) L[r_{j-2},\field{U}_{j}]
\end{split}\end{equation*}\normalsize
where,
\small\begin{equation*}
L[r_{j-k_1},\field{U}_{j-k_2}] = L[r_{j-k_1},u_{j-k_2} \rightarrow u_{j-(k_2+1)} \rightarrow \dots \rightarrow u_{j-k_1}] \textrm{  ,   }  k_2 \le k_1
\end{equation*}\normalsize
Expressing $f^{(k)}_{\pm} \left( u_{1}\atop{u_{2}} \right)$ as the convenient product,
\small\begin{equation*}\begin{split}
f^{(k)}_{\pm} \left( u_{1}\atop{u_{2}} \right) &=\frac{[1\mp(u_1-u_2)]}{[u_1-u_2]}\frac{[\left( \vec{a}_{(0)}+(N-1)\hat{e}_1+k\hat{e}_L \right)_{1,L}]}{[\left( \vec{a}_{(0)}+(N-1)\hat{e}_1+k\hat{e}_L  \right)_{1,L}+ 1 ]}\\
&= f^{\pm}_{(1)}\left( u_{1}\atop{u_{2}} \right) f_{(2)} \left( k \right)
\end{split}\end{equation*}\normalsize
we obtain the following suggestive form,
\small\begin{equation}\begin{split}
L[r_j,u_j] =\sum^{j-1}_{l=j-2}g^{(j-2)} \left( u_j \atop{u_l} \right) \frac{f_{(2)}(l-1) }{f_{(2)}(j-1)}  \left( \prod^{l-1}_{k=j-2} f^{(k-1)}_{+} \left( u_k \atop{u_l} \right) \right) \\
\times  \left( \prod^{j-1}_{k=l+1} f^{(k-1)}_{-} \left( u_k \atop{u_l} \right) \right) L[r_{j-2},\field{U}_l] + \left( \prod^{j-1}_{k=j-2} f^{(k-1)}_{+} \left( u_k \atop{u_j} \right) \right) L[r_{j-2},\field{U}_j]\\
= \sum^j_{l=j-2} \frac{g^{(j-2)}\left( u_j \atop{u_l} \right)}{f^{(j-2)}_{-}\left( u_j \atop{u_l} \right)} \frac{f_{(2)}(l-1) }{f_{(2)}(j-1)}  \left( \prod^{l-1}_{k=j-2} f^{(k-1)}_{+} \left( u_k \atop{u_l} \right) \right) \\
\times \left( \prod^{j}_{k=l+1} f^{(k-1)}_{-} \left( u_k \atop{u_l} \right) \right) L[r_{j-2},\field{U}_l] 
\label{suggestive}
\end{split}\end{equation}\normalsize
\textbf{Rolling Many Times.} The process of iterating this procedure many times is now straightforward. The obvious choice for the expression of $L[r_j,u_j]$ after a general number of rolling operations is,
\small\begin{equation}\begin{split}
L[r_j,u_j] = \sum^j_{l=j-n} \frac{g^{(j-2)}\left( u_j \atop{u_l} \right)}{f^{(j-2)}_{-}\left( u_j \atop{u_l} \right)} \frac{f_{(2)}(l-1) }{f_{(2)}(j-1)}  \left( \prod^{l-1}_{k=j-n} f^{(k-1)}_{+} \left( u_k \atop{u_l} \right) \right)\\
\times  \left( \prod^{j}_{k=l+1} f^{(k-1)}_{-} \left( u_k \atop{u_l} \right) \right) L[r_{j-n},\field{U}_l]
\label{generalPS}
\end{split}\end{equation}\normalsize
for $2 \le n \le j-1$. We now prove eq. \ref{generalPS} using induction. Obviously we know that it holds for $n=2$, assuming now that it holds for general $n=p-1$, we consider using the rolling procedure once more to obtain,
\small\begin{equation*}\begin{split} 
 \sum^j_{l=j-(p-1)} \frac{g^{(j-2)}\left( u_j \atop{u_l} \right)}{f^{(j-2)}_{-}\left( u_j \atop{u_l} \right)} \frac{f_{(2)}(l-1) }{f_{(2)}(j-1)}  \left( \prod^{l-1}_{k=j-p} f^{(k-1)}_{+} \left( u_k \atop{u_l} \right) \right)  \left( \prod^{j}_{k=l+1} f^{(k-1)}_{-} \left( u_k \atop{u_l} \right) \right)\\
\times L[r_{j-p},\field{U}_l] + \left\{ \left( \prod^{p-1}_{k=1} f^{(j-1-k)}_{-}\left( u_{j-k}\atop{u_j} \right) \right) g^{(j-1-p)}\left( u_{j}\atop{u_{j-p}} \right) + \left( \prod^{p-1}_{k=2} f_{(2)}(j-1-k) \right)\right.\\
\times \sum^{p-1}_{l=1} g^{(j-2)} \left( u_{j}\atop{u_{j-l}} \right)   \left. \left( \prod^{p-1}_{k=1\atop{\ne l}} f^{-}_{(1)}\left( u_{j-k}\atop{u_{j-l}} \right) \right) g^{(j-1-p)}\left( u_{j-l}\atop{u_{j-p}} \right) \right\} L[r_{j-p},\field{U}_{j-p}]
\end{split}\end{equation*}\normalsize 
which leads us to the following result,
\begin{proposition}
\label{PSlong}
\small \begin{equation*} 
\begin{split}
& \left\{ \left( \prod^{p-1}_{k=1} f^{(j-1-k)}_{-}\left( u_{j-k}\atop{u_j} \right) \right) g^{(j-1-p)}\left( u_{j}\atop{u_{j-p}} \right) + \left( \prod^{p-1}_{k=2} f_{(2)}(j-1-k) \right)\right. \\
&\times  \left. \sum^{p-1}_{l=1} g^{(j-2)} \left( u_{j}\atop{u_{j-l}} \right)  \left( \prod^{p-1}_{k=1\atop{\ne l}} f^{-}_{(1)}\left( u_{j-k}\atop{u_{j-l}} \right) \right) g^{(j-1-p)}\left( u_{j-l}\atop{u_{j-p}} \right) \right\} \\
=& g^{(j-2)}\left( u_j\atop{u_{j-p}} \right) \left\{ \prod^{p-1}_{k=1}f_{(2)}(j-(k+1)) f^-_{(1)} \left( u_{j-k}\atop{u_{j-p}} \right) \right\} 
\end{split}
\end{equation*}\normalsize
\end{proposition}
\noindent \textbf{A note on verifying the above result.} The method of verifying proposition \ref{PSlong} is exactly the same as verifying proposition \ref{bigprop} in the previous chapter. The right hand side of the above expression can be expressed as an equivalent diagram to fig. \ref{5.p}, where the rapidity labels stay the same and the state vectors can be easily inferred. We then apply the same Yang-Baxter procedure as explained in proposition \ref{bigprop} to obtain the left hand sum of $p$ terms. We then divide by the following factor,
\small\begin{equation*}
W^{(j-2)}_{B,+}\left(u_{j-p}-u_{j} \right)  \left\{ \prod^{p-1}_{k = 1} W^{(j-k-2)}_{B,+}\left(u_{j-k}-u_{j} \right)W^{(j-k-2)}_{B,+}\left(u_{j-p}-u_{j-p+k} \right) \right\}
\end{equation*}\normalsize
where $W^{(j-2)}_{B,+}\left(u \right) =W^{\vec{a}_{(0)}+(N-1) \hat{e}_1\atop{+(j-2)\hat{e}_L} }_{B,+}\left(u \right)$, to obtain the required expression. \\
\\
Using proposition \ref{PSlong} we instantly verify eq. \ref{generalPS} for general $n$. Taking $n = j-1$ we obtain,
\small \begin{equation}
\begin{split}
L[r_j,u_j] =&\sum^j_{l=1} \frac{g^{(j-2)}\left( u_j \atop{u_l} \right)}{f^{(j-2)}_{-}\left( u_j \atop{u_l} \right)} \frac{f_{(2)}(l-1) }{f_{(2)}(j-1)}  \left( \prod^{l-1}_{k=1} f^{(k-1)}_{+} \left( u_k \atop{u_l} \right) \right)  \left( \prod^{j}_{k=l+1} f^{(k-1)}_{-} \left( u_k \atop{u_l} \right) \right)\\
&\times \left( \prod^{N-1}_{k=1} W_{A,+}(u_j-v_k) \right)Z^{\vec{a}_{(0)}+\hat{e}_1 }_{N-1}(\vec{u},\vec{v},\hat{u}_j,\hat{v}_N)
\end{split}
\label{PS=1}
\end{equation}\normalsize
where we have applied the following obvious generalization to eq. \ref{L!!!},
\small\begin{equation*}
L[r_1,\field{U}_j] = \left( \prod^{N-1}_{k=1} W_{A,+}(u_j-v_k) \right)Z^{\vec{a}_{(0)}+\hat{e}_1 }_{N-1}(\vec{u},\vec{v},\hat{u}_j,\hat{v}_N)
\end{equation*}\normalsize
Thus, eq. \ref{L.1} becomes,
\small\begin{equation}
\begin{split}
Z^{\vec{a}_0}_{N }=\sum^N_{j=1} \sum^j_{l=1}\left(\prod^{j-1}_{k=1} W^{\vec{a}_0+(N-1)\hat{e}_1\atop{+(k-1)\hat{e}_L}}_{B,+}(u_k-v_N) \right) W^{\vec{a}_0+(N-1)\hat{e}_1\atop{+(j-1)\hat{e}_L}}_{C,+}(u_j-v_N) \\
\times\left(\prod^{N}_{k=j+1}W_{A,-}(u_k-v_N) \right) \frac{g^{(j-2)}\left( u_j \atop{u_l} \right)}{f^{(j-2)}_{-}\left( u_j \atop{u_l} \right)} \frac{f_{(2)}(l-1) }{f_{(2)}(j-1)}  \left( \prod^{l-1}_{k=1} f^{(k-1)}_{+} \left( u_k \atop{u_l} \right) \right) \\
\times \left( \prod^{j}_{k=l+1} f^{(k-1)}_{-} \left( u_k \atop{u_l} \right) \right) \left( \prod^{N-1}_{k=1} W_{A,+}(u_j-v_k) \right)Z^{\vec{a}_{(0)}+\hat{e}_1 }_{N-1}(\hat{u}_j,\hat{v}_N)
\end{split}
\label{PSL.1}\end{equation}\normalsize
The above expression contains two summations. We shall now proceed to use the same algorithm detailed in section \ref{STRL} to absorb one the the summations.\\
\\
\textbf{Further reducing the recurrence relation.} 
\begin{proposition}
\small\begin{equation} 
\begin{split}
Z^{\vec{a}_0}_{N } = \sum^{N}_{j=1}W^{\vec{a}_0+(N-1)\hat{e}_1\atop{+(N-1)\hat{e}_L}}_{C,+}(u_{j}-v_N) \left( \prod^{l-1}_{k=1} f^{(k-1)}_{+} \left( u_k \atop{u_l} \right)\right)  \left( \prod^{j}_{k=l+1} f^{(k-1)}_{-} \left( u_k \atop{u_l} \right) \right) \\
\times \left(\prod^{N}_{k=1\atop{\ne j}} W_{B}(u_k-v_N) \right)  \left( \prod^{N-1}_{k=1} W_{A,+}(u_j-v_k) \right)Z^{\vec{a}_{(0)}+\hat{e}_1 }_{N-1}(\hat{u}_j,\hat{v}_N)
\end{split}
\label{fifinal}
\end{equation}\normalsize
\end{proposition}
\noindent \textbf{A note on verifying the above result.} The method of verifying proposition \ref{fifinal} is exactly the same as verifying proposition \ref{N.25} in section \ref{STRL}. Minor changes appear in the state vectors of the necessary Yang-Baxter equation(s), but these can easily be derived as we already know which of the six necessary configurations the corresponding faces must be.\\
\\
\textbf{Generating the elliptic identity.} We now substitute the known product form for $Z^{\vec{a}_0}_{N }$ and $Z^{\vec{a}_{(0)}+\hat{e}_1 }_{N-1}(\hat{u}_j,\hat{v}_N)$ to obtain the following elliptic identity for general $N$,
\small\begin{equation}
\begin{split}
[\left(\vec{a}_0\right)_{1,L}+(N-1)-\sum^N_{j=1}(u_j-v_j)] \left(\prod^{N-1}_{j=1}[1-(v_j-v_N)] \right) \left(\prod_{1\le i<j \le N}[u_i-u_j] \right) \\
\times [\left(\vec{a}_0\right)_{1,L}+2(N-1)]=\sum^{N}_{k=1}(-1)^{N-k}[\left(\vec{a}_0\right)_{1,L}+2(N-1)-(u_{k}-v_N)] \left(\prod^{N}_{j=1\atop{\ne k}} [u_j-v_N] \right)\\
\times\left(\prod^{N-1}_{j=1} [1+u_{k}-v_j] \right) [\left(\vec{a}_0\right)_{1,L}+(N-1)-\sum^N_{j=1\atop{\ne k}}u_j+\sum^{N-1}_{j=1}v_j]\left( \prod_{1\le i<j \le N \atop{i,j \ne k}}[u_i-u_j]\right)
\end{split}
\label{ellipident}
\end{equation}\normalsize
Proving such an identity using theorem \ref{Louv} would obviously be a non trivial task. Nevertheless, a starting point would be to consider eq. \ref{simple}, which we know to be one of the simplest non trivial elliptic identities. Performing the following change of variables,
\small\begin{equation*}
\begin{split}
u \rightarrow &\frac{1}{2} \left(\vec{a}_0\right)_{1,L} + 1 + \frac{1}{2} u_1 -  \frac{1}{2} u_2 \\
 v \rightarrow &\frac{1}{2} \left(\vec{a}_0\right)_{1,L} + 1 - \frac{1}{2} u_1 +  \frac{1}{2} u_2\\
x \rightarrow &-\frac{1}{2} \left(\vec{a}_0\right)_{1,L} - 1 + \frac{1}{2} u_1 +  \frac{1}{2} u_2 -v_2 \\
 y \rightarrow &-\frac{1}{2} \left(\vec{a}_0\right)_{1,L}  + \frac{1}{2} u_1 +  \frac{1}{2} u_2-v_1
\end{split}
\end{equation*}\normalsize
we obtain the $N=2$ case of eq. \ref{ellipident}.
%%%%%%%%%%%%%%%%%%%%%%%%%%%%%%%%%%%%%%%%%%%%%%%%%%%%%%%%%%%%%%%%%%%%%%%%%%%%%%%%%%%%%%%%%%%%%%%%%%%%%%%%%%%%%%%%%%%%
\newpage
%%%%%%%%%%%%%%%%%%%%%%%%%%%%%%%%%%%%%%%%%%%%%%%%%%%%%%%%%%%%%%%%%%%%%%%%%%%%%%%%%%%%%%%%%%%%%%%%%%%%%%%%%%%%%%%%%%%%

%%%%%%%%%%%%%%%%%%%%%%%%%%%%%%%%%%%%%%%%%%%%%%%%%%%%%%%%%%%%%%%%%%%%%%%%%%%%%%%%%%%%%%%%%%%%%%%%%%%%%%%%%%%%%%%%%%%%
\bibliographystyle{plain}

\end{document}